%% file: global1.tex
\documentstyle[pifont,amsmath,psfig,12pt]{toto}

\input{tcilatex}
\begin{document}

\input{titre.tex}
\newpage
\setcounter{page}{1}
\input{remercim.tex}
\newpage
\tableofcontents
\newpage

\vfill\break \setcounter{page}{1}
\input{intro.tex}

\input{chapter2.tex}

\input{chapter3.tex}

\input{chapter4.tex}

\input{chapter5.tex}
\input{conclu.tex}

\headoff

\end{document}

%% file: titre.tex
\thispagestyle{empty}
\begin{center}
{\bf \huge LABORATOIRE KASTLER BROSSEL}

\vspace{2cm}

{\bf\large Th\`{e}se de Doctorat de l'Universit\'{e} Paris VI }

\bigskip

{\bf\large Sp\'{e}cialit\'{e} : Physique Quantique }

\vspace{2cm}

{\large Pr\'{e}sent\'{e}e par}

\smallskip

\LARGE Yassine Hadjar

\vspace{1cm}

{\large Pour obtenir le titre de Docteur de l'Universit\'{e} Paris
VI}

\vspace{2cm}

{\large Sujet de la th\`{e}se :}

\bigskip

\LARGE Etude du couplage optom\'{e}canique dans une cavit\'{e}\\
de grande finesse.\\
Observation du mouvement Brownien d'un miroir

\vspace{2cm}

\large Soutenue le 25 novembre 1998 devant le jury compos\'{e} de
:

\bigskip

\renewcommand \arraystretch {1.3}

\begin{tabular}{ll@{\extracolsep{2cm}}l}

 & M. Philippe TOURRENC &Pr\'{e}sident\\
 & M. Antoine HEIDMANN & Directeur de th\`{e}se\\
 & M. Ariel LEVENSON & Rapporteur\\
 & M. Michel PINARD & \\
 & M. Jacques ROBERT &\\
 & M. Jean-Yves VINET & Rapporteur
\end{tabular}

\end{center}

%% file: remercim.tex
\bigskip
{\it Le pr\'{e}sent travail a \'{e}t\'{e} effectu\'{e} au laboratoire
Kastler Brossel durant la p\'{e}riode 1994-1998. Je remercie Mich\`{e}le
Leduc de m'avoir accueilli dans son laboratoire et le directeur du
D\'{e}partement de Physique Serge Haroche que j'ai souvent sollicit\'{e}
dans les moments difficiles. Je voudrais aussi remercier Elisabeth Giacobino
et Claude Fabre pour m'avoir accueilli au sein du groupe d'Optique Quantique.%
}

{\it Cette th\`{e}se a \'{e}t\'{e} financ\'{e}e par le M.E.N.E.S.R. et j'ai
aussi b\'{e}n\'{e}fici\'{e} d'une bourse de l'Association Louis de Broglie
d'Aide \`{a} la Recherche. Je remercie les responsables de l'Association et
en particulier Jacques Robert pour l'int\'{e}r\^{e}t qu'il a port\'{e} \`{a}
ce travail et pour la sympathie qu'il m'a manifest\'{e}e.\bigskip }

{\it J'ai eu l'immense plaisir de travailler dans la ''jeune'' \'{e}quipe
''optom\'{e}canique'' dirig\'{e}e par Antoine Heidmann et Michel Pinard.
J'ai eu en particulier la chance de contribuer \`{a} la naissance de
l'exp\'{e}rience. J'ai pu remarquer et appr\'{e}cier l'efficacit\'{e}
redoutable avec laquelle Antoine et Michel abordaient les probl\`{e}mes
aussi bien th\'{e}oriques qu'exp\'{e}rimentaux. Par le soin qu'ils portaient
aux d\'{e}tails, il m'ont encourag\'{e} \`{a} faire toujours mieux, plus
efficace, plus pr\'{e}cis et ''plus compact'' dans mon travail. Je ne les
remercierai jamais assez pour tout ce qu'ils m'ont apport\'{e} durant ces
quatre derni\`{e}res ann\'{e}es et je pense sinc\`{e}rement que les mots ne
pourront pas exprimer ce que je leur dois et devrai toujours. Je tiens aussi
\`{a} remercier Antoine pour les nombreuses fois o\`{u} j'ai d\^{u} le
solliciter pour des probl\`{e}mes administratifs et personnels. Il a en
effet d\^{u} faire face non seulement aux difficult\'{e}s li\'{e}es \`{a}
l'exp\'{e}rience mais celles aussi li\'{e}es \`{a} mon statut
''d'\'{e}tranger''. Pour tout cela, Randa et moi tenons \`{a} le remercier
de tout coeur.}

{\it Pierre-Fran\c{c}ois Cohadon a commenc\'{e} sa th\`{e}se dans
l'\'{e}quipe en 1996. Sa gentillesse, sa disponibilit\'{e} et son
enthousiasme (sans oublier ses commentaires et ses notes d'humour \`{a} la
PF), ont beaucoup apport\'{e} \`{a} ce travail. Sa contribution m'a permis
de r\'{e}diger la th\`{e}se dans des conditions id\'{e}ales et je l'en
remercie vivement.\bigskip }

{\it Je voudrais remercier l'\'{e}quipe du professeur Claude Boccara de
l'E.S.P.C.I. qui a caract\'{e}ris\'{e} l'\'{e}tat de surface de nos
substrats, l'\'{e}quipe de Jean-Marie Mackowski du Service des Mat\'{e}riaux
Avanc\'{e}s de l'I.P.N. \`{a} Lyon qui a r\'{e}alis\'{e} le traitement de
haute r\'{e}flectivit\'{e} de nos miroirs, et Fran\c{c}ois Bondu qui nous a
fourni le programme CYPRES. Un grand merci \`{a} Fran\c{c}ois Biraben et Fran%
\c{c}ois Nez qui nous ont beaucoup aid\'{e}s pour la construction de notre
laser Titane Saphir. Merci \`{a} Bernard Rodriguez pour toutes les
pi\`{e}ces m\'{e}caniques (je n'ose pas les compter) qu'il a
r\'{e}alis\'{e}es avec une pr\'{e}cision digne d'un ''horloger suisse'' et
en particulier pour les supports de la cavit\'{e}.}

{\it Je remercie Jean Michel Courty qui m'a fait b\'{e}n\'{e}ficier de sa
grande culture g\'{e}n\'{e}rale \`{a} travers les nombreuses discussions que
nous avons eues. Je le remercie aussi pour ses critiques souvent tr\`{e}s
constructives qui m'ont \'{e}t\'{e} utiles pour la r\'{e}daction ainsi que
pour la pr\'{e}sentation orale. Je tiens \`{a} remercier Serge Reynaud qui,
malgr\'{e} ses nombreuses responsabilit\'{e}s, a toujours \'{e}t\'{e} tr\`{e}%
s disponible; sa pr\'{e}sence amicale m'a beaucoup aid\'{e} durant la
th\`{e}se (merci pour le email du 25-11 \`{a} 22h30).\bigskip }

{\it Je remercie Philippe Tourrenc, Ariel Levenson, Jean Yves Vinet et
Jacques Robert d'avoir accept\'{e} de faire partie du jury malgr\'{e} leurs
nombreuses occupations et de m'avoir montr\'{e} de ce fait l'int\'{e}r\^{e}t
qu'ils portent \`{a} ce travail.\bigskip }

{\it Durant ce travail de th\`{e}se j'ai b\'{e}n\'{e}fici\'{e} de la superbe
ambiance qui r\`{e}gne dans le groupe d'Optique Quantique et je remercie
particuli\`{e}rement Francesca Grassia, Astrid Lambrecht, Pascal El Khoury,
Antonio Zelaquett-Khoury, Paulo Souto Ribeiro, Laurent Vernac et Agn\`{e}s
Ma\^{\i}tre pour avoir pr\^{e}t\'{e} une oreille attentive (et m\^{e}me plus)
\`{a} tous mes probl\`{e}mes et mes inqui\'{e}tudes. Je leur en suis
tr\`{e}s reconnaissant.}

{\it Un grand merci collectif \`{a} : Thomas Coudreau, Catherine Schwob et
Alberto Bramati pour les nombreuses discussions, leurs encouragements, leur
aide et leur camaraderie; Ga\"{e}tan Messin, Hichem Eleuch, Jean-Pierre
Hermier, Cedric Begon, Katsuyuki Kasai, Ga\"{e}tan Hagel, St\'{e}phane
Boucard et Matthias Vaupel pour la chaleureuse ambiance qui r\`{e}gne dans
les couloirs du laboratoire.}

{\it Je voudrais remercier toute l'\'{e}quipe technique du laboratoire sans
qui ce travail n'aurait pas \'{e}t\'{e} possible : Jean-Pierre Plaut, Guy
Flory, Francis Tr\'{e}hin, Sylvain Pledel, Alexis Poizat, Jean-Claude
Bernard, Jean-Pierre Okpisz, Philippe Pace et Mohamed Boujrad. Je remercie
beaucoup Blandine Moutiers et Karine Gautier pour leur aide dans les
diff\'{e}rentes d\'{e}marches administratives ainsi que pour leur bonne
humeur et leur gentillesse. Un grand merci \`{a} Marie-Claire Rigolet pour
sa gentillesse et son travail formidable. \bigskip }

{\it Je voudrais exprimer ici ma reconnaissance \`{a} tous ceux qui m'ont
encourag\'{e} et aid\'{e} dans les moments difficiles de ma vie en France :
Kaci Hadjar qui a subvenu \`{a} mes besoins le temps que je m'adapte \`{a}
la vie parisienne; Hassib Khoury sans qui je n'aurais jamais pu obtenir ma
premi\`{e}re carte de s\'{e}jour; Laure Homberg et ses parents \`{a} qui je
dois beaucoup; Edwidge Ghazal pour les heures de cours gr\^{a}ce auxquelles
j'ai pu manger et payer le loyer; Regis Ledu qui a toujours r\'{e}pondu pr%
\'{e}sent \`{a} mes SOS; Laurent Vernac (tu sais pourquoi); Jean-Philippe
Poizat, Pierre-Fran\c{c}ois et Val\'{e}rie pour leur gentillesse; Francesca
et Arne pour leur soutien; Astrid et Pascal pour l'int\'{e}r\^{e}t qu'ils ont
toujours port\'{e} \`{a} mes probl\`{e}mes.}

{\it Je tiens \`{a} d\'{e}dier cette th\`{e}se \`{a} : tous mes amis d'Alger
(oulad el houma) et en particulier \`{a} ceux qui y sont toujours et y
vivent dans des conditions tr\`{e}s p\'{e}nibles, \`{a} Mohamed et Djouher
que je consid\`{e}re comme des parents sur qui je peux compter en toutes
circonstances et \`{a} Louness et Meryam qui ont toujours pr\'{e}serv\'{e}
la m\'{e}moire de ma m\`{e}re. Le dernier mot va \`{a} ma soeur Soraya et
mes fr\`{e}res Bela\"{\i}d et Omar, la seule vraie famille qui me reste.}

%% file: intro.tex
\chapter{INTRODUCTION}

\section{Pr\'{e}sentation g\'{e}n\'{e}rale\label{I-1}}

Les mesures optiques ont atteint aujourd'hui une tr\`{e}s grande
sensibilit\'{e}. Les avanc\'{e}es technologiques dans diff\'{e}rents
domaines de la physique (optique, physique des mat\'{e}riaux,
\'{e}lectronique . . .) ont permis de r\'{e}duire consid\'{e}rablement les
sources de bruit classique. Ainsi beaucoup de dispositifs optiques sont
limit\'{e}s par le bruit quantique de la lumi\`{e}re. La nature quantique de
la lumi\`{e}re impose en effet l'existence de fluctuations du champ
\'{e}lectromagn\'{e}tique. Si ces fluctuations sont connues depuis les
fondements de la m\'{e}canique quantique\cite{wheeler 1983}, ce n'est
qu'\`{a} partir des ann\'{e}es 1970 que les physiciens ont \'{e}t\'{e}
r\'{e}ellement confront\'{e}s aux limitations induites par ces fluctuations
sur la pr\'{e}cision d'une mesure. Depuis lors, un grand nombre
d'exp\'{e}riences s'est d\'{e}velopp\'{e} afin de modifier ou de
contr\^{o}ler ces fluctuations. Des progr\`{e}s marquants ont \'{e}t\'{e}
r\'{e}alis\'{e}s dans ce domaine gr\^{a}ce \`{a} la mise en \'{e}vidence
exp\'{e}rimentale d'\'{e}tats comprim\'{e}s ({\it squeezed states})\cite{réf
géné squeezing}, pour lesquels les fluctuations d'une quadrature du champ
sont inf\'{e}rieures \`{a} celles du vide, et gr\^{a}ce \`{a} la
r\'{e}alisation de mesures qui ne perturbent pas le signal mesur\'{e} ({\it %
Quantum Non Demolition Measurements})\cite{Braginsky,Leven La Porta Grang}.

Ces recherches ont \'{e}t\'{e} motiv\'{e}es en partie par les projets de
d\'{e}tection des ondes gravitationnelles \`{a} l'aide de grands
interf\'{e}rom\`{e}tres (projet franco-italien VIRGO ou LIGO aux USA)\cite
{Meystre 83 Bradaschia 90}. Les effets induits par le passage d'une onde
gravitationnelle \'{e}tant tr\`{e}s petits, ces interf\'{e}rom\`{e}tres
doivent avoir des bras tr\`{e}s longs (de l'ordre de quelques
kilom\`{e}tres) afin d'atteindre une sensibilit\'{e} suffisante. M\^{e}me si
les g\'{e}n\'{e}rations actuelles d'antennes gravitationnelles sont
essentiellement limit\'{e}es par le bruit thermique des suspensions et des
miroirs\cite{Meystre 83 Bradaschia 90,masse virgo ligo}, il n'est pas exclu
que les projets futurs de d\'{e}tecteurs soient confront\'{e}s au bruit
quantique de la lumi\`{e}re. La limite impos\'{e}e par ce bruit peut
cependant \^{e}tre contourn\'{e}e gr\^{a}ce \`{a} l'utilisation d'\'{e}tats
comprim\'{e}s, comme cela a \'{e}t\'{e} d\'{e}montr\'{e}
exp\'{e}rimentalement en 1987\cite{xiao grangier}.

Une seconde limitation li\'{e}e \`{a} la nature quantique de la lumi\`{e}re
est due \`{a} la pression de radiation exerc\'{e}e par la lumi\`{e}re sur
les miroirs de l'interf\'{e}rom\`{e}tre. Certains effets associ\'{e}s aux
fluctuations quantiques de la force de pression de radiation sont maintenant
bien connus. Par exemple, la pression de radiation exerc\'{e}e par les
fluctuations du vide est \`{a} l'origine de la force de Casimir, qui a
\'{e}t\'{e} observ\'{e}e \`{a} la fin des ann\'{e}es 50\cite{casimir}. Dans
le cas des mesures interf\'{e}rom\'{e}triques, la force de pression de
radiation exerc\'{e}e par la lumi\`{e}re d\'{e}place les miroirs et rend la
longueur des bras de l'interf\'{e}rom\`{e}tre sensible aux fluctuations
quantiques du champ. Il existe ainsi deux sources de bruit li\'{e}es \`{a}
la nature quantique de la lumi\`{e}re : le bruit propre du faisceau lumineux
qui diminue lorsque l'intensit\'{e} du champ augmente, et le bruit de
pression de radiation qui est proportionnel \`{a} l'intensit\'{e} lumineuse.
Un compromis entre ces deux effets conduit \`{a} une {\it limite quantique
standard} pour une mesure de position. Malgr\'{e} les travaux th\'{e}oriques
consacr\'{e}s \`{a} l'\'{e}tude de cette limite quantique standard\cite
{Unruh Yurke Caves}, aucune mise en \'{e}vidence exp\'{e}rimentale n'a
\'{e}t\'{e} r\'{e}alis\'{e}e jusqu'\`{a} pr\'{e}sent. Les fluctuations
quantiques de la pression de radiation agissent en effet tr\`{e}s peu sur un
objet macroscopique tel qu'un miroir, et une telle exp\'{e}rience est
difficile \`{a} mettre en oeuvre.

Plus r\'{e}cemment, des \'{e}tudes th\'{e}oriques ont montr\'{e} que la
pression de radiation devrait permettre de contr\^{o}ler les fluctuations
quantiques de la lumi\`{e}re, en utilisant une cavit\'{e} Fabry-Perot dont
un miroir est susceptible de se d\'{e}placer. Le d\'{e}placement du miroir
sous l'effet de la force de pression de radiation modifie le d\'{e}saccord
entre le champ et la cavit\'{e} : on obtient ainsi l'\'{e}quivalent d'un
effet Kerr d'origine purement m\'{e}canique, puisque le champ
intracavit\'{e} subit un d\'{e}phasage qui d\'{e}pend de son intensit\'{e}.
Cet effet non lin\'{e}aire a pour cons\'{e}quence de rendre la cavit\'{e}
bistable\cite{Lugiato prog opt 84}, comme cela a \'{e}t\'{e}
d\'{e}montr\'{e} exp\'{e}rimentalement \`{a} l'aide d'une cavit\'{e} dont un
miroir \'{e}tait attach\'{e} \`{a} un dispositif pendulaire\cite{Dorsel prl
1983}. Un milieu Kerr plac\'{e} dans une cavit\'{e} permet par ailleurs de
r\'{e}duire le bruit de photon de la lumi\`{e}re en dessous du bruit
quantique standard\cite{Kerr PRA 89}, comme l'ont montr\'{e} des
exp\'{e}riences utilisant des atomes comme milieu Kerr\cite{Squeez atom 85
96}. Dans le cas d'une cavit\'{e} vide \`{a} miroir mobile, il se produit un
effet de redistribution temporelle des photons par la cavit\'{e}, li\'{e}
\`{a} la variation de {\it longueur physique} de la cavit\'{e} induite par
la pression de radiation\cite{Fabre PRA 94}. Ce dispositif devrait donc se
comporter comme un {\it mangeur de bruit quantique}, le bruit de photon
\`{a} la sortie de la cavit\'{e} \'{e}tant r\'{e}duit en dessous du bruit de
photon standard.

La pression de radiation a aussi pour effet de cr\'{e}er des
corr\'{e}lations entre les fluctuations quantiques du champ et le mouvement
du miroir. L'une des applications de ces corr\'{e}lations consiste \`{a}
r\'{e}aliser une {\it mesure quantique non destructive}, c'est \`{a} dire
une mesure qui ne perturbe pas l'intensit\'{e}. Pour mesurer la position du
miroir, on peut utiliser soit une d\'{e}tection capacitive, soit une
d\'{e}tection optique\cite{QND group 95 97}. La premi\`{e}re m\'{e}thode
consiste \`{a} d\'{e}poser le miroir sur un cristal pi\'{e}zo\'{e}lectrique,
et \`{a} utiliser un circuit \'{e}lectrique r\'{e}sonnant pour d\'{e}tecter
la charge induite par la variation de longueur du cristal. Il appara\^{\i }t
ainsi des corr\'{e}lations quantiques entre l'intensit\'{e} du faisceau et
le courant \'{e}lectrique. La seconde m\'{e}thode consiste \`{a} envoyer un
second faisceau moins intense dans la cavit\'{e}, pour mesurer la position
du miroir. On cr\'{e}e alors des corr\'{e}lations quantiques entre
l'intensit\'{e} du premier faisceau et la phase du faisceau de mesure.

Nous pr\'{e}sentons dans ce m\'{e}moire un dispositif constitu\'{e} d'une
cavit\'{e} Fabry-Perot de grande finesse \`{a} une seule entr\'{e}e-sortie,
dont l'un des miroirs est susceptible de se d\'{e}placer sous l'effet de la
force de pression de radiation du champ intracavit\'{e}. La grande finesse
de cette cavit\'{e} permet de rendre le faisceau r\'{e}fl\'{e}chi par la
cavit\'{e} sensible \`{a} des tr\`{e}s petits d\'{e}placements du miroir
mobile. Par ailleurs, la r\'{e}ponse m\'{e}canique du miroir mobile est
optimis\'{e}e de fa\c{c}on \`{a} exalter les effets induits par les
fluctuations quantiques de la pression de radiation. En particulier, la
masse du miroir est choisie aussi petite que possible. Gr\^{a}ce \`{a} ces
caract\'{e}ristiques, il est possible d'\'{e}tudier de mani\`{e}re
g\'{e}n\'{e}rale les cons\'{e}quences du couplage optom\'{e}canique, tant
pour les fluctuations quantiques du champ que pour la dynamique du miroir.

Les objectifs principaux du dispositif exp\'{e}rimental d\'{e}crit dans ce
m\'{e}moire sont donc la mise en \'{e}vidence des effets quantiques du
couplage optom\'{e}canique sur le champ, et plus g\'{e}n\'{e}ralement sur le
syst\`{e}me coupl\'{e} champ-miroir mobile. Cependant, un tel dispositif est
capable de mesurer des d\'{e}placements du miroir mobile avec une
sensibilit\'{e} extr\^{e}me, du m\^{e}me ordre que celle des antennes
gravitationnelles. Cette sensibilit\'{e} peut \^{e}tre mise \`{a} profit
pour \'{e}tudier le mouvement Brownien du miroir mobile. Une \'{e}tude
exp\'{e}rimentale pr\'{e}cise de ce bruit thermique est importante car il
est responsable de la limite actuelle de sensibilit\'{e} pour les grands
interf\'{e}rom\`{e}tres de d\'{e}tection des ondes gravitationnelles\cite
{masse virgo ligo}.

Une autre application de la tr\`{e}s grande sensibilit\'{e} de la cavit\'{e}
est la mesure des d\'{e}placements produits par une onde gravitationnelle
sur une barre de Weber\cite{Weber 1960,optique weber}. Notre dispositif
devrait permettre d'am\'{e}liorer de fa\c{c}on significative la
sensibilit\'{e} de la mesure par rapport aux meilleurs dispositifs de
d\'{e}tection actuels, qui sont bas\'{e}s sur une mesure capacitive du
d\'{e}placement\cite{Bocko Onofrio 96}.

\section{Organisation de la th\`{e}se\label{I-2}}

L'objectif de ce m\'{e}moire est de pr\'{e}senter une \'{e}tude
g\'{e}n\'{e}rale des propri\'{e}t\'{e}s du couplage optom\'{e}canique et la
r\'{e}alisation exp\'{e}rimentale d'un dispositif en vue d'une mise en
\'{e}vidence des effets quantiques de ce couplage. Le chapitre II est
consacr\'{e} \`{a} une introduction g\'{e}n\'{e}rale sur le couplage
optom\'{e}canique. Nous pr\'{e}senterons dans un premier temps les concepts
de base en utilisant un syst\`{e}me simple, constitu\'{e} d'un miroir mobile
sur lequel se r\'{e}fl\'{e}chit un faisceau laser. Nous d\'{e}crirons
ensuite le dispositif qui permettra de mettre en \'{e}vidence les effets du
couplage optom\'{e}canique, c'est \`{a} dire une cavit\'{e} optique de
grande finesse dont un miroir est mobile. Nous pr\'{e}senterons les
applications de ce dispositif, telles que la production d'\'{e}tats
comprim\'{e}s et la mesure de petits d\'{e}placements. Cette \'{e}tude sera
r\'{e}alis\'{e}e dans le cadre d'un mod\`{e}le th\'{e}orique simple, o\`{u}
le champ est d\'{e}crit comme une onde plane et le miroir mobile comme un
oscillateur harmonique. Cette introduction nous permettra n\'{e}anmoins de
d\'{e}gager les param\`{e}tres essentiels du syst\`{e}me afin d'optimiser
les effets du couplage optom\'{e}canique.

Dans le chapitre III, nous d\'{e}crirons le miroir mobile utilis\'{e} dans
l'exp\'{e}rience. Ce miroir est form\'{e} de couches multidi\'{e}lectriques
d\'{e}pos\'{e}es sur la face plane d'un r\'{e}sonateur m\'{e}canique
constitu\'{e} d'un substrat plan-convexe en silice pure. Pour \'{e}tudier le
couplage entre la lumi\`{e}re et le mouvement m\'{e}canique du
r\'{e}sonateur, nous d\'{e}velopperons dans ce chapitre un mod\`{e}le
th\'{e}orique qui tient compte de la pr\'{e}sence de diff\'{e}rents modes
acoustiques dans le r\'{e}sonateur et de la structure tridimensionnelle du
r\'{e}sonateur et du faisceau lumineux. Nous montrerons qu'il est possible
de se ramener \`{a} une description monodimensionnelle, en int\'{e}grant la
structure spatiale dans une {\it susceptibilit\'{e} effective} qui
d\'{e}crit l'effet sur le champ de la r\'{e}ponse m\'{e}canique du
r\'{e}sonateur \`{a} la pression de radiation intracavit\'{e}. Nous
g\'{e}n\'{e}raliserons alors les r\'{e}sultats obtenus dans le chapitre II
au cas du r\'{e}sonateur plan-convexe, et nous montrerons l'importance de
l'adaptation spatiale entre le faisceau lumineux et les modes acoustiques.
Nous verrons en particulier que la {\it masse effective} du r\'{e}sonateur
qui d\'{e}crit l'amplitude du couplage optom\'{e}canique d\'{e}pend de la
section du faisceau et peut \^{e}tre beaucoup plus petite que la masse
totale du miroir.

Le chapitre IV est consacr\'{e} \`{a} la description du montage
exp\'{e}rimental qui est essentiellement constitu\'{e} de quatre parties :
la cavit\'{e} \`{a} miroir mobile,qui est une cavit\'{e} lin\'{e}aire de
grande finesse \`{a} une seule entr\'{e}e sortie, la source laser, qui est
constitu\'{e}e d'un laser titane saphir stabilis\'{e} en fr\'{e}quence et en
intensit\'{e}, le syst\`{e}me de d\'{e}tection homodyne, et enfin un
dispositif qui permet d'exciter optiquement les modes acoustiques du
r\'{e}sonateur afin de caract\'{e}riser la r\'{e}ponse m\'{e}canique du
miroir. Nous pr\'{e}senterons les caract\'{e}ristiques ainsi que les
performances obtenues pour chacun de ces \'{e}l\'{e}ments.

Dans le chapitre V nous pr\'{e}senterons les r\'{e}sultats exp\'{e}rimentaux
qui ont permis de caract\'{e}riser le couplage optom\'{e}canique dans notre
dispositif. Nous d\'{e}montrerons l'extr\^{e}me sensibilit\'{e} de la
cavit\'{e} \`{a} des petits d\'{e}placements du miroir mobile, gr\^{a}ce
\`{a} l'observation du mouvement Brownien du miroir. Nous d\'{e}terminerons
enfin cette sensibilit\'{e} et effectuerons une comparaison avec les
r\'{e}sultats th\'{e}oriques.

%% file: chapter2.tex
\chapter{LE COUPLAGE OPTOMECANIQUE}

\bigskip \bigskip

Nous pr\'{e}sentons dans ce chapitre les caract\'{e}ristiques
g\'{e}n\'{e}rales du couplage optom\'{e}canique. Nous introduisons les
concepts de base \`{a} l'aide d'un syst\`{e}me simple, constitu\'{e} d'un
miroir mobile sur lequel se r\'{e}fl\'{e}chit un faisceau laser (partie \ref
{II-1}). L'\'{e}tude de ce syst\`{e}me permet de comprendre ais\'{e}ment les
effets quantiques induits par la pression de radiation exerc\'{e}e par la
lumi\`{e}re. Nous pr\'{e}sentons ensuite le dispositif qui est le coeur du
montage exp\'{e}rimental, c'est-\`{a}-dire une cavit\'{e} optique de grande
finesse dont un miroir est mobile (partie \ref{II-2}). Les deux
derni\`{e}res parties sont consacr\'{e}es aux applications du couplage
optom\'{e}canique : la production d'\'{e}tats comprim\'{e}s (partie \ref
{II-3}) et la mesure de tr\`{e}s petits d\'{e}placements (partie 2.4).

\section{Syst\`{e}me avec un seul miroir mobile\label{II-1}}

\bigskip

\begin{figure}[tbp]
\centerline{\psfig{figure=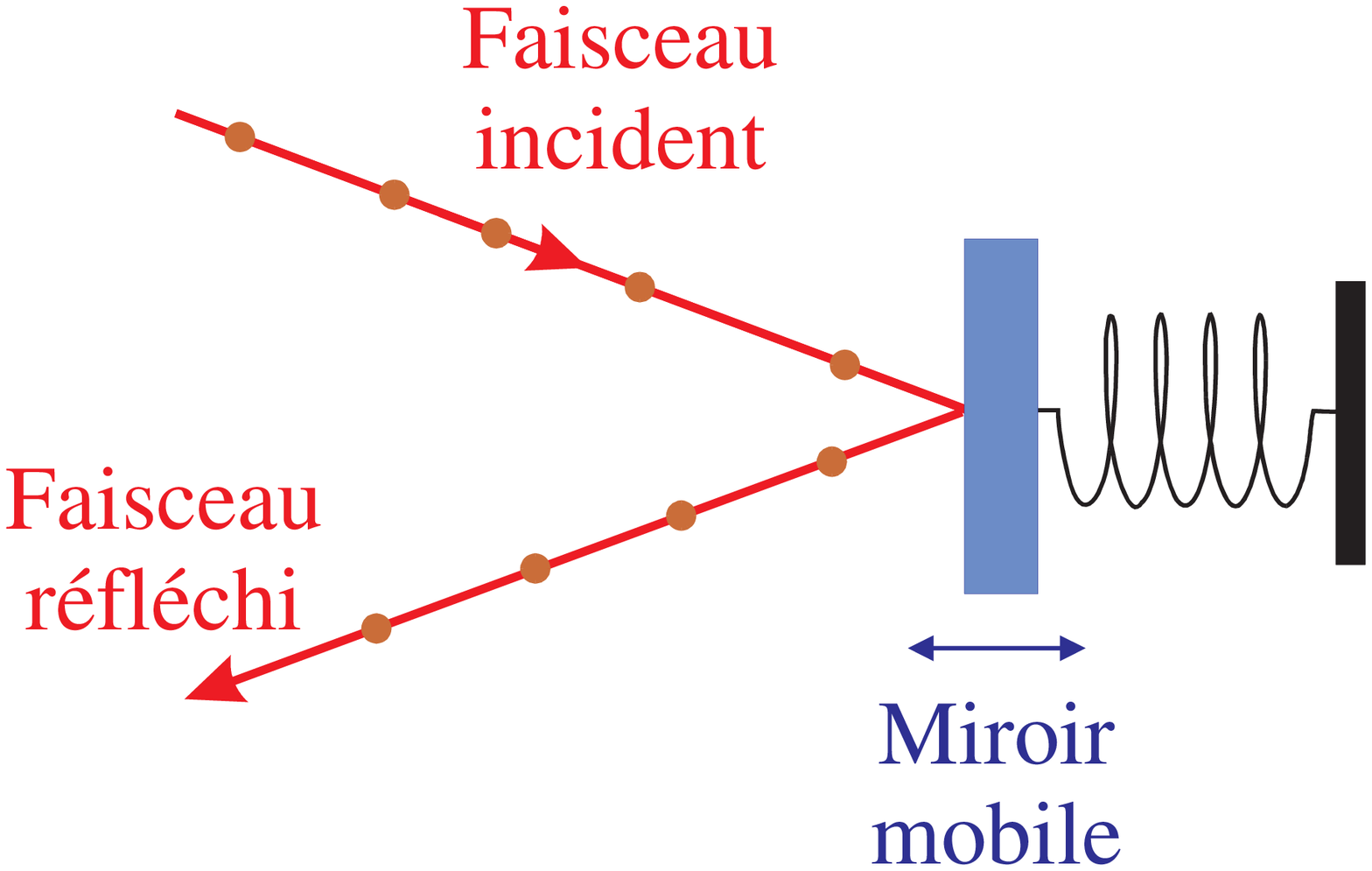,width=9cm}}
\caption{Syst\`{e}me simple o\`{u} un faisceau laser se refl\'{e}chit sur un
miroir mobile}
\label{Fig_1Mir}
\end{figure}

Les effets du couplage optom\'{e}canique peuvent \^{e}tre mis en
\'{e}vidence en consid\'{e}rant le syst\`{e}me simple o\`{u} un faisceau
lumineux se r\'{e}fl\'{e}chit sur un miroir mobile\cite{AH: miroir mobile}
(figure \ref{Fig_1Mir}). On utilisera dans cette partie une description
corpusculaire de la lumi\`{e}re. Ceci nous permettra de donner des images
physiques simples des ph\'{e}nom\`{e}nes qui interviennent au cours de
l'interaction. 
Nous allons tout d'abord d\'{e}crire le bruit de photon de la lumi\`{e}re
dans le cadre du mod\`{e}le corpusculaire. Nous pr\'{e}senterons ensuite les
deux aspects compl\'{e}mentaires du couplage optom\'{e}canique, c'est \`{a}
dire le d\'{e}placement du miroir sous l'effet de la force de pression de
radiation exerc\'{e}e par la lumi\`{e}re, et la modification du bruit
quantique de la lumi\`{e}re due \`{a} ce mouvement.

\subsection{Le bruit de photon\label{II-1-1}}

Le faisceau incident, issu d'une source laser coh\'{e}rente, est
caract\'{e}ris\'{e} par une distribution al\'{e}atoire des photons dans le
temps. On peut d\'{e}crire un tel faisceau \`{a} l'aide d'un mod\`{e}le
corpusculaire o\`{u} le flux de photons arrivant sur le miroir est
trait\'{e} comme un processus stochastique ponctuel. Chaque photon est
consid\'{e}r\'{e} comme un \'{e}v\`{e}nement discret, localis\'{e} dans le
temps, et l'intensit\'{e} du faisceau est d\'{e}finie comme le taux de
comptage de ces \'{e}v\'{e}nements. Plus pr\'{e}cis\'{e}ment,
l'intensit\'{e} peut s'\'{e}crire sous la forme: 
\begin{equation}
I(t)=\sum_{n}\delta (t-t_{n})  \label{2.1}
\end{equation}
o\`{u} $t_{n}$ correspond \`{a} l'instant d'arriv\'{e}e du $n$-i\`{e}me
photon sur le miroir. Pour un faisceau coh\'{e}rent, les temps $t_{n}$ sont
r\'{e}partis de fa\c{c}on al\'{e}atoire dans le temps.

On peut caract\'{e}riser la statistique de photon d'un faisceau coh\'{e}rent
de plusieurs mani\`{e}res \cite{Teich Loudon}. Par exemple, le nombre $N$ de
photons compt\'{e}s pendant un intervalle de temps $T$ est une variable
al\'{e}atoire dont la distribution de probabilit\'{e} $P_{T}(N)$ ob\'{e}it
\`{a} la loi de Poisson : 
\begin{equation}
P_{T}(N)=\frac{\overline{N}^{N}}{N!}~e^{-\overline{N}}  \label{2.2}
\end{equation}
o\`{u} $\overline{N}$ est le nombre moyen de photons compt\'{e}s dans
l'intervalle de temps $T$, reli\'{e} \`{a} l'intensit\'{e} moyenne $%
\overline{I}$ par: 
\begin{equation}
\overline{N}=\overline{I}T  \label{2.3}
\end{equation}

Il appara\^{\i }t ainsi que le nombre de photons compt\'{e}s sur un
intervalle de temps donn\'{e} n'est pas constant. En particulier, la
variance $\Delta N^{2}$ est \'{e}gale, pour une telle loi Poissonienne,
\`{a} $\overline{N}$ : 
\begin{equation}
\Delta N^{2}=\overline{N}  \label{2.4}
\end{equation}
Ces fluctuations sont \`{a} l'origine d'un bruit de ''grenaille'' lors de la
mesure de l'intensit\'{e} du faisceau lumineux. Ce bruit, qui est de nature
quantique car il est li\'{e} \`{a} la discr\'{e}tisation sous forme de
photons de la lumi\`{e}re, n'est autre que le bruit quantique standard, ou 
{\it shot noise}.

\begin{figure}[tbp]
\centerline{\psfig{figure=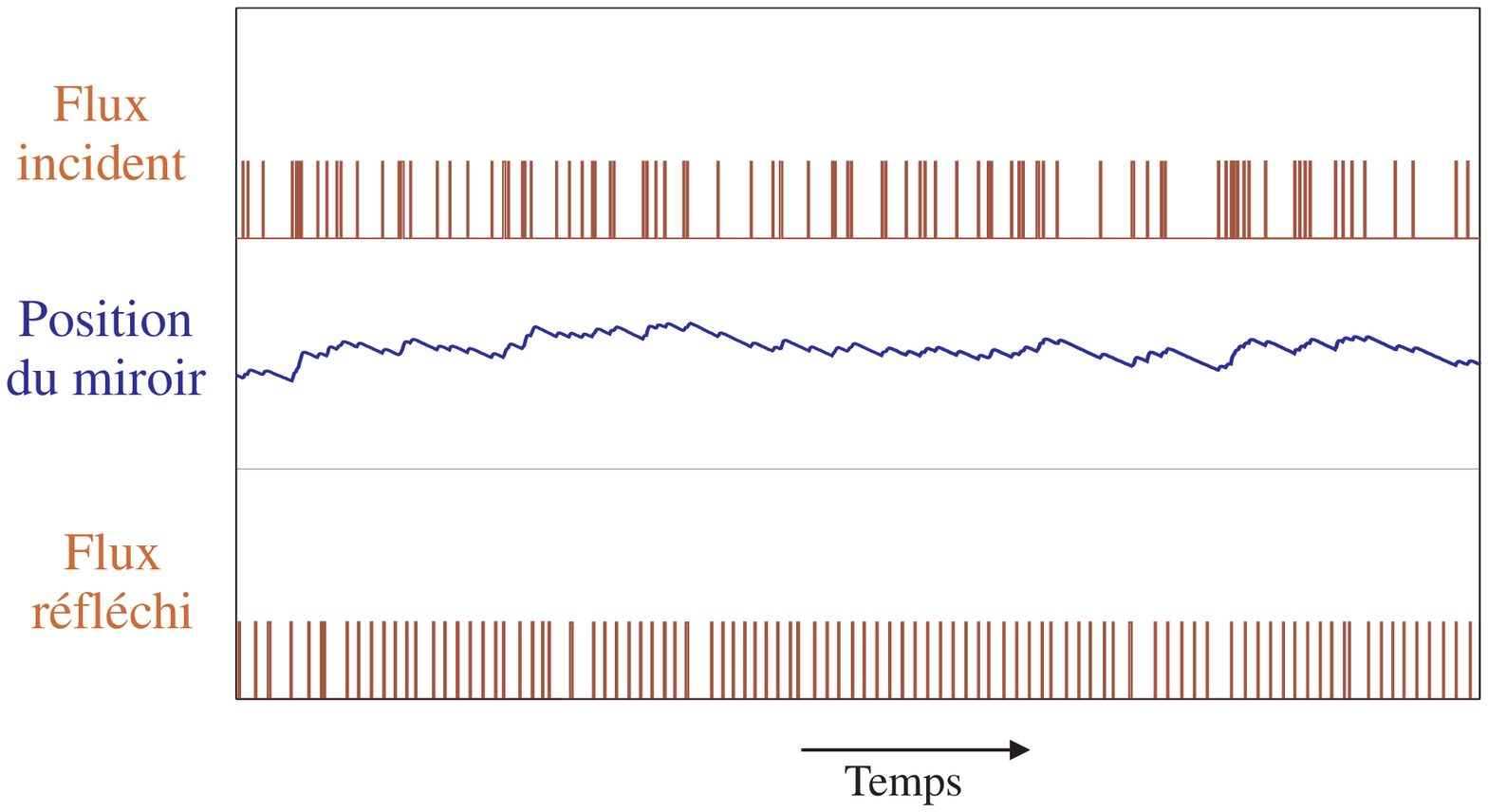,width=11cm}}
\caption{Simulation Monte-Carlo de l'\'{e}volution temporelle du syst\`{e}me
dans le cas d'un r\'{e}gime de faible flux de photons}
\label{Fig_MontCarl}
\end{figure}

Une autre caract\'{e}risation d'un faisceau coh\'{e}rent consiste \`{a}
utiliser la fonction d\'{e}lai entre deux photons successifs : ces
d\'{e}lais sont des variables al\'{e}atoires ind\'{e}pendantes ob\'{e}issant
\`{a} une loi de probabilit\'{e} exponentielle. On peut utiliser cette
propri\'{e}t\'{e} pour r\'{e}aliser une simulation num\'{e}rique du bruit de
photon. Les temps d'arriv\'{e}e des photons sont d\'{e}termin\'{e}s par
r\'{e}currence, en tirant num\'{e}riquement les d\'{e}lais entre photons
successifs \`{a} l'aide d'un g\'{e}n\'{e}rateur pseudo-al\'{e}atoire
ob\'{e}issant \`{a} une loi de distribution exponentielle. Le r\'{e}sultat
de la simulation est repr\'{e}sent\'{e} sur la premi\`{e}re courbe de la
figure \ref{Fig_MontCarl}. 
Chaque pulse repr\'{e}sente un temps d'arriv\'{e}e $t_{n}$ d'un photon sur
le miroir. Cette courbe simule tr\`{e}s bien le signal fourni par un
photomultiplicateur plac\'{e} en face d'un faisceau peu intense : chaque
photon d\'{e}tect\'{e} se traduit par un pulse \'{e}lectrique.
L'irr\'{e}gularit\'{e} dans les temps d'arriv\'{e}e des pulses est une
manifestation du bruit quantique de la lumi\`{e}re.

\subsection{Effet de la pression de radiation sur un miroir mobile\label%
{II-1-2}}

Chaque r\'{e}flexion d'un photon sur le miroir mobile se traduit par un
transfert de quantit\'{e} de mouvement \'{e}gal \`{a} $2\hbar k$ o\`{u} $k$
est le vecteur d'onde de la lumi\`{e}re. Ce transfert produit une force de
pression de radiation $F_{rad}(t)$ qui s'exerce sur le miroir. Cette force
est \'{e}gale \`{a} la quantit\'{e} de mouvement \'{e}chang\'{e}e par
photon, multipli\'{e}e par le nombre de photons r\'{e}fl\'{e}chis par
seconde: 
\begin{equation}
F_{rad}(t)=2\hbar k~I(t)  \label{2.5}
\end{equation}
Cette force induit un d\'{e}placement du miroir, et la position du miroir
est ainsi corr\'{e}l\'{e}e \`{a} l'intensit\'{e} lumineuse.

Dans le cadre du mod\`{e}le simple d\'{e}crit ici, on suppose que le
mouvement du miroir est harmonique, caract\'{e}ris\'{e} par une
fr\'{e}quence de r\'{e}sonance m\'{e}canique $\Omega _{M}$ (il s'agit par
exemple d'un miroir suspendu \`{a} un pendule). La simulation num\'{e}rique,
repr\'{e}sent\'{e}e par la seconde courbe de la figure \ref{Fig_MontCarl},
montre que le mouvement du miroir est sensible aux fluctuations
d'intensit\'{e} du faisceau incident. Lorsque le nombre de photons arrivant
sur le miroir est important, ce dernier a tendance \`{a} \^{e}tre
pouss\'{e}, alors qu'il tend \`{a} revenir vers sa position d'\'{e}quilibre
lorsque le flux diminue. La position du miroir reproduit ainsi les
fluctuations d'intensit\'{e} incidente, filtr\'{e}es par la r\'{e}ponse
m\'{e}canique du miroir. L'existence de ces corr\'{e}lations quantiques peut
\^{e}tre mise \`{a} profit pour r\'{e}aliser une mesure quantique de
l'intensit\'{e} du faisceau lumineux. Il faut pour cela disposer d'un
appareil de mesure capable de d\'{e}tecter les tr\`{e}s petits
d\'{e}placements du miroir sans pour autant perturber ce mouvement, et
extraire ensuite toute l'information sur les fluctuations d'intensit\'{e} du
faisceau incident. Nous d\'{e}crirons par la suite un syst\`{e}me capable de
r\'{e}aliser une telle mesure (section \ref{II-4-3}).

\subsection{Effet du mouvement du miroir sur le bruit de photon\label{II-1-3}
}

Le mouvement du miroir agit \`{a} son tour sur le bruit de photon en
modifiant le chemin optique suivi par la lumi\`{e}re. Les photons sont ainsi
plus ou moins retard\'{e}s lors de leur r\'{e}flexion sur le miroir. Comme
on peut le voir sur la derni\`{e}re courbe de la figure \ref{Fig_MontCarl},
la distribution des photons r\'{e}fl\'{e}chis est modifi\'{e}e, et dans
certains cas il est possible de r\'{e}guler le flux r\'{e}fl\'{e}chi.

Cette r\'{e}gulation peut s'interpr\'{e}ter de la mani\`{e}re suivante :
lorsque le flux incident est trop important, le miroir est pouss\'{e} vers
l'arri\`{e}re, ce qui a pour effet de retarder les photons. Le flux
r\'{e}fl\'{e}chi est donc r\'{e}duit par rapport au flux incident. L'effet
inverse se produit lorsque le flux incident est trop faible. Bien que ce
raisonnement ne soit pas tout \`{a} fait exact car il ne tient pas compte de
la dynamique du miroir, il permet de comprendre l'effet de r\'{e}gulation
temporelle du flux de photons.

Il appara\^{\i}t ainsi que le couplage optom\'{e}canique est bas\'{e} sur
deux effets compl\'{e}mentaires. Tout d'abord, le miroir se d\'{e}place sous
l'effet de la pression de radiation. La position du miroir effectue en
quelque sorte une ''mesure'' de l'intensit\'{e} de la lumi\`{e}re. Ensuite,
le mouvement modifie le chemin optique suivi par la lumi\`{e}re. La
combinaison de ces deux effets permet de contr\^{o}ler les fluctuations
quantiques de la lumi\`{e}re.

\subsection{R\'{e}gime de flux intense\label{II-1-4}}

\begin{figure}[tbp]
\centerline{\psfig{figure=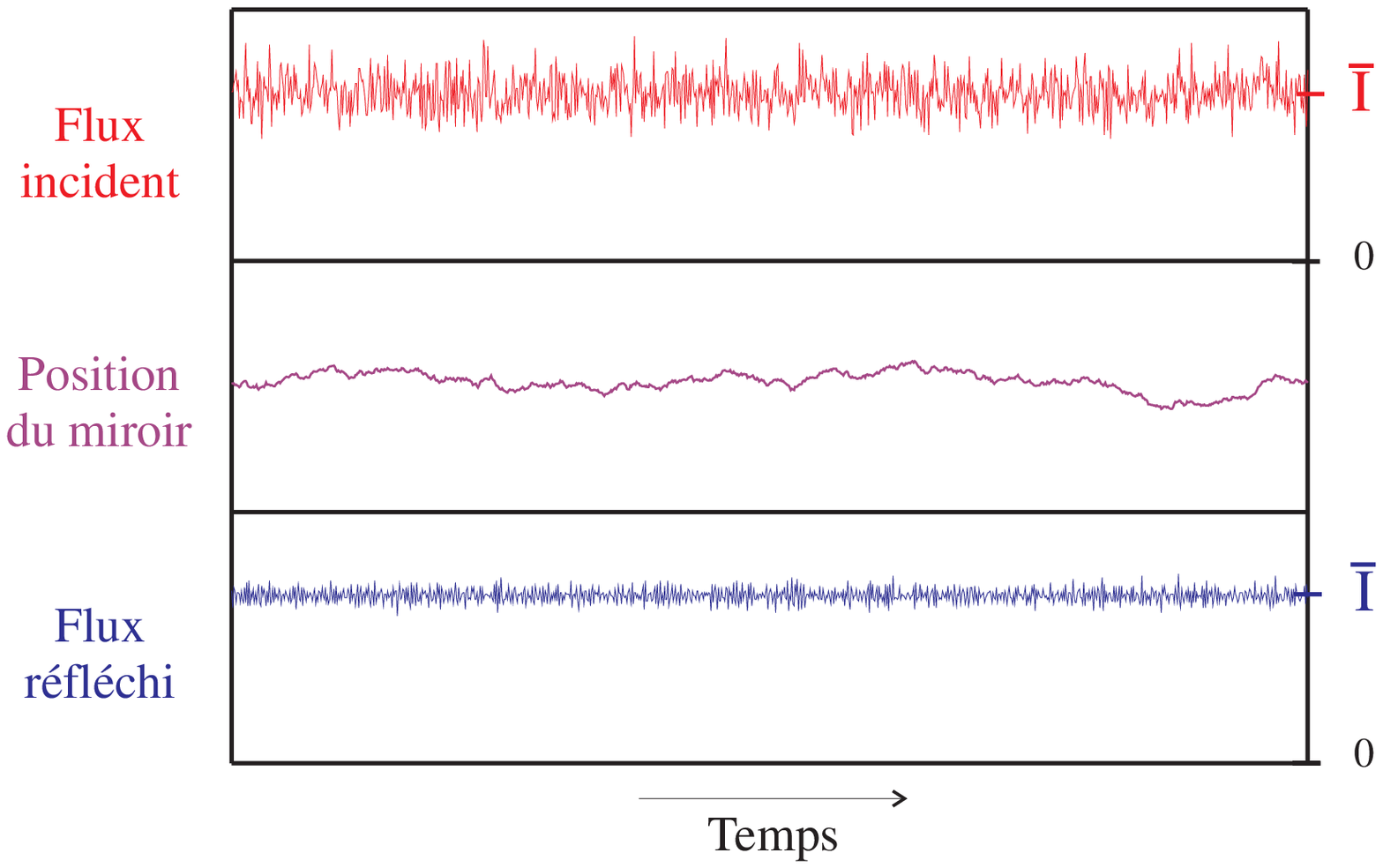,width=11cm}}
\caption{Simulation num\'{e}rique de l'\'{e}volution temporelle du
syst\`{e}me dans le cas d'un r\'{e}gime de flux intense. L'intensit\'{e}
apparait sous la forme d'un signal moyen non nul sur lequel se superpose des
fluctuations}
\label{Fig_1Mir_hf}
\end{figure}

Afin d'agir efficacement sur le bruit quantique, le flux incident doit
\^{e}tre suffisamment intense pour produire des d\'{e}placements importants
du miroir. La figure \ref{Fig_1Mir_hf} montre le r\'{e}sultat de la
simulation num\'{e}rique pour un r\'{e}gime de flux intense. On ne peut plus
dans ce cas repr\'{e}senter chaque photon par un pulse individuel. Afin de
visualiser le flux de photons, on divise le temps en petits canaux de
dur\'{e}e $\tau _{C}$ et on compte le nombre $N$ de photons dans chaque
canal. On obtient ainsi un signal d'intensit\'{e} moyenne non nulle $%
\overline{I}=\frac{\overline{N}}{\tau _{C}}$ auquel vient se rajouter des
fluctuations $\delta I(t)=\frac{\delta N(t)}{\tau _{C}}$, qui sont
associ\'{e}es \`{a} la variation $\delta N$ du nombre de photons par canal
au cours du temps. L'intensit\'{e} s'\'{e}crit:

\begin{equation}
I(t)=\overline{I}+\delta I(t)  \label{2.6}
\end{equation}

La premi\`{e}re courbe de la figure \ref{Fig_1Mir_hf} montre le flux de
photons incidents. Cette courbe simule tr\`{e}s bien le signal
\'{e}lectrique obtenu \`{a} l'aide d'une photodiode plac\'{e}e en face d'un
faisceau lumineux intense. On voit nettement sur la troisi\`{e}me courbe de
la figure \ref{Fig_1Mir_hf} la r\'{e}gulation temporelle du flux de photons,
qui se traduit par une r\'{e}duction du bruit d'intensit\'{e} du faisceau
r\'{e}fl\'{e}chi.

\subsection{Spectre de bruit\label{II-1-5}}

La repr\'{e}sentation temporelle que nous avons utilis\'{e}e jusqu'ici
pr\'{e}sente l'inconv\'{e}nient de ne pas mettre en \'{e}vidence la
dynamique du miroir mobile. On peut \'{e}tudier la r\'{e}ponse dynamique du
syst\`{e}me en se pla\c{c}ant dans l'espace des fr\'{e}quences (espace de
Fourier). Dans cet espace, le bruit d'intensit\'{e} est caract\'{e}ris\'{e}
par son spectre et le miroir mobile par sa r\'{e}ponse m\'{e}canique \`{a}
une force ext\'{e}rieure.

Le spectre de bruit d'intensit\'{e}, not\'{e} $S_{I}[\Omega ]$,
repr\'{e}sente la puissance de bruit de l'intensit\'{e} $I(t)$. Il est par
d\'{e}finition la transform\'{e}e de Fourier \`{a} la fr\'{e}quence $\Omega $
de la fonction d'autocorr\'{e}lation C$_{I}$(t) :

\begin{equation}
S_{I}\left[ \Omega \right] =\int_{-\infty }^{+\infty }C_{I}(t)e^{i\Omega t}dt%
\text{\qquad avec\qquad }C_{I}(t)=\left\langle \delta I(0)\delta
I(t)\right\rangle  \label{2.7}
\end{equation}
o\`{u} les crochets $\left\langle {}\right\rangle $ repr\'{e}sentent une
moyenne sur la statistique des photons. On peut aussi exprimer le spectre
d'intensit\'{e} en fonction de la transform\'{e}e de Fourier $\delta I\left[
\Omega \right] $ des fluctuations d'intensit\'{e} : 
\begin{equation}
\left\langle \delta I\left[ \Omega \right] \delta I[\Omega ^{\prime
}]\right\rangle =2\pi \text{ }\delta (\Omega +\Omega ^{^{\prime }})\text{ }%
S_{I}\left[ \Omega \right]  \label{2.8}
\end{equation}
Cette d\'{e}finition du spectre est tout \`{a} fait g\'{e}n\'{e}rale et
s'applique en fait \`{a} n'importe quelle variable al\'{e}atoire
stationnaire. On peut ainsi d\'{e}finir le spectre de bruit de position du
miroir mobile $S_{x}\left[ \Omega \right] $ en fonction des fluctuations de
position $\delta x\left[ \Omega \right] $ (transform\'{e}e de Fourier de $%
\delta x(t)$) : 
\begin{equation}
\left\langle \delta x\left[ \Omega \right] \delta x[\Omega ^{\prime
}]\right\rangle =2\pi \text{ }\delta (\Omega +\Omega ^{^{\prime }})\text{ }%
S_{x}\left[ \Omega \right]  \label{2.9}
\end{equation}

D'autre part, la r\'{e}ponse m\'{e}canique du miroir mobile permet de relier
la position du miroir \`{a} la force de pression de radiation. Dans le cas
de petits d\'{e}placements, la th\'{e}orie de la r\'{e}ponse lin\'{e}aire%
\cite{Cours Landau} montre qu'il existe une relation de proportionnalit\'{e}
dans l'espace de Fourier entre la position et la force appliqu\'{e}e: 
\begin{equation}
x\left[ \Omega \right] =\chi \left[ \Omega \right] F_{rad}\left[ \Omega
\right]  \label{2.10}
\end{equation}
o\`{u} $\chi \left[ \Omega \right] $ est la susceptibilit\'{e} m\'{e}canique
du miroir mobile. Pour un oscillateur harmonique de fr\'{e}quence de
r\'{e}sonance $\Omega _{M}$, cette susceptibilit\'{e} est d\'{e}crite par
une Lorent-\newline
zienne qui a la forme suivante: 
\begin{equation}
\chi \left[ \Omega \right] =\frac{1}{M(\Omega _{M}^{2}-\Omega ^{2}-i\Omega
\Omega _{M}/Q)}  \label{2.11}
\end{equation}
o\`{u} $M$ est la masse du miroir et $Q$ le facteur de qualit\'{e} de la
r\'{e}sonance. A l'aide de l'expression (\ref{2.10}) et celle de la force de
pression de radiation (\'{e}quation \ref{2.5}), on peut obtenir une relation
simple entre les spectres de position et d'intensit\'{e} incidente : 
\begin{equation}
S_{x}\left[ \Omega \right] =4\hbar ^{2}k^{2}\left| \chi \left[ \Omega
\right] \right| ^{2}~S_{I}\left[ \Omega \right]  \label{2.12}
\end{equation}
Cette relation montre clairement que la position du miroir reproduit les
fluctuations d'intensit\'{e}, filtr\'{e}es par la r\'{e}ponse m\'{e}canique
du miroir mobile.

\begin{figure}[tbp]
\centerline{\psfig{figure=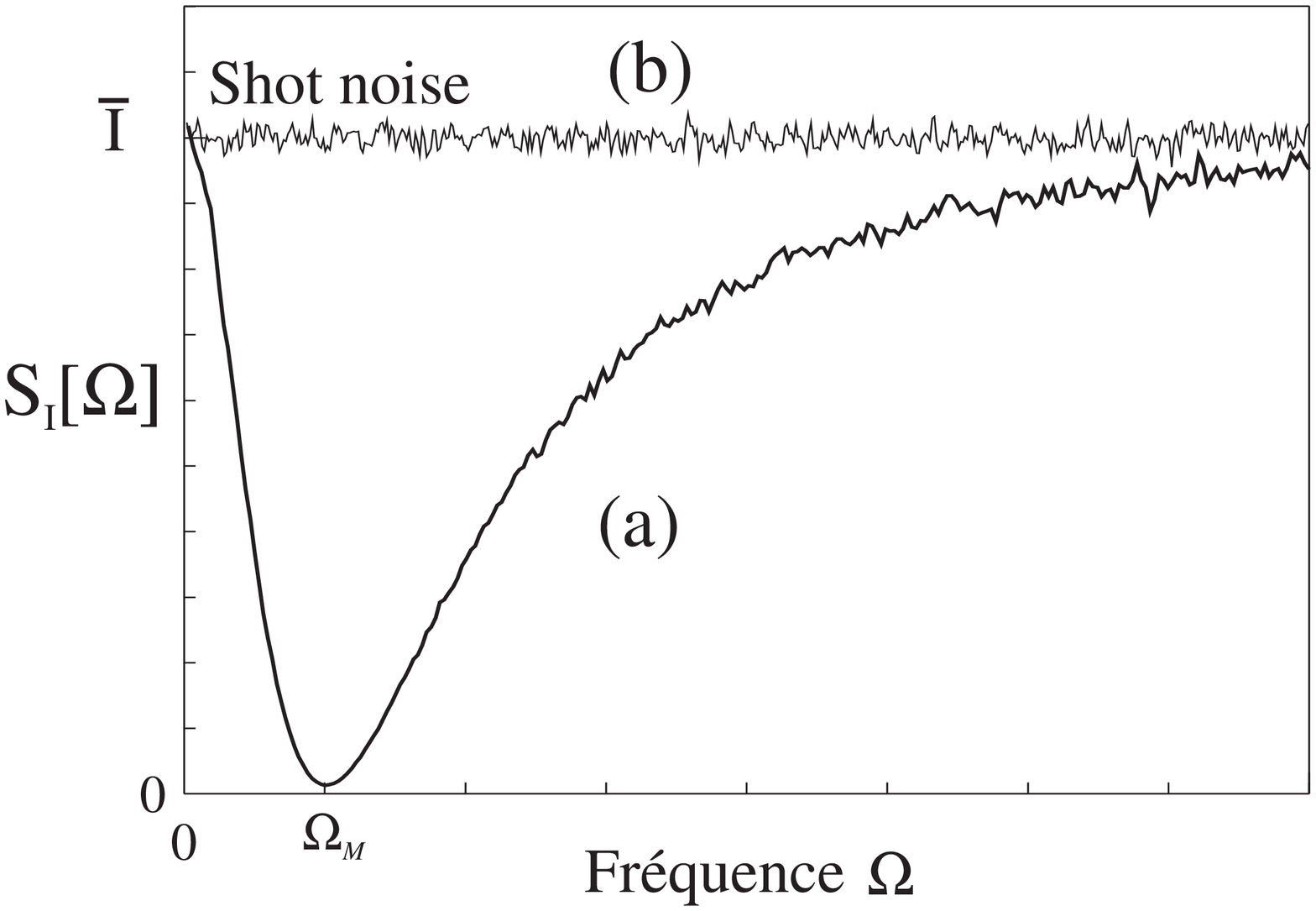,width=10cm}}
\caption{Simulation du spectre de bruit d'intensit\'{e} r\'{e}fl\'{e}chi
(a), qui est minimum \`{a} la fr\'{e}quence de r\'{e}sonance m\'{e}canique $%
\Omega _{M}$ du miroir. Le spectre de bruit du faisceau incident (b) est
plat en fr\'{e}quence (shot noise) }
\label{Fig_1Mir_sp}
\end{figure}

La simulation Monte Carlo permet de d\'{e}terminer les spectres de bruit
d'intensit\'{e}, par transform\'{e}e de Fourier num\'{e}rique des flux de
photons. La figure \ref{Fig_1Mir_sp} montre le r\'{e}sultat de cette
simulation, pour les spectres de bruit des faisceaux incident et
r\'{e}fl\'{e}chi\cite{AH: miroir mobile}. 
Le faisceau incident, caract\'{e}ris\'{e} par une statistique Poissonienne,
a un bruit plat en fr\'{e}quence (shot noise) : 
\begin{equation}
S_{I}\left[ \Omega \right] =\overline{I}  \label{2.13}
\end{equation}
La simulation num\'{e}rique montre que le spectre de bruit d'intensit\'{e}
du faisceau r\'{e}fl\'{e}chi n'est plus plat en fr\'{e}quence mais qu'il est
r\'{e}duit en dessous de la limite quantique standard. La r\'{e}duction
d\'{e}pend de la r\'{e}ponse en fr\'{e}quence du miroir, qui agit comme un
filtre pour les fluctuations quantiques de la lumi\`{e}re autour de la
fr\'{e}quence de r\'{e}sonance m\'{e}canique $\Omega _{M}$. Ce syst\`{e}me
permet ainsi de contr\^{o}ler les fluctuations quantiques de la lumi\`{e}re
et de produire des \'{e}tats non classiques, ou \'{e}tats comprim\'{e}s (%
{\it squeezed states})\cite{réf géné squeezing}. Nous d\'{e}crirons par la
suite plus en d\'{e}tail les caract\'{e}ristiques de ces \'{e}tats non
classiques.

\subsection{Quelques ordres de grandeurs\label{II-1-6}}

Nous allons maintenant donner quelques ordres de grandeurs concernant d'une
part les d\'{e}placements n\'{e}cessaires pour agir sur les fluctuations
quantiques de la lumi\`{e}re et d'autre part les d\'{e}placements du miroir
mobile produits par la force de pression de radiation. Nous allons utiliser
pour cela des raisonnements simples, certes incomplets, mais qui nous
permettront de garder une image physique des ph\'{e}nom\`{e}nes mis en jeu.

\subsubsection{D\'{e}placements n\'{e}cessaires pour corriger les
fluctuations quantiques de l'intensit\'{e} lumineuse\label{II-1-6-1}}

\begin{figure}[tbp]
\centerline{\psfig{figure=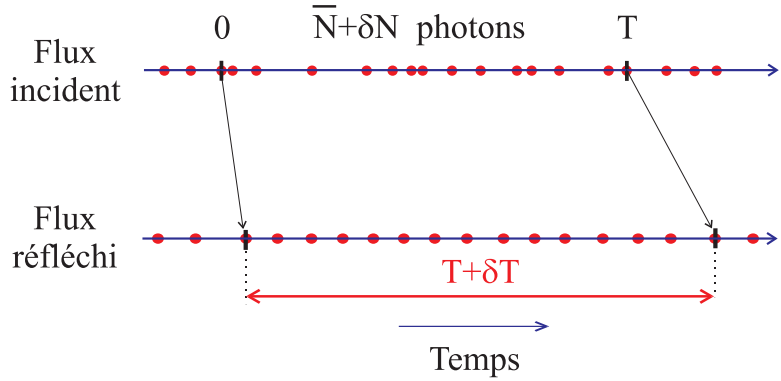,height=5cm}}
\caption{Le mouvement du miroir redistribue les $\overline{N}+\delta N$
photons incidents dans un intervalle de temps $T+\delta T$, ce qui a pour
effet de r\'{e}guler le flux de photons r\'{e}fl\'{e}chi }
\label{2Fig_Delai}
\end{figure}

On peut comprendre le ph\'{e}nom\`{e}ne de r\'{e}gulation temporelle \`{a}
partir de la figure \ref{2Fig_Delai}. 
Consid\'{e}rons un intervalle de temps $T$ au cours duquel on compte $%
\overline{N}+\delta N$ photons dans le flux incident. Le miroir, en se
d\'{e}pla\c{c}ant, retarde les photons. Les $\overline{N}+\delta N$ photons
se retrouvent ainsi dans le flux r\'{e}fl\'{e}chi r\'{e}partis sur un
intervalle de temps diff\'{e}rent, \'{e}gal \`{a} $T+\delta T$ o\`{u} $%
\delta T$ est proportionnel au d\'{e}placement du miroir : 
\begin{equation}
\delta T=2~\frac{x(T)-x(0)}{c}  \label{2.14}
\end{equation}
($c$ est la vitesse de la lumi\`{e}re). Le miroir r\'{e}gule le flux de
photons si la variation temporelle $\delta T$ est telle que le nombre $%
\overline{N}+\delta N$ de photons est proche du nombre moyen de photons
attendu dans l'intervalle de temps $T+\delta T$ : 
\begin{equation}
\frac{\overline{N}+\delta N}{T+\delta T}\approx \overline{I}  \label{2.15}
\end{equation}
On en d\'{e}duit que le miroir doit se d\'{e}placer au cours du temps $T$
d'une quantit\'{e} $\delta x_{T}$ \'{e}gale \`{a} : 
\begin{equation}
\delta x_{T}=x(T)-x(0)=\frac{c}{2\overline{I}}~\delta N  \label{2.16}
\end{equation}
On trouve ainsi l'\'{e}cart type $\Delta x_{T}$ des d\'{e}placements
n\'{e}cessaires \`{a} partir de la variance $\Delta N^{2}$ donn\'{e}e par
l'\'{e}quation (\ref{2.4}) : 
\begin{equation}
\Delta x_{T}=\frac{c}{2}\sqrt{\frac{T}{\overline{I}}}  \label{2.17}
\end{equation}
Le temps de mesure $T$ est typiquement de l'ordre de la p\'{e}riode de
r\'{e}sonance m\'{e}canique. C'est en effet au voisinage de la fr\'{e}quence
de r\'{e}sonance $\Omega _{M}$ que la r\'{e}duction de bruit est la plus
importante (figure \ref{Fig_1Mir_sp}). On arrive alors \`{a} l'expression
suivante pour les d\'{e}placements que doit effectuer le miroir pour
corriger les fluctuations quantiques de la lumi\`{e}re : 
\begin{equation}
\Delta x_{T}\approx \frac{c}{2\sqrt{\overline{I}\Omega _{M}}}  \label{2.18}
\end{equation}

Pour d\'{e}terminer un ordre de grandeur de ces d\'{e}placements, on peut
consid\'{e}rer un miroir fix\'{e} \`{a} un syst\`{e}me pendulaire. Dans ce
cas, la fr\'{e}quence de r\'{e}sonance m\'{e}canique $\Omega _{M}$ est au
plus de l'ordre de $10^{5}~rad/s$. L'intensit\'{e} moyenne $\overline{I}$
est reli\'{e}e \`{a} la puissance lumineuse $P$ par l'interm\'{e}diaire de
l'\'{e}nergie $\hbar kc$ d'un photon : 
\begin{equation}
P=\hbar kc~\overline{I}  \label{2.19}
\end{equation}
Pour une puissance lumineuse de $10~Watts$ et une longueur d'onde $\lambda
=2\pi /k$ de $1~\mu m$, on obtient une intensit\'{e} moyenne $\overline{I}$
\'{e}gale \`{a} $5~10^{19}~photons/s$. Ces chiffres conduisent \`{a} des
d\'{e}placements du miroir de l'ordre de $\Delta x_{T}=65~\mu m$, soit $%
65~\lambda $.

\subsubsection{D\'{e}placements dus \`{a} la force de pression de radiation%
\label{II-1-6-2}}

La force de pression de radiation s'\'{e}crit, d'apr\`{e}s les \'{e}quations
(\ref{2.5}) et (\ref{2.6}), comme la somme d'une force moyenne associ\'{e}e
\`{a} l'intensit\'{e} moyenne du faisceau et d'un terme fluctuant li\'{e} au
bruit de photon. Le miroir subit donc \`{a} la fois un recul moyen $%
\overline{x}$ et des fluctuations de position $\delta x$. Nous allons voir
que ces deux quantit\'{e}s correspondent en fait \`{a} de tr\`{e}s petits
d\'{e}placements.

Afin de d\'{e}terminer le d\'{e}placement moyen, on \'{e}crit qu'\`{a}
l'\'{e}quilibre, la force moyenne de rappel de l'oscillateur harmonique $%
M\Omega _{M}^{2}\overline{x}$ est \'{e}gale \`{a} la force de pression de
radiation moyenne $2\hbar k\overline{I}$. On obtient alors l'expression
suivante : 
\begin{equation}
\overline{x}=\frac{2\hbar k\overline{I}}{M\Omega _{M}^{2}}  \label{2.20}
\end{equation}
Pour une masse du miroir mobile \'{e}gale \`{a} $100~mg$, on trouve un
d\'{e}placement moyen beaucoup plus petit que la longueur d'onde, de l'ordre
de $7~10^{-8}\lambda $.

Pour \'{e}valuer \`{a} pr\'{e}sent le d\'{e}placement d\^{u} aux
fluctuations de la force de pression de radiation, nous allons
d\'{e}terminer la variance $\Delta x^{2}$ qui est par d\'{e}finition
l'int\'{e}grale du spectre de position $S_{x}\left[ \Omega \right] $.
D'apr\`{e}s les \'{e}quations (\ref{2.12}) et (\ref{2.13}), cette variance
s'exprime simplement en fonction de la susceptibilit\'{e} m\'{e}canique du
miroir, c'est-\`{a}-dire comme l'int\'{e}grale d'une Lorentzienne
(\'{e}quation \ref{2.11}). On obtient ainsi :

\begin{equation}
\Delta x=\overline{x}~\sqrt{\frac{\Omega _{M}Q}{2\overline{I}}}  \label{2.21}
\end{equation}
Pour un facteur de qualit\'{e} $Q=10^{6}$, ce d\'{e}placement est de l'ordre
de $\Delta x\approx 2~10^{-12}\lambda $.

Ce tr\`{e}s petit d\'{e}placement est bien s\^{u}r insuffisant pour agir
efficacement sur le bruit quantique, puisqu'il est environ $10^{14}$ fois
plus faible que le d\'{e}placement n\'{e}cessaire $\Delta x_{T}$\cite
{heidmann2}. Il semble par ailleurs difficile d'augmenter de fa\c{c}on
significative les d\'{e}placements produits par les fluctuations de pression
de radiation : ces fluctuations quantiques induisent sur un objet
macroscopique tel qu'un miroir des d\'{e}placements petits devant la
longueur d'onde.

Il faut cependant noter qu'il est envisageable de d\'{e}tecter d'aussi
petites variations de position, en particulier gr\^{a}ce aux progr\`{e}s
r\'{e}alis\'{e}s dans le domaine des mesures interf\'{e}rom\'{e}triques
destin\'{e}es \`{a} d\'{e}tecter des ondes gravitationnelles\cite{Meystre 83
Bradaschia 90}. En fait, la phase du champ \'{e}lectromagn\'{e}tique est
beaucoup plus sensible que l'intensit\'{e} \`{a} une variation de position
du miroir. S'il est n\'{e}cessaire de d\'{e}placer le miroir de $65\lambda $
pour changer l'intensit\'{e} d'une quantit\'{e} de l'ordre du bruit
quantique, un d\'{e}placement de l'ordre de $\lambda $ induit une variation
de phase de $2\pi $. Il suffit donc de d\'{e}placer le miroir d'une tr\`{e}s
petite fraction de longueur d'onde pour induire sur le faisceau une
variation de phase de l'ordre de ses fluctuations quantiques.

En conclusion, si le syst\`{e}me pr\'{e}sent\'{e} dans cette partie permet
de comprendre les principes de base du couplage optom\'{e}canique, il est
n\'{e}cessaire d'utiliser un dispositif plus complexe, o\`{u} la phase joue
un r\^{o}le, pour esp\'{e}rer agir sur le bruit quantique de la lumi\`{e}re.
Les parties suivantes de ce chapitre ont pour but de d\'{e}crire un tel
syst\`{e}me.

\section{Cavit\'{e} Fabry-Perot \`{a} miroir mobile\label{II-2}}

\bigskip

\begin{figure}[tbp]
\centerline{\psfig{figure=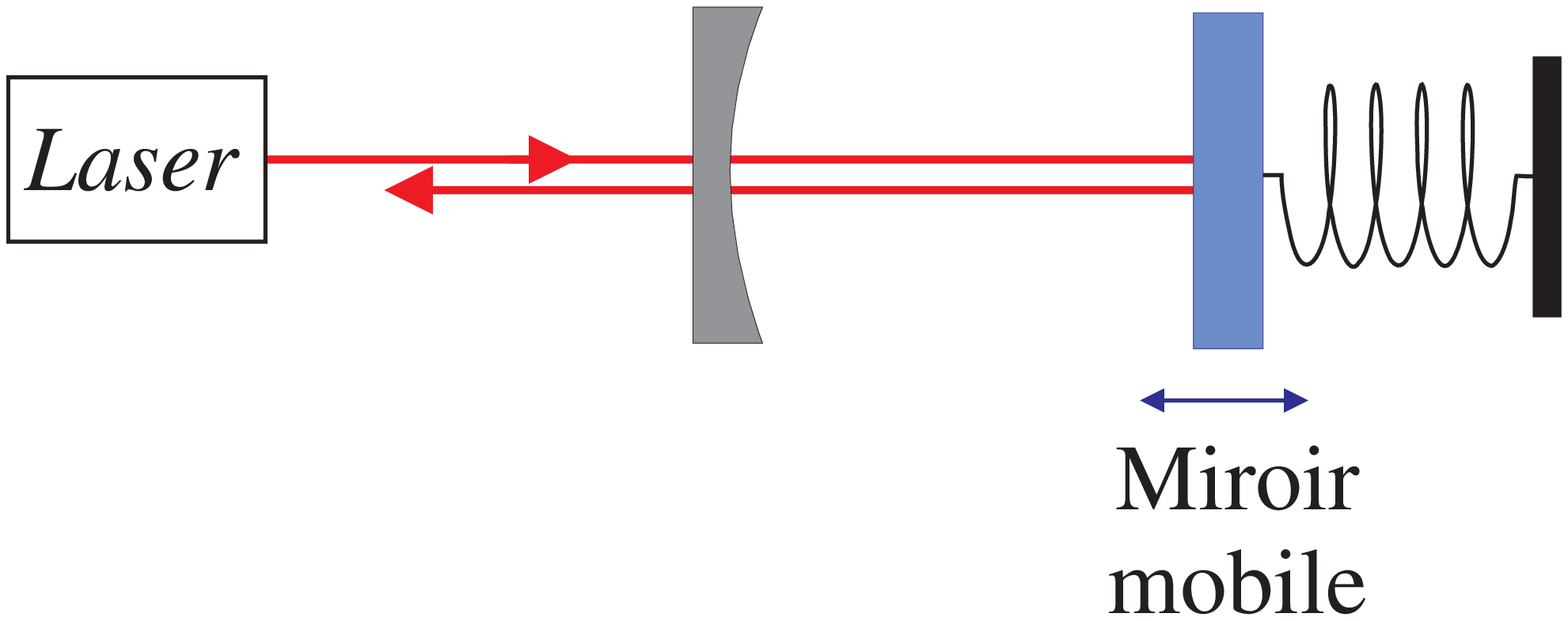,width=8cm}}
\caption{Cavit\'{e} Fabry Perot \`{a} miroir mobile}
\label{Fig_cavmob}
\end{figure}

Nous avons vu que les d\'{e}placements induits par les fluctuations de la
force de pression de radiation sont tr\`{e}s petits (de l'ordre de $\lambda
/10^{12}$). Il sont par cons\'{e}quent insuffisants pour agir directement
sur les fluctuations quantiques de l'intensit\'{e} lumineuse. Par ailleurs,
la phase du champ \'{e}lectromagn\'{e}tique est beaucoup plus sensible aux
d\'{e}placements du miroir mobile. Pour mettre en \'{e}vidence les effets
quantiques du couplage optom\'{e}canique, il faut donc disposer d'un
syst\`{e}me capable de coupler les fluctuations de phase \`{a} celles de
l'intensit\'{e}. Il suffit en fait de rajouter un miroir fixe de
transmission non nulle devant le miroir mobile de fa\c{c}on \`{a} former une
cavit\'{e} Fabry-Perot \`{a} une seule entr\'{e}e-sortie (figure \ref
{Fig_cavmob}). Nous allons d\'{e}crire dans cette partie les
propri\'{e}t\'{e}s g\'{e}n\'{e}rales de ce syst\`{e}me, les effets sur le
faisceau lumineux et quelques applications qui en d\'{e}coulent. Nous
\'{e}tudierons plus en d\'{e}tail ces applications dans les parties
suivantes. 

\subsection{Cavit\'{e} Fabry-Perot \`{a} une seule entr\'{e}e-sortie\label%
{II-2-1}}

\begin{figure}[tbp]
\centerline{\psfig{figure=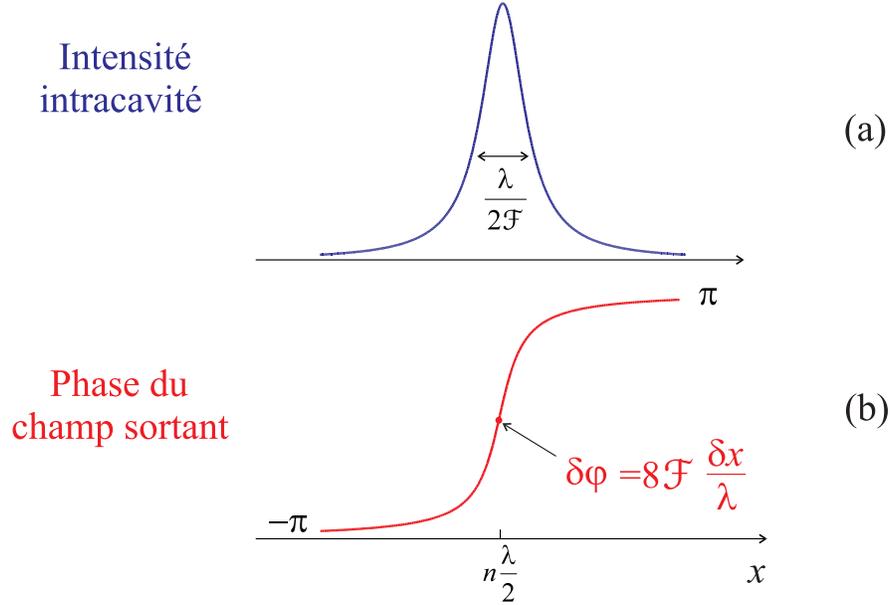,height=8cm}}
\caption{Variation de l'intensit\'{e} du champ \`{a} l'int\'{e}rieur de la
cavit\'{e} (a) et de la phase du faisceau r\'{e}fl\'{e}chi (b) en fonction
de la longueur de la cavit\'{e}}
\label{Fig_2intphas}
\end{figure}

On va s'int\'{e}resser au syst\`{e}me constitu\'{e} d'une cavit\'{e}
Fabry-Perot dont un miroir est mobile (figure \ref{Fig_cavmob}). Une telle
cavit\'{e} pr\'{e}sente des r\'{e}sonances pour des longueurs pr\'{e}cises
de la cavit\'{e}. Le champ intracavit\'{e} est en effet la somme d'une
infinit\'{e} d'ondes qui interf\`{e}rent de fa\c{c}on constructive pour des
longueurs multiples de la demi longueur d'onde lumineuse $\lambda /2$. Il
appara\^{\i }t de ce fait un maximum d'intensit\'{e} \`{a} chaque fois que
la longueur est un multiple de $\lambda /2$. Lorsqu'on s'\'{e}carte de ces
r\'{e}sonances en d\'{e}pla\c{c}ant par exemple le miroir mobile,
l'intensit\'{e} dans la cavit\'{e} diminue en d\'{e}crivant un pic d'Airy
qui a la forme d'une Lorentzienne pour une cavit\'{e} de grande finesse
(figure \ref{Fig_2intphas}a).

Un param\`{e}tre important d'une cavit\'{e} Fabry-Perot est sa finesse $%
{\cal F}$, d\'{e}finie par la largeur $\lambda /2{\cal F}$ \`{a} mi-hauteur
des pics d'Airy. La finesse ne d\'{e}pend que de la transmission et des
pertes des deux miroirs. Elle d\'{e}termine aussi l'amplification de
l'intensit\'{e} intracavit\'{e} moyenne $\bar{I}$ \`{a} r\'{e}sonance qui
est reli\'{e}e \`{a} l'intensit\'{e} moyenne incidente $\bar{I}^{in}$ par la
relation : 
\begin{equation}
\bar{I}=\frac{2}{\pi }{\cal F}\bar{I}^{in}  \label{2.22}
\end{equation}

Dans le cas d'une cavit\'{e} sans perte optique, tous les photons incidents
finissent par ressortir de la cavit\'{e} apr\`{e}s un certain temps de
stockage. Le faisceau r\'{e}fl\'{e}chi a donc la m\^{e}me intensit\'{e}
moyenne que le faisceau incident. Par contre il subit une variation de phase
qui d\'{e}pend de la longueur de la cavit\'{e}, et donc de la position du
miroir mobile. La figure \ref{Fig_2intphas}b montre que cette variation de
phase est de l'ordre de $2\pi $ pour un d\'{e}placement du miroir \'{e}gal
\`{a} la largeur du pic d'Airy. A r\'{e}sonance, la pente de la courbe est
maximale et vaut $8{\cal F}/\lambda $. Un petit d\'{e}placement $\delta x$
du miroir mobile autour de la r\'{e}sonance produit une variation de phase $%
\delta \varphi $ \'{e}gale \`{a}: 
\begin{equation}
\delta \varphi =8{\cal F}~\frac{\delta x}{\lambda }  \label{2.23}
\end{equation}

On peut comparer cette sensibilit\'{e} \`{a} celle obtenue avec un seul
miroir mobile. Pour le syst\`{e}me \'{e}tudi\'{e} dans la partie \ref{II-1},
un d\'{e}placement $\delta x$ du miroir induit une modification $2\delta x$
du chemin optique de la lumi\`{e}re. Le champ r\'{e}fl\'{e}chi subit donc un
d\'{e}phasage \'{e}gal \`{a} $4\pi \delta x/\lambda $. La pr\'{e}sence de la
cavit\'{e} augmente ainsi la sensibilit\'{e} de ce d\'{e}phasage par un
facteur de l'ordre de la finesse ${\cal F}$ de la cavit\'{e}.

Une cavit\'{e} de grande finesse devrait donc permettre de mesurer des
d\'{e}placements du miroir correspondants \`{a} une tr\`{e}s petite fraction
de la longueur d'onde. Nous \'{e}tudierons dans la partie \ref{II-2-3} la
possibilit\'{e} de r\'{e}aliser des mesures de tr\`{e}s faibles
d\'{e}placements, qu'ils soient dus aux fluctuations quantiques de la
pression de radiation, ou qu'ils soient li\'{e}s \`{a} d'autres sources,
comme le bruit thermique du miroir mobile.

Une autre application de la tr\`{e}s grande sensibilit\'{e} de la cavit\'{e}
\`{a} des petits d\'{e}placements consiste \`{a} contr\^{o}ler les
fluctuations quantiques du champ. Les fluctuations de phase du faisceau
r\'{e}fl\'{e}chi induites par le couplage optom\'{e}canique peuvent en effet
\^{e}tre du m\^{e}me ordre de grandeur que les fluctuations quantiques. Il
est ainsi possible de modifier ces fluctuations, et en particulier de les
r\'{e}duire de fa\c{c}on \`{a} obtenir des \'{e}tats comprim\'{e}s. Nous
\'{e}tudierons dans la partie suivante la possibilit\'{e} de produire de
tels \'{e}tats avec une cavit\'{e} \`{a} miroir mobile.

\subsection{Compression du champ par couplage optom\'{e}canique\label{II-2-2}
}

En \'{e}lectrodynamique quantique, le champ \'{e}lectromagn\'{e}tique se
d\'{e}compose en une somme de modes \'{e}quivalents \`{a} des oscillateurs
harmoniques ind\'{e}pendants. Nous allons supposer ici que le champ peut
\^{e}tre d\'{e}crit par un seul mode, caract\'{e}ris\'{e} par des
op\'{e}rateurs d'annihilation et de cr\'{e}ation $\hat{a}$ et $\hat{a}%
^{\dagger }$ ind\'{e}pendants du temps. Ces op\'{e}rateurs ob\'{e}issent
\`{a} la relation de commutation:

\begin{equation}
\left[ \hat{a},\hat{a}^{\dagger }\right] =1  \label{2.24}
\end{equation}
Comme dans le cas d'un oscillateur mat\'{e}riel caract\'{e}ris\'{e} par les
deux observables de position $\hat{x}$ et de quantit\'{e} de mouvement $\hat{%
p}$, on introduit pour le champ \'{e}lectromagn\'{e}tique deux observables $%
\hat{a}_{1}$ et $\hat{a}_{2}$, appel\'{e}es quadratures du champ, qui sont
associ\'{e}es aux parties r\'{e}elle et imaginaire du champ : 
\begin{subequations}
\begin{eqnarray}
\hat{a}_{1} &=&\hat{a}+\hat{a}^{\dagger }  \label{2.25.a} \\
\hat{a}_{2} &=&i(\hat{a}^{\dagger }-\hat{a})  \label{2.25.b}
\end{eqnarray}  \label{2.25}
\end{subequations}
Ces deux observables ne commutent pas et leurs variances $\Delta \hat{a}_{1}$
et $\Delta \hat{a}_{2}$ v\'{e}rifient une in\'{e}galit\'{e} de Heisenberg:

\begin{equation}
\Delta \hat{a}_{1}~\Delta \hat{a}_{2}\geq 1  \label{2.26}
\end{equation}
Cette in\'{e}galit\'{e} traduit l'existence de fluctuations quantiques. Pour
\'{e}tudier ces fluctuations, nous allons utiliser la m\'{e}thode
semi-classique qui va permettre de leur associer une repr\'{e}sentation
g\'{e}om\'{e}trique dans l'espace des phases.

\subsubsection{Repr\'{e}sentation semi-classique des fluctuations quantiques%
\label{II-2-2-1}}

La m\'{e}thode semi-classique consiste \`{a} repr\'{e}senter les
fluctuations quantiques par une distribution de quasi-probabilit\'{e} de
Wigner\cite{Wigner Phys Rev 1932}. On associe ainsi des variables classiques
al\'{e}atoires${\cal \ }\alpha $ et $\alpha ^{*}$ ($\alpha ^{*}$ \'{e}tant
le complexe conjugu\'{e} de $\alpha $) aux op\'{e}rateurs du champ $\hat{a}$
et $\hat{a}^{\dagger }$. La distribution de Wigner, qui est une fonction des
variables $\alpha $ et $\alpha ^{*}$, d\'{e}crit la loi de distribution de
ces variables. Ainsi toute moyenne quantique des op\'{e}rateurs $\hat{a}$ et 
$\hat{a}^{\dagger }$ rang\'{e}s dans l'ordre sym\'{e}trique est \'{e}gale
\`{a} la moyenne classique de la m\^{e}me combinaison des variables $\alpha $
et $\alpha ^{*}$, pond\'{e}r\'{e}e par la distribution de Wigner.

\begin{figure}[tbp]
\centerline{\psfig{figure=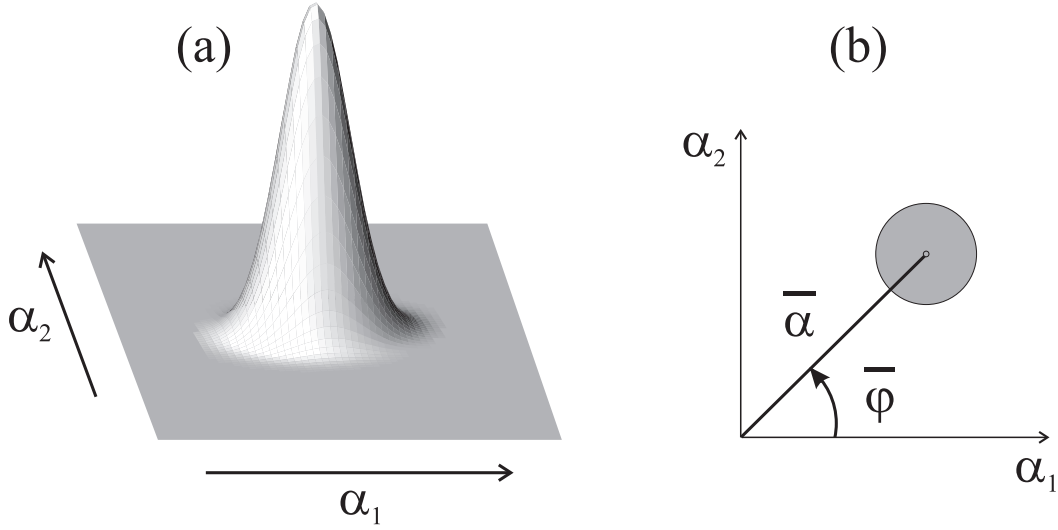,height=7cm}}
\caption{Repr\'{e}sentation dans l'espace des phases d'un champ
coh\'{e}rent. La distribution de Wigner est une Gaussienne (a) et sa
projection dans le plan \{$\alpha _{1};$ $\alpha _{2}$\} est un disque qui
d\'{e}crit les r\'{e}alisations les plus probables du champ. $\bar{\alpha}$
et $\bar{\varphi}$ repr\'{e}sentent respectivement l'amplitude et la phase
du champ moyen}
\label{Fig_2wigcoh}
\end{figure}

L'un des int\'{e}r\^{e}ts de la distribution de Wigner est qu'elle est
positive pour les \'{e}tats usuels du champ. Elle peut donc \^{e}tre
consid\'{e}r\'{e}e comme une v\'{e}ritable distribution de probabilit\'{e},
et les variables $\alpha $ et $\alpha ^{*}$ repr\'{e}sentent les valeurs
possibles du champ. La figure \ref{Fig_2wigcoh}a montre la distribution de
Wigner pour un \'{e}tat coh\'{e}rent. Le plan horizontal repr\'{e}sente
l'espace des phases dont les axes sont d\'{e}finis par les parties
r\'{e}elle $\alpha _{1}=\alpha +\alpha ^{*}$ et imaginaire $\alpha
_{2}=i(\alpha ^{*}-\alpha )$ du champ. La distribution de Wigner est une
Gaussienne dont la variance dans toutes les directions est \'{e}gale \`{a} $%
1 $.

Une repr\'{e}sentation plus commode \`{a} deux dimensions (figure \ref
{Fig_2wigcoh}b) est obtenue en faisant une projection de la distribution
dans le plan \{$\alpha _{1};\alpha _{2}$\}. Les fluctuations du champ sont
alors d\'{e}crites par un disque d\'{e}limit\'{e} par la courbe
d'isoprobabilit\'{e} \`{a} $1/e$ de la Gaussienne. Chaque point \`{a}
l'int\'{e}rieur du disque repr\'{e}sente une r\'{e}alisation possible du
champ. Le champ \'{e}lectromagn\'{e}tique peut ainsi s'\'{e}crire comme la
somme d'un champ moyen $\bar{\alpha}$ et d'un champ fluctuant $\delta \alpha 
$ qui d\'{e}crit les fluctuations quantiques: 
\begin{equation}
\alpha =\overline{\alpha }+\delta \alpha  \label{2.27}
\end{equation}
Le bruit sur chaque quadrature $\alpha _{1}$ et $\alpha _{2}$ est donn\'{e}
par la projection de la distribution sur les axes horizontal et vertical.
Pour un \'{e}tat coh\'{e}rent, on trouve que les variances $\Delta \alpha
_{1}$ et $\Delta \alpha _{2}$ sont toutes deux \'{e}gales \`{a} $1$. Un
\'{e}tat coh\'{e}rent est donc un \'{e}tat minimal pour lequel le produit de
ces variances a la valeur minimale autoris\'{e}e par l'in\'{e}galit\'{e} de
Heisenberg (\'{e}quation \ref{2.26}). 

De mani\`{e}re g\'{e}n\'{e}rale, on peut d\'{e}finir une quadrature
quelconque du champ par l'expression: 
\begin{equation}
\alpha _{\theta }=e^{-i\theta }\alpha +e^{i\theta }\alpha ^{*}  \label{2.28}
\end{equation}
Les quadratures $\alpha _{1}$ et $\alpha _{2}$ correspondent respectivement
\`{a} $\theta =0$ et $\theta =\pi /2$. Les fluctuations pour la quadrature $%
\alpha _{\theta }$ sont obtenues en projetant la distribution sur l'axe
faisant un angle $\theta $ avec l'horizontale dans l'espace des phases. Pour
un \'{e}tat coh\'{e}rent, toutes les quadratures ont donc le m\^{e}me bruit,
la variance $\Delta \alpha _{\theta }$ \'{e}tant \'{e}gale \`{a} $1$ quelle
que soit la valeur de $\theta $.

On peut aussi d\'{e}terminer les bruits d'intensit\'{e} et de phase. Pour un
champ monomode et ind\'{e}pendant du temps, l'intensit\'{e} est
caract\'{e}ris\'{e}e par le nombre $N$ de photons dans le mode. Ce nombre de
photons et la phase $\varphi $ sont reli\'{e}s au champ $\alpha $ par la
relation: 
\begin{equation}
\alpha =\sqrt{N}e^{i\varphi }  \label{2.29}
\end{equation}
En lin\'{e}arisant les fluctuations autour des valeurs moyennes, on trouve
que les fluctuations d'intensit\'{e} et de phase sont reli\'{e}es
respectivement \`{a} la quadrature d'amplitude $\alpha _{\overline{\varphi }%
} $ et \`{a} la quadrature orthogonale $\alpha _{\overline{\varphi }+\pi /2}$%
, o\`{u} $\overline{\varphi }$ est la phase du champ moyen: 
\begin{subequations}
\begin{eqnarray}
\delta N &=&\left| \overline{\alpha }\right| \delta \alpha _{\overline{%
\varphi }}  \label{2.30.a} \\
\delta \varphi &=&\frac{1}{2\left| \overline{\alpha }\right| }\delta \alpha
_{\overline{\varphi }+\pi /2}  \label{2.30.b}
\end{eqnarray}  \label{2.30}
\end{subequations}
Les bruits d'intensit\'{e} et de phase sont donc associ\'{e}s \`{a} la
projection de la distribution sur les axes parall\`{e}le et perpendiculaire
au champ moyen. Pour un \'{e}tat coh\'{e}rent, on trouve que le bruit
d'intensit\'{e} correspond au bruit quantique standard, puisque la variance $%
\Delta N^{2}$ est \'{e}gale au nombre moyen $\overline{N}$ de photons. Par
contre, la variance des fluctuations de phase est inversement
proportionnelle au nombre moyen de photons: 
\begin{equation}
\Delta \varphi ^{2}=\frac{1}{4\overline{N}}  \label{2.31}
\end{equation}
Ce r\'{e}sultat peut s'interpr\'{e}ter de la mani\`{e}re suivante : la
dispersion de phase $\Delta \varphi $ correspond \`{a} l'angle sous lequel
on voit la distribution depuis l'origine; elle est donc d'autant plus petite
que l'amplitude moyenne $\overline{\alpha }$ du champ est grande. Notons
enfin qu'il existe une relation de Heisenberg entre les fluctuations
d'intensit\'{e} et de phase: 
\begin{equation}
\Delta N~\Delta \varphi \geq \frac{1}{2}  \label{2.32}
\end{equation}
Un \'{e}tat coh\'{e}rent est un \'{e}tat minimal vis-\`{a}-vis de cette
in\'{e}galit\'{e}.

\subsubsection{La m\'{e}thode semi-classique\label{II-2-2-2}}

Nous venons d'associer une repr\'{e}sentation semi-classique aux
fluctuations quantiques. Ceci permet de d\'{e}crire les champs entrants dans
le syst\`{e}me \`{a} l'aide de variables classiques al\'{e}atoires. Afin de
d\'{e}terminer les fluctuations des champs sortants, il est n\'{e}cessaire
de d\'{e}crire les effets de l'interaction avec le syst\`{e}me.

Dans le cas d'un r\'{e}gime de champ intense o\`{u} les fluctuations $\delta
\alpha $ sont petites devant la valeur moyenne $\overline{\alpha }$ du
champ, on peut lin\'{e}ariser l'\'{e}quation d'\'{e}volution de la
distribution de Wigner. On obtient alors une \'{e}quation tout \`{a} fait
similaire \`{a} celle d\'{e}crivant l'\'{e}volution classique du syst\`{e}me%
\cite{Reynaud Fabre Ekert}. La m\'{e}thode semi-classique consiste donc
\`{a} remplacer les fluctuations quantiques des champs par une distribution
semi-classique dans l'espace des phases, puis \`{a} \'{e}tudier
l'\'{e}volution de cette distribution \`{a} l'aide des \'{e}quations
classiques d\'{e}crivant le syst\`{e}me.

Notons que la m\'{e}thode pr\'{e}sent\'{e}e ici correspond \`{a} une analyse
statique du syst\`{e}me: on s'int\'{e}resse \`{a} la variance des
fluctuations du champ, le syst\`{e}me \'{e}tant dans un \'{e}tat
stationnaire. On peut bien s\^{u}r g\'{e}n\'{e}raliser la m\'{e}thode
semi-classique pour tenir compte de la dynamique du syst\`{e}me. On
s'int\'{e}resse alors aux fluctuations $\delta \alpha \left[ \Omega \right] $
du champ \`{a} une fr\'{e}quence d'analyse $\Omega $. La m\'{e}thode
semi-classique permet d'obtenir une relation d'entr\'{e}e-sortie pour ces
fluctuations, relation qui fait intervenir la r\'{e}ponse dynamique du
syst\`{e}me. Cette relation permet de d\'{e}terminer les spectres de bruit
des fluctuations sortantes. Nous aurons l'occasion de pr\'{e}senter cette
approche dynamique dans la partie (2.3).

\subsubsection{Production d'\'{e}tats comprim\'{e}s par couplage
optom\'{e}canique\label{II-2-2-3}}

Nous allons expliquer de fa\c{c}on simple comment le couplage
optom\'{e}canique peut permettre de comprimer les fluctuations du champ. Une
\'{e}tude plus rigoureuse est pr\'{e}sent\'{e}e dans la partie \ref{II-3}.
Nous n\'{e}gligeons ici la dynamique du syst\`{e}me, ce qui revient \`{a} ne
consid\'{e}rer que les fluctuations \`{a} fr\'{e}quence nulle. Dans le cadre
de cette analyse statique, on peut consid\'{e}rer que le champ est monomode
et ind\'{e}pendant du temps.

\begin{figure}[tbp]
\centerline{\psfig{figure=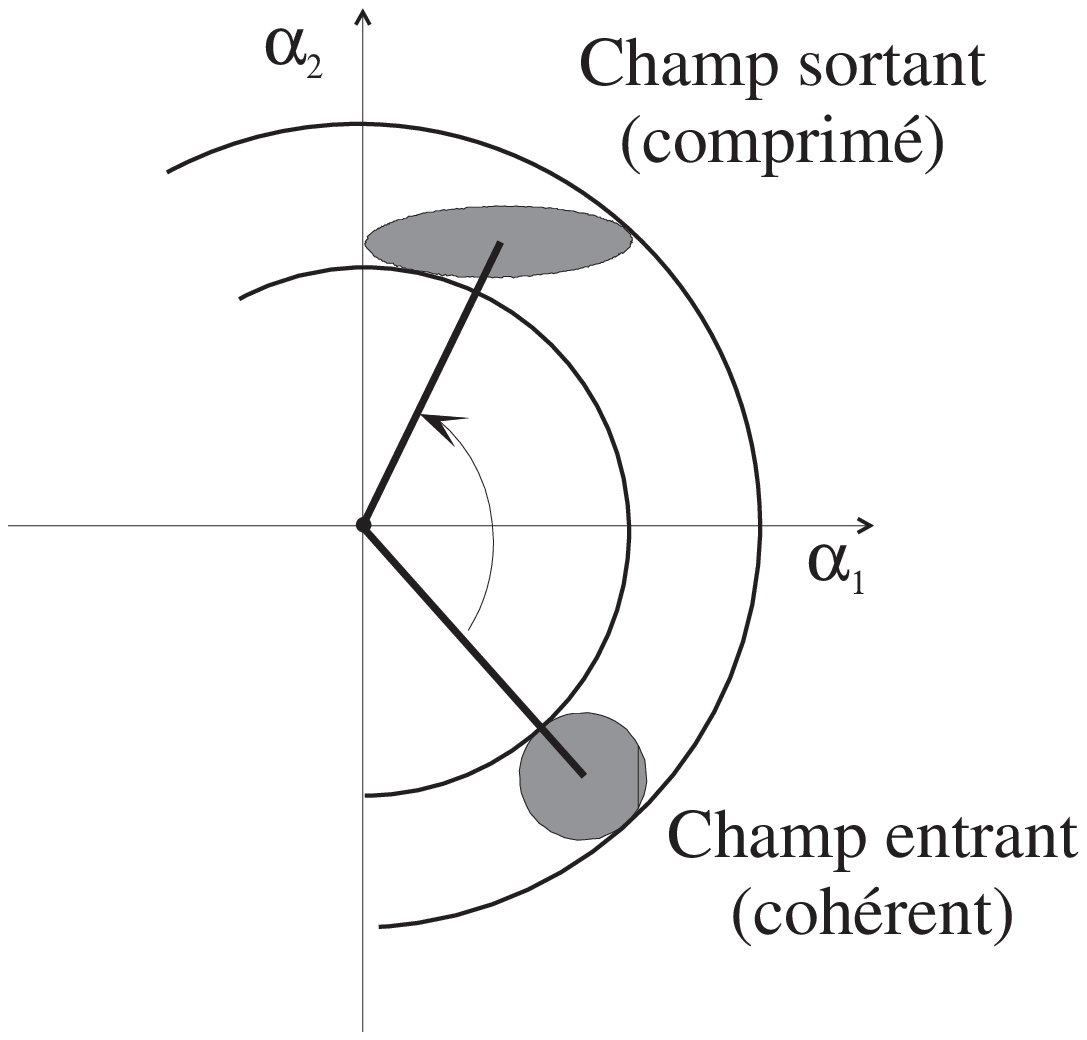,height=7cm}}
\caption{Repr\'{e}sentation dans l'espace des phases des champs entrant et
sortant de la cavit\'{e}. Chaque point de la distribution incidente subit
une rotation dont l'angle d\'{e}pend de l'amplitude du point}
\label{Fig_2sqzesph}
\end{figure}

Nous avons vu au d\'{e}but de cette partie que le champ r\'{e}fl\'{e}chi par
la cavit\'{e} subit un d\'{e}phasage qui d\'{e}pend de la longueur de la
cavit\'{e} (figure \ref{Fig_2intphas}b). Cette longueur est modifi\'{e}e par
le mouvement du miroir mobile sous l'effet de la force de pression de
radiation exerc\'{e}e par le champ intracavit\'{e}. Le champ \`{a} la sortie
subit donc un d\'{e}phasage qui d\'{e}pend de l'amplitude du champ. Dans
l'espace des phases, ce d\'{e}phasage se traduit par une rotation du champ
(figure \ref{Fig_2sqzesph}). La m\'{e}thode semi-classique permet alors
d'appliquer cette transformation classique \`{a} l'ensemble de la
distribution du champ coh\'{e}rent incident (disque de la figure \ref
{Fig_2sqzesph}). Chaque r\'{e}alisation possible du champ subit une rotation
qui d\'{e}pend de son amplitude, ce qui a pour effet de modifier la forme de
la distribution qui devient elliptique\cite{Kerr PRA 89}. On obtient ainsi
un \'{e}tat du champ dont le bruit sur la composante $\alpha _{2}$ est
r\'{e}duit ($\Delta \alpha _{2}<1$). Afin de conserver l'in\'{e}galit\'{e}
de Heisenberg, le bruit sur la composante $\alpha _{1}$ est augment\'{e}. 

Dans ce syst\`{e}me, le champ subit une transformation unitaire, et la
surface de la distribution est conserv\'{e}e. Le champ sortant de la
cavit\'{e} est donc dans un \'{e}tat minimal vis-\`{a}-vis de
l'in\'{e}galit\'{e} de Heisenberg. Un tel \'{e}tat est un \'{e}tat non
classique du champ appel\'{e} \'{e}tat comprim\'{e} ({\it squeezed state}).
Il est caract\'{e}ris\'{e} par la distribution de Wigner de la figure \ref
{Fig_2wigsq}. La distribution est toujours une Gaussienne, mais la variance
pour une quadrature quelconque varie selon la quadrature consid\'{e}r\'{e}e.

Nous venons de montrer, de mani\`{e}re simple et g\'{e}om\'{e}trique, que la
cavit\'{e} \`{a} miroir mobile est capable de produire un \'{e}tat
comprim\'{e} du champ\cite{Fabre PRA 94}. Nous exposerons dans la partie \ref
{II-3} une \'{e}tude plus d\'{e}taill\'{e}e de ce syst\`{e}me. En
particulier nous nous int\'{e}resserons au spectre de bruit du champ
r\'{e}fl\'{e}chi par la cavit\'{e}.

\subsection{Mesure de petits d\'{e}placements\label{II-2-3}}

La phase du champ r\'{e}fl\'{e}chi par la cavit\'{e} est sensible \`{a} de
tr\`{e}s petits d\'{e}placements du miroir mobile. L'\'{e}quation (\ref{2.23}%
) montre qu'un d\'{e}placement $\delta x$ du miroir mobile provoque une
variation de la phase du champ r\'{e}fl\'{e}chi. Les fluctuations de phase $%
\delta \varphi ^{out}$ du faisceau r\'{e}fl\'{e}chi refl\`{e}tent donc les
d\'{e}placements du miroir, auxquels se superposent le bruit propre du
faisceau. Nous verrons dans la partie 2.4 que pour une cavit\'{e}
r\'{e}sonnante avec le champ, ce bruit quantique est de l'ordre du bruit de
phase $\delta \varphi ^{in}$ du faisceau incident: 
\begin{figure}[tbp]
\centerline{\psfig{figure=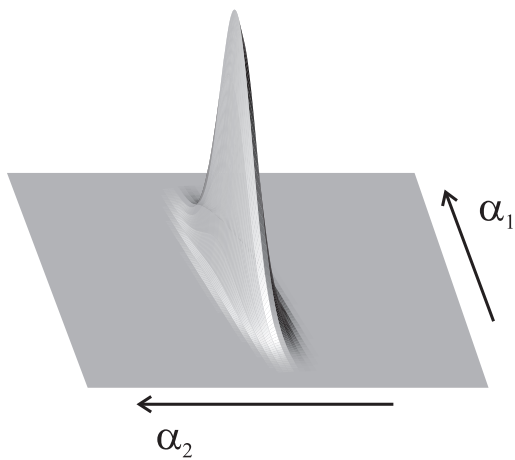,height=6cm}}
\caption{Repr\'{e}sentation dans l'espace des phases d'un \'{e}tat
comprim\'{e} selon la quadrature $\alpha _{2}$}
\label{Fig_2wigsq}
\end{figure}
\begin{equation}
\delta \varphi ^{out}\approx \delta \varphi ^{in}+\frac{8{\cal F}}{\lambda }%
~\delta x  \label{2.33}
\end{equation}
Le spectre $S_{\varphi }^{out}$ des fluctuations de phase du faisceau
r\'{e}fl\'{e}chi reproduit donc le spectre de position $S_{x}$ du miroir
mobile. La sensibilit\'{e} est limit\'{e}e par le bruit propre $S_{\varphi
}^{in}$ du faisceau incident. Pour un faisceau coh\'{e}rent ce bruit est
inversement proportionnel \`{a} l'intensit\'{e} moyenne (relation similaire
\`{a} l'\'{e}quation \ref{2.31}): 
\begin{equation}
S_{\varphi }^{in}\left[ \Omega \right] =\frac{1}{4\overline{I}^{in}}
\label{2.34}
\end{equation}
On trouve ainsi le plus petit d\'{e}placement mesurable $\delta x_{\min }$\ :

\begin{equation}
\delta x_{\min }\approx \frac{\lambda }{8{\cal F}}~\sqrt{S_{\varphi }^{in}}%
\approx \frac{\lambda }{16{\cal F}}\frac{1}{\sqrt{\overline{I}^{in}}}
\label{2.35}
\end{equation}
Pour une cavit\'{e} de finesse $3~10^{5}$ et une puissance incidente de $%
3~mW $ ($\overline{I}^{in}=2~10^{16}$ photons/s), on obtient une
sensibilit\'{e} $\delta x_{\min }$ \'{e}gale \`{a} $10^{-21}m/\sqrt{Hz}$.
L'unit\'{e} en $m/\sqrt{Hz}$ est celle utilis\'{e}e pour d\'{e}crire une
amplitude de bruit de position. Cette amplitude correspond en effet \`{a} la
racine carr\'{e}e de la puissance de bruit $S_{x}$, qui s'exprime en $%
m^{2}/Hz$ (puissance de bruit par bande spectrale d'analyse).

Cette sensibilit\'{e} est tout \`{a} fait remarquable puisqu'elle est
comparable, ou m\^{e}me meilleure, que celle des dispositifs les plus
sensibles \`{a} l'heure actuelle. L'essentiel des efforts concernant la
mesure de petits d\'{e}placements se concentre aujourd'hui autour des
projets de d\'{e}tection des ondes gravitationnelles, qu'il s'agisse des
projets optiques (projets VIRGO\cite{Bradaschia virgo 90}, LIGO\cite
{Abramovici science 92}), ou m\'{e}caniques (barres de Weber\cite{Weber 1960}%
).

La d\'{e}tection optique des ondes gravitationnelles est bas\'{e}e sur
l'utilisation d'un interf\'{e}rom\`{e}tre de Michelson. Chaque bras de
l'interf\'{e}rom\`{e}tre, d'environ $3~km$ de long, est constitu\'{e} d'une
cavit\'{e} Fabry-Perot de finesse $50$. Gr\^{a}ce \`{a} un dispositif de
recyclage de la lumi\`{e}re, la puissance lumineuse envoy\'{e}e dans
l'interf\'{e}rom\`{e}tre est de l'ordre de $1000~Watts$. Le passage d'une
onde gravitationnelle se traduit par une variation apparente de la longueur
des bras, \'{e}quivalente \`{a} un d\'{e}placement des miroirs des
cavit\'{e}s Fabry-Perot, qui induit un d\'{e}filement des franges
d'interf\'{e}rence \`{a} la sortie de l'appareil.

\begin{figure}[tbp]
\centerline{\psfig{figure=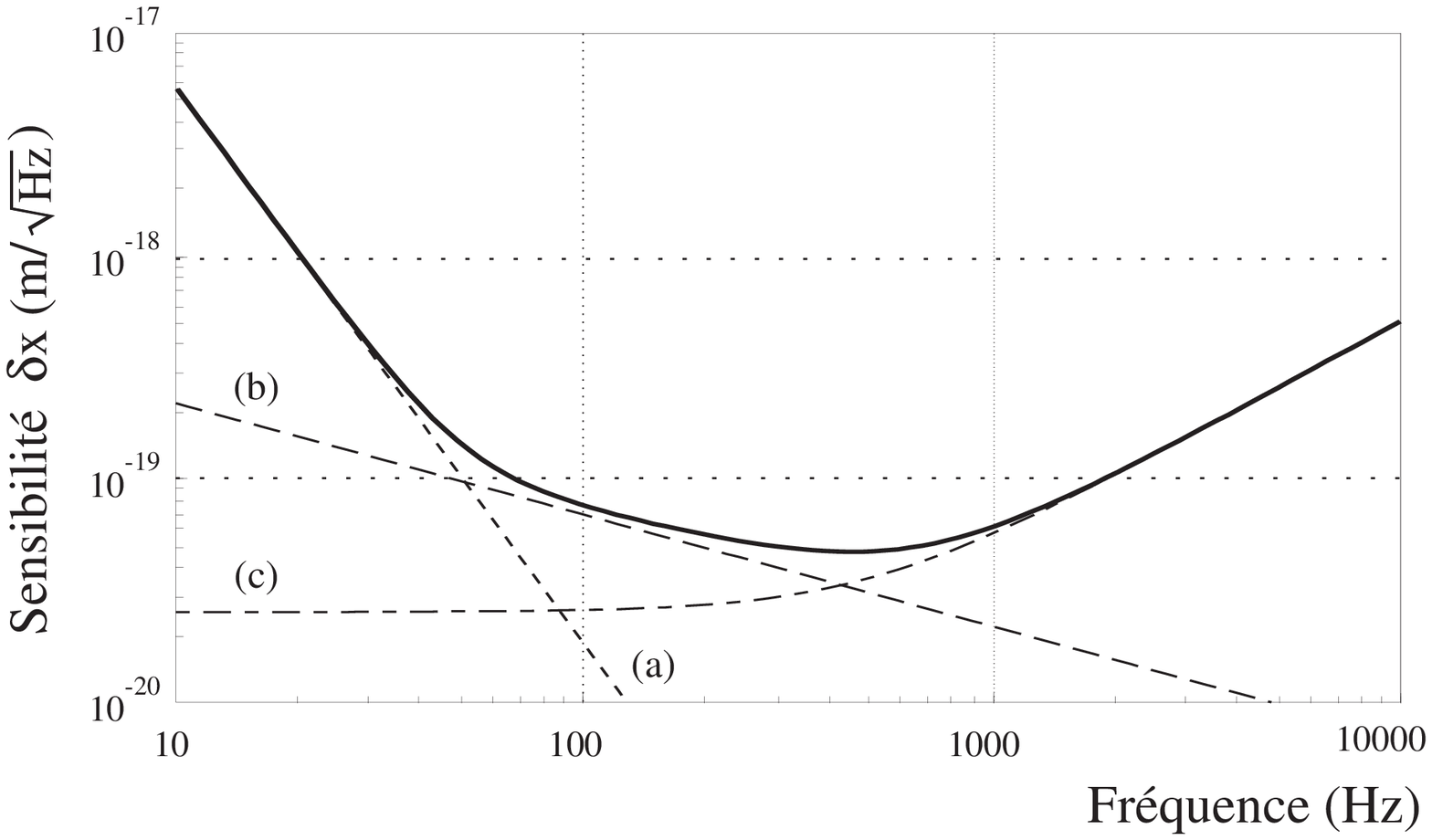,height=8cm}}
\caption{Sensibilit\'{e} de l'interf\'{e}rom\`{e}tre VIRGO pour une
variation $\delta x$ de la longueur des bras. La limite de sensibilit\'{e}
est essentiellement due au bruit thermique des suspensions des miroirs (a),
au bruit thermique de vibration des modes internes des miroirs (b) et au
bruit de photon (c)}
\label{Fig_2senvirg}
\end{figure}

La sensibilit\'{e} pr\'{e}vue pour VIRGO est indiqu\'{e}e sur la figure \ref
{Fig_2senvirg}\cite{Bondu thèse 1996}. On distingue essentiellement trois
sources de bruit. A basse fr\'{e}quence, la sensibilit\'{e} est limit\'{e}e
par le bruit thermique des suspensions des miroirs. Entre $50$ et $400~Hz$,
le bruit thermique des modes de vibration interne des miroirs devient
dominant. A haute fr\'{e}quence la limite est fix\'{e}e par le bruit de
photon, filtr\'{e} par la bande passante de $1~kHz$ des cavit\'{e}s
Fabry-Perot. La figure \ref{Fig_2senvirg} montre que la sensibilit\'{e} de
l'interf\'{e}rom\`{e}tre est au mieux de l'ordre de $5~10^{-20}m/\sqrt{Hz}$. 

La sensibilit\'{e} atteinte avec une cavit\'{e} de grande finesse peut aussi
\^{e}tre compar\'{e}e \`{a} celle obtenue par les dispositifs capacitifs
plac\'{e}s sur les barres de Weber. Une barre de Weber est constitu\'{e}e
d'un corps massif, g\'{e}n\'{e}ralement de forme cylindrique. Le
mat\'{e}riau utilis\'{e} peut \^{e}tre du niobium, du saphir, de l'aluminium
ou encore de la silice, et le poids de la barre peut atteindre plusieurs
tonnes. Elle est soigneusement isol\'{e}e du monde ext\'{e}rieur\ : elle est
suspendue par des fils dans le vide, \`{a} l'int\'{e}rieur d'un cryostat
\`{a} tr\`{e}s basse temp\'{e}rature (jusqu'\`{a} $100~mK$). Le passage
d'une onde gravitationnelle se traduit par une excitation d'un mode de
vibration m\'{e}canique de la barre. L'amplitude du mouvement attendu est
extr\^{e}mement faible. Le transducteur utilis\'{e} pour d\'{e}tecter ces
oscillations est constitu\'{e} d'un r\'{e}sonateur m\'{e}canique de faible
masse coupl\'{e} m\'{e}caniquement \`{a} la barre et capacitivement \`{a} un
circuit \'{e}lectrique r\'{e}sonnant. On atteint ainsi une sensibilit\'{e}
comprise entre $10^{-18}$ et $10^{-19}m/\sqrt{Hz}$\cite{Bocko Onofrio 96}.

Une cavit\'{e} de grande finesse devrait donc permettre de r\'{e}aliser une
mesure de d\'{e}placement avec une sensibilit\'{e} meilleure que celle des
dispositifs actuels\cite{optique weber}. Nous pr\'{e}senterons dans la suite
de cette section deux applications de cette sensibilit\'{e} extr\^{e}me: la
mesure du mouvement Brownien du miroir, et la mesure quantique non
destructive de l'intensit\'{e} de la lumi\`{e}re.

\subsubsection{Mesure du bruit thermique\label{II-2-3-1}}

Comme le montre la figure \ref{Fig_2senvirg}, le bruit thermique est la
limitation essentielle dans les dispositifs de d\'{e}tection
interf\'{e}rom\'{e}trique des ondes gravitationnelles. D'apr\`{e}s les
th\'{e}ories astrophysiques actuelles, le nombre attendu
d'\'{e}v\`{e}nements varie brutalement avec la sensibilit\'{e} atteinte par
les interf\'{e}rom\`{e}tres. Il est de ce fait important de d\'{e}terminer
avec pr\'{e}cision les diff\'{e}rentes sources de bruit pouvant limiter
cette sensibilit\'{e}.

Les m\'{e}canismes de dissipation thermique dans les solides sont cependant
mal connus. Une cavit\'{e} \`{a} miroir mobile, utilis\'{e}e \`{a}
temp\'{e}rature ambiante, devrait permettre de caract\'{e}riser l'amplitude
et la distribution spectrale du bruit thermique du miroir. Le principe de la
mesure consiste \`{a} envoyer un faisceau laser dans la cavit\'{e} \`{a}
miroir mobile, le laser \'{e}tant \`{a} r\'{e}sonance avec la cavit\'{e}.
Les fluctuations de phase du faisceau r\'{e}fl\'{e}chi refl\`{e}tent alors
les d\'{e}placements du miroir mobile, auxquels se superposent le bruit
propre du faisceau (\'{e}quation \ref{2.33}). A temp\'{e}rature ambiante, et
pour des puissances incidentes raisonnables, les effets de pression de
radiation sont n\'{e}gligeables devant le mouvement Brownien du miroir : le
d\'{e}placement $\delta x$ est donc pour l'essentiel d\^{u} au bruit
thermique du miroir mobile.

\begin{figure}[tbp]
\centerline{\psfig{figure=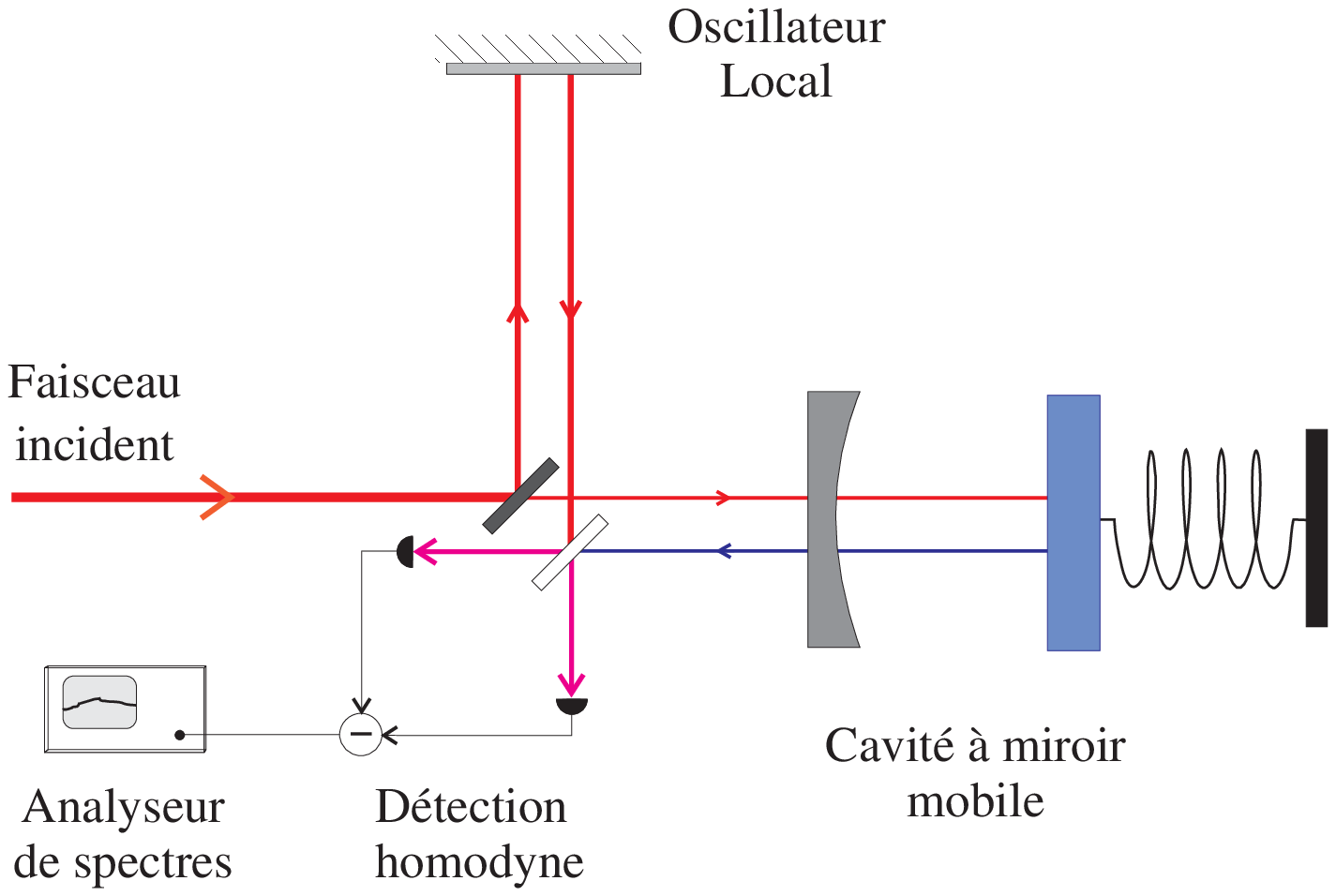,height=7cm}}
\caption{Repr\'{e}sentation sch\'{e}matique d'une d\'{e}tection homodyne. Le
spectre de bruit de phase du faisceau r\'{e}fl\'{e}chi est obtenu par
battement avec un faisceau de r\'{e}f\'{e}rence (oscillateur local), les
deux champs moyens \'{e}tant en quadrature de phase. En pratique, la
s\'{e}paration entre les faisceaux aller et retour peut \^{e}tre
r\'{e}alis\'{e}e \`{a} l'aide de cubes s\'{e}parateurs de polarisation}
\label{Fig_2prinhet}
\end{figure}

On peut mesurer exp\'{e}rimentalement le bruit du faisceau r\'{e}fl\'{e}chi
en utilisant une d\'{e}tection homodyne (figure \ref{Fig_2prinhet})\cite
{hétérodyn}. Une grande partie du faisceau incident est pr\'{e}lev\'{e}e
\`{a} l'aide d'une lame s\'{e}paratrice afin de fournir un faisceau de
r\'{e}f\'{e}rence (oscillateur local). Apr\`{e}s un aller et retour dans un
bras dont la longueur est soigneusement contr\^{o}l\'{e}e, ce faisceau est
m\'{e}lang\'{e} au niveau d'une seconde lame parfaitement
semi-r\'{e}fl\'{e}chissante avec le faisceau r\'{e}fl\'{e}chi par la
cavit\'{e}. Les faisceaux transmis et r\'{e}fl\'{e}chi par la lame sont
d\'{e}tect\'{e}s \`{a} l'aide de deux photodiodes de grande efficacit\'{e}
quantique. Lorsque la longueur du bras de l'oscillateur local est tel que le
champ r\'{e}fl\'{e}chi par la cavit\'{e} et le faisceau de r\'{e}f\'{e}rence
sont en quadrature de phase, la diff\'{e}rence des deux photocourants est
proportionnelle aux fluctuations de phase du faisceau r\'{e}fl\'{e}chi. On
obtient alors le spectre de bruit de phase \`{a} l'aide d'un analyseur de
spectres.

On peut noter la grande similitude du dispositif sch\'{e}matis\'{e} sur la
figure \ref{Fig_2prinhet} avec un interf\'{e}rom\`{e}tre de Michelson. La
diff\'{e}rence essentielle est la dissym\'{e}trie entre les deux bras de
l'interf\'{e}rom\`{e}tre. La pr\'{e}sence de la cavit\'{e} permet
d'amplifier l'effet du d\'{e}placement du miroir sur la lumi\`{e}re. D'autre
part, pour mesurer les fluctuations de la quadrature de phase du champ
r\'{e}fl\'{e}chi, il est n\'{e}cessaire que l'intensit\'{e} dans le bras de
l'oscillateur local soit grande devant celle du faisceau envoy\'{e} dans la
cavit\'{e}.

\subsubsection{Mesure quantique non destructive de l'intensit\'{e}\label%
{II-2-3-2}}

La d\'{e}tection des ondes gravitationnelles \`{a} l'aide d'une barre de
Weber est bas\'{e}e sur une mesure continue de la position du r\'{e}sonateur
m\'{e}canique. Pour d\'{e}tecter une onde gravitationnelle, il faut mesurer
la position du r\'{e}sonateur avec une grande pr\'{e}cision, pendant le
temps de passage de l'onde. Cette mesure se heurte \`{a} une limite
fondamentale impos\'{e}e par l'in\'{e}galit\'{e} de Heisenberg. En effet,
une mesure initiale de la position du r\'{e}sonateur avec une pr\'{e}cision $%
\Delta x_{0}$ perturbe in\'{e}vitablement son impulsion d'une quantit\'{e} $%
\Delta p_{0}\geq \frac{\hbar }{2\Delta x_{0}}$. Apr\`{e}s une \'{e}volution
libre pendant un temps $\tau $, la position du r\'{e}sonateur pr\'{e}sente
une dispersion $\Delta x_{1}=\tau \frac{\Delta p_{0}}{M}\geq \frac{\hbar
\tau }{2M\Delta x_{0}}$, o\`{u} $M$ est la masse de la barre. Cette
dispersion peut masquer l'effet du passage d'une onde gravitationnelle. Un
compromis entre les deux mesures conduit \`{a} une limite quantique standard 
$\Delta x_{LQS}$ pour la pr\'{e}cision d'une mesure de position du
r\'{e}sonateur : 
\begin{equation}
\Delta x_{LQS}=\sqrt{\frac{\hbar \tau }{2M}}  \label{2.36}
\end{equation}

Il est cependant possible de s'affranchir de cette limite en utilisant une
technique de mesure qui reporte tout le bruit de la mesure sur une grandeur
d\'{e}coupl\'{e}e de l'\'{e}volution libre de l'observable \`{a} mesurer\cite
{Braginsky}. On peut par exemple mesurer l'\'{e}nergie du r\'{e}sonateur
m\'{e}canique sans perturber son \'{e}volution temporelle, le bruit
associ\'{e} \`{a} la mesure \'{e}tant report\'{e} sur la phase (grandeur
conjugu\'{e}e de l'\'{e}nergie).

Ce concept de mesure quantique non destructive ({\it Quantum Non Demolition
Measurement}) peut se g\'{e}n\'{e}raliser \`{a} d'autres syst\`{e}mes que
les r\'{e}sonateurs m\'{e}caniques. La plupart des r\'{e}alisations
exp\'{e}rimentales ont d'ailleurs \'{e}t\'{e} effectu\'{e}es en optique\cite
{Leven La Porta Grang}. Deux crit\`{e}res doivent \^{e}tre satisfaits pour
qualifier une mesure de QND. La mesure ne doit tout d'abord pas alt\'{e}rer
l'observable \`{a} mesurer : l'exc\`{e}s de bruit li\'{e} \`{a} la mesure
est report\'{e} sur la composante conjugu\'{e}e. Il faut de plus que le
signal fourni par la mesure contienne le maximum d'informations sur
l'observable mesur\'{e}e, ce qui impose d'avoir de fortes corr\'{e}lations
quantiques entre l'appareil de mesure et l'observable mesur\'{e}e.

\begin{figure}[tbp]
\centerline{\psfig{figure=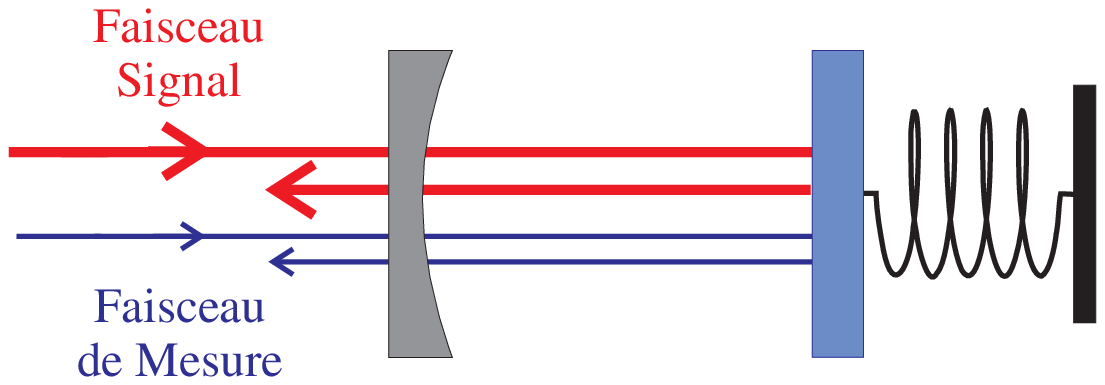,height=3cm}}
\caption{Principe d'une mesure QND de l'intensit\'{e}. L'intensit\'{e} du
faisceau signal agit sur le mouvement du miroir et modifie la phase du
faisceau de mesure}
\label{Fig_2qnd}
\end{figure}

Une cavit\'{e} \`{a} miroir mobile devrait permettre de r\'{e}aliser une
mesure QND de l'intensit\'{e} d'un faisceau lumineux\cite{QND group 95 97}.
Le principe de la mesure consiste \`{a} utiliser les corr\'{e}lations
quantiques qui existent entre l'intensit\'{e} du faisceau incident et la
position du miroir mobile de la cavit\'{e}. On envoie simultan\'{e}ment deux
faisceaux dans la cavit\'{e} comme le montre la figure \ref{Fig_2qnd}. Le
premier faisceau, le faisceau signal dont on veut mesurer l'intensit\'{e},
d\'{e}place le miroir sous l'effet de la pression de radiation. Le second
faisceau, le faisceau de mesure, est beaucoup moins intense de fa\c{c}on
\`{a} pouvoir n\'{e}gliger son influence sur le mouvement du miroir. Ce
faisceau est cependant sensible au d\'{e}placement du miroir provoqu\'{e}
par le faisceau signal. 

Pour satisfaire au premier crit\`{e}re QND, la mesure ne doit pas
alt\'{e}rer l'intensit\'{e} du signal. Ceci se produit lorsque le faisceau
signal est \`{a} r\'{e}sonance avec la cavit\'{e} : le point de
fonctionnement de la cavit\'{e} est alors au maximum du pic d'Airy, l\`{a}
o\`{u} l'intensit\'{e} du faisceau r\'{e}fl\'{e}chi n'est plus sensible
\`{a} des petites variations de longueur de la cavit\'{e}.

Le second crit\`{e}re QND impose un maximum de corr\'{e}lations quantiques
entre l'intensit\'{e} du signal et le faisceau de mesure. Pour cela, on met
\'{e}galement le faisceau de mesure \`{a} r\'{e}sonance avec la cavit\'{e},
puisque la phase du faisceau r\'{e}fl\'{e}chi est alors la plus sensible aux
variations de longueur de la cavit\'{e}. L'efficacit\'{e} de la mesure QND
est alors caract\'{e}ris\'{e}e par les corr\'{e}lations quantiques entre les
fluctuations d'intensit\'{e} du signal et les fluctuations de phase du
faisceau de mesure. Ces fluctuations de phase peuvent \^{e}tre
d\'{e}tect\'{e}es \`{a} l'aide d'un syst\`{e}me de mesure homodyne, comme
celui pr\'{e}sent\'{e} pour la mesure du bruit thermique du miroir mobile
(section \ref{II-2-3-1}). Bien s\^{u}r, l'observation de ces
corr\'{e}lations quantiques n\'{e}cessite de placer la cavit\'{e} \`{a}
basse temp\'{e}rature, de fa\c{c}on \`{a} rendre le bruit thermique
n\'{e}gligeable par rapport aux effets de pression de radiation. Une
\'{e}tude d\'{e}taill\'{e}e de ce dispositif est pr\'{e}sent\'{e}e dans la
partie 2.4.

\section{G\'{e}n\'{e}ration d'\'{e}tats comprim\'{e}s par couplage
optom\'{e}canique\label{II-3}}

\bigskip

L'existence de fluctuations quantiques est connue depuis le d\'{e}but de la
m\'{e}canique quantique. Mais ce n'est que depuis une vingtaine d'ann\'{e}es
que les physiciens se sont trouv\'{e}s confront\'{e}s aux limitations
impos\'{e}es par ces fluctuations sur la pr\'{e}cision d'une mesure. Le
d\'{e}veloppement des projets de d\'{e}tection des ondes gravitationnelles
ont amen\'{e} les physiciens \`{a} \'{e}tudier le bruit quantique et ses
cons\'{e}quences sur la sensibilit\'{e} d'une mesure. On s'est ainsi rendu
compte que le bruit quantique standard ne constitue pas une limite
fondamentale et qu'il est possible d'am\'{e}liorer la sensibilit\'{e} d'une
mesure en utilisant des \'{e}tats comprim\'{e}s\cite{Unruh Yurke Caves}.

Bien que le concept d'\'{e}tat comprim\'{e} soit tout \`{a} fait
g\'{e}n\'{e}ral, c'est dans le domaine de l'optique qu'a \'{e}t\'{e}
men\'{e}e la plupart des \'{e}tudes exp\'{e}rimentales. Les mesures optiques
pr\'{e}sentent en effet des caract\'{e}ristiques qui permettent d'atteindre
le niveau quantique plus facilement que dans d'autres domaines. Les signaux
optiques sont naturellement prot\'{e}g\'{e}s des perturbations
ext\'{e}rieures et la qualit\'{e} des dispositifs optiques permet
d'atteindre un bruit instrumental tr\`{e}s faible.

De nombreuses mises en \'{e}vidence exp\'{e}rimentales d'\'{e}tats
comprim\'{e}s ont \'{e}t\'{e} r\'{e}alis\'{e}es ces derni\`{e}res ann\'{e}es%
\cite{réf géné squeezing}. Des facteurs de r\'{e}duction du bruit quantique
de l'ordre de $90\%$ ont \'{e}t\'{e} obtenus aussi bien avec des
oscillateurs param\'{e}triques optiques \cite{Mertz Opt Lett 91} qu'avec des
diodes laser\cite{Squeez Diode 91}. La production d'un \'{e}tat comprim\'{e}
est g\'{e}n\'{e}ralement bas\'{e}e sur une interaction non lin\'{e}aire
entre la lumi\`{e}re et un milieu mat\'{e}riel. Ces processus non
lin\'{e}aires sont obtenus par m\'{e}lange \`{a} trois ou \`{a} quatre ondes
(nonlin\'{e}arit\'{e} de type $\chi ^{(2)}$ ou $\chi ^{(3)}$). Lorsqu'un
milieu non lin\'{e}aire de type $\chi ^{(3)}$ est plac\'{e} dans une
cavit\'{e}, la lumi\`{e}re subit un d\'{e}phasage non lin\'{e}aire puisque
l'indice du milieu d\'{e}pend de l'intensit\'{e} de la lumi\`{e}re\cite
{Squeez atom 85 96}. Le r\^{o}le principal jou\'{e} par le milieu est donc
de rendre la {\it longueur optique} de la cavit\'{e} d\'{e}pendante de
l'intensit\'{e} lumineuse. On peut obtenir le m\^{e}me effet en modifiant la 
{\it longueur physique} de la cavit\'{e}. Nous nous proposons dans cette
partie de montrer qu'une cavit\'{e} vide dont un miroir est mobile permet de
produire des \'{e}tats comprim\'{e}s\cite{Fabre PRA 94}. Apr\`{e}s avoir
d\'{e}crit la configuration du syst\`{e}me, nous d\'{e}terminerons les
\'{e}quations de base liant les champs entrant et sortant en fonction du
mouvement du miroir (sections \ref{II-3-1} et \ref{II-3-2}). Nous
analyserons ensuite les effets statiques (section \ref{II-3-3}) puis
dynamiques (sections \ref{II-3-4} et \ref{II-3-5}) du couplage
optom\'{e}canique sur le champ sortant de la cavit\'{e}.

\subsection{Evolution du champ dans la cavit\'{e}\label{II-3-1}}

Le syst\`{e}me consid\'{e}r\'{e} est constitu\'{e} d'une cavit\'{e}
Fabry-Perot \`{a} une seule entr\'{e}e-sortie dans laquelle est envoy\'{e}
un faisceau laser coh\'{e}rent de fr\'{e}quence $\omega _{0}$. Le miroir
d'entr\'{e}e a une tr\`{e}s grande r\'{e}flectivit\'{e} $r=1-\gamma $, avec $%
\gamma \ll 1$. Il est de plus suppos\'{e} sans perte, de sorte que sa
transmission $t$ est \'{e}gale \`{a} $\sqrt{2\gamma }$. On suppose que le
miroir mobile est totalement r\'{e}fl\'{e}chissant et sans perte. Dans ces
conditions, les pertes totales de la cavit\'{e} sont uniquement dues \`{a}
la transmission du miroir d'entr\'{e}e et la finesse de la cavit\'{e} est
d'autant plus grande que $\gamma $ est petit.

\begin{figure}[tbp]
\centerline{\psfig{figure=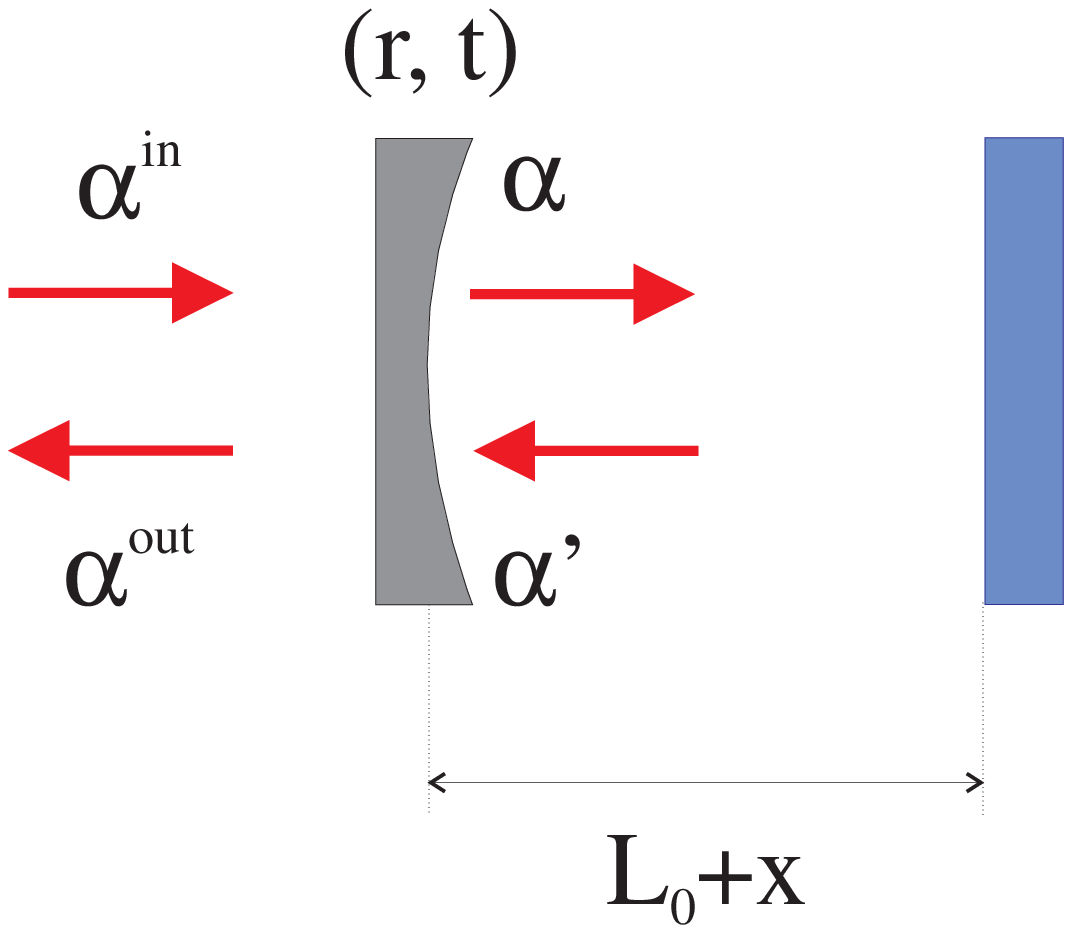,height=5cm}}
\caption{D\'{e}finition des champs dans une cavit\'{e} Fabry-Perot \`{a} une
seule entr\'{e}e-sortie }
\label{Fig_inoutcav}
\end{figure}

Nous nous limiterons dans cette partie \`{a} une description
monodimensionnelle du champ et de la cavit\'{e}. Le champ est trait\'{e}
comme une onde plane se propageant uniquement selon l'axe de la cavit\'{e}
(figure \ref{Fig_inoutcav}). 
Chaque onde lumineuse peut \^{e}tre d\'{e}crite par une amplitude complexe $%
\alpha (t)$, qui est une fonction lentement variable du temps $t$. La
cavit\'{e} est enti\`{e}rement d\'{e}termin\'{e}e par sa longueur $%
L(t)=L_{0}+x(t)$ o\`{u} $x(t)$ est le d\'{e}placement du miroir mobile.

La conservation de l'\'{e}nergie au niveau du miroir d'entr\'{e}e entra\^{\i %
}ne l'existence de relations lin\'{e}aires et unitaires entre les champs
incident $\alpha ^{in}$, r\'{e}fl\'{e}chi $\alpha ^{out}$ et les champs dans
la cavit\'{e} $\alpha $ et $\alpha ^{\prime }$ (figure \ref{Fig_inoutcav}): 
\begin{subequations}
\label{3.1}
\begin{eqnarray}
\alpha (t) &=&t\alpha ^{in}(t)+r\alpha ^{\prime }(t)  \label{3.1a} \\
\alpha ^{out}(t) &=&t\alpha ^{\prime }(t)-r\alpha ^{in}(t)  \label{3.1b}
\end{eqnarray}
La premi\`{e}re relation montre que le champ intracavit\'{e} $\alpha $ est
la somme du champ incident $\alpha ^{in}$ au m\^{e}me instant et du champ
intracavit\'{e} $\alpha ^{\prime }$ qui a effectu\'{e} un aller et retour
dans la cavit\'{e}. La seconde relation signifie que le champ sortant
r\'{e}sulte de l'interf\'{e}rence entre le champ incident directement
r\'{e}fl\'{e}chi et le champ intracavit\'{e} transmis par le miroir.

La propagation du champ dans la cavit\'{e} permet de relier le champ $\alpha
^{\prime }$ revenant sur le miroir au champ $\alpha $:

\end{subequations}
\begin{equation}
\alpha ^{\prime }(t)=\alpha (t-\tau )\ e^{i\Psi (t)}  \label{3.2}
\end{equation}
$\tau $ est le temps moyen mis par la lumi\`{e}re pour parcourir un aller et
retour dans la cavit\'{e} ($\tau =2L_{0}/c$) et $\Psi $ est le d\'{e}phasage
subi par le champ: 
\begin{equation}
\Psi (t)\equiv 2kL(t)~\left[ 2\pi \right]   \label{3.3}
\end{equation}
o\`{u} $k=\omega _{0}/c$ est le vecteur d'onde du champ. Pour \'{e}crire
l'\'{e}quation (\ref{3.2}), nous avons n\'{e}glig\'{e} les effets de retard
temporel subis par le champ\cite{Tourrenc Annal Phys 85} : pour des petits
d\'{e}placements du miroir mobile, le champ est essentiellement
d\'{e}phas\'{e} d'une quantit\'{e} $\Psi (t)$, et la variation du temps $%
\tau $ avec la position du miroir peut \^{e}tre n\'{e}glig\'{e}e.

Pour une cavit\'{e} quasi r\'{e}sonnante et de grande finesse ($\Psi (t)$, $%
\gamma \ll 1$), l'enveloppe du champ varie peu sur un tour. En combinant les
\'{e}quations (\ref{3.1}) et (\ref{3.2}), on obtient alors l'\'{e}quation
d'\'{e}volution du champ intracavit\'{e} et l'expression du champ sortant: 
\begin{subequations}
\label{3.4}
\begin{eqnarray}
\tau \frac{d\alpha (t)}{dt} &=&(-\gamma +i\Psi (t))~\alpha (t)+\sqrt{2\gamma 
}~\alpha ^{in}(t)  \label{3.4a} \\
\alpha ^{out}(t) &=&\sqrt{2\gamma }~\alpha (t)-\alpha ^{in}(t)  \label{3.4b}
\end{eqnarray}
Ces \'{e}quations sont identiques \`{a} celles obtenues pour une cavit\'{e}
usuelle \`{a} une seule entr\'{e}e-sortie, avec toutefois un d\'{e}phasage $%
\Psi (t)$ d\'{e}pendant du d\'{e}placement $x(t)$: 
\end{subequations}
\begin{equation}
\Psi (t)=\Psi _{0}+2kx(t)  \label{3.5}
\end{equation}
o\`{u} $\Psi _{0}\equiv \omega _{0}\tau ~\left[ 2\pi \right] $ est le
d\'{e}phasage entre le champ et la cavit\'{e} en absence de d\'{e}placement
du miroir.

\subsection{Mouvement du miroir mobile\label{II-3-2}}

Pour des petits d\'{e}placements, la th\'{e}orie de la r\'{e}ponse
lin\'{e}aire\cite{Cours Landau} permet de relier la transform\'{e}e de
Fourier $x\left[ \Omega \right] $ du d\'{e}placement \`{a} la force
appliqu\'{e}e: 
\begin{equation}
x\left[ \Omega \right] =\chi \left[ \Omega \right] (F_{rad}\left[ \Omega
\right] +F_{T}\left[ \Omega \right] )  \label{2.42}
\end{equation}
o\`{u} $\chi \left[ \Omega \right] $ est la susceptibilit\'{e} m\'{e}canique
du miroir. La force appliqu\'{e}e est la somme de la force $F_{rad}$ due
\`{a} la pression de radiation et de la force de Langevin $F_{T}$
d\'{e}crivant le couplage du miroir avec un bain thermique. Comme nous
l'avons vu dans la partie (2.1), la force de pression de radiation est
proportionnelle \`{a} l'intensit\'{e} du champ: 
\begin{equation}
F_{rad}(t)=2\hbar k~I(t)=2\hbar k\left| \alpha (t)\right| ^{2}  \label{2.43}
\end{equation}
La force de Langevin $F_{T}$ a une valeur moyenne nulle et son spectre de
bruit $S_{T}$ est reli\'{e} \`{a} la susceptibilit\'{e} m\'{e}canique par le
th\'{e}or\`{e}me fluctuation-dissipation\cite{Cours Landau}: 
\begin{equation}
S_{T}\left[ \Omega \right] =-\frac{2k_{B}T}{\Omega }{\cal I}m\left( \frac{1}{%
\chi \left[ \Omega \right] }\right)  \label{2.44}
\end{equation}
o\`{u} $T$ est la temp\'{e}rature du bain thermique et $k_{B}$ la constante
de Boltzmann.

\subsection{Etat stationnaire et bistabilit\'{e}\label{II-3-3}}

Les valeurs moyennes des champs et du d\'{e}placement du miroir mobile
s'obtiennent en consid\'{e}rant le r\'{e}gime stationnaire dans les
\'{e}quations d'\'{e}volution pr\'{e}c\'{e}dentes. Ainsi le d\'{e}placement
moyen $\overline{x}$ est donn\'{e} par l'\'{e}quation (\ref{2.42}) \`{a}
fr\'{e}quence nulle : 
\begin{equation}
\overline{x}=2\hbar k\chi \left[ 0\right] \overline{I}  \label{2.45}
\end{equation}
o\`{u} $\overline{I}=\left| \overline{\alpha }\right| ^{2}$ est
l'intensit\'{e} moyenne du champ intracavit\'{e}. Le d\'{e}phasage moyen $%
\overline{\Psi }$ du champ dans la cavit\'{e} est alors donn\'{e} par
l'\'{e}quation (\ref{3.5}): 
\begin{equation}
\overline{\Psi }=\Psi _{0}+\Psi _{NL}  \label{2.46}
\end{equation}
o\`{u} $\Psi _{NL}=2k\overline{x}$ est le d\'{e}phasage non lin\'{e}aire
li\'{e} au d\'{e}placement du miroir sous l'effet de la pression de
radiation moyenne: 
\begin{equation}
\Psi _{NL}=4\hbar k^{2}\chi \left[ 0\right] \overline{I}  \label{2.47}
\end{equation}
Ces relations montrent que l'effet de la pression de radiation moyenne
correspond \`{a} un effet Kerr, comme celui produit par un milieu non
lin\'{e}aire de type $\chi ^{(3)}$ plac\'{e} dans une cavit\'{e} : le champ
dans la cavit\'{e} subit un d\'{e}phasage proportionnel \`{a} son
intensit\'{e} moyenne $\overline{I}$. Comme nous le verrons par la suite, le
d\'{e}phasage non lin\'{e}aire $\Psi _{NL}$ est le param\`{e}tre essentiel
pour d\'{e}crire l'amplitude des effets optom\'{e}caniques : le couplage
optom\'{e}canique agit de fa\c{c}on appr\'{e}ciable sur les fluctuations
quantiques lorsque le d\'{e}phasage non lin\'{e}aire est de l'ordre des
pertes $2\gamma $ de la cavit\'{e}.

La valeur moyenne des champs s'obtient en posant $\frac{d\alpha (t)}{dt}=0$
dans l'\'{e}quation (\ref{3.4a}): 
\begin{subequations}
\label{2.48}
\begin{eqnarray}
\overline{\alpha } &=&\frac{\sqrt{2\gamma }}{\gamma -i\overline{\Psi }}~%
\overline{\alpha }^{in}  \label{2.48a} \\
\overline{\alpha }^{out} &=&\frac{\gamma +i\overline{\Psi }}{\gamma -i%
\overline{\Psi }}~\overline{\alpha }^{in}  \label{2.48b}
\end{eqnarray}
\begin{figure}[tbp]
\centerline{\psfig{figure=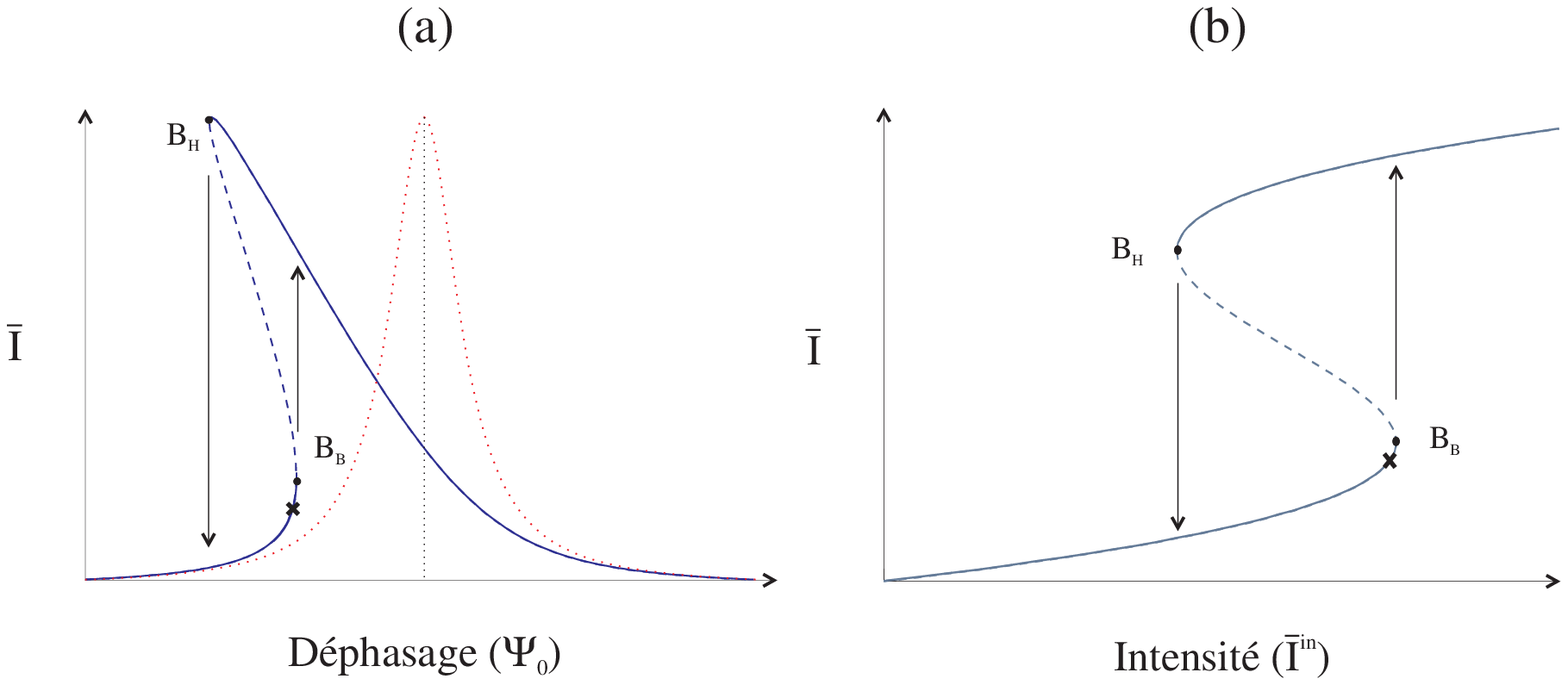,height=7cm}}
\caption{Comportement bistable de l'intensit\'{e} dans une cavit\'{e} \`{a}
miroir mobile, en fonction du d\'{e}saccord en fr\'{e}quence entre le laser
et la cavit\'{e} (a) ou de l'intensit\'{e} incidente (b). La partie en
traits tiret\'{e}s repr\'{e}sente la zone instable de la courbe de
bistabilit\'{e} et les fl\`{e}ches indiquent le cycle d'hyst\'{e}r\'{e}sis.
Les croix sur les deux figures repr\'{e}sentent le point de fonctionnement
pour lequel sont calcul\'{e}s les spectres de bruit (figures \ref
{Fig_2int_opt} et \ref{Fig_2specint})}
\label{Fig_2bist_fr}
\end{figure}
Si l'on oublie la variation du d\'{e}phasage total $\overline{\Psi }$ avec
l'intensit\'{e} intracavit\'{e}, on retrouve ici les expressions usuelles
des champs pour une cavit\'{e} Fabry-Perot \`{a} une seule
entr\'{e}e-sortie. La cavit\'{e} \'{e}tant suppos\'{e}e sans perte, le champ
r\'{e}fl\'{e}chi a la m\^{e}me intensit\'{e} que le champ incident
(\'{e}quation \ref{2.48b}). Le champ subit simplement un d\'{e}phasage qui
se traduit par une rotation dans l'espace des phases, avec un angle qui
d\'{e}pend de $\overline{\Psi }$. L'intensit\'{e} intracavit\'{e} $\overline{%
I}$ est donn\'{e}e par le module carr\'{e} de l'\'{e}quation (\ref{2.48a}).
Il appara\^{\i }t une r\'{e}sonance Lorentzienne autour de $\overline{\Psi }%
=0$. La largeur de ce pic d'Airy est \'{e}gale \`{a} $2\gamma $ et
l'intensit\'{e} intracavit\'{e} $\overline{I}$ \`{a} r\'{e}sonance est
\'{e}gale \`{a} l'intensit\'{e} incidente $\overline{I}^{in}$, amplifi\'{e}e
par un facteur $2/\gamma $. On d\'{e}duit de ces r\'{e}sultats l'expression
de la finesse ${\cal F}$ de la cavit\'{e}\ : 
\end{subequations}
\begin{equation}
{\cal F}=\frac{\pi }{\gamma }  \label{2.49}
\end{equation}
Notons enfin que tous les champs sont d\'{e}finis \`{a} une phase globale
pr\`{e}s. Dans toute la suite, on choisira cette phase de telle mani\`{e}re
que le champ intracavit\'{e} $\overline{\alpha }$ soit r\'{e}el. Si l'on
tient compte de la variation du d\'{e}phasage $\overline{\Psi }$ avec
l'intensit\'{e} intracavit\'{e}, la pr\'{e}sence du d\'{e}phasage non
lin\'{e}aire $\Psi _{NL}$ entra\^{\i }ne une d\'{e}formation du pic d'Airy.
La figure \ref{Fig_2bist_fr}a montre comment varie l'intensit\'{e}
intracavit\'{e} $\overline{I}$ lorsque le d\'{e}phasage $\Psi _{0}$ est
balay\'{e}, par exemple en modifiant la fr\'{e}quence optique $\omega _{0}$
du faisceau incident. Cet effet est responsable du comportement bistable de
la cavit\'{e}\cite{Lugiato prog opt 84} : pour certaines valeurs du
d\'{e}phasage, il existe plusieurs solutions pour l'intensit\'{e}
intracavit\'{e}.

Le ph\'{e}nom\`{e}ne de bistabilit\'{e} peut se comprendre de la fa\c{c}on
suivante. Lorsqu'on balaye la fr\'{e}quence du faisceau incident de
mani\`{e}re \`{a} augmenter $\Psi _{0}$, l'intensit\'{e} suit la branche
basse de la r\'{e}sonance jusqu'\`{a} atteindre le point tournant $B_{B}$. A
partir de ce point, l'intensit\'{e} augmente brusquement et passe sur la
branche haute. Lorsqu'on balaye $\Psi _{0}$ dans le sens inverse,
l'intensit\'{e} suit d'abord la branche haute de la r\'{e}sonance
jusqu'\`{a} atteindre le second point tournant $B_{H}$, \`{a} partir duquel
l'intensit\'{e} chute brusquement pour passer sur la branche basse. On
obtient ainsi un cycle d'hyst\'{e}r\'{e}sis caract\'{e}ristique des
syst\`{e}mes bistables.

On peut aussi observer la bistabilit\'{e} en faisant varier l'intensit\'{e}
incidente $\overline{I}^{in}$, pour un d\'{e}phasage $\Psi _{0}$ fix\'{e}
(figure \ref{Fig_2bist_fr}b). L'intensit\'{e} $\overline{I}$ est en effet
solution d'une \'{e}quation du troisi\`{e}me degr\'{e} que l'on obtient
\`{a} partir des \'{e}quations (\ref{2.46}) \`{a} (\ref{2.48a}) : 
\begin{equation}
\overline{I}~[\gamma ^{2}+(\Psi _{0}+4\hbar k^{2}\chi \left[ 0\right] 
\overline{I})^{2}]=2\gamma \overline{I}^{in}  \label{2.50}
\end{equation}
\begin{figure}[tbp]
\centerline{\psfig{figure=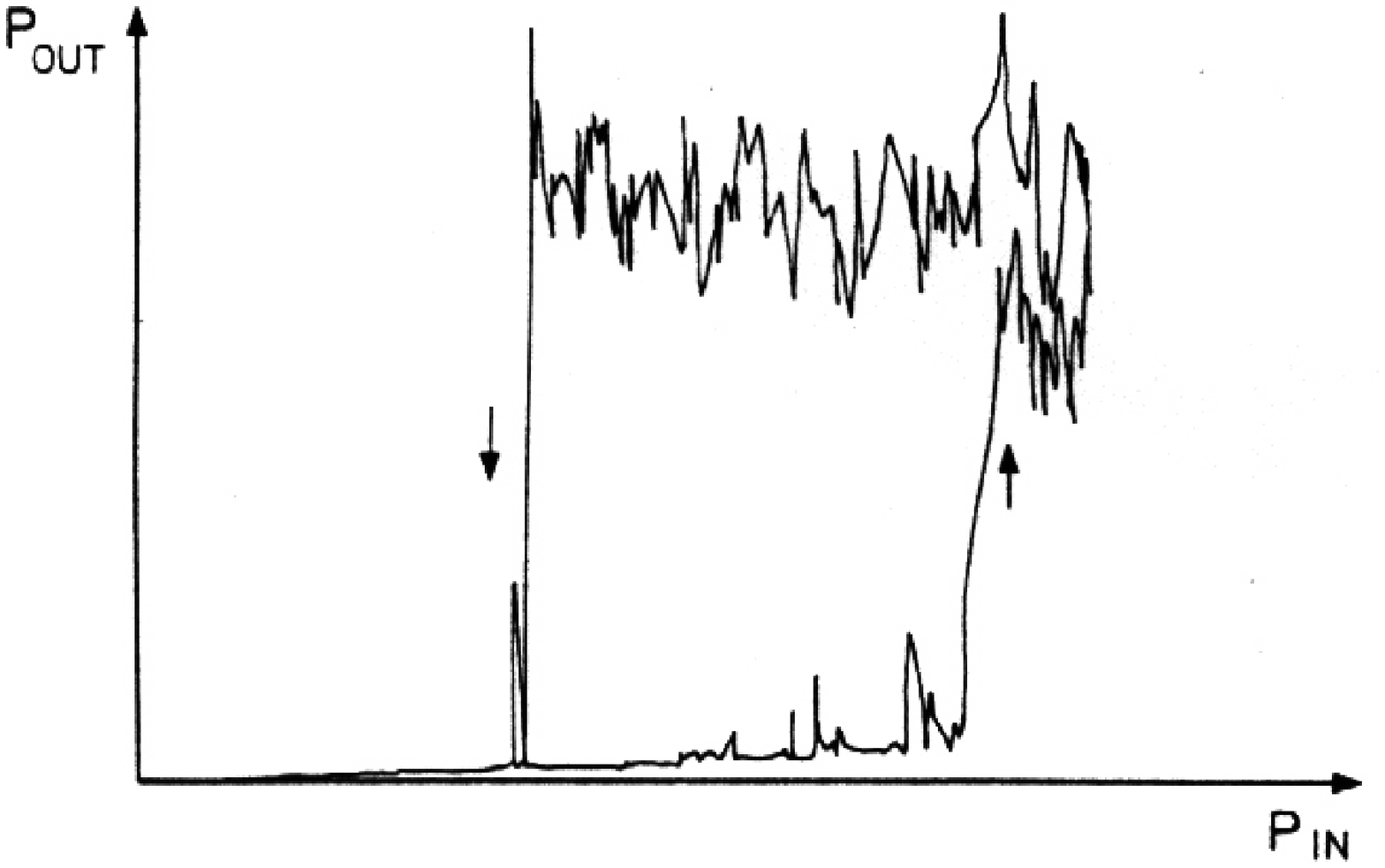,height=5cm}}
\caption{Cycle d'hyst\'{e}r\'{e}sis observ\'{e} par {\it Dorsel et al}%
\protect\cite{Dorsel prl 1983} en mesurant la puissance intracavit\'{e} en
fonction de la puissance incidente}
\label{Fig_2dorsel}
\end{figure}
Cette relation montre qu'il peut exister, pour une valeur donn\'{e}e de
l'intensit\'{e} incidente, trois solutions stationnaires de l'intensit\'{e}
intracavit\'{e}. L'une des solutions se trouve sur la branche instable qui
relie les deux points tournants de la courbe de bistabilit\'{e}\cite{Lugiato
prog opt 84}. Un raisonnement similaire \`{a} celui fait pour la figure (\ref
{Fig_2bist_fr}a), montre que le syst\`{e}me parcourt un cycle
d'hyst\'{e}r\'{e}sis lorsqu'on fait varier l'intensit\'{e} incidente. Cette
bistabilit\'{e} d'origine m\'{e}canique a d\'{e}j\`{a} \'{e}t\'{e}
observ\'{e}e exp\'{e}rimentalement\cite{Dorsel prl 1983}. Le montage
utilis\'{e} dans cette exp\'{e}rience \'{e}tait constitu\'{e} d'une
cavit\'{e} de finesse $15$ dont l'un des miroirs es d\'{e}pos\'{e} sur une
plaque en quartz, pesant $60~mg$ et suspendue par deux fils en
tungst\`{e}ne. La fr\'{e}quence de r\'{e}sonance m\'{e}canique de ce pendule
est de l'ordre de quelques Hertz. La figure \ref{Fig_2dorsel} montre le
cycle d'hyst\'{e}r\'{e}sis observ\'{e} exp\'{e}rimentalement, en faisant
varier la puissance incidente $P_{in}$. L'axe vertical repr\'{e}sente la
puissance $P_{out}$ de la lumi\`{e}re r\'{e}siduelle transmise par le second
miroir de la cavit\'{e}. Cette puissance est directement proportionnelle
\`{a} l'intensit\'{e} intracavit\'{e}. Dans cette exp\'{e}rience, les points
tournants de la bistabilit\'{e} correspondent \`{a} des puissances
incidentes de $1.1$ et $2.2~W$.

\subsection{Evolution des fluctuations quantiques\label{II-3-4}}

Pour \'{e}tudier les fluctuations quantiques, nous utilisons la m\'{e}thode
semi-classique dont les principes ont \'{e}t\'{e} expos\'{e}s dans la partie
2.2\cite{Reynaud Fabre Ekert}. Elle consiste \`{a} \'{e}crire le champ sous
la forme $\alpha (t)=\overline{\alpha }+\delta \alpha (t)$ o\`{u} $\overline{%
\alpha }$ est la valeur moyenne du champ et $\delta \alpha (t)$ est une
variable al\'{e}atoire classique, associ\'{e}e aux fluctuations quantiques
par l'interm\'{e}diaire de la distribution de Wigner. Plus
pr\'{e}cis\'{e}ment, la distribution de Wigner pour un champ d\'{e}pendant
du temps est une fonction d'un ensemble de variables $\alpha \left[ \Omega
\right] $ et $\alpha ^{*}\left[ \Omega \right] $, qui sont les
transform\'{e}es de Fourier des variables $\alpha (t)$ et $\alpha ^{*}(t)$: 
\begin{subequations}
\label{2.51}
\begin{eqnarray}
\alpha \left[ \Omega \right] &=&\int_{-\infty }^{+\infty }\alpha \left(
t\right) ~e^{i\Omega t}~dt  \label{2.51a} \\
\alpha ^{*}\left[ \Omega \right] &=&\int_{-\infty }^{+\infty }\alpha
^{*}\left( t\right) ~e^{i\Omega t}~dt  \label{2.51b}
\end{eqnarray}
Notons que $\alpha ^{*}\left[ \Omega \right] $ n'est pas le complexe
conjugu\'{e} de $\alpha \left[ \Omega \right] $, puisque les deux variables
\'{e}voluent \`{a} la m\^{e}me fr\'{e}quence $\Omega $. On a en fait la
relation: 
\end{subequations}
\begin{equation}
\alpha ^{*}\left[ \Omega \right] =\left( \alpha \left[ -\Omega \right]
\right) ^{*}  \label{2.52}
\end{equation}

La distribution de Wigner pour un \'{e}tat coh\'{e}rent est une Gaussienne
dont la largeur est d\'{e}finie par les valeurs moyennes\cite{Reynaud Fabre
Ekert}: 
\begin{equation}
\begin{array}{l}
\left\langle \delta \alpha \left[ \Omega \right] ~\delta \alpha \left[
\Omega ^{\prime }\right] \right\rangle =\left\langle \delta \alpha
^{*}\left[ \Omega \right] ~\delta \alpha ^{*}\left[ \Omega ^{\prime }\right]
\right\rangle =0 \\ 
\left\langle \delta \alpha \left[ \Omega \right] ~\delta \alpha ^{*}\left[
\Omega ^{\prime }\right] \right\rangle =\pi ~\delta \left( \Omega +\Omega
^{\prime }\right) 
\end{array}
\label{2.53}
\end{equation}
Les principes de la m\'{e}thode semi-classique pr\'{e}sent\'{e}s dans la
partie 2.2 se g\'{e}n\'{e}ralisent alors pour un champ d\'{e}pendant du
temps. Une quadrature quelconque du champ est d\'{e}finie par ses
composantes spectrales $\alpha _{\theta }\left[ \Omega \right] $,
reli\'{e}es aux variables $\alpha $ et $\alpha ^{*}$ par une relation
similaire \`{a} (\ref{2.28}): 
\begin{equation}
\alpha _{\theta }\left[ \Omega \right] =e^{-i\theta }~\alpha \left[ \Omega
\right] +e^{i\theta }~\alpha ^{*}\left[ \Omega \right]   \label{2.54}
\end{equation}
Le bruit de cette quadrature n'est plus d\'{e}fini par une variance mais par
un spectre $S_{\theta }\left[ \Omega \right] $: 
\begin{equation}
\left\langle \delta \alpha _{\theta }\left[ \Omega \right] ~\delta \alpha
_{\theta }\left[ \Omega ^{\prime }\right] \right\rangle =2\pi ~\delta \left(
\Omega +\Omega ^{\prime }\right) ~S_{\theta }\left[ \Omega \right] 
\label{2.55}
\end{equation}
A partir des relations (\ref{2.53}), on retrouve que pour un \'{e}tat
coh\'{e}rent, le spectre $S_{\theta }\left[ \Omega \right] $ est \'{e}gal
\`{a} $1$ quelque soit l'angle $\theta $.

Les bruits d'intensit\'{e} et de phase sont toujours reli\'{e}s aux
quadratures d'amplitude $\alpha _{\bar{\varphi}}$ et de phase $\alpha _{\bar{%
\varphi}+\pi /2}$, o\`{u} $\bar{\varphi}$ est la phase du champ moyen: 
\begin{subequations}
\label{2.56}
\begin{eqnarray}
\delta I\left[ \Omega \right] &=&\left| \overline{\alpha }\right| ~\delta
\alpha _{\bar{\varphi}}\left[ \Omega \right]  \label{2.56a} \\
\delta \varphi \left[ \Omega \right] &=&\frac{1}{2\left| \overline{\alpha }%
\right| }~\delta \alpha _{\bar{\varphi}+\pi /2}\left[ \Omega \right]
\label{2.56b}
\end{eqnarray}
Pour un \'{e}tat coh\'{e}rent, ces bruits sont ind\'{e}pendants de la
fr\'{e}quence et ne d\'{e}pendent que de l'intensit\'{e} moyenne $\overline{I%
}$: 
\end{subequations}
\begin{equation}
S_{I}\left[ \Omega \right] =\overline{I}\qquad ,\qquad S_{\varphi }\left[
\Omega \right] =\frac{1}{4\overline{I}}  \label{2.57}
\end{equation}

Ces diff\'{e}rentes relations permettent d'associer une repr\'{e}sentation
semi-classique aux fluctuations quantiques. Comme dans le cas d'un champ
ind\'{e}pendant du temps, l'\'{e}volution de ces fluctuations depuis
l'entr\'{e}e jusqu'\`{a} la sortie de la cavit\'{e} est obtenue en
lin\'{e}arisant les \'{e}quations classiques autour du point de
fonctionnement moyen. A partir des \'{e}quations (\ref{3.4}) \`{a} (\ref
{2.42}), on trouve pour la cavit\'{e} \`{a} miroir mobile:

\begin{subequations}
\label{2.58}
\begin{eqnarray}
(\gamma -i\overline{\Psi }-i\Omega \tau )~\delta \alpha \left[ \Omega
\right] &=&\sqrt{2\gamma }~\delta \alpha ^{in}\left[ \Omega \right] +i%
\overline{\alpha }~\delta \Psi \left[ \Omega \right]  \label{2.58a} \\
\delta \alpha ^{out}\left[ \Omega \right] &=&\sqrt{2\gamma }~\delta \alpha
\left[ \Omega \right] -\delta \alpha ^{in}\left[ \Omega \right]
\label{2.58b} \\
\overline{\alpha }~\delta \Psi \left[ \Omega \right] &=&\overline{\chi }%
\left[ \Omega \right] \Psi _{NL}~(\delta \alpha \left[ \Omega \right]
+\delta \alpha ^{*}\left[ \Omega \right] )+2k\overline{\alpha }\chi \left[
\Omega \right] ~F_{T}\left[ \Omega \right]  \label{2.58c}
\end{eqnarray}
o\`{u} $\overline{\chi }\left[ \Omega \right] =\chi \left[ \Omega \right]
/\chi \left[ 0\right] $ est la susceptibilit\'{e} normalis\'{e}e \`{a} $1$
\`{a} fr\'{e}quence nulle. On peut \'{e}liminer les variables $\delta \alpha
\left[ \Omega \right] $ et $\delta \Psi \left[ \Omega \right] $ dans ce
syst\`{e}me d'\'{e}quations. On obtient alors une relation
d'entr\'{e}e-sortie pour les fluctuations, qui donne les fluctuations
sortantes $\delta \alpha ^{out}\left[ \Omega \right] $ en fonction des
fluctuations entrantes $\delta \alpha ^{in}\left[ \Omega \right] $ et $%
F_{T}\left[ \Omega \right] $: 
\end{subequations}
\begin{equation}
\delta \alpha ^{out}\left[ \Omega \right] =c_{1}\left[ \Omega \right] \delta
\alpha ^{in}\left[ \Omega \right] +c_{2}\left[ \Omega \right] \delta \alpha
^{in*}\left[ \Omega \right] +c_{T}\left[ \Omega \right] F_{T}\left[ \Omega
\right]  \label{2.59}
\end{equation}
o\`{u} les coefficients $c_{1}\left[ \Omega \right] $, $c_{2}\left[ \Omega
\right] $ et $c_{T}\left[ \Omega \right] $ d\'{e}pendent des param\`{e}tres
du syst\`{e}me$:$%
\begin{subequations}
\label{2.60}
\begin{eqnarray}
c_{1}\left[ \Omega \right] &=&\frac{1}{\Delta }\left\{ \left( \Omega \tau
\right) ^{2}+\left( \gamma +i\overline{\Psi }\right) \left( \gamma +i%
\overline{\Psi }+2i\Psi _{NL}~\overline{\chi }\left[ \Omega \right] \right)
\right\}  \label{2.60a} \\
c_{2}\left[ \Omega \right] &=&\frac{2i}{\Delta }\gamma \Psi _{NL}~\overline{%
\chi }\left[ \Omega \right]  \label{2.60b} \\
c_{T}\left[ \Omega \right] &=&\frac{2i}{\Delta }\sqrt{2\gamma }k\overline{%
\alpha }\left( \gamma +i\overline{\Psi }-i\Omega \tau \right) ~\chi \left[
\Omega \right]  \label{2.60c}
\end{eqnarray}
avec: 
\end{subequations}
\begin{equation}
\Delta =\left( \gamma -i\Omega \tau \right) ^{2}+\overline{\Psi }^{2}+2%
\overline{\Psi }\,\Psi _{NL}~\overline{\chi }\left[ \Omega \right]
\label{2.61}
\end{equation}

\subsection{Spectre de bruit quantique\label{II-3-5}}

La relation d'entr\'{e}e-sortie des fluctuations \'{e}tablie dans la section
pr\'{e}c\'{e}dente (\'{e}quation \ref{2.59}) permet de relier le spectre $%
S_{\theta }^{out}\left[ \Omega \right] $ pour n'importe quelle quadrature du
champ r\'{e}fl\'{e}chi aux fluctuations entrantes $\delta \alpha ^{in}$ et $%
F_{T}$. Ces fluctuations \'{e}tant ind\'{e}pendantes, tous les termes
crois\'{e}s du type $\left\langle \delta \alpha ^{in}\left[ \Omega \right]
\delta F_{T}[\Omega ^{\prime }]\right\rangle $ sont nuls. On suppose d'autre
part que le champ incident est dans un \'{e}tat coh\'{e}rent. Les
fluctuations incidentes sont alors caract\'{e}ris\'{e}es par les fonctions
de corr\'{e}lations d'ordre deux donn\'{e}es par les \'{e}quations (\ref
{2.53}). Enfin, la force de Langevin $F_{T}$ est caract\'{e}ris\'{e}e par le
spectre de bruit thermique $S_{T}\left[ \Omega \right] $ (\'{e}quation \ref
{2.44}). On obtient alors: 
\begin{equation}
S_{\theta }^{out}\left[ \Omega \right] =\frac{1}{2}\left( {\cal C}%
^{in}\left[ \Omega \right] +{\cal C}^{in}\left[ -\Omega \right] \right) +%
{\cal C}_{T}\left[ \Omega \right] S_{T}\left[ \Omega \right]   \label{2.62}
\end{equation}
o\`{u} les quantit\'{e}s ${\cal C}^{in}\left[ \Omega \right] $ et ${\cal C}%
_{T}\left[ \Omega \right] $ s'expriment en fonction des coefficients $c_{1}$%
, $c_{2}$, et $c_{T}~:$

\begin{subequations}
\label{2.63}
\begin{eqnarray}
{\cal C}^{in}\left[ \Omega \right] &=&\left| e^{-i\theta }c_{1}\left[ \Omega
\right] +e^{i\theta }c_{2}^{*}\left[ -\Omega \right] \right| ^{2}
\label{2.63a} \\
{\cal C}_{T}\left[ \Omega \right] &=&\left| e^{-i\theta }c_{T}\left[ \Omega
\right] +e^{i\theta }c_{T}^{*}\left[ -\Omega \right] \right| ^{2}
\label{2.63b}
\end{eqnarray}
Cette expression permet de d\'{e}terminer le spectre de n'importe quelle
quadrature du champ sortant. Nous allons nous int\'{e}resser plus
particuli\`{e}rement au bruit d'intensit\'{e} et au spectre de bruit
optimum. Ce dernier correspond au spectre $S_{opt}\left[ \Omega \right] $
obtenu en choisissant, \`{a} chaque fr\'{e}quence $\Omega $, la quadrature $%
\alpha _{\theta }$ qui a le bruit minimum.

Les fluctuations d'intensit\'{e} sont reli\'{e}es \`{a} la quadrature
d'amplitude $\alpha _{\bar{\varphi}}$ (\'{e}quation \ref{2.56a}): 
\end{subequations}
\begin{equation}
S_{I}^{out}\left[ \Omega \right] =\overline{I}^{out}~S_{\bar{\varphi}%
}^{out}\left[ \Omega \right]  \label{2.64}
\end{equation}
o\`{u} $\overline{\varphi }$ est la phase moyenne du champ sortant,
donn\'{e}e par:

\begin{equation}
e^{i\overline{\varphi }}=\frac{\gamma +i\overline{\Psi }}{\sqrt{\gamma ^{2}+%
\overline{\Psi }^{2}}}  \label{2.65}
\end{equation}
A partir des \'{e}quations (\ref{2.62}) et (\ref{2.63}), on obtient alors: 
\begin{equation}
S_{I}^{out}\left[ \Omega \right] =\overline{I}^{out}\left( 1+{\cal S}\left[
\Omega \right] +{\cal T}\left[ \Omega \right] \right)   \label{2.66}
\end{equation}
Les quantit\'{e}s ${\cal S}\left[ \Omega \right] $ et ${\cal T}\left[ \Omega
\right] $ sont donn\'{e}es par: 
\begin{subequations}
\label{2.67}
\begin{eqnarray}
{\cal S}\left[ \Omega \right] &=&\frac{8\gamma ^{2}\overline{\Psi }\Omega
\tau }{\left| \Delta \left[ \Omega \right] \right| ^{2}}~\frac{\Psi _{NL}}{%
\gamma }~\left\{ \overline{\chi }_{I}\left[ \Omega \right] +\frac{\gamma
\Omega \tau }{\gamma ^{2}+\overline{\Psi }^{2}}\overline{\chi }_{R}\left[
\Omega \right] \right\}  \label{2.67a} \\
{\cal T}\left[ \Omega \right] &=&\frac{8\left( \gamma \overline{\Psi }\Omega
\tau \right) ^{2}}{\left( \gamma ^{2}+\overline{\Psi }^{2}\right) \left|
\Delta \left[ \Omega \right] \right| ^{2}}~\frac{\Psi _{NL}}{\gamma }~\left| 
\bar{\chi}\left[ \Omega \right] \right| ^{2}\frac{\chi \left[ 0\right] }{%
\hbar }~S_{T}\left[ \Omega \right]  \label{2.67b}
\end{eqnarray}
o\`{u} $\overline{\chi }_{R}\left[ \Omega \right] $ et $\overline{\chi }%
_{I}\left[ \Omega \right] $ sont respectivement les parties r\'{e}elle et
imaginaire de la susceptibilit\'{e} normalis\'{e}e $\overline{\chi }\left[
\Omega \right] $.

Lorsque ${\cal S}\left[ \Omega \right] $ et ${\cal T}\left[ \Omega \right] $
sont nuls, comme c'est le cas \`{a} fr\'{e}quence nulle, le spectre de bruit
d'intensit\'{e} est \'{e}gal \`{a} l'intensit\'{e} moyenne $\overline{I}%
^{out}$, c'est-\`{a}-dire au bruit quantique standard. Le fait que le bruit
d'intensit\'{e} n'est pas modifi\'{e} \`{a} fr\'{e}quence nulle peut se
comprendre \`{a} partir de la repr\'{e}sentation de la distribution de
Wigner dans l'espace des phases (figure \ref{Fig_2sqzesph}, page \pageref
{Fig_2sqzesph}). Chaque point de la distribution du faisceau incident subit
en effet une rotation autour de l'origine, et la projection de la
distribution sur l'axe du champ moyen n'est pas modifi\'{e}e. La
conservation de la distribution en intensit\'{e} \`{a} fr\'{e}quence nulle
est li\'{e}e \`{a} la conservation du nombre de photons sur des temps longs
par rapport au temps de stockage de la cavit\'{e}.

A fr\'{e}quence non nulle, ${\cal S}\left[ \Omega \right] $ et ${\cal T}%
\left[ \Omega \right] $ sont en g\'{e}n\'{e}ral non nuls, et le bruit de
photon du champ r\'{e}fl\'{e}chi n'est plus \'{e}gal au bruit quantique
standard. On peut remarquer que ${\cal S}\left[ \Omega \right] $ et ${\cal T}%
\left[ \Omega \right] $ sont proportionnels au rapport $\Psi _{NL}/\gamma $
entre le d\'{e}phasage non lin\'{e}aire et les pertes de la cavit\'{e} : le
bruit de photon n'est modifi\'{e} de mani\`{e}re appr\'{e}ciable que si ce
rapport est de l'ordre de $1$. D'autre part, le param\`{e}tre ${\cal T}%
\left[ \Omega \right] $ est associ\'{e} au bruit thermique du miroir mobile
et il est toujours positif : le mouvement Brownien du miroir mobile induit
toujours une augmentation du bruit de photon du faisceau r\'{e}fl\'{e}chi
par la cavit\'{e}. Par contre, le param\`{e}tre ${\cal S}\left[ \Omega
\right] $ peut \^{e}tre n\'{e}gatif, si le d\'{e}saccord $\overline{\Psi }$
est n\'{e}gatif. Comme nous allons le montrer, on peut alors trouver des
conditions de fonctionnement pour lesquels le bruit de photon est r\'{e}duit
en-dessous du bruit quantique standard. Pour cela, nous allons nous placer
dans le cas simple o\`{u} le mouvement du miroir est harmonique. La
susceptibilit\'{e} m\'{e}canique $\chi \left[ \Omega \right] $ est alors
donn\'{e}e par la relation (2.11).

\subsubsection{R\'{e}duction du bruit de photon \`{a} temp\'{e}rature nulle%
\label{II.3.5.1}}

Nous supposons que le miroir mobile est caract\'{e}ris\'{e} par les
m\^{e}mes param\`{e}tres que dans la partie 2.1 : fr\'{e}quence de
r\'{e}sonance m\'{e}canique $\Omega _{M}=10^{5}~rad/s$, masse $M=100~mg$ et
facteur de qualit\'{e} $Q=10^{6}$. Nous supposons d'autre part que la
cavit\'{e} a une finesse ${\cal F}=3~10^{5}$ (soit $\gamma =10^{-5}$) et une
bande passante $\Omega _{cav}=\gamma /\tau $ \'{e}gale \`{a} $\Omega _{M}/3$%
. L'efficacit\'{e} de la compression du champ d\'{e}pend beaucoup de
l'\'{e}cart entre le point de fonctionnement de la cavit\'{e} et les points
tournants de la bistabilit\'{e}. C'est en effet au voisinage de ces points
tournants que les effets non lin\'{e}aires, responsables de la modification
du bruit quantique, sont les plus importants. On peut montrer que le bruit
de photon \`{a} basse fr\'{e}quence est directement proportionnel \`{a} la
pente $\sigma =d\overline{I}^{in}/d\overline{I}$ de la courbe de
bistabilit\'{e}, pente qui s'annule aux points tournants (voir courbe \ref
{Fig_2bist_fr}b, page \pageref{Fig_2bist_fr})\cite{Kerr PRA 89}. A partir de
l'\'{e}quation (\ref{2.50}), on peut exprimer cette pente en fonction du
point de fonctionnement de la cavit\'{e} (d\'{e}phasage $\Psi _{0}$ et
d\'{e}phasage non lin\'{e}aire $\Psi _{NL}$): 
\end{subequations}
\begin{equation}
\sigma =\frac{1}{2\gamma }\left( 3\Psi _{NL}^{2}+4\Psi _{0}\Psi _{NL}+\gamma
^{2}+\Psi _{0}^{2}\right)   \label{2.68}
\end{equation}
La pente $\sigma $ peut s'annuler pour des valeurs positives du
d\'{e}phasage non lin\'{e}aire $\Psi _{NL}$ \`{a} condition que le
d\'{e}phasage $\Psi _{0}$ soit inf\'{e}rieur \`{a} $-\sqrt{3}\gamma $. On
choisit donc un d\'{e}phasage $\Psi _{0}$ \'{e}gal \`{a} $-3\gamma $. Dans
ces conditions, les deux points tournants de la bistabilit\'{e} (points $%
B_{B}$ et $B_{H}$ de la figure \ref{Fig_2bist_fr}b) correspondent \`{a} des
d\'{e}phasages non lin\'{e}aires $\Psi _{NL}$ \'{e}gaux \`{a} $1.2\gamma $
et $2.8\gamma $. On choisit alors un point de fonctionnement situ\'{e} sur
la branche basse de la courbe de bistabilit\'{e}, au voisinage du point
tournant $B_{B}$, en prenant $\Psi _{NL}=\gamma $. En d'autres termes, la
pression de radiation moyenne exerc\'{e}e sur le miroir mobile d\'{e}place
celui-ci d'une quantit\'{e} $\overline{x}$ \'{e}gale \`{a} la moiti\'{e} de
la largeur $\lambda /2{\cal F}$ de la r\'{e}sonance, soit environ $10^{-12}~m%
\grave{e}tre$.

Le choix du d\'{e}phasage $\Psi _{0}$ et du d\'{e}phasage non lin\'{e}aire $%
\Psi _{NL}$ fixe le point de fonctionnement sur la courbe de bistabilit\'{e}
(croix sur la figure \ref{Fig_2bist_fr}). On peut aussi d\'{e}finir le point
de fonctionnement par le d\'{e}phasage global $\overline{\Psi }$, \'{e}gal
\`{a} $-2\gamma $. Ces param\`{e}tres correspondent \`{a} une puissance
lumineuse incidente que l'on peut d\'{e}duire des \'{e}quations (\ref{2.47})
et (\ref{2.48a}): 
\begin{equation}
\Psi _{NL}=\frac{2\gamma }{\gamma ^{2}+\overline{\Psi }^{2}}~\frac{4\hbar
k^{2}}{M\Omega _{M}^{2}}~\overline{I}^{in}  \label{2.69}
\end{equation}
Avec les param\`{e}tres choisis, on trouve une puissance incidente $P_{in}$
de $3~mW$ ($\overline{I}^{in}\approx 2~10^{16}photons/s$).

\begin{figure}[tbp]
\centerline{\psfig{figure=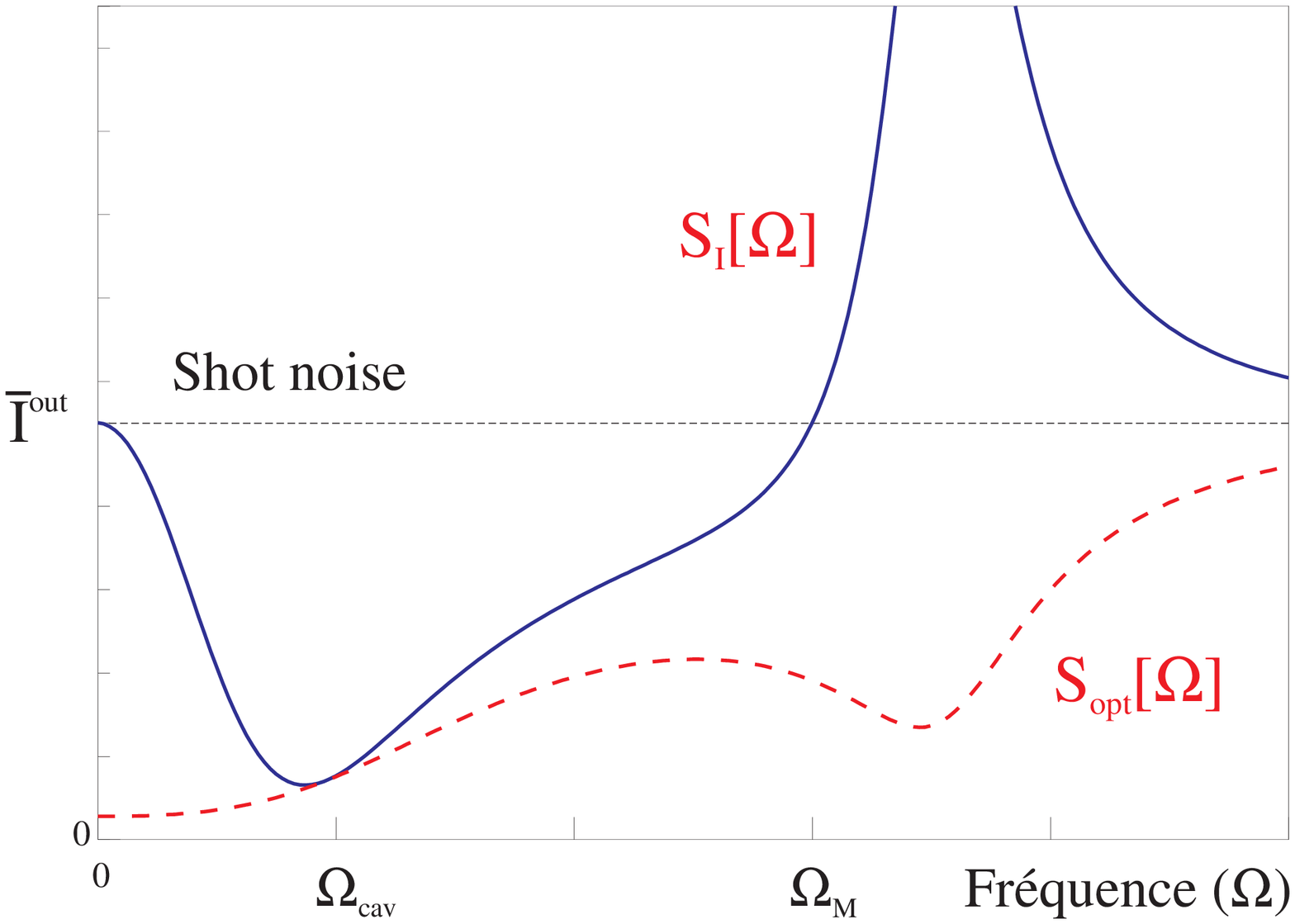,height=7cm}}
\caption{Spectre d'intensit\'{e} (trait plein) et spectre optimum (tirets)
en fonction de la fr\'{e}quence $\Omega $, \`{a} temp\'{e}rature nulle. Le
point de fonctionnement de la cavit\'{e} est fix\'{e} par le d\'{e}saccord $%
\overline{\Psi }=-2\gamma $ et par le d\'{e}phasage non lin\'{e}aire $\Psi
_{NL}=\gamma $}
\label{Fig_2int_opt}
\end{figure}

Le spectre d'intensit\'{e} et le spectre optimum obtenus \`{a}
temp\'{e}rature nulle sont repr\'{e}sent\'{e}s sur la figure \ref
{Fig_2int_opt}. 
On peut distinguer deux domaines de fr\'{e}quences : \`{a} basse
fr\'{e}quence ($\Omega \ll \Omega _{M}$) et au voisinage de la r\'{e}sonance
m\'{e}canique ($\Omega \approx \Omega _{M}$). A basse fr\'{e}quence, le
comportement du syst\`{e}me est similaire \`{a} celui d'un milieu Kerr
id\'{e}al plac\'{e} dans une cavit\'{e}\cite{Kerr PRA 89}. Dans cette plage
de fr\'{e}quence, la susceptibilit\'{e} m\'{e}canique peut en effet \^{e}tre
consid\'{e}r\'{e}e comme constante et \'{e}gale \`{a} sa valeur statique $%
1/M\Omega _{M}^{2}$. Le d\'{e}phasage non lin\'{e}aire subi par le champ est
alors tout \`{a} fait \'{e}quivalent \`{a} celui produit par un milieu Kerr
id\'{e}al.

\begin{figure}[tbp]
\centerline{\psfig{figure=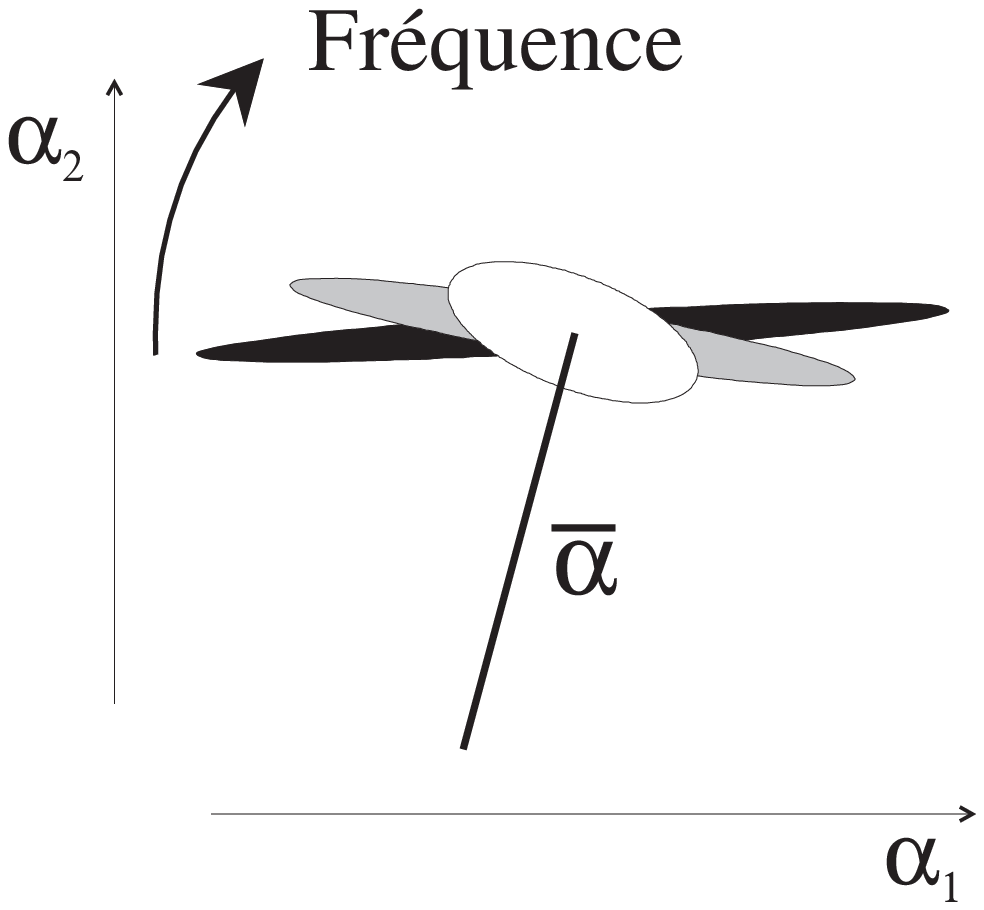,height=6cm}}
\caption{Evolution avec la fr\'{e}quence de la distribution du champ
r\'{e}fl\'{e}chi, dans l'espace des phases}
\label{Fig_2distfrq}
\end{figure}

On peut comprendre le comportement des spectres de la figure \ref
{Fig_2int_opt} \`{a} l'aide de la repr\'{e}sentation de la distribution de
Wigner dans l'espace des phases. La figure \ref{Fig_2distfrq} montre
l'\'{e}volution avec la fr\'{e}quence de la distribution du champ
r\'{e}fl\'{e}chi, depuis la fr\'{e}quence nulle (ellipse noire), jusqu'\`{a}
une fr\'{e}quence voisine de la bande passante de la cavit\'{e} $\Omega
_{cav}$ (ellipse blanche). Le premier effet est une diminution de
l'excentricit\'{e} de l'ellipse lorsque la fr\'{e}quence augmente. La
cavit\'{e} se comporte en effet comme un filtre passe bas pour les
fluctuations, et l'efficacit\'{e} de la non lin\'{e}arit\'{e} diminue au fur
et \`{a} mesure que la fr\'{e}quence augmente. Le bruit optimum $%
S_{opt}\left[ \Omega \right] $, qui est en fait reli\'{e} \`{a} la longueur
du petit axe de l'ellipse, cro\^{\i}t avec la fr\'{e}quence.

Le second effet est une rotation de l'ellipse avec la fr\'{e}quence. Le
petit axe de l'ellipse peut alors devenir parall\`{e}le au champ moyen
(ellipse grise sur la figure \ref{Fig_2distfrq}). Ainsi le bruit
d'intensit\'{e}, qui est reli\'{e} \`{a} la projection de la distribution
sur le champ moyen, est r\'{e}duit pour des fr\'{e}quences non nulles :
partant du bruit quantique standard \`{a} fr\'{e}quence nulle, le bruit de
photon diminue jusqu'\`{a} rejoindre le bruit optimum. Pour des
fr\'{e}quences sup\'{e}rieures, le petit axe n'est plus align\'{e} avec le
champ moyen et le bruit de photon remonte au dessus du bruit optimum. 

A haute fr\'{e}quence ($\Omega \gg \Omega _{cav}$), les deux spectres
rejoignent le bruit quantique standard, puisque la distribution tend vers un
disque identique \`{a} la distribution du champ incident. Ce comportement
est toutefois perturb\'{e} au voisinage de la r\'{e}sonance m\'{e}canique ($%
\Omega \approx \Omega _{M}$), o\`{u} la dynamique du miroir mobile joue un
r\^{o}le important. Dans cette r\'{e}gion, le bruit optimum est plus
r\'{e}duit qu'avec un milieu Kerr id\'{e}al. Cependant le spectre
d'intensit\'{e} pr\'{e}sente un fort exc\`{e}s de bruit.

\subsubsection{Effets du bruit thermique\label{II.3.5.2}}

A temp\'{e}rature non nulle, le bruit thermique du miroir mobile agit sur
les fluctuations quantiques du champ. On peut voir sur l'expression du bruit
d'intensit\'{e} (param\`{e}tre ${\cal T}\left[ \Omega \right] $ dans
l'\'{e}quation \ref{2.66}), mais aussi sur l'expression g\'{e}n\'{e}rale du
spectre $S_{\theta }^{out}\left[ \Omega \right] $ (dernier terme dans
l'\'{e}quation \ref{2.62}), que le bruit thermique augmente toujours le
bruit du faisceau r\'{e}fl\'{e}chi.

\begin{figure}[tbp]
\centerline{\psfig{figure=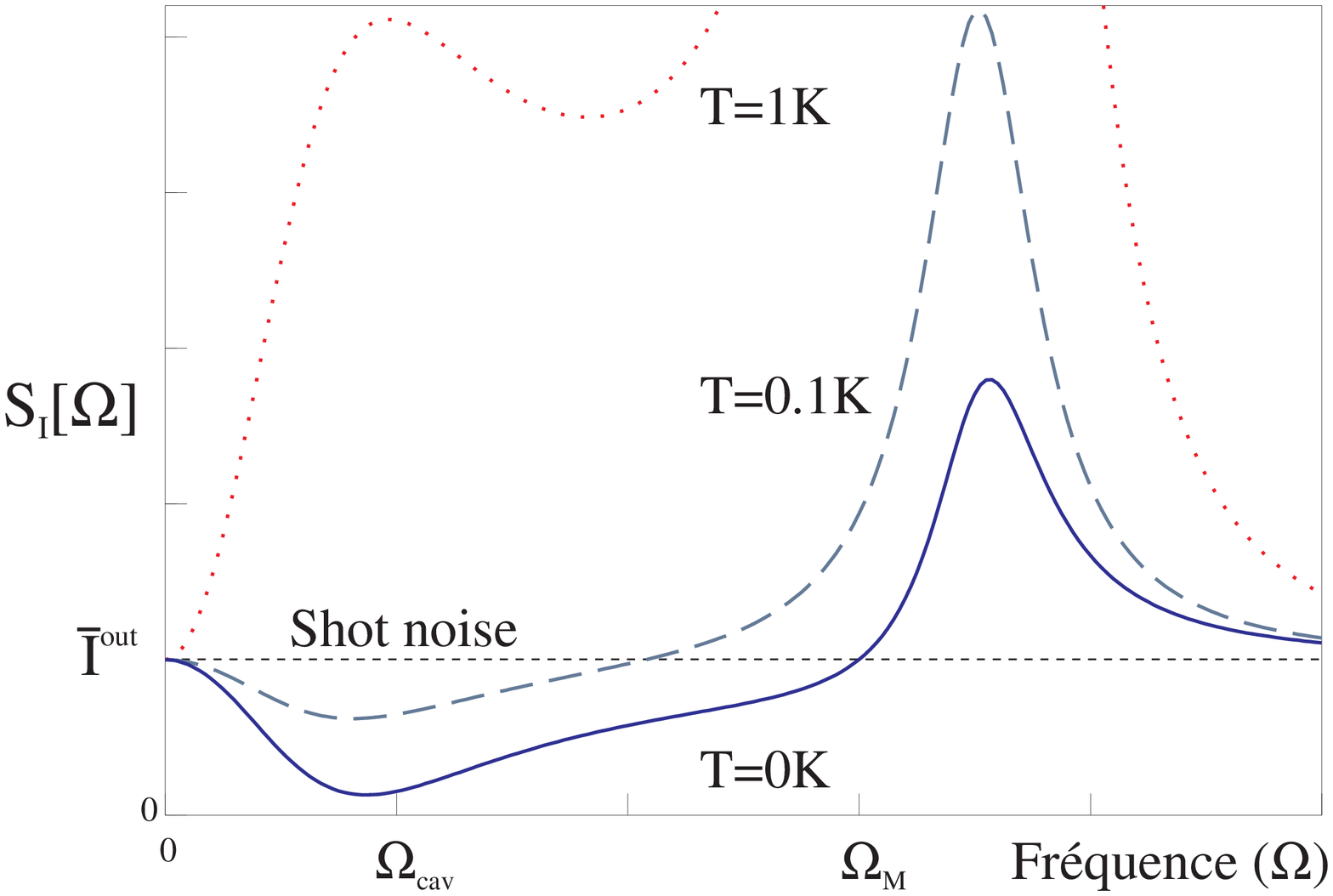,height=7cm}}
\caption{Effet du bruit thermique du miroir mobile sur le spectre de bruit
d'intensit\'{e} du champ r\'{e}fl\'{e}chi par la cavit\'{e}, \`{a} des
temp\'{e}ratures de 0.1 et 1 Kelvin}
\label{Fig_2specint}
\end{figure}

En utilisant l'expression du spectre de bruit thermique $S_{T}\left[ \Omega
\right] $ (\'{e}quation \ref{2.44}), on montre que la contribution du bruit
thermique au spectre d'intensit\'{e} est donn\'{e}e par : 
\begin{equation}
{\cal T}\left[ \Omega \right] =\frac{16\left( \gamma \overline{\Psi }\Omega
\tau \right) ^{2}}{\left( \gamma ^{2}+\overline{\Psi }^{2}\right) \left|
\Delta \left[ \Omega \right] \right| ^{2}}~\frac{\Psi _{NL}}{\gamma }~\left| 
\bar{\chi}\left[ \Omega \right] \right| ^{2}~\frac{n_{T}}{Q}  \label{2.70}
\end{equation}
o\`{u} $n_{T}$ repr\'{e}sente le nombre de phonons thermiques \`{a} la
fr\'{e}quence de r\'{e}sonance m\'{e}canique $\Omega _{M}$ : 
\begin{equation}
n_{T}=\frac{k_{B}T}{\hbar \Omega _{M}}  \label{2.71}
\end{equation}
Pour minimiser les effets du bruit thermique, il est n\'{e}cessaire de se
placer \`{a} basse temp\'{e}rature et d'utiliser un oscillateur
m\'{e}canique ayant un grand facteur de qualit\'{e} $Q$. La figure \ref
{Fig_2specint} montre l'effet du bruit thermique sur le spectre de bruit
d'intensit\'{e} du champ sortant, \`{a} des temp\'{e}ratures de $0.1$ et $%
1~Kelvin$. 
Pour des temp\'{e}ratures tr\`{e}s basses, inf\'{e}rieures \`{a} $0.1~K$,
l'exc\`{e}s de bruit \`{a} basse fr\'{e}quence reste mod\'{e}r\'{e} puisque
le bruit de photon est toujours r\'{e}duit en dessous du bruit quantique
standard. Cependant le bruit thermique masque compl\`{e}tement les effets
quantiques d\`{e}s que la temp\'{e}rature s'\'{e}l\`{e}ve : \`{a} une
temp\'{e}rature aussi basse que $1~Kelvin$, le bruit du faisceau
r\'{e}fl\'{e}chi est largement au dessus du bruit de photon standard.

Il est donc n\'{e}cessaire d'optimiser les param\`{e}tres du syst\`{e}me de
fa\c{c}on \`{a} r\'{e}duire l'influence du bruit thermique. L'\'{e}quation (%
\ref{2.71}) montre que le bruit thermique est inversement proportionnel
\`{a} la fr\'{e}quence de r\'{e}sonance m\'{e}canique $\Omega _{M}$. On peut
r\'{e}duire les effets du bruit thermique en augmentant cette fr\'{e}quence.
Ceci ne peut \^{e}tre fait avec un syst\`{e}me pendulaire, mais plut\^{o}t
en utilisant un r\'{e}sonateur m\'{e}canique, comme par exemple les
r\'{e}sonateurs pi\'{e}zo\'{e}lectriques en quartz. De tels r\'{e}sonateurs
ont des fr\'{e}quences de r\'{e}sonance m\'{e}canique bien sup\'{e}rieures
\`{a} $10^{5}rad/s$. 
\begin{figure}[tbp]
\centerline{\psfig{figure=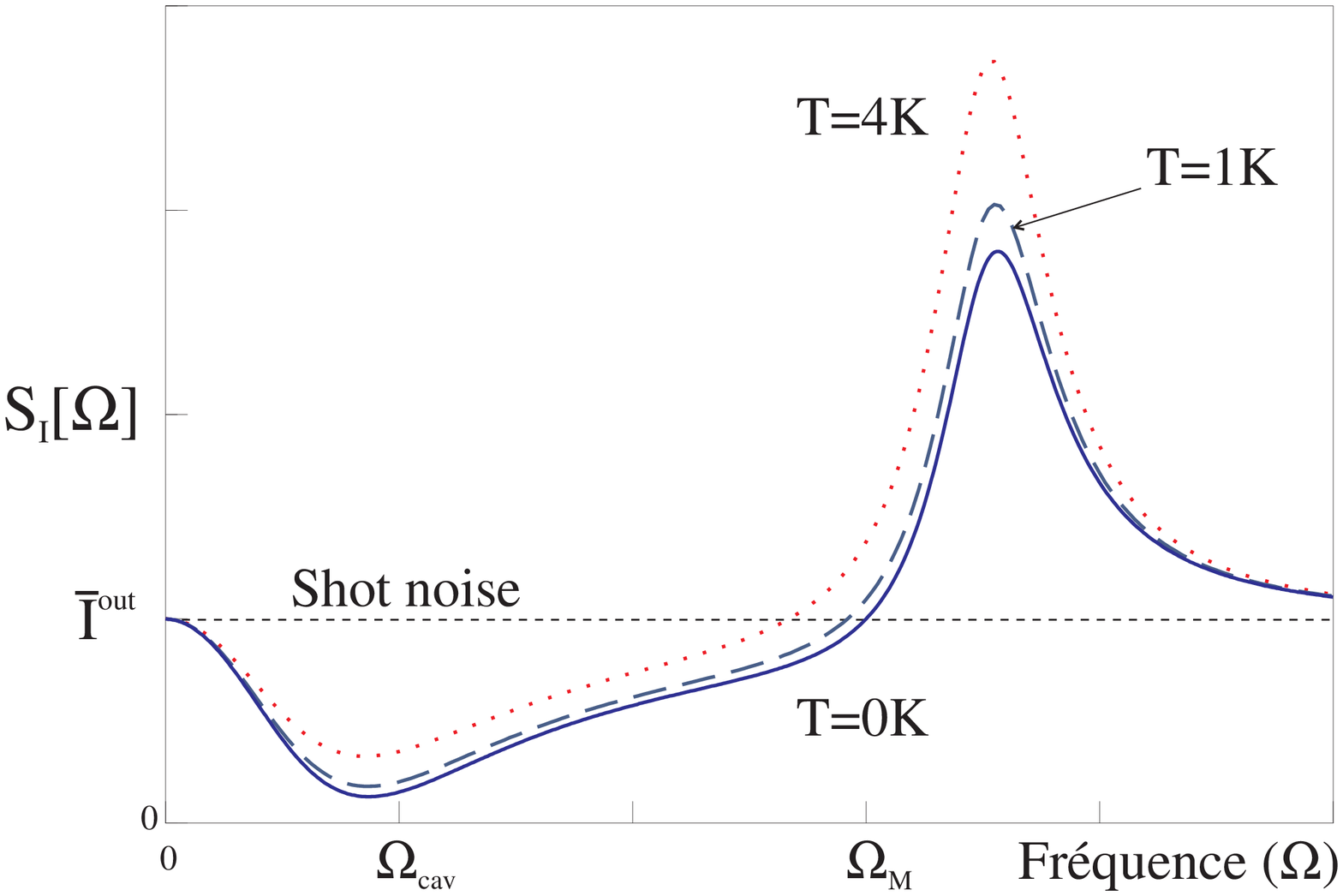,height=7cm}}
\caption{Effet du bruit thermique du miroir mobile sur le spectre de bruit
d'intensit\'{e} du champ r\'{e}fl\'{e}chi par la cavit\'{e}, pour une
fr\'{e}quence de r\'{e}sonance m\'{e}canique $\Omega _{M}$ du miroir
\'{e}gale \`{a} $10^{7}~rad/s$. Pour garder le m\^{e}me point de
fonctionnement que dans les figures \ref{Fig_2int_opt} et \ref{Fig_2specint}%
, le faisceau incident a une puissance $P_{in}$ \'{e}gale \`{a} $300~mW$ et
le miroir mobile a une masse $M$ de $1mg$}
\label{Fig_2spcint2}
\end{figure}

La figure \ref{Fig_2spcint2} montre les spectres de bruit obtenus pour une
fr\'{e}quence de r\'{e}sonance $\Omega _{M}$ \'{e}gale \`{a} $10^{7}~rad/s$,
les autres param\`{e}tres \'{e}tant identiques \`{a} ceux de la figure \ref
{Fig_2specint} ($\Psi _{NL}=\gamma $, $\overline{\Psi }=-2\gamma $). Le
bruit thermique est essentiellement concentr\'{e} au voisinage de la
r\'{e}sonance m\'{e}canique. A basse fr\'{e}quence, l'exc\`{e}s de bruit
reste mod\'{e}r\'{e} m\^{e}me \`{a} une temp\'{e}rature de $4~Kelvin$.

Augmenter la fr\'{e}quence de r\'{e}sonance permet donc de rendre les effets
quantiques dominants devant les effets thermiques pour des temp\'{e}ratures
raisonnables. Notons cependant que cela pr\'{e}sente plusieurs
inconv\'{e}nients. Tout d'abord, il est n\'{e}cessaire d'augmenter dans les
m\^{e}mes proportions la bande passante $\Omega _{cav}$ de la cavit\'{e} de
fa\c{c}on \`{a} limiter l'effet de filtrage produit par celle-ci. Etant
donn\'{e} les finesses mises en jeu, cela impose de construire une
cavit\'{e} tr\`{e}s courte, de l'ordre de $0.5~mm$ de longueur. D'autre
part, pour conserver la condition $\Psi _{NL}=\gamma $, il est
n\'{e}cessaire de diminuer la masse du r\'{e}sonateur et d'augmenter
l'intensit\'{e} lumineuse (voir \'{e}quation \ref{2.69}). Pour un
r\'{e}sonateur de masse $M$ \'{e}gale \`{a} $1~mg$, la puissance incidente
doit \^{e}tre de l'ordre de $300~mW$.

\section{Mesures de petits d\'{e}placements\label{II-4}}

\bigskip

A l'aide des r\'{e}sultats obtenus dans la partie pr\'{e}c\'{e}dente, nous
pouvons \`{a} pr\'{e}sent mener une \'{e}tude plus d\'{e}taill\'{e}e de la
limite de sensibilit\'{e} lors d'une mesure de d\'{e}placement du miroir de
la cavit\'{e}. Nous commencerons cette partie en \'{e}tablissant de
mani\`{e}re g\'{e}n\'{e}rale l'expression de la sensibilit\'{e} optimale du
dispositif (section 2.4.1). Nous appliquerons ensuite le r\'{e}sultat obtenu
\`{a} la mesure du bruit thermique du miroir mobile (section 2.4.2) et nous
pr\'{e}senterons les spectres de bruit pour deux mod\`{e}les de dissipation
thermique. Nous terminerons cette partie par une \'{e}tude du dispositif
permettant de r\'{e}aliser une mesure quantique non destructive de
l'intensit\'{e} d'un faisceau lumineux (section 2.4.3).

\subsection{Sensibilit\'{e} optimale d'une mesure optique de d\'{e}placement 
\label{II-4-1}}

Nous avons vu dans les sections 2.2.1 et 2.2.3 que la phase du champ
r\'{e}fl\'{e}chi par la cavit\'{e} est sensible aux d\'{e}placements du
miroir mobile. Cette sensibilit\'{e} atteint son maximum lorsque le faisceau
est \`{a} r\'{e}sonance avec la cavit\'{e}. Les champs moyens
intracavit\'{e} et r\'{e}fl\'{e}chi sont alors donn\'{e}s par les
\'{e}quations (\ref{2.48}) avec $\overline{\Psi }=0$: 
\begin{equation}
\overline{\alpha }^{out}=\overline{\alpha }^{in}=\sqrt{\frac{\gamma }{2}}~%
\overline{\alpha }  \label{4.1}
\end{equation}
Le champ intracavit\'{e} $\overline{\alpha }$ \'{e}tant choisi r\'{e}el, on
en d\'{e}duit qu'\`{a} r\'{e}sonance tous les champs sont r\'{e}els et le
champ r\'{e}fl\'{e}chi est \'{e}gal au champ incident.

Nous supposons que le miroir est soumis \`{a} une variation de position $%
\delta x\left[ \Omega \right] $. Nous allons maintenant d\'{e}terminer
comment ce d\'{e}placement se traduit sur le spectre du faisceau
r\'{e}fl\'{e}chi par la cavit\'{e}. Nous nous int\'{e}resserons dans la
suite aux deux composantes d'amplitude et de phase du champ $\alpha $, que
l'on notera respectivement $p$ et $q$. Dans le cas d'un champ r\'{e}el,
l'amplitude et la phase s'identifient aux deux quadratures $\alpha _{1}$ et $%
\alpha _{2}$ d\'{e}finies dans la section 2.2.2: 
\begin{equation}
p=\alpha +\alpha ^{*}\quad ,\quad q=i(\alpha ^{*}-\alpha )  \label{4.2}
\end{equation}
En substituant le d\'{e}phasage $\delta \Psi $ du champ intracavit\'{e} par $%
2k\delta x$ dans les \'{e}quations (\ref{2.58a}) et (\ref{2.58b}), on
obtient les fluctuations des champs intracavit\'{e} et r\'{e}fl\'{e}chi en
fonction des fluctuations du champ incident et de la position du miroir
mobile: 
\begin{subequations}
\label{4.5}
\begin{eqnarray}
\delta p\left[ \Omega \right] &=&\frac{\sqrt{2}}{\gamma -i\Omega \tau }%
~\delta p^{in}\left[ \Omega \right]  \label{4.5a} \\
\delta q\left[ \Omega \right] &=&\frac{\sqrt{2\gamma }}{\gamma -i\Omega \tau 
}~\delta q^{in}\left[ \Omega \right] +\frac{4}{\gamma -i\Omega \tau }~%
\overline{\alpha }k~\delta x\left[ \Omega \right]  \label{4.5b} \\
\delta p^{out}\left[ \Omega \right] &=&\frac{\gamma +i\Omega \tau }{\gamma
-i\Omega \tau }~\delta p^{in}\left[ \Omega \right]  \label{4.5c} \\
\delta q^{out}\left[ \Omega \right] &=&\frac{\gamma +i\Omega \tau }{\gamma
-i\Omega \tau }~\delta q^{in}\left[ \Omega \right] +\frac{4\sqrt{2\gamma }}{%
\gamma -i\Omega \tau }~\overline{\alpha }k~\delta x\left[ \Omega \right]
\label{4.5d}
\end{eqnarray}
Les \'{e}quations (\ref{4.5a}) et (\ref{4.5c}) montrent qu'\`{a}
r\'{e}sonance, les fluctuations d'amplitude ne sont coupl\'{e}es ni aux
fluctuations de phase, ni aux variations de position du miroir. En
particulier, le spectre de bruit d'amplitude du faisceau r\'{e}fl\'{e}chi $%
S_{p}^{out}\left[ \Omega \right] $ est \'{e}gal au spectre de bruit incident 
$S_{p}^{in}\left[ \Omega \right] $:

\end{subequations}
\begin{equation}
S_{p}^{out}\left[ \Omega \right] =S_{p}^{in}\left[ \Omega \right]
\label{4.6}
\end{equation}
Pour un champ incident coh\'{e}rent, le champ r\'{e}fl\'{e}chi a donc un
spectre de bruit d'amplitude \'{e}gal \`{a} $1$, c'est \`{a} dire un spectre
de bruit d'intensit\'{e} \'{e}gal au bruit quantique standard $\overline{I}%
^{out}$ (\'{e}quation 2.64). Nous verrons dans la section 2.4.3 que la
relation (\ref{4.6}) est indispensable si l'on veut r\'{e}aliser une mesure
quantique non destructive.

Par contre, les fluctuations de phase en sortie sont li\'{e}es non seulement
au bruit de phase incident mais aussi aux variations de position du miroir
(\'{e}quation \ref{4.5d}). Pour d\'{e}terminer le spectre de bruit de phase $%
S_{q}^{out}$ du faisceau r\'{e}fl\'{e}chi, il est n\'{e}cessaire de faire
des hypoth\`{e}ses suppl\'{e}mentaires sur les corr\'{e}lations entre les
variations de position $\delta x$ du miroir et les fluctuations de phase $%
\delta q^{in}$ du champ incident. Nous supposons ici que ces fluctuations
sont d\'{e}corr\'{e}l\'{e}es. Nous verrons que cette hypoth\`{e}se est
justifi\'{e}e dans les deux configurations que nous \'{e}tudions (mesure du
bruit thermique du miroir, mesure QND de l'intensit\'{e}). On obtient alors
l'expression suivante pour le spectre de bruit de phase du faisceau
r\'{e}fl\'{e}chi: 
\begin{equation}
S_{q}^{out}\left[ \Omega \right] =S_{q}^{in}\left[ \Omega \right] +256\frac{1%
}{1+\left( \Omega /\Omega _{cav}\right) ^{2}}\frac{{\cal F}^{2}\overline{I}%
^{in}}{\lambda ^{2}}~S_{x}\left[ \Omega \right]  \label{4.7}
\end{equation}
o\`{u} $S_{q}^{in}\left[ \Omega \right] $ est le spectre de bruit de phase
du champ incident, \'{e}gal \`{a} $1$ pour un \'{e}tat coh\'{e}rent, et $%
S_{x}\left[ \Omega \right] $ d\'{e}signe le spectre de position du miroir
mobile. Le spectre de phase en sortie appara\^{\i }t donc comme la somme
d'un terme de ''bruit'', \'{e}gal au bruit de la composante de phase du
champ incident, et d'un ''signal'', li\'{e} au spectre des d\'{e}placements
du miroir, et filtr\'{e} par la bande passante $\Omega _{cav}=\gamma /\tau $
de la cavit\'{e}.

On peut \`{a} pr\'{e}sent donner une estimation de la sensibilit\'{e} $%
\delta x_{\min }$ d'une mesure de position. $\delta x_{\min }$ est
l'amplitude de bruit de position du miroir qui fournit un signal du m\^{e}me
ordre de grandeur que le bruit $S_{q}^{in}$ (rapport signal \`{a} bruit
\'{e}gal \`{a} $1$). Pour un faisceau incident coh\'{e}rent ($%
S_{q}^{in}\left[ \Omega \right] =1$) et des fr\'{e}quences petites devant la
bande passante de la cavit\'{e} ($\Omega \ll \Omega _{cav}$), on obtient une
relation qui donne, en $m/\sqrt{Hz}$, le plus petit d\'{e}placement
d\'{e}tectable du miroir mobile: 
\begin{equation}
\delta x_{\min }=\frac{\lambda }{16{\cal F}}\frac{1}{\sqrt{\overline{I}^{in}}%
}  \label{4.9}
\end{equation}
On retrouve ici l'expression obtenue \`{a} partir de raisonnements physiques
simples dans la partie 2.2 (\'{e}quation \ref{2.35}, page \pageref{2.35}).
On obtient ainsi une sensibilit\'{e} optimale $\delta x_{\min }\approx
10^{-21}m/\sqrt{Hz}$ pour une finesse ${\cal F}=3~10^{5}$ et une puissance
incidente $P_{in}=3~mW$.

Notons enfin que nous avons suppos\'{e} que le mouvement du miroir est
d\'{e}corr\'{e}l\'{e} des fluctuations de phase du faisceau incident. Si ce
n'est pas le cas, il est possible de r\'{e}duire le bruit du faisceau
r\'{e}fl\'{e}chi en produisant un \'{e}tat comprim\'{e} comme nous l'avons
montr\'{e} dans la partie 2.3. Ceci peut permettre d'augmenter encore la
sensibilit\'{e} de la mesure. Nous reviendrons sur ce point dans la section
2.4.3.3.

\subsection{Mesure du bruit thermique\label{II-4-2}}

Nous nous proposons dans cette section de montrer que la sensibilit\'{e} de
la cavit\'{e} est suffisante pour mesurer le spectre des fluctuations
thermiques du miroir. Afin d'obtenir une expression analytique du spectre,
nous nous placerons dans le cas simple o\`{u} le mouvement du miroir est
d\'{e}crit par un oscillateur harmonique amorti et nous pr\'{e}senterons
deux mod\`{e}les d'amortissement : l'amortissement visqueux et
l'amortissement interne. Nous donnerons l'allure des spectres de bruit qui
correspondent \`{a} chacun de ces deux mod\`{e}les. Nous nous placerons
\`{a} temp\'{e}rature ambiante et nous supposerons l'intensit\'{e} incidente
suffisamment faible pour pouvoir n\'{e}gliger les effets de pression de
radiation sur le miroir mobile : le d\'{e}placement du miroir correspond au
mouvement Brownien associ\'{e} au couplage avec le bain thermique. Ce
d\'{e}placement est donc d\'{e}corr\'{e}l\'{e} des fluctuations du champ
incident et le spectre de bruit de phase du faisceau r\'{e}fl\'{e}chi est
donn\'{e} par l'\'{e}quation (\ref{4.7}).

\subsubsection{Th\'{e}or\`{e}me fluctuation-dissipation\label{II-4-2-1}}

Les fluctuations de position li\'{e}es au bruit thermique peuvent \^{e}tre
d\'{e}crites \`{a} l'aide d'une force de Langevin $F_{T}$ appliqu\'{e}e au
miroir mobile. Les fluctuations de position $\delta x\left[ \Omega \right] $
sont alors \'{e}gales au produit de la susceptibilit\'{e} $\chi \left[
\Omega \right] $ par la force de Langevin, dont le spectre de bruit $%
S_{T}\left[ \Omega \right] $ est donn\'{e} par le th\'{e}or\`{e}me
fluctuation-dissipation (\'{e}quation 2.44). Le spectre des fluctuations
thermiques de position du miroir $S_{x}\left[ \Omega \right] $ s'\'{e}crit
alors: 
\begin{equation}
S_{x}\left[ \Omega \right] =\left| \chi \left[ \Omega \right] \right|
^{2}~S_{T}\left[ \Omega \right] =\frac{2k_{B}T}{\Omega }~{\cal I}m\left(
\chi \left[ \Omega \right] \right)  \label{4.10}
\end{equation}

Le th\'{e}or\`{e}me fluctuation-dissipation permet de relier le spectre de
bruit thermique du miroir \`{a} la partie dissipative (partie imaginaire) de
la susceptibilit\'{e} qui d\'{e}crit l'amortissement. Malheureusement, il
n'existe pas de mod\`{e}le th\'{e}orique satisfaisant capable de d\'{e}crire
l'ensemble des effets de dissipation thermique dans les solides. Nous allons
dans la suite consid\'{e}rer le cas d'un miroir harmonique amorti, ce qui
nous permettra de donner une expression simple du spectre de bruit de phase
du faisceau r\'{e}fl\'{e}chi.

\subsubsection{Oscillateur harmonique amorti : amortissements visqueux et
interne \label{II-4-2-2}}

L'\'{e}quation qui r\'{e}git le mouvement d'un oscillateur harmonique libre
de masse $M$, ayant une force de rappel \'{e}gale \`{a} $-k_{0}~x$,
s'\'{e}crit: 
\begin{equation}
M~\ddot{x}(t)=F(t)-k_{0}~x(t)  \label{4.11}
\end{equation}
o\`{u} $F(t)$ est une force ext\'{e}rieure appliqu\'{e}e \`{a} l'oscillateur
harmonique. La dynamique de ce dernier est alors d\'{e}crite par la
susceptibilit\'{e} m\'{e}canique $\chi _{0}\left[ \Omega \right] $: 
\begin{equation}
\chi _{0}\left[ \Omega \right] =\frac{1}{M\left( \Omega _{M}^{2}-\Omega
^{2}\right) }  \label{4.12}
\end{equation}
o\`{u} la fr\'{e}quence de r\'{e}sonance $\Omega _{M}$ est d\'{e}finie par $%
\Omega _{M}=\sqrt{\frac{k_{0}}{M}}$. Lorsque l'oscillateur harmonique est
faiblement coupl\'{e} \`{a} son environnement, il est n\'{e}cessaire de
rajouter \`{a} la susceptibilit\'{e} $\chi _{0}\left[ \Omega \right] $ un
terme imaginaire pour tenir compte de l'amortissement issu de ce couplage.
La susceptibilit\'{e} d'un oscillateur harmonique amorti peut donc
s'\'{e}crire sous la forme: 
\begin{equation}
\chi \left[ \Omega \right] =\frac{1}{M\left( \Omega _{M}^{2}-\Omega
^{2}-i\Omega _{M}^{2}\Phi \left[ \Omega \right] \right) }  \label{4.13}
\end{equation}
o\`{u} $\Phi \left[ \Omega \right] $ est une fonction a priori quelconque de
la fr\'{e}quence, qui doit n\'{e}anmoins s'annuler \`{a} fr\'{e}quence nulle
car la susceptibilit\'{e} $\chi \left( t\right) $ est r\'{e}elle. On
supposera dans la suite que l'amortissement est faible, c'est \`{a} dire que 
$\Phi $ est petit devant $1$. En comparant les d\'{e}nominateurs des deux
expressions (\ref{4.12}) et (\ref{4.13}), on voit qu'il suffit de rajouter
un terme imaginaire \`{a} la constante de raideur de l'oscillateur libre
pour tenir compte de la dissipation: 
\begin{equation}
k\left[ \Omega \right] =k_{0}~\left( 1-i\Phi \left[ \Omega \right] \right)
\label{4.14}
\end{equation}
Si on applique \`{a} l'oscillateur harmonique une force sinuso\"{\i }dale,
on montre alors que $\Phi \left[ \Omega \right] $ est li\'{e} \`{a} la
fraction d'\'{e}nergie dissip\'{e}e durant chaque cycle, d'o\`{u}
l'appellation usuelle d{\it 'angle de perte} pour $\Phi $. Notons aussi que
dans le cas o\`{u} l'angle de perte varie peu au voisinage de la
r\'{e}sonance m\'{e}canique, on peut relier l'angle de perte \`{a}
r\'{e}sonance $\Phi \left[ \Omega _{M}\right] $ au facteur de qualit\'{e} $Q$%
: 
\begin{equation}
\Phi \left[ \Omega _{M}\right] =\frac{1}{Q}  \label{4.15}
\end{equation}

La d\'{e}pendance en fr\'{e}quence de l'angle de perte n'est pas simple
\`{a} d\'{e}terminer car elle d\'{e}pend des nombreux processus de
dissipation qui se manifestent lorsqu'un corps est coupl\'{e} \`{a} son
environnement. Si l'on consid\`{e}re, par exemple, un miroir mobile fix\'{e}
\`{a} un syst\`{e}me pendulaire, le mouvement harmonique du pendule est
amorti de la m\^{e}me mani\`{e}re qu'une particule Brownienne plong\'{e}e
dans un liquide : le pendule est soumis \`{a} une force de friction
proportionnelle \`{a} sa vitesse. Ceci revient \`{a} rajouter au
d\'{e}nominateur de la susceptibilit\'{e} $\chi _{0}$ une partie imaginaire
proportionnelle \`{a} la fr\'{e}quence. Dans le cadre de ce mod\`{e}le
d'amortissement visqueux\cite{Landau Nav Stoks}, l'angle de perte est donc
une fonction lin\'{e}aire de la fr\'{e}quence: 
\begin{equation}
\Phi _{vis}\left[ \Omega \right] =\frac{\Omega }{Q\Omega _{M}}  \label{4.17}
\end{equation}

Ce mod\`{e}le ne permet pas cependant de d\'{e}crire de fa\c{c}on
satisfaisante la dissipation dans un solide. Plusieurs processus de
dissipation peuvent coexister dans un solide. Citons par exemple la
dissipation thermo\'{e}lastique qui est due aux effets coupl\'{e}s des
d\'{e}formations et des gradients de temp\'{e}rature. Lors d'une excitation
d'un mode de vibration interne, les zones dilat\'{e}es du solide se
refroidissent et les zones contract\'{e}es se r\'{e}chauffent. La
thermalisation entre ces r\'{e}gions entra\^{\i }ne un effet de dissipation.
Un autre type de dissipation est li\'{e} \`{a} la propagation de
dislocations dues \`{a} la pr\'{e}sence d'impuret\'{e}s dans le solide. Ces
impuret\'{e}s absorbent une partie de l'\'{e}nergie apport\'{e}e par une
excitation d'un mode de vibration interne du solide. On peut enfin citer la
dissipation par pertes de recul qui est due au contact du solide avec son
support. En g\'{e}n\'{e}ral la masse du support n'est pas infinie et une
partie de l'\'{e}nergie stock\'{e}e dans le solide peut \^{e}tre
dissip\'{e}e sous forme d'une excitation du support. Il existe encore
d'autres processus de dissipation, et la diversit\'{e} de ces processus rend
difficile l'identification de la source principale de dissipation pour un
mat\'{e}riau ou pour une g\'{e}om\'{e}trie donn\'{e}e d'un solide.

Les mod\`{e}les th\'{e}oriques actuels ne permettent donc pas de d\'{e}crire
de fa\c{c}on compl\`{e}te les diff\'{e}rent processus de dissipation. Il
existe cependant une approximation raisonnable, bas\'{e}e sur des
r\'{e}sultats exp\'{e}rimentaux, qui consiste \`{a} d\'{e}crire l'ensemble
des processus de dissipation par un angle de perte qui varie peu avec la
fr\'{e}quence\cite{Gillespie Saulson}. Un mod\`{e}le simple bas\'{e} sur la
th\'{e}orie an\'{e}lastique dans les solides d\'{e}crit assez bien ce
comportement de la dissipation\cite{Saulson Zener}. Cette approximation
pr\'{e}dit donc un angle de perte constant sur une large bande de
fr\'{e}quence, et \'{e}gal d'apr\`{e}s l'\'{e}quation (\ref{4.15}) \`{a}: 
\begin{equation}
\Phi _{cst}\left[ \Omega \right] =1/Q  \label{4.20}
\end{equation}

\subsubsection{Spectre de bruit thermique\label{II-4-2-3}}

Pour un oscillateur harmonique amorti, il est possible d'exprimer de
mani\`{e}re simple le spectre de bruit de phase du champ r\'{e}fl\'{e}chi $%
S_{q}^{out}\left[ \Omega \right] $. En utilisant les relations (\ref{4.7}), (%
\ref{4.10}) et (\ref{4.13}) on trouve: 
\begin{equation}
S_{q}^{out}\left[ \Omega \right] =S_{q}^{in}\left[ \Omega \right] +16\frac{%
\left| \overline{\chi }\left[ \Omega \right] \right| ^{2}}{1+\left( \Omega
/\Omega _{cav}\right) ^{2}}\frac{\Psi _{NL}}{\gamma }\frac{\Phi \left[
\Omega \right] /\Omega }{\Phi \left[ \Omega _{M}\right] /\Omega _{M}}~\frac{%
n_{T}}{Q}  \label{4.21}
\end{equation}
On obtient une expression qui ressemble beaucoup \`{a} celles obtenues dans
la partie pr\'{e}c\'{e}dente, telles que l'\'{e}quation (\ref{2.70}) qui
d\'{e}crit la contribution du bruit thermique au spectre d'intensit\'{e} du
faisceau r\'{e}fl\'{e}chi. On retrouve ici les m\^{e}mes param\`{e}tres : le
rapport entre le d\'{e}phasage non lin\'{e}aire $\Psi _{NL}$ et les pertes $%
\gamma $ de la cavit\'{e}, et le rapport entre le nombre $n_{T}$ de phonons
thermiques et le facteur de qualit\'{e} $Q$. Ces deux param\`{e}tres fixent
l'amplitude globale de la contribution du bruit thermique au bruit de phase
du faisceau r\'{e}fl\'{e}chi. La d\'{e}pendance en fr\'{e}quence est pour
l'essentiel li\'{e}e \`{a} la susceptibilit\'{e} normalis\'{e}e $\overline{%
\chi }\left[ \Omega \right] $ et \`{a} un terme de filtrage d\^{u} \`{a} la
bande passante $\Omega _{cav}$ de la cavit\'{e}. Le dernier param\`{e}tre
fait intervenir explicitement l'angle de perte $\Phi \left[ \Omega \right] $
et d\'{e}pend crucialement du mod\`{e}le de dissipation choisi. 
\begin{figure}[tbp]
\centerline{\psfig{figure=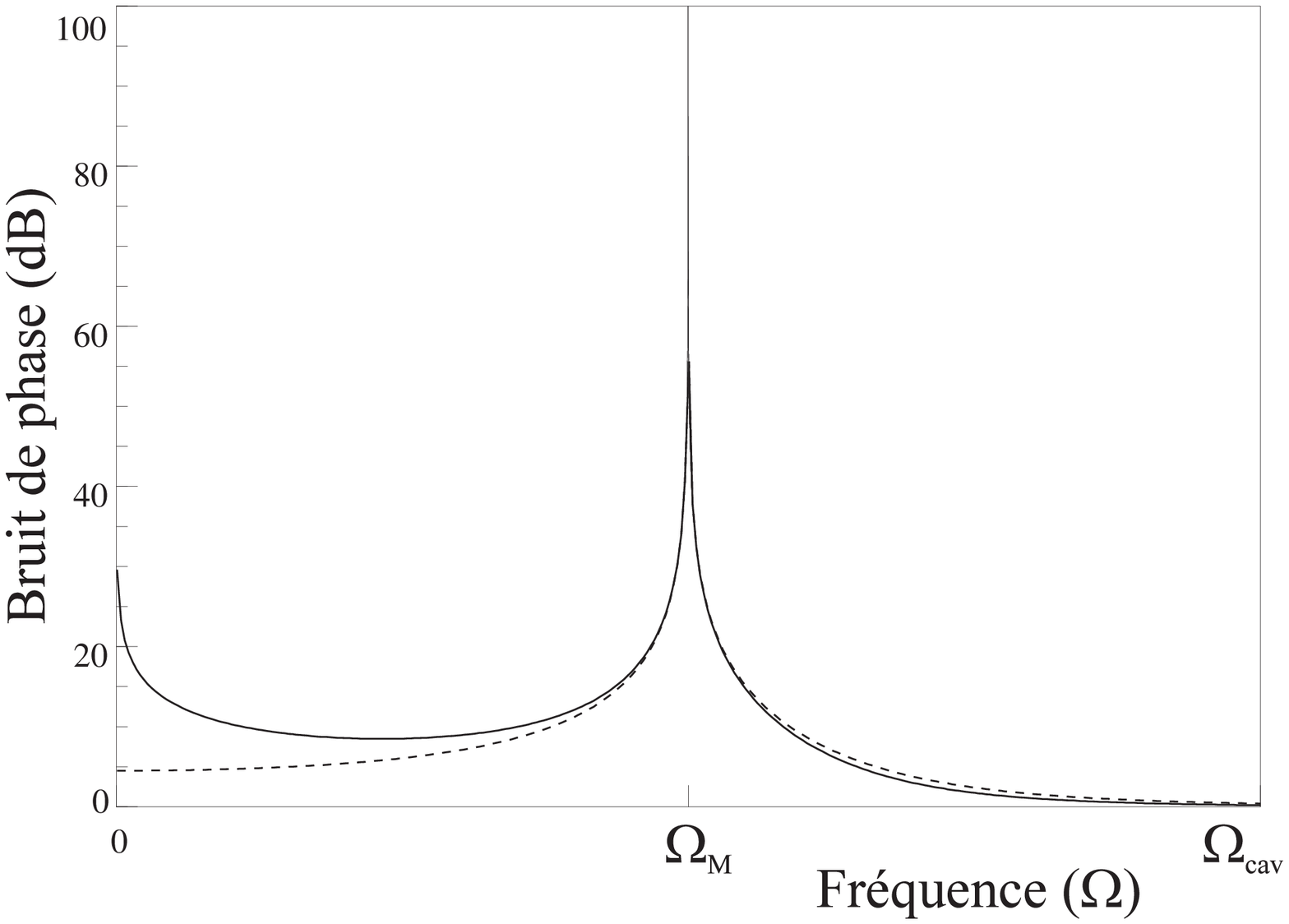,height=8cm}}
\caption{Spectre de bruit de phase du faisceau r\'{e}fl\'{e}chi \`{a}
temp\'{e}rature ambiante. En dehors de la r\'{e}sonance m\'{e}canique, le
bruit thermique d\'{e}pend du mod\`{e}le d'amortissement que l'on adopte :
amortissement visqueux (tirets) et mod\`{e}le $\Phi $ constant (trait plein)}
\label{Fig_2brthoh}
\end{figure}

La figure \ref{Fig_2brthoh} montre les spectres de bruit $S_{q}^{out}\left[
\Omega \right] $ obtenus \`{a} temp\'{e}rature ambiante pour les deux
mod\`{e}les de dissipation (amortissement visqueux et $\Phi $ constant). Les
param\`{e}tres sont identiques \`{a} ceux choisis dans la partie
pr\'{e}c\'{e}dente (fr\'{e}quence de r\'{e}sonance $\Omega _{M}=10^{7}~rad/s$%
, masse $M=1~mg$, facteur de qualit\'{e} $Q=10^{6}$, finesse de la
cavit\'{e} ${\cal F}=3~10^{5}$). Pour pouvoir n\'{e}gliger les effets de
pression de radiation devant les effets thermiques, le d\'{e}phasage non
lin\'{e}aire $\Psi _{NL}$ est \'{e}gal \`{a} $\gamma /20$ ce qui correspond
\`{a} une puissance incidente $P_{in}$ de $3~mW$ (\'{e}quation \ref{2.69}).
Enfin la bande passante $\Omega _{cav}$ de la cavit\'{e} est choisie
\'{e}gale \`{a} $2\Omega _{M}$ de fa\c{c}on \`{a} r\'{e}duire l'effet de
filtrage du bruit thermique.

Du fait de la tr\`{e}s grande dynamique du bruit thermique (le terme $\left| 
\overline{\chi }\left[ \Omega \right] \right| ^{2}$ dans l'\'{e}quation (\ref
{4.21}) prend des valeurs comprises entre $1$ \`{a} basse fr\'{e}quence et $%
Q^{2}$ \`{a} r\'{e}sonance), les spectres sont repr\'{e}sent\'{e}s en
\'{e}chelle logarithmique (d\'{e}cibel). L'axe vertical repr\'{e}sente $%
10\log S_{q}^{out}\left[ \Omega \right] $ et la valeur $0~dB$ correspond au
bruit de photon standard ($S_{q}^{out}=1$). On voit que le bruit thermique
est essentiellement concentr\'{e} au voisinage de la r\'{e}sonance
m\'{e}canique. Il est toutefois possible d'observer le bruit thermique
m\^{e}me tr\`{e}s loin de la r\'{e}sonance m\'{e}canique, puisqu'\`{a} basse
fr\'{e}quence le niveau de bruit est sup\'{e}rieur d'au moins $5~dB$ par
rapport au bruit de photon. Notons que les techniques de mesure homodyne
permettent de mesurer des \'{e}carts par rapport au bruit de photon standard
avec une pr\'{e}cision de l'ordre de $0.1~dB$.

On voit que les deux mod\`{e}les de dissipation se distinguent l'un de
l'autre au niveau des ailes de la r\'{e}sonance : \`{a} basse fr\'{e}quence,
le mod\`{e}le visqueux pr\'{e}voit un ''fond'' de bruit thermique beaucoup
plus faible que le mod\`{e}le $\Phi $ constant. Il appara\^{\i }t ainsi
qu'une cavit\'{e} \`{a} miroir mobile atteint une sensibilit\'{e} suffisante
pour \'{e}tudier quantitativement le niveau et la distribution spectrale du
bruit thermique du miroir mobile.

\subsection{Mesure quantique non destructive de l'intensit\'{e} lumineuse 
\label{II-4-3}}

Nous avons montr\'{e} dans la partie pr\'{e}c\'{e}dente que le spectre de
phase d'un faisceau interagissant avec la cavit\'{e} de mani\`{e}re
r\'{e}sonnante reproduit le bruit thermique du miroir mobile lorsque ce
dernier est \`{a} temp\'{e}rature ambiante. On peut donc penser utiliser
cette grande sensibilit\'{e} pour sonder le mouvement du miroir mobile
lorsqu'il est soumis \`{a} une force ext\'{e}rieure autre que la force de
Langevin. Cette force ext\'{e}rieure peut \^{e}tre produite par les
fluctuations quantiques de la pression de radiation d'un second faisceau (%
{\it faisceau signal}) qui interagit aussi avec la cavit\'{e} (voir figure 
\ref{Fig_2qnd}, page \pageref{Fig_2qnd}).

Pour un faisceau signal suffisamment intense, le mouvement du miroir
reproduit les fluctuations d'intensit\'{e} du faisceau. On cr\'{e}e ainsi
des corr\'{e}lations quantiques entre l'intensit\'{e} du faisceau signal et
la phase du faisceau de mesure. Nous avons vu dans la section 2.2.3.2 que ce
syst\`{e}me permet de r\'{e}aliser une mesure QND de l'intensit\'{e} sous
certaines conditions. Il faut d'une part que les deux faisceaux soient
r\'{e}sonnants avec la cavit\'{e} et d'autre part que toutes les sources de
bruit telles que le bruit thermique et le bruit de pression de radiation du
faisceau de mesure soient n\'{e}gligeables devant les fluctuations de la
pression de radiation du faisceau signal.

Nous allons dans cette section \'{e}tudier en d\'{e}tail ce dispositif. Nous
examinerons successivement les deux crit\`{e}res d'une mesure QND, \`{a}
savoir les perturbations induites par la mesure sur le faisceau signal et
les corr\'{e}lations entre les faisceaux signal et mesure. La fin de la
section est consacr\'{e}e \`{a} une situation particuli\`{e}re o\`{u} le
faisceau de mesure, tout en assurant la d\'{e}tection de l'intensit\'{e} du
faisceau signal, est comprim\'{e} par la cavit\'{e}.

\subsubsection{Perturbations du faisceau signal\label{II-4-3-1}}

Les faisceaux signal et mesure \'{e}tant r\'{e}sonnants avec la cavit\'{e},
leurs quadratures d'amplitude et de phase v\'{e}rifient les relations
\'{e}tablies au d\'{e}but de cette partie (\'{e}quations \ref{4.1} et 2.74).
On trouve en particulier que les quadratures d'amplitude ne sont pas
modifi\'{e}es : une cavit\'{e} r\'{e}sonnante se comporte comme un
dispositif transparent pour l'intensit\'{e} du champ. Aussi bien
l'intensit\'{e} moyenne que le spectre de bruit d'intensit\'{e} du faisceau
signal ne sont pas modifi\'{e}s par la cavit\'{e}: 
\begin{equation}
\left| \overline{\alpha }_{s}^{out}\right| ^{2}=\left| \overline{\alpha }%
_{s}^{in}\right| ^{2}\qquad ,\qquad S_{p,s}^{out}\left[ \Omega \right]
=S_{p,s}^{in}\left[ \Omega \right]  \label{4.21bis}
\end{equation}
L'appareil de mesure ne perturbe donc pas le signal mesur\'{e}. En fait,
tout le bruit de la mesure est report\'{e} sur la variable conjugu\'{e}e,
c'est \`{a} dire la quadrature de phase du champ r\'{e}fl\'{e}chi. On trouve
en effet \`{a} partir des \'{e}quations (2.74) que les quadratures $q_{s/m}$
des deux champs signal et mesure v\'{e}rifient la relation: 
\begin{equation}
\delta q_{s/m}^{out}\left[ \Omega \right] =\frac{\gamma +i\Omega \tau }{%
\gamma -i\Omega \tau }~\delta q_{s/m}^{in}\left[ \Omega \right] +\frac{4%
\sqrt{2\gamma }}{\gamma -i\Omega \tau }~k~\overline{\alpha }_{s/m}~\delta
x\left[ \Omega \right]  \label{4.21ter}
\end{equation}

Le d\'{e}placement $\delta x$ du miroir d\'{e}pend de la pr\'{e}sence des
deux champs dans la cavit\'{e}. Le miroir est soumis \`{a} trois forces :
les forces de pression de radiation de chacun des deux champs et la force de
Langevin associ\'{e}e au bruit thermique. Le mouvement du miroir est donc
d\'{e}crit par la relation: 
\begin{equation}
\delta x\left[ \Omega \right] =\chi \left[ \Omega \right] ~\left( 2\hbar k~%
\overline{\alpha }_{s}~\delta p_{s}\left[ \Omega \right] +2\hbar k~\overline{%
\alpha }_{m}~\delta p_{m}\left[ \Omega \right] +F_{T}\left[ \Omega \right]
\right)  \label{4.22}
\end{equation}
Les quantit\'{e}s $\overline{\alpha }_{s/m}~\delta p_{s/m}\left[ \Omega
\right] $ repr\'{e}sentent les fluctuations d'intensit\'{e} $\delta
I_{s/m}\left[ \Omega \right] $ des deux champs intracavit\'{e} (\'{e}quation 
\ref{2.56a}). Nous avons suppos\'{e} d'autre part que les deux champs
intracavit\'{e} n'interf\`{e}rent pas entre eux, par exemple en choisissant
des polarisations orthogonales. A partir de l'\'{e}quation (\ref{4.5a}), on
peut relier le d\'{e}placement du miroir aux fluctuations incidentes $\delta
p_{s/m}^{in}\left[ \Omega \right] $ et $F_{T}\left[ \Omega \right] $: 
\begin{equation}
\delta x\left[ \Omega \right] =\frac{1}{2k}\frac{\sqrt{2\gamma }}{\gamma
-i\Omega \tau }\overline{\chi }\left[ \Omega \right] ~\left( \frac{\Psi _{s}%
}{\overline{\alpha }_{s}}~\delta p_{s}^{in}\left[ \Omega \right] +\frac{\Psi
_{m}}{\overline{\alpha }_{m}}~\delta p_{m}^{in}\left[ \Omega \right] \right)
+\chi \left[ \Omega \right] ~F_{T}\left[ \Omega \right]  \label{4.23}
\end{equation}
$\Psi _{s}$ et $\Psi _{m}$ sont les d\'{e}phasages non lin\'{e}aires
produits respectivement par les pressions de radiation moyennes des
faisceaux signal et mesure (\'{e}quation \ref{2.47}): 
\begin{equation}
\Psi _{s/m}=4\hbar k^{2}~\overline{\alpha }_{s/m}^{2}~\chi \left[ 0\right]
\label{4.23bis}
\end{equation}
Les relations (\ref{4.21ter}) et (\ref{4.23}) montrent que les quadratures
de phase des faisceaux r\'{e}fl\'{e}chis sont modifi\'{e}es par la
cavit\'{e}. En ce qui concerne le faisceau de mesure, cela correspond au
processus m\^{e}me de mesure, puisque cela conduit \`{a} des
corr\'{e}lations entre le signal $\delta p_{s}^{in}$ et la phase $\delta
q_{m}^{out}$ du faisceau de mesure. Pour le faisceau signal, ces relations
traduisent la perturbation provoqu\'{e}e par le processus de mesure : tout
le bruit est report\'{e} sur la phase du faisceau r\'{e}fl\'{e}chi. Le
dispositif est donc non destructif pour l'intensit\'{e} du faisceau signal.

\subsubsection{Corr\'{e}lations signal-mesure\label{II-4-3-2}}

L'efficacit\'{e} quantique de la mesure d\'{e}pend du niveau de
corr\'{e}lation entre les fluctuations $\delta q_{m}^{out}\left[ \Omega
\right] $ de la quadrature de phase du faisceau de mesure r\'{e}fl\'{e}chi
et les fluctuations d'amplitude $\delta p_{s}^{in}\left[ \Omega \right] $ du
faisceau signal incident. A partir des \'{e}quations (\ref{4.21ter}) et (\ref
{4.23}), on voit appara\^{\i }tre dans l'expression de $\delta
q_{m}^{out}\left[ \Omega \right] $ un terme en $\delta p_{s}^{in}\left[
\Omega \right] $ qui est \`{a} l'origine de ces corr\'{e}lations. Les autres
termes apparaissent comme des sources de bruit qui tendent \`{a} r\'{e}duire
l'efficacit\'{e} de la mesure. Le premier terme de l'\'{e}quation (\ref
{4.21ter}) correspond au bruit de phase du faisceau de mesure incident. Les
deux derniers termes de l'\'{e}quation (\ref{4.23}) d\'{e}crivent les bruits
li\'{e}s au mouvement du miroir mobile sous l'effet des fluctuations de la
pression de radiation du faisceau de mesure et des fluctuations thermiques.

Nous allons \`{a} pr\'{e}sent estimer l'efficacit\'{e} de la mesure en
exprimant la fonction de corr\'{e}lation entre l'intensit\'{e} du signal et
la phase du faisceau de mesure. Cette fonction de corr\'{e}lation ${\cal C}%
_{sm}\left[ \Omega \right] $ est d\'{e}finie par: 
\begin{equation}
\left\langle \delta p_{s}^{in}\left[ \Omega \right] ~\delta
q_{m}^{out}\left[ \Omega ^{\prime }\right] \right\rangle =2\pi \delta \left(
\Omega +\Omega ^{\prime }\right) ~\sqrt{S_{p,s}^{in}\left[ \Omega \right]
S_{q,m}^{out}\left[ \Omega \right] }~{\cal C}_{sm}\left[ \Omega \right]
\label{4.25}
\end{equation}
La fonction $\left| {\cal C}_{sm}\left[ \Omega \right] \right| ^{2}$ varie
de $0$ lorsque les fluctuations ne sont pas corr\'{e}l\'{e}es, \`{a} $1$
pour des fluctuations parfaitement corr\'{e}l\'{e}es. Puisque les champs
incidents sont dans des \'{e}tats coh\'{e}rents, toutes les sources de bruit
qui apparaissent dans l'expression de $\delta q_{m}^{out}$ sont
ind\'{e}pendantes, avec un spectre de bruit \'{e}gal \`{a} $1$ pour les
champs incidents et \`{a} $S_{T}\left[ \Omega \right] $ pour les
fluctuations thermiques (\'{e}quation \ref{2.44}). Le spectre $S_{q,m}^{out}$
des fluctuations de phase du faisceau de mesure r\'{e}fl\'{e}chi et la
fonction de corr\'{e}lation $\left| {\cal C}_{sm}\left[ \Omega \right]
\right| ^{2}$ peuvent ainsi s'\'{e}crire sous la forme: 
\begin{subequations}
\label{4.26}
\begin{eqnarray}
S_{q,m}^{out}\left[ \Omega \right] &=&{\cal S}\left[ \Omega \right] +{\cal N}%
\left[ \Omega \right]  \label{4.26a} \\
\left| {\cal C}_{sm}\left[ \Omega \right] \right| ^{2} &=&\frac{{\cal S}%
\left[ \Omega \right] }{{\cal S}\left[ \Omega \right] +{\cal N}\left[ \Omega
\right] }  \label{4.26b}
\end{eqnarray}
o\`{u} le ''signal'' ${\cal S}\left[ \Omega \right] $ d\'{e}crit la
contribution des fluctuations d'intensit\'{e} du signal et ${\cal N}\left[
\Omega \right] $ celle des diff\'{e}rentes sources de bruit: 
\end{subequations}
\begin{subequations}
\label{4.27}
\begin{eqnarray}
{\cal S}\left[ \Omega \right] &=&16\frac{\gamma ^{4}}{\left( \gamma
^{2}+\Omega ^{2}\tau ^{2}\right) ^{2}}~\left| \overline{\chi }\left[ \Omega
\right] \right| ^{2}~\frac{\Psi _{s}\Psi _{m}}{\gamma ^{2}}%
~S_{p,s}^{in}\left[ \Omega \right]  \label{4.27a} \\
{\cal N}\left[ \Omega \right] &=&S_{q,m}^{in}\left[ \Omega \right] +16\frac{%
\gamma ^{4}}{\left( \gamma ^{2}+\Omega ^{2}\tau ^{2}\right) ^{2}}~\left| 
\overline{\chi }\left[ \Omega \right] \right| ^{2}~\frac{\Psi _{m}^{2}}{%
\gamma ^{2}}~S_{p,m}^{in}\left[ \Omega \right]  \label{4.27b}
\end{eqnarray}
\end{subequations}
\[
\qquad \qquad \qquad \quad ~+16\frac{\gamma ^{2}}{\left( \gamma ^{2}+\Omega
^{2}\tau ^{2}\right) }~{\cal I}m\left( \overline{\chi }\left[ \Omega \right]
\right) ~\frac{\Psi _{m}}{\gamma }~\frac{k_{B}T}{\hbar \Omega } 
\]
Les relations(\ref{4.26}) montrent que la fonction $\left| {\cal C}%
_{sm}\left[ \Omega \right] \right| ^{2}$ est \'{e}gale \`{a} la contribution
relative du signal ${\cal S}$ au spectre total $S_{q,m}^{out}\left[ \Omega
\right] $ du faisceau de mesure. Le signal ${\cal S}\left[ \Omega \right] $
est proportionnel au produit des intensit\'{e}s moyennes des deux champs
(terme $\Psi _{s}\Psi _{m}$) alors que le bruit ${\cal N}\left[ \Omega
\right] $ ne d\'{e}pend pas de $\Psi _{s}$. On peut donc obtenir des
corr\'{e}lations arbitrairement grandes en prenant un faisceau signal
suffisamment intense. On retrouve dans l'expression du bruit ${\cal N}\left[
\Omega \right] $ la contribution des trois sources de bruit qui
interviennent dans le processus de mesure. Le premier terme est associ\'{e}
au bruit quantique standard du faisceau de mesure incident et les deux
derniers termes sont associ\'{e}s aux fluctuations de position du miroir
dues respectivement au bruit de pression de radiation du faisceau de mesure
et au bruit thermique.

\begin{figure}[tbp]
\centerline{\psfig{figure=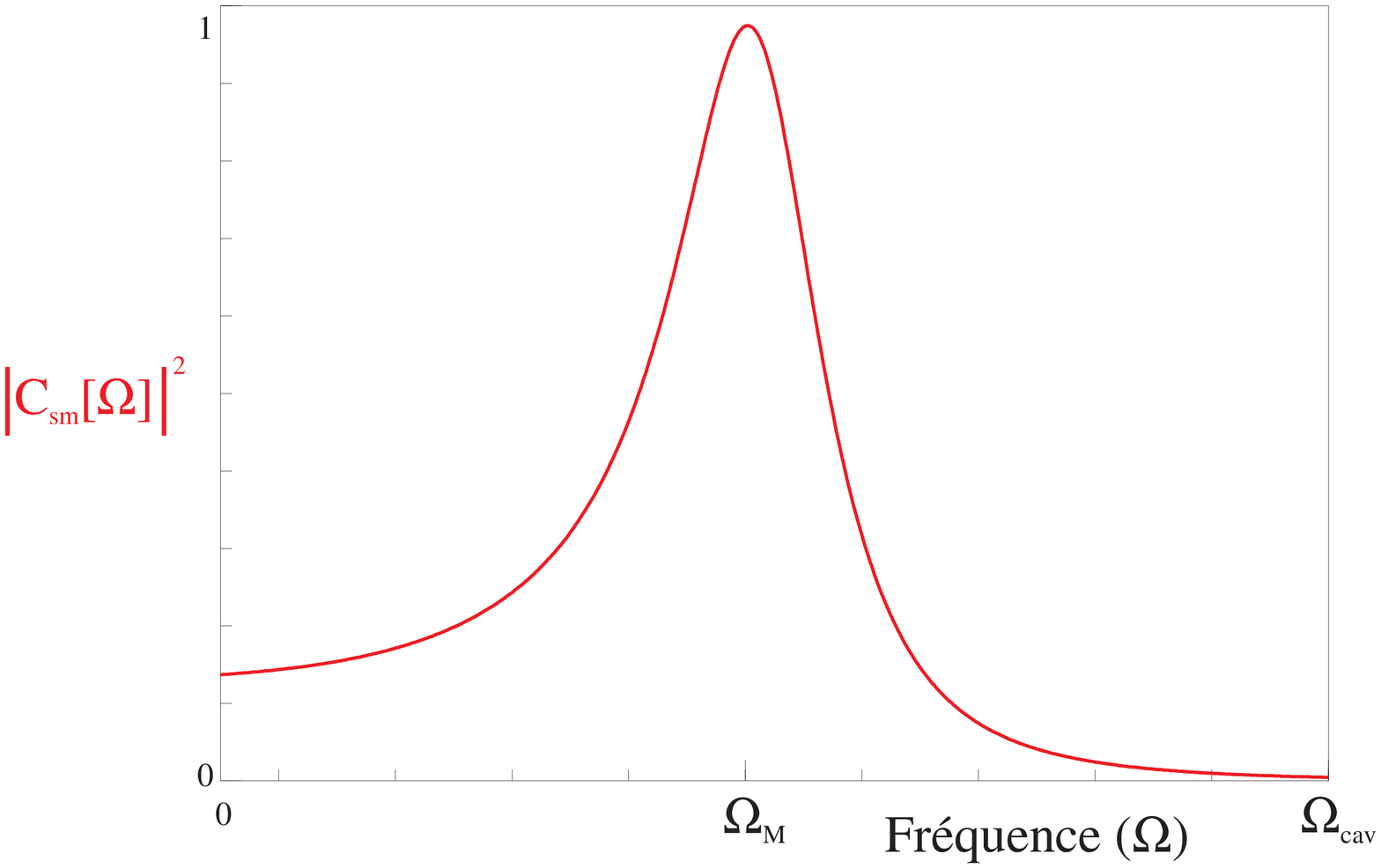,height=7cm}}
\caption{Spectre des corr\'{e}lations quantiques entre la quadrature
d'amplitude du signal et la quadrature de phase du faisceau de mesure. La
cavit\'{e} refroidie \`{a} une temp\'{e}rature de $1K$ est r\'{e}sonnante
avec les deux faisceaux. Les d\'{e}phasages non lin\'{e}aires des deux
faisceaux sont $\Psi _{s}=\gamma $ et $\Psi _{m}=\gamma /100$ }
\label{Fig_2csmint}
\end{figure}

La figure \ref{Fig_2csmint} montre l'\'{e}volution en fr\'{e}quence de la
fonction de corr\'{e}lation, en supposant que le mouvement du miroir peut
\^{e}tre d\'{e}crit par un oscillateur harmonique et pour des param\`{e}tres
similaires \`{a} ceux choisis pour la mesure du bruit thermique
(fr\'{e}quence de r\'{e}sonance $\Omega _{M}=10^{7}~rad/s$, masse $M=1~mg$,
facteur de qualit\'{e} $Q=10^{6}$, finesse de la cavit\'{e} ${\cal F}%
=3~10^{5}$, bande passante de la cavit\'{e} $\Omega _{cav}=2\Omega _{M}$).
Afin de r\'{e}duire les sources de bruit, on suppose la temp\'{e}rature
\'{e}gale \`{a} $1~Kelvin$ (nombre de phonons thermiques $n_{T}=10^{4}$).
Les corr\'{e}lations sont d'autant plus importantes que le faisceau signal
est intense. On choisit donc un d\'{e}phasage non lin\'{e}aire $\Psi _{s}$
pour le signal \'{e}gal \`{a} $\gamma $. Les corr\'{e}lations d\'{e}pendent
aussi de l'intensit\'{e} du faisceau de mesure. Si le d\'{e}phasage non
lin\'{e}aire $\Psi _{m}$ est trop petit, le bruit de phase du faisceau de
mesure (terme $S_{q,m}^{in}\left[ \Omega \right] =1$ dans ${\cal N}$)
devient dominant devant le signal, qui est proportionnel \`{a} $\Psi _{m}$.
Par contre, si $\Psi _{m}$ est trop grand, la pression de radiation du
faisceau de mesure (terme en $\Psi _{m}^{2}$ dans ${\cal N}$) n'est plus
n\'{e}gligeable, ce qui conduit \`{a} une diminution des corr\'{e}lations.
La figure \ref{Fig_2csmint} a \'{e}t\'{e} trac\'{e}e pour un d\'{e}phasage $%
\Psi _{m}$ \'{e}gal \`{a} $\gamma /100$. Pour une masse de $1~mg$, ces
d\'{e}phasages correspondent \`{a} des puissances incidentes de $60~mWatt$
pour le faisceau signal et de $600~\mu Watt$ pour le faisceau de mesure. 

Les corr\'{e}lations signal-mesure reproduisent la d\'{e}pendance en
fr\'{e}quence de la r\'{e}ponse m\'{e}canique du miroir. Elles sont
maximales \`{a} r\'{e}sonance, o\`{u} des valeurs sup\'{e}rieures \`{a} $%
95\% $ sont atteintes. On voit d'autre part que la largeur des
corr\'{e}lations est beaucoup plus grande que la largeur $\Omega
_{M}/Q=10^{-6}\Omega _{M}$ de la r\'{e}sonance m\'{e}canique. Ceci peut
s'expliquer par la d\'{e}pendance en fr\'{e}quence des diff\'{e}rentes
sources de bruit. Si l'on n\'{e}glige l'effet de filtrage de la cavit\'{e}
qui n'est pas significatif au voisinage de la r\'{e}sonance m\'{e}canique,
les expressions de ${\cal S}$ et ${\cal N}$ (\'{e}quations \ref{4.27}) se
simplifient: 
\begin{subequations}
\label{4.28}
\begin{eqnarray}
{\cal S}\left[ \Omega \right] &\approx &16~\left| \overline{\chi }\left[
\Omega \right] \right| ^{2}~\frac{\Psi _{s}\Psi _{m}}{\gamma ^{2}}
\label{4.28a} \\
{\cal N}\left[ \Omega \right] &\approx &1+16~\left| \overline{\chi }\left[
\Omega \right] \right| ^{2}~\frac{\Psi _{m}^{2}}{\gamma ^{2}}+16~\left| 
\overline{\chi }\left[ \Omega \right] \right| ^{2}~\frac{\Psi _{m}}{\gamma }~%
\frac{n_{T}}{Q}  \label{4.28b}
\end{eqnarray}
o\`{u} $n_{T}$ est le nombre de phonons thermiques donn\'{e} par
l'\'{e}quation (\ref{2.71}). Notons que l'on retrouve ici les param\`{e}tres
essentiels du couplage optom\'{e}canique : les effets de pression de
radiation d\'{e}pendent du rapport entre les d\'{e}phasages non
lin\'{e}aires et les pertes $\gamma $ de la cavit\'{e}, tandis que les
effets thermiques sont proportionnels au rapport $n_{T}/Q$.

Les bruits associ\'{e}s au mouvement du miroir (deuxi\`{e}me et
troisi\`{e}me termes dans l'\'{e}quation \ref{4.28b}) ont la m\^{e}me
d\'{e}pendance en fr\'{e}quence que le signal ${\cal S}\left[ \Omega \right] 
$ et repr\'{e}sentent une petite perturbation au signal (environ $2\%$ du
signal pour les param\`{e}tres utili-\newline
s\'{e}s). En dehors de la r\'{e}sonance
m\'{e}canique, les corr\'{e}lations sont donc essentiellement limit\'{e}es
par le bruit de phase du faisceau de mesure incident (premier terme de
l'\'{e}quation \ref{4.28b}). Les corr\'{e}lations sont importantes si le
signal est sup\'{e}rieur au bruit du faisceau de mesure, c'est \`{a} dire $%
{\cal S}\geq 1$, ou encore $\left| \overline{\chi }\left[ \Omega \right]
\right| \geq 2$. Cette condition peut \^{e}tre satisfaite sur une bande de
fr\'{e}quence beaucoup plus grande que la largeur de la r\'{e}sonance
m\'{e}canique, d\'{e}finie par $\left| \overline{\chi }\left[ \Omega \right]
\right| \geq \left| \overline{\chi }\left[ \Omega _{M}\right] \right| /\sqrt{%
2}=Q/\sqrt{2}$.

\subsubsection{Mesure QND avec un faisceau auto-comprim\'{e}\label{II-4-3-3}}

Nous venons de montrer que les corr\'{e}lations signal-mesure peuvent
\^{e}tre tr\`{e}s importantes au voisinage de la fr\'{e}quence de
r\'{e}sonance m\'{e}canique. En dehors de cette plage de fr\'{e}quence, les
corr\'{e}lations sont limit\'{e}es par le bruit propre du faisceau de
mesure, qui est \'{e}gal au bruit du faisceau incident lorsque celui-ci est
r\'{e}sonnant avec la cavit\'{e}. Cette condition de r\'{e}sonance appara\^{\i }t en fait
comme une configuration optimale dans le cas o\`{u} l'intensit\'{e} du
faisceau de mesure est faible. Dans un r\'{e}gime de forte intensit\'{e}
(d\'{e}phasage $\Psi _{m}$ de l'ordre de $\gamma $), on sait que le couplage
optom\'{e}canique peut produire un faisceau de mesure dans un \'{e}tat
comprim\'{e}, c'est \`{a} dire un faisceau r\'{e}fl\'{e}chi dont le bruit
propre est r\'{e}duit (voir partie 2.3). Un choix judicieux du d\'{e}saccord
de la cavit\'{e} $\overline{\Psi }$ et de la quadrature utilis\'{e}e pour
r\'{e}aliser la mesure devrait donc permettre d'am\'{e}liorer les corr\'{e}lations
signal-mesure. Bien s\^{u}r, le faisceau signal doit quant \`{a} lui rester r%
\'{e}sonnant avec la cavit\'{e}, de fa\c{c}on \`{a} ce que son intensit\'{e}
ne soit pas perturb\'{e}e par la mesure.

Comme nous l'avons vu dans la partie 2.3, on peut exprimer le spectre de
bruit du faisceau r\'{e}fl\'{e}chi pour n'importe quelle quadrature $\delta
\alpha _{\theta }^{out}$ (\'{e}quation \ref{2.62}) en fonction des diff\'{e}rentes
sources de bruit. Il faut tenir compte ici de la pr\'{e}sence d'une source
de bruit suppl\'{e}mentaire due \`{a} la pression de radiation du faisceau
signal. L'action de cette force fluctuante sur le faisceau de mesure est
tout \`{a} fait similaire \`{a} celle exerc\'{e}e par la force de Langevin $%
F_{T}$. Il est donc possible de faire une \'{e}quivalence entre le
syst\`{e}me \`{a} un seul faisceau, \'{e}tudi\'{e} dans la partie 2.3, et
celui \`{a} deux faisceaux. Pour d\'{e}terminer l'expression du spectre de
bruit de la composante $\alpha _{\theta ,m}^{out}$ du faisceau de mesure
r\'{e}fl\'{e}chi, il suffit de substituer le terme $S_{T}$ dans la relation
(\ref{2.62}) par le terme $S_{T}+S_{F}$ o\`{u} le spectre $S_{F}$ d\'{e}crit le
bruit de pression de radiation du faisceau signal, que l'on peut d\'{e}duire
de l'\'{e}quation (\ref{4.23}): 
\end{subequations}
\begin{equation}
S_{F}\left[ \Omega \right] =\frac{2\gamma }{\gamma ^{2}+\Omega ^{2}\tau ^{2}}%
\frac{\hbar \Psi _{s}}{\chi \left[ 0\right] }~S_{p,s}^{in}\left[ \Omega
\right]  \label{4.29}
\end{equation}
On obtient ainsi l'expression du spectre de bruit du faisceau de mesure $%
S_{\theta ,m}^{out}\left[ \Omega \right] $ en fonction des fluctuations de
la force de pression de radiation du signal $S_{F}\left[ \Omega \right] $,
du bruit thermique $S_{T}\left[ \Omega \right] $ et des coefficients ${\cal C%
}^{in}\left[ \Omega \right] $ et ${\cal C}_{T}\left[ \Omega \right] $
donn\'{e}s par les \'{e}quations (\ref{2.63}): 
\begin{equation}
S_{\theta ,m}^{out}\left[ \Omega \right] =\frac{1}{2}\left( {\cal C}%
^{in}\left[ \Omega \right] +{\cal C}^{in}\left[ -\Omega \right] \right) +%
{\cal C}_{T}\left[ \Omega \right] ~S_{T}\left[ \Omega \right] +{\cal C}%
_{T}\left[ \Omega \right] ~S_{F}\left[ \Omega \right]  \label{4.30}
\end{equation}

Le premier terme repr\'{e}sente le bruit propre du faisceau
r\'{e}fl\'{e}chi, \'{e}quivalent aux deux premiers termes de l'expression de 
${\cal N}$ trouv\'{e}e dans la section pr\'{e}c\'{e}dente (\'{e}quation \ref
{4.28b}). Selon le point de fonctionnement choisi et la quadrature $\theta $
mesur\'{e}e, ce bruit propre peut \^{e}tre plus petit que $1$. Le
deuxi\`{e}me terme repr\'{e}sente l'effet du bruit thermique (\'{e}quivalent
au dernier terme de ${\cal N}$). Le dernier terme est \`{a} l'origine des
corr\'{e}lations quantiques signal-mesure. La contribution relative de ce
terme au spectre $S_{\theta ,m}^{out}$ est par d\'{e}finition \'{e}gale
\`{a} la fonction de corr\'{e}lation $\left| {\cal C}_{sm}\left[ \Omega
\right] \right| ^{2}$: 
\begin{equation}
\left| {\cal C}_{sm}\left[ \Omega \right] \right| ^{2}=\frac{{\cal C}%
_{T}\left[ \Omega \right] ~S_{F}\left[ \Omega \right] }{S_{\theta
,m}^{out}\left[ \Omega \right] }  \label{4.31}
\end{equation}

\begin{figure}[tbp]
\centerline{\psfig{figure=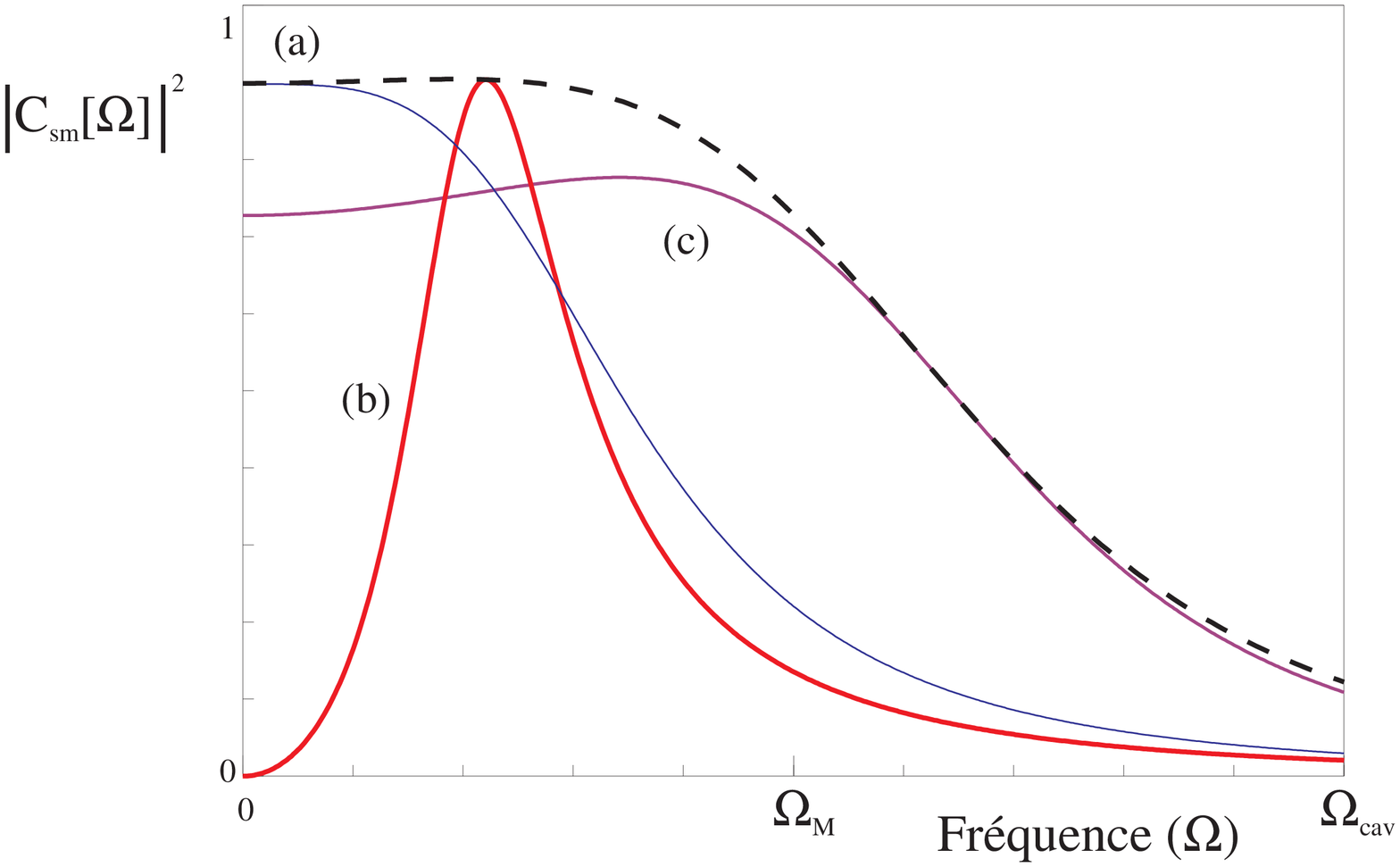,height=8cm}}
\caption{Corr\'{e}lations quantiques entre l'intensit\'{e} du signal et
diff\'{e}rentes quadratures du faisceau de mesure r\'{e}fl\'{e}chi. Les
courbes (a), (b) et (c) sont obtenues respectivement pour des quadratures
d'angle $\theta $ \'{e}gal \`{a} $15^{\circ }$, $0^{\circ }$ et $-60^{\circ
} $. La courbe en tirets repr\'{e}sente le maximum des corr\'{e}lations
atteintes lorsque $\theta $ varie. Le point de fonctionnement est d\'{e}fini
par le d\'{e}saccord du faisceau de mesure $\overline{\Psi }=-2\gamma $, les
d\'{e}phasages $\Psi _{s}=\Psi _{m}=\gamma $ et une temp\'{e}rature $%
T=1~Kelvin$}
\label{Fig_2csmteta}
\end{figure}

La figure \ref{Fig_2csmteta} montre la fonction de corr\'{e}lation $\left| 
{\cal C}_{sm}\left[ \Omega \right] \right| ^{2}$ obtenue pour les m\^{e}mes
param\`{e}tres que dans le cas r\'{e}sonant, sauf pour l'intensit\'{e} du
faisceau de mesure qui est \'{e}gale \`{a} l'intensit\'{e} du signal ($\Psi
_{m}=\Psi _{s}=\gamma $), et pour le d\'{e}saccord de la cavit\'{e} $%
\overline{\Psi }$ \'{e}gal \`{a} $-2\gamma $. Ces valeurs correspondent au
m\^{e}me point de fonctionnement de la cavit\'{e} que celui choisi dans la
partie 2.3. 
Les trois courbes sont obtenues pour diff\'{e}rentes quadratures du faisceau
de mesure r\'{e}fl\'{e}chi.

\begin{figure}[tbp]
\centerline{\psfig{figure=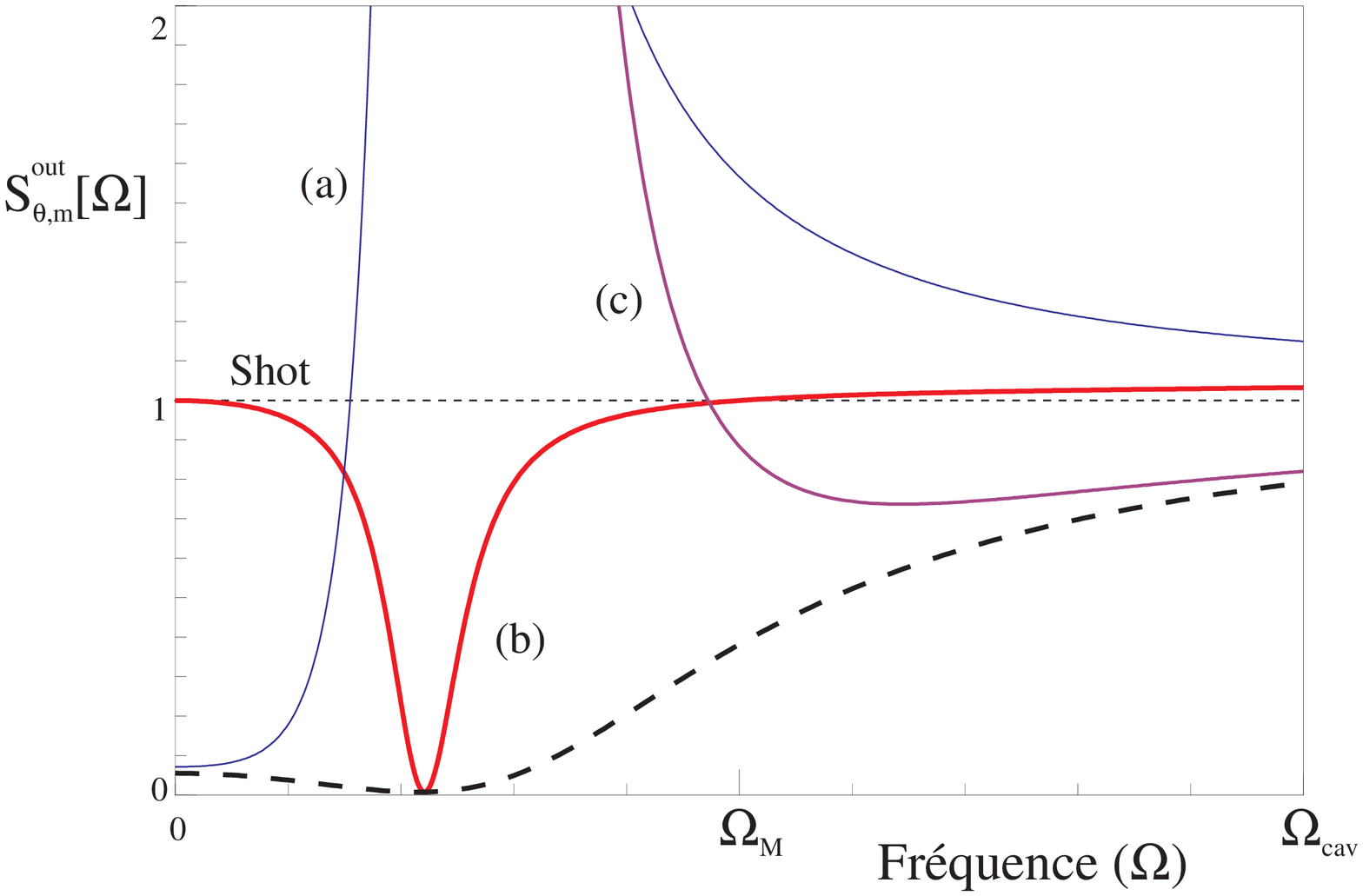,height=8cm}}
\caption{Spectre de bruit propre du faisceau de mesure r\'{e}fl\'{e}chi pour
diff\'{e}rentes quadratures, \`{a} temp\'{e}rature nulle et en absence de
faisceau signal. Le point de fonctionnement de la cavit\'{e} est fix\'{e}
par le d\'{e}saccord $\overline{\Psi }=-2\gamma $ et le d\'{e}phasage $\Psi
_{m}=\gamma $. Les courbes (a), (b) et (c) sont obtenues respectivement pour
des quadratures d'angle $\theta $ \'{e}gal \`{a} $15^{\circ }$, $0^{\circ }$
et $-60^{\circ }$. La courbe en tirets repr\'{e}sente le spectre optimum qui
d\'{e}crit le niveau de bruit minimum atteint lorsque $\theta $ varie}
\label{Fig_2spcteta}
\end{figure}

On peut comparer ces courbes \`{a} celles de la figure \ref{Fig_2spcteta}
qui repr\'{e}sentent le spectre de bruit $S_{\theta ,m}^{out}$ pour les
m\^{e}mes quadratures, en absence de signal et \`{a} temp\'{e}rature nulle.
La figure \ref{Fig_2spcteta} montre en fait le bruit propre du faisceau
r\'{e}fl\'{e}chi, donn\'{e} par le terme $\frac{1}{2}\left( {\cal C}%
^{in}\left[ \Omega \right] +{\cal C}^{in}\left[ -\Omega \right] \right) $
dans l'expression de $S_{\theta ,m}^{out}$ (\'{e}quation \ref{4.30}). Les
fr\'{e}quences pour lesquelles le bruit propre est inf\'{e}rieur au bruit
quantique standard d\'{e}pendent de la quadrature que l'on consid\`{e}re. Il
est de ce fait possible d'atteindre un niveau de bruit propre du faisceau de
mesure inf\'{e}rieur \`{a} $1$ sur l'ensemble de la bande passante de la
cavit\'{e} en variant l'angle $\theta $ (courbe en tirets de la figure \ref
{Fig_2spcteta}). En comparant les deux figures, on constate que les
corr\'{e}lations sont maximales lorsque le bruit propre du faisceau
r\'{e}fl\'{e}chi est minimum.

On parcourt ainsi la fonction de corr\'{e}lation optimale (courbe en tirets
de la figure \ref{Fig_2csmteta}) lorsqu'on choisit \`{a} chaque
fr\'{e}quence la quadrature qui pr\'{e}sente le bruit minimal. 
Contrairement \`{a} la configuration utilisant un faisceau de mesure
r\'{e}sonnant et de faible intensit\'{e} (figure \ref{Fig_2csmint}), o\`{u}
les corr\'{e}lations sont essentiellement concentr\'{e}es autour de la
r\'{e}sonance m\'{e}canique, on peut obtenir ici de fortes corr\'{e}lations
sur toute la plage de fr\'{e}quence d\'{e}finie par la bande passante de la
cavit\'{e}.

\section{Conclusion\label{II-5}}

\bigskip

Nous avons pr\'{e}sent\'{e} dans ce chapitre les propri\'{e}t\'{e}s
g\'{e}n\'{e}rales du couplage optom\'{e}canique. Nous avons montr\'{e}
qu'une cavit\'{e} de grande finesse dont un miroir est mobile peut \^{e}tre
utilis\'{e}e pour mettre en \'{e}vidence les effets quantiques dus \`{a} la
pression de radiation. Il est ainsi possible de contr\^{o}ler les
fluctuations de la lumi\`{e}re en produisant un \'{e}tat comprim\'{e}, ou
encore de cr\'{e}er des corr\'{e}lations quantiques entre la position du
miroir mobile et l'intensit\'{e} lumineuse. D'autre part, une telle
cavit\'{e} permet de mesurer de tr\`{e}s petits d\'{e}placements du miroir
mobile. Il devrait ainsi \^{e}tre possible de mesurer le bruit thermique du
miroir mobile ou encore de r\'{e}aliser une mesure quantique non destructive
de l'intensit\'{e} de la lumi\`{e}re.

Nous avons cherch\'{e} \`{a} d\'{e}gager les param\`{e}tres physiques
importants. Malgr\'{e} le grand nombre de caract\'{e}ristiques du
syst\`{e}me pouvant intervenir (propri\'{e}t\'{e}s optiques de la cavit\'{e}
et caract\'{e}ristiques m\'{e}caniques du miroir mobile), l'efficacit\'{e}
du couplage optom\'{e}canique d\'{e}pend essentiellement de deux
param\`{e}tres :

{\bf -} Les effets li\'{e}s \`{a} la pression de radiation sont
significatifs lorsque le d\'{e}phasage non lin\'{e}aire $\Psi _{NL}$ est de
l'ordre des pertes $\gamma $ de la cavit\'{e}, c'est \`{a} dire lorsque le
d\'{e}placement moyen $\overline{x}$ du miroir produit par la pression de
radiation est de l'ordre de la largeur $\lambda /2{\cal F}$ de la
r\'{e}sonance optique.

{\bf -} Les effets thermiques sont proportionnels au rapport $n_{T}/Q$ entre
le nombre de phonons thermiques \`{a} la fr\'{e}quence de r\'{e}sonance
m\'{e}canique et le facteur de qualit\'{e} de la r\'{e}sonance.

Afin de r\'{e}duire les effets thermiques, le r\'{e}sonateur doit avoir un
grand facteur de qualit\'{e} ($Q\geq 10^{6}$). Il est aussi n\'{e}cessaire
de travailler \`{a} basse temp\'{e}rature et avec un miroir mobile dont la
fr\'{e}quence de r\'{e}sonance est \'{e}lev\'{e}e : pour une temp\'{e}rature
de $1~Kelvin$ et une fr\'{e}quence de r\'{e}sonance de $10^{7}~rad/s$, le
nombre $n_{T}$ de phonons thermiques est de l'ordre de $10^{4}$.

La condition $\Psi _{NL}\approx \gamma $ est plus difficile \`{a}
\'{e}valuer. Elle fait intervenir \`{a} la fois les caract\'{e}ristiques
optiques (finesse de la cavit\'{e}, puissance lumineuse incidente) et
m\'{e}caniques (r\'{e}ponse \`{a} basse fr\'{e}quence du r\'{e}sonateur).
Nous avons jusqu'\`{a} pr\'{e}sent utilis\'{e} un mod\`{e}le d'oscillateur
harmonique pour d\'{e}crire la r\'{e}ponse m\'{e}canique du r\'{e}sonateur.
Dans le cadre de ce mod\`{e}le, la r\'{e}ponse \`{a} basse fr\'{e}quence
d\'{e}pend essentiellement de la masse du r\'{e}sonateur, qui doit \^{e}tre
aussi petite que possible ($M\leq 1~mg$).

Ces diff\'{e}rentes contraintes nous ont amen\'{e}s \`{a} choisir un
r\'{e}sonateur m\'{e}canique cons- titu\'{e} d'un substrat en silice de
structure plan-convexe, plut\^{o}t qu'un syst\`{e}me pendulaire. Ceci doit
permettre d'atteindre des fr\'{e}quences de r\'{e}sonance \'{e}lev\'{e}es et
une faible masse. Le prochain chapitre est consacr\'{e} \`{a} l'\'{e}tude
th\'{e}orique de ce r\'{e}sonateur. Nous verrons en particulier que la
complexit\'{e} de la r\'{e}ponse m\'{e}canique d'un tel r\'{e}sonateur peut
\^{e}tre int\'{e}gr\'{e}e dans la d\'{e}finition d'une susceptibilit\'{e}
effective. On peut alors d\'{e}finir une masse effective qui caract\'{e}rise
la r\'{e}ponse \`{a} basse fr\'{e}quence du r\'{e}sonateur. Cette masse
effective d\'{e}pend de la g\'{e}om\'{e}trie du syst\`{e}me et peut \^{e}tre
inf\'{e}rieure au milligramme. \newpage

%% file: chapter3.tex
\chapter{LE RESONATEUR MECANIQUE}

\bigskip \bigskip

Nous avons pr\'{e}sent\'{e} dans le chapitre pr\'{e}c\'{e}dent les
caract\'{e}ristiques g\'{e}n\'{e}rales du couplage optom\'{e}canique, en
supposant que le miroir subit un mouvement d'ensemble cara-ct\'{e}ris\'{e}
par un d\'{e}placement global $x$. Ceci nous a permis de simplifier la
description du syst\`{e}me puisqu'\`{a} la fois le champ et la cavit\'{e}
\`{a} miroir mobile peuvent \^{e}tre d\'{e}crits dans le cadre d'un
mod\`{e}le monodimensionnel. 
\begin{figure}[tbp]
\centerline{\psfig{figure=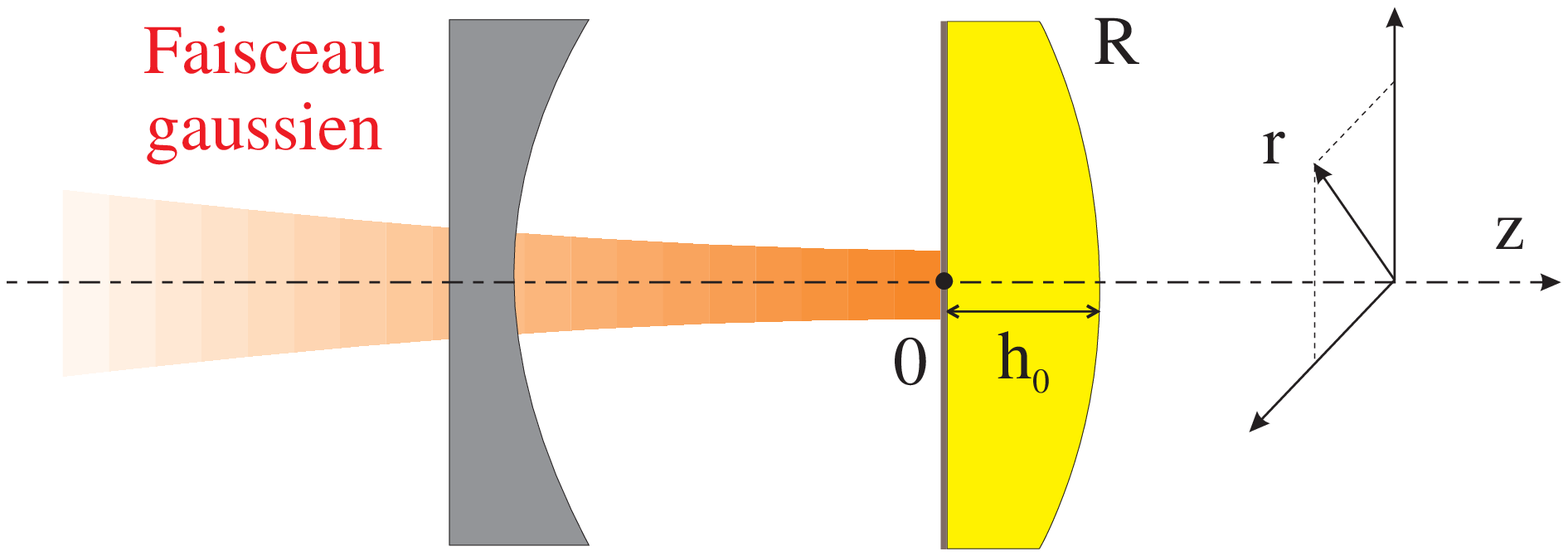,height=5cm}}
\caption{G\'{e}ometrie du r\'{e}sonateur m\'{e}canique plan-convexe. Le
miroir mobile est d\'{e}pos\'{e} sur la face plane du r\'{e}sonateur
caract\'{e}ris\'{e} par une \'{e}paisseur au centre $h_{0}$ et un rayon de
courbure $R$ de la face convexe. Le faisceau lumineux a une structure
gaussienne correspondant au mode fondamental $TEM_{00}$ de la cavit\'{e}}
\label{Fig_2crisp_c}
\end{figure}

Nous avons montr\'{e} cependant que la principale limitation \`{a}
l'observation d'effets quantiques est due au mouvement thermique du miroir,
m\^{e}me \`{a} tr\`{e}s basse temp\'{e}rature. Afin de r\'{e}duire ces
effets, il est n\'{e}cessaire d'utiliser un syst\`{e}me m\'{e}canique ayant
de grands facteurs de qualit\'{e} et des fr\'{e}quences de r\'{e}sonance
\'{e}lev\'{e}es, de l'ordre du m\'{e}gahertz. Des fr\'{e}quences aussi
\'{e}lev\'{e}es ne peuvent pas correspondre \`{a} un mouvement d'ensemble du
miroir : ces fr\'{e}quences correspondent aux r\'{e}sonances des modes
acoustiques internes du r\'{e}sonateur, qui induisent des d\'{e}formations
de la surface du miroir.

Nous utilisons dans l'exp\'{e}rience un r\'{e}sonateur m\'{e}canique
constitu\'{e} d'un substrat en silice tr\`{e}s pure, de structure
plan-convexe (figure \ref{Fig_2crisp_c}). Le miroir est form\'{e} de couches
multidi\'{e}lectriques d\'{e}pos\'{e}es sur la face plane du r\'{e}sonateur.
Le mouvement du miroir est d\^{u} aux d\'{e}formations de la face plane du
r\'{e}sonateur produites par les modes de vibrations acoustiques internes du
substrat. Le choix de la g\'{e}om\'{e}trie plan-convexe permet un
confinement radial des modes acoustiques au centre du r\'{e}sonateur. Cette
propri\'{e}t\'{e} permet d'obtenir des facteurs de qualit\'{e}
m\'{e}caniques \'{e}lev\'{e}s, ind\'{e}pendamment de la fa\c{c}on dont le
r\'{e}sonateur est tenu sur son bord ext\'{e}rieur cylindrique.

Ce chapitre est consacr\'{e} \`{a} l'\'{e}tude du dispositif
repr\'{e}sent\'{e} sur la figure \ref{Fig_2crisp_c}. Nous allons montrer
qu'il est possible de se ramener \`{a} une description monodimensionnelle,
en int\'{e}grant la structure spatiale dans une susceptibilit\'{e} effective
qui d\'{e}crit l'effet sur le champ de la r\'{e}ponse m\'{e}canique du
miroir \`{a} la pression de radiation du faisceau lumineux. Nous
commencerons par \'{e}tudier l'effet d'une d\'{e}formation du r\'{e}sonateur
sur le champ (partie 3.1). Nous montrerons que le d\'{e}phasage subi par le
champ ne d\'{e}pend que du d\'{e}placement du miroir moyenn\'{e} sur la
section du faisceau lumineux. Nous \'{e}tudierons ensuite le mouvement du
r\'{e}sonateur lorsque sa face plane est soumise \`{a} une force
ext\'{e}rieure, en d\'{e}composant ce mouvement sur l'ensemble des modes
acoustiques (partie 3.2). Nous d\'{e}finirons alors la susceptibilit\'{e}
effective qui d\'{e}crit la r\'{e}ponse m\'{e}canique du r\'{e}sonateur sous
l'effet de la pression de radiation du champ intracavit\'{e} (partie 3.3).
Nous pourrons alors g\'{e}n\'{e}raliser les r\'{e}sultats obtenus dans le
chapitre pr\'{e}c\'{e}dent au cas du r\'{e}sonateur plan-convexe, et relier
le comportement du r\'{e}sonateur \`{a} basse fr\'{e}quence \`{a} sa masse
effective (partie 3.4).

\section{Effet d'une d\'{e}formation du r\'{e}sonateur sur le champ \label%
{III-1}}

\bigskip

Nous allons d\'{e}crire dans cette partie l'effet d'une d\'{e}formation
quelconque de la face plane du r\'{e}sonateur sur le champ dans la
cavit\'{e}. Etant donn\'{e} la sym\'{e}trie du syst\`{e}me, on utilise des
coordonn\'{e}es cylindriques d'axe $Oz$ dont l'origine $z=0$ se situe sur la
face plane du r\'{e}sonateur (figure \ref{Fig_2crisp_c}). Une
d\'{e}formation longitudinale de cette face peut \^{e}tre d\'{e}crite par un
d\'{e}placement $u(r,z=0,t)$ selon $Oz$ en tout point $r$ de la surface.

La structure spatiale du champ \'{e}lectromagn\'{e}tique se d\'{e}duit de
l'\'{e}quation de pro-\newline
pagation du champ \'{e}lectrique dans le vide \`{a} laquelle on impose des
conditions aux limites li\'{e}es \`{a} la pr\'{e}sence des miroirs de la
cavit\'{e} optique. Pour cela nous nous pla\c{c}ons dans l'approximation
paraxiale : nous supposons que l'onde se propage essentiellement suivant $Oz$
et on utilise la condition de transversalit\'{e} du champ qui permet de
n\'{e}gliger la composante du champ selon $Oz$. Dans ces conditions, les
modes de la cavit\'{e} sont les modes {\it gaussiens} $v_{pql}\left( \vec{r}%
\right) $ qui forment une base orthonorm\'{e}e sur laquelle on peut
d\'{e}composer le champ intracavit\'{e}\cite{Kogelnik}. Contrairement \`{a}
une onde plane, un faisceau gaussien est d\'{e}fini par une taille
transversale finie et un rayon de courbure du front d'onde, qui
d\'{e}pendent de la position $z$. Les nombres radial $q$ et angulaire $l$
rep\`{e}rent les diff\'{e}rents modes transverses du champ alors que le
nombre $p$, qui est associ\'{e} \`{a} la condition de p\'{e}riodicit\'{e}
sur la longueur de la cavit\'{e}, rep\`{e}re les diff\'{e}rents modes
longitudinaux du champ.

En pratique, nous supposons que l'\'{e}cart entre les modes longitudinaux
est tr\`{e}s grand compar\'{e} \`{a} celui entre les modes transverses, et
que le champ incident sur la cavit\'{e} est parfaitement adapt\'{e} \`{a} un
mode $TEM_{00}$, c'est \`{a} dire \`{a} un mode gaussien fondamental $%
v_{p00} $ o\`{u} l'indice $p$ est fix\'{e} et les indices $q$ et $l$ sont
nuls. Ce mode fondamental est caract\'{e}ris\'{e} au niveau de la face plane
du r\'{e}sonateur (en $z=0$) par un rayon de courbure infini et un {\it col} 
$w_{0}$ ({\it waist} en Anglais) qui d\'{e}termine l'extension radiale
minimale du faisceau. La structure spatiale du mode est donn\'{e}e dans ce
plan par la gaussienne\cite{Kogelnik}: 
\begin{equation}
v_{0}(r)=\frac{\sqrt{2/\pi }}{w_{o}}~e^{-r^{2}/w_{o}^{2}}  \label{3.1.1}
\end{equation}
o\`{u} le col $w_{0}$ d\'{e}pend des param\`{e}tres g\'{e}om\'{e}triques de
la cavit\'{e}, c'est \`{a} dire de sa longueur $L$ et du rayon de courbure $%
R_{cav}$ du miroir d'entr\'{e}e: 
\begin{equation}
w_{o}^{2}=\frac{\lambda }{\pi }\sqrt{L\left( R_{cav}-L\right) }
\label{3.1.2}
\end{equation}

Nous allons d\'{e}terminer l'effet sur ce mode d'une d\'{e}formation $%
u(r,z=0,t)$ du miroir mobile. Au niveau du r\'{e}sonateur, dans le plan $z=0$%
, le champ avant r\'{e}flexion s'\'{e}crit: 
\begin{equation}
E\left( r,t\right) =v_{0}(r)\alpha (t)~e^{-i\omega _{0}t}  \label{3.1.3}
\end{equation}
o\`{u} $\alpha \left( t\right) $ est l'amplitude lentement variable du champ
et $\omega _{0}$ la fr\'{e}quence optique, reli\'{e}e \`{a} l'indice $p$ et
\`{a} la longueur $L$ de la cavit\'{e} par : 
\begin{equation}
\omega _{0}=p~\frac{\pi c}{L}  \label{3.1.3bis}
\end{equation}
En suivant les trajets optiques, on constate qu'en tout point $r$ le champ
subit un d\'{e}phasage proportionnel au d\'{e}placement $u\left(
r,z=0,t\right) $ du miroir. Apr\`{e}s r\'{e}flexion totale, il devient donc: 
\begin{equation}
E^{\prime }\left( r,t\right) =v_{0}(r)\alpha (t)~e^{-i\omega
_{0}t}~e^{2iku\left( r,z=0,t\right) }  \label{3.1.4}
\end{equation}
Cette expression montre que le champ r\'{e}fl\'{e}chi ne se r\'{e}duit plus
uniquement \`{a} sa composante fondamentale $v_{0}$ mais qu'il pr\'{e}sente
des composantes sur l'ensemble des modes $v_{p^{\prime }q^{\prime }l^{\prime
}}$ de la cavit\'{e}. On peut en effet \'{e}crire le champ r\'{e}fl\'{e}chi
sous la forme: 
\begin{equation}
E^{\prime }\left( r,t\right) =\stackunder{p^{\prime },q^{\prime },l^{\prime }%
}{\sum }\left\langle v_{0}~e^{2iku},v_{p^{\prime }q^{\prime }l^{\prime
}}\right\rangle ~v_{p^{\prime }q^{\prime }l^{\prime }}\left( r\right)
~\alpha (t)~e^{-i\omega _{0}t}  \label{3.1.4bis}
\end{equation}
o\`{u} les crochets $\left\langle ~,~\right\rangle $ repr\'{e}sentent
l'int\'{e}grale de recouvrement dans le plan $z=0$: 
\begin{equation}
\left\langle f,g\right\rangle =\stackunder{z=0}{\int }dr^{2}~f\left(
r\right) g\left( r\right)  \label{3.1.4ter}
\end{equation}
Ce ph\'{e}nom\`{e}ne de diffusion du mode fondamental dans les autres modes
propres de la cavit\'{e} est li\'{e} \`{a} la d\'{e}formation du front
d'onde du champ, qui reproduit apr\`{e}s r\'{e}flexion la forme de la face
du r\'{e}sonateur. Cette diffusion devient cependant n\'{e}gligeable pour
une cavit\'{e} non d\'{e}g\'{e}n\'{e}r\'{e}e et de grande finesse. Dans ce
cas, l'\'{e}cart entre les fr\'{e}quences de r\'{e}sonance des
diff\'{e}rents modes est grand par rapport \`{a} la bande passante de la
cavit\'{e}. Tous les modes apparaissant dans la somme (\ref{3.1.4bis})
\'{e}voluent \`{a} des fr\'{e}quences voisines de la r\'{e}sonance
fondamentale $\omega _{0}$ de la cavit\'{e}, c'est \`{a} dire \`{a} des
fr\'{e}quences tr\`{e}s \'{e}loign\'{e}es de leur propre fr\'{e}quence de
r\'{e}sonance (on suppose que les fr\'{e}quences d'\'{e}volution des
d\'{e}placements $u\left( r,z=0,t\right) $ du miroir restent de l'ordre de
grandeur de la bande passante de la cavit\'{e}). Ces modes sont donc
fortement filtr\'{e}s par la cavit\'{e} et ne peuvent pas se propager dans
celle-ci. La cavit\'{e} inhibe ainsi la diffusion dans les autres modes et
seul le mode fondamental $v_{p00} $ subsiste dans la somme (\ref{3.1.4bis})%
\cite{JMC PRA 96}: 
\begin{equation}
E^{\prime }\left( r,t\right) =\left\langle e^{2iku},v_{0}^{2}\right\rangle
~E\left( r,t\right)  \label{3.1.5}
\end{equation}
Pour des petits d\'{e}placements, on obtient \`{a} l'ordre le plus bas:

\begin{equation}
E^{\prime }\left( r,t\right) =\left[ 1+2ik\left\langle
u,v_{0}^{2}\right\rangle \right] ~E\left( r,t\right)  \label{3.1.6}
\end{equation}

Le champ subit essentiellement un d\'{e}phasage lorsqu'il se
r\'{e}fl\'{e}chit sur le r\'{e}sonateur. Ce d\'{e}phasage est \'{e}gal au
d\'{e}phasage $2ku(r,z=0,t)$ en tout point $r$ du r\'{e}sonateur,
pond\'{e}r\'{e} par la structure spatiale $v_{0}^{2}(r)$ du mode
fondamental. Le d\'{e}saccord $\Psi \left( t\right) $ entre le champ et la
cavit\'{e} d\'{e}fini dans le chapitre pr\'{e}c\'{e}dent est alors donn\'{e}
par une \'{e}quation similaire \`{a} (2.41), \`{a} condition de remplacer le
d\'{e}placement monodimensionnel $x\left( t\right) $ par un d\'{e}placement $%
\hat{u}(t)$, \'{e}gal \`{a} la moyenne du d\'{e}placement sur la section du
faisceau lumineux: 
\begin{subequations}
\label{3.1.7}
\begin{eqnarray}
\Psi \left( t\right)  &=&\Psi _{0}+2k~\hat{u}(t)  \label{3.1.7a} \\
\hat{u}(t) &=&\left\langle u(r,z=0,t),v_{0}^{2}(r)\right\rangle 
\label{3.1.7b}
\end{eqnarray} 
En conclusion, l'analyse monodimensionnelle des effets du mouvement du
miroir sur le champ expos\'{e}e dans le chapitre pr\'{e}c\'{e}dent peut se
g\'{e}n\'{e}raliser au cas du r\'{e}sonateur plan-convexe : il suffit pour
cela de moyenner le d\'{e}placement du r\'{e}sonateur sur la structure
transverse du champ. Par exemple le bruit de phase $S_{q}^{out}\left[ \Omega
\right] $ du champ r\'{e}fl\'{e}chi, lorsque celui-ci est r\'{e}sonnant avec
la cavit\'{e}, est donn\'{e} par une expression similaire \`{a}
l'\'{e}quation (2.76) en rempla\c{c}ant le spectre de position $S_{x}\left[
\Omega \right] $ par le spectre $S_{\hat{u}}\left[ \Omega \right] $.
\end{subequations}

\section{Mouvement du r\'{e}sonateur\label{III-2}}

\bigskip

Nous allons maintenant d\'{e}terminer le mouvement du r\'{e}sonateur
lorsqu'il est \`{a} l'\'{e}quilibre thermodynamique ou lorsqu'il est soumis
\`{a} la force de pression de radiation du faisceau lumineux. Nous
commencerons par \'{e}tudier les modes acoustiques du r\'{e}sonateur, c'est
\`{a} dire le mouvement du r\'{e}sonateur libre, en absence de force
ext\'{e}rieure. Tout mouvement du r\'{e}sonateur peut se d\'{e}composer sur
ces modes, et nous \'{e}tudierons l'\'{e}volution de ces modes lorsqu'une
force est appliqu\'{e}e sur la face plane du r\'{e}sonateur.

\subsection{Modes acoustiques\label{III-2-1}}

Le r\'{e}sonateur \'{e}tant en silice pure, on consid\`{e}re que le milieu
est \'{e}lastique et isotrope. Pour des petites d\'{e}formations du
r\'{e}sonateur, les mouvements consid\'{e}r\'{e}s dans la th\'{e}orie de
l'\'{e}lasticit\'{e} sont des petites vibrations \'{e}lastiques ou {\it %
ondes acoustiques}. Une onde acoustique quelconque est d\'{e}finie par son
vecteur de d\'{e}formation $\vec{u}(\vec{r},t)$, qui ob\'{e}it \`{a} une
\'{e}quation de propagation ainsi qu'\`{a} des conditions aux limites
li\'{e}es aux contraintes ext\'{e}rieures appliqu\'{e}es au r\'{e}sonateur.
Il y a dans le r\'{e}sonateur deux types d'ondes, les ondes de compression,
qui provoquent des d\'{e}placements selon la direction $Oz$, et les ondes de
cisaillement, qui induisent des d\'{e}placements transverses de la face
plane du r\'{e}sonateur. Comme le faisceau n'est sensible qu'aux
d\'{e}placements longitudinaux du r\'{e}sonateur, nous nous
int\'{e}resserons seulement aux ondes de compression. L'\'{e}quation de
propagation d'une onde de compression dans le r\'{e}sonateur est donn\'{e}e
par\cite{landau propagation}: 
\begin{equation}
\frac{\partial ^{2}\vec{u}}{\partial t^{2}}(\vec{r},t)-c_{l}^{2}~\Delta \vec{%
u}\left( \vec{r},t\right) =0  \label{3.2.1}
\end{equation}
o\`{u} la vitesse de propagation de l'onde est: 
\begin{equation}
c_{l}=\sqrt{\frac{\lambda +2\mu }{\rho }}  \label{3.2.2}
\end{equation}
$\lambda $ et $\mu $ \'{e}tant les constantes de Lam\'{e} du r\'{e}sonateur
et $\rho $ sa masse volumique.

Le mouvement doit aussi satisfaire les conditions aux limites qui
s'\'{e}crivent, en l'absence de force sur les faces du r\'{e}sonateur: 
\begin{equation}
\stackunder{j}{\sum }\sigma _{ij}\left( \vec{r},t\right) ~n_{j}=0
\label{3.2.3}
\end{equation}
en tout point $\vec{r}$ de la surface o\`{u} le vecteur normal \`{a} la
surface est $\vec{n}$, $\left[ \sigma \right] $ \'{e}tant le tenseur des
contraintes dans le r\'{e}sonateur, li\'{e} au tenseur de d\'{e}formation $%
\left[ u\right] $ par la loi de Hooke\cite{landau propagation}: 
\begin{subequations}
\label{3.2.4}
\begin{eqnarray}
\sigma _{ij} &=&2\mu ~u_{ij}+\lambda ~(\vec{\nabla}\cdot \vec{u})
\label{3.2.4a} \\
u_{ij} &=&\frac{1}{2}\left( \partial _{i}u_{j}+\partial _{j}u_{i}\right) 
\label{3.2.4b}
\end{eqnarray}

Puisque les ondes de compression v\'{e}rifient une \'{e}quation de
propagation identique \`{a} celle d'un champ \'{e}lectromagn\'{e}tique se
propageant dans le vide, on peut chercher une solution de la forme $\vec{u}%
\left( \vec{r},t\right) =\vec{u}\left( \vec{r}\right) e^{-i\Omega t}$ o\`{u} 
$\Omega $ est la fr\'{e}quence d'\'{e}volution de l'onde. En reportant cette
expression dans l'\'{e}quation de propagation (\ref{3.2.1}), on obtient: 
\end{subequations}
\begin{equation}
\Delta \vec{u}\left( \vec{r}\right) =-\frac{\Omega ^{2}}{c_{l}^{2}}~\vec{u}%
\left( \vec{r}\right)  \label{3.2.5}
\end{equation}
Les solutions de cette \'{e}quation avec les conditions aux limites (\ref
{3.2.3}) sont les modes propres du r\'{e}sonateur. L'ensemble $\left\{ \vec{u%
}_{n}\left( \vec{r}\right) \right\} $ de ces modes propres forme une base
orthogonale de l'ensemble des solutions du syst\`{e}me d'\'{e}quations (\ref
{3.2.3}) et (\ref{3.2.5}), le produit scalaire \'{e}tant d\'{e}fini comme
l'int\'{e}grale sur tout le volume du r\'{e}sonateur. Comme nous le verrons
dans la partie \ref{III-4}, un avantage du r\'{e}sonateur plan-convexe est
qu'il est possible d'obtenir des expressions analytiques pour ces modes
propres, expressions en tout point similaires \`{a} celles des modes
gaussiens en optique. Cependant, l'\'{e}tude que nous menons ici ne
d\'{e}pend pas de la forme explicite de ces modes et peut \^{e}tre
appliqu\'{e}e \`{a} d'autres g\'{e}om\'{e}tries du r\'{e}sonateur. Le seul
point important pour l'\'{e}tude du mouvement du r\'{e}sonateur est
l'existence d'une base de modes propres $\left\{ \vec{u}_{n}\left( \vec{r}%
\right) \right\} $, chaque mode \'{e}tant caract\'{e}ris\'{e} par une
fr\'{e}quence d'\'{e}volution $\Omega _{n}$.

Cette base permet de d\'{e}composer tout d\'{e}placement $\vec{u}\left( \vec{%
r},t\right) $ sous la forme:

\begin{equation}
\vec{u}\left( \vec{r},t\right) =\stackunder{n}{\sum }\vec{u}_{n}\left( \vec{r%
}\right) ~a_{n}(t)  \label{3.2.13}
\end{equation}
o\`{u} $a_{n}(t)$ d\'{e}signe l'amplitude du mode acoustique $\vec{u}_{n}$.
A l'aide de cette d\'{e}composition, nous allons d\'{e}terminer
l'\'{e}volution de chaque mode lorsqu'une force ext\'{e}rieure est
appliqu\'{e}e sur la face plane du r\'{e}sonateur.

\subsection{Mouvement du r\'{e}sonateur en absence de dissipation\label%
{III-2-2}}

Nous nous int\'{e}ressons tout d'abord au cas d'un r\'{e}sonateur libre,
soumis \`{a} aucune contrainte ext\'{e}rieure. Afin de caract\'{e}riser le
mouvement, on peut calculer l'\'{e}nergie ${\cal E}$ associ\'{e}e au
d\'{e}placement $\vec{u}\left( \vec{r},t\right) $. Cette \'{e}nergie est
\'{e}gale \`{a} la somme des int\'{e}grales sur tout le volume du
r\'{e}sonateur des densit\'{e}s d'\'{e}nergie cin\'{e}tique et potentielle.
L'\'{e}nergie cin\'{e}tique s'\'{e}crit: 
\begin{equation}
{\cal E}_{c}=\frac{1}{2}\int d^{3}r~\rho \left[ \frac{\partial \vec{u}}{%
\partial t}(\vec{r},t)\right] ^{2}  \label{3.214}
\end{equation}
En utilisant la d\'{e}composition (\ref{3.2.13}) et la propri\'{e}t\'{e}
d'orthogonalit\'{e} des modes propres $\vec{u}_{n}\left( \vec{r}\right) $,
on obtient alors: 
\begin{equation}
{\cal E}_{c}=\stackunder{n}{\sum }\frac{1}{2}M_{n}~\left[ \frac{da_{n}}{dt}%
(t)\right] ^{2}  \label{3.2.15}
\end{equation}
o\`{u} le param\`{e}tre $M_{n}$ est la masse du mode acoustique $\vec{u}_{n}$%
, donn\'{e}e par: 
\begin{equation}
M_{n}=\rho \int d^{3}r~\vec{u}_{n}^{2}\left( \vec{r}\right)  \label{3.2.16}
\end{equation}
Cette masse appara\^{\i }t comme le produit de la masse volumique $\rho $
par le {\it volume} du {\it mode acoustique}, c'est \`{a} dire le volume de
la partie du r\'{e}sonateur mise en mouvement lors d'une excitation du mode
acoustique.

L'\'{e}nergie potentielle est reli\'{e}e aux constantes de Lam\'{e} $\lambda 
$ et $\mu $ du r\'{e}sonateur et au tenseur des d\'{e}formations $\left[
u\right] $ (\'{e}quation \ref{3.2.4b})\cite{landau elasticité}:

\begin{equation}
{\cal E}_{p}=\frac{1}{2}\int d^{3}r~\left\{ \lambda \left[ \vec{\nabla}\cdot 
\vec{u}\left( \vec{r},t\right) \right] ^{2}+2\mu \stackunder{i,j}{\sum }%
\left[ u_{ij}\left( \vec{r},t\right) \right] ^{2}\right\}  \label{3.2.18}
\end{equation}
Cette expression peut \^{e}tre transform\'{e}e par int\'{e}gration par
partie. On trouve pour une onde de compression: 
\begin{equation}
{\cal E}_{p}=-\frac{1}{2}\rho c_{l}^{2}\int d^{3}r~\vec{u}\left( \vec{r}%
,t\right) \cdot \Delta \vec{u}\left( \vec{r},t\right)  \label{3.2.19}
\end{equation}
En utilisant la d\'{e}composition en modes propres (\ref{3.2.13}) et
l'\'{e}quation de propagation (\ref{3.2.5}), on trouve que l'\'{e}nergie
potentielle s'\'{e}crit: 
\begin{equation}
{\cal E}_{p}=\stackunder{n}{\sum }\frac{1}{2}M_{n}\Omega _{n}^{2}\left[
a_{n}(t)\right] ^{2}  \label{3.2.20}
\end{equation}
o\`{u} $\Omega _{n}$ est la fr\'{e}quence d'\'{e}volution du mode propre $%
u_{n}$. En additionnant les deux contributions des \'{e}nergies
cin\'{e}tique et potentielle, on obtient finalement l'expression suivante
pour l'\'{e}nergie totale ${\cal E}$: 
\begin{equation}
{\cal E}=\stackunder{n}{\sum }\frac{1}{2}M_{n}~\left\{ \left[ \frac{da_{n}}{%
dt}(t)\right] ^{2}+\Omega _{n}^{2}\left[ a_{n}(t)\right] ^{2}\right\}
\label{3.2.21}
\end{equation}
On reconna\^{\i }t dans cette expression la somme des \'{e}nergies
d'oscillateurs harmoniques non amortis de masse $M_{n}$ et de pulsation
propre $\Omega _{n}$. Le mouvement du r\'{e}sonateur libre en absence de
dissipation interne peut donc se d\'{e}composer sur l'ensemble des modes
propres acoustiques $\vec{u}_{n}$ dont les amplitudes sont d\'{e}crites
comme des oscillateurs harmoniques ind\'{e}pendants. Cette d\'{e}composition
est tout \`{a} fait g\'{e}n\'{e}rale et correspond \`{a} une description du
mouvement en terme de phonons.

\subsection{Effet d'une force ext\'{e}rieure\label{III-2-3}}

Nous allons \`{a} pr\'{e}sent d\'{e}terminer l'\'{e}quation du mouvement du
r\'{e}sonateur en pr\'{e}sence d'une force ext\'{e}rieure. On
s'int\'{e}ressera au cas d'une force appliqu\'{e}e sur la face plane du
r\'{e}sonateur (en $z=0$) et parall\`{e}le \`{a} l'axe $Oz$. L'\'{e}nergie
totale du r\'{e}sonateur est donn\'{e}e par la relation (\ref{3.2.21}) \`{a}
laquelle on rajoute un terme suppl\'{e}mentaire li\'{e} au travail des
forces de contraintes internes qui s'opposent \`{a} la force ext\'{e}rieure.
Ce travail est \'{e}gal et oppos\'{e} au travail accompli par la force
ext\'{e}rieure sur toute la surface plane du r\'{e}sonateur: 
\begin{equation}
{\cal W}=-\stackunder{z=0}{\int }d^{2}r~\vec{F}(\vec{r},t)\cdot \vec{u}(\vec{%
r},t)  \label{3.2.22}
\end{equation}
o\`{u} $\vec{F}(\vec{r},t)$ est la force par unit\'{e} de surface
appliqu\'{e}e au point $\vec{r}$. En utilisant la d\'{e}composition en modes
propres (\ref{3.2.13}), on obtient: 
\begin{equation}
{\cal W}=-\stackunder{n}{\sum }\left\langle \vec{F}(t),\vec{u}%
_{n}\right\rangle ~a_{n}(t)  \label{3.2.23}
\end{equation}
o\`{u} les crochets repr\'{e}sentent l'int\'{e}grale de recouvrement dans le
plan $z=0$ (\'{e}quation \ref{3.1.4ter}). L'\'{e}nergie totale s'\'{e}crit
alors : 
\begin{equation}
{\cal E}+{\cal W}=\stackunder{n}{\sum }\left\{ \frac{1}{2}M_{n}\left[ \frac{%
da_{n}}{dt}(t)\right] ^{2}+\frac{1}{2}M_{n}\Omega _{n}^{2}\left[
a_{n}(t)\right] ^{2}-\left\langle \vec{F}(t),\vec{u}_{n}\right\rangle
~a_{n}(t)\right\}  \label{3.2.24}
\end{equation}
Cette \'{e}nergie est \'{e}quivalente \`{a} celle d'un ensemble
d'oscillateurs harmoniques ind\'{e}pendants. Chaque oscillateur est soumis
\`{a} une force $\left\langle \vec{F}(t),\vec{u}_{n}\right\rangle $,
\'{e}gale \`{a} la projection spatiale de la force ext\'{e}rieure sur le
mode consid\'{e}r\'{e}. En d'autres termes, le mouvement du r\'{e}sonateur
se d\'{e}compose sur l'ensemble des modes acoustiques, l'amplitude $a_{n}(t)$
de chaque mode \'{e}tant r\'{e}gie par l'\'{e}quation du mouvement d'un
oscillateur harmonique forc\'{e}:

\begin{equation}
\frac{d^{2}a_{n}}{dt^{2}}(t)+\Omega _{n}^{2}~a_{n}(t)=\frac{1}{M_{n}}%
~\left\langle \vec{F}(t),\vec{u}_{n}\right\rangle  \label{3.2.25}
\end{equation}
Dans l'espace de Fourier, l'amplitude de chaque mode peut s'\'{e}crire sous
la forme: 
\begin{equation}
a_{n}\left[ \Omega \right] =\chi _{n}\left[ \Omega \right] ~\left\langle 
\vec{F}\left[ \Omega \right] ,\vec{u}_{n}\right\rangle  \label{3.2.26}
\end{equation}
o\`{u} $\chi _{n}\left[ \Omega \right] =1/M_{n}\left( \Omega _{n}^{2}-\Omega
^{2}\right) $ est la susceptibilit\'{e} m\'{e}canique d'un oscillateur
harmonique non amorti, de masse $M_{n}$ et de fr\'{e}quence propre $\Omega
_{n}$.

\subsection{Mouvement Brownien du r\'{e}sonateur m\'{e}canique\label{III-2-4}
}

Jusqu'\`{a} pr\'{e}sent, nous avons suppos\'{e} le r\'{e}sonateur non
amorti. On peut tenir compte de l'amortissement et du couplage avec le bain
thermique en g\'{e}n\'{e}ralisant l'approche utilis\'{e}e dans le chapitre
pr\'{e}c\'{e}dent (section 2.4.2). Chaque mode acoustique est en effet
\'{e}quivalent \`{a} un oscillateur harmonique et on peut d\'{e}crire le
couplage avec un bain thermique par des forces de Langevin appliqu\'{e}es
\`{a} chaque oscillateur et par un angle de perte. Pour simplifier, nous
supposerons que l'angle de perte $\Phi \left[ \Omega \right] $ ne d\'{e}pend
pas du mode acoustique consid\'{e}r\'{e}. Ceci d\'{e}coule en fait du
mod\`{e}le de Navier-Stokes ou du mod\`{e}le $\Phi $ constant, et cette
hypoth\`{e}se semble justifi\'{e}e lorsque les m\'{e}canismes de dissipation
sont internes (les m\'{e}canismes de dissipation par contact avec le support
peuvent d\'{e}pendre de l'extension spatiale des modes acoustiques). En
pr\'{e}sence d'amortissement les susceptibilit\'{e}s $\chi _{n}\left[ \Omega
\right] $ des modes acoustiques s'\'{e}crivent alors: 
\begin{equation}
\chi _{n}\left[ \Omega \right] =\frac{1}{M_{n}\left( \Omega _{n}^{2}-\Omega
^{2}-i\Omega _{n}^{2}\Phi \left[ \Omega \right] \right) }  \label{3.2.27}
\end{equation}

Comme nous l'avons vu dans le chapitre pr\'{e}c\'{e}dent, les m\'{e}canismes
de dissipation dans les solides sont mal connus et on ne dispose pas de
mod\`{e}le th\'{e}orique satisfaisant pour d\'{e}crire le comportement de
l'angle de perte en fonction de la fr\'{e}quence. Les deux approches que
nous consid\'{e}rons sont les mod\`{e}les de Navier-Stokes, bas\'{e} sur une
dissipation de type visqueuse, et le mod\`{e}le $\Phi $ constant qui associe
\`{a} tout processus de dissipation un angle de perte constant. Dans le
cadre du mod\`{e}le de Navier-Stokes\cite{Landau Nav Stoks}, l'angle de
perte $\Phi _{vis}$ est donn\'{e} par une relation similaire \`{a} (2.84): 
\begin{equation}
\Phi _{vis}\left[ \Omega \right] =\frac{\Omega }{Q\Omega _{M}}
\label{3.2.28}
\end{equation}
o\`{u} $Q$ et $\Omega _{M}$ sont respectivement le facteur de qualit\'{e} et
la fr\'{e}quence de r\'{e}sonance du mode acoustique fondamental du
r\'{e}sonateur.

Le second mod\`{e}le, bas\'{e} sur des constatations exp\'{e}rimentales,
pr\'{e}dit un angle de perte $\Phi _{cst}$ constant en fr\'{e}quence. En
utilisant le r\'{e}sultat de la section 2.4.2.2 (\'{e}quation 2.85), on
obtient: 
\begin{equation}
\Phi _{cst}\left[ \Omega \right] =\frac{1}{Q}  \label{3.2.29}
\end{equation}

Une autre cons\'{e}quence du couplage avec le bain thermique est la
pr\'{e}sence de forces de Langevin $F_{T,n}$ pour chaque mode acoustique.
Ces forces sont ind\'{e}pendantes les unes des autres et v\'{e}rifient le
th\'{e}or\`{e}me fluctuation-dissipation. Le spectre $S_{T,n}\left[ \Omega
\right] $ de la force $F_{T,n}$ est donc reli\'{e} \`{a} la partie
imaginaire de la susceptibilit\'{e} $\chi _{n}$ par: 
\begin{equation}
S_{T,n}\left[ \Omega \right] =-\frac{2k_{B}T}{\Omega }~{\cal I}m\left\{ 
\frac{1}{\chi _{n}\left[ \Omega \right] }\right\}  \label{3.2.31}
\end{equation}

En conclusion, le mouvement du r\'{e}sonateur en pr\'{e}sence d'une force
ext\'{e}rieure $F$ est d\'{e}crit par l'ensemble des amplitudes $\left\{
a_{n}\right\} $ des modes acoustiques, reli\'{e}es \`{a} la force
appliqu\'{e}e par la relation:

\begin{equation}
a_{n}\left[ \Omega \right] =\chi _{n}\left[ \Omega \right] ~\left(
\left\langle \vec{F}\left[ \Omega \right] ,\vec{u}_{n}\right\rangle
+F_{T,n}\left[ \Omega \right] \right)  \label{3.2.30}
\end{equation}
Ce r\'{e}sultat pr\'{e}sente certaines analogies avec la description
monodimensionnelle du mouvement pr\'{e}sent\'{e} dans le chapitre
pr\'{e}c\'{e}dent (\'{e}quation 2.42). Les dif\'{e}rences sont li\'{e}es
\`{a} la pr\'{e}sence de multiples modes acoustiques et au fait que la force
appliqu\'{e}e pour chaque mode est la projection de la force sur la
structure spatiale du mode.

\section{Effet de la pression de radiation et susceptibilit\'{e} effective%
\label{III-3}}

\bigskip

Pour d\'{e}terminer l'effet du faisceau lumineux sur le mouvement du
r\'{e}sonateur, il suffit d'identifier la force ext\'{e}rieure $F$ \`{a} la
pression de radiation $F_{rad}$ exerc\'{e}e par le champ intracavit\'{e}.
Cette pression est proportionnelle \`{a} l'intensit\'{e} du champ dont
l'amplitude pr\'{e}sente une distribution spatiale gaussienne (\'{e}quation 
\ref{3.1.3}). La force par unit\'{e} de surface qu'exerce le champ sur le
r\'{e}sonateur est donc une gaussienne qui s'\'{e}crit: 
\begin{equation}
F_{rad}(r,t)=2\hbar k\left| E\left( r,t\right) \right| ^{2}=2\hbar k~\left[
v_{0}(r)\right] ^{2}I(t)  \label{3.2.32}
\end{equation}
o\`{u} $v_{0}(r)$ est donn\'{e} par la relation (\ref{3.1.1}) et $I(t)$ est
l'intensit\'{e} du champ (nombre de photons par seconde int\'{e}gr\'{e} sur
toute la section du faisceau), reli\'{e}e \`{a} l'amplitude $\alpha \left(
t\right) $ du champ par: 
\begin{equation}
I(t)=\stackunder{z=0}{\int }d^{2}r~\left| E\left( r,t\right) \right|
^{2}=\left| \alpha \left( t\right) \right| ^{2}  \label{3.2.33}
\end{equation}

Nous allons \`{a} pr\'{e}sent d\'{e}terminer l'expression du d\'{e}placement
moyen $\hat{u}$ (\'{e}quation \ref{3.1.7b}) auquel est sensible le champ
dans la cavit\'{e}. Ce d\'{e}placement est obtenu en utilisant la
d\'{e}composition (\ref{3.2.13}) de $u$ en fonction des amplitudes $a_{n}$
et l'\'{e}quation du mouvement (\ref{3.2.30}) de chaque mode. Dans l'espace
de Fourier ce d\'{e}placement est donn\'{e} par: 
\begin{equation}
\hat{u}\left[ \Omega \right] =\stackunder{n}{\sum }\left\langle
v_{0}^{2},u_{n}\right\rangle ~\chi _{n}\left[ \Omega \right] \left( 2\hbar
k~I\left[ \Omega \right] \left\langle v_{0}^{2},u_{n}\right\rangle
+F_{T,n}\left[ \Omega \right] \right)  \label{3.2.34}
\end{equation}
\ Cette relation peut \^{e}tre \'{e}crite de fa\c{c}on condens\'{e}e, sous
une forme similaire \`{a} celle pr\'{e}sent\'{e}e dans le cadre de l'analyse
monodimensionnelle (\'{e}quations 2.42 et 2.43): 
\begin{equation}
\hat{u}\left[ \Omega \right] =\chi _{eff}\left[ \Omega \right] \left( 2\hbar
k~I\left[ \Omega \right] +F_{T}\left[ \Omega \right] \right)  \label{3.2.35}
\end{equation}
o\`{u} la susceptibilit\'{e} effective $\chi _{eff}$ est \'{e}gale \`{a} la
somme des susceptibilit\'{e}s $\chi _{n}$ de tous les modes acoustiques du
r\'{e}sonateur, pond\'{e}r\'{e}es par leur recouvrement avec le mode
optique: 
\begin{equation}
\chi _{eff}\left[ \Omega \right] =\stackunder{n}{\sum }\left\langle
v_{0}^{2},u_{n}\right\rangle ^{2}~\chi _{n}\left[ \Omega \right]
\label{3.2.36}
\end{equation}
Dans la relation (\ref{3.2.35}), $F_{T}$ repr\'{e}sente une force de
Langevin effective qui s'exprime en fonction des forces de Langevin $F_{T,n}$
de tous les modes acoustiques: 
\begin{equation}
F_{T}\left[ \Omega \right] =\stackunder{n}{\sum }\left\langle
v_{0}^{2},u_{n}\right\rangle \frac{\chi _{n}\left[ \Omega \right] }{\chi
_{eff}\left[ \Omega \right] }~F_{T,n}\left[ \Omega \right]  \label{3.2.37}
\end{equation}

On peut d\'{e}terminer le spectre de $F_{T}$ en utilisant l'expression des
spectres de chaque force de Langevin $F_{T,n}$ (\'{e}quation \ref{3.2.31}).
On trouve alors que la force $F_{T}$ v\'{e}rifie le th\'{e}or\`{e}me
fluctuation-dissipation:

\begin{equation}
S_{T}\left[ \Omega \right] =-\frac{2k_{B}T}{\Omega }{\cal I}m\left\{ \frac{1%
}{\chi _{eff}\left[ \Omega \right] }\right\}  \label{3.2.38}
\end{equation}
En d'autres termes, le r\'{e}sonateur dans son ensemble est \`{a}
l'\'{e}quilibre thermodynamique en absence de faisceau lumineux.

En conclusion, nous avons d\'{e}fini une susceptibilit\'{e} effective $\chi
_{eff}$ qui permet de relier le d\'{e}placement moyen $\hat{u}$ du miroir
mobile \`{a} l'intensit\'{e} lumineuse et \`{a} une force de Langevin
(\'{e}quation \ref{3.2.35}) de la m\^{e}me mani\`{e}re que dans le cas d'un
oscillateur harmonique (\'{e}quation 2.42) : le couplage optom\'{e}canique
entre le r\'{e}sonateur et le champ gaussien intracavit\'{e} est
compl\`{e}tement d\'{e}crit par la susceptibilit\'{e} effective $\chi _{eff}$
qui int\`{e}gre l'ensemble des propri\'{e}t\'{e}s spatiales du couplage.
Cette susceptibilit\'{e} tient compte de la contribution de tous les modes
acoustiques du r\'{e}sonateur ainsi que de leur adaptation spatiale avec le
faisceau lumineux. Tous les r\'{e}sultats obtenus dans le cadre de
l'approche monodimensionnelle sont donc valables \`{a} condition de
remplacer la susceptibilit\'{e} $\chi $ d'un seul oscillateur harmonique par
la susceptibilit\'{e} effective $\chi _{eff}$.

\section{Couplage optom\'{e}canique avec un r\'{e}sonateur plan-convexe\label%
{III-4}}

\bigskip

La d\'{e}marche qui a \'{e}t\'{e} suivie dans le chapitre 2 pour mesurer le
bruit thermique et pour r\'{e}aliser une mesure QND de l'intensit\'{e}
lumineuse peut se g\'{e}n\'{e}raliser au cas d'un r\'{e}sonateur
plan-convexe. Nous allons tout d'abord d\'{e}terminer la structure des modes
acoustiques pour un tel r\'{e}sonateur, afin d'expliciter les diff\'{e}rents
param\`{e}tres intervenant dans la susceptibilit\'{e} effective (section
3.4.1). Nous pr\'{e}senterons ensuite le spectre du bruit thermique du
r\'{e}sonateur plan-convexe (section 3.4.2) ainsi que la fonction de
corr\'{e}lation d'une mesure QND (section 3.4.3).

\subsection{Modes acoustiques d'un r\'{e}sonateur plan-convexe\label{III-4-1}
}

Les modes acoustiques longitudinaux sont solutions de l'\'{e}quation de
propagation (\ref{3.2.5}) avec les conditions aux limites (\ref{3.2.3}).
Pour un r\'{e}sonateur plan-convexe, ces \'{e}quations sont similaires \`{a}
celles obtenues en optique pour une cavit\'{e} Fabry-Perot constitu\'{e}e
d'un miroir plan et d'un miroir courbe. On peut alors obtenir des
expressions analytiques approch\'{e}es pour les modes acoustiques, valides
dans le cadre de l'approximation paraxiale\cite{Wilson 74}. Chaque mode est
d\'{e}fini par trois indices ($p$,$q$,$l$) et le d\'{e}placement est
caract\'{e}ris\'{e} par une amplitude $u_{pql}\left( \vec{r}\right) $ selon
l'axe $Oz$ dont l'expression est similaire \`{a} celle des modes gaussiens
en optique: 
\begin{equation}
u_{pql}\left( r,\theta ,z\right) =e^{-\left( \frac{r}{w_{p}}\right)
^{2}}\left( \sqrt{2}\frac{r}{w_{p}}\right) ^{l}L_{q}^{l}\left( \frac{2r^{2}}{%
w_{p}^{2}}\right) ~e^{i\left[ \frac{p\pi z}{h(r)}+l\theta \right] }
\label{mod ac1}
\end{equation}
o\`{u} $L_{q}^{l}$ d\'{e}signe un polyn\^{o}me de Laguerre
g\'{e}n\'{e}ralis\'{e} et $h\left( r\right) $ est l'\'{e}paiseur du
r\'{e}sonateur \`{a} une distance radiale $r$, reli\'{e}e \`{a}
l'\'{e}paisseur au centre $h_{0}$ et au rayon de courbure $R$ de la face
convexe par la relation: 
\begin{equation}
h(r)\simeq h_{0}-\frac{r^{2}}{2R}  \label{mod ac2}
\end{equation}
lorsque $r\ll R$. Le col acoustique $w_{p}$ est donn\'{e} par: 
\begin{equation}
w_{p}^{2}=2\frac{h_{0}}{p\pi }\sqrt{Rh_{0}}=\frac{w_{1}^{2}}{p}
\label{mod ac3}
\end{equation}
o\`{u} $w_{1}$ d\'{e}signe le col du mode fondamental. L'\'{e}volution
temporelle du mode acoustique $u_{pql}$ est fix\'{e}e par la fr\'{e}quence
propre $\Omega _{pql}$ que l'on d\'{e}duit de la condition de
p\'{e}riodicit\'{e} sur l'\'{e}paisseur du r\'{e}sonateur: 
\begin{equation}
\Omega _{pql}^{2}=c_{l}^{2}~\left[ \frac{p^{2}\pi ^{2}}{h_{0}^{2}}+\left(
2q+l+1\right) \frac{4}{w_{p}^{2}}\right]  \label{mod ac4}
\end{equation}
On voit ainsi que l'indice $p$ rep\`{e}re les diff\'{e}rentes harmoniques,
le nombre radial $q$ donne le nombre de z\'{e}ros de la fonction radiale,
tandis que $l$ est le nombre angulaire. On peut aussi exprimer les
fr\'{e}quences propres en fonction de la fr\'{e}quence caract\'{e}ristique $%
\Omega _{M}=\pi \left( c_{l}/h_{0}\right) $ qui repr\'{e}sente la
fr\'{e}quence longitudinale fondamentale du r\'{e}sonateur: 
\begin{equation}
\Omega _{pql}^{2}=\Omega _{M}^{2}~\left[ p^{2}+\left( 2q+l+1\right) \frac{2p%
}{\pi }\sqrt{\frac{h_{0}}{R}}\right]  \label{mod ac5}
\end{equation}

Ces modes propres acoustiques sont tr\`{e}s similaires aux modes gaussiens
optiques\cite{Kogelnik}. Ils diff\`{e}rent cependant par la condition de
validit\'{e} de l'approximation paraxiale. En r\`{e}gle g\'{e}n\'{e}rale,
cette approximation est valable lorsque la composante transverse $k_{\bot }$
du vecteur d'onde est tr\`{e}s petite devant la composante longitudinale $%
k_{z}$. En optique, cette condition est v\'{e}rifi\'{e}e d\`{e}s que la
longueur de la cavit\'{e} est grande compar\'{e}e \`{a} la longueur d'onde,
ce qui est g\'{e}n\'{e}ralement le cas. En acoustique, cette condition n'est
pas r\'{e}alis\'{e}e. Les composantes du vecteur d'onde sont donn\'{e}es
par: 
\begin{equation}
k_{z}=\frac{p\pi }{h_{0}}\quad ,\quad k_{\bot }=\sqrt{\left( 2q+l+1\right) 
\frac{2p\pi }{h_{0}\sqrt{Rh_{0}}}}  \label{3.2.11}
\end{equation}
Les ordres des diff\'{e}rentes harmoniques (nombre $p$) pouvant \^{e}tre du
m\^{e}me ordre de grandeur que les nombres $q$ et $l$ associ\'{e}s aux modes
transverses , l'approximation para-\newline
xiale impose de prendre $R\gg h_{0}$.

Parmis tous les modes acoustiques $u_{pql}$, seuls nous int\'{e}ressent ceux
qui contribuent au d\'{e}placement moyen $\hat{u}$. Pour des raisons de
sym\'{e}trie, les modes concern\'{e}s sont invariants par rotation autour de
l'axe de propagation $Oz$. On ne consid\`{e}re donc que les modes $l=0$ dont
le d\'{e}placement est non nul dans le plan $z=0$, modes que l'on peut
\'{e}crire sous la forme: 
\begin{equation}
u_{pq}\left( r,\theta ,z\right) =e^{-\left( \frac{r}{w_{p}}\right)
^{2}}L_{q}\left( \frac{2r^{2}}{w_{p}^{2}}\right) \cos \left[ \frac{p\pi z}{%
h(r)}\right]  \label{mod ac6}
\end{equation}
Ces modes sont caract\'{e}ris\'{e}s par des fr\'{e}quences de r\'{e}sonance $%
\Omega _{pq}$ donn\'{e}es par l'\'{e}quation (\ref{mod ac5}) avec $l=0$.

La forme explicite de l'amplitude $u_{pq}\left( \vec{r}\right) $ du mode
acoustique permet de d\'{e}terminer les param\`{e}tres intervenant dans la
susceptibilit\'{e} effective, c'est \`{a} dire la masse $M_{pq}$ du mode et
le recouvrement $\left\langle v_{0}^{2},u_{pq}\right\rangle $ avec le
faisceau lumineux. On peut calculer la masse $M_{pq}$ \`{a} l'ordre le plus
bas en $h_{0}/R$ (approximation paraxiale), en rempla\c{c}ant
l'\'{e}paisseur $h\left( r\right) $ au point $r$ par l'\'{e}paisseur au
centre $h_{0}$. On se ram\`{e}ne ainsi \`{a} une int\'{e}grale simple de
polynomes de Laguerre, qui donne\cite{Gradshteyn}: 
\begin{equation}
M_{pq}=\frac{\pi }{4}\rho h_{0}~w_{p}^{2}  \label{3.2.17}
\end{equation}
Cette relation montre bien que la masse $M_{pq}$ est reli\'{e}e au {\it %
volume} du mode, \'{e}gal \`{a} l'\'{e}paisseur du r\'{e}sonateur par la
section du mode. Cette masse est plus petite que la masse totale du
r\'{e}sonateur, et d\'{e}croit lorsque $p$ augmente.

Le recouvrement $\left\langle v_{0}^{2},u_{pq}\right\rangle $ entre le
faisceau lumineux et le mode acoustique $u_{pq}$ peut \^{e}tre calcul\'{e}
en utilisant les expressions des modes (\'{e}quations \ref{3.1.1} et \ref
{mod ac6}) et la relation $\int_{0}^{\infty }dx~e^{-bx}L_{q}\left( x\right)
=\left( b-1\right) ^{q}/b^{q+1}$\cite{Gradshteyn}. On obtient: 
\begin{equation}
\left\langle v_{0}^{2},u_{pq}\right\rangle =\frac{\left( \frac{w_{p}}{w_{o}}%
\right) ^{2}}{\left( \frac{w_{p}}{w_{o}}\right) ^{2}+\frac{1}{2}}\left[ 
\frac{\left( \frac{w_{p}}{w_{o}}\right) ^{2}-\frac{1}{2}}{\left( \frac{w_{p}%
}{w_{o}}\right) ^{2}+\frac{1}{2}}\right] ^{q}  \label{3.2.42}
\end{equation}
Le recouvrement spatial d\'{e}pend uniquement du rapport $w_{p}/w_{o}$ entre
les cols acoustique et optique.

\subsection{Mesure du bruit thermique du r\'{e}sonateur plan-convexe\label%
{III-4-2}}

Pour mesurer le spectre des fluctuations thermiques du r\'{e}sonateur, on
utilise un faisceau de mesure r\'{e}sonnant avec la cavit\'{e} et
d'intensit\'{e} suffisamment faible pour n\'{e}g-\newline
liger les effets de pression de radiation (voir section 2.4.2). Le spectre
de bruit de phase du faisceau r\'{e}fl\'{e}chi est alors donn\'{e} par la
relation (2.76) en rempla\c{c}ant le spectre $S_{x}\left[ \Omega \right] $
par $S_{\hat{u}}\left[ \Omega \right] $. Les fluctuations de position $\hat{u%
}$ du miroir sont reli\'{e}es uniquement \`{a} la force de Langevin $F_{T}$
(\'{e}quation \ref{3.2.35}) et le spectre $S_{\hat{u}}\left[ \Omega \right] $
est donn\'{e} par: 
\begin{equation}
S_{\hat{u}}\left[ \Omega \right] =\left| \chi _{eff}\left[ \Omega \right]
\right| ^{2}S_{T}\left[ \Omega \right] =\frac{2k_{B}T}{\Omega }~{\cal I}%
m\left\{ \chi _{eff}\left[ \Omega \right] \right\}  \label{3.2.39}
\end{equation}
On obtient finalement pour le spectre des fluctuations de phase du faisceau
r\'{e}fl\'{e}chi $S_{q}^{out}\left[ \Omega \right] $: 
\begin{equation}
S_{q}^{out}\left[ \Omega \right] =S_{q}^{in}\left[ \Omega \right] +16\frac{1%
}{1+\left( \Omega /\Omega _{cav}\right) ^{2}}\frac{\hat{\Psi}_{NL}}{\gamma }%
\frac{k_{B}T}{\hbar \Omega }~{\cal I}m\left\{ \bar{\chi}_{eff}\left[ \Omega
\right] \right\}  \label{3.2.40}
\end{equation}
o\`{u} $\bar{\chi}_{eff}\left[ \Omega \right] =\chi _{eff}\left[ \Omega
\right] /\chi _{eff}\left[ 0\right] $ est la susceptibilit\'{e} effective
normalis\'{e}e \`{a} $1$ \`{a} fr\'{e}quence nulle et $\hat{\Psi}_{NL}$ est
le d\'{e}phasage non lin\'{e}aire li\'{e} au d\'{e}placement de la face
plane du r\'{e}sonateur sous l'effet de la pression de radiation moyenne. Ce
d\'{e}phasage s'\'{e}crit, d'apr\`{e}s les relations (2.47), (\ref{3.2.27})
et (\ref{3.2.36}):

\begin{equation}
\hat{\Psi}_{NL}=4\hbar k^{2}\overline{I}~\chi _{eff}\left[ 0\right] =4\hbar
k^{2}\overline{I}~\stackunder{p,q}{\sum }\frac{\left\langle
v_{0}^{2},u_{pq}\right\rangle ^{2}}{M_{pq}\Omega _{pq}^{2}}  \label{3.2.41}
\end{equation}
Par rapport au cas simple de l'oscillateur harmonique, l'expression du
d\'{e}phasage non lin\'{e}aire est plus difficile \`{a} \'{e}valuer
puisqu'elle fait intervenir ici la contribution d'une infinit\'{e}
d'oscillateurs harmoniques pond\'{e}r\'{e}s par leur recouvrement spatial $%
\left\langle v_{0}^{2},u_{pq}\right\rangle $ avec le mode optique. Nous
reviendrons sur ce point dans la section 3.4.4.

On peut relier le spectre de bruit de phase $S_{q}^{out}\left[ \Omega
\right] $ \`{a} l'angle de perte $\Phi $ qui d\'{e}crit la dissipation dans
le r\'{e}sonateur: 
\begin{equation}
S_{q}^{out}\left[ \Omega \right] =S_{q}^{in}\left[ \Omega \right] +\frac{16}{%
1+\left( \Omega /\Omega _{cav}\right) ^{2}}\frac{\hat{\Psi}_{NL}}{\gamma }%
\frac{\Phi \left[ \Omega \right] /\Omega }{\Phi \left[ \Omega _{M}\right]
/\Omega _{M}}\frac{n_{T}}{Q}  \label{3.2.43}
\end{equation}
\[
\qquad \qquad \quad \times \stackunder{pq}{\sum }\left\langle
v_{0}^{2},u_{pq}\right\rangle ^{2}\frac{\chi _{pq}\left[ 0\right] }{\chi
_{eff}\left[ 0\right] }\left| \bar{\chi}_{pq}\left[ \Omega \right] \right|
^{2} 
\]
o\`{u} $n_{T}$ est le nombre de phonons thermiques \`{a} la fr\'{e}quence
fondamentale $\Omega _{M}$ (\'{e}quation 2.71). On obtient ainsi une
expression similaire \`{a} celle obtenue dans la section 2.4.2 (\'{e}quation
2.86). La contribution du bruit thermique au spectre de bruit de phase du
faisceau r\'{e}fl\'{e}chi est caract\'{e}ris\'{e}e par les m\^{e}mes
param\`{e}tres, \`{a} savoir le rapport $\hat{\Psi}_{NL}/\gamma $ qui
d\'{e}crit l'amplitude du couplage optom\'{e}canique, le rapport $n_{T}/Q$
qui d\'{e}crit l'amplitude du bruit thermique et enfin le terme faisant
intervenir l'angle de perte $\Phi \left[ \Omega \right] $ qui d\'{e}crit la
dissipation dans le r\'{e}sonateur. La seule diff\'{e}rence avec
l'oscillateur harmonique est le terme de somme dans lequel appara\^{\i }t la
contribution de tous les modes acoustiques (terme $\left| \bar{\chi}%
_{pq}\left[ \Omega \right] \right| ^{2}$). Cette contribution est
pond\'{e}r\'{e}e par un terme qui tient compte de la contribution relative
du mode consid\'{e}r\'{e} \`{a} la susceptibilit\'{e} effective \`{a}
fr\'{e}quence nulle.

\begin{figure}[tbp]
\centerline{\psfig{figure=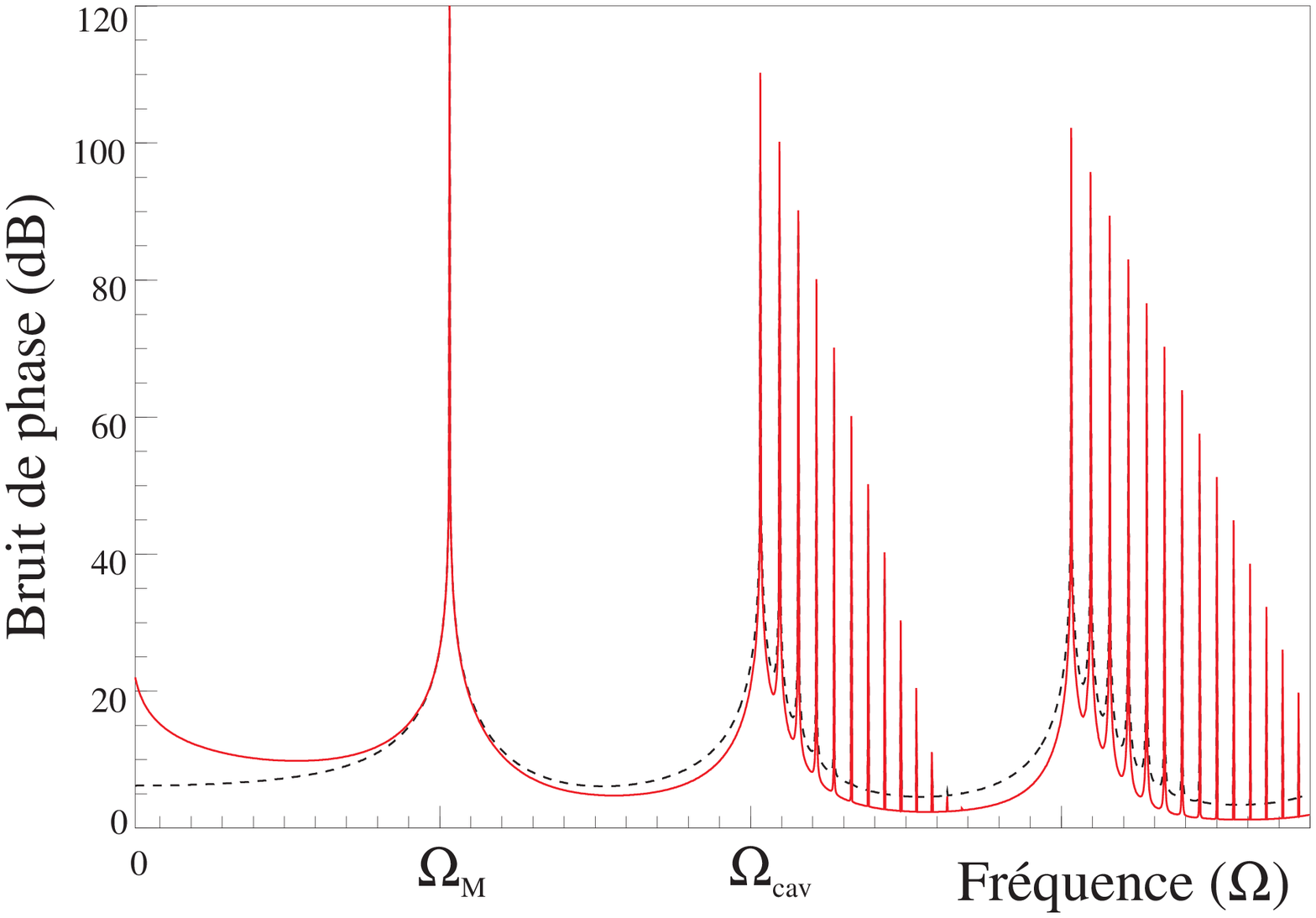,height=8cm}}
\caption{Spectres de bruit de phase en sortie lorsque le faisceau,
r\'{e}sonnant avec la cavit\'{e}, est adapt\'{e} spatialement au mode
acoustique fondamental ($w_{o}=\protect\sqrt{2}w_{1}$). La courbe en tirets
repr\'{e}sente le bruit thermique pour un amortissement de type visqueux,
alors que la courbe en trait plein d\'{e}crit un amortissement \`{a} angle
de perte constant}
\label{Fig_3brthew1}
\end{figure}

Le spectre de phase en sortie $S_{q}^{out}\left[ \Omega \right] $ peut
\^{e}tre d\'{e}termin\'{e} en calculant num\'{e}riquement la double somme
dans l'\'{e}quation (\ref{3.2.43}). La figure \ref{Fig_3brthew1} montre les
spectres de bruit $S_{q}^{out}\left[ \Omega \right] $ obtenus \`{a}
temp\'{e}rature ambiante pour les deux mod\`{e}les de dissipation
(amortissement visqueux et $\Phi $ constant). Les param\`{e}tres optiques de
la cavit\'{e} sont identiques \`{a} ceux utilis\'{e}s dans la section 2.4.2
: finesse ${\cal F}=3~10^{5}$ et bande passante $\Omega _{cav}=2\Omega _{M}$%
. Pour pouvoir n\'{e}gliger les effets de pression de radiation devant les
effets thermiques, on garde aussi le m\^{e}me d\'{e}phasage non lin\'{e}aire 
$\hat{\Psi}_{NL}=\gamma /20$. Le r\'{e}sonateur m\'{e}canique,
caract\'{e}ris\'{e} par une \'{e}paisseur au centre $h_{0}=1.5~mm$ et un
rayon de courbure $R=150~mm$, est constitu\'{e} d'un substrat en silice
fondue de masse volumique $\rho =2.2~g.cm^{-3}$ et de vitesse de propagation
longitudinale du son $c_{l}=5970~m/s$. Le col $w_{1}$ du mode acoustique
fondamental est ainsi \'{e}gal \`{a} $3.8~mm$ (\'{e}quation \ref{mod ac3})
et la fr\'{e}quence de r\'{e}sonance $\Omega _{M}$ est \'{e}gale \`{a} $%
1.2~10^{7}rad/s$. Le facteur de qualit\'{e} $Q$ du fondamental est choisi
\'{e}gal \`{a} $10^{6}$. 

La figure \ref{Fig_3brthew1} montre que le spectre de bruit thermique
pr\'{e}sente plusieurs pics associ\'{e}s aux modes acoustiques du
r\'{e}sonateur : chaque mode contribue au spectre de bruit thermique par une
r\'{e}ponse Lorentzienne \`{a} la fr\'{e}quence $\Omega _{pq}$. On distingue
sur cette figure les diff\'{e}rents modes longitudinaux du r\'{e}sonateur,
correspondant \`{a} la fr\'{e}quence $\Omega _{M}$ et \`{a} ses harmoniques.
Chaque mode longitudinal $\left( p,q=0\right) $ est suivi d'une s\'{e}rie de
modes transverses $\left( p,q\geq 1\right) $, except\'{e} le mode
fondamental. La contribution de chaque mode d\'{e}pend en fait de la fa\c{c}%
on dont ils sont adapt\'{e}s au mode optique. L'expression du recouvrement
spatial $\left\langle v_{0}^{2},u_{pq}\right\rangle $ (\'{e}quation \ref
{3.2.42}) montre qu'un mode acoustique longitudinal ($p$, $q=0$) est
parfaitement adapt\'{e} au mode optique si la relation $w_{o}=\sqrt{2}w_{p}$
est satisfaite : le terme de recouvrement est alors maximum. Dans ce cas,
l'int\'{e}grale de recouvrement est nulle pour les modes transverses
associ\'{e}s \`{a} ce mode longitudinal. Les spectres de la figure \ref
{Fig_3brthew1} sont trac\'{e}s dans le cas o\`{u} le faisceau incident est
adapt\'{e} au mode acoustique fondamental, c'est-\`{a}-dire pour $w_{o}=%
\sqrt{2}w_{1}$. On voit que les modes acoustiques transverses du fondamental
ne sont pas coupl\'{e}s au faisceau lumineux et ne contribuent pas au
spectre. Par contre, les autres modes acoustiques longitudinaux ne sont pas
parfaitement adapt\'{e}s au faisceau lumineux et le spectre fait appara\^{\i
}tre pour ces modes un peigne de modes transverses d'amplitude
d\'{e}croissante : on peut montrer \`{a} partir de l'\'{e}quation (\ref
{3.2.42}) que le recouvrement des modes transverses est d'autant plus faible
que l'ordre $q$ du mode est \'{e}lev\'{e}.

\begin{figure}[tbp]
\centerline{\psfig{figure=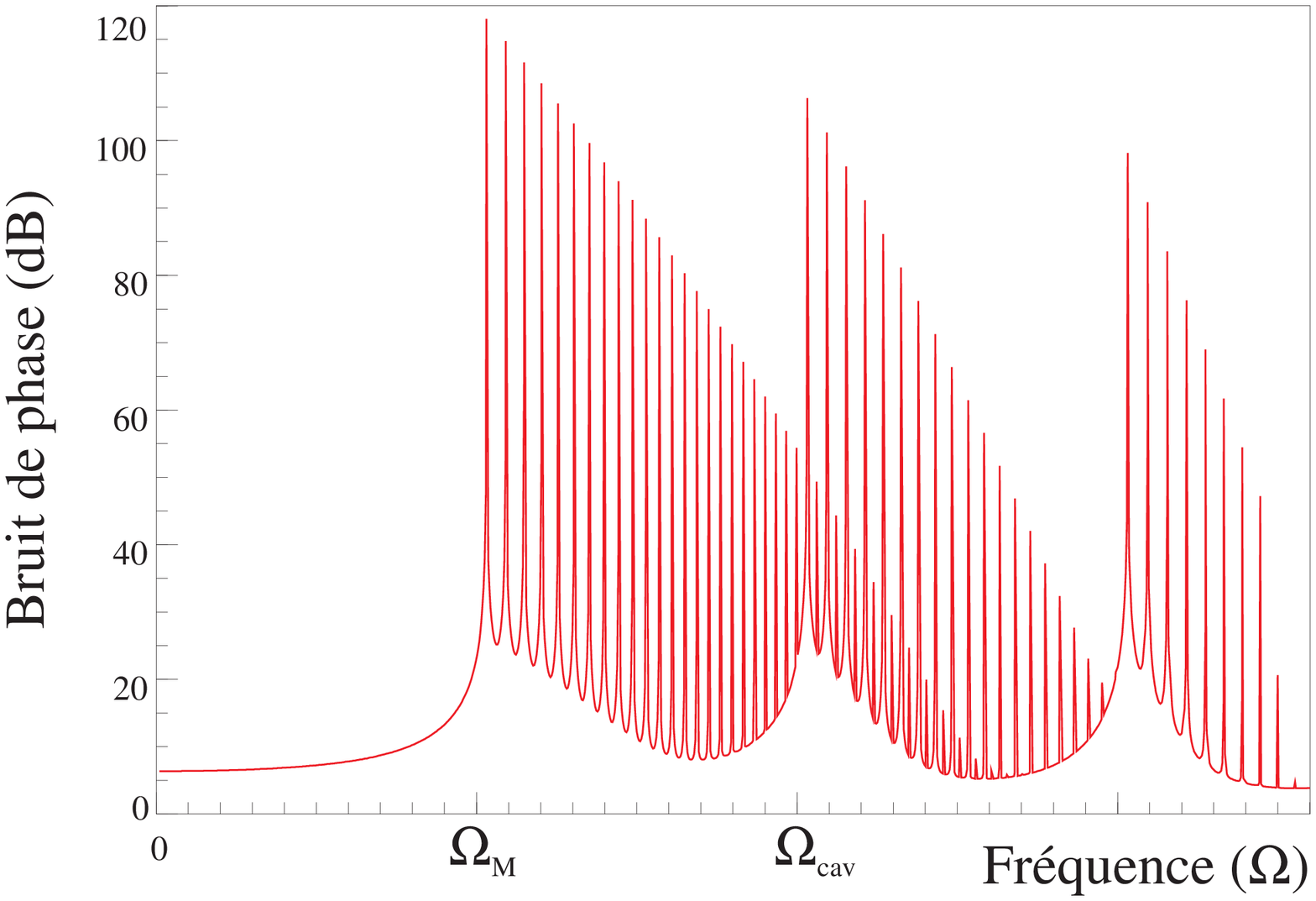,height=8cm}}
\caption{Spectre de bruit de phase en sortie pour un col optique $w_{o}=%
\frac{1}{2}w_{1}$. Le spectre est calcul\'{e} dans le cas d'un amortissement
visqueux. Les modes longitudinaux n'\'{e}tant pas adapt\'{e}s spatialement
au mode optique, on voit appara\^{\i }tre sur le signal des s\'{e}ries
importantes de modes transverses qui se recouvrent}
\label{Fig_3brthew2}
\end{figure}

L'allure du spectre de bruit de phase d\'{e}pend donc beaucoup de
l'adaptation spatiale entre les modes acoustiques et optique. La figure \ref
{Fig_3brthew2} montre le spectre de bruit de phase en sortie dans le cas
d'un amortissement visqueux, pour un col optique $w_{o}$ plus petit,
\'{e}gal \`{a} $w_{1}/2$. Le faisceau lumineux n'est plus adapt\'{e}
spatialement au mode fondamental et il est ainsi coupl\'{e} \`{a} un grand
nombre de ses modes transverses, si bien que les s\'{e}ries de modes
transverses associ\'{e}es aux diff\'{e}rents modes longitudinaux se
recouvrent. 

En conclusion, nous avons montr\'{e} que le spectre de bruit du faisceau
r\'{e}fl\'{e}chi par la cavit\'{e} peut fournir des informations sur le
mouvement Brownien du r\'{e}sonateur plan-convexe. Cette \'{e}tude montre
d'autre part l'importance de l'adaptation spatiale entre le faisceau
lumineux et les modes acoustiques du r\'{e}sonateur. En modifiant la taille
du faisceau lumineux, il est possible de modifier le couplage avec les modes
acoustiques, et ainsi de modifier l'influence dans le signal optique du
bruit thermique de certains modes acoustiques.

\subsection{Mesure QND de l'intensit\'{e} lumineuse\label{III-4-3}}

Le principe de la mesure QND d\'{e}crit dans la partie 2.4 reste valable
pour un r\'{e}sonateur plan-convexe. Le faisceau signal et le faisceau de
mesure sont tous deux r\'{e}sonnants avec la cavit\'{e} et adapt\'{e}s
spatialement \`{a} son mode fondamental $TEM_{00}$. Dans ces conditions
l'intensit\'{e} du signal n'est pas d\'{e}grad\'{e}e par la mesure et elle
est corr\'{e}l\'{e}e \`{a} la quadrature de phase du faisceau de mesure
r\'{e}fl\'{e}chi. Les deux faisceaux \'{e}tant gaussiens, le d\'{e}placement 
$\hat{u}\left[ \Omega \right] $ auquel est sensible la phase $\delta
q_{m}^{out}$ du faisceau de mesure r\'{e}fl\'{e}chi est donn\'{e} par
l'\'{e}quation (\ref{3.2.35}) dans laquelle l'intensit\'{e} intracavit\'{e}
est \'{e}gale \`{a} la somme des intensit\'{e}s des deux faisceaux (on
suppose les deux faisceaux polaris\'{e}s orthogonalement). L'expression de $%
\delta q_{m}^{out}$ est obtenue en rempla\c{c}ant dans la relation (2.88) le
d\'{e}placement $\delta x$ par $\hat{u}$, ce qui revient \`{a} substituer
dans les r\'{e}sultats de la section 2.4.3 la susceptibilit\'{e} harmonique $%
\chi $ par la susceptibilit\'{e} effective $\chi _{eff}$ qui prend en compte
la structure spatiale du syst\`{e}me. 
\begin{figure}[tbp]
\centerline{\psfig{figure=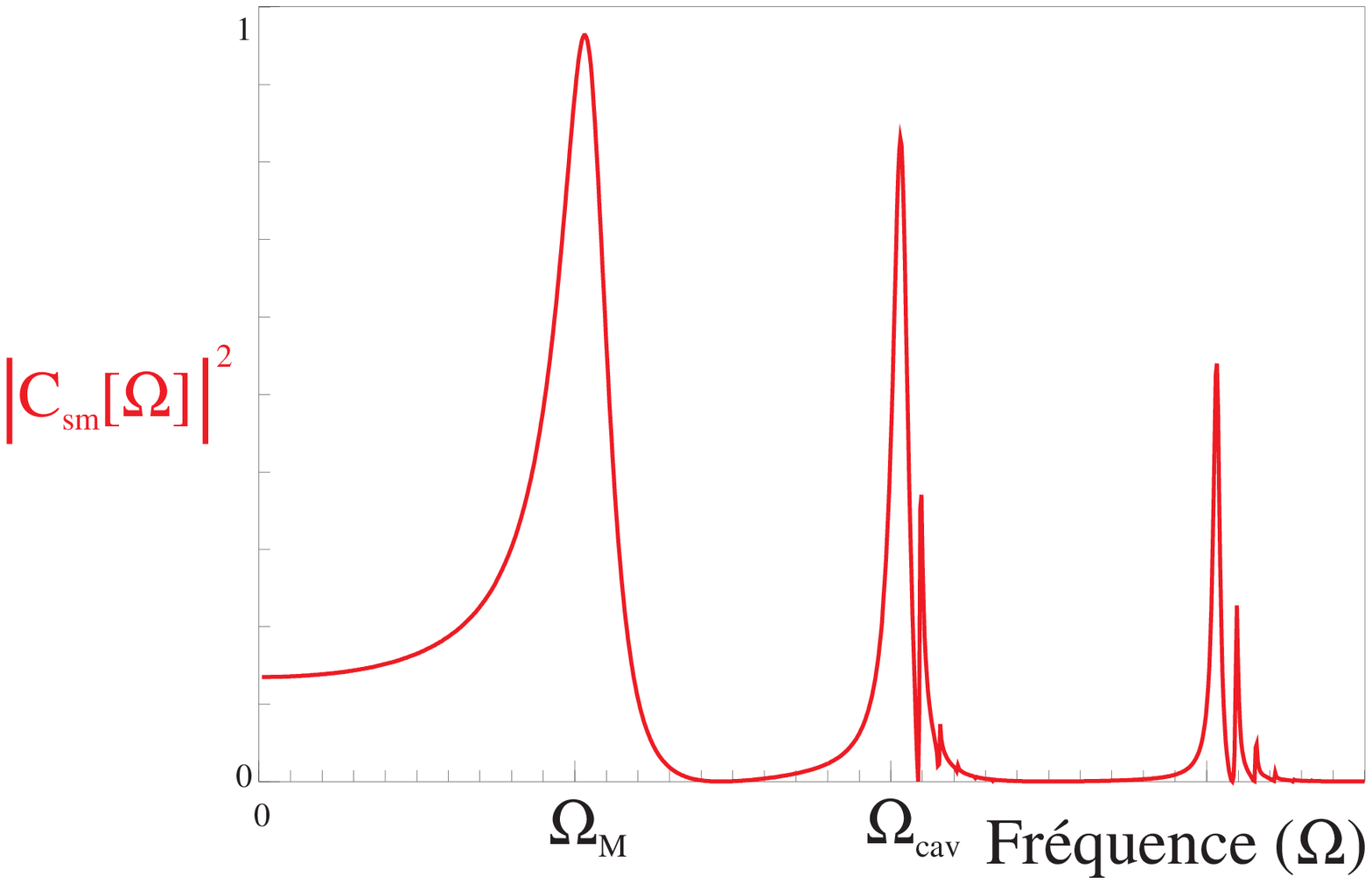,height=8cm}}
\caption{Spectre des corr\'{e}lations quantiques entre la quadrature
d'amplitude du signal et la quadrature de phase du faisceau de mesure. Les
deux faisceaux sont r\'{e}sonnants avec la cavit\'{e} et adapt\'{e}s
spatialement au mode acoustique fondamental du r\'{e}sonateur m\'{e}canique
plan-convexe ($w_{0}=\protect\sqrt{2}w_{1}$)}
\label{Fig_3csmint}
\end{figure}

L'expression de la fonction de corr\'{e}lation signal-mesure $\left| {\cal C}%
_{sm}\left[ \Omega \right] \right| ^{2}$ est donc donn\'{e}e par la relation
(2.93b) en rempla\c{c}ant la susceptibilit\'{e} harmonique $\overline{\chi }$
dans les \'{e}quations (2.94) par la susceptibilit\'{e} $\bar{\chi}_{eff}$.
Le spectre des corr\'{e}lations est ensuite obtenu en calculant
num\'{e}riquement la double somme qui appara\^{\i }t dans l'expression de $%
\chi _{eff}$. La figure \ref{Fig_3csmint} montre le r\'{e}sultat du calcul
dans le cas o\`{u} les deux faisceaux sont adapt\'{e}s spatialement au mode
acoustique fondamental ($w_{0}=\sqrt{2}w_{1}$). Les param\`{e}tres de la
cavit\'{e} sont les m\^{e}mes que dans la section 2.4.3; en particulier les
d\'{e}phasages non lin\'{e}aires $\hat{\Psi}_{s}$ et $\hat{\Psi}_{m}$ sont
\'{e}gaux \`{a} $\gamma $ et $\gamma /100$. Les caract\'{e}ristiques du
r\'{e}sonateur m\'{e}canique sont les m\^{e}mes que dans la section
pr\'{e}c\'{e}dente.

Comme dans le cas du mod\`{e}le harmonique (figure \ref{Fig_2csmint}, page 
\pageref{Fig_2csmint}), les corr\'{e}lations quantiques reproduisent la
d\'{e}pendance en fr\'{e}quence de la r\'{e}ponse m\'{e}canique du miroir
mobile. Dans le cas du r\'{e}sonateur plan-convexe, la r\'{e}ponse est
caract\'{e}ris\'{e}e par plusieurs r\'{e}sonances et les corr\'{e}lations
signal-mesure sont maximales aux fr\'{e}quences de r\'{e}sonances
acoustiques du r\'{e}sonateur. Comme dans le cas monodimensionnel, on
observe l'effet de filtrage de la cavit\'{e} qui limite les corr\'{e}lations
pour des fr\'{e}quences sup\'{e}rieures \`{a} la bande passante de la
cavit\'{e}.

On retrouve dans le cas du r\'{e}sonateur plan-convexe les param\`{e}tres
essentiels du couplage optom\'{e}canique que nous avions pr\'{e}cis\'{e}s
dans le chapitre pr\'{e}c\'{e}dent. L'efficacit\'{e} du couplage d\'{e}pend
des rapports $\hat{\Psi}_{s/m}/\gamma $ et $n_{T}/Q$ qui d\'{e}crivent
respectivement les effets de pression de radiation et du bruit thermique. A
basse temp\'{e}rature, les effets thermiques sont r\'{e}duits gr\^{a}ce aux
valeurs \'{e}lev\'{e}es de la fr\'{e}quence de r\'{e}sonance fondamentale $%
\Omega _{M}$ et du facteur de qualit\'{e} du r\'{e}sonateur. Les effets
quantiques li\'{e}s \`{a} la pression de radiation sont significatifs
lorsque les d\'{e}phasages non lin\'{e}aire $\hat{\Psi}_{s/m}$ sont de
l'ordre des pertes $\gamma $ de la cavit\'{e}.

\subsection{Masse effective\label{III-4-4}}

Le d\'{e}phasage non lin\'{e}aire est un param\`{e}tre important qui
intervient dans la caract\'{e}risation du couplage optom\'{e}canique. Nous
allons \'{e}tudier maintenant comment il est reli\'{e} aux param\`{e}tres
physiques du syst\`{e}me. Nous allons montrer en particulier que le
d\'{e}phasage non lin\'{e}aire produit par un r\'{e}sonateur plan-convexe
est beaucoup plus grand que dans le cas d'un simple oscillateur harmonique
de m\^{e}me masse totale.

Le d\'{e}phasage non lin\'{e}aire d\'{e}pend de la r\'{e}ponse m\'{e}canique
\`{a} fr\'{e}quence nulle $\chi _{eff}\left[ 0\right] $ du r\'{e}sonateur
(\'{e}quation \ref{3.2.41}). Cette r\'{e}ponse est reli\'{e}e \`{a} la
r\'{e}ponse de l'ensemble des modes acoustiques: 
\begin{equation}
\chi _{eff}\left[ 0\right] =\stackunder{p,q}{\sum }\frac{\left\langle
v_{0}^{2},u_{pq}\right\rangle ^{2}}{M_{pq}\Omega _{pq}^{2}}  \label{3.2.44}
\end{equation}
En fait, l'effet du r\'{e}sonateur \`{a} basse fr\'{e}quence est similaire
\`{a} celui d'un simple oscillateur harmonique de fr\'{e}quence de
r\'{e}sonance \'{e}gale \`{a} la fr\'{e}quence fondamentale $\Omega _{M}$ du
r\'{e}sonateur, mais de {\it masse effective} diff\'{e}rente de la masse
totale du r\'{e}sonateur. On peut d\'{e}finir cette masse effective $M_{eff}$
en identifiant la r\'{e}ponse \`{a} basse fr\'{e}quence $\chi _{eff}\left[
0\right] $ \`{a} celle d'un seul oscillateur harmonique:

\begin{equation}
\frac{1}{M_{eff}\Omega _{M}^{2}}=\stackunder{p,q}{\sum }\frac{\left\langle
v_{0}^{2},u_{pq}\right\rangle ^{2}}{M_{pq}\Omega _{pq}^{2}}  \label{3.2.45}
\end{equation}
La masse effective est donc reli\'{e}e aux masses $M_{pq}$ des modes
acoustiques et d\'{e}pend du recouvrement spatial avec le faisceau lumineux.

\begin{figure}[tbp]
\centerline{\psfig{figure=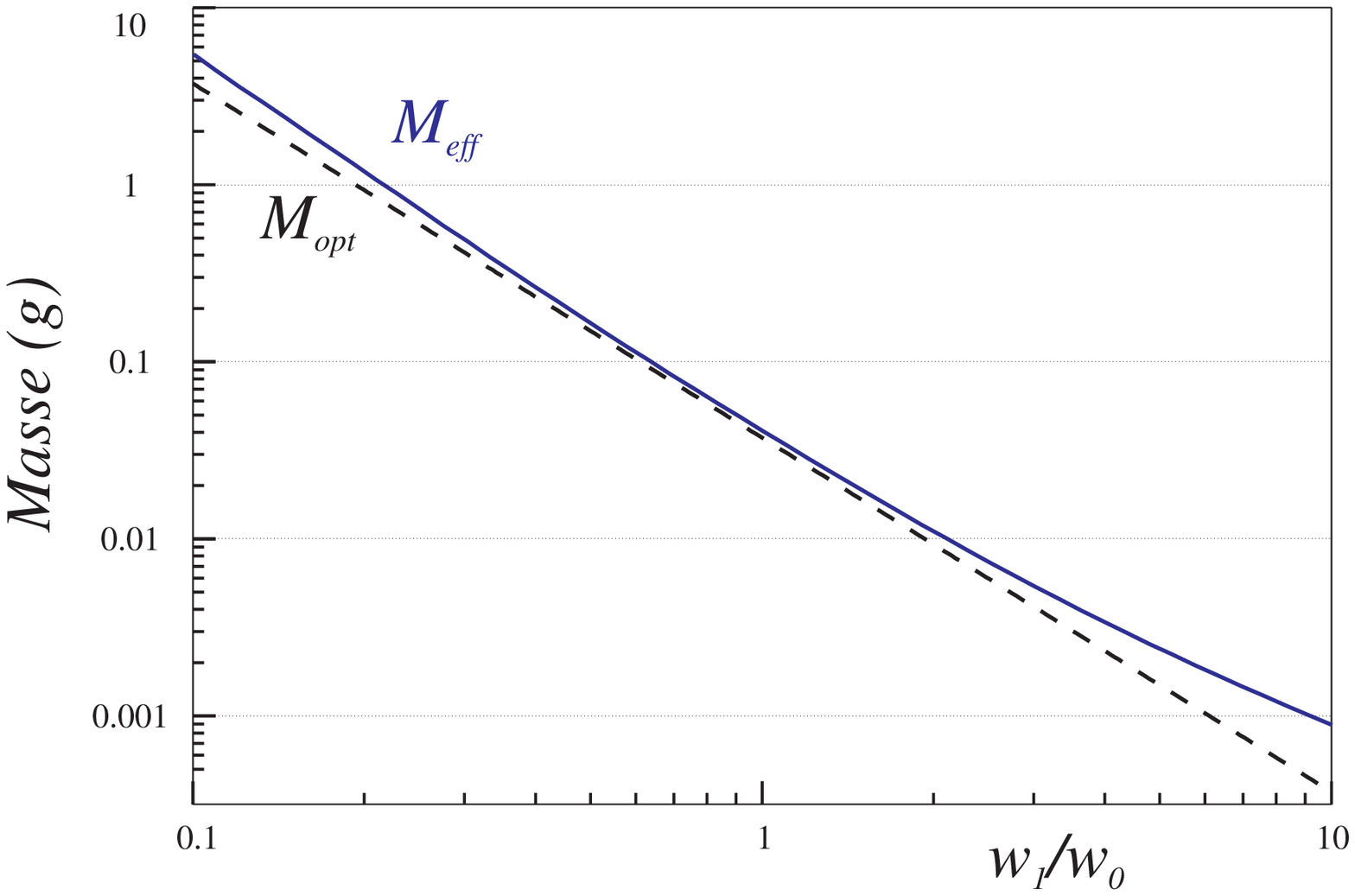,height=9cm}}
\caption{Variation des masses effective et optique en fonction du col du
faisceau lumineux}
\label{Fig_3masseff}
\end{figure}

A partir des expressions du col $w_{p}$, de la fr\'{e}quence de
r\'{e}sonance $\Omega _{pq}$, de la masse $M_{pq}$ et du recouvrement
spatial $\left\langle v_{0}^{2},u_{pq}\right\rangle $ (\'{e}quations \ref
{mod ac3}, \ref{mod ac4}, \ref{3.2.17}, \ref{3.2.42}), on peut exprimer la
masse effective en fonction de la masse du mode acoustique fondamental $M_{1}
$ et du rapport $r=w_{1}/w_{0}$ entre les cols du mode acoustique
fondamental et du mode optique: 
\begin{equation}
\frac{1}{M_{eff}}=\frac{1}{M_{1}}\stackunder{p,q}{\sum }\frac{4r^{4}}{\left(
2r^{2}+p\right) ^{2}}\left( \frac{2r^{2}-p}{2r^{2}+p}\right) ^{2q}\frac{1}{p+%
\frac{2}{\pi }\sqrt{\frac{h_{0}}{R}}\left( 2q+1\right) }  \label{3.2.46}
\end{equation}
La courbe en trait plein de la figure \ref{Fig_3masseff} montre
l'\'{e}volution de la masse effective en fonction du rapport $r$, en gardant
les m\^{e}mes param\`{e}tres g\'{e}om\'{e}triques du r\'{e}sonateur que ceux
utilis\'{e}s dans la partie pr\'{e}c\'{e}dente ($h_{0}=1.5~mm$ et $R=150~mm$%
). Le mode fondamental a alors une masse $M_{1}=37~mg$ et un col $%
w_{1}=3.8~mm$. La masse effective est d'autant plus petite que le col
optique $w_{0}$ est petit devant le col acoustique $w_{1}$ et elle peut
atteindre une valeur de $1~mg$ seulement pour un col optique $w_{0}$
\'{e}gal \`{a} $w_{1}/10$, soit $380~\mu m$. La masse effective peut donc
\^{e}tre deux \`{a} trois ordres de grandeur plus petite que la masse totale
du r\'{e}sonateur, qui est de l'ordre de $400~mg$, augmentant dans les
m\^{e}mes proportions les effets du couplage optom\'{e}canique. Pour une
masse effective de $1~mg$, il suffit d'une puissance lumineuse incidente de $%
60~mW$ pour avoir un d\'{e}phasage non lin\'{e}aire $\hat{\Psi}_{NL}$
\'{e}gal \`{a} $\gamma $.

On peut par ailleurs donner une estimation simple, not\'{e}e $M_{opt}$, de
la masse effective \`{a} la limite $h_{0}\ll R$. On peut alors n\'{e}gliger
la d\'{e}pendance en $q$ dans le dernier terme de l'expression (\ref{3.2.46}%
). Cela revient \`{a} supposer tous les modes transverses
d\'{e}g\'{e}n\'{e}r\'{e}s ($\Omega _{pq}=\Omega _{p0}$). La somme sur $p$
est alors une simple s\'{e}rie g\'{e}om\'{e}trique, et on obtient: 
\begin{equation}
\frac{1}{M_{opt}}=\frac{r^{2}}{2M_{1}}\stackunder{p}{\sum }\frac{1}{p^{2}}=%
\frac{\pi ^{2}}{12}\frac{r^{2}}{M_{1}}
\end{equation}
En utilisant l'expression de $M_{1}$ (\'{e}quation \ref{3.2.17}) on trouve
finalement: 
\begin{equation}
M_{opt}=\frac{12}{\pi^{2} }~\left( \frac{\pi }{4}\rho h_{0}w_{0}^{2}\right)
\end{equation}
Le terme entre parenth\`{e}ses repr\'{e}sente la {\it masse optique}, qui
correspond \`{a} la masse du volume du r\'{e}sonateur d\'{e}limit\'{e} par
le faisceau lumineux. La courbe en traits tiret\'{e}s de la figure \ref
{Fig_3masseff} d\'{e}crit la variation de la masse $M_{opt}$ en fonction du
col du faisceau dans la cavit\'{e}. Comme on peut le voir, cette masse
optique fournit une bonne estimation de la masse effective $M_{eff}$.

\section{Conclusion\label{III-5}}

\bigskip

Nous avons \'{e}tudi\'{e} dans ce chapitre les caract\'{e}ristiques du
couplage optom\'{e}canique entre un miroir mobile d\'{e}pos\'{e} sur un
r\'{e}sonateur plan-convexe et un faisceau lumineux gaussien. Nous avons
montr\'{e} qu'une d\'{e}formation du miroir est \'{e}quivalente pour le
champ \`{a} un d\'{e}placement suivant l'axe de la cavit\'{e},
d\'{e}placement \'{e}gal \`{a} la moyenne spatiale de la d\'{e}formation sur
la section du faisceau lumineux.

Nous avons ensuite d\'{e}termin\'{e} le d\'{e}placement du miroir lorsque le
r\'{e}sonateur est soumis \`{a} la force de pression de radiation du
faisceau lumineux tout en \'{e}tant en contact avec un bain thermique. Pour
cela nous avons utilis\'{e} l'\'{e}quivalence entre le r\'{e}sonateur
m\'{e}canique et un ensemble d'oscillateurs harmoniques associ\'{e}s aux
modes acoustiques propres du r\'{e}sonateur. Chaque mode acoustique
contribue au mouvement du miroir dans une proportion qui d\'{e}pend de sa
susceptibilit\'{e} m\'{e}canique et du recouvrement spatial entre le mode
acoustique et le faisceau lumineux.

Nous avons montr\'{e} qu'il est possible de rendre compte de la
complexit\'{e} de la r\'{e}ponse m\'{e}canique du r\'{e}sonateur en
d\'{e}finissant une susceptibilit\'{e} effective $\chi _{eff}$. Cette
susceptibilit\'{e} permet de g\'{e}n\'{e}raliser les r\'{e}sultats obtenus
dans le chapitre pr\'{e}c\'{e}dent pour un seul oscillateur harmonique. Nous
avons ainsi \'{e}tudi\'{e} les possibilit\'{e}s de mesurer le mouvement
Brownien du r\'{e}sonateur et de r\'{e}aliser une mesure quantique non
destructive. Les sp\'{e}cificit\'{e}s li\'{e}es \`{a} l'utilisation d'un
r\'{e}sonateur plan-convexe apparaissent essentiellement dans la
pr\'{e}sence de multiples fr\'{e}quences de r\'{e}sonance acoustiques et
dans l'importance de l'adaptation spatiale entre la lumi\`{e}re et les modes
acoustiques.

Les param\`{e}tres physiques essentiels du couplage optom\'{e}canique
restent cependant les m\^{e}mes. Les effets li\'{e}s \`{a} la pression de
radiation sont significatifs lorsque le d\'{e}phasage non lin\'{e}aire $\hat{%
\Psi}_{NL}$ est de l'ordre des pertes $\gamma $ de la cavit\'{e}. Les effets
thermiques sont proportionnels au rapport $n_{T}/Q$ entre le nombre de
phonons thermiques \`{a} la fr\'{e}quence de r\'{e}sonance fondamentale du
r\'{e}sonateur et le facteur de qualit\'{e} de cette r\'{e}sonance.

Nous avons enfin d\'{e}fini une masse effective qui permet de d\'{e}crire
les effets du couplage optom\'{e}canique \`{a} basse fr\'{e}quence. Un
r\'{e}sonateur plan-convexe est \'{e}quivalent \`{a} un oscillateur
harmonique dont la masse effective est de l'ordre de la {\it masse optique},
c'est \`{a} dire la masse du volume du r\'{e}sonateur {\it \'{e}clair\'{e}}
par le faisceau lumineux. On obtient ainsi une masse effective tr\`{e}s
petite, inf\'{e}rieure au milligramme, lorsque la section du faisceau
lumineux est suffisamment petite. Ceci devrait permettre d'observer les
effets du couplage optom\'{e}canique pour des puissances lumineuses
incidentes raisonnables, de l'ordre de quelques milliwatts.

Notons enfin que cette \'{e}tude n'est pas limit\'{e}e \`{a} la seule
g\'{e}om\'{e}trie plan-convexe. Le choix d'une g\'{e}om\'{e}trie pour le
r\'{e}sonateur modifie simplement la structure spatiale des modes
acoustiques. Des \'{e}tudes similaires \`{a} celle pr\'{e}sent\'{e}e ici ont
\'{e}t\'{e} r\'{e}alis\'{e}es pour caract\'{e}riser le bruit thermique des
modes de vibrations internes des miroirs des interf\'{e}rom\`{e}tres VIRGO
et LIGO\cite{masse virgo ligo}. Ces miroirs ont une structure cylindrique et
il est n\'{e}cessaire d'utiliser des m\'{e}thodes de calcul num\'{e}rique
par \'{e}l\'{e}ments finis pour d\'{e}terminer la structure spatiale des
modes acoustiques ainsi que leur fr\'{e}quences propres. On trouve alors que
la masse effective est $5$ \`{a} $10$ fois plus petite que la masse totale
du miroir. C'est finalement gr\^{a}ce au confinement radial des modes
acoustiques qu'il est possible d'obtenir avec un r\'{e}sonateur plan-convexe
une masse effective beaucoup plus petite. \newpage

%% file: chapter4.tex
\chapter{LE MONTAGE EXPERIMENTAL}

\bigskip \bigskip

L'exp\'{e}rience que nous allons d\'{e}crire dans ce chapitre a
d\'{e}but\'{e} en d\'{e}cembre 1994. Nous avons r\'{e}alis\'{e} au cours de
ces derni\`{e}res ann\'{e}es le montage exp\'{e}rimental qui doit permettre
de mettre en \'{e}vidence les effets quantiques du couplage
optom\'{e}canique pr\'{e}sent\'{e}s dans les chapitres pr\'{e}c\'{e}dents.
Ce montage est essentiellement compos\'{e} de quatre parties comme le montre
la figure \ref{Fig_4manip1} : 
\begin{figure}[tbp]
\centerline{\psfig{figure=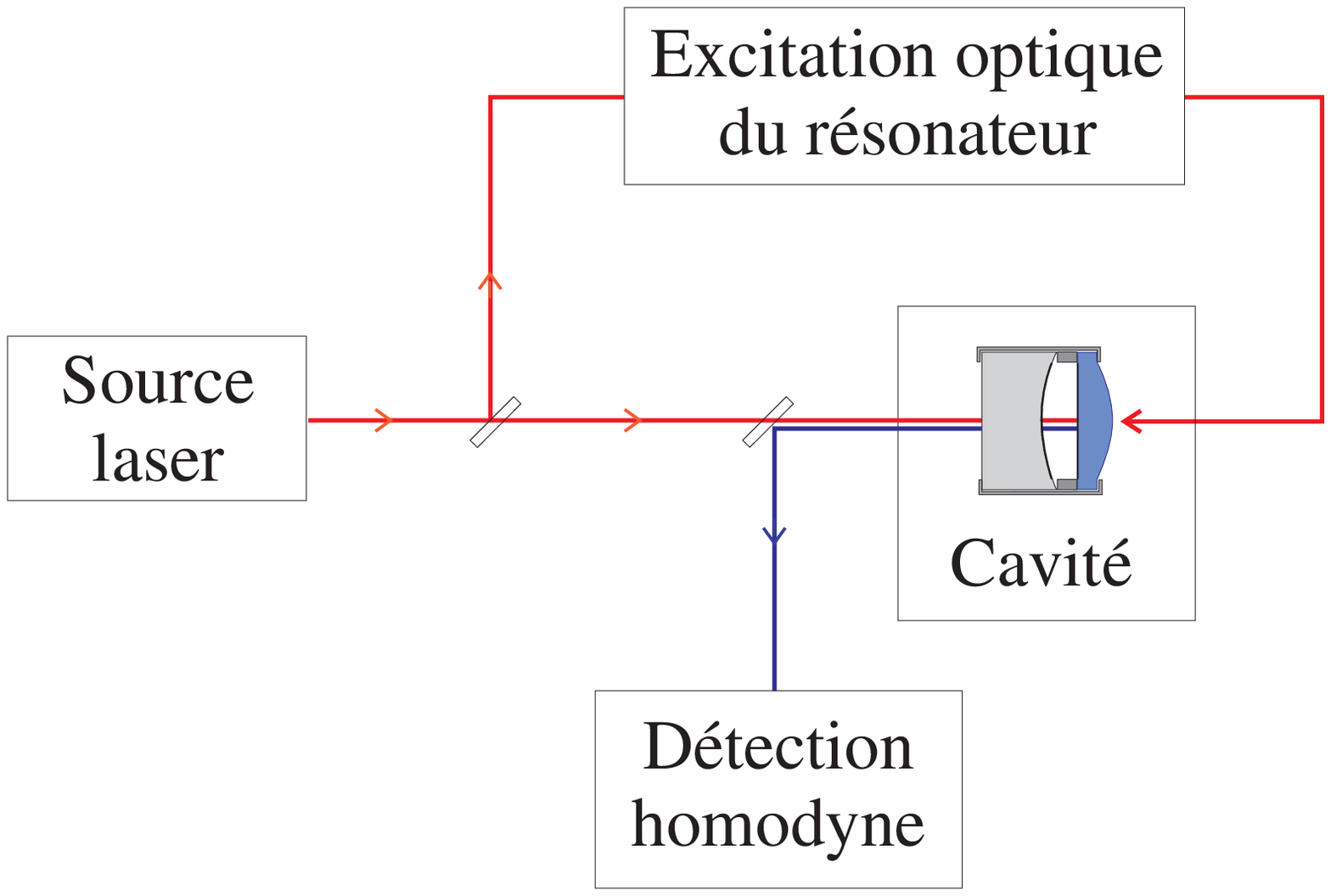,height=7cm}}
\caption{Sch\'{e}ma g\'{e}n\'{e}ral de l'exp\'{e}rience. Le montage est
compos\'{e} de quatre parties : la source laser, la cavit\'{e} \`{a} miroir
mobile, un syst\`{e}me de d\'{e}tection homodyne, et un dispositif
permettant d'exciter optiquement les modes acoustiques du r\'{e}sonateur}
\label{Fig_4manip1}
\end{figure}

{\bf -} La cavit\'{e} \`{a} miroir mobile constitue le coeur de
l'exp\'{e}rience car ses caract\'{e}ristiques imposent des contraintes
s\'{e}v\`{e}res pour le reste du montage. La cavit\'{e} est compos\'{e}e
d'un miroir qui sert de coupleur d'entr\'{e}e-sortie et du miroir mobile de
haute r\'{e}flectivit\'{e} d\'{e}pos\'{e} sur un r\'{e}sonateur
m\'{e}canique de structure plan-convexe. L'ensemble forme une cavit\'{e}
lin\'{e}aire de grande finesse \`{a} une seule entr\'{e}e-sortie.

{\bf -} La source laser est constitu\'{e}e d'un laser titane saphir
stabilis\'{e} en fr\'{e}quence et en intensit\'{e}. Etant donn\'{e} la
grande finesse de la cavit\'{e} \`{a} miroir mobile, la source laser doit
pr\'{e}senter une tr\`{e}s grande stabilit\'{e} en fr\'{e}quence et en
intensit\'{e}. Ceci a n\'{e}cessit\'{e} la r\'{e}alisation de syst\`{e}mes
de stabilisation tr\`{e}s performants. D'autre part, le faisceau \`{a} la
sortie du laser pr\'{e}sente un certain astigmatisme. Pour le supprimer, le
faisceau laser est filtr\'{e} spatialement \`{a} l'aide d'une cavit\'{e}
r\'{e}sonnante, ce qui permet d'adapter pr\'{e}cis\'{e}ment le faisceau au
mode fondamental de la cavit\'{e} \`{a} miroir mobile.

{\bf -} Pour mesurer les fluctuations du faisceau r\'{e}fl\'{e}chi par la
cavit\'{e}, on utilise un syst\`{e}me de d\'{e}tection homodyne qui permet
de mesurer le bruit de n'importe quelle quadrature du faisceau
r\'{e}fl\'{e}chi, \`{a} l'\'{e}chelle des fluctuations quantiques.

{\bf -} Le dernier \'{e}l\'{e}ment est un dispositif permettant d'exciter
les modes acoustiques du r\'{e}sonateur, afin de caract\'{e}riser la
r\'{e}ponse m\'{e}canique du miroir mobile. On utilise une partie du
faisceau laser, modul\'{e}e en intensit\'{e} \`{a} une fr\'{e}quence
variable. Ce faisceau se r\'{e}fl\'{e}chit par l'arri\`{e}re sur le miroir
mobile et exerce ainsi une force de pression de radiation dont l'amplitude
est modul\'{e}e.

Dans la suite nous allons d\'{e}crire ces diff\'{e}rentes parties et
pr\'{e}senter pour chacune d'entre elles les performances obtenues sur notre
montage.

\section{La cavit\'{e} \`{a} miroir mobile\label{IV-1}}

\bigskip

\begin{figure}[tbp]
\centerline{\psfig{figure=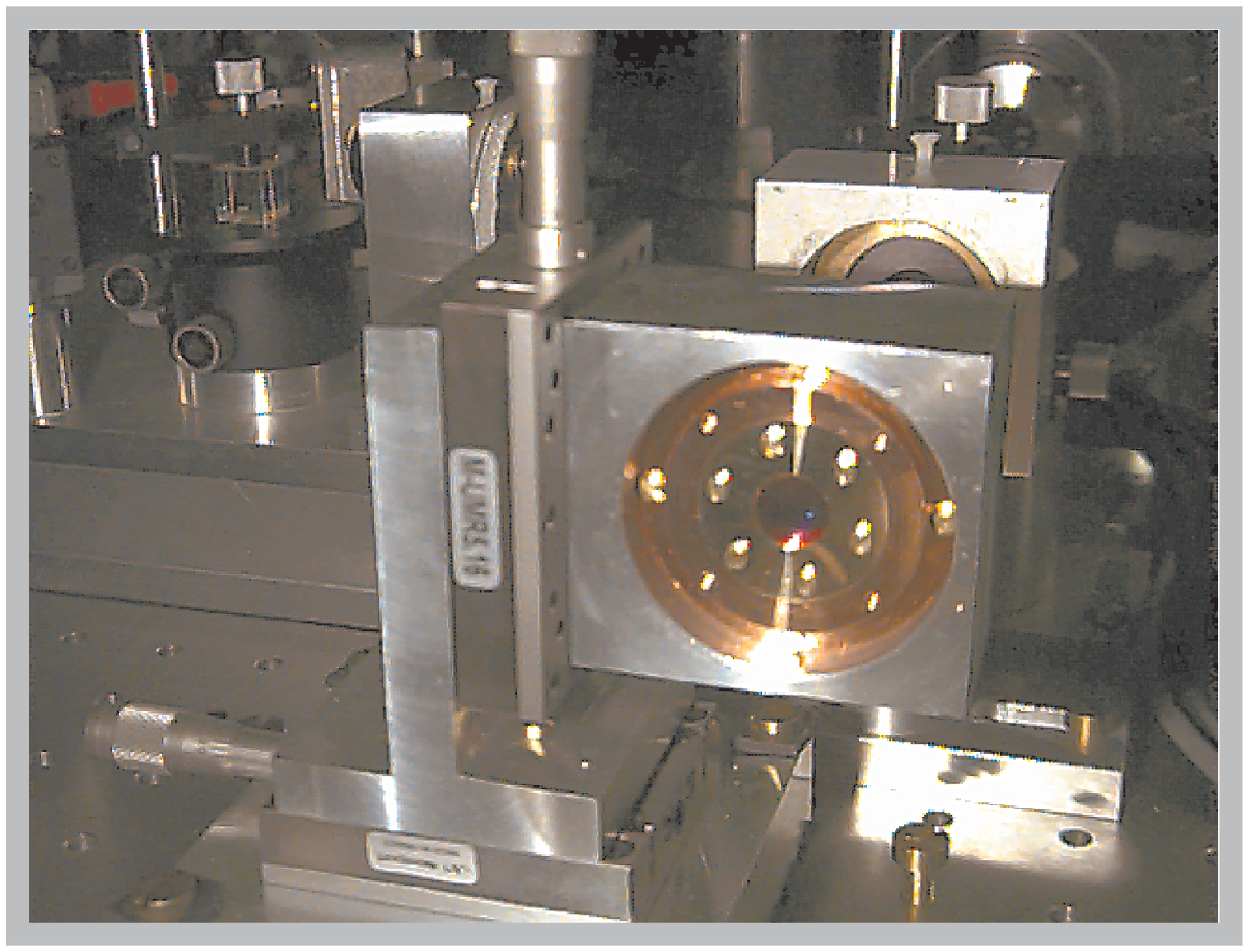,height=7cm}}
\caption{Vue arri\`{e}re de la cavit\'{e} \`{a} miroir mobile mont\'{e}e
dans son support}
\label{Fig_4photfpm}
\end{figure}

La cavit\'{e} \`{a} miroir mobile est une cavit\'{e} lin\'{e}aire d'une
longueur comprise entre $0.5$ et $1~mm$. La longueur de la cavit\'{e} ainsi
que le parall\'{e}lisme et le centrage des deux miroirs sont fix\'{e}s
gr\^{a}ce \`{a} deux pi\`{e}ces cylindriques en cuivre parfaitement
centr\'{e}es, dans lesquelles viennent se loger les miroirs. Pour \'{e}viter
d'endommager les miroirs, l'ensemble est serr\'{e} par des lames de
chrysocale de forme cylindrique qui appliquent une contrainte uniforme sur
la p\'{e}riph\'{e}rie des miroirs. La figure \ref{Fig_4photfpm} montre une
photo de la cavit\'{e} vue du c\^{o}t\'{e} du miroir mobile. On distingue la
lame de chrysocale fix\'{e}e par six vis au support en cuivre. L'ensemble
est plac\'{e} dans un support en dural, mont\'{e} sur platines de
translation.

Le miroir mobile (au premier plan sur la figure \ref{Fig_4photfpm}) a
\'{e}t\'{e} r\'{e}alis\'{e} par le Service des Mat\'{e}riaux Avanc\'{e}s de
l'Institut de Physique Nucl\'{e}aire \`{a} Lyon. Cette \'{e}quipe travaille
\`{a} la r\'{e}alisation des miroirs destin\'{e}s \`{a}
l'interf\'{e}rom\`{e}tre Virgo. Le miroir de tr\`{e}s haute
r\'{e}flectivit\'{e} et \`{a} faible perte est d\'{e}pos\'{e} sur un
substrat plan-convexe en silice fondue tr\`{e}s pure, de petite dimension ($%
1.5~mm$ d'\'{e}paisseur et $14~mm$ de diam\`{e}tre). Le miroir qui sert de
coupleur d'entr\'{e}e pour la cavit\'{e} pr\'{e}sente une transmission plus
importante mais est aussi \`{a} faible perte.

Nous allons pr\'{e}senter dans cette partie les caract\'{e}ristiques de la
cavit\'{e}. Nous commencerons par d\'{e}crire le miroir mobile en
pr\'{e}sentant les sp\'{e}cifications fournies par les fabricants (section 
\ref{IV-1-1}). Nous d\'{e}crirons ensuite le coupleur d'entr\'{e}e (section 
\ref{IV-1-2}), la cavit\'{e} optique (section \ref{IV-1-3}) et les
caract\'{e}ristiques (bande passante, intervalle entre ordres, finesse,
perte et transmission des miroirs) que nous avons mesur\'{e}es (section \ref
{IV-1-4}).

\subsection{Le miroir mobile\label{IV-1-1}}

Le miroir mobile est r\'{e}alis\'{e} en deux \'{e}tapes distinctes : la
fabrications du substrat et le d\'{e}p\^{o}t des couches
multidi\'{e}lectriques. En ce qui concerne le substrat, ses
caract\'{e}ristiques (mat\'{e}riau utilis\'{e}, g\'{e}om\'{e}trie, \'{e}tat
de surface) d\'{e}terminent non seulement la qualit\'{e} optique du miroir
mais aussi les propri\'{e}t\'{e}s m\'{e}caniques du r\'{e}sonateur. Nous
avons choisi d'utiliser des substrats en silice fondue. Ce mat\'{e}riau
synth\'{e}tique pr\'{e}sente plusieurs avantages, tant sur le plan optique
que m\'{e}canique. Etant utilis\'{e} depuis de nombreuses ann\'{e}es comme
substrat pour la r\'{e}alisation de miroirs, les techniques de polissage et
de d\'{e}pot de couches multidi\'{e}lectriques sur ce mat\'{e}riau sont
parfaitement ma\^{\i }tris\'{e}es. Ceci permet d'obtenir des miroirs ayant
des pertes de l'ordre du $ppm$ (une part par million) et une excellente
tenue au flux. Les caract\'{e}ristiques m\'{e}caniques de ce mat\'{e}riau
sont aussi tr\`{e}s bonnes, puisque les facteurs de qualit\'{e}
intrins\`{e}que de la silice \`{a} basse temp\'{e}rature sont sup\'{e}rieurs
\`{a} $10^{8}$.

Le substrat a une g\'{e}om\'{e}trie plan-convexe. Gr\^{a}ce au confinement
des modes acoustiques au centre du r\'{e}sonateur, cette structure devrait
permettre d'obtenir des facteurs de qualit\'{e} m\'{e}canique proche de sa
valeur intrins\`{e}que. Ce type de g\'{e}om\'{e}trie a d\'{e}j\`{a}
\'{e}t\'{e} utilis\'{e}e dans le cas de r\'{e}sonateurs en quartz pour
\'{e}tudier les propri\'{e}t\'{e}s des modes de cisaillement. Les
exp\'{e}riences r\'{e}alis\'{e}es ont montr\'{e} un excellent accord avec la
th\'{e}orie tant en ce qui concerne la fr\'{e}quence de r\'{e}sonance que la
structure spatiale des modes. En particulier, des exp\'{e}riences de
diffraction de rayons X ont permis de visualiser la structure gaussienne de
ces modes et leur confinement\cite{zarka 1992}. D'autres exp\'{e}riences ont
montr\'{e} que le facteur de qualit\'{e} de ces r\'{e}sonances peut
atteindre sous vide et \`{a} basse temp\'{e}rature des valeurs proches de $%
10^{8}$\cite{el hatbi 1993}.

\begin{figure}[tbp]
\centerline{\psfig{figure=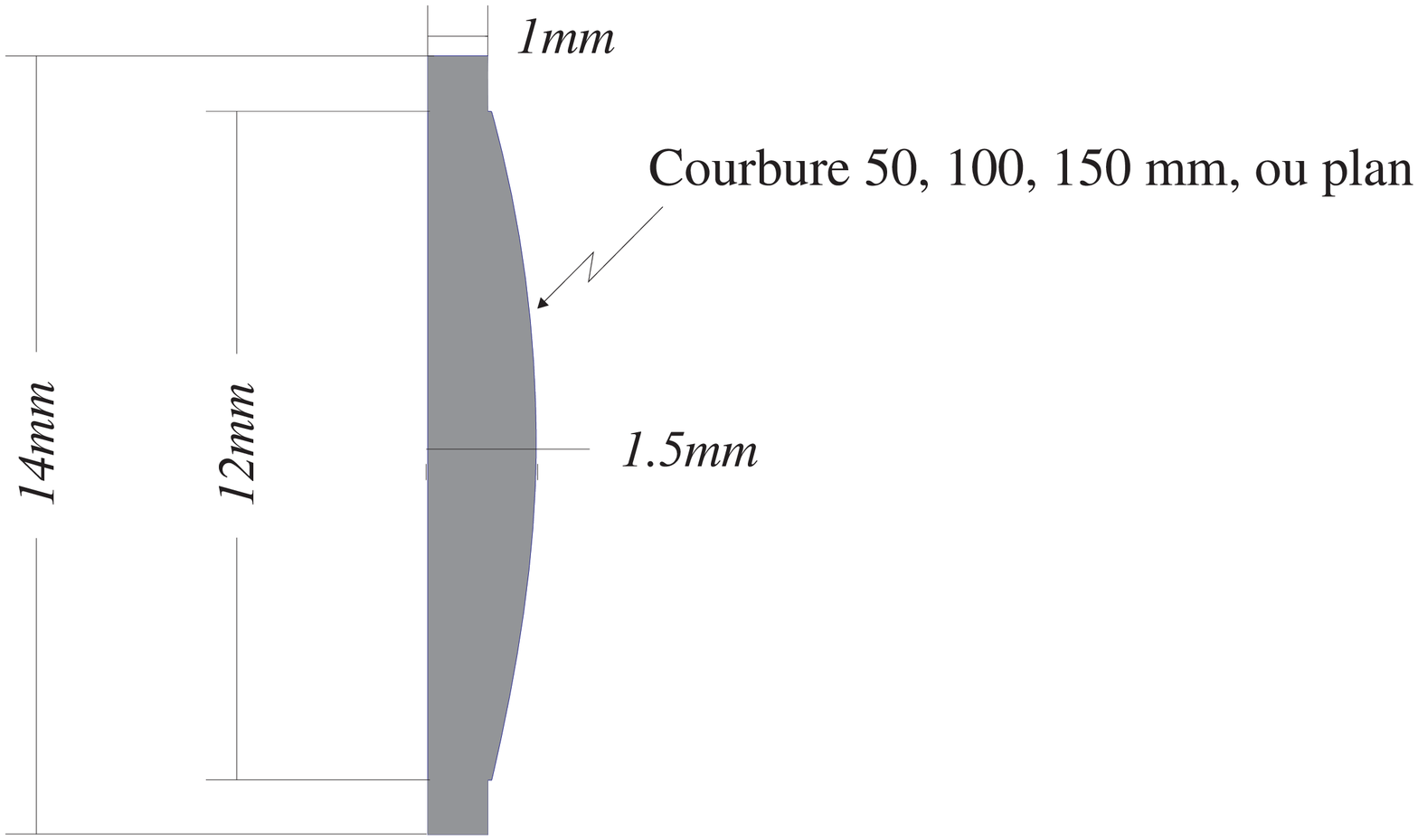,height=6cm}}
\caption{Sch\'{e}ma de la structure plan-convexe des substrats en silice
fondue}
\label{Fig_4substra}
\end{figure}

Nos substrats ont \'{e}t\'{e} r\'{e}alis\'{e}s par la soci\'{e}t\'{e} Maris
Delfour. Nous disposons ainsi d'un ensemble de substrats d'\'{e}paisseur au
centre $h_{0}=1.5~mm$ et de diff\'{e}rents rayons de courbure ($50$, $100$
et $150~mm$). Nous avons aussi des substrats plan-plan pour \'{e}tudier les
caract\'{e}ristiques des modes acoustiques dans le cas o\`{u} les modes ne
sont pas confin\'{e}s. La figure \ref{Fig_4substra} montre la
g\'{e}om\'{e}trie pr\'{e}cise de nos substrats. La partie
p\'{e}riph\'{e}rique du substrat plan-convexe de $12~mm$ de diam\`{e}tre est
prolong\'{e}e de $1~mm$ de fa\c{c}on \`{a} former un anneau de $1~mm$
d'\'{e}paisseur. Cet anneau permet de tenir le r\'{e}sonateur sans
endommager le traitement d\'{e}pos\'{e} sur la face plane. En tenant ainsi
le miroir sur les bords, on diminue aussi l'effet des contraintes sur les
caract\'{e}ristiques des modes acoustiques (fr\'{e}quence de r\'{e}sonance
et facteur de qualit\'{e}).

\subsubsection{Etat de surface des substrats\label{IV-1-1-1}}

Les caract\'{e}ristiques optiques du miroir mobile qui sont importantes en
vue de la r\'{e}alisation d'une cavit\'{e} de grande finesse sont la
transmission r\'{e}siduelle, les pertes et la tenue au flux. Les deux
premiers param\`{e}tres peuvent \^{e}tre d\'{e}finis par des coefficients $T$
et $P$ \'{e}gaux respectivement aux rapports entre l'\'{e}nergie lumineuse
transmise ou perdue et l'\'{e}nergie lumineuse incidente. Par conservation
de l'\'{e}nergie, ces param\`{e}tres sont reli\'{e}s au coefficient de
r\'{e}flexion en intensit\'{e} $R$ par: 
\begin{equation}
R+T+P=1  \label{0}
\end{equation}
Le coefficient de perte $P$ est lui-m\^{e}me la somme des pertes par
absorption $A$ et des pertes par diffusion $D$: 
\begin{equation}
P=A+D  \label{0bis}
\end{equation}
Enfin, la tenue au flux est caract\'{e}ris\'{e}e par la puissance maximale
admissible par unit\'{e} de surface.

La plupart de ces param\`{e}tres d\'{e}pendent de la qualit\'{e} des couches
multi-di\'{e}lectriques. Cependant, les pertes par diffusion sont pour
l'essentiel li\'{e}es \`{a} l'\'{e}tat de surface des substrats. On
d\'{e}finit la rugosit\'{e} $\sigma $ du substrat comme la moyenne des
\'{e}carts quadratiques des irr\'{e}gularit\'{e}s de la surface du substrat
par rapport \`{a} une surface id\'{e}ale. Le coefficient de diffusion $D$
est alors donn\'{e} par\cite{Bennet 61 (rugosité)}: 
\begin{equation}
D=\left( \frac{4\pi \sigma }{\lambda }\right) ^{2}  \label{1}
\end{equation}
La soci\'{e}t\'{e} Maris Delfour qui a r\'{e}alis\'{e} nos substrats
garantit un poli de la surface \`{a} traiter correspondant \`{a} une
rugosit\'{e} de l'ordre de $0.2~\AA $, ce qui induit des pertes par
diffusion inf\'{e}rieures \`{a} $1~ppm$, pour une longueur d'onde de $800~nm$%
.

\begin{figure}[tbp]
\centerline{\psfig{figure=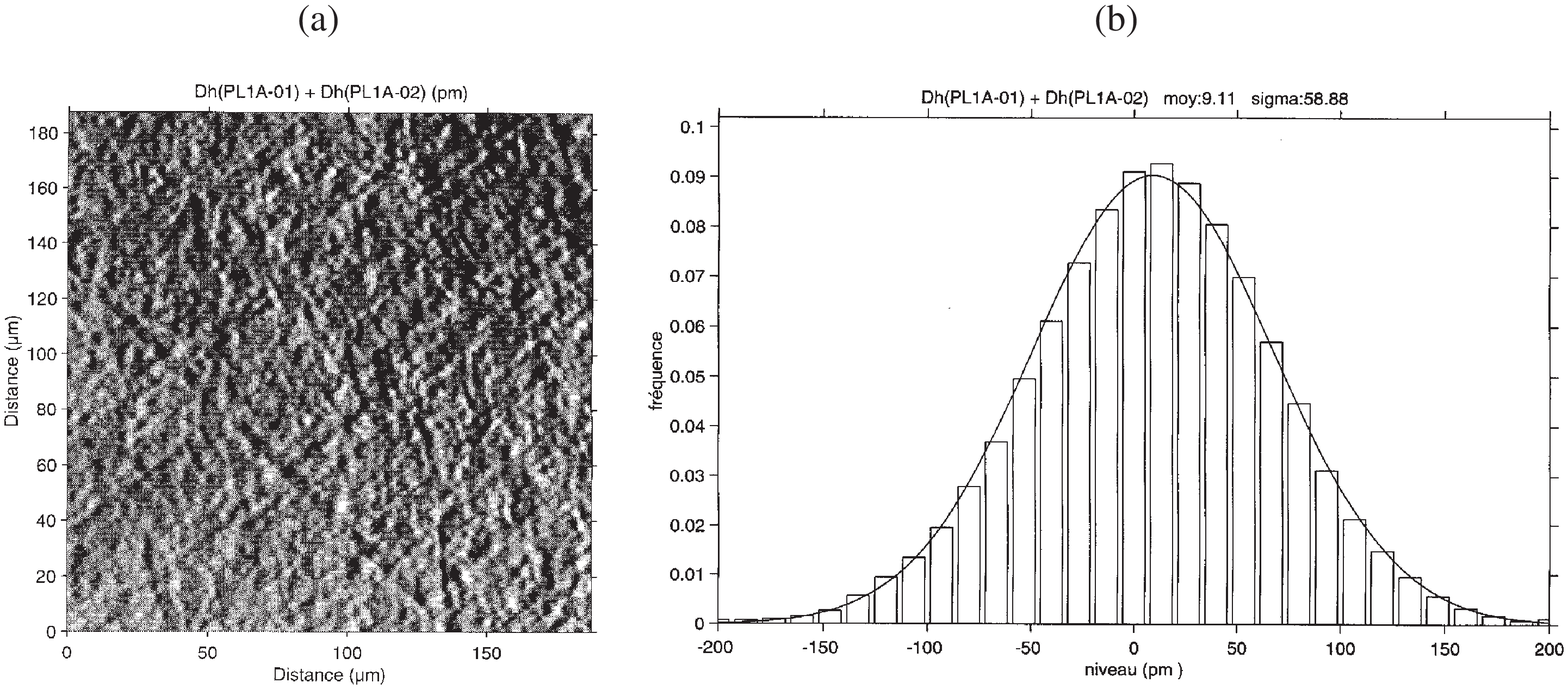,height=7cm}}
\caption{Caract\'{e}risation de l'\'{e}tat de surface de nos substrats en
silice fondue. La figure (a) repr\'{e}sente une topographie au centre du
substrat obtenue par une m\'{e}thode d'interf\'{e}rom\'{e}trie
diff\'{e}rentielle d\'{e}velopp\'{e}e \`{a} l'ESPCI. Ces r\'{e}sultats
permettent d'obtenir un histogramme du relief de la surface
\'{e}chantillonn\'{e}e (b), \`{a} partir duquel est d\'{e}duite la
rugosit\'{e} $\sigma $}
\label{Fig_4rugosit}
\end{figure}

Nous avons caract\'{e}ris\'{e} la rugosit\'{e} de nos substrats au
Laboratoire de l'Ecole Sup\'{e}rieure de Physico-Chimie Industrielle
(E.S.P.C.I.) en utilisant un dispositif d\'{e}velopp\'{e} par P. Gleyzes et
C. Boccara\cite{Boccara 1990}. Ce dispositif permet de mesurer le
d\'{e}phasage entre deux faisceaux r\'{e}fl\'{e}chis par deux points de
l'\'{e}chantillon, mesure qui donne acc\`{e}s \`{a} la d\'{e}nivellation
entre ces deux points. Le principe du dispositif est le suivant : un
faisceau de lumi\`{e}re issu d'une diode laser traverse un prisme de
Wollaston qui le s\'{e}pare angulairement en deux faisceaux de m\^{e}me
intensit\'{e} et de polarisations orthogonales. Ces deux faisceaux sont
ensuite focalis\'{e}s en deux points de l'\'{e}chantillon \'{e}cart\'{e}s
d'un micron. Apr\`{e}s r\'{e}flexion sur l'\'{e}chantillon, les deux
faisceaux sont recombin\'{e}s par le Wollaston. A cause du d\'{e}phasage
induit par le cube et par la diff\'{e}rence de marche des deux points
analys\'{e}s, le faisceau r\'{e}sultant a une polarisation elliptique. La
mesure de l'ellipticit\'{e} permet donc de d\'{e}terminer la diff\'{e}rence
de hauteur entre les deux points. Afin d'am\'{e}liorer le rapport signal
\`{a} bruit, cette mesure utilise une technique de modulation de la
polarisation qui permet de mesurer des d\'{e}nivellations inf\'{e}rieures
\`{a} $10~picom\grave{e}tres$. Une microplatine permet de d\'{e}placer
l'\'{e}chantillon et d'acqu\'{e}rir des valeurs du profil diff\'{e}rentiel
sur des intervalles de quelques centaines de microns avec un pas
d'\'{e}chantillonnage \'{e}gal \`{a} $0.5~\mu m$. Ces valeurs sont
stock\'{e}e puis trait\'{e}es afin d'en d\'{e}duire le profil r\'{e}el de la
surface \'{e}chantillonn\'{e}e. La figure \ref{Fig_4rugosit} (a) montre la
topographie au centre de l'un de nos substrats : les points sombres et
clairs repr\'{e}sentent des ''bosses'' ou des ''creux''. La figure \ref
{Fig_4rugosit} (b) repr\'{e}sente l'histogramme des valeurs du profil obtenu
avec le m\^{e}me substrat. On obtient pour ce substrat une rugosit\'{e} de
l'ordre de $0.6~\AA $, soit une valeur trois fois plus \'{e}lev\'{e}e que
celle annonc\'{e}e par le fabriquant. 

Dans l'ensemble, cette caract\'{e}risation a montr\'{e} que la rugosit\'{e}
de nos meilleurs substrats est plut\^{o}t de l'ordre de l'\AA ngstrom et que
les surfaces pr\'{e}sentent des d\'{e}fauts importants, sous forme de
longues rayures ou bosses. Les pertes par diffusion attendues seront donc au
moins de l'ordre de quelques $ppm$. A terme, si l'on veut mettre en
\'{e}vidence les effets quantiques du couplage optom\'{e}canique, il sera
n\'{e}cessaire de disposer de substrats ayant un meilleur \'{e}tat de
surface. La qualit\'{e} de ces substrats est cependant suffisante pour la
premi\`{e}re \'{e}tape de l'exp\'{e}rience qui consiste \`{a}
caract\'{e}riser le couplage optom\'{e}canique et mesurer le bruit thermique
du r\'{e}sonateur. Nous verrons que les pertes des miroirs r\'{e}alis\'{e}s
avec ces substrats sont suffisamment faibles pour construire une cavit\'{e}
optique de grande finesse.

L'\'{e}tat de surface de la face convexe intervient aussi au niveau du
facteur de qualit\'{e} des r\'{e}sonances acoustiques du r\'{e}sonateur
m\'{e}canique. Cette face est polie optiquement \`{a} $\lambda /10$ ce qui
est largement suffisant pour \'{e}viter toute d\'{e}gradation du facteur de
qualit\'{e} m\'{e}canique.

\subsubsection{Les traitements multidi\'{e}lectriques\label{IV-1-1-2}}

Les pertes par absorption dans les couches multidi\'{e}lectriques peuvent
\^{e}tre limit\'{e}es par un choix appropri\'{e} des mat\'{e}riaux et aussi
par la m\'{e}thode utilis\'{e}e pour d\'{e}poser les couches. Nos miroirs
ont \'{e}t\'{e} r\'{e}alis\'{e}s et caract\'{e}ris\'{e}s par l'\'{e}quipe de
J. M. Mackowski du Service des Mat\'{e}riaux Avanc\'{e}s de l'Institut de
Physique Nucl\'{e}aire de Lyon qui r\'{e}alise les miroirs de Virgo dont les
sp\'{e}cifications sont tr\`{e}s s\'{e}v\`{e}res. La technique de
d\'{e}p\^{o}t utilis\'{e}e est la pulv\'{e}risation r\'{e}active par double
faisceau d'ions (Dual Ion Beam Sputtering)\cite{dibs 1995}. Cette technique
de d\'{e}p\^{o}t permet d'obtenir des couches minces tr\`{e}s denses et
parfaitement isotropes.

Dans ce proc\'{e}d\'{e}, on utilise un faisceau \'{e}nerg\'{e}tique d'ions
argon $Ar^{+}$ pour bombarder une cible constitu\'{e}e du mat\'{e}riau que
l'on veut d\'{e}poser sur les substrats. Les esp\`{e}ces pulv\'{e}ris\'{e}es
sont \'{e}mises dans le demi-espace face \`{a} la cible et viennent se
condenser sur le substrat pour former le film mince. Tel quel, le film
pr\'{e}sente des microstructures qui peuvent constituer des sources
d'absorption importantes, limitant ainsi la tenue au flux du traitement.
C'est pourquoi le substrat est soumis \`{a} un autre faisceau d'ions
d'oxyg\`{e}ne $O_{2}^{+}$ peu \'{e}nerg\'{e}tique afin d'assister la
croissance du film. On obtient ainsi des couches tr\`{e}s compactes avec un
niveau d'absorption tr\`{e}s bas (inf\'{e}rieur au $ppm$) assurant une tenue
au flux sup\'{e}rieure \`{a} $10~kW/cm^{2}$. Le nombre de couches
multidi\'{e}lectriques, qui est \'{e}gal \`{a} $43$, est tel que le niveau
de transmission est pratiquement nul. 
\begin{figure}[tbp]
\centerline{\psfig{figure=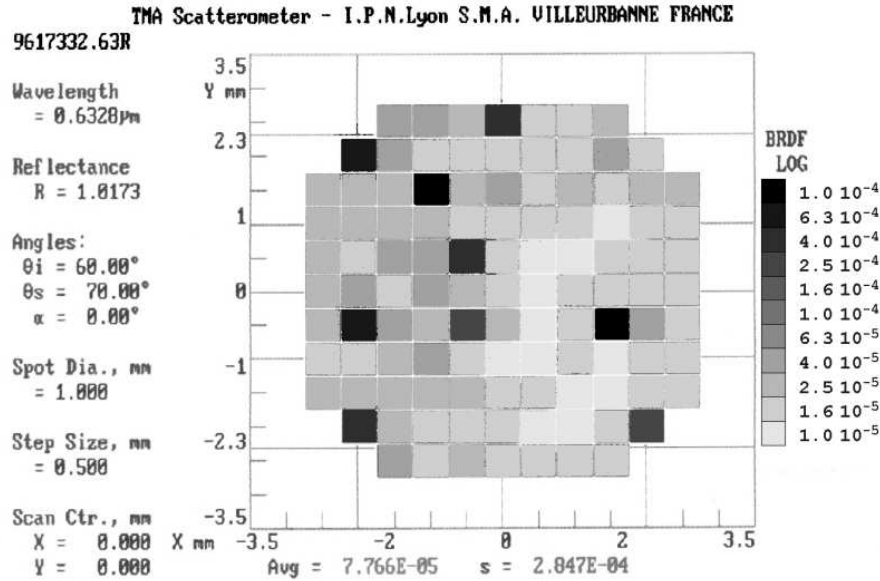,height=8cm}}
\caption{R\'{e}sultat d'une mesure de diffusion au centre d'un miroir \`{a}
l'aide du diffusom\`{e}tre de l'IPN \`{a} Lyon}
\label{Fig_4diffmac}
\end{figure}

La qualit\'{e} du traitement d\'{e}pend beaucoup de l'\'{e}tat de surface de
la face trait\'{e}e du substrat. En fait, les pertes totales de nos miroirs
sont essentiellement dues \`{a} la diffusion li\'{e}e \`{a} la rugosit\'{e}
relativement importante des substrats qui impose la reproduction des
d\'{e}fauts couche apr\`{e}s couche. Le syst\`{e}me utilis\'{e} \`{a}
l'I.P.N. pour mesurer les pertes par diffusion est un diffusom\`{e}tre de
type C.A.S.I. (complete angle scan instrument)\cite{amra 1990}. Le miroir,
mont\'{e} sur un support orientable, est \'{e}clair\'{e} par un faisceau
laser de section $1~mm^{2}$ et la lumi\`{e}re diffus\'{e}e est
d\'{e}tect\'{e}e \`{a} l'aide d'un photomultiplicateur port\'{e} par le bras
mobile d'un goniom\`{e}tre. On peut ainsi faire varier l'angle d'incidence $%
\theta _{i}$ du faisceau incident par rapport au miroir et l'angle de
diffusion $\theta _{s}$ correspondant au d\'{e}tecteur.

Ce dispositif donne acc\`{e}s \`{a} la fonction BRDF (bidirectional
reflectance distribution function), d\'{e}finie comme le rapport entre
l'\'{e}nergie totale diffus\'{e}e par unit\'{e} d'angle solide et le produit
de l'\'{e}nergie incidente par le cosinus de l'angle de diffusion $\theta
_{s}$. En d'autres termes, le produit $BRDF\times \cos \left( \theta
_{s}\right) $ est le flux diffus\'{e} par unit\'{e} de surface et d'angle
solide, norm\'{e} au flux incident. Ainsi, en d\'{e}pla\c{c}ant le miroir
dans son plan on obtient la fonction BRDF en plusieurs points de sa surface
pour une direction de diffusion donn\'{e}e. La figure \ref{Fig_4diffmac}
montre une cartographie de diffusion d'une surface d'environ $16~mm^{2}$ au
centre du miroir pour des angles $\theta _{i}$ et $\theta _{s}$ fix\'{e}s.
Ce type de cartographie permet seulement d'avoir une id\'{e}e
g\'{e}n\'{e}rale concernant les pertes par diffusion auquel on doit
s'attendre dans notre cavit\'{e} \`{a} miroir mobile. En effet, pour une
position donn\'{e}e, le taux de diffusion obtenu est en fait une moyenne sur
la section du miroir \'{e}clair\'{e}e par le faisceau, section qui est
relativement importante ($1~mm^{2}$). D'autre part, la diffusion dans une
direction donn\'{e}e n'est pas n\'{e}cessairement repr\'{e}sentative de la
diffusion totale.

Les r\'{e}sultats obtenus sont r\'{e}sum\'{e}s dans le tableau suivant
o\`{u} sont rep\'{e}r\'{e}s les miroirs que nous avons utilis\'{e} (lignes
gris\'{e}es). La troisi\`{e}me colonne indique la valeur de la fonction BRDF
pour une direction de diffusion donn\'{e}e, et moyenn\'{e}e sur toute la
surface analys\'{e}e. La quatri\`{e}me colonne donne la valeur de la
fonction au centre du miroir (section $1~mm^{2}$). Enfin, les pertes totales
par diffusion (cinqui\`{e}me colonne) sont obtenues en int\'{e}grant la
courbe angulaire de la fonction BRDF dans le demi-espace, en supposant que
la diffusion est la m\^{e}me dans toutes les directions. Ceci suppose des
propri\'{e}t\'{e}s isotropes du traitement multidi\'{e}lectrique. Les
r\'{e}sultats obtenus montrent que les pertes par diffusion de nos miroirs
sont importantes, ce qui est probablement li\'{e} \`{a} l'\'{e}tat de
surface des substrats qui ne permet d'atteindre qu'au mieux des pertes de
l'ordre de $10~ppm$.\bigskip \medskip

\centerline{\psfig{figure=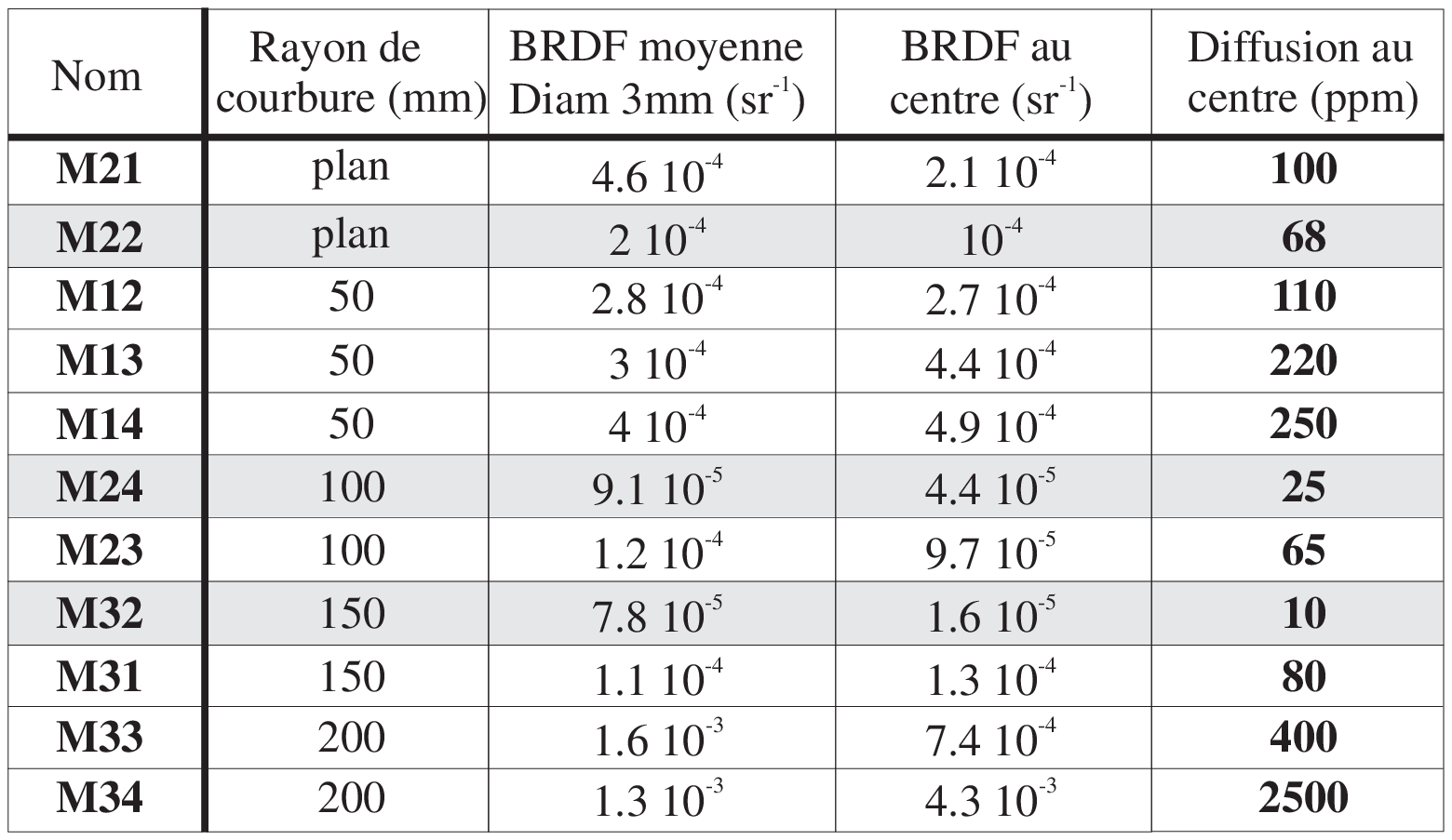,height=8cm}}\label{4tabppm}\bigskip
\medskip

Il est important de noter que cette caract\'{e}risation ne nous permet pas
de fixer une valeur assez pr\'{e}cise pour la diffusion puisque dans le cas
de notre exp\'{e}rience, la taille du faisceau au niveau du miroir mobile
est au plus de l'ordre de $100~\mu m$. D'autre part, \'{e}tant donn\'{e}
l'\'{e}tat de surface de nos substrats, la condition d'isotropie du
traitement n'est probablement pas satisfaite ce qui peut induire un
\'{e}cart non n\'{e}gligeable par rapport aux valeurs obtenues. Nous verrons
que nos mesures sur la cavit\'{e} \`{a} miroir mobile permettent de donner
des valeurs plus pr\'{e}cises pour les pertes de nos miroirs.

Notons pour terminer que ces miroirs sont tr\`{e}s fragiles : le
d\'{e}p\^{o}t de poussi\`{e}re augmente tr\`{e}s rapidement les pertes par
diffusion, jusqu'\`{a} des valeurs sup\'{e}rieures \`{a} $100~ppm$. D'autre
part, ces miroirs ne peuvent pas \^{e}tre nettoy\'{e}s par les techniques
usuelles employ\'{e}es pour des miroirs moins performants. C'est pourquoi
nous avons install\'{e} l'ensemble de l'exp\'{e}rience sous un flux
laminaire qui assure une propret\'{e} au niveau de la table optique
correspondant \`{a} une salle blanche de classe $100$.

\subsection{Le coupleur d'entr\'{e}e\label{IV-1-2}}

La cavit\'{e} optique est constitu\'{e}e du miroir mobile et d'un miroir qui
sert de coupleur d'entr\'{e}e et qui pr\'{e}sente une transmission plus
grande, mais n\'{e}anmoins assez faible pour assurer une finesse
\'{e}lev\'{e}e. Nous avons choisi d'utiliser des miroirs commerciaux
fabriqu\'{e}s par Micro-controle Newport sous la d\'{e}nomination de
''SuperMirrors'' de s\'{e}rie haute finesse. D'apr\`{e}s les
sp\'{e}cifications du constructeur, ces miroirs pr\'{e}sentent une
transmission de $50~ppm$, des pertes (diffusion et absorption) du m\^{e}me
ordre et une tenue au flux sup\'{e}rieure \`{a} $1~kW/cm^{2}$. La taille de
ces miroirs est beaucoup plus grande que celle des miroirs mobiles
puisqu'ils ont une \'{e}paisseur de $6.35~mm$ et un diam\`{e}tre de $25.4~mm$%
. Le rayon de courbure $R_{N}$ de la face trait\'{e}e est de $1~m$.

Les pertes totales $2\gamma $ de la cavit\'{e} sont \'{e}gales \`{a} la
somme des transmission $T_{M}$, $T_{N}$ et des pertes $P_{M}$, $P_{N}$ du
miroir mobile et du coupleur Newport: 
\begin{equation}
2\gamma =T_{M}+T_{N}+P_{M}+P_{N}  \label{2}
\end{equation}
D'apr\`{e}s les sp\'{e}cifications des miroirs fournies par les fabricants,
ont devrait donc obtenir des cavit\'{e}s dont les pertes $2\gamma $ sont
comprises entre $50$ et $150~ppm$ environ (soit des finesses comprises entre 
$2~10^{4}$ et $6~10^{4}$). Nous verrons dans les sections suivantes comment
on peut d\'{e}terminer ces pertes en mesurant les caract\'{e}ristiques de la
cavit\'{e} telles que la bande passante, l'intervalle spectral libre et le
coefficient de r\'{e}flexion.

\subsection{La cavit\'{e} optique\label{IV-1-3}}

\begin{figure}[tbp]
\centerline{\psfig{figure=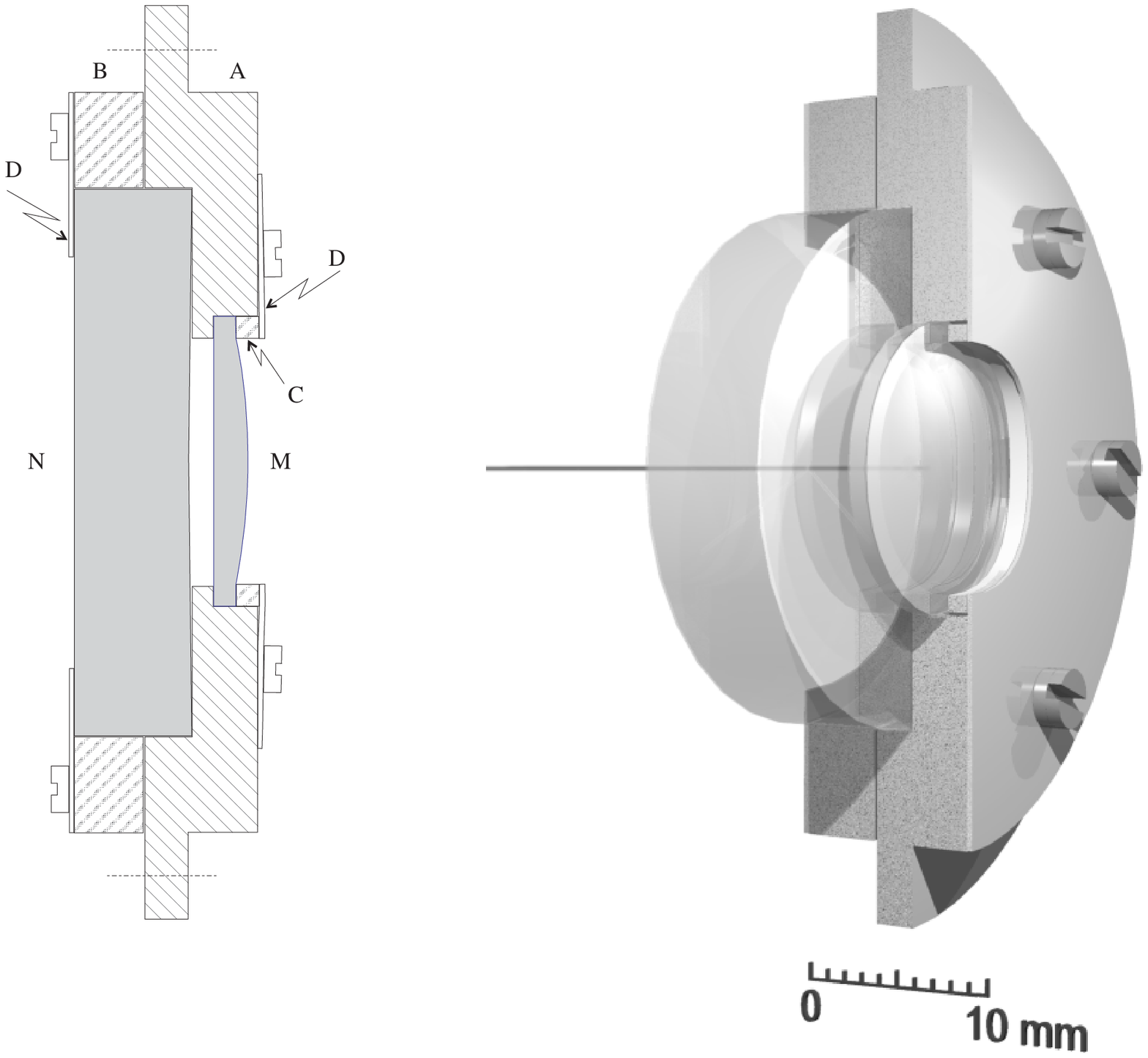,height=12cm}}
\caption{Sch\'{e}ma g\'{e}n\'{e}ral de la cavit\'{e} form\'{e}e par le
miroir mobile $M$ et le coupleur Newport $N$, s\'{e}par\'{e}s par un
espaceur en cuivre de $1~mm$ d'\'{e}paisseur usin\'{e} dans la pi\`{e}ce $A$%
. Le support constitu\'{e} des pi\`{e}ces en cuivre $A$, $B$ et $C$ assure
le centrage et le parall\'{e}lisme de la cavit\'{e} et l'ensemble est tenu
gr\^{a}ce \`{a} deux lames de chrysocale $D$}
\label{Fig_4cavcuiv}
\end{figure}

Les deux miroirs de la cavit\'{e} Fabry Perot sont mont\'{e}s dans un
support en cuivre qui assure le centrage et le parall\'{e}lisme des miroirs.
Pour fixer la longueur de la cavit\'{e}, nous avons utilis\'{e} un espaceur
en cuivre, solidaire du support, de $1~mm$ d'\'{e}paisseur. La figure \ref
{Fig_4cavcuiv} montre comment sont mont\'{e}s les diff\'{e}rents
\'{e}l\'{e}ments de la cavit\'{e}. Les \'{e}l\'{e}ments $A$, $B$ et $C$ sont
des pi\`{e}ces en cuivre qui servent de support aux deux miroirs $M$ (miroir
mobile) et $N$ (coupleur d'entr\'{e}e). L'ensemble est tenu gr\^{a}ce \`{a}
des lames de chrysocale $D$ qui exercent une contrainte uniforme sur les
bords des miroirs, assurant ainsi une fixation stable des miroirs sans pour
autant risquer de les endommager lors du serrage.

Une attention toute particuli\`{e}re doit \^{e}tre port\'{e}e au
parall\'{e}lisme des deux faces du support $A$ sur lesquelles reposent les
deux miroirs. En effet, un d\'{e}faut de parall\'{e}lisme d'un angle $\alpha 
$ se traduit par un d\'{e}calage de l'axe optique de la cavit\'{e} d'une
quantit\'{e} $\alpha ~R_{N}$ par rapport au centre des miroirs ($R_{N}=1~m$
est le rayon de courbure du coupleur d'entr\'{e}e). Ceci a pour effet de
modifier le recouvrement spatial entre le mode optique et les modes
acoustiques du r\'{e}sonateur. Pour minimiser cet effet, le d\'{e}calage
doit \^{e}tre petit devant la taille des cols des premiers modes
acoustiques, en particulier le col $w_{1}$ du mode fondamental qui est
\'{e}gal \`{a} $3.8~mm$ pour un rayon de courbure du r\'{e}sonateur de $%
150~mm$ (\'{e}quation 3.43). On se fixe donc une tol\'{e}rance sur le
d\'{e}calage inf\'{e}rieure au millim\`{e}tre, ce qui impose un
parall\'{e}lisme entre les surface de la pi\`{e}ce $A$ meilleur que $1~mrad$%
, c'est \`{a} dire une variation d'\'{e}paisseur sur le diam\`{e}tre du
miroir mobile inf\'{e}rieure \`{a} $10~\mu m$. Nous n'avons pas constat\'{e}
de d\'{e}centrage notable de l'axe optique de la cavit\'{e} avec les
pi\`{e}ces r\'{e}alis\'{e}es par B. Rodriguez \`{a} l'atelier de
m\'{e}canique du laboratoire, mais cette contrainte rend difficile l'usinage
d'une pi\`{e}ce en cuivre avec un espaceur d'\'{e}paisseur inf\'{e}rieure au
millim\`{e}tre.

Pour obtenir une cavit\'{e} de longueur $0.5~mm$ avec un parall\'{e}lisme
correct entre les deux miroirs, nous avons construit un autre support de
cavit\'{e} dans lequel on utilise un espaceur en silice de $0.5~mm$
d'\'{e}paisseur et de m\^{e}me diam\`{e}tre que le miroir mobile. Dans cette
configuration, les deux miroirs sont s\'{e}par\'{e}s par l'espaceur qui est
en contact direct avec les deux miroirs. Selon les sp\'{e}cifications du
fabriquant Maris Delfour, on obtient ainsi un parall\'{e}lisme inf\'{e}rieur
\`{a} $1~mrad$. Notons cependant que le montage de cette cavit\'{e} est plus
d\'{e}licat du fait de la faible \'{e}paisseur de l'espaceur en silice qui
peut facilement se casser lors du montage de l'ensemble de la cavit\'{e}.

Nous avons enfin construit un troisi\`{e}me support pour une cavit\'{e}
utilisant deux miroirs Newport courbes. L'espaceur est en cuivre, formant
ainsi une cavit\'{e} bi-convexe ($R_{N}=1~m$) de $1~mm$ d'\'{e}paisseur.
Comme nous le verrons plus loin, l'\'{e}tude de cette cavit\'{e} permet de
d\'{e}terminer s\'{e}par\'{e}ment les caract\'{e}ristiques des coupleurs
Newport et du miroir mobile.

\subsection{Caract\'{e}ristiques de la cavit\'{e}\label{IV-1-4}}

Nous allons maintenant pr\'{e}senter les mesures qui nous ont permis de
d\'{e}terminer les caract\'{e}ristiques de la cavit\'{e}. Nous avons
utilis\'{e} trois miroirs mobiles (not\'{e}s $M22$, $M24$, $M32$ sur le
tableau de la page \pageref{4tabppm}) et deux miroirs Newport de rayon de
courbure d'un m\`{e}tre (not\'{e}s $N1$ et $N2$). Cet ensemble de miroirs
nous a permis de r\'{e}aliser plusieurs cavit\'{e}s en combinant un coupleur
Newport avec l'un des miroirs mobiles, ainsi qu'une cavit\'{e} utilisant les
deux miroirs Newport. Pour chaque cavit\'{e} nous avons mesur\'{e} la bande
passante, l'intervalle spectral libre et le coefficient de r\'{e}flexion
\`{a} r\'{e}sonance. Ces mesures nous ont permis de d\'{e}terminer la
finesse et la longueur de chaque cavit\'{e}, les pertes totales des miroirs
et la transmission du coupleur d'entr\'{e}e. La combinaison de ces
r\'{e}sultats permet en outre de d\'{e}terminer pour chaque miroir les
pertes et le coefficient de transmission.

\subsubsection{Alignement et adaptation de la cavit\'{e}\label{IV-1-4-1}}

La cavit\'{e} pr\'{e}sente des r\'{e}sonances longitudinales auxquelles sont
associ\'{e}es une s\'{e}rie de r\'{e}sonances transversales. Dans
l'approximation paraxiale la structure spatiale des modes propres est
gaussienne et le mode fondamental $TEM_{00}$ est caract\'{e}ris\'{e} par un
col $w_{0}$ situ\'{e} au niveau du miroir plan et dont la taille est
donn\'{e}e par\cite{Kogelnik (a)}: 
\begin{equation}
w_{0}^{2}=\frac{\lambda }{\pi }\sqrt{L\left( R_{N}-L\right) }  \label{3}
\end{equation}
o\`{u} $R_{N}=1~m$ est le rayon de courbure du miroir Newport et $L$ est la
longueur de la cavit\'{e}, \'{e}gale \`{a} $0.5~mm$ ou $1~mm$. Selon la
longueur de la cavit\'{e}, la taille $w_{0}$ du col est comprise entre $75$
et $90~\mu m$. La longueur de la cavit\'{e} d\'{e}finit l'\'{e}cart en
fr\'{e}quence entre deux r\'{e}sonances longitudinales successives. Cet
intervalle spectral libre $\nu _{ISL}$ est donn\'{e} par: 
\begin{equation}
\nu _{ISL}=\frac{c}{2L}  \label{4}
\end{equation}
L'intervalle spectrale libre est de $300~GHz$ pour une longueur de $0.5~mm$
et de $150~GHz$ pour une longueur de $1~mm$. La position relative en
fr\'{e}quence des modes transverses $TEM_{pl}$ associ\'{e}s au fondamental $%
TEM_{00}$ d\'{e}pend non seulement de la longueur $L$ mais aussi du rayon de
courbure $R_{N}$\cite{Kogelnik (a)}: 
\begin{equation}
\nu _{pl}=\frac{\nu _{ISL}}{\pi }\left( p+l\right) ~\cos ^{-1}\left( \sqrt{1-%
\frac{L}{R_{N}}}\right)  \label{5}
\end{equation}
Cette relation montre que les modes transverses correspondant \`{a} des
valeurs diff\'{e}rentes de ($p+l$) ont des fr\'{e}quences de r\'{e}sonance
diff\'{e}rentes. La d\'{e}g\'{e}n\'{e}rescence en ($p+l$) est li\'{e}e \`{a}
la sym\'{e}trie cylindrique de la cavit\'{e}.

Comme nous le verrons dans la section 4.2.4, le faisceau issu de la source
laser est filtr\'{e} spatialement de fa\c{c}on \`{a} ce que sa structure
spatiale soit parfaitement gaussienne $TEM_{00}$. On dispose d'un
syst\`{e}me \`{a} deux lentilles de focales $40$ et $150~mm$ qui permet de
transformer le faisceau pour que son col soit \'{e}gal \`{a} $w_{0}$ et
situ\'{e} au niveau du miroir mobile de la cavit\'{e}. On utilise enfin un
syst\`{e}me de deux miroirs mont\'{e}s dans des supports r\'{e}glables pour
aligner le faisceau sur la cavit\'{e}. On r\'{e}alise un pr\'{e}-alignement
en observant la structure de la lumi\`{e}re transmise par la cavit\'{e}.
Etant donn\'{e}e la faible intensit\'{e} transmise par le miroir mobile, on
utilise une cam\'{e}ra infrarouge directement plac\'{e}e derri\`{e}re la
cavit\'{e} : une t\^{a}che unique centr\'{e}e sur le miroir indique un
alignement correct.

Pour optimiser l'adaptation spatiale, il est n\'{e}cessaire de balayer le
d\'{e}saccord entre le faisceau incident et la cavit\'{e}. Pour des raisons
de stabilit\'{e}, les cavit\'{e}s que nous utilisons sont rigides et ne
disposent pas de cale pi\'{e}zo\'{e}lectrique permettant de balayer la
longueur de la cavit\'{e}. Il est donc n\'{e}cessaire de balayer la
fr\'{e}quence de la source laser sur une plage d'au moins $300~GHz$ pour
trouver les r\'{e}sonances de la cavit\'{e}. Comme nous le verrons dans la
partie suivante consacr\'{e}e \`{a} la source laser, nous disposons de trois
dispositifs de balayage : le bilame du laser titane saphir permet de faire
un balayage continu sur plusieurs dizaines de gigahertz, tandis que des
sauts de fr\'{e}quence de $19$ ou $150~GHz$ peuvent \^{e}tre
r\'{e}alis\'{e}s avec l'\'{e}talon mince ou le filtre de Lyot. Un
lambdam\`{e}tre de r\'{e}solution \'{e}gale \`{a} $10^{-3}~nm$ permet de se
rep\'{e}rer au cours du balayage. On contr\^{o}le l'adaptation spatiale du
faisceau incident en mesurant l'intensit\'{e} transmise \`{a} l'aide d'une
photodiode plac\'{e}e derri\`{e}re la cavit\'{e}. En balayant continuement
la fr\'{e}quence du laser, on visualise les pics de transmission
associ\'{e}s aux diff\'{e}rents modes transverses de la cavit\'{e}.

\begin{figure}[tbp]
\centerline{\psfig{figure=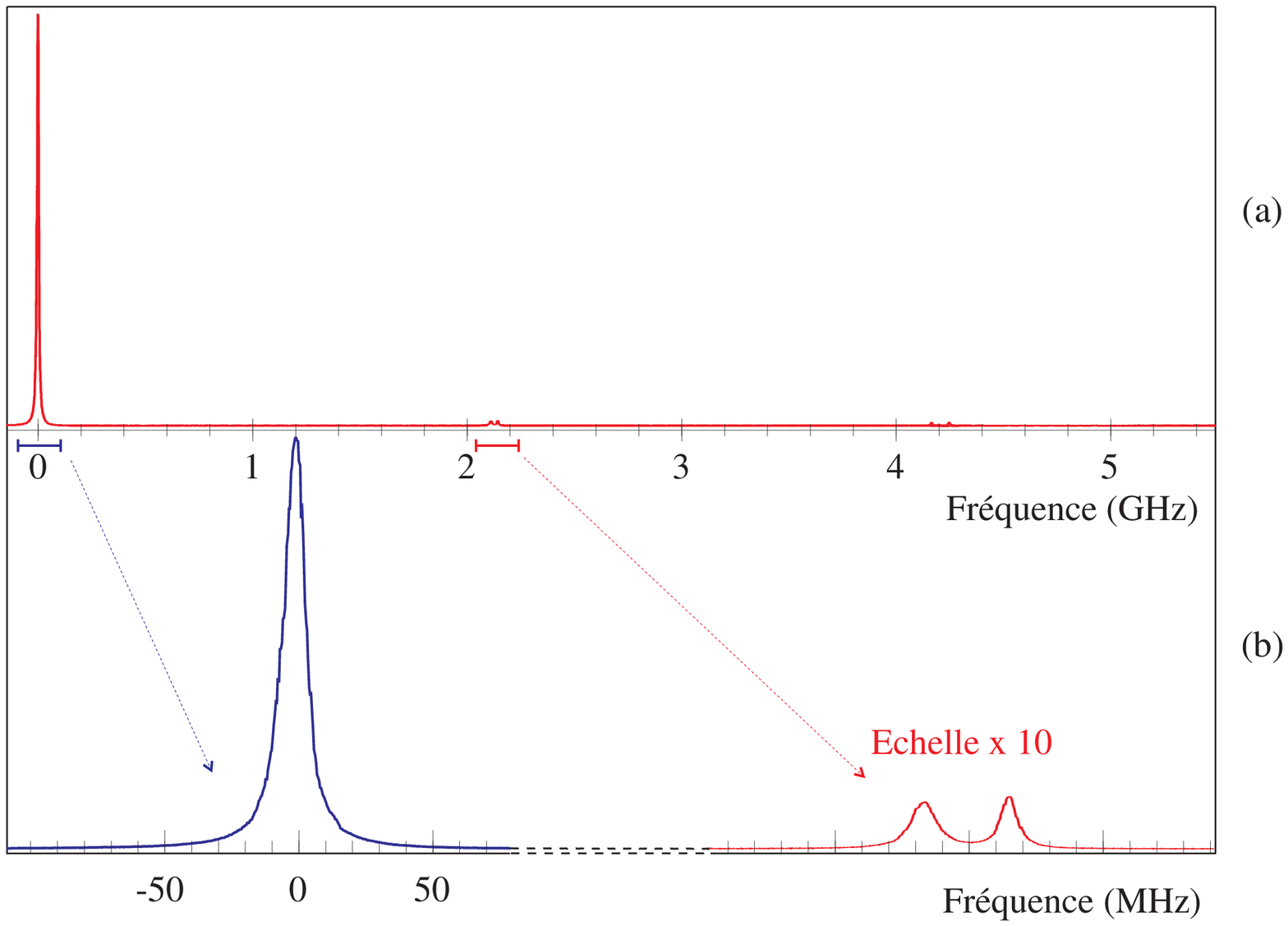,height=10cm}}
\caption{Transmission en intensit\'{e} de la cavit\'{e} \`{a} miroir mobile $%
\left\{ N1,~M24\right\} $ de longueur 0.5 mm. Un balayage de la
fr\'{e}quence du laser permet de visualiser le pic de transmission du mode
fondamental de la cavit\'{e} ainsi que les premiers modes transverses}
\label{Fig_4bilame}
\end{figure}

La figure \ref{Fig_4bilame} montre le r\'{e}sultat obtenu avec la cavit\'{e}
constitu\'{e}e par le miroir mobile $M24$ (voir tableau page \pageref
{4tabppm}) et le miroir Newport $N1$, s\'{e}par\'{e}s par l'espaceur en
silice de $0.5~mm$. On observe un grand pic qui correspond au mode
fondamental $TEM_{00}$ de la cavit\'{e} suivi des premiers modes transverses
fortement att\'{e}nu\'{e}s du fait de l'adaptation entre le faisceau et le
mode fondamental. Le balayage en fr\'{e}quence a \'{e}t\'{e} r\'{e}alis\'{e}
en appliquant une rampe lin\'{e}aire sur le bilame du laser titane saphir.
Ce balayage a \'{e}t\'{e} \'{e}talonn\'{e} en utilisant une cavit\'{e} de
r\'{e}f\'{e}rence (cavit\'{e} de filtrage de la source laser). On observe
que l'\'{e}cart entre le fondamental et les deux modes transverses $TEM_{10}$
et $TEM_{01}$ est de $2.1~GHz$, ce qui correspond bien au r\'{e}sultat
obtenu en utilisant la relation (\ref{5}). Un agrandissement autour des pics
permet de visualiser la forme des r\'{e}sonances, et en particulier la
lev\'{e}e de d\'{e}g\'{e}n\'{e}rescence des modes transverses li\'{e}e sans
doute \`{a} une dissym\'{e}trie du rayon de courbure du coupleur. Par
ailleurs, le rapport entre la hauteur des modes transverses et celle du
fondamental permet d'\'{e}valuer l'adaptation spatiale entre le faisceau
incident et le mode fondamental. Seuls les deux premi\`{e}res harmoniques
transverses sont visibles et leur hauteur est att\'{e}nu\'{e}e d'un facteur
sup\'{e}rieur \`{a} $100$ par rapport au fondamental. La puissance lumineuse
coupl\'{e}e au mode fondamental de la cavit\'{e} est donc de l'ordre de $98\%
$ de la puissance incidente.

La figure \ref{Fig_4bilame} permet par ailleurs d'\'{e}valuer la bande
passante de la cavit\'{e}. Il s'agit ici simplement d'une \'{e}valuation, le
pic \'{e}tant l\'{e}g\`{e}rement dissym\'{e}trique, sans doute \`{a} cause
d'effets thermiques dans la cavit\'{e} lors du passage \`{a} r\'{e}sonance.
Nous verrons plus loin comment la bande passante peut \^{e}tre
d\'{e}termin\'{e}e plus pr\'{e}cis\'{e}ment. La largeur \`{a} mi-hauteur du
pic d'Airy sur la figure \ref{Fig_4bilame} est de l'ordre de $5$ \`{a} $%
10~MHz$, correspondant \`{a} une bande passante $\nu _{BP}$ de $2.5$ \`{a} $%
5~MHz$. On peut alors estimer la finesse ${\cal F}=\nu _{ISL}/2\nu _{BP}$
sachant que l'intervalle spectral libre $\nu _{ISL}$ est \'{e}gal \`{a} $%
300~GHz$. On trouve une finesse comprise entre $30000$ et $60000$. Une
d\'{e}termination plus pr\'{e}cise est pr\'{e}sent\'{e}e dans la section
suivante.

Nous avons par ailleurs cherch\'{e} \`{a} caract\'{e}riser la tenue au flux
des miroirs. La cavit\'{e} \{$N1$, $M24$\} supporte une puissance incidente $%
P_{in}$ de $300~\mu W$, sans d\'{e}gradation des caract\'{e}ristiques ni
dommage pour les miroirs. La puissance intracavit\'{e} est dans ce cas de
l'ordre de quelques Watts et le flux de photon est de l'ordre de quelques
dizaines de kilowatts par centim\`{e}tre carr\'{e}. La tenue au flux est
donc bien meilleure que les valeurs annoc\'{e}es par le fabricant ($%
1~kW/cm^{2}$ pour le coupleur Newport).

\subsubsection{Mesure de la bande passante de la cavit\'{e}\label{IV-1-4-2}}

L'intensit\'{e} intracavit\'{e} est maximum lorsque la fr\'{e}quence du
faisceau incident co\"{\i }ncide avec une r\'{e}sonance de la cavit\'{e}.
Lorsqu'on s'\'{e}carte de part et d'autre de cette position en modulant par
exemple la fr\'{e}quence du laser, l'intensit\'{e} diminue en d\'{e}crivant
une lorentzienne de largeur \`{a} mi-hauteur \'{e}gale \`{a} $2\nu _{BP}$,
o\`{u} $\nu _{BP}$ est la bande passante de la cavit\'{e}. On retrouve ce
comportement dans l'intensit\'{e} transmise par la cavit\'{e} puisque
celle-ci est proportionnelle \`{a} l'intensit\'{e} intracavit\'{e}. Par
ailleurs, la bande passante est reli\'{e}e \`{a} la finesse de la cavit\'{e}
par la relation: 
\begin{equation}
{\cal F}=\frac{\pi }{\gamma }=\frac{\nu _{ISL}}{2\nu _{BP}}  \label{6}
\end{equation}
Cette relation permet de d\'{e}terminer la finesse \`{a} partir de la mesure
de la bande passante et de la longueur de la cavit\'{e}. Pour mesurer la
bande passante, on utilise une cavit\'{e} Fabry-Perot de r\'{e}f\'{e}rence
dont on conna\^{\i }t la bande passante et on compare les deux largeurs des
r\'{e}sonances pour en d\'{e}duire celle de la cavit\'{e} \`{a} miroir
mobile. Nous verrons dans la section \ref{IV-2-4} comment nous avons
mesur\'{e} la bande passante de la cavit\'{e} de r\'{e}f\'{e}rence, qui est
en fait la cavit\'{e} de filtrage spatial de la source laser. Le
r\'{e}sultat de la mesure donne une bande passante \'{e}gale \`{a} $5.8~MHz$
pour cette cavit\'{e} de r\'{e}f\'{e}rence.

\begin{figure}[tbp]
\centerline{\psfig{figure=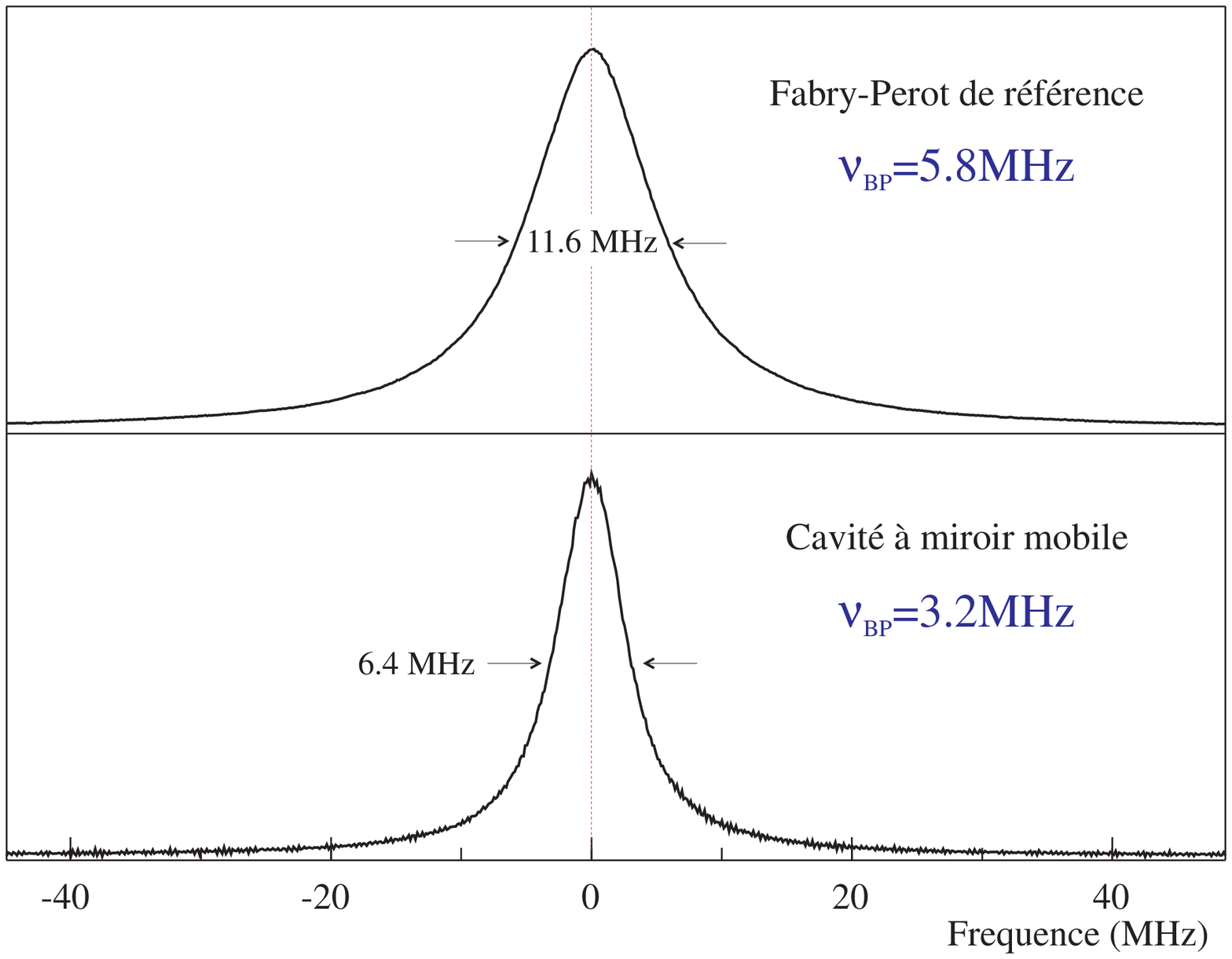,height=10cm}}
\caption{Profil de l'intensit\'{e} transmise par la cavit\'{e} de
r\'{e}f\'{e}rence et par la cavit\'{e} \`{a} miroir mobile $\left\{ N1\text{%
, }M24\right\} $. Connaissant la bande passante du Fabry-Perot de
r\'{e}f\'{e}rence, on en d\'{e}duit celle de la cavit\'{e}}
\label{Fig_4bandpas}
\end{figure}

On observe les r\'{e}sonances de la cavit\'{e} en modulant la fr\'{e}quence
du laser \`{a} $100~Hz$ et en d\'{e}tectant l'intensit\'{e} transmise par la
cavit\'{e} de r\'{e}f\'{e}rence puis celle transmise par la cavit\'{e} \`{a}
miroir mobile. La premi\`{e}re mesure permet de calibrer la modulation de
fr\'{e}quence du laser. La figure \ref{Fig_4bandpas} montre d'une part la
r\'{e}sonance de la cavit\'{e} de r\'{e}f\'{e}rence de largeur \`{a}
mi-hauteur \'{e}gale \`{a} $2\times 5.8=11.6~MHz$ et la r\'{e}sonance de la
cavit\'{e} \`{a} miroir mobile constitu\'{e}e du couple de miroirs \{$N1$, $%
M24$\} s\'{e}par\'{e}s par l'entretoise en silice de $0.5~mm$
d'\'{e}paisseur. On obtient une bande passante $\nu _{BP}$ \'{e}gale \`{a} $%
3.2~MHz$, correspondant \`{a} une finesse ${\cal F}$ de $47000$. Nous avons
repris la m\^{e}me proc\'{e}dure de mesure en utilisant diff\'{e}rents
miroirs mobiles tout en gardant le m\^{e}me coupleur $N1$, et on obtient
globalement des finesses comprises entre $30000$ et $47000$. Nous avons
aussi mesur\'{e} la finesse de la cavit\'{e} form\'{e}e des deux miroirs
Newport \{$N1$, $N2$\} et on obtient une finesse de $33500$. 

Cette mesure de la finesse permet d'\'{e}valuer les pertes totales dans la
cavit\'{e}, c'est \`{a} dire la somme des transmissions et des pertes des
deux miroirs (\'{e}quation \ref{2}). A ce stade, il n'est pas possible de
discriminer entre les transmissions et les pertes de chaque miroir. Si on
veut d\'{e}terminer s\'{e}par\'{e}ment les caract\'{e}ristiques des miroirs,
il est n\'{e}cessaire de r\'{e}aliser d'autres mesures, en particulier la
mesure du coefficient de r\'{e}flexion \`{a} r\'{e}sonance de la cavit\'{e}.

\subsubsection{Mesure du coefficient de r\'{e}flexion de la cavit\'{e}\label%
{IV-1-4-3}}

Nous avons vu dans la partie 2.3 que le champ r\'{e}fl\'{e}chi par une
cavit\'{e} \`{a} une seule entr\'{e}e-sortie a la m\^{e}me intensit\'{e} que
le champ incident. Ceci n'est vrai que pour une cavit\'{e} sans perte : en
pr\'{e}sence de pertes, l'intensit\'{e} r\'{e}fl\'{e}chie pr\'{e}sente un
pic d'Airy en absorption au voisinage de la r\'{e}sonance. On peut calculer
l'intensit\'{e} r\'{e}fl\'{e}chie en d\'{e}crivant chaque miroir par un
coefficient de r\'{e}flexion $r$ en amplitude, un coefficient de
transmission $t$ et un coefficient de perte $p$ (les carr\'{e}s de ces
coefficients sont \'{e}gaux aux param\`{e}tres $R$, $T$, $P$ introduits dans
la section \ref{IV-1-1}). La conservation de l'\'{e}nergie au niveau des
deux miroirs se traduit par la relation: 
\begin{equation}
r_{i}^{2}+t_{i}^{2}+p_{i}^{2}=1\qquad ,\quad i=1,2  \label{7}
\end{equation}
Les champs moyens sont reli\'{e}s par des \'{e}quations similaires aux
relations (2.37) et (2.38): 
\begin{subequations}
\label{7bis}
\begin{eqnarray}
\bar{\alpha} &=&t_{1}~\bar{\alpha}^{in}+r_{1}~\bar{\alpha}^{\prime }
\label{7bisa} \\
\bar{\alpha}^{out} &=&t_{1}~\bar{\alpha}^{\prime }-r_{1}~\bar{\alpha}^{in}
\label{7bisb} \\
\bar{\alpha}^{\prime } &=&r_{2}~e^{i\overline{\Psi }}\bar{\alpha}
\label{7bisc}
\end{eqnarray}
\begin{figure}[tbp]
\centerline{\psfig{figure=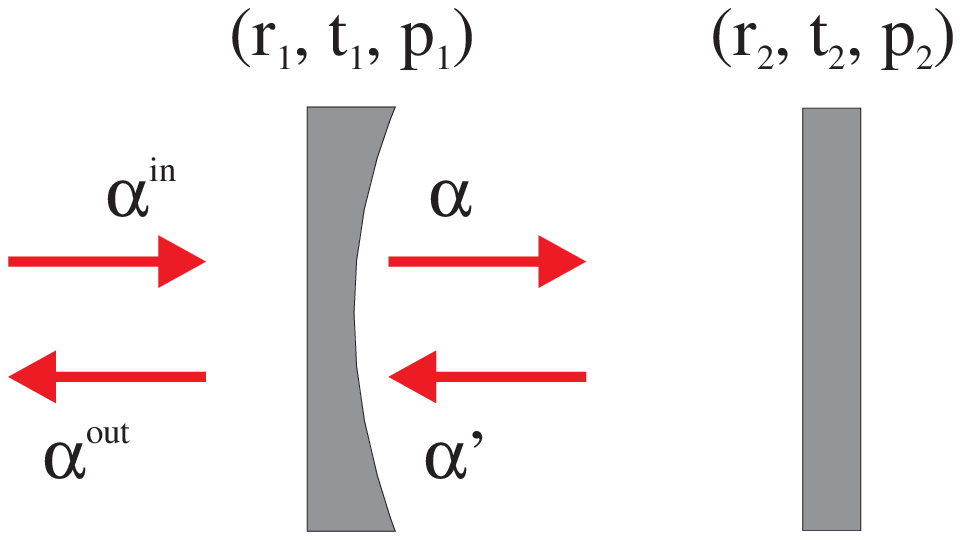,height=4cm}}
\caption{Cavit\'{e} Fabry-Perot avec pertes}
\label{Fig_4fpmcav}
\end{figure}
o\`{u} $\bar{\alpha}^{\prime }$ est le champ intracavit\'{e} revenant sur le
miroir d'entr\'{e}e (figure \ref{Fig_4fpmcav}) et $\overline{\Psi }$ le
d\'{e}phasage moyen du champ dans la cavit\'{e}. En \'{e}liminant les champs
intracavit\'{e} $\bar{\alpha}$ et $\bar{\alpha}^{\prime }$, on obtient la
relation d'entr\'{e}e-sortie pour le champ moyen: 
\end{subequations}
\begin{equation}
\overline{\alpha }^{out}=\frac{T_{1}-\gamma +i\overline{\Psi }}{\gamma -i%
\overline{\Psi }}~\overline{\alpha }^{in}  \label{8}
\end{equation}
o\`{u} $T_{1}=t_{1}^{2}$ est le coefficient de transmission en intensit\'{e}
du miroir d'entr\'{e}e et $2\gamma $ repr\'{e}sente les pertes totales de la
cavit\'{e}, reli\'{e}es aux coefficients de transmission et de perte en
intensit\'{e} des deux miroirs par l'\'{e}quation (\ref{2}).

On obtient ainsi l'expression du coefficient de r\'{e}flexion en
intensit\'{e} \`{a} r\'{e}sonance ${\cal R}_{0}=\overline{I}^{out}/\overline{%
I}^{in}$, que l'on peut exprimer en fonction de la finesse et du coefficient
de transmission du coupleur d'entr\'{e}e: 
\begin{equation}
{\cal R}_{0}=\left[ \frac{T_{1}-\gamma }{\gamma }\right] ^{2}=\left[ 1-\frac{%
{\cal F}~T_{1}}{\pi }\right] ^{2}  \label{9}
\end{equation}
Ce r\'{e}sultat montre que l'on peut d\'{e}terminer la transmission $T_{1}$
du coupleur si on mesure \`{a} la fois la finesse et le coefficient de
r\'{e}flexion de la cavit\'{e}.

Le coefficient de r\'{e}flexion ${\cal R}_{0}$ est obtenu
exp\'{e}rimentalement en faisant le rapport entre l'intensit\'{e}
r\'{e}fl\'{e}chie $\overline{I}_{res}^{out}$ lorsque la cavit\'{e} est \`{a}
r\'{e}sonance avec le faisceau incident et l'intensit\'{e} incidente ${\bar{I%
}}^{in}$. En dehors de la r\'{e}sonance, le faisceau incident est totalement
r\'{e}fl\'{e}chi par la cavit\'{e} et dans ce cas l'intensit\'{e}
r\'{e}fl\'{e}chie $\bar{I}_{off}^{out}$ est \'{e}gale \`{a} l'intensit\'{e}
incidente. Il suffit donc de balayer la fr\'{e}quence du laser autour de la
r\'{e}sonance et de mesurer l'intensit\'{e} r\'{e}fl\'{e}chie \`{a}
r\'{e}sonance et loin de la r\'{e}sonance pour en d\'{e}duire le coefficient
de r\'{e}flexion ${\cal R}_{0}$. Cette m\'{e}thode permet en outre de
s'affranchir des pertes subies par le faisceau incident, comme par exemple
la r\'{e}flexion sur la face avant du miroir d'entr\'{e}e, et des pertes
dans le syst\`{e}me de d\'{e}tection.

Par contre, cette expression du coefficient de r\'{e}flexion ne tient pas
compte de l'adaptation spatiale entre le faisceau et la cavit\'{e}. En fait,
le faisceau incident ne se projette pas seulement sur le mode fondamental de
la cavit\'{e}, mais sur l'ensemble des modes propres que l'on peut
caract\'{e}riser par une base orthonorm\'{e}e \{$v_{n}\left( r\right) $\}:

\begin{equation}
\bar{\alpha}^{in}\left( r\right) =\stackunder{n}{\sum }c_{n}~v_{n}\left(
r\right)  \label{10}
\end{equation}
L'adaptation est d\'{e}finie par le param\`{e}tre $\eta =\left| c_{0}\right|
^{2}/\stackunder{n}{\sum }\left| c_{n}\right| ^{2}$ qui est \'{e}gal \`{a}
la proportion de la puissance lumineuse incidente qui se projette sur le
mode fondamental.

Au voisinage de la r\'{e}sonance du mode fondamental, seul le terme $n=0$
dans la somme (\ref{10}) est modifi\'{e}, les autres termes ($n\geq 1$)
\'{e}tant totalement r\'{e}fl\'{e}chis. On trouve ainsi pour les
intensit\'{e}s r\'{e}fl\'{e}chies \`{a} r\'{e}sonance et hors r\'{e}sonance
les expressions suivantes: 
\begin{figure}[tbp]
\centerline{\psfig{figure=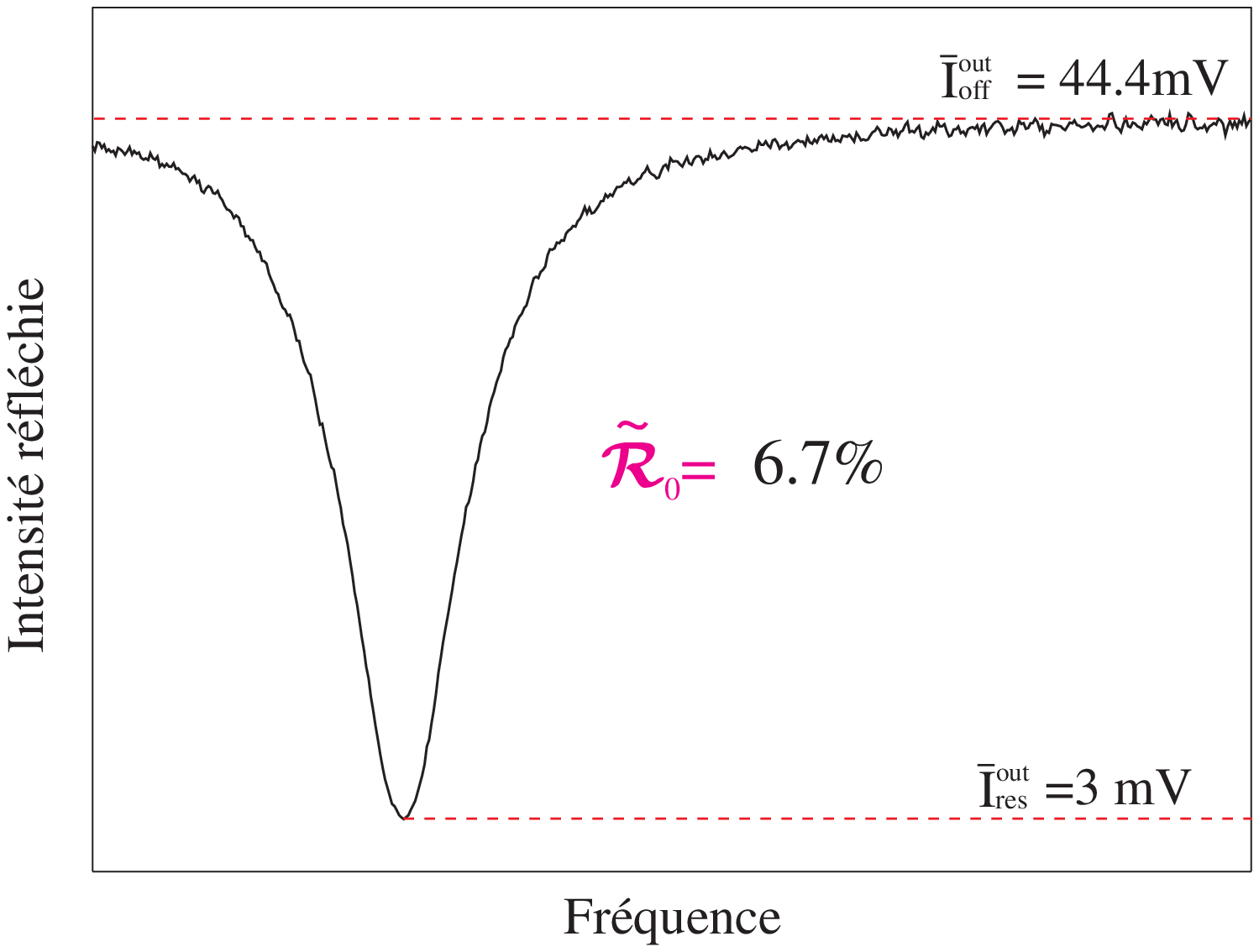,height=10cm}}
\caption{Variation de l'intensit\'{e} r\'{e}fl\'{e}chie par la cavit\'{e} $%
\left\{ N1\text{, }M24\right\} $ lorsque la fr\'{e}quence du faisceau
incident est balay\'{e}e autour de la r\'{e}sonance. Le coefficient ${\cal 
\tilde{R}}_{0}$ de r\'{e}flexion en intensit\'{e} est \'{e}gal au rapport
entre l'intensit\'{e} \`{a} r\'{e}sonance et l'intensit\'{e} loin de
r\'{e}sonance}
\label{Fig_4reflfpm}
\end{figure}

\begin{subequations}
\label{10}
\begin{eqnarray}
\bar{I}_{res}^{out} &=&{\cal R}_{0}~\left| c_{0}\right| ^{2}+\stackunder{%
n\geq 1}{\sum }\left| c_{n}\right| ^{2}  \label{10a} \\
\bar{I}_{off}^{out} &=&\left| c_{0}\right| ^{2}+\stackunder{n\geq 1}{\sum }%
\left| c_{n}\right| ^{2}  \label{10b}
\end{eqnarray}
Le coefficient de r\'{e}flexion \`{a} r\'{e}sonance ${\cal \tilde{R}}_{0}=%
\bar{I}_{res}^{out}/\bar{I}_{off}^{out}$ que l'on mesure
exp\'{e}rimentalement s'\'{e}crit alors en fonction du coefficient ${\cal R}%
_{0}$ (\'{e}quation \ref{9}) et du coefficient d'adaptation spatiale $\eta $%
: 
\end{subequations}
\begin{equation}
{\cal \tilde{R}}_{0}=1-\eta \left( 1-{\cal R}_{0}\right)  \label{11}
\end{equation}
Pour une adaptation parfaite ($\eta =1$), l'expression de ${\cal \tilde{R}}%
_{0}$ se r\'{e}duit \`{a} celle de ${\cal R}_{0}$. Par contre, d\`{e}s que $%
\eta $ est inf\'{e}rieur \`{a} $1$, la valeur mesur\'{e}e ${\cal \tilde{R}}%
_{0}$ est diff\'{e}rente de la valeur th\'{e}orique ${\cal R}_{0}$
donn\'{e}e par l'\'{e}quation (\ref{9}). Etant donn\'{e} la qualit\'{e} de
l'adaptation spatiale r\'{e}alis\'{e}e dans l'exp\'{e}rience ($\eta \approx
98\%$), cet \'{e}cart se traduit par une modification de l'\'{e}valuation du
coefficient de transmission $T_{1}$ de quelques $ppm$ seulement. Nous avons
donc tenu compte de l'adaptation spatiale \`{a} l'aide de l'\'{e}quation (%
\ref{11}), mais sans chercher \`{a} d\'{e}terminer plus pr\'{e}cis\'{e}ment
le coefficient $\eta $.

La figure \ref{Fig_4reflfpm} montre l'intensit\'{e} r\'{e}fl\'{e}chie par la
cavit\'{e} $\{N1$, $M24\}$ lorsqu'on balaye la fr\'{e}quence du faisceau
incident. On mesure \`{a} partir de ce r\'{e}sultat les intensit\'{e}s $\bar{%
I}_{off}^{out}=44.4~mV$ et $\overline{I}_{res}^{out}=3~mV$, qui
correspondent \`{a} un coefficient de r\'{e}flexion ${\cal \tilde{R}}_{0}$
\'{e}gal \`{a} $6.7\%$ et un coefficient th\'{e}orique ${\cal R}_{0}$ de $%
4.8\%$. En utilisant l'expression (\ref{9}) et la valeur obtenue
pr\'{e}c\'{e}demment pour la finesse (${\cal F}=47000$), on trouve que le
coefficient de transmission $T_{N1}$ du coupleur Newport $N1$ est \'{e}gal
\`{a} $52~ppm$. Nous avons r\'{e}p\'{e}t\'{e} cette mesure pour
diff\'{e}rentes cavit\'{e}s, constitu\'{e}es de diff\'{e}rents miroirs
mobiles mais toujours avec le m\^{e}me coupleur d'entr\'{e}e $N_{1}$. Nous
avons obtenu des valeurs de la transmission du coupleur $N_{1}$ comprises
entre $48$ et $52~ppm$. Nous avons aussi mesur\'{e} la transmission du
coupleur Newport $N_{2}$ et nous avons trouv\'{e} une transmission de $%
58~ppm $. Ces r\'{e}sultats sont en tr\`{e}s bon accord avec les valeurs
annonc\'{e}es par le fabriquant.

La mesure de la finesse et du coefficient de r\'{e}flexion \`{a}
r\'{e}sonance permettent donc de d\'{e}terminer la transmission du coupleur
d'entr\'{e}e. A partir de l'\'{e}quation (\ref{2}), on peut aussi
d\'{e}terminer la somme $T_{M}+P_{M}+P_{N}$ des pertes totales (transmission
plus pertes) du miroir mobile et des pertes du coupleur Newport. Ainsi pour
la cavit\'{e} \{$N1$, $M24$\}, on trouve: 
\begin{equation}
T_{M24}+P_{M24}+P_{N1}=81~ppm  \label{12}
\end{equation}
Ces mesures ne permettent pas cependant de s\'{e}parer les pertes du miroir
mobile de celles du coupleur d'entr\'{e}e.

\subsubsection{Mesure des pertes du miroir mobile\label{IV-1-4-4}}

Il est possible de s\'{e}parer les pertes du miroir mobile de celles du
coupleur d'entr\'{e}e en r\'{e}p\'{e}tant les mesures pr\'{e}c\'{e}dentes
pour diff\'{e}rentes cavit\'{e}s. On utilise en fait trois miroirs, le
miroir mobile $M32$ et les deux coupleurs Newport $N1$ et $N2$. On peut
ainsi former les trois cavit\'{e}s \{$N1$, $M32$\}, \{$N2$, $M32$\} et \{$N1$%
, $N2$\}, et mesurer leur bande passante $\nu _{BP}$ et leur coefficient de
r\'{e}flexion \`{a} r\'{e}sonance ${\cal \tilde{R}}_{0}$. Le tableau suivant
r\'{e}sume les mesures effectu\'{e}es:\bigskip \medskip

\centerline{\psfig{figure=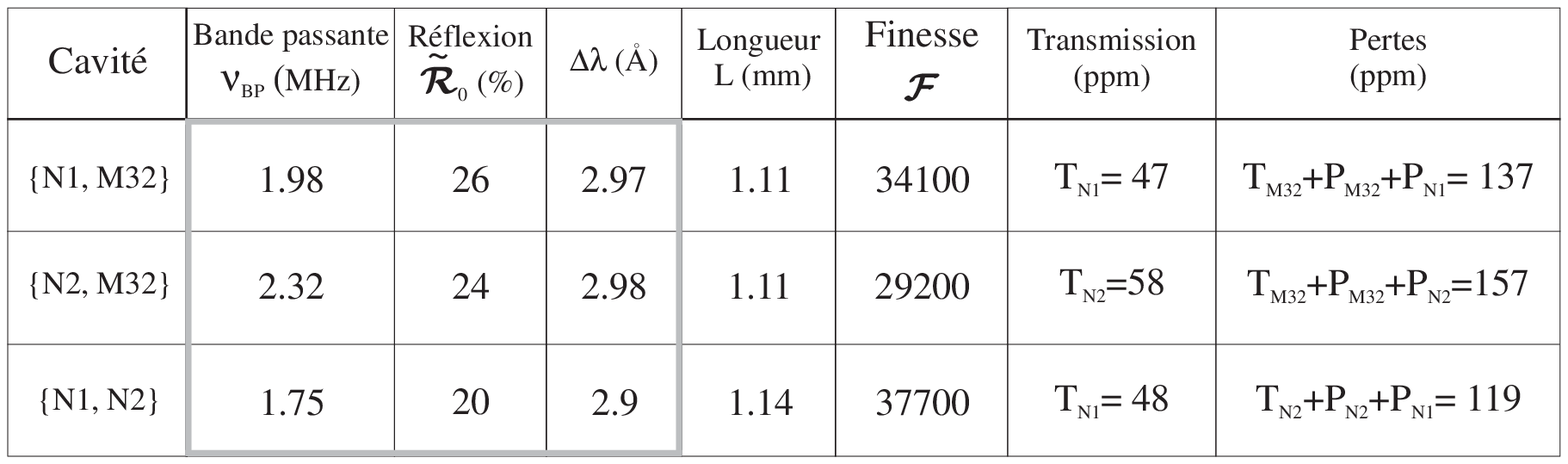,height=45mm}}\label{4tabpert}\bigskip
\medskip

Nous avons utilis\'{e} pour ces mesures les supports de cavit\'{e} avec
espaceur en cuivre, dont la longueur est environ $1~mm$. Pour d\'{e}terminer
pr\'{e}cis\'{e}ment la longueur de la cavit\'{e}, nous avons mesur\'{e}
l'intervalle spectral libre $\nu _{ISL}$ de chaque cavit\'{e}. Pour cela on
fait varier la longueur d'onde $\lambda $ de la source laser de fa\c{c}on
\`{a} passer d'un mode $TEM_{00}$ de la cavit\'{e} au suivant. On parcourt
ainsi un intervalle spectral libre $\nu _{ISL}$ de la cavit\'{e} et on peut
d\'{e}duire la longueur de la cavit\'{e} $L=\lambda ^{2}/2\Delta \lambda $
en mesurant la variation de longueur d'onde $\Delta \lambda $ \`{a} l'aide
d'un lambdam\`{e}tre dont la pr\'{e}cision est de $10^{-2}~\AA $ (la
longueur d'onde $\lambda $ est de l'ordre de $812~nm$).

Les trois premi\`{e}res colonnes du tableau pr\'{e}sentent les r\'{e}sultats
des mesures concernant la bande passante de la cavit\'{e} $\nu _{BP}$
d\'{e}termin\'{e}e par comparaison avec la cavit\'{e} de r\'{e}f\'{e}rence,
le coefficient de r\'{e}flexion \`{a} r\'{e}sonance ${\cal \tilde{R}}_{0}$
et la variation de la longueur d'onde $\Delta \lambda $ correspondant \`{a}
un intervalle spectral libre. La quatri\`{e}me colonne donne la longueur $L$
de la cavit\'{e} d\'{e}duite de la mesure de $\Delta \lambda $. Notons que
les valeurs obtenues sont en bon accord avec une longueur de l'espaceur de $%
1\pm 0.1~mm$, \`{a} condition de tenir compte de la courbure des miroirs
Newport. La distance selon l'axe de la cavit\'{e} entre le centre du miroir
Newport et son bord est en effet de $80~\mu m$.

La cinqui\`{e}me colonne du tableau pr\'{e}sente la finesse de la cavit\'{e}
obtenue \`{a} partir des \'{e}quations (\ref{4}) et (\ref{6}), en utilisant
les valeurs de la bande passante $\nu _{BP}$ et la longueur $L$ de la
cavit\'{e}. Les deux derni\`{e}res colonnes pr\'{e}sentent la transmission
du coupleur d'entr\'{e}e et la somme des pertes du miroir arri\`{e}re et du
coupleur, d\'{e}duites de la finesse ${\cal F}$ et du coefficient de
r\'{e}flexion \`{a} r\'{e}sonance ${\cal \tilde{R}}_{0}$ mesur\'{e} avec une
adaptation spatiale $\eta $ de $98\%$.

On voit que la derni\`{e}re colonne fournit des valeurs pour des
combinaisons lin\'{e}aires diff\'{e}rentes des pertes $T_{M32}+P_{M32}$, $%
P_{N1}$ et $P_{N2}$ des trois miroirs, \`{a} partir desquelles on peut
d\'{e}terminer les valeurs de chacune de ces pertes. On trouve ainsi que les
pertes totales $T_{M32}+P_{M32}$ du miroir mobile $M32$ sont \'{e}gales
\`{a} $116~ppm$, alors que les pertes des coupleurs Newport sont
respectivement $P_{N1}=21~ppm$ et $P_{N2}=41~ppm$. Le tableau suivant
pr\'{e}sente l'ensemble des r\'{e}sultats de nos mesures pour tous les
miroirs que nous avons utilis\'{e}s:\bigskip \medskip

\centerline{\psfig{figure=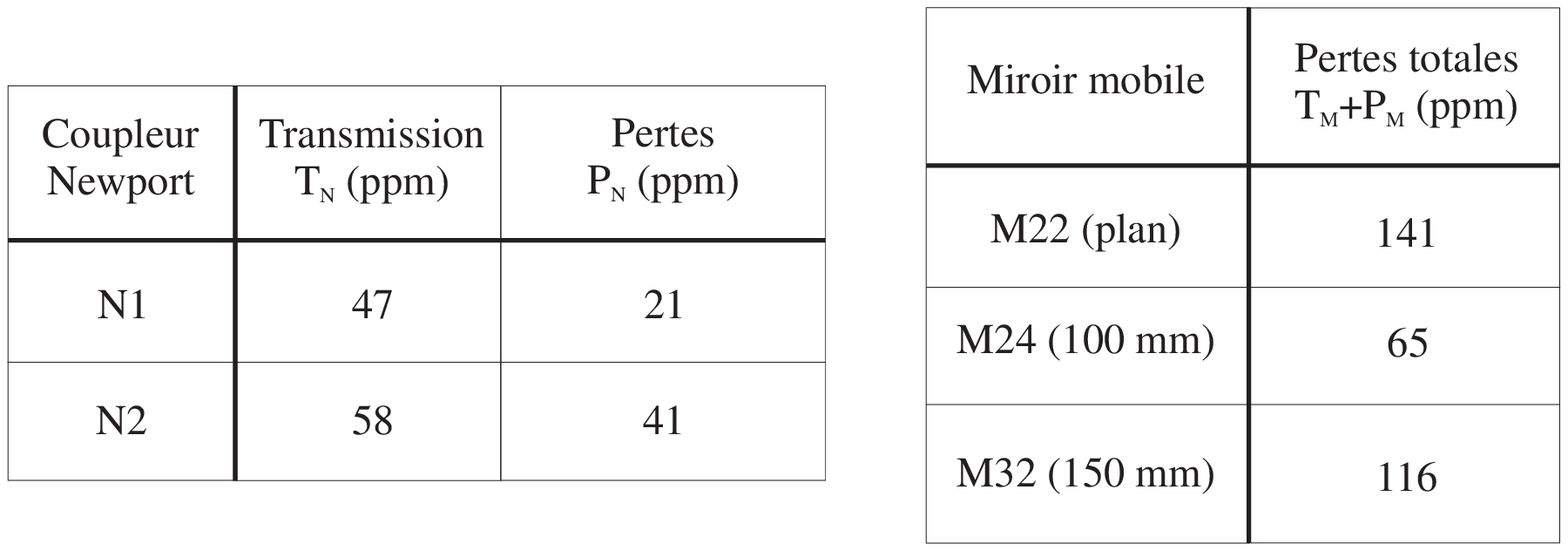,height=5cm}}\label{4tabresu}\bigskip
\medskip

On constate que les caract\'{e}ristiques des miroirs Newport sont conformes
aux sp\'{e}cifications du constructeur. Par contre les pertes des miroirs
mobiles sont plus importantes que pr\'{e}vu. Ceci pourrait \^{e}tre d\^{u}
\`{a} un mauvais \'{e}tat de surface des substrats, qui induirait des
d\'{e}formations des couches multidi\'{e}lectriques. Ces d\'{e}formations
peuvent entra\^{\i }ner une absorption plus importante dans les couches
multidi\'{e}lectriques et une augmentation de la transmission de l'ensemble
des couches.

Notons que le d\'{e}saccord apparent entre les valeurs mesur\'{e}es ici et
les pertes par diffusion mesur\'{e}es \`{a} l'Institut de Physique
Nucl\'{e}aire (tableau page \pageref{4tabppm}) peut s'expliquer en comparant
le principe des deux m\'{e}thodes : ici on mesure les pertes totales du
miroir pour un faisceau lumineux de rayon inf\'{e}rieur \`{a} $100~\mu m$,
alors que le diffusom\`{e}tre mesure les pertes diffus\'{e}es dans une
direction donn\'{e}e, pour un faisceau incident de section \'{e}gale \`{a} $%
1~mm$.

On peut enfin estimer la pr\'{e}cision de nos mesures. Les r\'{e}sultats
n'ont de sens que si la qualit\'{e} des miroirs est suffisamment
homog\`{e}ne au voisinage du centre des miroirs. Le col du faisceau lumineux
\'{e}tant inf\'{e}rieur \`{a} $100~\mu m$, la zone du miroir sur laquelle se
r\'{e}fl\'{e}chit la lumi\`{e}re peut \^{e}tre diff\'{e}rente d'une
cavit\'{e} \`{a} l'autre. Nous n'avons pas cependant constat\'{e} de
variations appr\'{e}ciables lors de nos diff\'{e}rents essais de cavit\'{e}.
A partir de toutes les mesures que nous avons effectu\'{e}es et des
recoupements que l'on peut faire entre elles, on peut conclure que
l'incertitude sur la mesure des pertes est de l'ordre de $\pm 10~ppm$.

\section{La source laser\label{IV-2}}

\bigskip

Du fait des contraintes impos\'{e}es par la cavit\'{e} \`{a} miroir mobile,
nous avons \'{e}t\'{e} amen\'{e}s \`{a} concevoir une source laser
adapt\'{e}e \`{a} nos besoins. Cette source doit pouvoir fonctionner selon
trois r\'{e}gimes distincts :

{\bf -} Pour trouver la r\'{e}sonance de la cavit\'{e} \`{a} miroir mobile,
il est n\'{e}cessaire de balayer la fr\'{e}quence de la cavit\'{e} sur une
plage d'au moins $300~GHz$. Ce balayage n'a cependant pas besoin d'\^{e}tre
effectu\'{e} en continu. Lors de la proc\'{e}dure de r\'{e}glage de
l'adaptation spatiale du faisceau sur la cavit\'{e}, celle-ci pr\'{e}sente
un peigne large de r\'{e}sonances correspondants aux diff\'{e}rents modes
transverses. Il suffit en fait de pouvoir balayer en continu quelques
intervalles entre modes transverses (typiquement $10~GHz$), et de faire des
sauts de fr\'{e}quence discontinus pour explorer l'ensemble du peigne. On
dispose aussi d'un lambdam\`{e}tre de fa\c{c}on \`{a} rep\'{e}rer \`{a} tout
instant la longueur d'onde du faisceau laser.

{\bf -} Une fois l'adaptation spatiale r\'{e}alis\'{e}e, on utilise une
rampe de modulation de la fr\'{e}quence du laser afin de parcourir le pic
d'Airy de la r\'{e}sonance. Ceci permet en particulier de d\'{e}terminer la
finesse de la cavit\'{e}. Cette rampe doit avoir une excursion plus large
que la bande passante de la cavit\'{e} (typiquement $20$ \`{a} $30~MHz$) et
une fr\'{e}quence de modulation de l'ordre de $100~Hz$.

{\bf -} En fonctionnement normal, la fr\'{e}quence du laser doit \^{e}tre
contr\^{o}l\'{e}e de fa\c{c}on \`{a} correspondre \`{a} la r\'{e}sonance de
la cavit\'{e} \`{a} miroir mobile. Il est ainsi n\'{e}cessaire de
r\'{e}duire les fluctuations de fr\'{e}quence du laser \`{a} une valeur
petite devant la bande passante de la cavit\'{e}, et de contr\^{o}ler les
d\'{e}rives lentes entre la fr\'{e}quence du laser et celle de la cavit\'{e}
\`{a} miroir mobile. D'autre part, nous avons vu dans la partie \ref{II-3}
que la cavit\'{e} est un syst\`{e}me bistable : \`{a} cause du terme de
d\'{e}phasage non lin\'{e}aire, le d\'{e}saccord entre la r\'{e}sonance de
la cavit\'{e} et la fr\'{e}quence du laser d\'{e}pend de l'intensit\'{e}
incidente. Il faut par cons\'{e}quent contr\^{o}ler l'intensit\'{e} de la
source laser, de fa\c{c}on \`{a} pouvoir ajuster l'intensit\'{e} moyenne
\`{a} une valeur pr\'{e}cise et r\'{e}duire ses fluctuations basse
fr\'{e}quence. Enfin le faisceau doit \^{e}tre parfaitement adapt\'{e}
spatialement au mode fondamental de la cavit\'{e}.

Ces contraintes nous ont amen\'{e}s \`{a} choisir un laser titane saphir. Ce
laser est facilement balayable, de mani\`{e}re r\'{e}p\'{e}titive et
contr\^{o}l\'{e}e. La longueur d'onde de notre laser se situe aux alentours
de $810~nm$, valeur pour laquelle on sait faire d'excellents miroirs ayant
une bonne tenue au flux. Ce laser fournit une puissance importante, de
l'ordre du Watt, ce qui permet de l'utiliser aussi pour l'excitation optique
des modes acoustiques du r\'{e}sonateur. Enfin le laser titane saphir
fournit un faisceau dont le bruit technique reste mod\'{e}r\'{e} : les
fluctuations d'intensit\'{e} et de phase se r\'{e}duisent aux fluctuations
quantiques pour des fr\'{e}quences sup\'{e}rieures \`{a} $2~MHz$, et
l'exc\`{e}s de bruit \`{a} plus basse fr\'{e}quence n'exc\`{e}de pas $30$
\`{a} $40~dB$. La situation aurait \'{e}t\'{e} diff\'{e}rente avec des
diodes lasers par exemple, qui pr\'{e}sentent un bruit de phase important
m\^{e}me \`{a} haute fr\'{e}quence. Ce bruit aurait pu \^{e}tre
pr\'{e}judiciable pour une mesure pr\'{e}cise de la quadrature de phase du
faisceau r\'{e}fl\'{e}chi par la cavit\'{e} \`{a} miroir mobile. Il est par
ailleurs important pour l'\'{e}tude des effets quantiques du couplage
optom\'{e}canique que l'intensit\'{e} du faisceau incident soit au niveau du
bruit de photon standard aux fr\'{e}quences d'analyse. Nous verrons dans la
partie \ref{IV-5} comment la cavit\'{e} de filtrage spatial du faisceau
laser peut \^{e}tre utilis\'{e}e pour r\'{e}duire le bruit technique basse
fr\'{e}quence du faisceau.

\begin{figure}[tbp]
\centerline{\psfig{figure=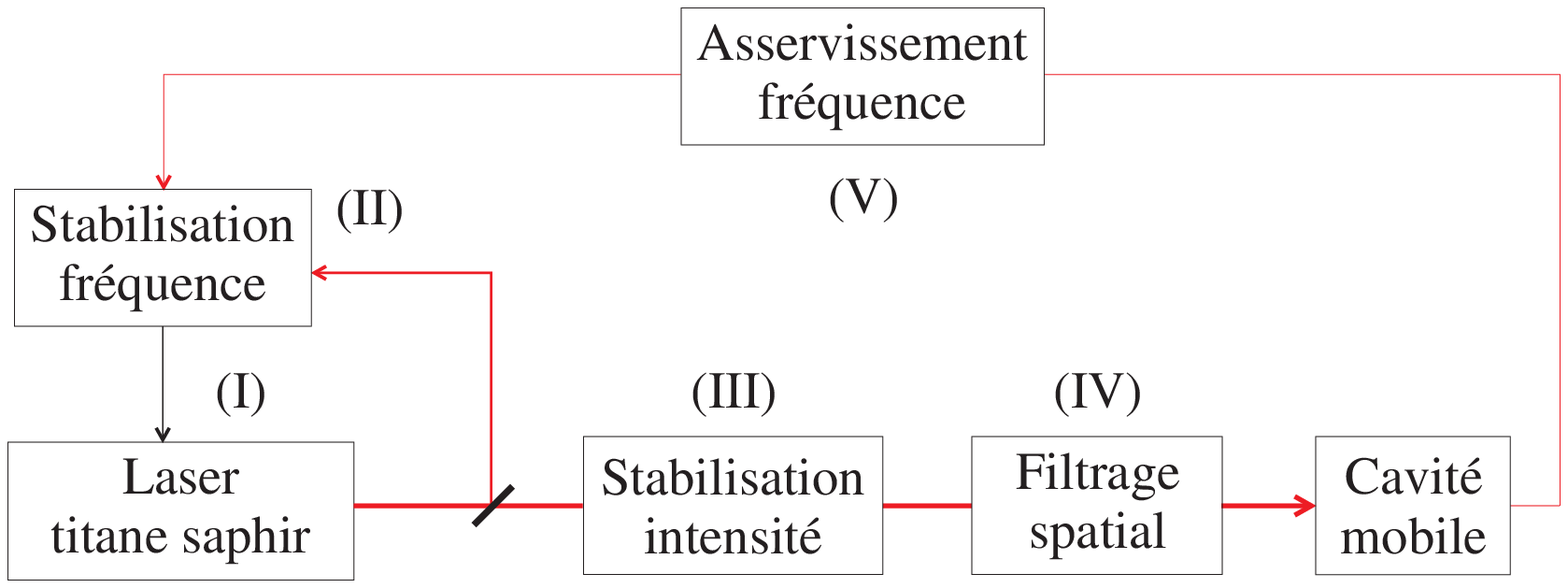,height=5cm}}
\caption{Sch\'{e}ma g\'{e}n\'{e}ral de la source laser}
\label{Fig_4laser1}
\end{figure}

L'ensemble de notre source lumineuse comporte essentiellement cinq
\'{e}l\'{e}ments comme le montre la figure \ref{Fig_4laser1}.
L'\'{e}l\'{e}ment principal est le laser titane saphir (I) qui produit un
faisceau lumineux monomode, balayable, et de forte puissance. Les fluc-%
\newline
tuations de fr\'{e}quence du mode laser (jitter) \'{e}tant importantes, on
utilise un syst\`{e}me de stabilisation en fr\'{e}quence (II) qui r\'{e}duit
consid\'{e}rablement ces fluctuations. Les deux \'{e}l\'{e}ments (III) et
(IV) assurent respectivement la stabilit\'{e} en intensit\'{e} du faisceau
et le filtrage spatial. On obtient ainsi \`{a} l'entr\'{e}e de la cavit\'{e}
\`{a} miroir mobile un faisceau $TEM_{00}$ monomode balayable, stable en
fr\'{e}quence et en intensit\'{e}. Enfin, une boucle d'asservissement
suppl\'{e}mentaire (V) vient se rajouter au syst\`{e}me de stabilisation en
fr\'{e}quence pour maintenir le laser \`{a} r\'{e}sonance avec la cavit\'{e}
\`{a} miroir mobile. Nous allons pr\'{e}senter plus en d\'{e}tail dans les
sections suivantes ces cinq \'{e}l\'{e}ments.

\subsection{Le laser titane saphir\label{IV-2-1}}

\begin{figure}[tbp]
\centerline{\psfig{figure=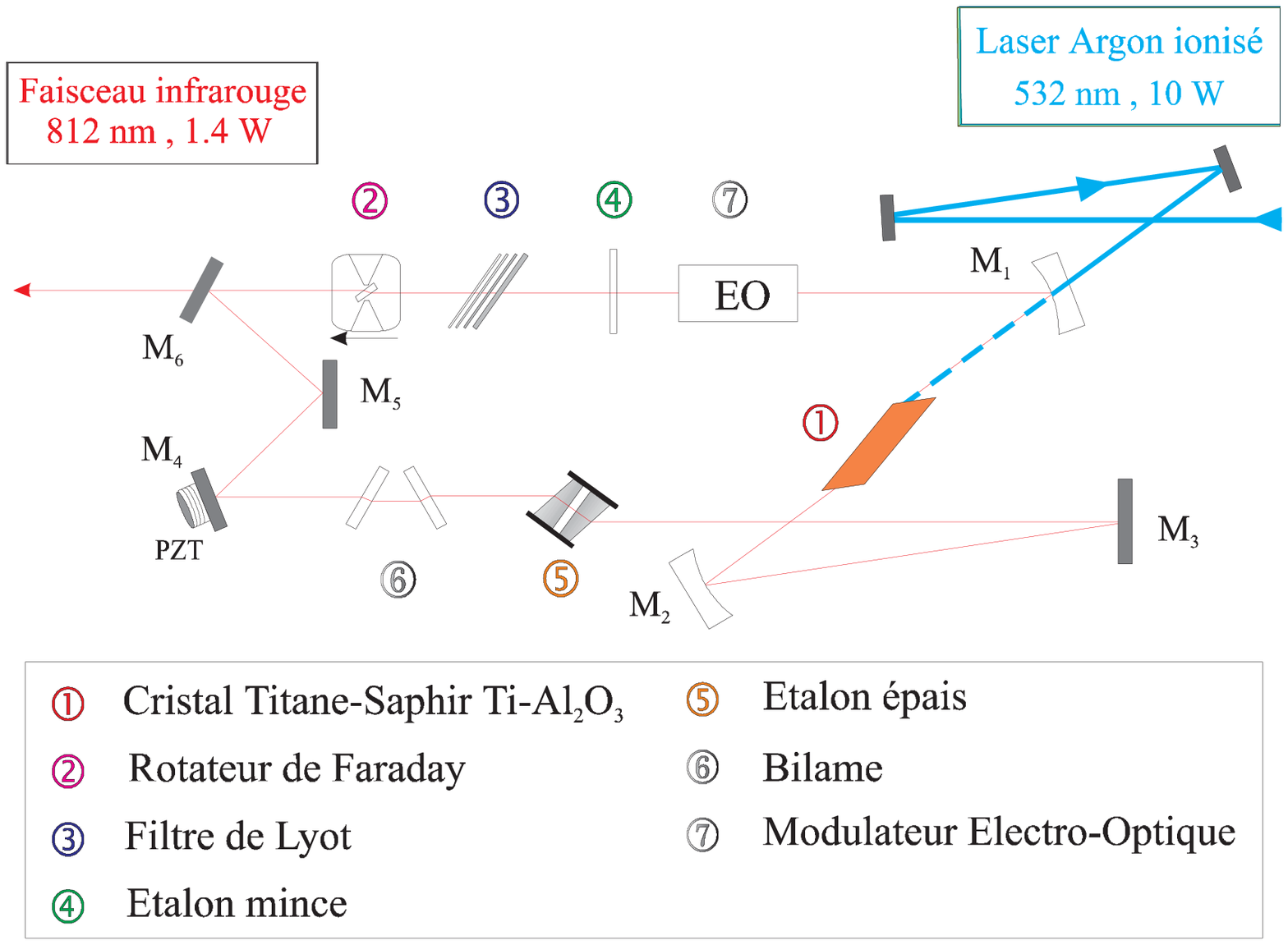,height=115mm}}
\caption{Sch\'{e}ma du laser titane saphir pomp\'{e} par un laser \`{a}
Argon commercial}
\label{Fig_4tisa}
\end{figure}

Nous avons construit le laser titane saphir selon le mod\`{e}le
d\'{e}velopp\'{e} au laboratoire par F. Biraben\cite{biraben 82}. Comme le
montre la figure \ref{Fig_4tisa}, il s'agit d'une cavit\'{e} en anneau dans
laquelle sont dispos\'{e}s les diff\'{e}rent \'{e}l\'{e}ments optiques. La
cavit\'{e} a une longueur de $1.6~m$, ce qui correspond \`{a} un intervalle
spectral libre de $187~MHz$. Les miroirs $M_{1}$ \`{a} $M_{5}$ sont
totalement r\'{e}fl\'{e}chissants ($R_{\max }$) alors que le miroir $M_{6}$
sert de coupleur de sortie et a une transmission de $4\%$. Le miroir $M_{4}$
est mont\'{e} sur une cale pi\'{e}zo\'{e}lectrique utilis\'{e}e pour des
corrections lentes de l'asservissement en fr\'{e}quence. Le cristal
Titane-Saphir $Ti$-$Al_{2}O_{3}$ (\ding{172} sur la figure \ref{Fig_4tisa}),
taill\'{e} \`{a} l'angle de Brewster, est pomp\'{e} par un laser \`{a} Argon
ionis\'{e} continu (Coherent INNOVA 400), utilis\'{e} en mode multiraies
avec une puissance de $10~W$. Les miroirs sph\'{e}riques $M_{1}$ et $M_{2}$ (%
$150~mm$ de rayon de courbure) sont dichro\"{\i }ques avec un coefficient de
transmission pour les raies de l'Argon \'{e}gal \`{a} $97\%$. Le rayon de
courbure des miroirs est choisi de fa\c{c}on \`{a} focaliser le faisceau au
niveau du cristal ce qui permet d'augmenter l'\'{e}mission stimul\'{e}e par
rapport \`{a} l'\'{e}mission spontan\'{e}e. 

Parmi les \'{e}l\'{e}ments plac\'{e}s dans la cavit\'{e}, on trouve le
rotateur de Faraday (\ding{173} sur la figure \ref{Fig_4tisa}) constitu\'{e}
d'une lame de verre Hoya $FR5$ ayant une forte constante de Verdet,
d'\'{e}paisseur $4~mm$ et d'un aimant cr\'{e}ant un champ magn\'{e}tique de $%
5~kG$ environ. Le r\^{o}le du rotateur de Faraday est de forcer le sens de
propagation du faisceau lumineux. Dans un laser en anneau, il y a a priori
deux ondes progressives se propageant en sens inverse. Le rotateur de
Faraday fait tourner la polarisation du faisceau d'un angle ind\'{e}pendant
du sens de propagation du faisceau. Il est associ\'{e} \`{a} un syst\`{e}me
form\'{e} des miroirs non coplanaires $M_{4}$, $M_{5}$ et $M_{6}$ ($M_{5}$
est situ\'{e} au dessus du plan du laser) pour assurer un fonctionnement
unidirectionnel du laser : ces miroirs compensent la rotation de
polarisation due \`{a} l'effet Faraday pour un sens de propagation et
l'augmentent pour l'autre. Du fait de la pr\'{e}sence d'\'{e}l\'{e}ments
polarisants (lames \`{a} incidence de Brewster), le faisceau se propageant
dans le second sens subit des pertes. La diff\'{e}rence des pertes entre les
deux sens de propagation est suffisante pour que la comp\'{e}tition entre
modes assure un fonctionnement monodirectionnel stable, dans la direction
indiqu\'{e}e par une fl\`{e}che sur la figure \ref{Fig_4tisa}.

Les autres \'{e}l\'{e}ments servent essentiellement \`{a} la s\'{e}lection
de la fr\'{e}quence d'\'{e}mission du laser. La courbe de fluorescence du
cristal Titane-Saphir est tr\`{e}s large compar\'{e}e \`{a} l'intervalle
entre les modes longitudinaux de la cavit\'{e} laser, ce qui se traduit par
une \'{e}mission laser multimode. Pour rendre ce comportement monomode, on
utilise trois \'{e}l\'{e}ments s\'{e}lectifs en fr\'{e}quence. Le premier
est le filtre de Lyot (\ding{174} sur la figure \ref{Fig_4tisa}) qui
s\'{e}lectionne une bande de fr\'{e}quence de quelque centaines de
gigahertz. Ce filtre est constitu\'{e} de quatre lames bir\'{e}fringentes
plac\'{e}es \`{a} incidence de Brewster. Afin que toutes les lames
s\'{e}lectionnent la m\^{e}me longueur d'onde, leurs axes optiques sont
parall\`{e}les et leur \'{e}paisseur sont choisies dans les rapports 1, 2,
4, 16, la lame la plus mince faisant $410~\mu m$ d'\'{e}paisseur. Ces lames
sont plac\'{e}es dans un support qui peut tourner dans le plan des lames
gr\^{a}ce \`{a} un moteur \'{e}lectrique, modifiant ainsi la zone de
fr\'{e}quence s\'{e}lectionn\'{e}e.

Le second \'{e}l\'{e}ment s\'{e}lectif est l'\'{e}talon mince (\ding{175}
sur la figure \ref{Fig_4tisa}) qui est constitu\'{e} d'une lame en silice
d'indice $n=1.43$, non trait\'{e}e optiquement et de $0.7~mm$
d'\'{e}paisseur. Cet \'{e}l\'{e}ment joue le r\^{o}le d'une cavit\'{e}
Fabry-Perot d'intervalle spectral libre \'{e}gal \`{a} $150~GHz$. La lame
est plac\'{e}e sur un support que l'on peut incliner par rapport au faisceau
laser gr\^{a}ce \`{a} un syst\`{e}me de bras de levier command\'{e} par un
moteur. On modifie ainsi l'\'{e}paisseur optique de la lame et donc la
fr\'{e}quence s\'{e}lectionn\'{e}e. Lorsqu'on effectue un balayage de la
fr\'{e}quence du laser, l'intensit\'{e} transmise par la lame diminue tandis
que l'intensit\'{e} r\'{e}fl\'{e}chie augmente. Pour maintenir la lame \`{a}
r\'{e}sonance (transmission maximale) durant un tel balayage, on utilise un
asservissement qui pilote le moteur de fa\c{c}on \`{a} minimiser
l'intensit\'{e} r\'{e}fl\'{e}chie. Le signal d'erreur est obtenu en faisant
la diff\'{e}rence entre l'intensit\'{e} d\'{e}tect\'{e}e en r\'{e}flexion
sur la lame et celle d\'{e}tect\'{e}es \`{a} la sortie du laser. Un
\'{e}quilibrage \'{e}lectronique entre ces deux intensit\'{e}s permet de
fixer le point de fonctionnement de l'asservissement au voisinage du minimum
d'intensit\'{e} r\'{e}fl\'{e}chie.

Le dernier \'{e}l\'{e}ment s\'{e}lectif en fr\'{e}quence est l'\'{e}talon
\'{e}pais qui est en fait une cavit\'{e} Fabry-Perot (\ding{176} sur la
figure \ref{Fig_4tisa}). Il est form\'{e} par deux prismes dont les faces en
regard sont trait\'{e}es pour avoir un coefficient de r\'{e}flexion
d'environ $30\%$. L'\'{e}paisseur de ce Fabry-Perot est d'environ $8~mm$,
correspondant \`{a} un intervalle spectral libre de $19~GHz$. L'un des deux
prismes est mont\'{e} sur une cale pi\'{e}zo\'{e}lectrique afin de modifier
la fr\'{e}quence de r\'{e}sonance. Cette fr\'{e}quence est asservie sur la
fr\'{e}quence du mode laser gr\^{a}ce \`{a} une d\'{e}tection synchrone. Le
principe de cet asservissement consiste \`{a} moduler la longueur du
Fabry-Perot \`{a} une fr\'{e}quence de $3~kHz$. Ceci induit une modulation
de l'intensit\'{e} en sortie du laser, dont l'amplitude est proportionnelle
au d\'{e}saccord entre la r\'{e}sonance de l'\'{e}talon \'{e}pais et la
fr\'{e}quence du mode laser s\'{e}lectionn\'{e}. Ce signal est
d\'{e}tect\'{e} par une photodiode plac\'{e}e \`{a} la sortie du laser et
envoy\'{e} dans une d\'{e}tection synchrone. Le signal d'erreur pilote la
cale pi\'{e}zo\'{e}lectrique de l'\'{e}talon \'{e}pais.

Le fonctionnement monomode du laser est assur\'{e} lorsque ces trois
\'{e}l\'{e}ments s\'{e}lectifs sont centr\'{e}s sur un mode de la cavit\'{e}
laser. Lorsque ces \'{e}l\'{e}ments sont r\'{e}gl\'{e}s et que les
asservissements du Fabry-Perot \'{e}pais et de l'\'{e}talon mince sont
verrouill\'{e}s, on obtient \`{a} la sortie du laser titane saphir un
faisceau monomode d'une puissance de $1.4~W$ sur une plage de longueur
d'onde comprise entre $800$ et $850~nm$. Les caract\'{e}ristiques spatiales
du faisceau, mesur\'{e}es \`{a} l'aide d'un analyseur de modes (mode master
Coherent), correspondent \`{a} un col $w_{L}$ \'{e}gal \`{a} $0.6~mm$
situ\'{e} \`{a} environ $180~mm$ du miroir de sortie du laser et \`{a} une
longueur de Raighley $z_{R}=\pi w_{L}^{2}/\lambda $ \'{e}gale \`{a} $1.4~m$.
Notons aussi que le faisceau pr\'{e}sente un astigmatisme relativement
important (environ $15\%$) li\'{e} essentiellement \`{a} des effets de
lentille dus \`{a} l'\'{e}chauffement du cristal Titane-Saphir qui absorbe
une partie importante du faisceau pompe. Du fait de cet astigmatisme, le
faisceau \`{a} la sortie du laser n'est pas parfaitement gaussien et nous
verrons par la suite comment obtenir un faisceau gaussien en utilisant une
cavit\'{e} de filtrage.

Le dernier \'{e}l\'{e}ment (\ding{177} sur la figure \ref{Fig_4tisa}) est un
bilame qui permet de balayer de mani\`{e}re continue la fr\'{e}quence du
laser. Ce dispositif est constitu\'{e} de deux lames de $10~mm$
d'\'{e}paisseur plac\'{e}es au voisinage de l'incidence de Brewster. Un
syst\`{e}me m\'{e}canique motoris\'{e} permet de faire tourner les lames de
mani\`{e}re sym\'{e}trique, ce qui a pour effet de varier la longueur
optique de la cavit\'{e} et donc la fr\'{e}quence du laser, sans
d\'{e}placer le faisceau. Les asservissements de la cavit\'{e} permettent un
balayage continu de la fr\'{e}quence du laser sur plusieurs dizaines de
gigahertz. On peut par ailleurs effectuer des excursions de $19$~$GHz$ par
saut de modes de l'\'{e}talon \'{e}pais en inclinant l'\'{e}talon mince, ou
encore de $150$~$GHz$ par saut de modes de l'\'{e}talon mince en tournant
les lames du Lyot. Pour se rep\'{e}rer lors de ces sauts de modes, on mesure
la longueur d'onde du faisceau \`{a} l'aide d'un lambdam\`{e}tre. Cette
proc\'{e}dure permet de parcourir facilement un intervalle spectral libre de
la cavit\'{e} \`{a} miroir mobile, qui est de $300~GHz$ pour une longueur de 
$0.5~mm$.

\subsection{Stabilisation en fr\'{e}quence\label{IV-2-2}}

Le faisceau issu du laser titane saphir est monomode et balayable en
fr\'{e}quence, mais il pr\'{e}sente d'importantes fluctuations de
fr\'{e}quence ({\it jitter}) li\'{e}es aux vibrations m\'{e}caniques des
diff\'{e}rents \'{e}l\'{e}ments optiques de la cavit\'{e} laser. Ces
fluctuations de fr\'{e}quence ont une amplitude de l'ordre du m\'{e}gahertz
et elles ne permettent pas de d\'{e}finir un point de fonctionnement
pr\'{e}cis de la cavit\'{e} \`{a} miroir mobile. Il est donc imp\'{e}ratif
de r\'{e}duire ces fluctuations de telle sorte qu'elles soient
n\'{e}gligeables devant la bande passante de la cavit\'{e} qui est de
l'ordre de $2~MHz$. On utilise pour cela une cavit\'{e} ext\'{e}rieure
tr\`{e}s stable m\'{e}caniquement et on asservit la fr\'{e}quence du laser
sur la r\'{e}sonance de cette cavit\'{e}.

\subsubsection{Cavit\'{e} Fabry-Perot externe\label{IV-2-2-1}}

La stabilisation en fr\'{e}quence du laser est r\'{e}alis\'{e}e \`{a} l'aide
d'un asservissement utilisant la m\'{e}thode des bandes lat\'{e}rales\cite
{drever 1983} en r\'{e}flexion sur une cavit\'{e} Fabry-Perot externe (FPE).
Cette cavit\'{e} est form\'{e}e d'un coupleur d'entr\'{e}e plan fix\'{e} sur
une cale pi\'{e}zo\'{e}lectrique et d'un miroir de fond sph\'{e}rique ($%
R=1~m $) de grande r\'{e}flectivit\'{e}. Les coefficients de r\'{e}flexion
des deux miroirs sont respectivement \'{e}gaux \`{a} $98\%$ et $99.9\%$. La
cavit\'{e} est mont\'{e}e sur un barreau en invar (de coefficient de
dilatation tr\`{e}s faible) de $29~cm$ de long. Elle est isol\'{e}e des
vibrations m\'{e}caniques de la table gr\^{a}ce \`{a} une boite cylindrique
en laiton \`{a} l'int\'{e}rieur de laquelle la cavit\'{e} est suspendue par
des tiges \'{e}lastiques. L'intervalle spectral libre de la cavit\'{e} est
de $513~MHz$ et sa finesse est \'{e}gale \`{a} $290$.

Une petite partie du faisceau est pr\'{e}lev\'{e}e \`{a} la sortie du laser
par une lame pour \^{e}tre envoy\'{e}e dans la cavit\'{e} FPE (figure \ref
{Fig_4fpeasse}). Pour adapter spatialement le faisceau laser au mode
fondamental $TEM_{00}$ de la cavit\'{e} FPE, on utilise un syst\`{e}me de
deux lentilles convergentes. Le but de l'adaptation est de transformer le
col du laser $w_{L}=0.6~mm$ en un col $w_{FPE}$ qui correspond au mode
fondamental de la cavit\'{e} FPE. D'apr\`{e}s les caract\'{e}ristiques
g\'{e}om\'{e}triques de la cavit\'{e} FPE, ce col se trouve au niveau du
miroir plan et il est \'{e}gal \`{a} $0.34~mm$ (\'{e}quation \ref{3}). En
fonction de l'emplacement de la cavit\'{e} FPE par rapport au laser titane
saphir, nous avons choisi une premi\`{e}re lentille de focale $75~mm$
plac\'{e}e \`{a} $400~mm$ du col $w_{L}$. La seconde lentille de focale $%
105~mm$ est plac\'{e}e $200~mm$ plus loin sur un support microm\'{e}trique
qui permet d'ajuster la position de la lentille selon l'axe de propagation.
En variant la distance entre les deux lentilles, on change \`{a} la fois la
taille et la position du col image. Cependant, le choix des focales est tel
que seul la taille du col image varie de fa\c{c}on appr\'{e}ciable lorsqu'on
translate la seconde lentille. L'alignement du faisceau sur la cavit\'{e}
est r\'{e}alis\'{e} \`{a} l'aide de deux miroirs mont\'{e}s sur des supports
microm\'{e}triques ($M_{1}$, $M_{2}$ sur la figure \ref{Fig_4fpeasse}). En
jouant sur la distance entre les deux lentilles et l'alignement du faisceau,
on arrive \`{a} adapter assez pr\'{e}cis\'{e}ment le faisceau sur la
cavit\'{e}, puisque le recouvrement du faisceau avec le mode fondamental de
la cavit\'{e} est de l'ordre de $99\%$.

Avant d'\^{e}tre envoy\'{e} dans la cavit\'{e} FPE, le faisceau traverse un
dispositif constitu\'{e} d'un cube s\'{e}parateur de polarisation et d'une
lame $\lambda /4$ (figure \ref{Fig_4fpeasse}). Le faisceau incident sur la
cavit\'{e} a ainsi une polarisation circulaire. Le faisceau r\'{e}fl\'{e}chi
retourne sur le cube avec une polarisation lin\'{e}aire orthogonale \`{a} la
polarisation incidente et il est \'{e}ject\'{e} par le cube. Ce dispositif
permet donc de s\'{e}parer le faisceau r\'{e}fl\'{e}chi du faisceau
incident, afin de mesurer l'intensit\'{e} r\'{e}fl\'{e}chie \`{a} l'aide
d'une photodiode.

\subsubsection{Signal d'erreur produit par la m\'{e}thode des bandes
lat\'{e}rales\label{IV-2-2-2}}

\begin{figure}[tbp]
\centerline{\psfig{figure=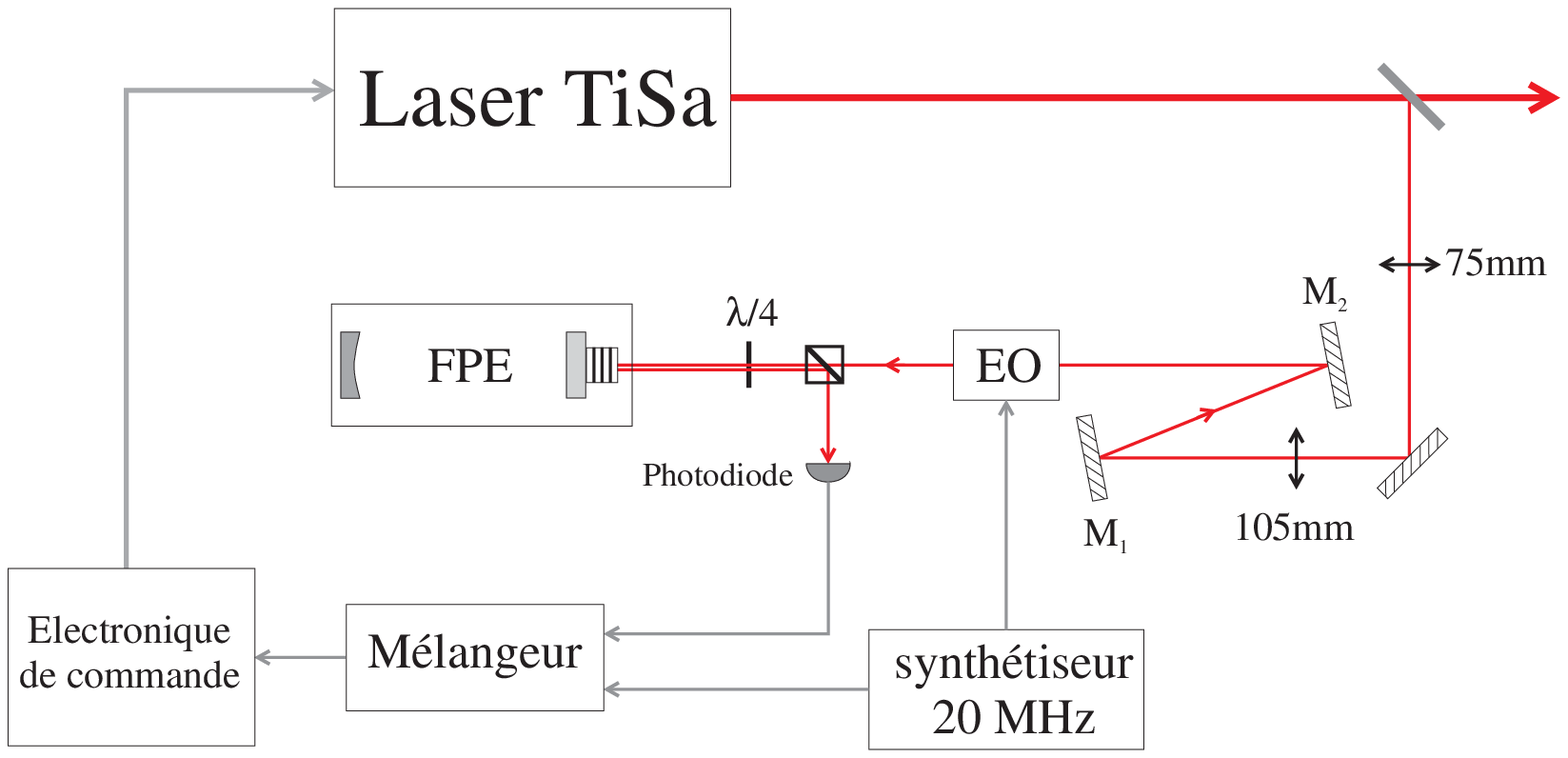,height=7cm}}
\caption{Sch\'{e}ma de l'asservissement en fr\'{e}quence du laser titane
saphir sur la cavit\'{e} FPE}
\label{Fig_4fpeasse}
\end{figure}

Le principe de l'asservissement par bandes lat\'{e}rales est
sch\'{e}matis\'{e} sur la figure \ref{Fig_4fpeasse}. Il consiste \`{a}
cr\'{e}er une courbe en dispersion ayant \`{a} la fois une pente importante
au voisinage de la r\'{e}sonance et une large plage de capture en
fr\'{e}quence. Le faisceau incident traverse un modulateur
\'{e}lectro-optique (EO) pilot\'{e} par un synth\'{e}tiseur afin de moduler
la phase du faisceau (de fr\'{e}quence optique $\omega _{L}$) \`{a} une
fr\'{e}quence $\omega _{m}$ \'{e}gale \`{a} $20~MHz$. La fr\'{e}quence de
modulation $\omega _{m}$ est choisie de fa\c{c}on \`{a} \^{e}tre grande par
rapport \`{a} la bande passante de la cavit\'{e} FPE et petite devant
l'intervalle entre modes transverses de la cavit\'{e} ($90~MHz$). Lorsque
l'amplitude de modulation est suffisamment faible, on obtient \`{a}
l'entr\'{e}e de la cavit\'{e} un champ qui pr\'{e}sente deux raies
lat\'{e}rales de fr\'{e}quences $\omega _{L}\pm \omega _{m}$ en quadrature
de phase par rapport \`{a} la porteuse de fr\'{e}quence $\omega _{L}$ ($%
\alpha _{0}$ et $\alpha _{1}$ sont suppos\'{e}s r\'{e}els): 
\begin{equation}
\bar{\alpha}^{in}(t)=\alpha _{0}~e^{i\omega _{L}t}+i~\alpha _{1}~\left\{
e^{i\left( \omega _{L}+\omega _{m}\right) t}+e^{i\left( \omega _{L}-\omega
_{m}\right) t}\right\}  \label{4.2.1}
\end{equation}
Au voisinage de la r\'{e}sonance, la porteuse est r\'{e}fl\'{e}chie avec un
coefficient de r\'{e}flexion $r\left( \bar{\Psi}\right) $ qui d\'{e}pend de
son d\'{e}saccord $\bar{\Psi}$ avec la cavit\'{e} et dont l'expression est
donn\'{e}e par l'\'{e}quation (\ref{8}). Les deux bandes lat\'{e}rales sont
quand \`{a} elles directement r\'{e}fl\'{e}chies puisqu'elles se trouvent en
dehors de la r\'{e}sonance. Le champ r\'{e}fl\'{e}chi par la cavit\'{e}
s'\'{e}crit alors: 
\begin{equation}
\bar{\alpha}^{out}(t)=r\left( \bar{\Psi}\right) ~\alpha _{0}~e^{i\omega
_{L}t}-i~\alpha _{1}~\left\{ e^{i\left( \omega _{L}+\omega _{m}\right)
t}+e^{i\left( \omega _{L}-\omega _{m}\right) t}\right\}  \label{4.2.2}
\end{equation}

Le faisceau r\'{e}fl\'{e}chi est d\'{e}tect\'{e} par une photodiode qui
fournit un signal proportionnel \`{a} l'intensit\'{e} du champ $\bar{\alpha}%
^{out}$. Le signal est amplifi\'{e} puis filtr\'{e}, pour \'{e}liminer les
\'{e}ventuelles harmoniques de la modulation \`{a} $20~MHz$, avant
d'\^{e}tre envoy\'{e} sur un m\'{e}langeur qui d\'{e}module le signal \`{a}
la fr\'{e}quence $\omega _{m}$ en utilisant comme signal de
r\'{e}f\'{e}rence la modulation utilis\'{e}e pour piloter
l'\'{e}lectro-optique (figure \ref{Fig_4fpeasse}). La phase relative entre
les deux entr\'{e}es du m\'{e}langeur est ajust\'{e}e en modifiant la
longueur du c\^{a}ble coaxial reliant le synth\'{e}tiseur au m\'{e}langeur.
Le signal \`{a} la sortie du m\'{e}langeur est alors proportionnel \`{a} la
composante de fr\'{e}quence $\omega _{m}$ de l'intensit\'{e} du faisceau
r\'{e}fl\'{e}chi, c'est \`{a} dire au battement entre les bandes
lat\'{e}rales et la porteuse: 
\begin{equation}
\bar{I}^{out}\left[ \omega _{m}\right] =4\alpha _{0}\alpha _{1}~{\cal I}%
m\left\{ r\left( \bar{\Psi}\right) \right\} =4\alpha _{0}\alpha _{1}\frac{%
\bar{\Psi}T_{1}}{\gamma ^{2}+\bar{\Psi}^{2}}  \label{4.2.3}
\end{equation}
o\`{u} $T_{1}$ et $2\gamma $ sont respectivement la transmission du coupleur
d'entr\'{e}e et les pertes totales de la cavit\'{e}. Cette expression montre
que le signal d'erreur reproduit en fonction du d\'{e}saccord $\bar{\Psi}$
une courbe de dispersion qui s'annule \`{a} r\'{e}sonance en changeant de
signe. Le signal d'erreur obtenu \`{a} la sortie du m\'{e}langeur est \`{a}
nouveau filtr\'{e} (filtre passe-bas \`{a} $5~MHz$) pour \'{e}liminer tout
r\'{e}sidu \`{a} la fr\'{e}quence de modulation $\omega _{m}$.

\begin{figure}[tbp]
\centerline{\psfig{figure=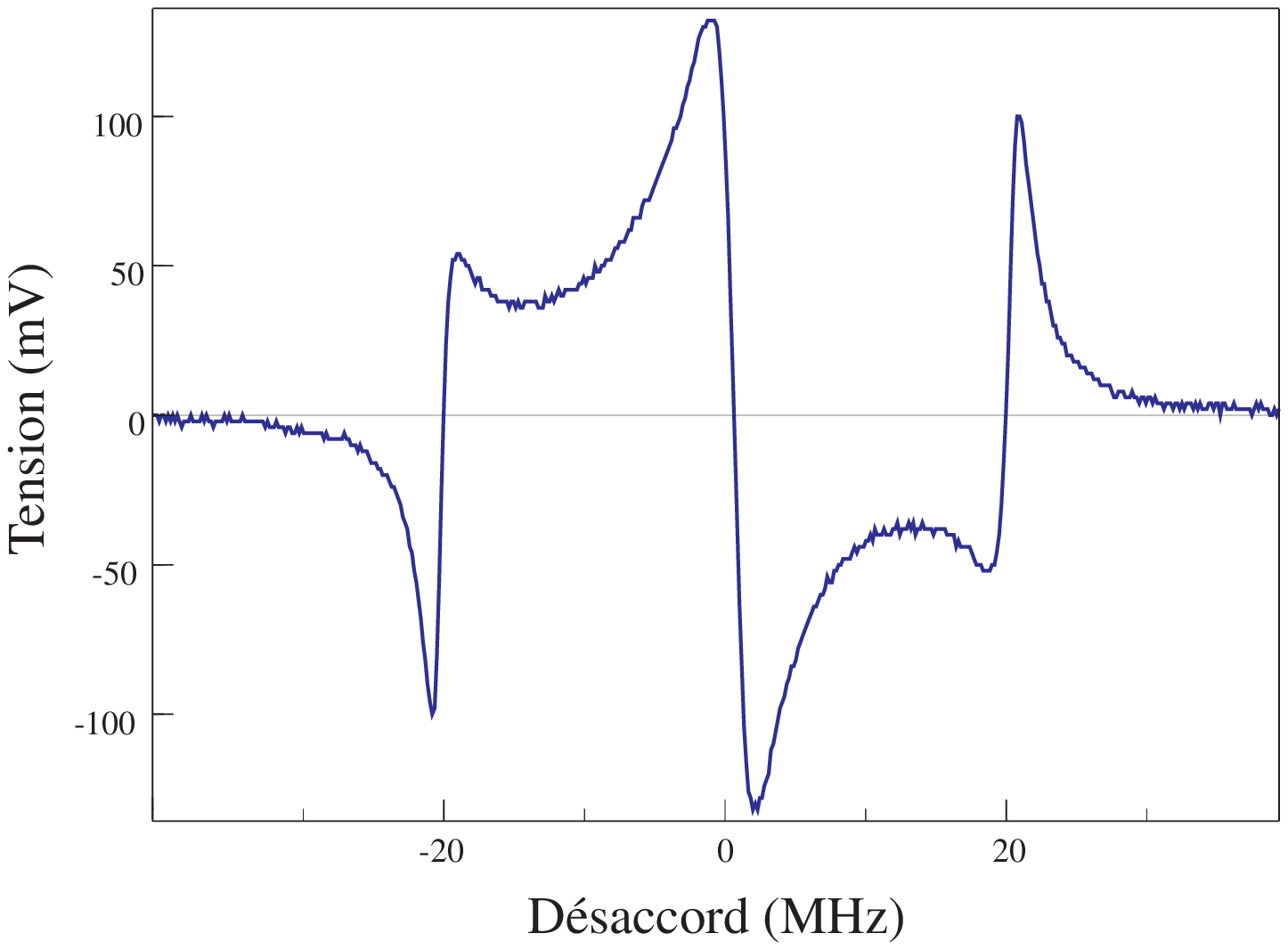,height=8cm}}
\caption{Signal d'erreur obtenu en r\'{e}flexion sur la cavit\'{e} FPE}
\label{Fig_4signerr}
\end{figure}

En pratique, le modulateur \'{e}lectro-optique est un mod\`{e}le New Focus $%
4001M$ r\'{e}sonnant \`{a} $20~MHz$, mont\'{e} sur une platine de
positionnement New Focus $9071$ de fa\c{c}on \`{a} l'aligner
pr\'{e}cis\'{e}ment sur le faisceau. Le synth\'{e}tiseur et le m\'{e}langeur
ont \'{e}t\'{e} r\'{e}alis\'{e}s au laboratoire \`{a} partir
d'\'{e}l\'{e}ments Mini Circuit. Le synth\'{e}tiseur fournit une puissance
de modulation de $0~dBm$ sur chacune de ses voies, la voie pilotant
l'\'{e}lectro-optique \'{e}tant r\'{e}glable de fa\c{c}on \`{a} ajuster la
profondeur de modulation. Le bloc photodiode est un syst\`{e}me rapide et
\`{a} faible bruit compos\'{e} d'une photodiode $FND100$ ($EG\&G$) suivi par
un pr\'{e}amplificateur transimp\'{e}dance architectur\'{e} autour d'un $%
CLC425$. Ce dispositif est similaire \`{a} ceux utilis\'{e}s pour la
d\'{e}tection du faisceau r\'{e}fl\'{e}chi par la cavit\'{e} \`{a} miroir
mobile; il est pr\'{e}sent\'{e} plus en d\'{e}tail dans la section \ref
{IV-3-3}.

La figure \ref{Fig_4signerr} repr\'{e}sente le signal \`{a} la sortie du
m\'{e}langeur lorsque la longueur de la cavit\'{e} FPE est balay\'{e}e
autour de la r\'{e}sonance, \`{a} l'aide d'une rampe de tension
appliqu\'{e}e sur la cale pi\'{e}zo\'{e}lectrique du miroir plan.
L'\'{e}cart de $2\times 20~MHz$ entre les bandes lat\'{e}rales permet
d'\'{e}talonner l'axe horizontal. On constate sur cette figure que le signal
d'erreur pr\'{e}sente une tr\`{e}s grande pente au voisinage de la
r\'{e}sonance (environ $1~mV$ pour $7~kHz$). Ainsi le signal d'erreur est
tr\`{e}s sensible \`{a} des petites fluctuations de fr\'{e}quence autour de
la r\'{e}sonance. D'autre part, la plage de capture est tr\`{e}s large, de
l'ordre de $\pm 20~MHz$ autour de la r\'{e}sonance. On est ainsi assur\'{e}
que l'asservissement sera capable de ramener la fr\'{e}quence du laser \`{a}
r\'{e}sonance avec la cavit\'{e} FPE d\`{e}s que le d\'{e}saccord est
inf\'{e}rieur \`{a} $20~MHz$.

\subsubsection{Stabilisation en fr\'{e}quence du laser\label{IV-2-2-3}}

Le signal d'erreur est utilis\'{e} pour agir sur la fr\'{e}quence du laser
titane saphir de fa\c{c}on \`{a} maintenir celle-ci \`{a} r\'{e}sonance avec
la cavit\'{e} FPE. On agit en fait sur la longueur de la cavit\'{e} laser en
pilotant la cale pi\'{e}zo\'{e}lectrique du miroir $M_{4}$ (figure \ref
{Fig_4tisa}). Pour corriger des fluctuations \`{a} haute fr\'{e}quence, on
agit aussi sur l'\'{e}lectro-optique interne du laser (\ding{178} sur la
figure \ref{Fig_4tisa}), ce qui a pour effet de modifier la longueur optique
de la cavit\'{e}.

\begin{figure}[tbp]
\centerline{\psfig{figure=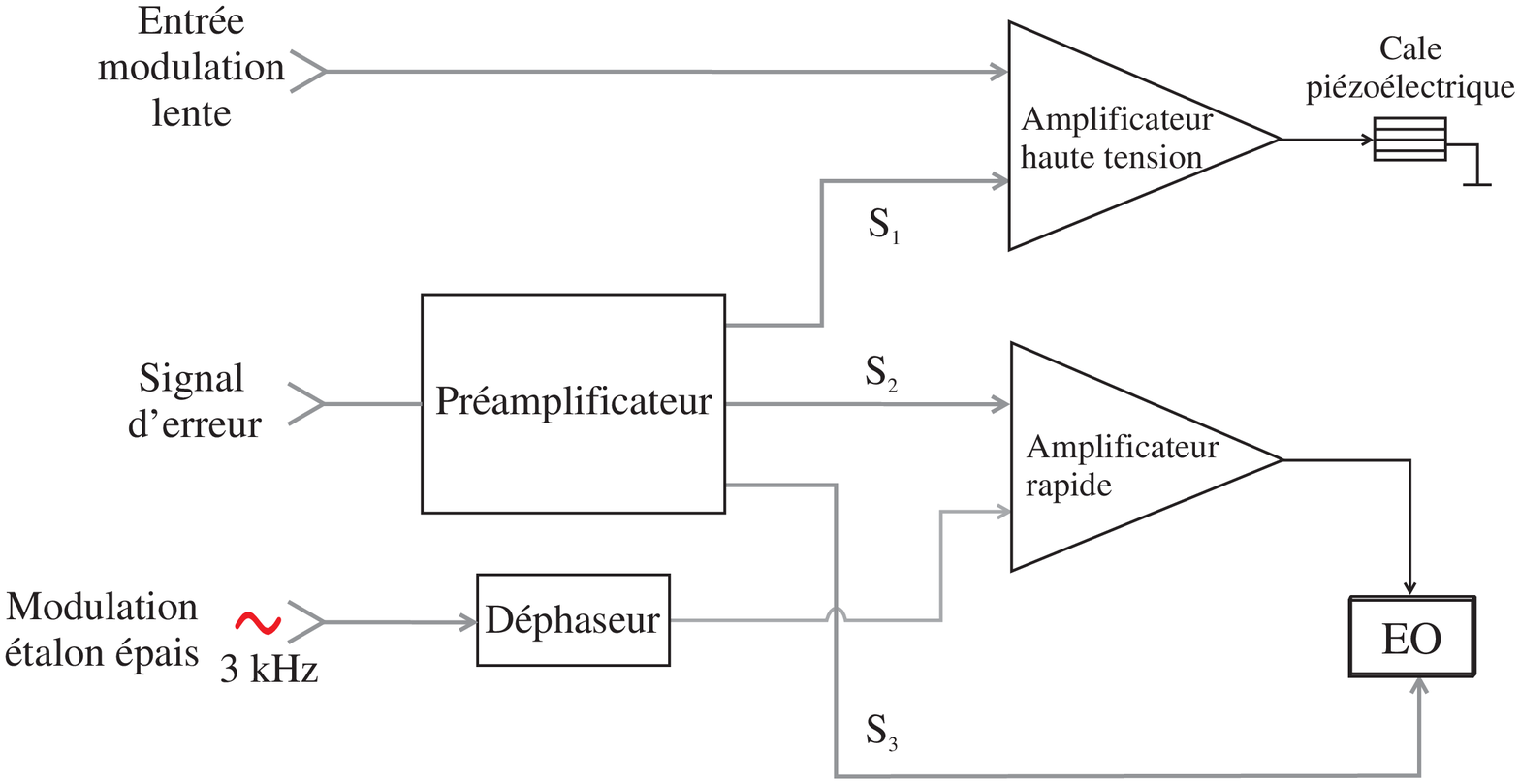,height=8cm}}
\caption{Sch\'{e}ma de principe de l'asservissement en fr\'{e}quence du
laser titane saphir. Pour contr\^{o}ler la fr\'{e}quence du laser, on agit
sur le laser en modifiant directement sa longueur par l'interm\'{e}diaire de
la cale pi\'{e}zo\'{e}lectrique du miroir $M_{4}$ (signal $S_{1}$). On
modifie aussi sa longueur optique en agissant sur les deux voies de
l'\'{e}lectro-optique interne du titane saphir (signaux $S_{2}$ et $S_{3}$)}
\label{Fig_4prinass}
\end{figure}

Afin de r\'{e}aliser un asservissement efficace sur une large plage de
fr\'{e}quence, nous avons construit un dispositif \`{a} trois boucles
d'asservissement en parall\`{e}le. Un sch\'{e}ma g\'{e}n\'{e}ral de
l'\'{e}lectronique de commande est pr\'{e}sent\'{e} sur la figure \ref
{Fig_4prinass}. Le signal d'erreur obtenu \`{a} l'aide de la m\'{e}thode des
bandes lat\'{e}rales est envoy\'{e} dans un pr\'{e}amplificateur qui fournit
trois signaux $S_{1}$, $S_{2}$ et $S_{3}$. La voie lente $S_{1}$ pilote la
cale pi\'{e}zo\'{e}lectrique du miroir $M_{4}$ du titane saphir, par
l'interm\'{e}diaire d'un amplificateur haute tension capable de fournir de
grandes amplitudes ($0$-$1000~volts$). Du fait du poids du miroir $M_{4}$,
la fr\'{e}quence de r\'{e}sonance de cette voie se situe aux alentours du
kilohertz. Cette boucle d'asservissement permet donc de corriger \`{a} basse
fr\'{e}quence d'importantes variations de longueur de la cavit\'{e} laser.
Cet amplificateur a aussi une entr\'{e}e qui permet de moduler \`{a} faible
vitesse la fr\'{e}quence du laser (entr\'{e}e modulation lente sur la figure 
\ref{Fig_4prinass}). Comme expliqu\'{e} dans l'introduction de cette partie,
cette entr\'{e}e est utilis\'{e}e pour parcourir le pic d'Airy de la
cavit\'{e} \`{a} miroir mobile. Bien s\^{u}r, cette entr\'{e}e n'est
utilis\'{e}e que lorsque le laser n'est pas asservi sur la cavit\'{e} FPE. 

La voie interm\'{e}diaire $S_{2}$ est envoy\'{e}e sur un amplificateur
rapide qui pilote l'une des deux voies de l'\'{e}lectro-optique du titane
saphir. Cet amplificateur est capable de r\'{e}aliser des excursions en
tension de $\pm 200~V$ \`{a} des fr\'{e}quences allant jusqu'\`{a} $100~kHz$%
. Cet asservissement permet aussi de compenser la modulation de
fr\'{e}quence induite par l'asservissement de l'\'{e}talon \'{e}pais. Nous
avons vu (section \ref{IV-2-1}) que la longueur de l'\'{e}talon \'{e}pais
est modul\'{e}e \`{a} une fr\'{e}quence de $3~kHz$ afin de produire une
modulation d'intensit\'{e} \`{a} la sortie du laser. Ce signal pilote
l'asservissement de l'\'{e}talon \'{e}pais et s'annule lorsqu'il est \`{a}
r\'{e}sonance avec le laser. Il reste n\'{e}anmoins une petite modulation de
fr\'{e}quence du laser \`{a} $3~kHz$ car la modulation de longueur de
l'\'{e}talon \'{e}pais est en fait \'{e}quivalente \`{a} une variation de la
longueur optique de la cavit\'{e} laser. Cet effet est compens\'{e} en
appliquant sur l'\'{e}lectro-optique la modulation de r\'{e}f\'{e}rence
\`{a} $3~kHz$ fourni par la d\'{e}tection synchrone. Ce signal est
auparavant pr\'{e}cis\'{e}ment ajust\'{e} en phase et en amplitude \`{a}
l'aide d'un d\'{e}phaseur de gain variable.

Enfin, la voie rapide $S_{3}$ pilote directement la seconde voie de
l'\'{e}lectro-optique. Elle permet de contr\^{o}ler des fluctuations de
faible amplitude ($\pm 10~V$ maximum \`{a} l'entr\'{e}e de
l'\'{e}lectro-optique) jusqu'\`{a} des fr\'{e}quences de l'ordre du
m\'{e}gahertz.

\begin{figure}[tbp]
\centerline{\psfig{figure=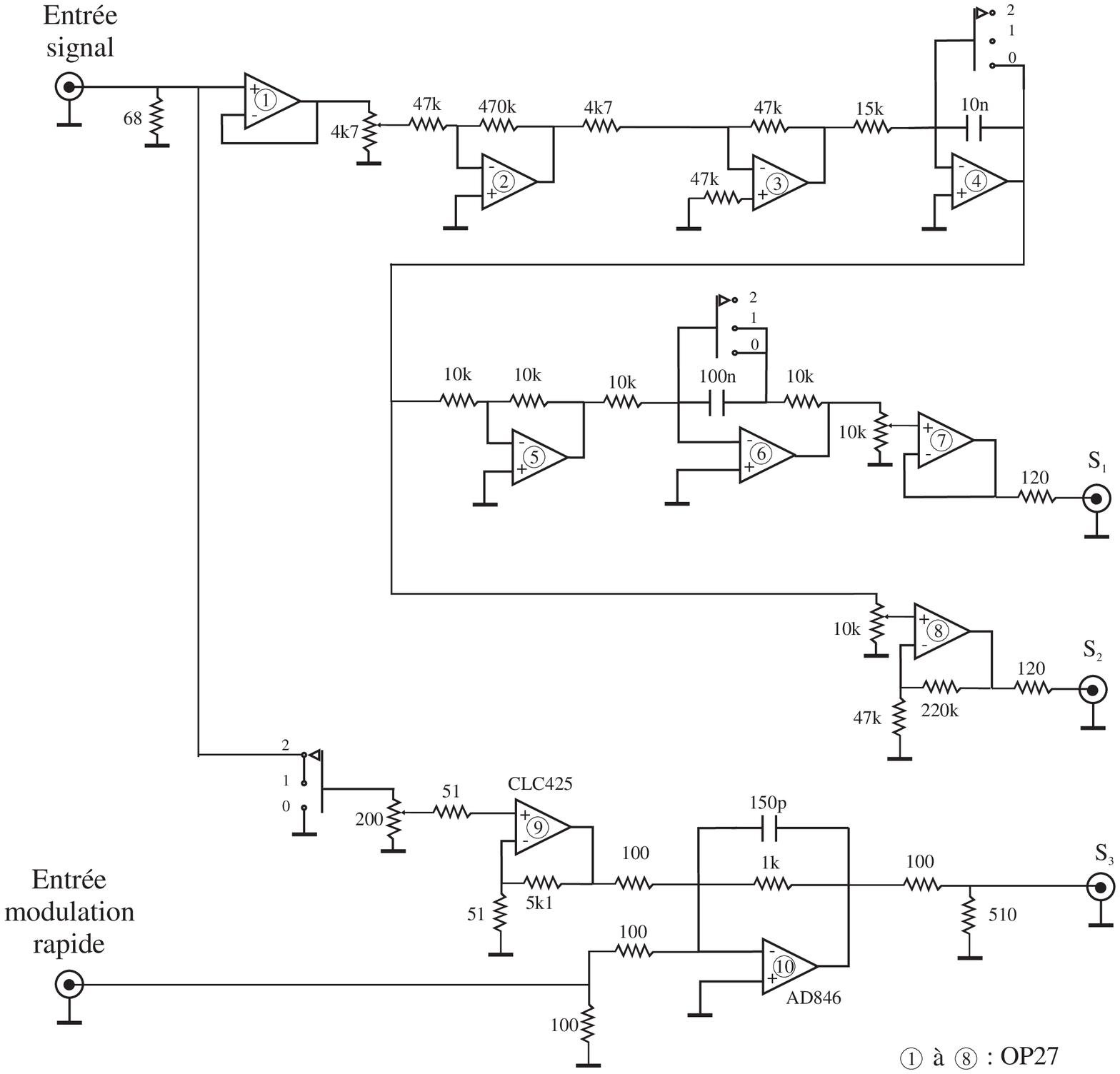,height=145mm}}
\caption{Sch\'{e}ma du syst\`{e}me de pr\'{e}amplification qui fournit,
\`{a} partir du signal d'erreur, trois signaux $S_{1}$, $S_{2}$ et $S_{3}$
qui constituent respectivement les voies lente, interm\'{e}diaire et rapide
de l'asservissement en fr\'{e}quence du laser titane saphir. Une seconde
entr\'{e}e est pr\'{e}vue pour appliquer une rampe de modulation rapide de
la fr\'{e}quence du laser}
\label{Fig_4alhp}
\end{figure}

Le pr\'{e}amplificateur d\'{e}termine les fonctions de transfert de chacune
des trois boucles d'asservissement. La figure \ref{Fig_4alhp} montre le
sch\'{e}ma du syst\`{e}me de pr\'{e}amplification. La voie rapide $S_{3}$
est obtenue en amplifiant le signal d'erreur \`{a} l'aide de deux \'{e}tages
amplificateurs. Le premier \'{e}tage utilise un $CLC425$ (\ding{180}) qui
assure une amplification de gain $100$ \`{a} faible bruit et sur une large
bande de fr\'{e}quence. Cet amplificateur op\'{e}rationnel pr\'{e}sente
n\'{e}anmoins l'inconv\'{e}nient de saturer pour de faibles tensions. C'est
pourquoi nous utilisons dans le second \'{e}tage un $AD846$ (\ding{181}) en
gain $10$ qui permet d'obtenir \`{a} la sortie des tensions suffisamment
\'{e}lev\'{e}es pour piloter l'\'{e}lectro-optique du laser. Ces deux
\'{e}tages ont un gain plat en fr\'{e}quence jusqu'\`{a} des fr\'{e}quences
de l'ordre du m\'{e}gahertz. La fonction de transfert de la voie rapide se
comporte toutefois comme un filtre passe-bas puisque l'\'{e}lectro-optique
est \'{e}quivalent \`{a} un condensateur de $3~nF$. La fr\'{e}quence de
coupure de ce filtre est de l'ordre de $600~kHz$. Ceci permet de r\'{e}gler
le gain global de cette voie de mani\`{e}re \`{a} ce que l'asservissement
agisse au dessous de $600~kHz$, en \'{e}vitant l'entr\'{e}e en oscillation
aux alentours du m\'{e}gahertz.

Pour les deux autres voies, le signal d'erreur est amplifi\'{e} (\'{e}tages %
\ding{172} \`{a} \ding{174}) puis int\'{e}gr\'{e} (\'{e}tage \ding{175}), ce
qui assure une pente constante de $-6~dB/octave$. La voie lente subit en
plus un filtrage passe-bas \`{a} partir de $150~Hz$, du fait de la
r\'{e}sistance de sortie de l'amplificateur haute tension ($100~k\Omega $)
et de la capacit\'{e} de la cale pi\'{e}zo\'{e}lectrique ($10~nF$). Ces
diff\'{e}rentes r\'{e}ponses en fr\'{e}quence et les potentiom\`{e}tres
plac\'{e}s dans le pr\'{e}amplificateur permettent d'assurer un
fonctionnement correct des trois voies de l'asservissement : la voie lente
est dominante \`{a} basse fr\'{e}quence ($0$-$200~Hz$), la voie
interm\'{e}diaire agit de $200~Hz$ \`{a} $20~kHz$ et la voie rapide est
pr\'{e}pond\'{e}rante au del\`{a} de $20~kHz$. 

Un commutateur double \`{a} trois positions permet de commander
l'asservissement. En position $0$, les trois boucles sont ouvertes : le
laser est non asservi. En position $1$, toutes les boucles sont ferm\'{e}es.
La position $2$ de l'interrupteur rajoute au signal $S_{1}$ un
int\'{e}grateur agissant jusqu'\`{a} $150~Hz$. La pr\'{e}sence de cet
int\'{e}grateur assure un bon fonctionnement de l'asservissement \`{a} basse
fr\'{e}quence en augmentant le gain de la boucle lente.

Notons enfin la pr\'{e}sence d'une entr\'{e}e modulation sur la voie rapide
(figure \ref{Fig_4alhp}). Cette entr\'{e}e permet de moduler la
fr\'{e}quence du laser sur une petite amplitude (typiquement quelques
kilohertz) mais \`{a} grande vitesse : comme le signal est directement
envoy\'{e} sur l'\'{e}lectro-optique du laser, il est possible d'appliquer
une modulation dont la fr\'{e}quence est de quelques m\'{e}gahertz. Nous
verrons dans le chapitre $5$ que cette entr\'{e}e est utilis\'{e}e pour
calibrer les mesures de petits d\'{e}placements r\'{e}alis\'{e}es par la
cavit\'{e} \`{a} miroir mobile.

\begin{figure}[tbp]
\centerline{\psfig{figure=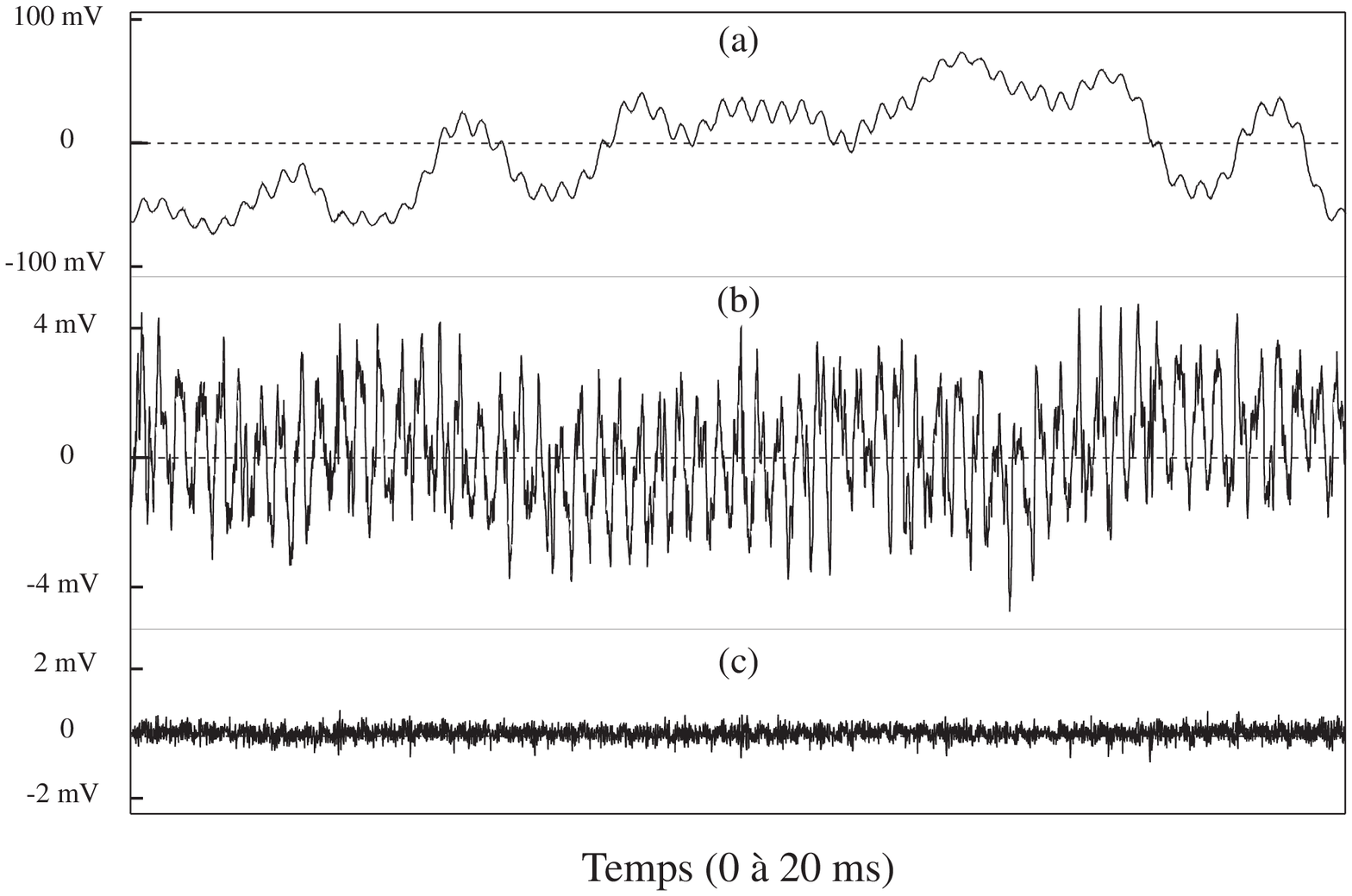,height=8cm}}
\caption{ Evolution temporelle du signal d'erreur. Le signal d'erreur (a)
est obtenu au voisinage de la r\'{e}sonance du FPE lorsque l'asservissement
est d\'{e}sactiv\'{e}. Les fluctuations du laser sont de l'ordre de $300~kHz$
rms. Le signal (b) est obtenu en activant les voies lente et
interm\'{e}diaire, ce qui r\'{e}duit les fluctuations \`{a} $30~kHz$ rms.
Lorsque la voie rapide est activ\'{e}e (courbe c), le signal d'erreur
fluctue tr\`{e}s peu (fluctuations de l'ordre de $4~kHz$ rms)}
\label{Fig_4signer2}
\end{figure}

L'efficacit\'{e} de l'asservissement est illustr\'{e}e sur la figure \ref
{Fig_4signer2}. La courbe (a) montre le signal d'erreur \`{a} r\'{e}sonance
lorsque les trois voies de l'asservissement sont d\'{e}sactiv\'{e}es.
L'amplitude de variation du signal traduit les fluctuations de fr\'{e}quence
du laser ({\it jitter}) qui peuvent \^{e}tre \'{e}valu\'{e}es \`{a} partir
de la sensibilit\'{e} du signal d'erreur qui est de $7~kHz/mV$ : ces
fluctuations sont de l'ordre de $300~kHz$ $rms$. L'oscillation visible sur
la courbe (a) est due \`{a} la modulation \`{a} $3~kHz$ de l'\'{e}talon
\'{e}pais. La courbe (b) montre le signal d'erreur lorsque les voies lente
et moyenne sont activ\'{e}es. On voit que le bruit est notablement
r\'{e}duit puisque l'amplitude du signal d'erreur reste inf\'{e}rieure \`{a} 
$5~mV$ $rms$. Ceci correspond \`{a} des fluctuations de fr\'{e}quence de
l'ordre de $30~kHz$ $rms$. Notons par ailleurs qu'une partie importante du
signal correspond au r\'{e}sidu de la modulation de l'\'{e}talon \'{e}pais
\`{a} $3~kHz$, qui n'est pas parfaitement compens\'{e}e par le signal
appliqu\'{e} sur l'\'{e}lectro-optique \`{a} travers le d\'{e}phaseur
(figure \ref{Fig_4prinass}). 

Le gain global est limit\'{e} par la r\'{e}sonance \`{a} $100~kHz$ de la
voie interm\'{e}diaire et il n'est pas possible de l'augmenter pour
am\'{e}liorer l'efficacit\'{e} de l'asservissement. La mise en service de la
voie rapide permet d'augmenter nettement le gain global des voies lente et
interm\'{e}diaire puisque la fr\'{e}quence d'oscillation de l'asservissement
est repouss\'{e} \`{a} $1~MHz$. Lorsque les gains des trois boucles sont
optimis\'{e}s, on obtient le signal d'erreur repr\'{e}sent\'{e} sur la
courbe (c) de la figure \ref{Fig_4signer2}. L'amplitude des fluctuations est
de l'ordre de $0.5~mV$ $rms$, ce qui correspond \`{a} des fluctuations de
fr\'{e}quence inf\'{e}rieures \`{a} $4~kHz$ $rms$. On constate aussi que
l'asservissement supprime totalement la modulation de l'\'{e}talon \'{e}pais
qui \'{e}tait encore pr\'{e}sente sur la courbe (b).

\subsection{Stabilisation en intensit\'{e}\label{IV-2-3}}

Le faisceau issu du laser titane saphir pr\'{e}sente d'importantes
fluctuations d'intensit\'{e} \`{a} basse fr\'{e}quence. Celles-ci sont
essentiellement li\'{e}es aux vibrations m\'{e}caniques des diff\'{e}rents
\'{e}l\'{e}ments optiques du laser, bien que l'ensemble du laser soit
rigidement fix\'{e} \`{a} une dalle en marbre pos\'{e}e sur la table optique
par l'interm\'{e}diaire d'amortisseurs en caoutchouc. Comme nous l'avons
soulign\'{e} au d\'{e}but de cette partie, le point de fonctionnement de la
cavit\'{e} \`{a} miroir mobile d\'{e}pend de l'intensit\'{e} incidente. Il
est donc important de contr\^{o}ler pr\'{e}cis\'{e}ment cette intensit\'{e}.
C'est le r\^{o}le de l'asservissement d'intensit\'{e} qui permet de fixer
l'intensit\'{e} moyenne \`{a} une valeur donn\'{e}e et de r\'{e}duire les
fluctuations \`{a} basse fr\'{e}quence.

\subsubsection{R\'{e}gulation de l'intensit\'{e} lumineuse \`{a} basse
fr\'{e}quence\label{IV-2-3-1}}

Le principe de l'asservissement consiste \`{a} utiliser un att\'{e}nuateur
variable pilot\'{e} par une boucle \'{e}lectronique de contre-r\'{e}action
qui contr\^{o}le l'intensit\'{e} transmise par l'att\'{e}nuateur. Le
faisceau \`{a} la sortie du laser \'{e}tant polaris\'{e} lin\'{e}airement,
on utilise un \'{e}lectro-optique dont les lignes neutres sont tourn\'{e}s
de $45{{}^{\circ }}$ par rapport \`{a} la polarisation incidente. Les deux
composantes du champ qui correspondent aux projections sur les deux lignes
neutres subissent des d\'{e}phasages diff\'{e}rents, qui d\'{e}pendent de la
tension appliqu\'{e}e sur l'\'{e}lectro-optique. Le faisceau transmis par
l'\'{e}lectro-optique a donc une polarisation elliptique dont
l'ellipticit\'{e} d\'{e}pend de la tension appliqu\'{e}e. Pour r\'{e}aliser
un att\'{e}nuateur variable, il suffit de faire suivre l'\'{e}lectro-optique
par un polariseur parall\`{e}le \`{a} la polarisation du faisceau incident.
L'intensit\'{e} transmise par ce dispositif\ est donn\'{e}e par: 
\begin{equation}
I^{out}\left( V\right) =\frac{I^{in}}{2}~\left[ 1+\cos \left( \epsilon
\right) \right]  \label{4.2.4}
\end{equation}
o\`{u} $\epsilon $ est la diff\'{e}rence des d\'{e}phasages subis par les
polarisations propres, qui d\'{e}pend lin\'{e}airement de la tension
appliqu\'{e}e. Nous utilisons un \'{e}lectro-optique Gz\"{a}nger $LM202$,
mont\'{e} sur un support New Focus $9082$ qui permet d'aligner
pr\'{e}cis\'{e}ment l'\'{e}lectro-optique sur le faisceau. On obtient
l'extinction ($I^{out}\approx 0$) pour une tension appliqu\'{e}e de $-130~V$
et la transmission maximale (de l'ordre de $80\%$) pour une tension de $%
+150~V$. Ces valeurs permettent de piloter l'\'{e}lectro-optique avec une
amplificateur rapide d'excursion $\pm 200~V$, identique \`{a} celui
utilis\'{e} dans la voie interm\'{e}diaire de l'asservissement en
fr\'{e}quence du laser. On est ainsi assur\'{e} de pouvoir agir sur les
fluctuations d'intensit\'{e} jusqu'\`{a} des fr\'{e}quences de plusieurs
dizaines de kilohertz.

\begin{figure}[tbp]
\centerline{\psfig{figure=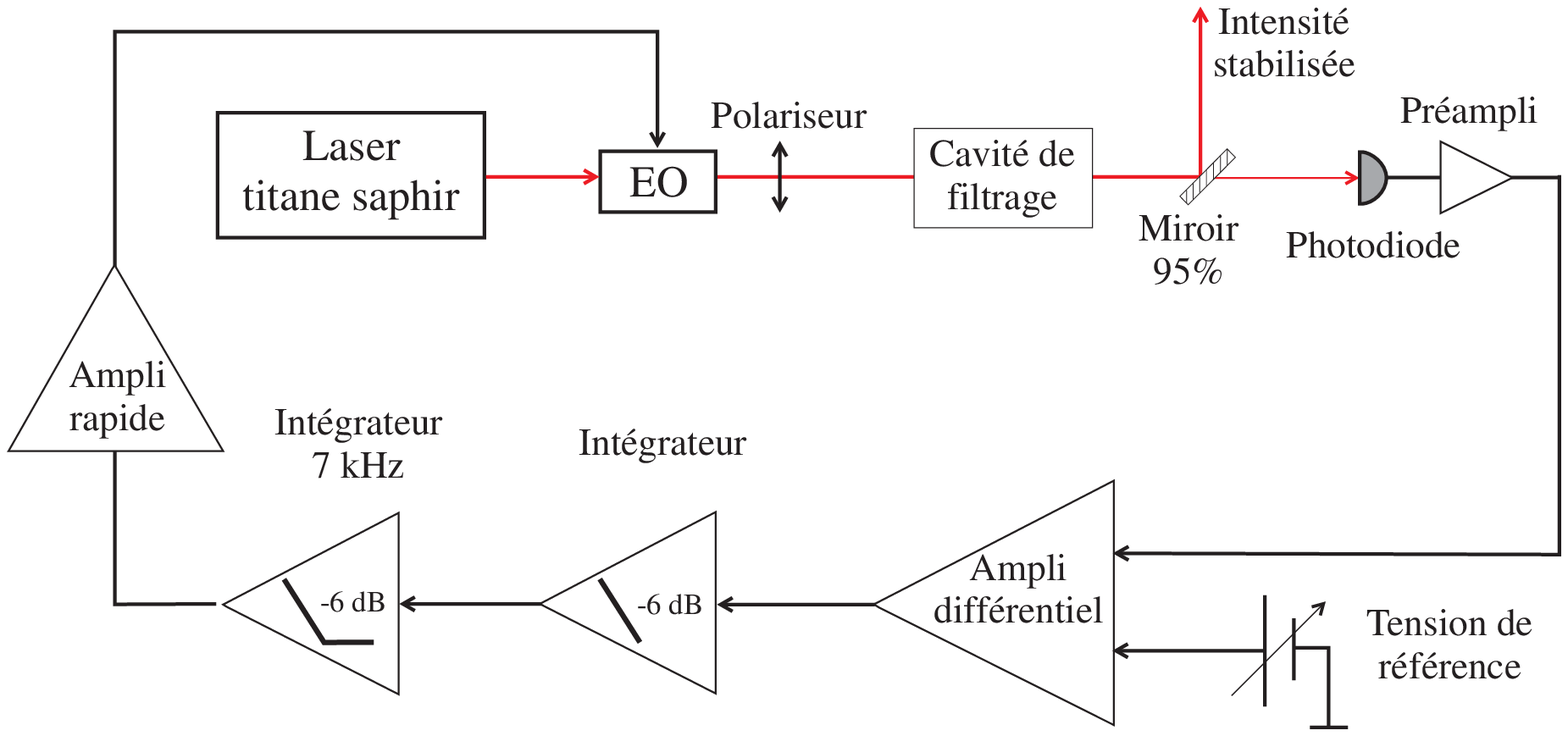,height=7cm}}
\caption{Sch\'{e}ma de l'asservissement de l'intensit\'{e} du laser titane
saphir}
\label{Fig_4assint}
\end{figure}

Le sch\'{e}ma g\'{e}n\'{e}ral du dispositif de r\'{e}gulation
d'intensit\'{e} est repr\'{e}sent\'{e} sur la figure \ref{Fig_4assint}. 
Ce dispositif comporte une partie optique, constitu\'{e}e de
l'\'{e}lectro-optique associ\'{e} \`{a} un polariseur et une partie
\'{e}lectronique pour piloter l'\'{e}lectro-optique \`{a} l'aide d'une
boucle de contre-r\'{e}action. Le signal d'erreur est fourni par une
photodiode $BPW34$ qui mesure l'intensit\'{e} d'une petite partie du
faisceau pr\'{e}lev\'{e}e \`{a} la sortie du dispositif par un miroir de
r\'{e}flexion $95\%$. A l'aide d'un amplificateur diff\'{e}rentiel, ce
signal est compar\'{e} \`{a} une tension de r\'{e}f\'{e}rence tr\`{e}s
stable dont on peut faire varier la valeur : ceci permet de modifier
l'intensit\'{e} moyenne du faisceau r\'{e}fl\'{e}chi par le miroir. La
fonction de transfert de l'\'{e}lectronique est d\'{e}termin\'{e}e par un
premier int\'{e}grateur de pente globale $-6~dB/octave$, suivi par un
deuxi\`{e}me int\'{e}grateur pour des fr\'{e}quences inf\'{e}rieures \`{a} $%
7~kHz$, qui permet d'augmenter le gain \`{a} basse fr\'{e}quence.

On peut noter sur la figure \ref{Fig_4assint} que la cavit\'{e} de filtrage,
d\'{e}crite dans la section suivante, est plac\'{e}e entre l'att\'{e}nuateur
variable et le miroir $95\%$. Ainsi l'asservissement contr\^{o}le
l'intensit\'{e} directement \`{a} la sortie du dispositif et rend celle-ci
ind\'{e}pendante des perturbations qui pourraient \^{e}tre induites par la
cavit\'{e} de filtrage.

Pour avoir un gain global suffisant de la boucle de contre-r\'{e}action, il
est pr\'{e}f\'{e}rable de faire travailler l'\'{e}lectro-optique au
voisinage de la mi-transmission. C'est en effet autour de ce point de
fonctionnement que la transmission est la plus sensible \`{a} une variation
de la tension appliqu\'{e}e (\'{e}quation \ref{4.2.4}). Ce choix du point de
fonctionnement d\'{e}pend de l'intensit\'{e} incidente sur
l'\'{e}lectro-optique et du point de consigne de l'intensit\'{e} transmise
d\'{e}termin\'{e}e par la tension de r\'{e}f\'{e}rence. Le fait de
travailler \`{a} mi-transmission pr\'{e}sente l'inconv\'{e}nient de perdre
la moiti\'{e} de la puissance lumineuse dans l'asservissement. Ceci n'est
cependant pas tr\`{e}s g\^{e}nant dans notre cas puisque la puissance
n\'{e}cessaire \`{a} la sortie du dispositif est de l'ordre de quelques
dizaines de milliwatts. Elle est donc bien inf\'{e}rieure \`{a} la puissance
disponible \`{a} la sortie du laser. D'autre part ce choix permet de faire
fonctionner l'amplificateur rapide $\pm 200~V$ dans des conditions optimales
puisque la tension moyenne de sortie reste voisine de $0~Volt$.

\begin{figure}[tbp]
\centerline{\psfig{figure=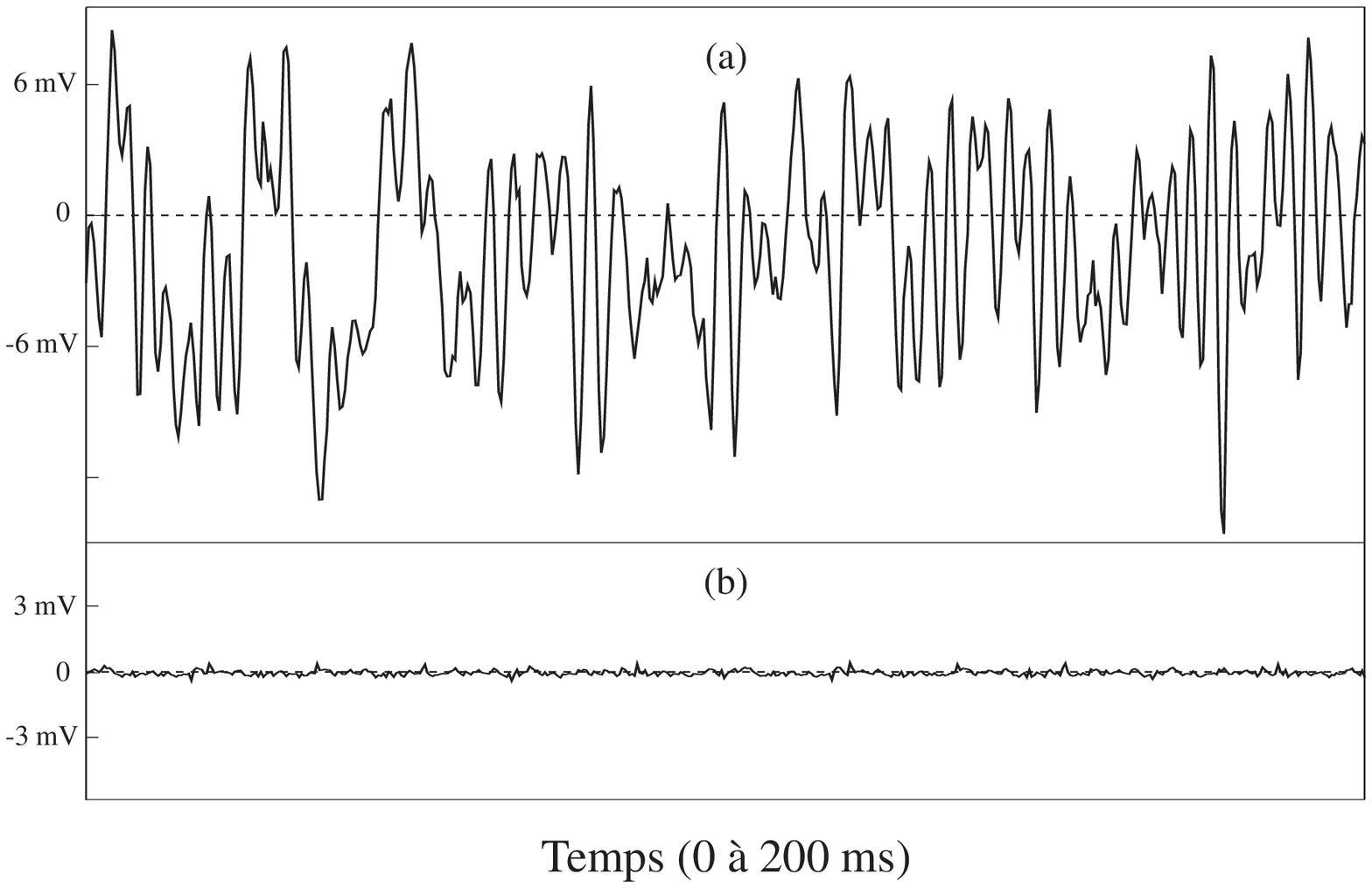,height=8cm}}
\caption{Evolution temporelle des fluctuations de l'intensit\'{e} transmise
sans asservissement (a) et avec asservissement (b). Ces traces
pr\'{e}sentent la partie alternative du signal; le niveau continu,
proportionnel \`{a} l'intensit\'{e} moyenne, est \'{e}gal \`{a} $630~mV$}
\label{Fig_4assint2}
\end{figure}

La figure \ref{Fig_4assint2} montre les fluctuations de l'intensit\'{e}
transmise autour de sa valeur moyenne, sans asservissement (trace a) et avec
asservissement (trace b). Sachant que l'intensit\'{e} moyenne transmise $%
\bar{I}$ correspond \`{a} une tension moyenne de $630~mV$, on peut estimer
les fluctuations relatives d'intensit\'{e} $\delta I/\bar{I}$ du faisceau.
On trouve que l'asservissement permet de r\'{e}duire ces fluctuations
relatives de quelques pourcent \`{a} environ $2~^{o}/_{oo}$.

\begin{figure}[tbp]
\centerline{\psfig{figure=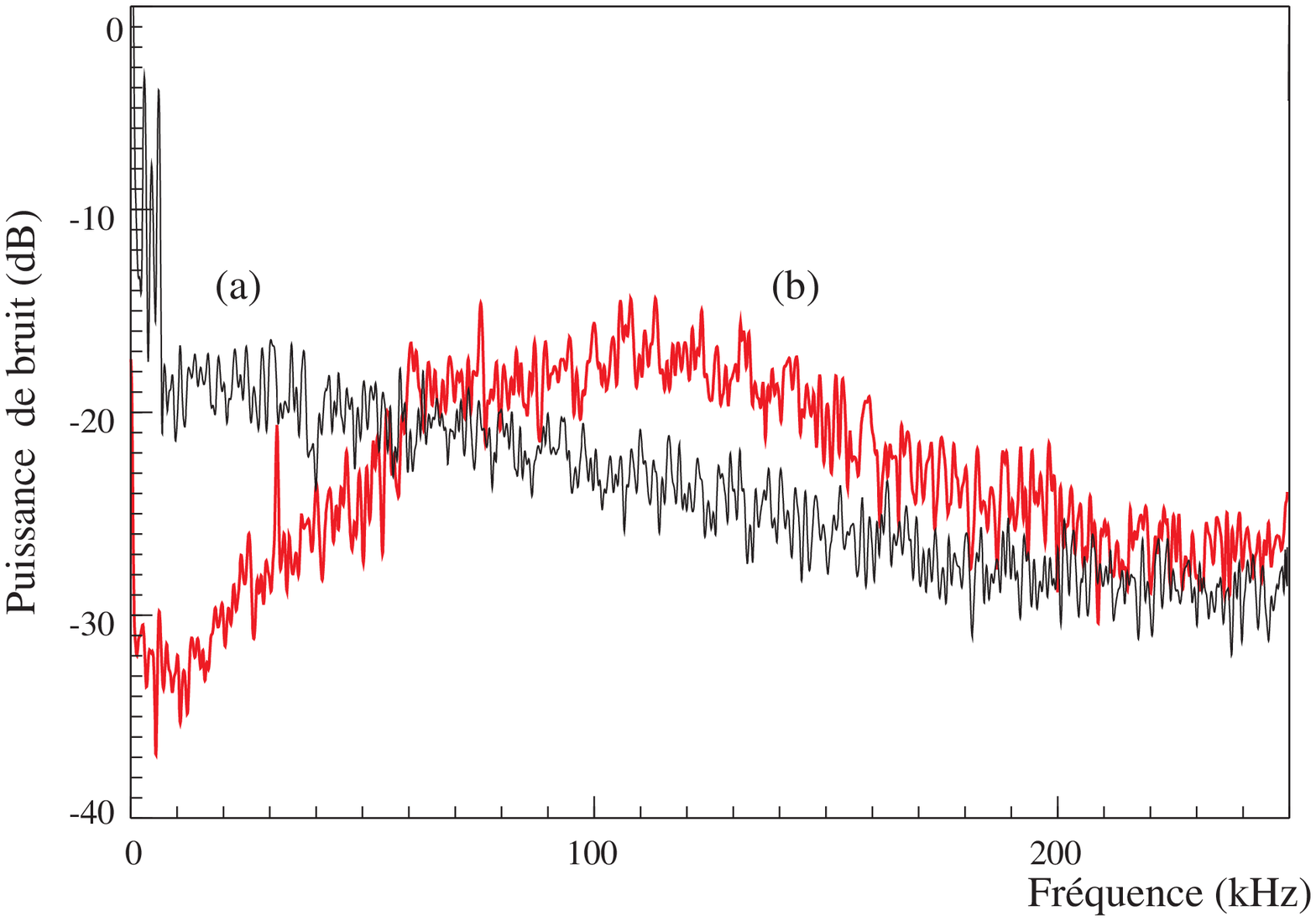,height=8cm}}
\caption{Spectres de bruit obtenus en faisant la transform\'{e}e de fourier
num\'{e}rique de l'\'{e}volution temporelle de l'intensit\'{e} transmise,
sans asservissement (a) et avec asservissement (b)}
\label{Fig_4asintdb}
\end{figure}

La figure \ref{Fig_4asintdb} montre l'action de l'asservissement en fonction
de la fr\'{e}quence. Plus pr\'{e}cis\'{e}ment, les deux courbes
repr\'{e}sentent les spectres de puissance de bruit de l'intensit\'{e}, avec
et sans asservissement. Ces courbes sont obtenues par transform\'{e}e de
Fourier num\'{e}rique ($FFT$) de l'\'{e}volution temporelle de
l'intensit\'{e} transmise. Sans asservissement (trace $a$), les fluctuations
d'intensit\'{e} sont tr\`{e}s importantes en dessous de $10~kHz$. La trace ($%
b$) montre que l'asservissement r\'{e}duit tr\`{e}s efficacement ce bruit
basse fr\'{e}quence: la r\'{e}duction est de l'ordre de $25~dB$ jusqu'\`{a} $%
10~kHz$. D'autre part, l'asservissement est efficace jusqu'\`{a} des
fr\'{e}quences sup\'{e}rieures \`{a} $50~kHz$. On observe un l\'{e}ger
exc\`{e}s de bruit au voisinage de $100~kHz$. Ce bruit est difficile \`{a}
supprimer; il est sans doute li\'{e} \`{a} des d\'{e}phasages dans la boucle
de contre-r\'{e}action qui inversent l'effet de l'asservissement dans cette
plage de fr\'{e}quence o\`{u} le gain est encore important. Le gain optimal
de l'asservissement est choisi de fa\c{c}on \`{a} r\'{e}duire efficacement
le bruit \`{a} basse fr\'{e}quence tout en ayant un exc\`{e}s de bruit
raisonnable au voisinage de $100~kHz$. On remarquera aussi que
l'asservissement ne rajoute pas de bruit \`{a} des fr\'{e}quences
sup\'{e}rieures \`{a} $200~kHz$. 

\subsubsection{Bruit technique du laser\label{IV-2-3-2}}

L'un des points importants en vue de l'\'{e}tude des effets quantiques du
couplage optom\'{e}canique est la qualit\'{e} du faisceau laser \`{a} la
fr\'{e}quence d'analyse. Comme nous l'avons soulign\'{e} au d\'{e}but de
cette section, l'intensit\'{e} du faisceau \`{a} la sortie du titane saphir
pr\'{e}sente d'importantes fluctuations \`{a} basse fr\'{e}quence li\'{e}es
\`{a} des sources de bruit technique du laser. A des fr\'{e}quences plus
\'{e}lev\'{e}es, ce bruit technique devient de moins en moins important et
les fluctuations d'intensit\'{e} se r\'{e}duisent aux fluctuations
quantiques (shot noise). Pour mettre en \'{e}vidence les corr\'{e}lations
quantiques entre l'intensit\'{e} lumineuse et la position du miroir mobile
(voir les sections 2.4.3 et 3.4.3), il est indispensable que le bruit
d'intensit\'{e} du faisceau incident sur la cavit\'{e} soit \'{e}gal au
bruit quantique standard pour des fr\'{e}quences voisines de la
fr\'{e}quence de r\'{e}sonance m\'{e}canique fondamentale (typiquement $%
2~MHz $).

\begin{figure}[tbp]
\centerline{\psfig{figure=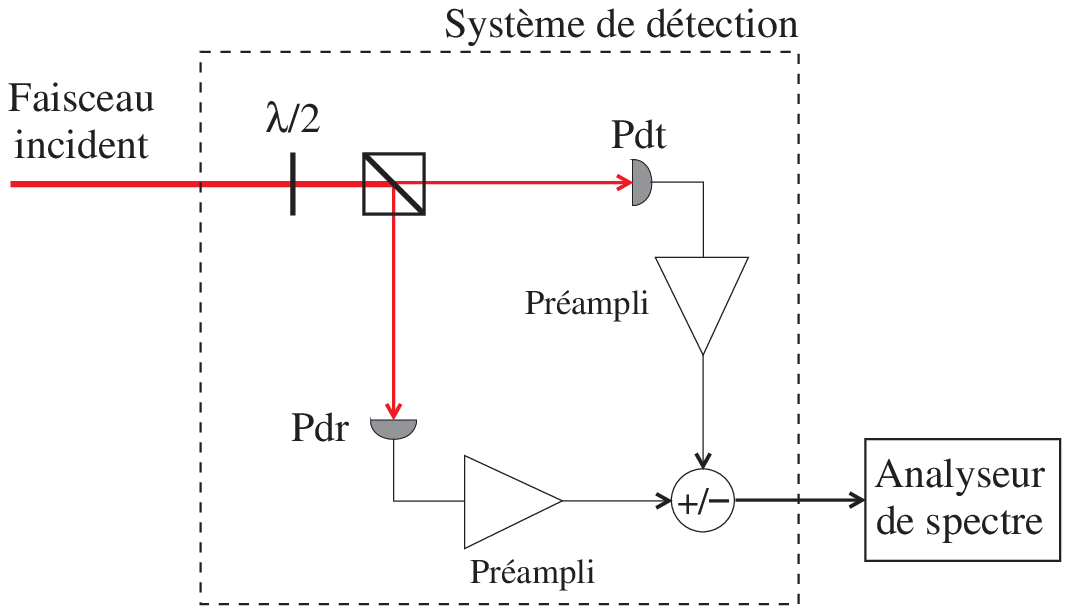,height=7cm}}
\caption{Principe de la mesure du bruit d'intensit\'{e} du faisceau laser.
Le syst\`{e}me de d\'{e}tection est constitu\'{e} d'une lame
semi-r\'{e}fl\'{e}chissante (lame demi-onde et cube s\'{e}parateur de
polarisation), de deux photodiodes et deux pr\'{e}amplificateurs
parfaitement \'{e}quilibr\'{e}s. Un dernier \'{e}tage permet d'ajouter ou de
soustraire les signaux fournis par les photodiodes. On mesure ainsi sur
l'analyseur de spectre alternativement le bruit d'intensit\'{e} total du
faisceau laser et le niveau du bruit de photon standard associ\'{e}}
\label{Fig_4brtechn}
\end{figure}

\begin{figure}[tbp]
\centerline{\psfig{figure=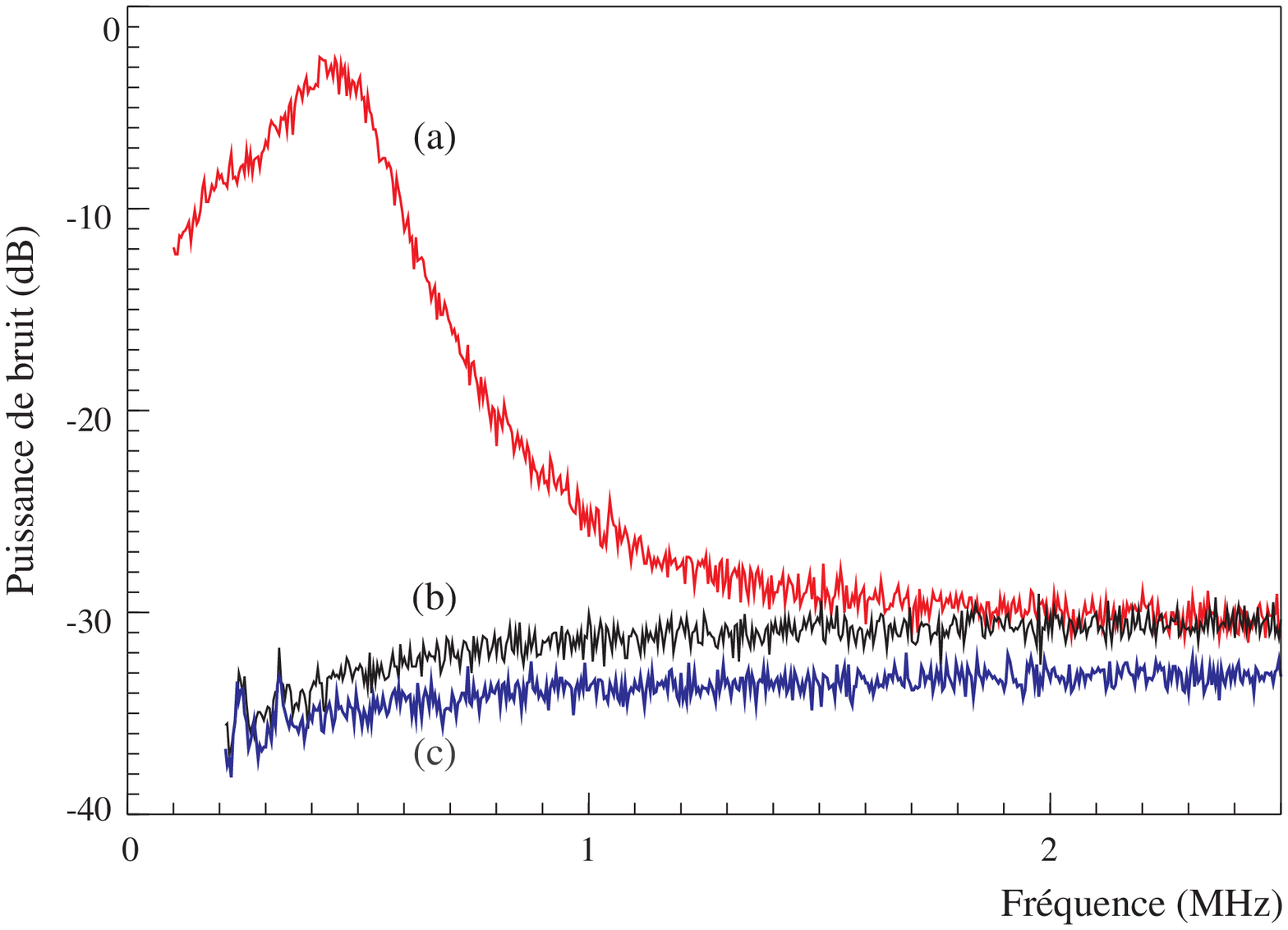,height=8cm}}
\caption{Spectre de bruit d'intensit\'{e} du faisceau laser (a) compar\'{e}
au bruit de photon standard (b) et au bruit \'{e}lectronique du syst\`{e}me
de d\'{e}tection (c), pour une puissance lumineuse de $500~\mu W$}
\label{Fig_4shot}
\end{figure}

Nous avons donc cherch\'{e} \`{a} caract\'{e}riser le bruit technique du
laser. Pour cela, nous avons utilis\'{e} le syst\`{e}me de d\'{e}tection
plac\'{e} normalement sur le faisceau r\'{e}fl\'{e}chi par la cavit\'{e}
\`{a} miroir mobile, en modifiant l\'{e}g\`{e}rement son fonctionnement de fa%
\c{c}on \`{a} mesurer l'intensit\'{e} et non la quadrature de phase du
faisceau. Les \'{e}l\'{e}ments constitutifs du dispositif sont cependant les
m\^{e}mes et ils sont d\'{e}crits en d\'{e}tail dans la section 4.3.3. Le
principe de la mesure est repr\'{e}sent\'{e} sur la figure \ref{Fig_4brtechn}%
. Le faisceau issu de la source laser est divis\'{e} en deux parties de
m\^{e}me intensit\'{e} \`{a} l'aide d'une lame demi-onde et d'un cube
s\'{e}parateur de polarisation. L'intensit\'{e} de chacun des deux faisceaux
est mesur\'{e}e \`{a} l'aide de deux photodiodes \'{e}quilibr\'{e}es. Les
signaux sont pr\'{e}amplifi\'{e}s puis ils peuvent \^{e}tre additionn\'{e}s
ou soustraits avant d'\^{e}tre envoy\'{e}s sur un analyseur de spectre. 
En position sommateur, on mesure la somme des intensit\'{e}s des deux
faisceaux. En d'autres termes, l'analyseur de spectre fournit le spectre
d'intensit\'{e} du faisceau incident sur le cube, que l'on peut \'{e}crire: 
\begin{equation}
S_{I+}\left[ \Omega \right] =\bar{I}~\left( 1+Q\left[ \Omega \right] \right)
\label{4.2.4bis}
\end{equation}
o\`{u} le facteur de Mandel $Q\left[ \Omega \right] $ repr\'{e}sente la
proportion d'exc\`{e}s de bruit par rapport au bruit de photon standard\cite
{mandel facteur}. En position soustracteur, on mesure en fait les
corr\'{e}lations quantiques entre les deux faisceaux issus du cube. Du fait
de la r\'{e}partition statistique al\'{e}atoire des photons induite par le
cube, ces corr\'{e}lations sont nulles et l'analyseur de spectre fournit le
spectre de bruit de photon standard\cite{lame semi plus vide}: 
\begin{equation}
S_{I-}\left[ \Omega \right] =\bar{I}  \label{4.2.4ter}
\end{equation}
La comparaison des deux spectres (\'{e}quations \ref{4.2.4bis} et \ref
{4.2.4ter}) permet donc de d\'{e}terminer le facteur de Mandel $Q$, c'est
\`{a} dire le bruit technique de notre source laser. La figure \ref
{Fig_4shot} montre le r\'{e}sultat de la mesure. Les traces ($a $) et ($b$)
repr\'{e}sentent les spectres $S_{I+}\left[ \Omega \right] $ et $%
S_{I-}\left[ \Omega \right] $ alors que la trace ($c$) donne le bruit
\'{e}lectronique du syst\`{e}me de d\'{e}tection, obtenu en coupant le
faisceau lumineux. L'intensit\'{e} pr\'{e}sente un exc\`{e}s de bruit
important ($20$ \`{a} $30~dBm$) pour des fr\'{e}quences inf\'{e}rieures
\`{a} $700~kHz$. Par contre, le bruit d'intensit\'{e} rejoint le bruit de
photon standard \`{a} partir de $2~MHz$ environ. Notons que ce r\'{e}sultat
d\'{e}pend de l'intensit\'{e} moyenne du faisceau laser. En effet, si cette
intensit\'{e} est att\'{e}nu\'{e}e par un facteur $\eta $, le spectre
d'intensit\'{e} $S_{I+}\left[ \Omega \right] $ devient\cite{mandel facteur}: 
\begin{equation}
S_{I+}\left[ \Omega \right] =\eta \bar{I}~\left( 1+\eta ~Q\left[ \Omega
\right] \right)  \label{4.2.5bis}
\end{equation}
Ainsi le facteur de Mandel est r\'{e}duit dans les m\^{e}mes proportions que
l'intensit\'{e} moyenne. Les r\'{e}sultats de la figure \ref{Fig_4shot} ont
\'{e}t\'{e} obtenus pour une puissance lumineuse de $500~\mu W$. 
Ces r\'{e}sultats sont donc satisfaisants pour observer les effets du
couplage optom\'{e}canique \`{a} des fr\'{e}quences d'analyse voisines de $%
2~MHz$. Comme nous le verrons dans la partie 4.5, ces caract\'{e}ristiques
peuvent \^{e}tre am\'{e}lior\'{e}es en utilisant une cavit\'{e} de filtrage
spatial de grande finesse. Une telle cavit\'{e} a pour effet de filtrer le
bruit technique du laser pour des fr\'{e}quences sup\'{e}rieures \`{a} sa
bande passante.

\subsection{Filtrage spatial\label{IV-2-4}}

L'adaptation spatiale entre le faisceau lumineux et la cavit\'{e} \`{a}
miroir mobile joue un r\^{o}le important dans l'observation des effets
quantiques du couplage optom\'{e}canique. Dans les chapitres
pr\'{e}c\'{e}dents nous avons suppos\'{e} une adaptation parfaite. Si ce
n'est pas le cas, seule la partie du faisceau incident qui se projette sur
le mode fondamental de la cavit\'{e} interagit avec celle-ci. La partie
restante du faisceau se r\'{e}fl\'{e}chit directement sur le miroir
d'entr\'{e}e et n'interagit pas avec le miroir mobile. Comme le dispositif
de mesure d\'{e}tecte l'ensemble du faisceau r\'{e}fl\'{e}chi,
l'inadaptation spatiale est \'{e}quivalente \`{a} des pertes pour le
faisceau r\'{e}fl\'{e}chi puisqu'elle r\'{e}duit les corr\'{e}lations entre
le faisceau et le miroir mobile. La mise en \'{e}vidence des
corr\'{e}lations quantiques n\'{e}cessite donc une adaptation spatiale aussi
bonne que possible.

Le faisceau issu du laser pr\'{e}sente un astigmatisme non n\'{e}gligeable,
li\'{e} \`{a} des effets de lentille thermique dans le cristal Titane-Saphir
et \`{a} la pr\'{e}sence dans la cavit\'{e} de nombreux \'{e}l\'{e}ments
optiques plac\'{e}s \`{a} incidence de Brewster. Ainsi la position et la
taille des cols du faisceau dans les directions horizontale et verticale
sont diff\'{e}rentes. L'\'{e}cart $\Delta z$ entre les positions des deux
cols est de l'ordre de $20~cm$, soit $15\%$ de la longueur de Rayleigh $%
z_{R} $. De m\^{e}me, l'assym\'{e}trie entre les tailles des cols est de
l'ordre de $10\%$. L'astigmatisme d\'{e}pend en plus des conditions de
fonctionnement du laser et peut varier d'un jour \`{a} l'autre. Le faisceau
a ainsi un profil elliptique qu'il est difficile d'adapter correctement
\`{a} la cavit\'{e} \`{a} miroir mobile. C'est pourquoi nous avons
r\'{e}alis\'{e} un dispositif de filtrage spatial qui donne au faisceau un
profil gaussien et cylindrique, plus facilement adaptable au mode
fondamental de la cavit\'{e} \`{a} miroir mobile.

Ce dispositif est constitu\'{e} d'une cavit\'{e} Fabry-Perot non
d\'{e}g\'{e}n\'{e}r\'{e}e, dans laquelle est envoy\'{e} le faisceau laser.
La longueur de la cavit\'{e} est contr\^{o}l\'{e}e de telle mani\`{e}re que
la fr\'{e}quence de r\'{e}sonance d'un mode $TEM_{00}$ de la cavit\'{e}
co\"{\i}ncide avec la fr\'{e}quence du laser. La cavit\'{e} transmet alors
la partie du faisceau incident qui est adapt\'{e}e \`{a} son mode propre
fondamental et elle r\'{e}fl\'{e}chit toutes les autres composantes du
faisceau incident. Ce dispositif transmet donc un faisceau dont la structure
est uniquement d\'{e}termin\'{e}e par la g\'{e}om\'{e}trie de la cavit\'{e}.

Nous allons pr\'{e}senter dans cette section les diff\'{e}rents
\'{e}l\'{e}ments du dispositif de filtrage. Nous commencerons par
d\'{e}crire et caract\'{e}riser la cavit\'{e} Fabry Perot de filtrage (FPF).
Nous pr\'{e}senterons ensuite les diff\'{e}rents \'{e}l\'{e}ments de
l'asservissement, qui permet de maintenir la cavit\'{e} FPF \`{a}
r\'{e}sonance avec le faisceau incident. Nous terminerons en d\'{e}crivant
la mesure de la bande passante de la cavit\'{e} qui, comme nous l'avons vu
dans la section 4.1.4, sert de r\'{e}f\'{e}rence pour d\'{e}terminer la
finesse de la cavit\'{e} \`{a} miroir mobile.

\subsubsection{La cavit\'{e} de filtrage\label{IV-2-4-1}}

La cavit\'{e} de filtrage spatial est constitu\'{e}e d'un miroir
d'entr\'{e}e plan et d'un miroir de sortie courbe de rayon de courbure
\'{e}gal \`{a} $1~m$. La cavit\'{e} a deux entr\'{e}es-sorties, les miroirs
ayant le m\^{e}me coefficient de r\'{e}flexion en intensit\'{e} \'{e}gal
\`{a} $95\%$. Ceci permet d'avoir une transmission de la cavit\'{e} maximale
\`{a} r\'{e}sonance. On montre en effet que la transmission ${\cal T}_{0}$
\`{a} r\'{e}sonance d'une cavit\'{e} \`{a} deux entr\'{e}es-sorties est: 
\begin{equation}
{\cal T}_{0}=\frac{T_{1}T_{2}}{\left( 1-\sqrt{R_{1}R_{2}}\right) ^{2}}
\label{4.2.5}
\end{equation}
o\`{u} les coefficients de transmission $T_{1,2}$ et de r\'{e}flexion $%
R_{1,2}$ des deux miroirs ob\'{e}issent \`{a} la relation de conservation $%
T_{1,2}+R_{1,2}=1$, \`{a} condition que les pertes soient n\'{e}gligeables.
On voit alors que ${\cal T}_{0}$ est maximal lorsque $R_{1}=R_{2}$ et $%
T_{1}=T_{2}$.

La longueur de la cavit\'{e} est de $12~cm$. Le support de la cavit\'{e} est
constitu\'{e} d'un barreau cylindrique en invar, le miroir plan \'{e}tant
mont\'{e} sur une cale pi\'{e}zo\'{e}lectrique. Deux fen\^{e}tres plan-plan
et plan-convexe (de courbure \'{e}gale \`{a} $1~m$) sont fix\'{e}es
respectivement \`{a} l'entr\'{e}e et \`{a} la sortie de la cavit\'{e}, afin
d'une part d'isoler la cavit\'{e} et d'autre part de compenser l'effet de
lentille divergente d\^{u} au miroir courbe.

L'adaptation spatiale du faisceau laser sur la cavit\'{e} de filtrage est
r\'{e}alis\'{e}e par un dispositif similaire \`{a} celui utilis\'{e} pour la
cavit\'{e} FPE de l'asservissement en fr\'{e}quence (section 4.2.2). Un jeu
de deux lentilles permet de modifier la taille du col du faisceau. La
premi\`{e}re lentille, plac\'{e}e \`{a} $500~mm$ du col du faisceau laser, a
une focale \'{e}gale \`{a} $30~mm$, alors que la deuxi\`{e}me lentille,
mont\'{e}e sur une platine de translation, est plac\'{e}e \`{a} $60~mm$ de
la premi\`{e}re et a une focale de $28~mm$. On obtient ainsi un col de $%
0.29~mm$ \`{a} l'entr\'{e}e de la cavit\'{e} FPF, que l'on peut ajuster
pr\'{e}cis\'{e}ment au col du mode fondamental de la cavit\'{e} \`{a} l'aide
de la platine de translation. Un jeu de deux miroirs mont\'{e}s sur des
supports microm\'{e}triques permet ensuite d'aligner le faisceau sur la
cavit\'{e}. 
\begin{figure}[tbp]
\centerline{\psfig{figure=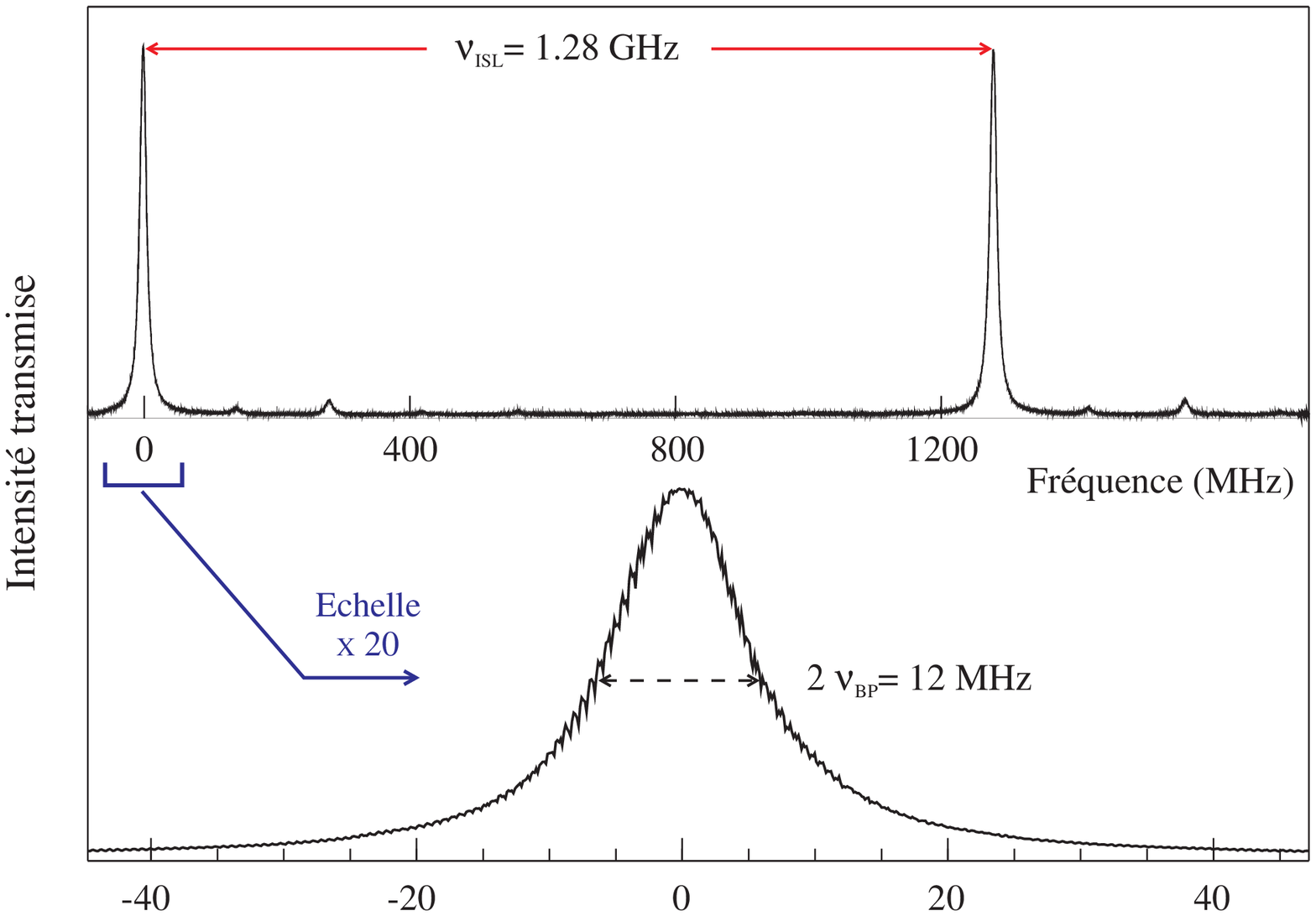,height=10cm}}
\caption{Intensit\'{e} transmise par la cavit\'{e} FPF lorsqu'on balaye la
fr\'{e}quence du laser titane saphir \`{a} l'aide du bilame sur un
intervalle spectral libre $\nu _{ISL}$. Un ajustement de la premi\`{e}re
r\'{e}sonance par une Lorentzienne donne une bande passante $\nu _{BP}$ de
l'ordre de $6~MHz$}
\label{Fig_4fpfisl1}
\end{figure}

Le r\'{e}sultat de l'adaptation spatiale appara\^{\i }t sur la courbe du
haut de la figure \ref{Fig_4fpfisl1}, obtenue en balayant la fr\'{e}quence
du laser. On peut voir deux pics d'Airy associ\'{e}s \`{a} deux modes $%
TEM_{00}$ successifs de la cavit\'{e}, s\'{e}par\'{e}s par un intervalle
spectral libre $\nu _{ISL}$ qui, pour une longueur de la cavit\'{e} de $%
12~cm $, est \'{e}gal \`{a} $1.28~GHz$. Les petits pics correspondent \`{a}
des modes transverses fortement att\'{e}nu\'{e}s du fait de l'adaptation
spatiale. L'intensit\'{e} transmise \`{a} r\'{e}sonance repr\'{e}sente
environ $80\%$ de l'intensit\'{e} incidente. Ces pertes peuvent \^{e}tre
attribu\'{e}es \`{a} l'astigmatisme du faisceau incident, mais aussi aux
pertes des miroirs de la cavit\'{e}. L'\'{e}quation (\ref{4.2.5}) n'est en
effet valable que dans le cas d'une adaptation spatiale parfaite et pour des
miroirs sans perte. Dans le cas contraire, la transmission ${\cal T}_{0}$
n'est plus \'{e}gale \`{a} $1$.

La figure \ref{Fig_4fpfisl1} permet aussi de d\'{e}terminer la bande
passante de la cavit\'{e}. Nous utilisons pour balayer la fr\'{e}quence du
faisceau incident une rampe de tension lin\'{e}aire appliqu\'{e}e sur le
moteur du bilame du laser titane saphir, l'asservissement en fr\'{e}quence
\'{e}tant d\'{e}sactiv\'{e}. L'intensit\'{e} transmise par la cavit\'{e} FPF
est d\'{e}tect\'{e}e puis envoy\'{e}e vers un oscilloscope digital Tektronix
TDS420 qui permet d'acqu\'{e}rir le signal avec un nombre de points
\'{e}lev\'{e} ($15000$ points). Connaissant l'intervalle spectrale libre $%
\nu _{ISL}$, on peut \'{e}talonner l'axe horizontal en fr\'{e}quence. On
\'{e}largit alors d'un facteur $20$ la zone autour de la premi\`{e}re
r\'{e}sonance. On obtient la courbe du bas sur la figure \ref{Fig_4fpfisl1}
qui permet de d\'{e}terminer la bande passante $\nu _{BP}$ de la cavit\'{e}
de filtrage en r\'{e}alisant un ajustement lorentzien de la r\'{e}sonance.
On trouve que la demi-largeur de la Lorentzienne est comprise entre $5.8$ et 
$6~MHz$. Il est difficile d'obtenir une valeur plus pr\'{e}cise \'{e}tant
donn\'{e}e la l\'{e}g\`{e}re dissym\'{e}trie de la r\'{e}sonance. Celle-ci
est sans doute due \`{a} des effets thermiques transitoires au niveau des
miroirs lorsque la cavit\'{e} passe \`{a} r\'{e}sonance.

\subsubsection{Asservissement de la cavit\'{e} FPF sur la fr\'{e}quence du
laser\label{IV-2-4-2}}

\begin{figure}[tbp]
\centerline{\psfig{figure=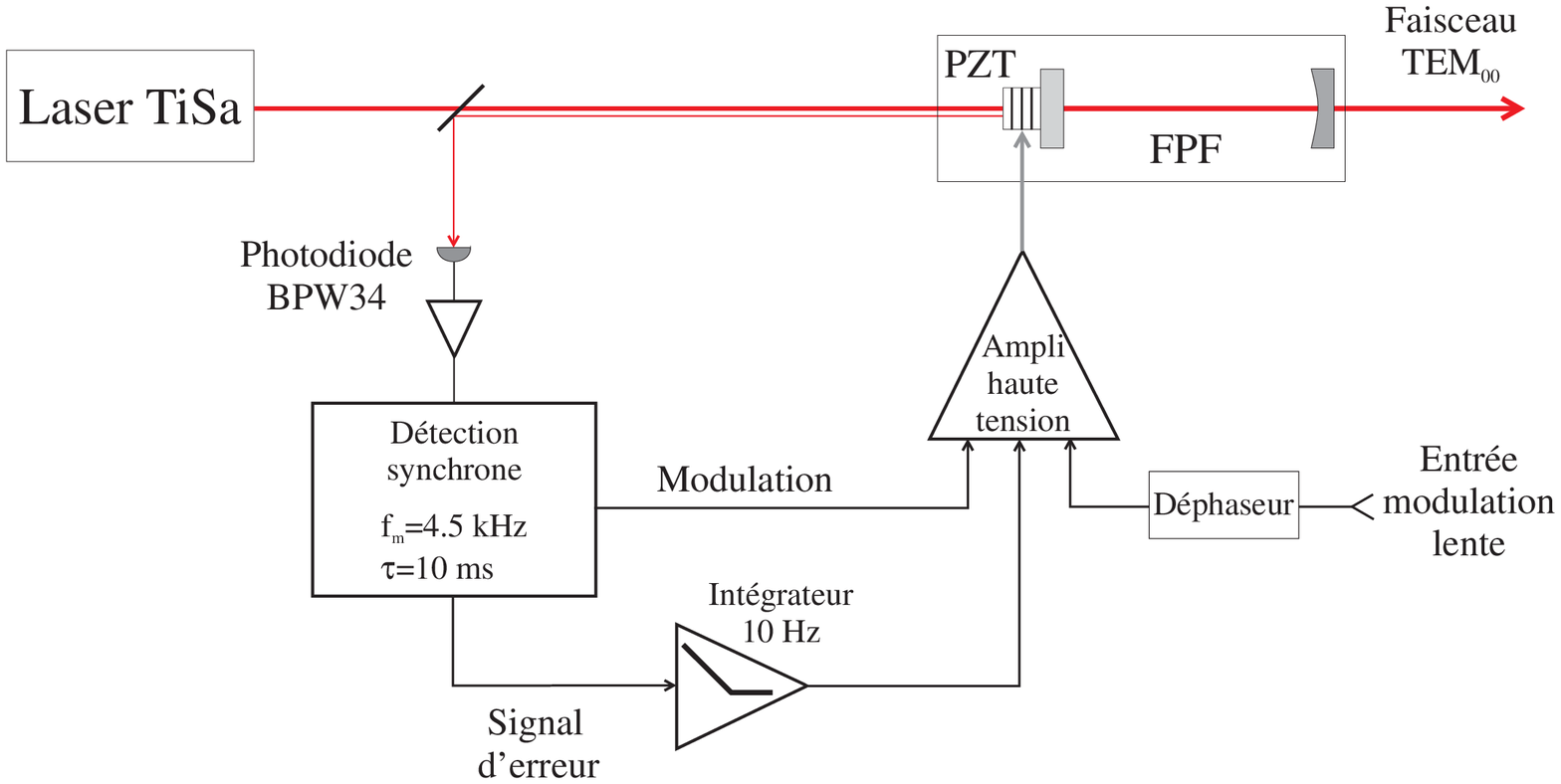,height=8cm}}
\caption{Repr\'{e}sentation sch\'{e}matique de l'asservissement de la
cavit\'{e} de filtrage FPF. La modulation appliqu\'{e}e sur la longueur de
la cavit\'{e} induit une modulation d'intensit\'{e} du faisceau
r\'{e}fl\'{e}chi. Le signal d'erreur qui pilote la cale
pi\'{e}zo\'{e}lectrique (PZT) du FPF est alors proportionnel au
d\'{e}saccord entre la fr\'{e}quence du faisceau et la r\'{e}sonance de la
cavit\'{e}}
\label{Fig_4fpfass}
\end{figure}

L'asservissement que nous avons mis au point pour maintenir la cavit\'{e}
FPF \`{a} r\'{e}sonance avec le faisceau laser repose sur la technique de
d\'{e}tection synchrone d\'{e}j\`{a} utilis\'{e}e pour l'asservissement de
l'\'{e}talon \'{e}pais du laser titane saphir. Le sch\'{e}ma de principe est
repr\'{e}sent\'{e} sur la figure \ref{Fig_4fpfass}. Contrairement \`{a} la
technique standard qui consiste \`{a} d\'{e}tecter l'intensit\'{e} du
faisceau transmis, modul\'{e} en intensit\'{e} par la modulation de longueur
de la cavit\'{e}, on d\'{e}tecte ici l'intensit\'{e} du faisceau
r\'{e}fl\'{e}chi. Le choix de cette configuration est li\'{e} au fait que la
cavit\'{e} de filtrage se trouve dans la boucle d'asservissement
d'intensit\'{e} (voir figure \ref{Fig_4assint}). Cet asservissement a pour
effet de supprimer la modulation d'intensit\'{e} du faisceau transmis. 

Plus pr\'{e}cis\'{e}ment, on utilise une d\'{e}tection synchrone commerciale 
$EG\&G$~$5205$ dont la fr\'{e}quence de modulation est \'{e}gale \`{a} $%
4.5~kHz$. La sortie modulation de la d\'{e}tection synchrone est
appliqu\'{e}e \`{a} la cale pi\'{e}zo\'{e}lectrique (PZT) de la cavit\'{e}
FPF par l'interm\'{e}diaire d'un amplificateur haute tension $0-1000~V$. La
variation de longueur de la cavit\'{e} induite par cette modulation est
\'{e}quivalente \`{a} une variation $\delta \Psi \left[ \omega _{m}\right] $
du d\'{e}phasage \`{a} la fr\'{e}quence de modulation $\omega _{m}$, autour
du d\'{e}phasage moyen $\bar{\Psi}$. On obtient ainsi une modulation des
intensit\'{e}s aussi bien en transmission qu'en r\'{e}flexion. En effet, si
la fr\'{e}quence $\omega _{m}$ est petite devant la bande passante de la
cavit\'{e}, ces deux modulations s'\'{e}crivent: 
\begin{subequations}
\label{4.2.6}
\begin{eqnarray}
\delta I^{t}\left[ \omega _{m}\right] &=&\bar{I}^{in}~\frac{\partial T_{c}}{%
\partial \bar{\Psi}}~\delta \Psi \left[ \omega _{m}\right]  \label{4.2.6a} \\
\delta I^{r}\left[ \omega _{m}\right] &=&\bar{I}^{in}~\frac{\partial R_{c}}{%
\partial \bar{\Psi}}~\delta \Psi \left[ \omega _{m}\right]  \label{4.2.6b}
\end{eqnarray}
o\`{u} $T_{c}$ et $R_{c}$ sont les coefficients de transmission et de
r\'{e}flexion de la cavit\'{e}, reli\'{e}s pour une cavit\'{e} sans perte et
sym\'{e}trique au coefficient de transmission $T$ des miroirs et au
d\'{e}phasage moyen $\bar{\Psi}$: 
\end{subequations}
\begin{equation}
T_{c}=\frac{T^{2}}{T^{2}+\bar{\Psi}^{2}}\qquad ,\qquad R_{c}=1-T_{c}
\label{4.2.7}
\end{equation}
On voit ainsi que la modulation d'intensit\'{e} du faisceau r\'{e}fl\'{e}chi
est simplement en opposition de phase par rapport \`{a} celle du faisceau
transmis.

Notons par ailleurs que l'asservissement d'intensit\'{e} ajoute une
modulation $\delta I^{in}\left[ \omega _{m}\right] $ du faisceau incident de
fa\c{c}on \`{a} supprimer la modulation $\delta I^{t}\left[ \omega
_{m}\right] $ transmise: 
\begin{equation}
\delta I^{t}\left[ \omega _{m}\right] =\bar{I}^{in}~\frac{\partial T_{c}}{%
\partial \bar{\Psi}}~\delta \Psi \left[ \omega _{m}\right] +T_{c}~\delta
I^{in}\left[ \omega _{m}\right] =0  \label{4.2.8}
\end{equation}
On constate alors que cela a pour effet d'augmenter la modulation du
faisceau r\'{e}fl\'{e}chi, par un facteur $1/T_{c}$: 
\begin{equation}
\delta I^{r}\left[ \omega _{m}\right] =-\frac{1}{T_{c}}~\bar{I}^{in}~\frac{%
\partial T_{c}}{\partial \bar{\Psi}}~\delta \Psi \left[ \omega _{m}\right]
\label{4.2.9}
\end{equation}

Une partie du faisceau r\'{e}fl\'{e}chi est d\'{e}tect\'{e}e par une
photodiode $BPW34$ puis amplifi\'{e}e (voir figure \ref{Fig_4fpfass}). Le
signal obtenu est envoy\'{e} dans la d\'{e}tection synchrone dont la
constante de temps est \'{e}gale \`{a} $10~ms$. Le signal est
d\'{e}modul\'{e} par la d\'{e}tection synchrone ce qui permet d'obtenir un
signal d'erreur proportionnel au d\'{e}saccord entre le laser et la
r\'{e}sonance de la cavit\'{e}. Entre la sortie de la d\'{e}tection
synchrone et l'amplificateur haute tension qui pilote la cale
pi\'{e}zoelectrique de la cavit\'{e} FPF, le signal d'erreur est
int\'{e}gr\'{e} pour des fr\'{e}quences inf\'{e}rieures \`{a} $10~Hz$, ce
qui permet d'agir efficacement sur les d\'{e}rives lentes du d\'{e}saccord.

Lorsque l'asservissement est verrouill\'{e}, la r\'{e}sonance de la
cavit\'{e} FPF est cal\'{e}e sur la fr\'{e}quence du laser et le faisceau
transmis est bien $TEM_{00}$. Nous avons en effet mesur\'{e} \`{a} l'aide
d'un analyseur de mode (Mode Master Coherent), les caract\'{e}ristiques
spatiales du faisceau transmis qui sont, \`{a} quelques pourcents pr\`{e}s
(la marge d'erreur du mode master \'{e}tant de cet ordre), celles d'un mode
parfaitement gaussien $TEM_{00}$. Notons que l'asservissement utilis\'{e}
ici est relativement simple et que nous n'avons pas eu besoin
d'am\'{e}liorer ses caract\'{e}ristiques. Les fluctuations \`{a} corriger
sont en effet assez faibles du fait de la stabilit\'{e} de la fr\'{e}quence
du laser et de la cavit\'{e} de filtrage. D'autre part, les imperfections de
l'asservissement de la cavit\'{e} sont automatiquement compens\'{e}es par
l'asservissement d'intensit\'{e}. Un d\'{e}saccord non nul entre la
fr\'{e}quence du laser et la r\'{e}sonance de la cavit\'{e} FPF ne modifie
pas ses capacit\'{e}s de filtrage spatial. Par contre, cela se traduit par
une modification du point de fonctionnement sur le pic d'Airy de la
r\'{e}sonance et donc par une variation de l'intensit\'{e} transmise par la
cavit\'{e}. Mais cette variation est corrig\'{e}e par l'asservissement
d'intensit\'{e} qui modifie la puissance incidente sur la cavit\'{e} de fa\c{%
c}on \`{a} ce que l'intensit\'{e} transmise soit constante.

On remarque enfin sur la figure \ref{Fig_4fpfass} la pr\'{e}sence d'une
entr\'{e}e modulation lente suivie d'un d\'{e}phaseur qui pilote la cale
pi\'{e}zo\'{e}lectrique de la cavit\'{e}. Cette entr\'{e}e est utilis\'{e}e
lorsqu'on applique une modulation sur l'entr\'{e}e du laser titane saphir
(figure \ref{Fig_4prinass}). La rampe de modulation de la fr\'{e}quence du
laser a typiquement une excursion de $20$ \`{a} $30~MHz$, et une
fr\'{e}quence de modulation de l'ordre de $100~Hz$. L'asservissement de la
cavit\'{e} FPF n'est pas capable de compenser une variation de la
fr\'{e}quence du laser aussi large (plusieurs fois la largeur du pic d'Airy)
et aussi rapide (de l'ordre de la constante de temps de la d\'{e}tection
synchrone). C'est pourquoi il est n\'{e}cessaire d'assister l'asservissement
afin que la r\'{e}sonance du FPF suive de fa\c{c}on synchrone la modulation
du laser. On applique la m\^{e}me rampe de modulation sur l'entr\'{e}e
correspondante de l'asservissement (figure \ref{Fig_4fpfass}) et on
r\`{e}gle l'amplitude et la phase du d\'{e}phaseur de fa\c{c}on \`{a} ce que
la fr\'{e}quence du laser reste sur le pic d'Airy de la r\'{e}sonance du
FPF. L'asservissement d'intensit\'{e} se charge alors de supprimer toute
variation r\'{e}siduelle de l'intensit\'{e} du faisceau transmis. On obtient
de cette mani\`{e}re un faisceau transmis parfaitement $TEM_{00}$,
modul\'{e} en fr\'{e}quence et dont l'intensit\'{e} est constante.

\subsubsection{Mesure de la bande passante de la cavit\'{e} de filtrage\label%
{IV-2-4-3}}

\begin{figure}[tbp]
\centerline{\psfig{figure=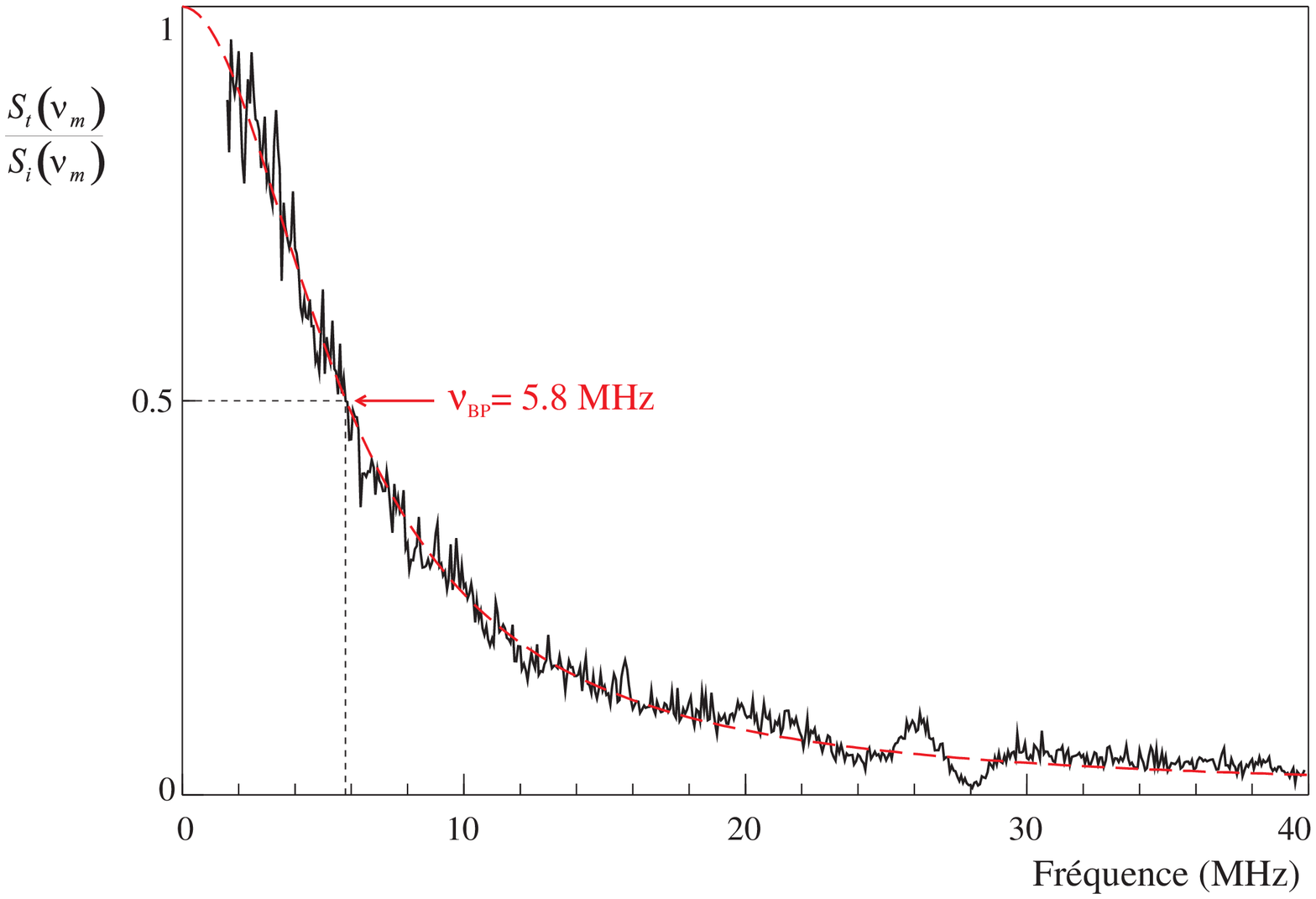,height=8cm}}
\caption{Fonction de transfert de la cavit\'{e} FPF obtenue en faisant le
rapport entre les spectres de modulation transmis $S_{t}\left( \nu
_{m}\right) $ et incident $S_{i}\left( \nu _{m}\right) $. L'ajustement
Lorentzien en tirets permet de d\'{e}terminer la bande passante $\nu _{BP}$
de la cavit\'{e} qui est \'{e}gale \`{a} $5.8~MHz$}
\label{Fig_4fpfbpas}
\end{figure}

La mesure de la bande passante $\nu _{BP}$ de la cavit\'{e} de filtrage sert
de r\'{e}f\'{e}rence de fr\'{e}quence pour d\'{e}terminer la bande passante
de la cavit\'{e} \`{a} miroir mobile (voir section 4.1.3.1). Nous avons
d'ores et d\'{e}j\`{a} d\'{e}termin\'{e} cette bande passante en balayant la
fr\'{e}quence du laser (section \ref{IV-1-4-2}). Cette m\'{e}thode
pr\'{e}sente n\'{e}anmoins des d\'{e}fauts, li\'{e}s \`{a} des effets
thermiques lors du passage \`{a} r\'{e}sonance. Ces d\'{e}fauts rendent
difficile une d\'{e}termination pr\'{e}cise de la bande passante, et nous
avons \'{e}t\'{e} amen\'{e}s \`{a} utiliser une m\'{e}thode de mesure plus
performante.

La mesure repose sur une technique de modulation en intensit\'{e} du
faisceau incident, la cavit\'{e} \'{e}tant asservie \`{a} r\'{e}sonance avec
le faisceau. L'intensit\'{e} du faisceau incident est modul\'{e}e en
appliquant sur l'\'{e}lectro-optique utilis\'{e} pour l'asservissement
d'intensit\'{e} une tension sinuso\"{\i }dale de fr\'{e}quence $\nu _{m}$
variable. La cavit\'{e} se comporte alors comme un filtre passe-bas de
fr\'{e}quence de coupure \'{e}gale \`{a} $\nu _{BP}$. Le signal $S_{t}\left(
\nu _{m}\right) $ correspondant \`{a} la puissance de modulation du faisceau
transmis est reli\'{e} au signal incident $S_{i}\left( \nu _{m}\right) $
par: 
\begin{equation}
S_{t}\left( \nu _{m}\right) =\frac{1}{1+\left( \nu _{m}/\nu _{BP}\right) ^{2}%
}~S_{i}\left( \nu _{m}\right)  \label{4.2.10}
\end{equation}

On d\'{e}termine ces deux signaux en pla\c{c}ant une photodiode rapide avant
et apr\`{e}s la cavit\'{e} FPF. Le montage utilis\'{e}, constitu\'{e} d'une
photodiode $FND100$ de $EG\&G$ et d'un pr\'{e}amplificateur rapide
transimp\'{e}dance, est similaire \`{a} celui utilis\'{e} dans la
d\'{e}tection homodyne (section \ref{IV-3-3}). Les signaux obtenus sont
envoy\'{e}s vers un analyseur de spectre ($HP~8560E$), et on balaye la
fr\'{e}quence de modulation $\nu _{m}$ de fa\c{c}on \`{a} obtenir les
spectres de modulation incident $S_{i}\left( \nu _{m}\right) $ et transmis $%
S_{t}\left( \nu _{m}\right) $. Le rapport des deux spectres donne la
fonction de transfert, en \'{e}liminant l'influence des r\'{e}ponses en
fr\'{e}quence du modulateur et de la d\'{e}tection. Le r\'{e}sultat est
repr\'{e}sent\'{e} sur la figure \ref{Fig_4fpfbpas}. Un ajustement
Lorentzien de la fonction de transfert obtenue permet de d\'{e}terminer la
bande passante $\nu _{BP}$ de la cavit\'{e} FPF que l'on trouve \'{e}gale
\`{a} $5.8~MHz$. Cette valeur est en tr\`{e}s bon accord avec celle obtenue
par la m\'{e}thode de balayage en fr\'{e}quence du laser.

\subsection{Asservissement de la fr\'{e}quence sur la cavit\'{e} \`{a}
miroir mobile\label{IV-2-5}}

Nous savons qu'une mesure de petits d\'{e}placements du miroir mobile ainsi
qu'une mesure QND de l'intensit\'{e} sont optimales lorsque le faisceau de
mesure incident est \`{a} r\'{e}sonance avec la cavit\'{e}. Malgr\'{e} la
stabilit\'{e} du faisceau laser et de la cavit\'{e} \`{a} miroir mobile, on
observe une d\'{e}rive lente du d\'{e}saccord entre la fr\'{e}quence du
faisceau incident et celle de la cavit\'{e}. Pour maintenir la fr\'{e}quence
du laser \`{a} r\'{e}sonance, nous avons r\'{e}alis\'{e} un asservissement
\`{a} basse fr\'{e}quence qui agit non pas sur la cavit\'{e} \`{a} miroir
mobile, qui est une cavit\'{e} rigide, mais sur la fr\'{e}quence du laser
titane saphir. Plus pr\'{e}cis\'{e}ment, l'asservissement pilote la cale
pi\'{e}zo\'{e}lectrique de la cavit\'{e} FPE sur laquelle le laser est
asservi (figure \ref{Fig_4fpeasse}). L'avantage de ce contr\^{o}le indirect
de la fr\'{e}quence du laser est que l'asservissement des fluctuations
propres du laser et la compensation des d\'{e}rives lentes sont
d\'{e}coupl\'{e}s. L'asservissement en fr\'{e}quence du laser est efficace
car la cavit\'{e} FPE est tr\`{e}s stable m\'{e}caniquement et il est
possible d'agir jusqu'\`{a} des fr\'{e}quences tr\`{e}s \'{e}lev\'{e}es. Par
contre, le pilotage du FPE \`{a} partir de la cavit\'{e} \`{a} miroir mobile
ne peut \^{e}tre aussi efficace : il s'agit d'une boucle de
contre-r\'{e}action globale sur l'ensemble du dispositif exp\'{e}rimental,
qui contr\^{o}le la source laser \`{a} partir d'un signal issu de la
cavit\'{e} \`{a} miroir mobile. Cet asservissement ne peut agir que sur les
d\'{e}rives lentes car tous les \'{e}l\'{e}ments interm\'{e}diaires du
montage (cavit\'{e} de filtrage, asservissement d'intensit\'{e}, oscillateur
local,...) doivent \^{e}tre capable de supporter cette boucle de
contre-r\'{e}action globale.

\begin{figure}[tbp]
\centerline{\psfig{figure=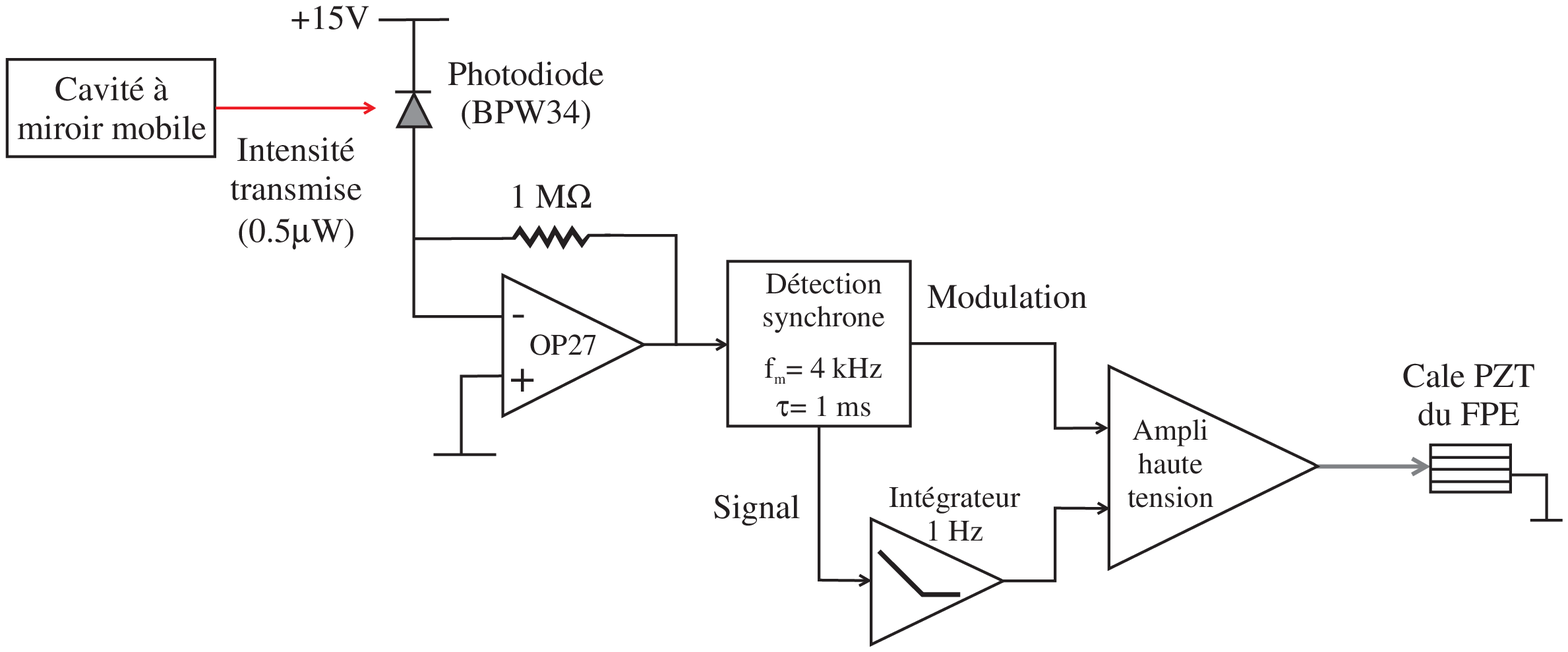,height=6cm}}
\caption{Principe de l'asservissement de la fr\'{e}quence du laser sur la
cavit\'{e} \`{a} miroir mobile. On utilise l'intensit\'{e} transmise par la
cavit\'{e} pour agir sur la cale pi\'{e}zo\'{e}lectrique de la cavit\'{e}
FPE via une chaine d'amplification et une d\'{e}tection synchrone
commerciale }
\label{Fig_4fpefpm}
\end{figure}

Le principe de l'asservissement est tout \`{a} fait similaire \`{a} celui
utilis\'{e} pour maintenir la cavit\'{e} FPF \`{a} r\'{e}sonance avec le
faisceau lumineux. Le sch\'{e}ma de l'asservissement est repr\'{e}sent\'{e}
sur la figure \ref{Fig_4fpefpm}. On module la fr\'{e}quence du laser et on
mesure \`{a} l'aide d'une d\'{e}tection synchrone la modulation induite sur
la lumi\`{e}re transmise par la cavit\'{e} \`{a} miroir mobile. Le signal
d'erreur ainsi produit contr\^{o}le la cale pi\'{e}zo\'{e}lectrique de la
cavit\'{e} FPE, par l'interm\'{e}diaire d'un amplificateur haute tension $%
0-1000~V$.

Le choix de la fr\'{e}quence de modulation est assez critique. Cette
fr\'{e}quence doit \^{e}tre suffisamment \'{e}lev\'{e}e pour ne pas
r\'{e}duire la plage de fr\'{e}quence o\`{u} l'asservissement agit. Elle
doit \^{e}tre suffisamment \'{e}loign\'{e}e de la fr\'{e}quence de
modulation de la cavit\'{e} de filtrage ($4.5~kHz$) pour ne pas perturber
son asservissement. Enfin, le faisceau incident sur la cavit\'{e} \`{a}
miroir mobile doit \^{e}tre exempt de toute modulation d'intensit\'{e}
synchrone avec la modulation de fr\'{e}quence du laser. En effet, la
modulation d'intensit\'{e} d\'{e}tect\'{e}e \`{a} la sortie de la cavit\'{e}
ne doit provenir que du d\'{e}saccord de la r\'{e}sonance de la cavit\'{e}
par rapport \`{a} la fr\'{e}quence du laser : toute modulation incidente se
traduirait comme une perturbation pour le signal d'erreur. Malheureusement
le faisceau traverse la cavit\'{e} de filtrage et celle-ci peut induire une
modulation d'intensit\'{e} si sa r\'{e}sonance n'est pas parfaitement
cal\'{e}e sur la fr\'{e}quence du laser. La fr\'{e}quence de modulation doit
donc \^{e}tre suffisamment basse pour que l'asservissement d'intensit\'{e}
puisse corriger toute modulation d'intensit\'{e} \`{a} l'entr\'{e}e de la
cavit\'{e} \`{a} miroir mobile. Pour ces diff\'{e}rentes raisons, nous avons
choisi une fr\'{e}quence de modulation de $4~kHz$. 

Un autre point d\'{e}licat est la d\'{e}tection du faisceau transmis par la
cavit\'{e} \`{a} miroir mobile. M\^{e}me \`{a} r\'{e}sonance, cette
intensit\'{e} est tr\`{e}s faible, de l'ordre de $0.5~\mu W$ pour une
intensit\'{e} incidente de $100$~$\mu W$. Notons par ailleurs qu'une partie
de la lumi\`{e}re d\'{e}tect\'{e}e \`{a} l'arri\`{e}re de la cavit\'{e}
provient de la lumi\`{e}re diffus\'{e}e par les miroirs. En utilisant une
cam\'{e}ra infrarouge, on peut voir au centre du miroir mobile un point
brillant qui correspond au faisceau directement transmis par la cavit\'{e}.
On observe aussi un anneau de lumi\`{e}re \`{a} la p\'{e}riph\'{e}rie du
miroir, qui provient de la lumi\`{e}re diffus\'{e}e qui traverse, apr\`{e}s
de multiples r\'{e}flexions, le substrat du miroir mobile au niveau de sa
partie non trait\'{e}e.

L'intensit\'{e} transmise est d\'{e}tect\'{e}e par une photodiode $BPW34$.
Le courant obtenu \'{e}tant tr\`{e}s petit (typiquement $300~nA$), on
utilise un montage en transimp\'{e}dance form\'{e} d'un amplificateur $OP27$
et d'une grande r\'{e}sistance de contre-r\'{e}action de $1~M\Omega $ pour
obtenir une tension en sortie raisonnable. Le montage transimp\'{e}dance
pr\'{e}sente l'avantage de r\'{e}duire l'influence de la capacit\'{e}
parasite de la photodiode : ceci permet d'atteindre des bandes passantes
suffisamment \'{e}lev\'{e}es malgr\'{e} la grande valeur de la
r\'{e}sistance de charge. Le signal est ensuite envoy\'{e} dans une
d\'{e}tection synchrone identique \`{a} celle utilis\'{e}e pour
l'asservissement de la cavit\'{e} FPF, dont la constante de temps est
\'{e}gale \`{a} $1~ms$. Le signal d'erreur produit est int\'{e}gr\'{e} pour
des fr\'{e}quences inf\'{e}rieures \`{a} $1~Hz$, ce qui permet d'augmenter
le gain global de l'asservissement \`{a} basse fr\'{e}quence.

\begin{figure}[tbp]
\centerline{\psfig{figure=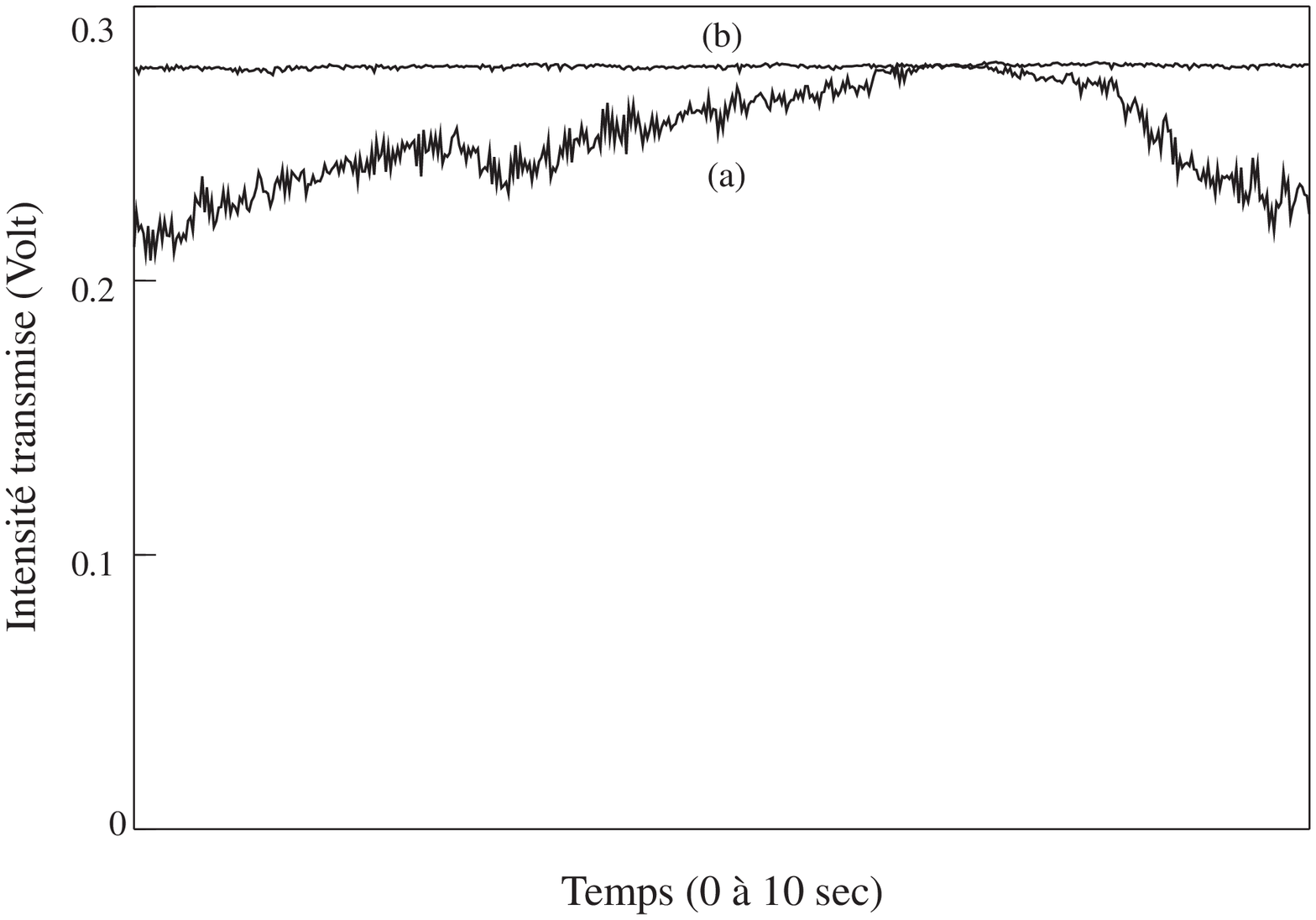,height=8cm}}
\caption{Intensit\'{e} transmise par la cavit\'{e} \`{a} miroir mobile sans
asservissement (trace a) et lorsque l'asservissement est activ\'{e} (trace
b) }
\label{Fig_4assfpm}
\end{figure}

On peut voir sur la figure \ref{Fig_4assfpm} le r\'{e}sultat de
l'asservissement sur l'intensit\'{e} transmise par la cavit\'{e} \`{a}
miroir mobile. La trace (a) montre l'\'{e}volution temporelle de
l'intensit\'{e} sans asservissement. Le d\'{e}saccord entre la fr\'{e}quence
du laser et la r\'{e}sonance de la cavit\'{e} d\'{e}rive lentement, ce qui
se traduit par une variation de l'intensit\'{e} transmise. On constate que
la d\'{e}rive est lente, puisque le d\'{e}saccord varie d'une quantit\'{e}
inf\'{e}rieure \`{a} la largeur du pic d'Airy en $10$ secondes. D'autre part
les fluctuations sur le signal restent mod\'{e}r\'{e}es : cela est d\^{u}
\`{a} la grande stabilit\'{e} du laser et de la cavit\'{e} \`{a} miroir
mobile. 

La trace (b) montre l'intensit\'{e} transmise lorsque l'asservissement est
activ\'{e}. La fr\'{e}quence du laser est contr\^{o}l\'{e}e de telle
mani\`{e}re que le faisceau reste \`{a} r\'{e}sonance avec la cavit\'{e}
(maximum du pic d'Airy). Cette courbe d\'{e}montre l'efficacit\'{e} de
l'asservissement et plus g\'{e}n\'{e}ralement la qualit\'{e} de notre source
laser. En effet, le bruit r\'{e}siduel de l'intensit\'{e} transmise est
d\^{u} \`{a} la fois aux fluctuations d'intensit\'{e} du faisceau incident,
aux fluctuations de fr\'{e}quence du laser et \'{e}ventuellement aux
fluctuations de longueur de la cavit\'{e} \`{a} miroir mobile. La
stabilit\'{e} de la source laser est telle que les fluctuations relatives $%
\Delta I/\bar{I}$ du faisceau transmis par la cavit\'{e} \`{a} miroir mobile
sont inf\'{e}rieures \`{a} $2^{\circ }/_{\circ \circ }$.

\subsection{Vue d'ensemble de la source laser\label{IV-2-6}}

\begin{figure}[tbp]
\centerline{\psfig{figure=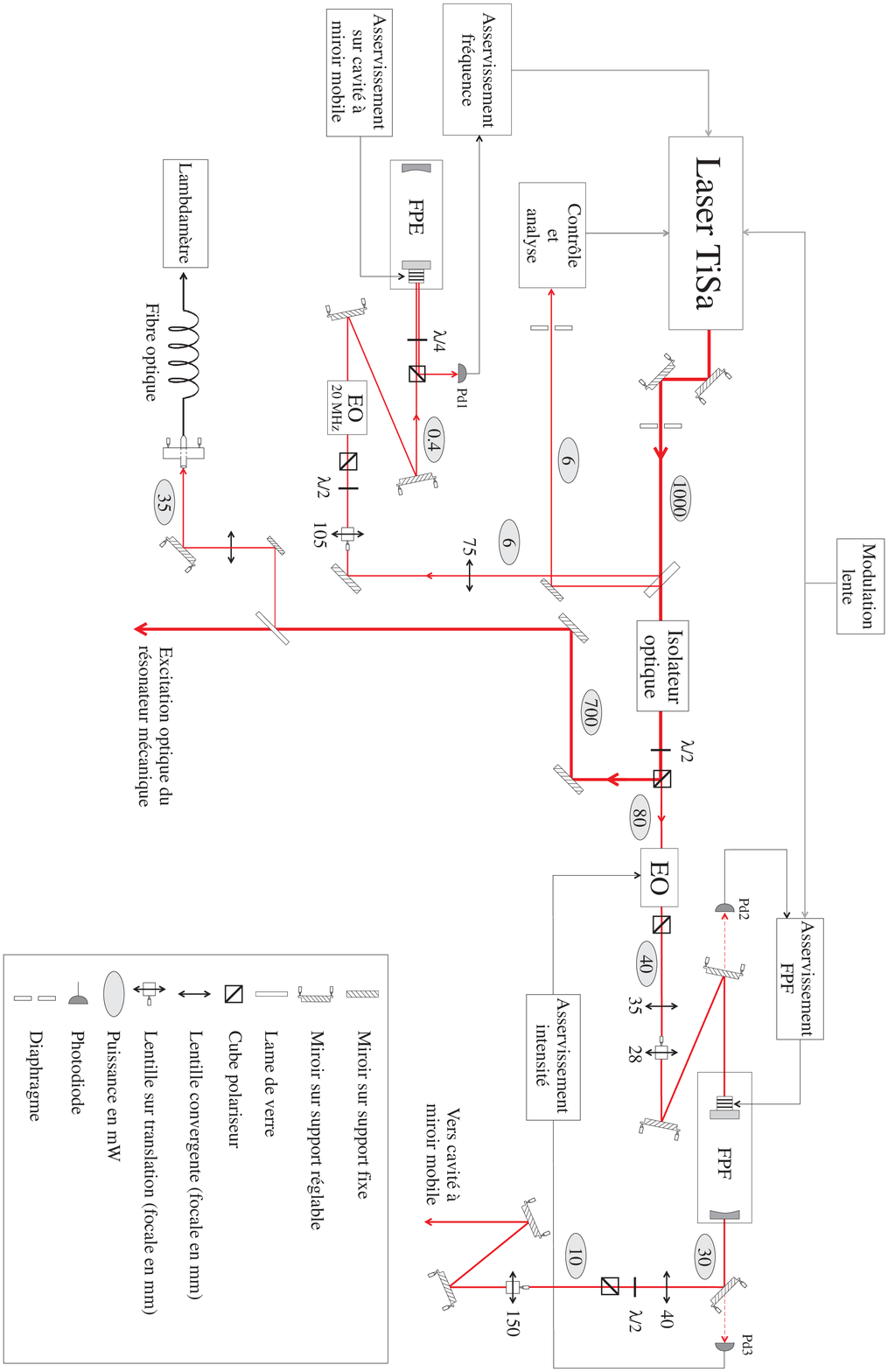,height=22cm}}
\caption{Repr\'{e}sentation sch\'{e}matique de la source laser
stabilis\'{e}e en fr\'{e}quence, en intensit\'{e} et filtr\'{e}e
spatialement }
\label{Fig_4tisasta}
\end{figure}

Nous avons pr\'{e}sent\'{e} dans les sections pr\'{e}c\'{e}dentes les
diff\'{e}rents \'{e}l\'{e}ments de la source laser. Nous allons maintenant
d\'{e}crire rapidement comment ces \'{e}l\'{e}ments sont interconnect\'{e}s.
Le sch\'{e}ma g\'{e}n\'{e}ral de la source laser est repr\'{e}sent\'{e} sur
la figure \ref{Fig_4tisasta}.

Le laser titane saphir se trouve en haut \`{a} gauche sur ce sch\'{e}ma. Il
d\'{e}livre un faisceau \`{a} $812~nm$, avec une puissance comprise entre $1$
et $1.4~Watt$. Une lame de verre \`{a} la sortie du laser permet de
pr\'{e}lever, par r\'{e}flexion sur ses deux faces, deux faisceaux de faible
puissance ($6~mW$ chacun). Le premier faisceau est envoy\'{e} vers un
dispositif de contr\^{o}le et d'analyse constitu\'{e} des deux photodiodes
utilis\'{e}es par les asservissements de l'\'{e}talon \'{e}pais et de
l'\'{e}talon mince, ainsi que d'un Fabry-Perot confocal qui permet de
v\'{e}rifier que le laser est monomode.

Le second faisceau est envoy\'{e} vers la cavit\'{e} FPE afin d'asservir en
fr\'{e}quence le laser. Il traverse tout d'abord un syst\`{e}me de deux
lentilles qui permet d'adapter le col du faisceau \`{a} celui du mode
fondamental de la cavit\'{e}, puis un att\'{e}nuateur r\'{e}glable
constitu\'{e} d'une lame demi-onde suivie d'un cube s\'{e}parateur de
polarisation. Cet att\'{e}nuateur est r\'{e}gl\'{e} de mani\`{e}re \`{a}
envoyer une puissance raisonnable dans la cavit\'{e}, de l'ordre de $400~\mu
W$. Le faisceau traverse ensuite l'\'{e}lectro-optique qui cr\'{e}e les
bandes lat\'{e}rales \`{a} $20~MHz$ et un syst\`{e}me \`{a} deux miroirs qui
permet l'alignement du faisceau sur la cavit\'{e}. Un cube s\'{e}parateur de
polarisation et une lame quart d'onde plac\'{e}s avant la cavit\'{e}
permettent de renvoyer le faisceau r\'{e}fl\'{e}chi vers la photodiode $Pd1$%
. Le signal obtenu est trait\'{e} par l'\'{e}lectronique d'asservissement
dont les diff\'{e}rentes sorties pilotent la cale pi\'{e}zo\'{e}lectrique et
l'\'{e}lectro-optique interne du laser.

Sur le trajet du faisceau principal nous avons plac\'{e} un isolateur
optique qui \'{e}vite tout retour de lumi\`{e}re vers le laser titane
saphir. Le faisceau traverse ensuite un att\'{e}nuateur variable
constitu\'{e} d'une lame demi-onde et d'un cube polariseur qui
pr\'{e}l\`{e}ve l'essentiel du faisceau (environ $700~mW$) pour le
dispositif d'excitation optique du r\'{e}sonateur m\'{e}canique, d\'{e}crit
dans la partie \ref{IV-4}. Une lame est plac\'{e}e sur le trajet de ce
faisceau de fa\c{c}on \`{a} pr\'{e}lever un faisceau de $35~mW$, envoy\'{e}
par fibre optique vers un lambdam\`{e}tre situ\'{e} sur une autre table
optique. Ce lambdam\`{e}tre permet de suivre la longueur d'onde du laser
lorsqu'on effectue un balayage de sa fr\'{e}quence. Il permet aussi de
retrouver la fr\'{e}quence de r\'{e}sonance de la cavit\'{e} \`{a} miroir
mobile. 

La partie du faisceau principal transmise par l'att\'{e}nuateur variable n'a
plus qu'une puissance de $80~mW$ et constitue le faisceau qui va interagir
avec la cavit\'{e} \`{a} miroir mobile. Il traverse l'\'{e}lectro-optique et
le cube polariseur utilis\'{e}s par l'asservissement en intensit\'{e} comme
att\'{e}nuateur pilotable. Le point de fonctionnement de cet att\'{e}nuateur
est situ\'{e} \`{a} mi-transmission : lorsque l'asservissement est
activ\'{e}, la puissance lumineuse transmise est \'{e}gale \`{a} $40~mW$.
Comme on peut le voir sur la figure \ref{Fig_4tisasta}, cet asservissement
contr\^{o}le l'intensit\'{e} apr\`{e}s la cavit\'{e} de filtrage spatial
(FPF) puisque la photodiode $Pd3$ d\'{e}tecte le faisceau transmis par le
miroir de r\'{e}flexion $95\%$ situ\'{e} apr\`{e}s cette cavit\'{e}. Cette
configuration permet de r\'{e}duire d'\'{e}ventuelles fluctuations
d'intensit\'{e} li\'{e}es \`{a} la pr\'{e}sence de la cavit\'{e} FPF.
D'autre part, cela \'{e}vite de placer l'\'{e}lectro-optique apr\`{e}s la
cavit\'{e} de filtrage, ce qui risquerait de modifier la structure spatiale
du faisceau apr\`{e}s le filtrage.

Comme pour la cavit\'{e} FPE, un dispositif de deux lentilles et deux
miroirs permet d'adapter et d'aligner le faisceau sur la cavit\'{e} FPF.
L'un des deux miroirs a un coefficient de r\'{e}flexion \'{e}gal \`{a} $95\%$%
, ce qui permet de d\'{e}tecter une partie du faisceau r\'{e}fl\'{e}chi par
la cavit\'{e}. Le signal d\'{e}livr\'{e} par la photodiode $Pd2$ est
utilis\'{e} par l'asservissement qui contr\^{o}le la r\'{e}sonance de la
cavit\'{e} FPF. Lorsque tous les asservissements sont verrouill\'{e}s, on
dispose \`{a} la sortie de la cavit\'{e} FPF d'un faisceau de $30~mW$
parfaitement stable en fr\'{e}quence, en intensit\'{e} et de structure
spatiale $TEM_{00}$.

Ce faisceau traverse un att\'{e}nuateur variable qui permet de r\'{e}gler
pr\'{e}cis\'{e}ment la puissance lumineuse envoy\'{e}e vers la cavit\'{e}
\`{a} miroir mobile et dans le dispositif de d\'{e}tection homodyne (voir
partie \ref{IV-3}). On trouve aussi \`{a} la sortie de la source laser deux
lentilles et deux miroirs qui permettent d'adapter et d'aligner le faisceau
sur la cavit\'{e} \`{a} miroir mobile.

On peut noter enfin la pr\'{e}sence sur la figure \ref{Fig_4tisasta} de
l'asservissement sur la cavit\'{e} \`{a} miroir mobile qui pilote la cale
pi\'{e}zo\'{e}lectrique de la cavit\'{e} FPE (en bas \`{a} gauche sur la
figure), ainsi que le dispositif de modulation lente (en haut sur la
figure). Ce dispositif est simplement constitu\'{e} d'un g\'{e}n\'{e}rateur
sinuso\"{\i }dal \`{a} $100~Hz$ dont l'amplitude de modulation est
r\'{e}glable. Il pilote \`{a} la fois le laser titane saphir et la
cavit\'{e} de filtrage, comme nous l'avons expliqu\'{e} dans les sections 
\ref{IV-2-2-3} et \ref{IV-2-4-2}.

L'un des points remarquables de notre source laser est sa stabilit\'{e}
\`{a} long terme. L'intensit\'{e} envoy\'{e}e vers la cavit\'{e} \`{a}
miroir mobile est uniquement d\'{e}termin\'{e}e par la tension de
r\'{e}f\'{e}rence de l'asservissement d'intensit\'{e}. La structure
spatiale, le point\'{e} et stabilit\'{e} de point\'{e} sont li\'{e}s \`{a}
la g\'{e}om\'{e}trie de la cavit\'{e} de filtrage. Ces caract\'{e}ristiques
sont donc compl\`{e}tement ind\'{e}pendantes du laser titane saphir et de
son \'{e}tat de fonctionnement. On est ainsi assur\'{e} de retrouver d'un
jour \`{a} l'autre les m\^{e}mes caract\'{e}ristiques.

Les op\'{e}rations de maintenance de la source laser sont de ce fait
tr\`{e}s simples. Il suffit d'optimiser la puissance du laser en
r\'{e}ajustant la cavit\'{e} laser (\`{a} l'aide du miroir $M_{3}$ de la
figure \ref{Fig_4tisa}) et l'\'{e}talon \'{e}pais. On s'assure ensuite de
l'alignement du faisceau sur les divers \'{e}l\'{e}ments constituant la
source laser. On dispose pour cela de deux miroirs mont\'{e}s dans des
supports microm\'{e}triques directement \`{a} la sortie du titane saphir, et
de deux diaphragmes qui permettent de rep\'{e}rer le bon alignement. Une
fois ces op\'{e}rations r\'{e}alis\'{e}es, il reste \`{a} choisir la
longueur d'onde du laser en s'aidant du lambdam\`{e}tre et d'activer tous
les asservissements pour retrouver les m\^{e}mes caract\'{e}ristiques du
faisceau de sortie.

\section{La d\'{e}tection homodyne\label{IV-3}}

\bigskip

Afin de mesurer les fluctuations de la lumi\`{e}re au niveau du bruit
quantique, un soin tout particulier doit \^{e}tre port\'{e} au syst\`{e}me
de d\'{e}tection. Le dispositif que nous allons d\'{e}crire dans cette
partie repose sur une technique d'homodynage entre le faisceau
r\'{e}fl\'{e}chi par la cavit\'{e} \`{a} miroir mobile, dont on veut mesurer
les fluctuations, et un faisceau de r\'{e}f\'{e}rence appel\'{e} {\it %
oscillateur local}. Le principe de ce type de mesure consiste \`{a} utiliser
un oscillateur local issu de la m\^{e}me source laser que le faisceau signal
de fa\c{c}on \`{a} fixer une r\'{e}f\'{e}rence de phase entre les deux
champs. Cette phase relative peut \^{e}tre contr\^{o}l\'{e}e en variant le
chemin optique suivi par l'oscillateur local. Le m\'{e}lange homodyne est
r\'{e}alis\'{e} en divisant en deux parties \'{e}gales les deux faisceaux et
en recombinant chacune des parties sur deux photodiodes de haut rendement
quantique parfaitement identiques. Les photocourants obtenus sont
amplifi\'{e}s \`{a} l'aide de deux cha\^{\i}nes d'amplification
soigneusement \'{e}quilibr\'{e}es et \`{a} faibles bruit. En utilisant un
oscillateur local beaucoup plus intense que le faisceau r\'{e}fl\'{e}chi et
en faisant la diff\'{e}rence des photocourants, on obtient un signal
proportionnel aux fluctuations de la quadrature du champ r\'{e}fl\'{e}chi en
phase avec l'oscillateur local. La visualisation et l'acquisition des
signaux est r\'{e}alis\'{e}e \`{a} l'aide d'un analyseur de spectre dont les
donn\'{e}es sont trait\'{e}es par un ordinateur.

Dans les exp\'{e}riences usuelles d'optique quantique telles que
l'observation d'\'{e}tats comprim\'{e}s du champ, la longueur du bras de
l'oscillateur local est modul\'{e}e de fa\c{c}on \`{a} parcourir l'ensemble
des quadratures du champ. Dans notre exp\'{e}rience, nous d\'{e}sirons
mesurer en continu une composante pr\'{e}cise du champ r\'{e}fl\'{e}chi par
la cavit\'{e}, en g\'{e}n\'{e}ral la quadrature de phase. C'est pourquoi
nous avons r\'{e}alis\'{e} un asservissement de la longueur du bras de
l'oscillateur local de fa\c{c}on \`{a} contr\^{o}ler la quadrature
mesur\'{e}e.

\begin{figure}[tbp]
\centerline{\psfig{figure=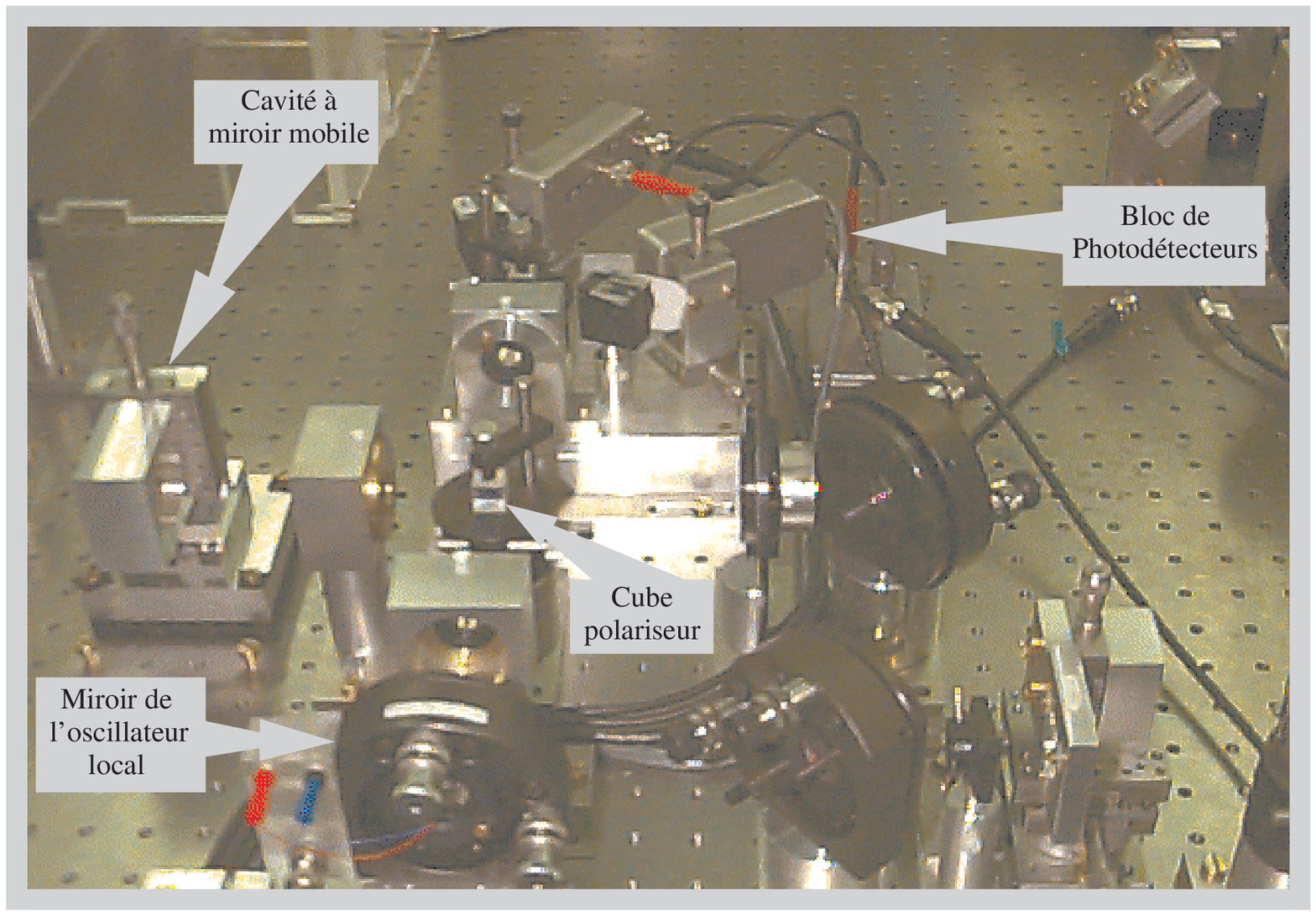,height=10cm}}
\caption{Photographie du dispositif de d\'{e}tection homodyne. On distingue
la cavit\'{e} \`{a} miroir mobile, le miroir de renvoi de l'oscillateur
local et le cube s\'{e}parateur de polarisation qui renvoie les deux
faisceaux sur le dispositif de d\'{e}tection constitu\'{e} d'un
s\'{e}parateur optique et de deux photodiodes}
\label{Fig_4phodet2}
\end{figure}

La figure \ref{Fig_4phodet2} montre une photo de l'ensemble du dispositif.
On reconna\^{\i}t sur la droite les derniers \'{e}l\'{e}ments de la source
laser d\'{e}crite dans la partie pr\'{e}c\'{e}dente : la seconde lentille
d'adaptation mont\'{e}e dans un support qui peut \^{e}tre translat\'{e} dans
les trois directions, et deux miroirs d'alignement mont\'{e}s dans des
supports Microcontrole. Le faisceau traverse ensuite une lame demi-onde et
un cube s\'{e}parateur de polarisation qui permettent de s\'{e}parer
l'oscillateur local du faisceau incident sur la cavit\'{e} \`{a} miroir
mobile. L'oscillateur local est renvoy\'{e} vers le cube \`{a} l'aide d'un
miroir mont\'{e} sur une cale pi\'{e}zo\'{e}lectrique. Deux lames quart
d'onde plac\'{e}es dans chacun des deux bras constituent avec le cube
polariseur un circulateur optique qui envoie l'oscillateur local et le
faisceau r\'{e}fl\'{e}chi par la cavit\'{e} \`{a} miroir mobile vers le
dispositif de d\'{e}tection constitu\'{e} d'un s\'{e}parateur optique et de
deux photodiodes. 

Nous allons pr\'{e}senter dans cette partie les caract\'{e}ristiques et les
performances de ce dispositif de d\'{e}tection homodyne. Nous commencerons
par une description du principe de la d\'{e}tection homodyne (section \ref
{IV-3-1}). Nous d\'{e}crirons ensuite les deux \'{e}l\'{e}ments
compl\'{e}mentaires du dispositif, c'est \`{a} dire l'oscillateur local
(section \ref{IV-3-2}) et le syst\`{e}me de d\'{e}tection \'{e}quilibr\'{e}e
(section \ref{IV-3-3}).

\subsection{Principe de la mesure homodyne\label{IV-3-1}}

\begin{figure}[tbp]
\centerline{\psfig{figure=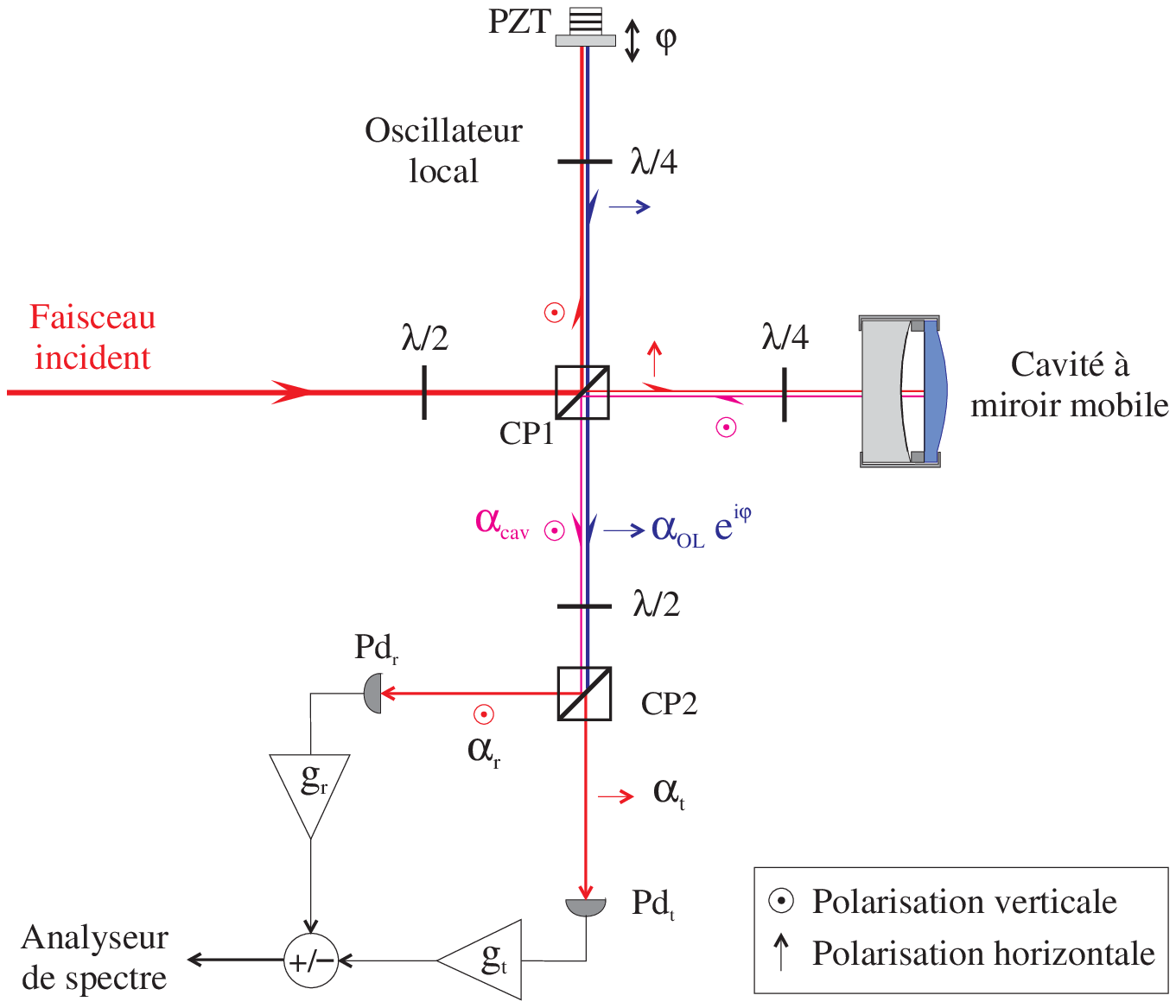,height=12cm}}
\caption{Sch\'{e}ma du dispositif de d\'{e}tection homodyne permettant de
mesurer les fluctuations quantiques du faisceau $\alpha _{cav}$
r\'{e}fl\'{e}chi par la cavit\'{e}. Le syst\`{e}me constitu\'{e} de la $%
\lambda /2$ et du cube polariseur $CP1$ permet de transmettre une petite
partie du faisceau vers la cavit\'{e}, le reste du faisceau incident
\'{e}tant r\'{e}fl\'{e}chi pour former l'oscillateur local dont le chemin
optique est control\'{e} par une cale pi\'{e}zo\'{e}lectrique (PZT). Les
lames $\lambda /4$ permettent de former un circulateur optique qui envoie
les deux faisceaux, avec des polarisations orthogonales, vers un syst\`{e}me
constitu\'{e} d'une $\lambda /2$ et du cube $CP2$, qui recombine les deux
champs au niveau des photodiodes $Pd_{t}$ et $Pd_{r}$. Les deux
photocourants sont amplifi\'{e}s et on visualise sur un analyseur de spectre
leur diff\'{e}rence ou leur somme}
\label{Fig_4detcdyn}
\end{figure}

Le sch\'{e}ma de principe de la mesure homodyne r\'{e}alis\'{e}e dans notre
exp\'{e}rience est repr\'{e}sent\'{e} sur la figure \ref{Fig_4detcdyn}. Le
faisceau incident, de polarisation lin\'{e}aire, traverse un dispositif
constitu\'{e} d'une lame demi-onde ($\lambda /2$) et d'un cube
s\'{e}parateur de polarisation ($CP1$). Ce dispositif permet, par rotation
de la $\lambda /2$ autour de son axe, de faire varier les intensit\'{e}s
relatives de l'oscillateur local et du faisceau qui interagit avec la
cavit\'{e} \`{a} miroir mobile. En pratique, la puissance du faisceau
incident est r\'{e}gl\'{e}e par l'att\'{e}nuateur variable situ\'{e} dans la
source laser, \`{a} une valeur de l'ordre de $10~mW$, et seulement $100~\mu
W $ sont envoy\'{e}s dans la cavit\'{e}. Ainsi, l'oscillateur local a une
puissance $100$ fois plus grande que celle du faisceau qui interagit avec la
cavit\'{e}. 

Avant d'\^{e}tre envoy\'{e} dans la cavit\'{e} \`{a} miroir mobile, le
faisceau transmis traverse une lame quart d'onde ($\lambda /4$) tourn\'{e}e
d'un angle de $45{{}^{\circ }}$par rapport \`{a} l'horizontale. Le faisceau
incident sur la cavit\'{e} a ainsi une polarisation circulaire. Le faisceau
r\'{e}fl\'{e}chi par la cavit\'{e} retourne alors sur le cube avec une
polarisation lin\'{e}aire verticale, et il est totalement r\'{e}fl\'{e}chi
par le cube $CP1$. De la m\^{e}me mani\`{e}re l'oscillateur local traverse
une $\lambda /4$, se r\'{e}fl\'{e}chit sur un miroir et retourne sur le cube
avec une polarisation horizontale. Il est donc totalement transmis par le
cube $CP1$. Le miroir de l'oscillateur local est mont\'{e} sur une cale
pi\'{e}zo\'{e}lectrique qui permet de contr\^{o}ler le d\'{e}phasage relatif 
$\varphi $ entre les deux faisceaux. On obtient ainsi \`{a} la sortie du
cube $CP1$ deux champs : le champ r\'{e}fl\'{e}chi par la cavit\'{e} \`{a}
miroir mobile $\alpha _{cav}$ de polarisation verticale et l'oscillateur
local $\alpha _{OL}~e^{i\varphi }$ de polarisation horizontale (les
amplitudes moyennes $\bar{\alpha}_{cav}$ et $\bar{\alpha}_{OL}$ de ces deux
champs sont choisies r\'{e}elles).

Ces deux champs sont envoy\'{e}s dans un second dispositif constitu\'{e}
d'une lame demi-onde et d'un cube s\'{e}parateur de polarisation ($CP2$). La 
$\lambda /2$ est tourn\'{e}e d'un angle de $22.5{^{\circ }}$ par rapport
\`{a} l'horizontale, ce qui a pour effet de tourner de $45{^{\circ }}$ la
polarisation des deux champs. Le cube $CP2$ s\'{e}pare alors chacun des
faisceaux en deux parties d'\'{e}gale intensit\'{e}, l'une \'{e}tant
transmise avec une polarisation horizontale et l'autre \'{e}tant
r\'{e}fl\'{e}chie avec une polarisation verticale. Notons que les
polarisations de l'oscillateur local et du faisceau r\'{e}fl\'{e}chi par la
cavit\'{e} sont identiques dans les deux voies de sortie du cube et que ces
champs peuvent maintenant interf\'{e}rer. Les champs transmis et
r\'{e}fl\'{e}chis par le cube $CP2$ s'\'{e}crivent: 
\begin{subequations}
\label{4.3.1}
\begin{eqnarray}
\alpha _{t}\left( t\right) &=&\frac{1}{\sqrt{2}}~\left[ \alpha _{cav}\left(
t\right) +\alpha _{OL}\left( t\right) ~e^{i\varphi }\right]  \label{4.3.1a}
\\
\alpha _{r}\left( t\right) &=&\frac{1}{\sqrt{2}}~\left[ \alpha _{cav}\left(
t\right) -\alpha _{OL}\left( t\right) ~e^{i\varphi }\right]  \label{4.3.1b}
\end{eqnarray}

L'ensemble $\lambda /2+CP2$ se comporte en fait comme une lame
semi-r\'{e}fl\'{e}chissante puisqu'il s\'{e}pare et m\'{e}lange les deux
faisceaux incidents. Cependant, la lame m\'{e}lange les deux champs de
m\^{e}me polarisation entrants par les deux faces de la lame. Ici, le
dispositif m\'{e}lange les deux champs entrants par le m\^{e}me c\^{o}t\'{e}
du cube et de polarisations orthogonales\cite{cube vide}. De mani\`{e}re plus g\'{e}n\'{e}rale, l'ensemble du
dispositif repr\'{e}sent\'{e} sur la figure \ref{Fig_4detcdyn} est similaire
au syst\`{e}me de mesure interf\'{e}rom\'{e}trique sch\'{e}matis\'{e} sur la
figure \ref{Fig_2prinhet}, page \pageref{Fig_2prinhet}, o\`{u} les lames
s\'{e}paratrices sont remplac\'{e}es par des cubes s\'{e}parateurs de
polarisation.

Les photocourants issus des photodiodes $Pd_{t}$ et $Pd_{r}$ plac\'{e}es
dans les deux voies de sortie du cube $CP2$ sont proportionnels aux
intensit\'{e}s $I_{t}=\alpha _{t}\alpha _{t}^{*}$ et $I_{r}=\alpha
_{r}\alpha _{r}^{*}$. A partir des relations (\ref{4.3.1}), on obtient les
expressions suivantes pour la diff\'{e}rence $I_{-}=I_{t}-I_{r}$ et la somme 
$I_{+}=I_{t}+I_{r}$ des intensit\'{e}s: 
\end{subequations}
\begin{subequations}
\label{4.3.2}
\begin{eqnarray}
I_{-}\left( t\right) &=&\alpha _{cav}\left( t\right) ~\alpha _{OL}^{*}\left(
t\right) ~e^{-i\varphi }+\alpha _{cav}^{*}\left( t\right) ~\alpha
_{OL}\left( t\right) ~e^{i\varphi }  \label{4.3.2a} \\
I_{+}\left( t\right) &=&\alpha _{cav}\left( t\right) ~\alpha
_{cav}^{*}\left( t\right) +\alpha _{OL}\left( t\right) ~\alpha
_{OL}^{*}\left( t\right)  \label{4.3.2b}
\end{eqnarray}
Ces relations montrent que l'intensit\'{e} $I_{-}$ n'est autre que le terme
d'interf\'{e}rence entre les deux champs alors que $I_{+}$ est \'{e}gal
\`{a} la somme des intensit\'{e}s $I_{cav}$ et $I_{OL}$. La partie continue (%
$DC$) des voies de d\'{e}tection permet d'acc\'{e}der aux valeurs moyennes
des intensit\'{e} $I_{-}$ et $I_{+}$: 
\end{subequations}
\begin{subequations}
\label{4.3.3}
\begin{eqnarray}
\bar{I}_{-} &=&2~\sqrt{\bar{I}_{cav}~\bar{I}_{OL}}~\cos \left( \varphi
\right)  \label{4.3.3a} \\
\bar{I}_{+} &=&\bar{I}_{cav}+\bar{I}_{OL}  \label{4.3.3b}
\end{eqnarray}
o\`{u} $\bar{I}_{cav}$ et $\bar{I}_{OL}$ sont respectivement les
intensit\'{e}s moyennes du champ r\'{e}fl\'{e}chi par la cavit\'{e} \`{a}
miroir mobile et de l'oscillateur local. La relation (\ref{4.3.3a}) montre
que la diff\'{e}rence des intensit\'{e}s moyennes des champs d\'{e}crit des
franges d'interf\'{e}rences lorsqu'on varie le d\'{e}phasage relatif $%
\varphi $ entre les deux champs. En particulier, l'intensit\'{e} $\bar{I}%
_{-} $ est nulle lorsque les deux champs sont en quadrature de phase ($%
\varphi =\pi /2$), et elle est maximale en valeur absolue lorsque les champs
sont en phase ou en opposition de phase ($\varphi =0$ ou $\pi $). Nous
verrons dans la section suivante (\ref{IV-3-2}) comment on utilise les
intensit\'{e}s moyennes $\bar{I}_{-}$ et $\bar{I}_{+}$ pour contr\^{o}ler la
longueur du bras de l'oscillateur local, ce qui permet de choisir la
quadrature du champ \`{a} mesurer.

Les fluctuations des photocourants sont li\'{e}es aux fluctuations $\delta
\alpha _{cav}$ et $\delta \alpha _{OL}$ des champs. En lin\'{e}arisant les
relations (\ref{4.3.2}) autour des valeurs moyennes, on trouve l'expression
suivante pour les fluctuations de la diff\'{e}rence des intensit\'{e}s $%
\delta I_{-}$: 
\end{subequations}
\begin{equation}
\begin{array}{r}
\delta I_{-}\left( t\right) =\bar{\alpha}_{OL}~\left[ \delta \alpha
_{cav}\left( t\right) ~e^{-i\varphi }+\delta \alpha _{cav}^{*}\left(
t\right) ~e^{i\varphi }\right] \\ 
\\ 
+\bar{\alpha}_{cav}~\left[ \delta \alpha _{OL}\left( t\right) ~e^{i\varphi
}+\delta \alpha _{OL}^{*}\left( t\right) ~e^{-i\varphi }\right]
\end{array}
\label{4.3.4}
\end{equation}
Les termes entre crochets repr\'{e}sentent les fluctuations des quadratures
d'angle $\pm \varphi $ des champs $\alpha _{cav}$ et $\alpha _{OL}$, que
nous avons d\'{e}finies dans la partie 2.2 du chapitre $2$ (voir
\'{e}quation 2.28). La quadrature de chacun des deux champs est
pond\'{e}r\'{e}e par un terme qui n'est autre que l'amplitude moyenne de
l'autre champ. Lorsque l'intensit\'{e} de l'oscillateur local est grande par
rapport \`{a} celle du champ r\'{e}fl\'{e}chi par la cavit\'{e}, le second
terme dans l'\'{e}quation (\ref{4.3.4}) est n\'{e}gligeable devant le
premier. Le signal $\delta I_{-}$ est ainsi proportionnel aux fluctuations
de la quadrature d'angle $\varphi $ du champ r\'{e}fl\'{e}chi $\alpha _{cav}$%
, et le spectre $S_{-}\left[ \Omega \right] $ des fluctuations $\delta I_{-}$
s'\'{e}crit:

\begin{equation}
S_{-}\left[ \Omega \right] =\bar{I}_{OL}~S_{\varphi }^{cav}\left[ \Omega
\right]  \label{4.3.5}
\end{equation}
o\`{u} $S_{\varphi }^{cav}\left[ \Omega \right] $ est le spectre de bruit de
la quadrature d'angle $\varphi $ du champ $\alpha _{cav}$ sortant de la
cavit\'{e} \`{a} miroir mobile. La d\'{e}tection homodyne permet ainsi
d'avoir acc\`{e}s au spectre de bruit de n'importe quelle quadrature du
champ $\alpha _{cav}$, en faisant varier la longueur du bras de
l'oscillateur local. Ce spectre est d'autre part amplifi\'{e} par
l'intensit\'{e} moyenne de l'oscillateur local. La calibration du spectre $%
S_{\varphi }^{cav}\left[ \Omega \right] $ est r\'{e}alis\'{e}e simplement en
coupant le faisceau r\'{e}fl\'{e}chi par la cavit\'{e} \`{a} l'aide d'un
cache. Le champ $\alpha _{cav}$ est alors remplac\'{e} par le vide et on
mesure le bruit de photon standard $S_{-}\left[ \Omega \right] =\bar{I}_{OL}$%
.

Notons que les r\'{e}sultats pr\'{e}sent\'{e}s ici supposent un parfait
\'{e}quilibrage entre les deux voies de d\'{e}tection, aussi bien au niveau
de la s\'{e}paration optique des faisceaux r\'{e}fl\'{e}chis et transmis par
le cube $CP2$, que pour le rendement quantique des photodiodes ou les gains
\'{e}lectroniques. Tout d\'{e}s\'{e}quilibre se traduit par une
contamination du signal mesur\'{e} $S_{-}\left[ \Omega \right] $ par le
bruit d'intensit\'{e} de l'oscillateur local. De plus, un
d\'{e}s\'{e}quilibre optique est \'{e}quivalent \`{a} des pertes optiques
qui ont tendance \`{a} ramener le signal mesur\'{e} vers le bruit de photon
standard. Nous \'{e}tudierons en d\'{e}tail l'\'{e}quilibrage du syst\`{e}me
de d\'{e}tection dans la section \ref{IV-3-3}.

Notons pour terminer que l'on dispose en fait d'un dispositif
soustracteur-sommateur qui permet d'envoyer dans l'analyseur de spectre les
signaux $I_{-}$ et $I_{+}$ (voir figure \ref{Fig_4detcdyn}). Seul le signal $%
I_{-}$ est utile pour la d\'{e}tection homodyne. Mais l'ensemble des
\'{e}l\'{e}ments constitu\'{e}s de la lame $\lambda /2$, du cube $CP2$ et
des deux voies de d\'{e}tection peut \^{e}tre consid\'{e}r\'{e} comme un
dispositif de d\'{e}tection \'{e}quilibr\'{e}e, utilisable \`{a} d'autres
fins que la mesure homodyne. C'est pourquoi nous avons con\c{c}u pour ces
\'{e}l\'{e}ments un dispositif autonome et compact (d\'{e}crit en d\'{e}tail
dans la section \ref{IV-3-3}), capable de fournir aussi bien la
diff\'{e}rence $I_{-}$ que la somme $I_{+}$ des intensit\'{e}s. Ce
dispositif nous a servi par exemple pour mesurer le bruit d'intensit\'{e} du
faisceau issu de la source laser (voir section \ref{IV-2-3-2}, page \pageref
{IV-2-3-2}). Dans ce cas, un seul faisceau est envoy\'{e} dans le
syst\`{e}me de d\'{e}tection. Le signal $I_{+}$ fournit le bruit
d'intensit\'{e} du faisceau, alors que $I_{-}$ donne le bruit de photon
standard associ\'{e}.

\subsection{L'oscillateur local\label{IV-3-2}}

L'efficacit\'{e} de la mesure repose sur un effet d'interf\'{e}rence entre
le champ sortant de la cavit\'{e} et l'oscillateur local. Il est de ce fait
indispensable de bien contr\^{o}ler deux param\`{e}tres : le d\'{e}phasage
relatif entre les deux faisceaux et leur recouvrement spatial au niveau des
d\'{e}tecteurs. Le d\'{e}phasage relatif $\varphi $ d\'{e}finit, comme nous
l'avons vu, la quadrature du champ que l'on mesure \`{a} l'aide de la
d\'{e}tection homodyne. Pour contr\^{o}ler pr\'{e}cis\'{e}ment ce
d\'{e}phasage, nous avons r\'{e}alis\'{e} un dispositif d'asservissement qui
agit sur le chemin optique suivi par l'oscillateur local.

Un mauvais recouvrement spatial entre les deux faisceaux se traduit par des
pertes au niveau de la mesure, puisque seule la partie du faisceau
r\'{e}fl\'{e}chi qui se projette sur l'oscillateur local interf\`{e}re pour
donner un signal. Il faut donc optimiser \`{a} la fois la position et
l'inclinaison du miroir de renvoi de l'oscillateur local.

\subsubsection{Asservissement de la longueur du bras\label{IV-3-2-1}}

Les fluctuations du d\'{e}phasage $\varphi $ sont essentiellement li\'{e}es
aux fluctuations d'indice de l'air dans lequel se propagent les deux
faisceaux et aux vibrations m\'{e}caniques des diff\'{e}rents supports sur
lesquels sont mont\'{e}s les \'{e}l\'{e}ments optiques. On observe aussi une
d\'{e}rive lente du d\'{e}phasage due \`{a} des effets de dilatation
thermique de ces supports. Pour compenser ces fluctuations de phase
relative, nous avons r\'{e}alis\'{e} un asservissement qui agit sur la
longueur du bras de l'oscillateur local.

Le signal d'erreur est obtenu en utilisant les tensions issues des voies
basse fr\'{e}quence ($DC$) des deux cha\^{\i}nes de d\'{e}tection. On
dispose ainsi de deux tensions $V_{t}$ et $V_{r}$ proportionnelles aux
intensit\'{e}s moyennes transmise et r\'{e}fl\'{e}chie par le cube CP2, avec
le m\^{e}me coefficient de proportionnalit\'{e} not\'{e} $g_{DC}$ lorsque
les deux voies sont \'{e}quilibr\'{e}es. On peut alors \'{e}crire la
diff\'{e}rence et la somme de ces tensions sous une forme similaire aux
relations (\ref{4.3.3}):

\begin{subequations}
\label{4.3.6}
\begin{eqnarray}
V_{-} &=&2g_{DC}\sqrt{\bar{I}_{cav}\bar{I}_{OL}}~\cos \left( \varphi \right)
\label{4.3.6a} \\
V_{+} &=&g_{DC}~\left( \bar{I}_{cav}+\bar{I}_{OL}\right)  \label{4.3.6b}
\end{eqnarray}
On voit que pour fixer le point de fonctionnement de
l'interf\'{e}rom\`{e}tre d\'{e}fini par le d\'{e}phasage $\varphi $, il
suffit de fixer la tension $V_{-}$. En fait, on compare $V_{-}$ \`{a} une
fraction de $V_{+}$, le signal d'erreur \'{e}tant de la forme $V_{-}-\kappa
~V_{+}$ o\`{u} $\kappa $ peut varier de $-1$ \`{a} $1$. L'avantage de cette
approche est qu'elle rend le fonctionnement de l'asservissement
ind\'{e}pendant d'\'{e}ventuelles fluctuations d'intensit\'{e} du faisceau
incident. En effet le d\'{e}phasage relatif $\varphi $ obtenu par
l'annulation du signal d'erreur s'\'{e}crit:

\end{subequations}
\begin{equation}
\cos \left( \varphi \right) =\kappa ~\frac{\bar{I}_{cav}+\bar{I}_{OL}}{2%
\sqrt{\bar{I}_{cav}\bar{I}_{OL}}}  \label{4.3.7}
\end{equation}

\begin{figure}[tbp]
\centerline{\psfig{figure=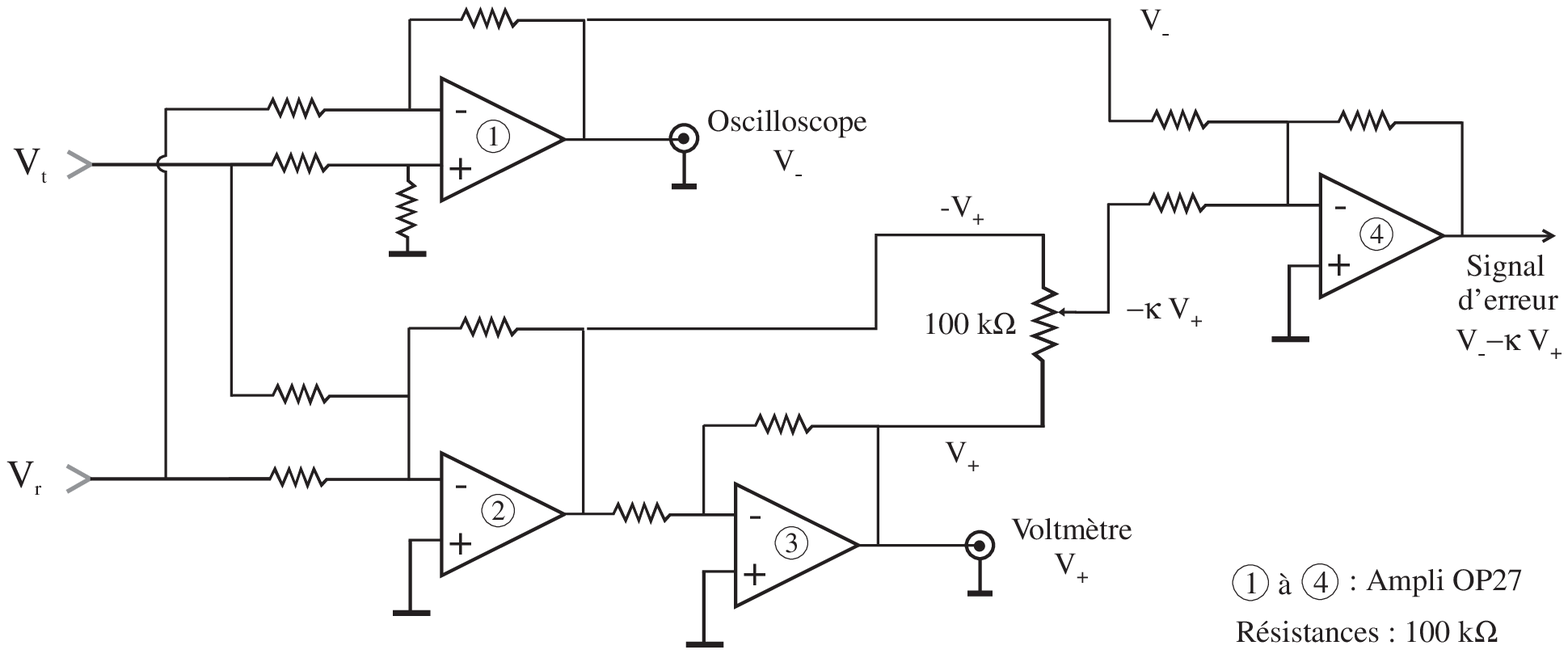,height=65mm}}
\caption{Sch\'{e}ma du pr\'{e}amplificateur qui fournit, \`{a} partir des
tensions DC des deux voies de d\'{e}tection, le signal d'erreur de
l'asservissement de la longueur du bras de l'oscillateur local. Deux voies
de sortie permettent de visualiser les franges d'interf\'{e}rences ($%
V_{-}=V_{t}-V_{r}$) sur un oscilloscope ainsi que l'intensit\'{e} totale ($%
V_{+}=V_{t}+V_{r}$) arrivant sur les deux d\'{e}tecteurs $Pd_{t}$ et $Pd_{r}$%
}
\label{Fig_4interse}
\end{figure}

Pour obtenir un tel signal d'erreur, on utilise le montage
repr\'{e}sent\'{e} sur la figure \ref{Fig_4interse}. Le signal
d'interf\'{e}rence $V_{-}$ est obtenu en utilisant un amplificateur (%
\ding{172}) qui fonctionne en soustracteur. Ce signal est utilis\'{e} pour
visualiser l'amplitude des franges d'interf\'{e}rences sur un oscilloscope.
A l'aide des amplificateurs \ding{173} et \ding{174}, qui fonctionnent
respectivement en additionneur et en inverseur, on obtient les tensions $%
-V_{+}$ et $V_{+}$. La tension $V_{+}$ est envoy\'{e}e sur un voltm\`{e}tre,
ce qui permet de contr\^{o}ler l'intensit\'{e} moyenne totale arrivant sur
les photodiodes. Le potentiom\`{e}tre connect\'{e} aux potentiels $V_{+}$ et 
$-V_{+}$ permet d'obtenir une tension $-\kappa ~V_{+}$ o\`{u} le coefficient 
$\kappa $ varie de $-1$ \`{a} $1$. Ce signal de r\'{e}f\'{e}rence est alors
compar\'{e} au signal d'interf\'{e}rence $V_{-}$ \`{a} l'aide de
l'amplificateur \ding{175} qui fonctionne en additionneur. On obtient alors
le signal d'erreur voulu, c'est \`{a} dire $V_{-}-\kappa ~V_{+}$.

Afin de r\'{e}aliser un contr\^{o}le efficace de la phase relative, nous
avons construit un asservissement \`{a} deux boucles en parall\`{e}le. La
premi\`{e}re boucle agit sur une petite cale pi\'{e}zo\'{e}lectrique de $%
1~mm $ d'\'{e}paisseur sur laquelle est coll\'{e}e un miroir de petite
taille ($7.75~mm$ de diam\`{e}tre et $4~mm$ d'\'{e}paisseur). L'ensemble
pr\'{e}sente une r\'{e}ponse dynamique relativement rapide puisque la
fr\'{e}quence de r\'{e}sonance se situe aux alentours de $5~kHz$. Cette
boucle permet de contr\^{o}ler la longueur du bras de l'oscillateur local
avec une amplitude maximale de l'ordre de $0.2~\mu m$ (soit $\lambda /4$).
La seconde boucle d'asservissement permet de compenser les d\'{e}rives
lentes du d\'{e}phasage en agissant sur un v\'{e}rin pi\'{e}zo\'{e}lectrique
Newport $ESA133OPT-01$ adapt\'{e} \`{a} la platine de translation sur
laquelle repose le support du miroir. Ce v\'{e}rin permet d'effectuer des
excursions de $30~\mu m$. Bien s\^{u}r, \'{e}tant donn\'{e} le poids de
l'ensemble du support et du miroir mont\'{e} sur la platine de translation,
ce dispositif ne contr\^{o}le la phase relative qu'\`{a} des fr\'{e}quences
basses, inf\'{e}rieures \`{a} quelques Hertz. 
\begin{figure}[tbp]
\centerline{\psfig{figure=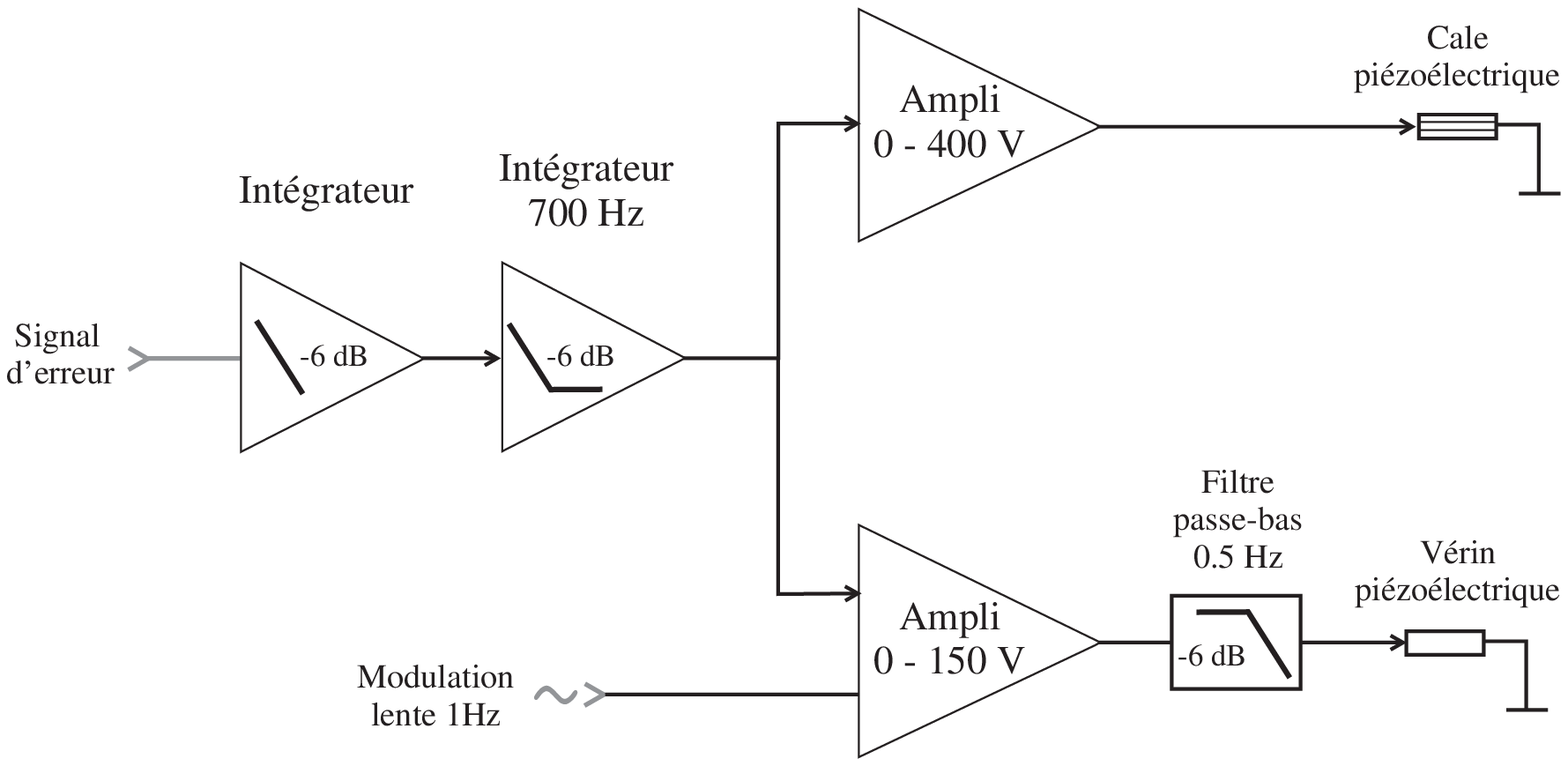,height=7cm}}
\caption{Dispositif d'asservissement du bras de l'oscillateur local. Le
signal d'erreur, apr\`{e}s int\'{e}gration, est divis\'{e} en deux pour
piloter d'une part la cale pi\'{e}zo\'{e}lectrique via un amplificateur $0$-$%
400~V$ et d'autre part un v\'{e}rin pi\'{e}zo\'{e}lectrique via un
amplificateur $0$-$150~V$ et un filtre passe-bas. Une rampe de modulation de 
$1~Hz$ permet de visualiser les franges d'interf\'{e}rences}
\label{Fig_4interas}
\end{figure}

Le sch\'{e}ma du dispositif d'asservissement est repr\'{e}sent\'{e} sur la
figure \ref{Fig_4interas}. Le signal d'erreur obtenu \`{a} l'aide du montage
de la figure \ref{Fig_4interse} est amplifi\'{e} \`{a} basse fr\'{e}quence
par un int\'{e}grateur de pente globale $-6~dB/octave$ et par un
int\'{e}grateur pour des fr\'{e}quences inf\'{e}rieures \`{a} $700~Hz$. Ce
signal pilote la petite cale pi\'{e}zo\'{e}lectrique par l'interm\'{e}diaire
d'un amplificateur $0$-$400~V$. Ce signal contr\^{o}le aussi le v\'{e}rin
pi\'{e}zo\'{e}lectrique par l'interm\'{e}diaire d'un amplificateur haute
tension $0$-$1000~V$ brid\'{e} \`{a} $150~V$ de fa\c{c}on \`{a} ne pas
endommager le v\'{e}rin. Etant donn\'{e} les gains des amplificateurs et la
diff\'{e}rence de sensibilit\'{e} entre la petite cale et le v\'{e}rin, le
gain de la boucle pilotant le v\'{e}rin est environ $100$ fois plus grand
\`{a} basse fr\'{e}quence que celui de la boucle agissant sur la petite
cale. Ainsi l'excursion de la cale reste mod\'{e}r\'{e}e \`{a} basse
fr\'{e}quence et les d\'{e}rives lentes de la phase relative sont
contr\^{o}l\'{e}es par le v\'{e}rin. Un filtre passe-bas de fr\'{e}quence de
coupure \'{e}gale \`{a} $0.5~Hz$ est ins\'{e}r\'{e} entre l'amplificateur et
le v\'{e}rin. La petite cale assure ainsi le contr\^{o}le de la phase
relative pour les fr\'{e}quences sup\'{e}rieures \`{a} $20~Hz$ environ. 
\begin{figure}[tbp]
\centerline{\psfig{figure=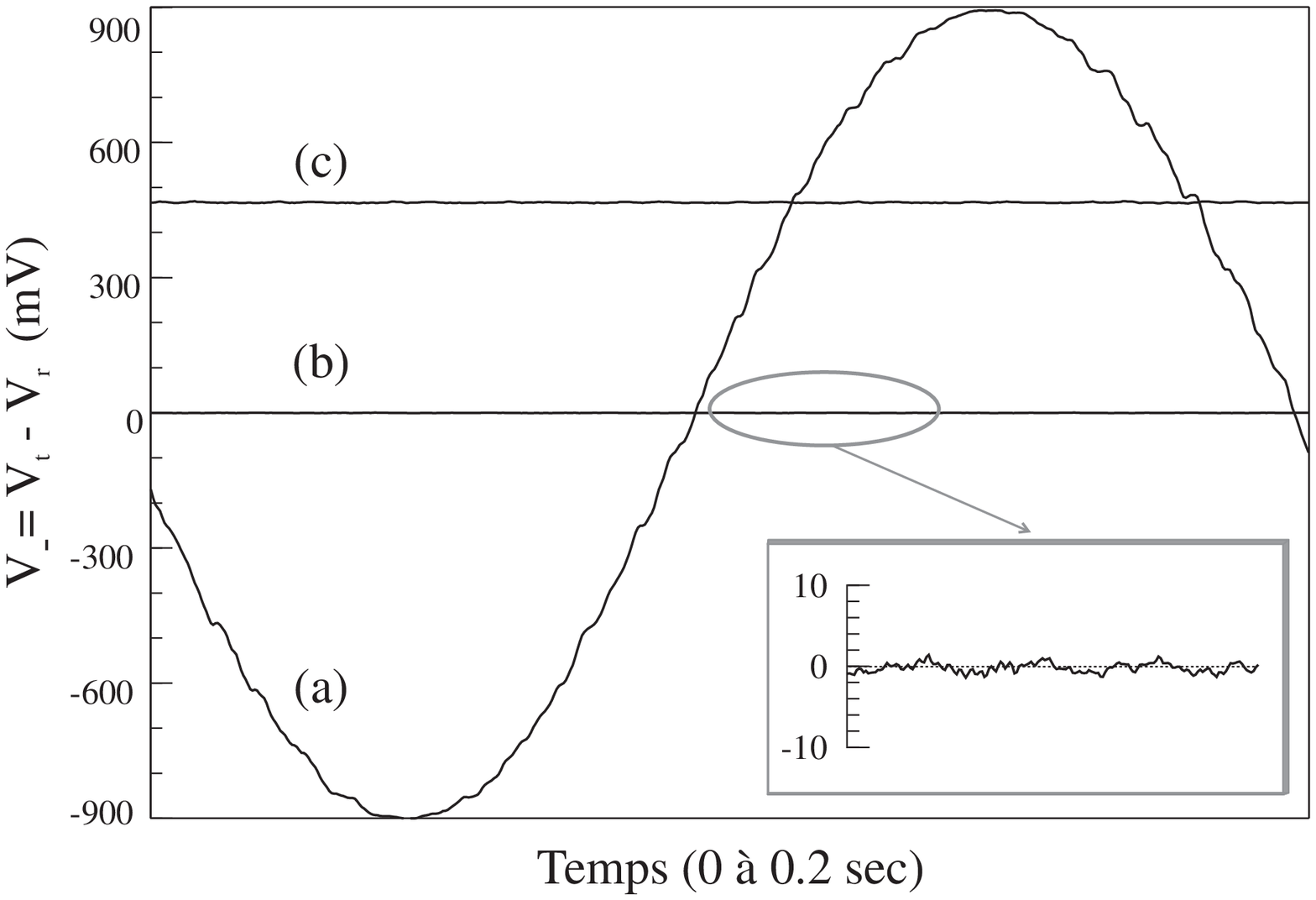,height=8cm}}
\caption{Evolution temporelle du signal d'interf\'{e}rence $%
V_{-}=V_{t}-V_{r} $. La trace (a) repr\'{e}sente le signal lorsqu'on
applique une rampe de modulation sur le v\'{e}rin pi\'{e}zo\'{e}lectrique.
Les traces (b) et (c) repr\'{e}sentent le signal d'interf\'{e}rence lorsque
l'asservissement est activ\'{e} pour des valeurs de la tension de
r\'{e}f\'{e}rence $\kappa ~V_{+} $ \'{e}gales \`{a} $0$ et $490~mV$.
L'insert correspond \`{a} un agrandissement vertical de la courbe (b) par un
facteur $50$. Ces traces correspondent \`{a} des valeurs des intensit\'{e}s
(en volts) de l'oscillateur local et du faisceau r\'{e}fl\'{e}chi \'{e}gales
respectivement \`{a} $5.68~V$ et $38~mV$, la tension cr\^{e}te-cr\^{e}te des
franges d'interf\'{e}rences \'{e}tant \'{e}gale \`{a} $1.8~V$}
\label{Fig_4inter3}
\end{figure}

Notons enfin la pr\'{e}sence d'une entr\'{e}e modulation sur l'amplificateur
pilotant le v\'{e}rin pi\'{e}zo\'{e}lectrique. Cette entr\'{e}e permet de
moduler \`{a} tr\`{e}s basse fr\'{e}quence (typiquement de l'ordre du Hertz)
la longueur du bras de l'oscillateur local afin de balayer les franges
d'interf\'{e}rences que l'on visualise gr\^{a}ce au signal $%
V_{-}=V_{t}-V_{r} $ (\'{e}quation \ref{4.3.6a}).

La figure \ref{Fig_4inter3} montre le r\'{e}sultat de l'asservissement sur
le signal d'interf\'{e}rence $V_{-}$. La trace (a) repr\'{e}sente les
franges d'interf\'{e}rences lorsqu'on module le v\'{e}rin
pi\'{e}zo\'{e}lectrique. On parcourt en une oscillation toutes les
quadratures du champ r\'{e}fl\'{e}chi. L'amplitude cr\^{e}te-cr\^{e}te de
l'oscillation d\'{e}termine, comme nous le verrons par la suite, le taux de
recouvrement spatial au niveau des photodiodes entre l'oscillateur local et
le faisceau r\'{e}fl\'{e}chi par la cavit\'{e}. Les r\'{e}sultats de la
figure \ref{Fig_4inter3} ont \'{e}t\'{e} obtenus apr\`{e}s avoir
soigneusement optimis\'{e} le recouvrement spatial, ce qui correspond \`{a}
une amplitude cr\^{e}te-cr\^{e}te de l'oscillation \'{e}gale \`{a} $1.8~V$.
Les intensit\'{e}s de l'oscillateur local et du faisceau r\'{e}fl\'{e}chi
sont mesur\'{e}es en utilisant le voltm\`{e}tre \`{a} la sortie de la voie $%
V_{+}$ du dispositif d\'{e}crit sur la figure \ref{Fig_4interse}. On obtient 
$V_{OL}=5.68~V$ et $V_{cav}=37~mV$ en masquant respectivement le faisceau
r\'{e}fl\'{e}chi par la cavit\'{e} puis l'oscillateur local. Les traces (b)
et (c) de la figure \ref{Fig_4inter3} repr\'{e}sentent la tension $V_{-}$
lorsque l'asservissement est activ\'{e}, pour des valeurs de la tension de
r\'{e}f\'{e}rence $\kappa ~V_{+}$ \'{e}gales respectivement \`{a} $0$ et $%
490~mV$ (d\'{e}phasage relatif $\varphi $ de $\pi /2$ et $\pi /3$ environ).
On voit que l'asservissement permet de fixer pr\'{e}cis\'{e}ment le
d\'{e}phasage $\varphi $, et par cons\'{e}quent la quadrature mesur\'{e}e du
champ r\'{e}fl\'{e}chi. Afin d'\'{e}valuer l'efficacit\'{e} de
l'asservissement, la trace (b) est agrandie selon l'axe vertical par un
facteur $50$ dans l'insert en bas \`{a} droite de la figure. On mesure alors
une tension r\'{e}siduelle de $3~mV$ ce qui correspond, d'apr\`{e}s la
relation (\ref{4.3.6a}), \`{a} des fluctuations maximales de la quadrature
de phase ($\varphi =\pi /2$) de l'ordre de $0.2$ degr\'{e}.

En variant la tension de r\'{e}f\'{e}rence $\kappa ~V_{+}$, il est possible
de s'asservir \`{a} diff\'{e}rentes valeurs du d\'{e}phasage $\varphi $,
comme le montrent les traces (b) et (c). On ne peut pas cependant
s'approcher aussi pr\`{e}s que l'on veut des extrema de la figure
d'interf\'{e}rence (trace a) puisqu'en ces points la pente du signal
d'erreur est nulle. Cependant, ces positions correspondent \`{a} un
d\'{e}phasage $\varphi $ \'{e}gal \`{a} $0$ ou $\pi $, c'est-\`{a}-dire
\`{a} une mesure de la quadrature d'amplitude du champ r\'{e}fl\'{e}chi par
la cavit\'{e}. Comme nous l'avons vu dans la section \ref{IV-2-3-2}, on peut
r\'{e}aliser une mesure directe de ce bruit d'intensit\'{e} en supprimant
l'oscillateur local et en utilisant la d\'{e}tection \'{e}quilibr\'{e}e en
position sommateur.

\subsubsection{Recouvrement spatial\label{IV-3-2-2}}

Le recouvrement spatial entre le faisceau r\'{e}fl\'{e}chi par la cavit\'{e}
\`{a} miroir mobile et l'oscillateur local d\'{e}finit la proportion de
l'intensit\'{e} lumineuse des deux faisceaux qui interf\`{e}re au niveau des
photodiodes pour donner le signal. Un mauvais recouvrement se traduit par
des pertes pour le faisceau r\'{e}fl\'{e}chi et induit une d\'{e}gradation
de la mesure de la quadrature. Il se traduit aussi par une r\'{e}duction des
interf\'{e}rences entre les deux faisceaux. La relation (\ref{4.3.6a}) qui
donne le signal d'interf\'{e}rence $V_{-}$ ne tient pas compte du
recouvrement entre les deux faisceaux. $V_{-}$ s'\'{e}crit en fait: 
\begin{equation}
V_{-}=\gamma \times 2~g_{DC}~\sqrt{\bar{I}_{cav}\bar{I}_{OL}}~\cos \left(
\varphi \right)  \label{4.3.8}
\end{equation}
o\`{u} le param\`{e}tre $\gamma $ est \'{e}gal \`{a} l'int\'{e}grale de
recouvrement spatial entre les deux champs.

La proc\'{e}dure d'adaptation consiste \`{a} positionner le miroir de renvoi
de l'oscillateur local exactement au niveau du col du faisceau incident de fa%
\c{c}on \`{a} adapter les structures des champs, et \`{a} orienter ce miroir
pour maximiser les interf\'{e}rences. La platine de translation et le
support Microcontrole sur lesquels est fix\'{e} le miroir permettent
d'optimiser ces param\`{e}tres. La proc\'{e}dure de r\'{e}glage est
r\'{e}alis\'{e}e en appliquant une rampe de tension sur le v\'{e}rin de
translation, et en jouant sur la position et l'orientation du miroir de
mani\`{e}re \`{a} maximiser l'amplitude cr\^{e}te-cr\^{e}te du signal
d'interf\'{e}rence (trace $a$ de la figure \ref{Fig_4inter3}).

Afin de d\'{e}terminer le coefficient de recouvrement $\gamma $, on mesure
cette amplitude cr\^{e}te-cr\^{e}te dont l'expression est donn\'{e}e par $%
\gamma \times 4~g_{DC}~\sqrt{\bar{I}_{cav}\bar{I}_{OL}}$. On mesure aussi
\`{a} l'aide du voltm\`{e}tre les tensions $V_{cav}=g_{DC}~\bar{I}_{cav}$ et 
$V_{OL}=g_{DC}~\bar{I}_{OL}$ en masquant respectivement l'oscillateur local
et le faisceau r\'{e}fl\'{e}chi par la cavit\'{e}. Lorsque l'adaptation
spatiale est optimis\'{e}e, on obtient d'apr\`{e}s la courbe (a) de la
figure \ref{Fig_4inter3} une amplitude cr\^{e}te-cr\^{e}te de $1.8~V$ pour
une tension $V_{cav}$ de $38~mV$ et une tension $V_{OL}$ de $5.68~V$. Ces
valeurs correspondent \`{a} un recouvrement spatial $\gamma $ \'{e}gal \`{a} 
$97\%$.

\subsection{La d\'{e}tection \'{e}quilibr\'{e}e\label{IV-3-3}}

Comme nous l'avons indiqu\'{e} au d\'{e}but de cette partie, nous avons con%
\c{c}u un dispositif de d\'{e}tection \'{e}quilibr\'{e}e autonome et
compact. Ceci pr\'{e}sente l'avantage de pouvoir utiliser ce dispositif pour
d'autres applications que la mesure homodyne. D'autre part ceci permet
d'obtenir un alignement pr\'{e}cis et stable des diff\'{e}rents
\'{e}l\'{e}ments optiques. La compacit\'{e} et la rigidit\'{e} de l'ensemble
conf\`{e}re au dispositif une bonne stabilit\'{e} et un bon contr\^{o}le des
param\`{e}tres tels que l'\'{e}quilibrage optique et \'{e}lectronique des
deux voies de d\'{e}tection. 
\begin{figure}[tbp]
\centerline{\psfig{figure=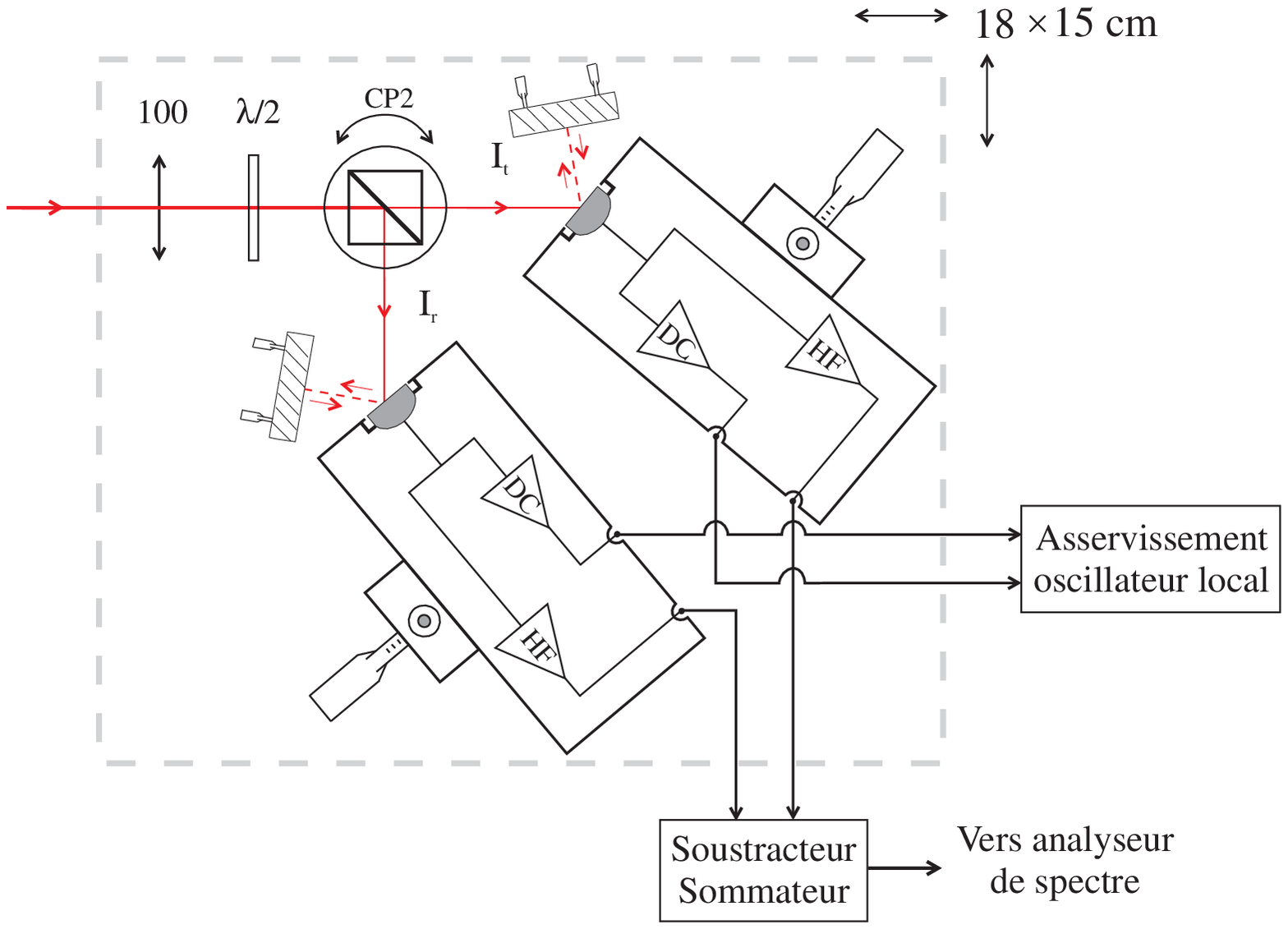,height=10cm}}
\caption{Sch\'{e}ma d'ensemble du bloc de d\'{e}tection \'{e}quilibr\'{e}e.
Tous les \'{e}l\'{e}ments optiques (lentille, lame demi-onde, cube
polariseur, miroirs de renvoi) et les deux d\'{e}tecteurs mont\'{e}s sur
translations sont fix\'{e}s sur une plaque en dural de dimension $18\times
15~cm$. Des c\^{a}bles BNC permettent d'envoyer les signaux $DC$ et $HF$
vers le dispositif d'asservissement du bras de l'oscillateur local et vers
le syst\`{e}me soustracteur-sommateur}
\label{Fig_4blocpd}
\end{figure}

Le sch\'{e}ma du montage est repr\'{e}sent\'{e} sur la figure \ref
{Fig_4blocpd}. Tous les \'{e}l\'{e}ments optiques de la d\'{e}tection
\'{e}quilibr\'{e}e sont fix\'{e}s sur une plaque en dural de dimension $%
18\times 15~cm$. Le faisceau incident traverse tout d'abord une lentille de
focale $100~mm$ qui permet de faire converger le faisceau au niveau des
surfaces photosensibles des deux photodiodes. Il traverse ensuite le
dispositif de s\'{e}paration optique constitu\'{e} d'une lame $\lambda /2$
et du cube $CP2$ qui divise le faisceau incident en deux parties d'\'{e}gale
intensit\'{e}. La lame demi-onde est plac\'{e}e dans un support tournant qui
permet de r\'{e}gler pr\'{e}cis\'{e}ment les axes de la lame par rapport aux
polarisations incidentes de l'oscillateur local et du faisceau
r\'{e}fl\'{e}chi par la cavit\'{e}. Le cube s\'{e}parateur de polarisation $%
CP2$ est fix\'{e} sur un support qui permet de l'aligner sur le faisceau
incident selon trois axes de rotation. Lorsque le cube est bien align\'{e},
il pr\'{e}sente un taux d'extinction de l'ordre de $3~10^{-4}$ (soit $%
300~ppm $).

Chaque photodiode et son pr\'{e}amplificateur sont plac\'{e}s dans un
bo\^{\i }tier m\'{e}tallique mont\'{e} sur deux translations verticale et
horizontale afin de centrer les photodiodes sur le faisceau. Les photodiodes
sont log\'{e}es dans des pi\`{e}ces en laiton qui assurent un excellent
contact de masse en reliant directement la structure m\'{e}tallique de la
photodiode au bo\^{\i }tier. Ceci nous a permis de r\'{e}duire tr\`{e}s
nettement les sources de bruit parasite \`{a} haute fr\'{e}quence. Ces
pi\`{e}ces m\'{e}talliques permettent d'autre part de fixer de mani\`{e}re
identique les photodiodes par rapport \`{a} leur bo\^{\i }tier. Les deux
bo\^{\i }tiers sont inclin\'{e}s par rapport au faisceau de fa\c{c}on \`{a}
r\'{e}cup\'{e}rer l'intensit\'{e} r\'{e}fl\'{e}chie par la surface
photosensible. Une part importante des pertes optiques au niveau des
photod\'{e}tecteurs est en effet due \`{a} la r\'{e}flexion du faisceau sur
la photodiode. On utilise donc des miroirs mont\'{e}s sur des supports
r\'{e}glables qui renvoient la lumi\`{e}re r\'{e}fl\'{e}chie sur la
photodiode. Les pr\'{e}amplificateurs sont constitu\'{e}s d'une voie basse
fr\'{e}quence ($DC$) qui fournit en particulier les valeurs moyennes des
intensit\'{e}s utilis\'{e}es par l'asservissement de l'oscillateur local, et
une voie haute fr\'{e}quence ($HF$). Les deux voies $HF$ sont envoy\'{e}es
dans le dispositif soustracteur-sommateur.

Nous allons dans la suite pr\'{e}senter plus en d\'{e}tail les
\'{e}l\'{e}ments constitutifs de la d\'{e}tection \'{e}quilibr\'{e}e : les
photodiodes, les pr\'{e}amplificateurs et le soustracteur-sommateur. Nous
\'{e}tudierons ensuite le probl\`{e}me de l'\'{e}quilibrage des deux voies
de d\'{e}tection et pr\'{e}senterons les caract\'{e}ristiques du dispositif.

\subsubsection{Les photodiodes\label{IV-3-3-1}}

Les principales qualit\'{e}s que doivent pr\'{e}senter les photodiodes sont
une excellente efficacit\'{e} quantique et une r\'{e}ponse en fr\'{e}quence
suffisamment large pour observer sans att\'{e}nuation les fluctuations
d'intensit\'{e} aux fr\'{e}quences d'analyse. Les photodiodes que nous avons
utilis\'{e}es sont des $FND100$ de $EG\&G$. Les $FND100$ sont des
photodiodes de tr\`{e}s large bande passante ($350~MHz$), de faible
capacit\'{e} interne ($8$ \`{a} $10~pF$), et dont la surface photosensible
est un disque de diam\`{e}tre $1~mm$. Leur domaine spectral de
fonctionnement optimal est compris entre $800$ et $850~nm$, pour une tension
de polarisation de $70~V$. Elles pr\'{e}sentent aussi une tr\`{e}s grande
efficacit\'{e} quantique, c'est-\`{a}-dire un taux de conversion
photon-\'{e}lectron sup\'{e}rieur \`{a} $80\%$, avec un bruit propre en
courant tr\`{e}s faible (de l'ordre de $40~fA/\sqrt{Hz}$). Ce rendement
quantique peut \^{e}tre augment\'{e} en retirant la fen\^{e}tre qui
prot\`{e}ge la surface photosensible et en utilisant un miroir de renvoi qui
permet de r\'{e}cup\'{e}rer le faisceau directement r\'{e}fl\'{e}chi par la
surface du d\'{e}tecteur (r\'{e}flexion sp\'{e}culaire).

La lentille convergente plac\'{e}e \`{a} l'entr\'{e}e du syst\`{e}me de
d\'{e}tection (figure \ref{Fig_4blocpd}) assure une focalisation des
faisceaux au niveau des d\'{e}tecteurs sur un diam\`{e}tre de l'ordre de $%
0.5~mm$. Cette valeur correspond \`{a} un bon compromis entre la taille de
la surface photosensible et les probl\`{e}mes de saturation qui pourraient
appara\^{\i }tre lorsque la focalisation est trop importante.

\subsubsection{Les pr\'{e}amplificateurs \label{IV-3-3-2}}

\begin{figure}[tbp]
\centerline{\psfig{figure=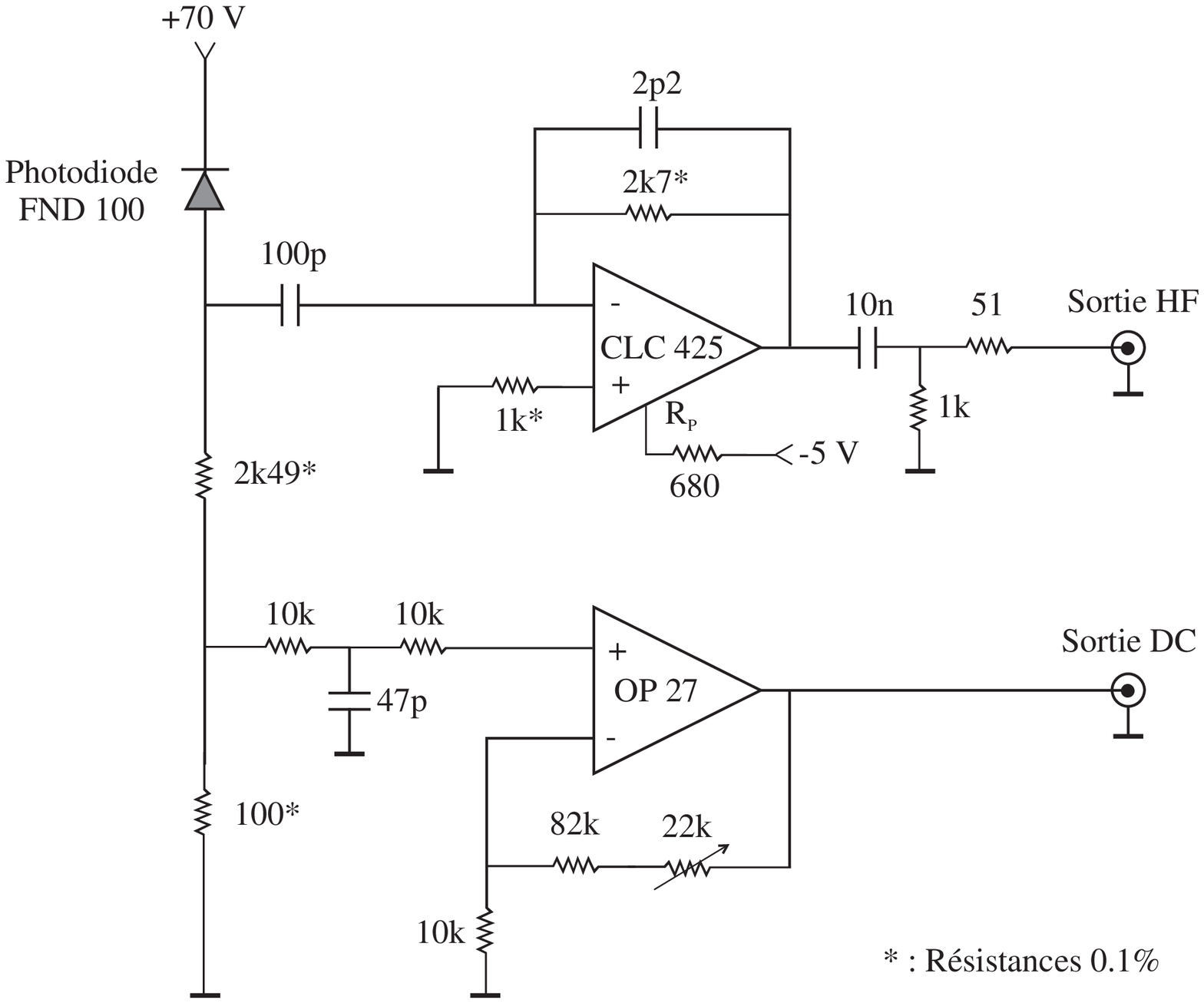,height=10cm}}
\caption{Sch\'{e}ma du pr\'{e}amplificateur pour chaque photodiode $FND100$.
Les deux \'{e}tages fournissent les signaux basse fr\'{e}quence (sortie $DC$%
) et haute fr\'{e}quence (sortie $HF$)}
\label{Fig_4amphdio}
\end{figure}

Les photodiodes se comportant comme des g\'{e}n\'{e}rateurs de courant, le
r\^{o}le du pr\'{e}amplificateur consiste \`{a} transformer le photocourant
en une tension mesurable sur un domaine de fr\'{e}quence \'{e}tendu,
incluant le continu. Les qualit\'{e}s requises pour les amplificateurs sont
une large bande passante et un faible bruit, afin de ne pas d\'{e}grader la
mesure des fluctuations quantiques de l'intensit\'{e} lumineuse. Le
sch\'{e}ma du montage pr\'{e}amplificateur est repr\'{e}sent\'{e} sur la
figure \ref{Fig_4amphdio}. 

Le pr\'{e}amplificateur est form\'{e} de deux \'{e}tages distincts, l'un
pour les signaux basse fr\'{e}quence (voie $DC$), l'autre pour les hautes
fr\'{e}quences (voie $HF$). La s\'{e}paration entre les deux voies est
r\'{e}alis\'{e}e par un filtre $RC$ constitu\'{e} par la r\'{e}sistance de
charge de la photodiode ($2.49~k\Omega $ et $100\Omega $ en s\'{e}rie) et
par la capacit\'{e} d'entr\'{e}e de l'\'{e}tage $HF$ ($100~pF$). La
fr\'{e}quence de s\'{e}paration est de l'ordre de $600~kHz$. La
r\'{e}sistance de charge est constitu\'{e}e de r\'{e}sistances \`{a} couches
m\'{e}talliques \`{a} faible bruit (r\'{e}sistances de pr\'{e}cision $0.1\%$%
). D'autre part, la valeur de la r\'{e}sistance est assez \'{e}lev\'{e}e de
fa\c{c}on \`{a} limiter son bruit thermique (bruit Johnson-Nyquist). Le
bruit de courant de la r\'{e}sistance vient en effet directement se
superposer au photocourant d\'{e}livr\'{e} par la photodiode et peut
perturber le signal mesur\'{e} pour de faibles intensit\'{e}s lumineuses.

La valeur de la r\'{e}sistance est cependant limit\'{e}e par le photocourant
d\'{e}bit\'{e} en continu par la photodiode. En effet si la r\'{e}sistance
est trop grande, la tension \`{a} ses bornes peut devenir assez
\'{e}lev\'{e}e pour r\'{e}duire de fa\c{c}on appr\'{e}ciable la polarisation
de la photodiode. Avec les valeurs choisies, la tension aux bornes de la
r\'{e}sistance de charge reste inf\'{e}rieure \`{a} $15~V$ pour des
puissances lumineuses inf\'{e}rieures \`{a} $10~mW$, et la tension de
polarisation reste toujours sup\'{e}rieure \`{a} $50~V$. 
\begin{figure}[tbp]
\centerline{\psfig{figure=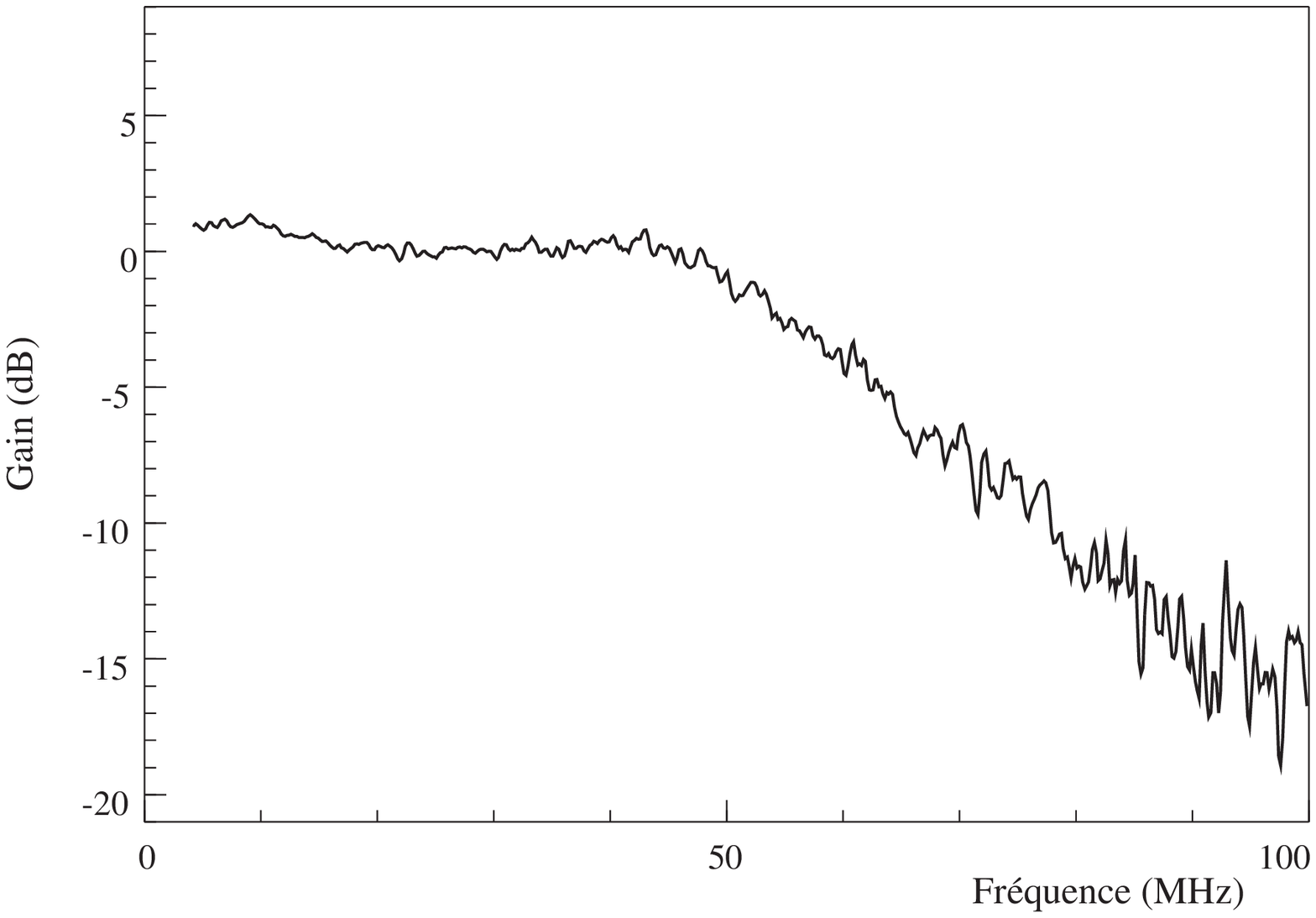,height=8cm}}
\caption{R\'{e}ponse en fr\'{e}quence de la voie $HF$, obtenue \`{a} partir
du bruit de photon d'un faisceau de $10~mW$ incident sur la photodiode. La
bande passante est sup\'{e}rieure \`{a} $50~MHz$}
\label{Fig_4bphdiod}
\end{figure}

La voie $DC$ pr\'{e}l\`{e}ve une partie de la tension aux bornes de la
r\'{e}sistance de charge. Elle est constitu\'{e}e d'un filtre en T qui
\'{e}vite tout retour de signaux parasites vers la voie $HF$ (fr\'{e}quence
de coupure $300~kHz$), suivi par un amplificateur en tension de gain $10$.
Le gain est ajustable \`{a} l'aide d'une r\'{e}sistance variable plac\'{e}e
en contre-r\'{e}action de fa\c{c}on \`{a} assurer l'\'{e}quilibrage des
voies $DC$ des deux d\'{e}tecteurs (voir section \ref{IV-3-3-4}) et
d'obtenir un taux de conversion global courant-tension de $1~V/mA$.

La voie $HF$ est bas\'{e}e sur un montage transimp\'{e}dance. Ce type de
circuit permet d'\'{e}liminer l'influence de la capacit\'{e} de la
photodiode. En effet, la photodiode fonctionne \`{a} haute fr\'{e}quence
avec une diff\'{e}rence de potentiel constante, puisque sa borne
n\'{e}gative se comporte comme une masse virtuelle. La capacit\'{e} parasite
en parall\`{e}le avec la photodiode ne limite donc pas la bande passante du
circuit. L'amplificateur utilis\'{e} (CLC425) est un amplificateur rapide
(produit gain-bande \'{e}gal \`{a} $1.7~GHz$) et faible bruit ($1~nV/\sqrt{Hz%
}$ et $1.6~pA/\sqrt{Hz}$). La valeur de la r\'{e}sistance de
contre-r\'{e}action ($2.7~k\Omega $) r\'{e}sulte d'un compromis entre la
bande passante de l'\'{e}tage et le bruit thermique ajout\'{e}. Une petite
capacit\'{e} plac\'{e}e en parall\`{e}le permet de modifier la r\'{e}ponse
dynamique de l'\'{e}tage de fa\c{c}on \`{a} obtenir une r\'{e}ponse en
fr\'{e}quence la plus constante possible. Un filtre passe-haut est plac\'{e}
en sortie afin d'\'{e}liminer toute composante continue (fr\'{e}quence de
coupure de $200~kHz$ pour une imp\'{e}dance en sortie de $50~\Omega $). La
r\'{e}sistance de $680~\Omega $ plac\'{e}e entre la broche $R_{p}$ du CLC425
et la tension d'alimentation n\'{e}gative permet d'optimiser la r\'{e}ponse
en fr\'{e}quence de l'amplificateur.

Le taux de conversion courant-tension de la voie $HF$ est de l'ordre de $%
1.3~V/mA$ pour une imp\'{e}dance de charge de $50~\Omega $. La figure \ref
{Fig_4bphdiod} montre la r\'{e}ponse en fr\'{e}quence du
pr\'{e}amplificateur. La courbe repr\'{e}sente en fait la puissance de bruit
en sortie de la voie $HF$ lorsqu'un faisceau de $10~mW$ est incident sur la
photodiode. Le bruit \'{e}lectronique, obtenu \`{a} partir du bruit en
sortie sans faisceau incident, a \'{e}t\'{e} retranch\'{e} de la courbe. Le
bruit d'intensit\'{e} du laser est \'{e}gal au bruit de photon au del\`{a}
de quelques m\'{e}gahertz. Il est donc plat en fr\'{e}quence et la figure
repr\'{e}sente en fait la fonction de transfert du dispositif. La bande
passante \`{a} $-3~dB$ est sup\'{e}rieure \`{a} $50~MHz$.

\subsubsection{Syst\`{e}me soustracteur-sommateur\label{IV-3-3-3}}

Le dispositif soustracteur-sommateur est le dernier \'{e}l\'{e}ment de la
cha\^{\i }ne de d\'{e}tection \'{e}quilibr\'{e}e. C'est un syst\`{e}me actif
qui dispose de deux entr\'{e}es pour recevoir les signaux envoy\'{e}s par
les voies $HF$ des deux d\'{e}tecteurs, la sortie fournissant la somme ou la
diff\'{e}rence des signaux d'entr\'{e}e. Un soin particulier doit \^{e}tre
pris pour la r\'{e}alisation de ce syst\`{e}me puisqu'il doit avoir un
faible bruit, une large bande passante et il doit \^{e}tre con\c{c}u de
mani\`{e}re \`{a} pr\'{e}senter le meilleurs \'{e}quilibrage possible sur
une large bande de fr\'{e}quence. 
\begin{figure}[tbp]
\centerline{\psfig{figure=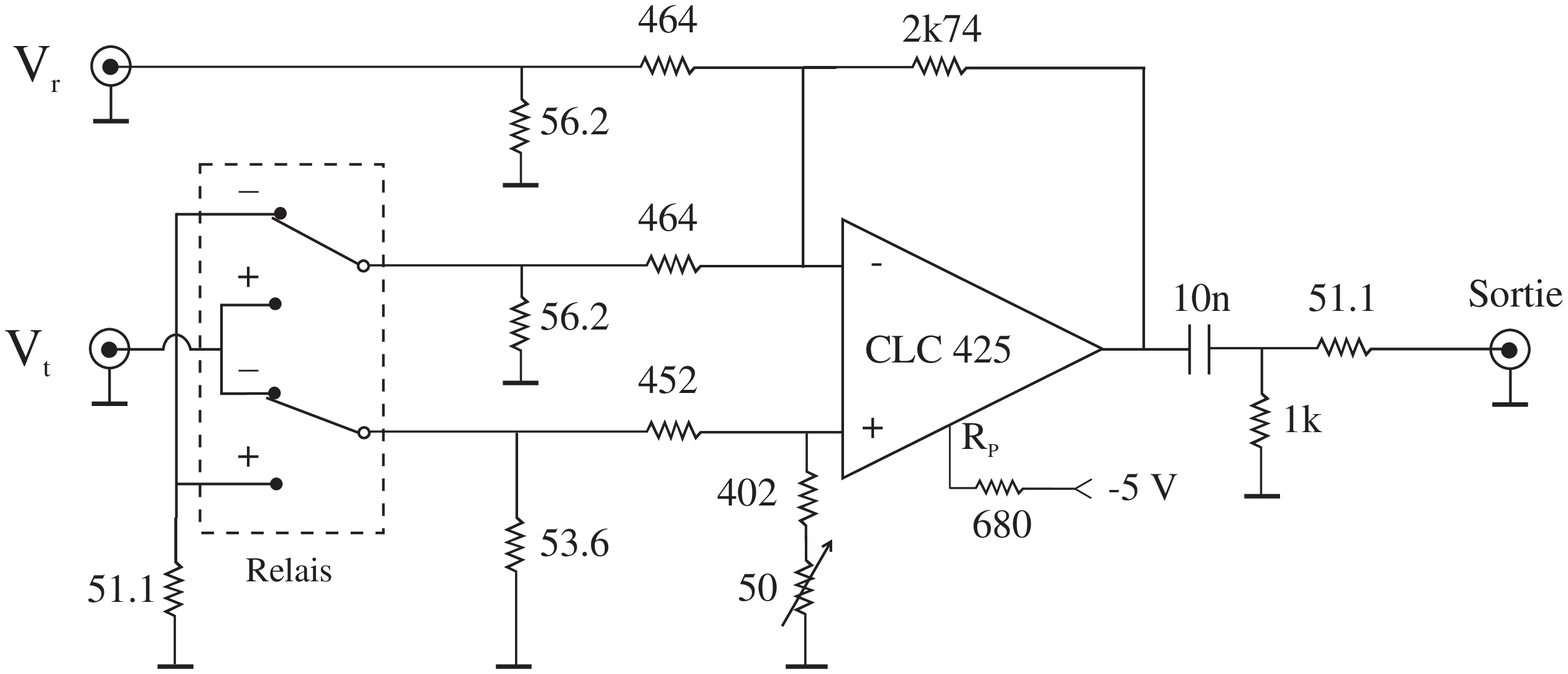,height=7cm}}
\caption{Sh\'{e}ma de principe du dispositif soustracteur-sommateur. Toutes
les r\'{e}sistances utilis\'{e}es ont une pr\'{e}cision de 0.1\%}
\label{Fig_4soustra}
\end{figure}
Le sch\'{e}ma \'{e}lectronique du soustracteur-sommateur est
repr\'{e}sent\'{e} sur la figure \ref{Fig_4soustra}. Il est construit autour
d'un amplificateur rapide et faible bruit CLC425 qui ajoute ou soustrait les
deux entr\'{e}es selon la position des commutateurs du relais. Nous avons
utilis\'{e} un relais \`{a} contact dor\'{e} plut\^{o}t qu'un simple
commutateur car il introduit moins de perturbations pour les signaux hautes
fr\'{e}quences (capacit\'{e}s parasites, isolation...). Le montage est con\c{%
c}u de telle mani\`{e}re que la gain \'{e}lectronique soit le m\^{e}me en
valeur absolue pour les deux voies, quelque soit la position du relais.
Ainsi, en position sommateur l'\'{e}quilibrage est assur\'{e} par la
construction sym\'{e}trique de la boucle de contre-r\'{e}action
vis-\`{a}-vis des deux voies d'entr\'{e}e, alors qu'en position soustracteur
l'\'{e}quilibrage est ajust\'{e} par la r\'{e}sistance variable sur
l'entr\'{e}e $+$ de l'amplificateur (valeur th\'{e}orique $50~\Omega $). De
m\^{e}me, les valeurs des r\'{e}sistances sont telles que l'imp\'{e}dance
d'entr\'{e}e pour les deux voies est toujours \'{e}gale \`{a} $50.3\pm
0.2~\Omega $, et les imp\'{e}dances aux entr\'{e}es $+$ et $-$ de
l'amplificateur sont \'{e}gales \`{a} $225~\Omega $. Comme nous le verrons
dans la section \ref{IV-3-3-4}, ceci assure un \'{e}quilibre des gains de
l'ordre de $2^{o}/_{oo}$ sur une plage de fr\'{e}quence allant jusqu'\`{a} $%
20~MHz$.

L'amplificateur est suivi par un filtre passe-haut similaire \`{a} celui
pr\'{e}sent\'{e} dans les pr\'{e}amplificateurs des photodiodes (figure \ref
{Fig_4amphdio}). Le gain global du soustracteur-sommateur est \'{e}gal \`{a} 
$3$ pour une imp\'{e}dance de charge de $50~\Omega $.

\subsubsection{Equilibrage des deux voies de la d\'{e}tection\label{IV-3-3-4}
}

L'\'{e}quilibrage des deux voies de d\'{e}tection est particuli\`{e}rement
important pour assurer le bon fonctionnement du dispositif, que ce soit dans
le cadre d'une mesure homodyne ou dans celui d'une mesure directe du bruit
d'intensit\'{e}. Un d\'{e}s\'{e}quilibre entre les deux voies peut \^{e}tre
aussi bien d'origine optique (s\'{e}paration en deux parties du faisceau
incident, rendement quantique des photodiodes) qu'\'{e}lectronique (gains
des voies $DC$ et $HF$, dispositif soustracteur-sommateur). Dans la suite
nous allons montrer comment l'\'{e}quilibrage de ces diff\'{e}rents
\'{e}l\'{e}ments est r\'{e}alis\'{e}.

\paragraph{Appariement des photodiodes\newline
}

\medskip

Les deux photodiodes doivent pr\'{e}senter une r\'{e}ponse aussi similaire
que possible. Parmi les photodiodes dont nous disposons, on en
s\'{e}lectionne deux qui pr\'{e}sentent les efficacit\'{e}s quantiques les
plus proches. La proc\'{e}dure suivie consiste \`{a} choisir une photodiode
de r\'{e}f\'{e}rence que l'on place par exemple en transmission du cube $CP2$
(figure \ref{Fig_4detcdyn}) alors qu'une seconde photodiode est plac\'{e}e
en r\'{e}flexion. On mesure alors l'\'{e}cart relatif des tensions $DC$, $%
\Delta V/V_{ref}$, qui permet d'\'{e}valuer l'\'{e}cart entre les rendements
quantiques des deux photodiodes en s'affranchissant des \'{e}ventuelles
fluctuations de l'intensit\'{e} lumineuse incidente sur le syst\`{e}me de
d\'{e}tection. On r\'{e}p\`{e}te cette mesure en rempla\c{c}ant \`{a} chaque
fois la photodiode en r\'{e}flexion et on s\'{e}lectionne finalement les
deux photodiodes qui pr\'{e}sentent les variations relatives les plus
proches. La r\'{e}ponse des photodiodes que nous avons choisies d'utiliser
sont identique \`{a} $0.3\%$ pr\`{e}s.

\paragraph{S\'{e}paration des faisceaux et gain des voies DC\newline
}

\medskip

\begin{figure}[tbp]
\centerline{\psfig{figure=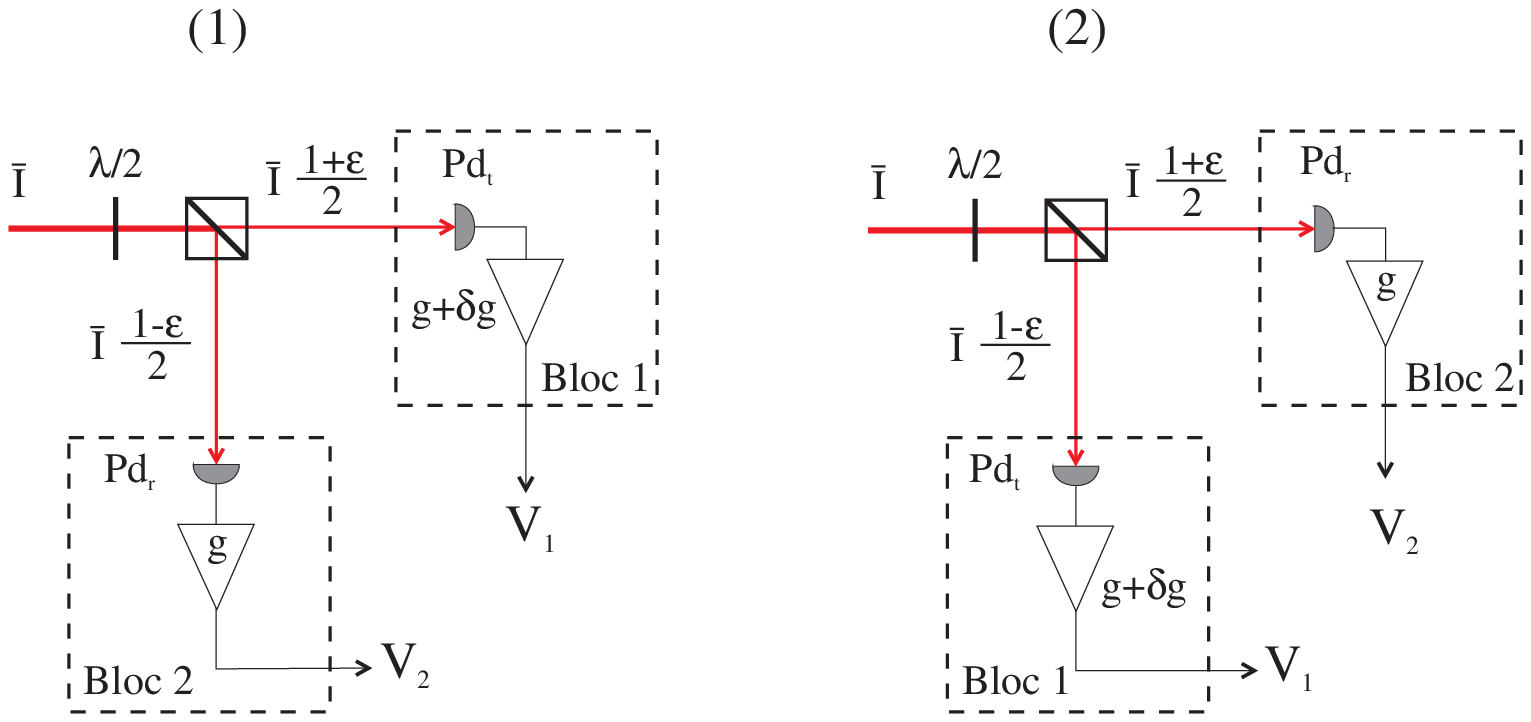,height=7cm}}
\caption{En intervertissant les deux blocs de d\'{e}tection (photodiodes et
pr\'{e}amplificateurs), on peut d\'{e}terminer le d\'{e}s\'{e}quilibre
optique $\varepsilon $ li\'{e} \`{a} un mauvais r\'{e}glage de la lame
demi-onde et le d\'{e}s\'{e}quilibre $\delta g$ des deux voies $DC$ des
pr\'{e}amplificateurs}
\label{Fig_4equilib}
\end{figure}

L'\'{e}quilibrage de la s\'{e}paration optique des faisceaux \`{a} l'aide de
la lame demi-onde et du cube $CP2$ (figure \ref{Fig_4detcdyn}), et
l'\'{e}quilibrage des gains \'{e}lectroniques des voies $DC$ sont
r\'{e}alis\'{e}s en intervertissant les deux blocs de d\'{e}tection comme le
montre la figure \ref{Fig_4equilib}. En notant $\varepsilon $ le
d\'{e}s\'{e}quilibre entre les intensit\'{e}s transmise et
r\'{e}fl\'{e}chie, et $\delta g$ l'\'{e}cart de gain entre les deux voies $%
DC $, on trouve que les tensions $V_{1}$ et $V_{2}$ \`{a} la sortie $DC$ des
deux blocs s'\'{e}crivent pour les deux configurations (\`{a} l'ordre $1$ en 
$\varepsilon $ et $\delta g$): 
\begin{equation}
(1)\left\{ 
\begin{array}{l}
V_{1}=g~\frac{\bar{I}}{2}\left( 1+\varepsilon +\frac{\delta g}{g}\right) \\ 
V_{2}=g~\frac{\bar{I}}{2}\left( 1-\varepsilon \right)
\end{array}
\right\} \quad ;\quad (2)\left\{ 
\begin{array}{l}
V_{1}=g~\frac{\bar{I}}{2}\left( 1-\varepsilon +\frac{\delta g}{g}\right) \\ 
V_{2}=g~\frac{I}{2}\left( 1+\varepsilon \right)
\end{array}
\right\}  \label{4.3.10bis}
\end{equation}
Pour d\'{e}terminer les facteurs $\varepsilon $ et $\delta g$, on mesure les
\'{e}carts relatifs $\left( V_{1}-V_{2}\right) /V_{1}$ dans les deux
configurations $R_{1}$ et $R_{2}$ dont les expressions sont donn\'{e}es par: 
\begin{subequations}
\label{4.3.10ter}
\begin{eqnarray}
R_{1} &=&2\varepsilon +\frac{\delta g}{g}  \label{4.3.10tera} \\
R_{2} &=&-2\varepsilon +\frac{\delta g}{g}  \label{4.3.10terb}
\end{eqnarray}
A partir de ces expressions, il est facile de voir que l'\'{e}quilibrage
optique, correspondant \`{a} $\varepsilon =0$, est r\'{e}alis\'{e} en
tournant la lame demi-onde jusqu'\`{a} ce que les valeurs mesur\'{e}es pour $%
R_{1}$ et $R_{2}$ soient \'{e}gales et de m\^{e}me signe. Par contre, un 
\'{e}quilibrage parfait au niveau des d\'{e}tecteurs ($\delta g=0$)
s'obtient lorsque $R_{1}$ et $R_{2}$ sont \'{e}gaux en valeur absolue mais
de signe contraire.

En pratique, pour \'{e}viter d'intervertir de nombreuses fois les deux blocs
d\'{e}tecteurs, on commence par \'{e}quilibrer les deux voies $DC$ en
modifiant la valeur de la r\'{e}sistance variable de l'un des
d\'{e}tecteurs, plac\'{e}e en contre-r\'{e}action sur l'amplificateur $OP27$
(figure \ref{Fig_4amphdio}). On mesure tout d'abord \`{a} l'aide d'un
voltm\`{e}tre les \'{e}carts relatifs $R_{1}$ et $R_{2}$. Ceci permet de
d\'{e}terminer la valeur du d\'{e}s\'{e}quilibre des gains $\delta g$ qui
est \'{e}gal, d'apr\`{e}s les relations (\ref{4.3.10ter}), \`{a} $g/2~\left(
R_{1}+R_{2}\right) $, o\`{u} le gain $g$ est fix\'{e} initialement \`{a} une
valeur voisine de $10$. On compense ce d\'{e}s\'{e}quilibre en modifiant la
valeur de la r\'{e}sistance variable d'une quantit\'{e} \'{e}gale \`{a} $%
\delta g\times 10~k\Omega $. Pour v\'{e}rifier le bon \'{e}quilibrage des
gains, on mesure une seconde fois les \'{e}carts relatifs $R_{1}$ et $R_{2}$
qui, dans le cas d'un \'{e}quilibre parfait, sont \'{e}gaux et de signes
contraires. Ceci permet de nous assurer que l'on ne compense pas un
d\'{e}s\'{e}quilibre optique par un d\'{e}s\'{e}quilibre \'{e}lectronique.
On obtient finalement un \'{e}quilibrage des voies $DC$ meilleur que $%
2^{o}/_{oo}$.

Une fois ces op\'{e}rations r\'{e}alis\'{e}es, l'\'{e}quilibrage de la
s\'{e}paration optique des faisceaux s'effectue simplement en annulant la
diff\'{e}rence $V_{1}-V_{2}$ entre les tensions $DC$. L'\'{e}quilibrage
obtenu est essentiellement limit\'{e} \`{a} environ $2^{o}/_{oo}$ par la
pr\'{e}cision de rotation de la lame $\lambda /2$.

\paragraph{Gain des voies HF et dispositif soustracteur-sommateur\newline}
\protect\medskip

Pour tester et r\'{e}gler le fonctionnement du syst\`{e}me \`{a} haute
fr\'{e}quence, on module l'intensit\'{e} du faisceau incident \`{a} une
fr\'{e}quence variable \`{a} l'aide d'un \'{e}lectro-optique suivi d'un
polariseur. Pour ajuster les gains des voies $HF$, nous avons utilis\'{e} un
soustracteur Mini-Circuit pour mesurer sur l'analyseur de spectre la
diff\'{e}rence des deux voies. La pr\'{e}cision du r\'{e}glage d\'{e}pend
bien s\^{u}r de l'\'{e}quilibrage du soustracteur qui est de l'ordre de $1\%$%
. Notons qu'un d\'{e}s\'{e}quilibre r\'{e}siduel des voies $HF$ sera
compens\'{e} par l'\'{e}quilibrage du dispositif soustracteur-sommateur. On
compense dans ce cas un gain \'{e}lectronique par un autre gain
\'{e}lectronique, et seul compte pour la qualit\'{e} du syst\`{e}me
l'\'{e}quilibrage global du pr\'{e}amplificateur suivi du
soustracteur-sommateur.

Nous avons r\'{e}alis\'{e} ce r\'{e}glage en modulant l'intensit\'{e} du
faisceau \`{a} une fr\'{e}quence de $3~MHz$. On corrige le gain de la voie
pr\'{e}sentant le plus fort gain en pla\c{c}ant une r\'{e}sistance de valeur
\'{e}lev\'{e}e en parall\`{e}le avec la r\'{e}sistance de $2.7~k\Omega $ en
contre-r\'{e}action du CLC425 (figure \ref{Fig_4amphdio}). Nous avons
ainsi obtenu un \'{e}quilibrage des voies $HF$ \`{a} la fr\'{e}quence de
modulation meilleur que $0.5\%$.

En ce qui concerne le dispositif soustracteur-sommateur, nous avons tout
d'abord test\'{e} son fonctionnement en appliquant la m\^{e}me modulation
\'{e}lectrique sur ses deux voies d'entr\'{e}es. La pr\'{e}cision de ce test
est cependant limit\'{e}e par le splitter Mini-Circuit utilis\'{e} pour
envoyer le m\^{e}me signal sur les deux entr\'{e}es avec une imp\'{e}dance
de $50~\Omega $. Nous avons donc repris ce test en connectant le
soustracteur-sommateur aux sorties des voies $HF$ et en appliquant une
modulation d'intensit\'{e} au faisceau incident. La r\'{e}sistance variable
sur l'entr\'{e}e $+$ du CLC425 (figure \ref{Fig_4soustra}) permet de
r\'{e}gler finement l'\'{e}quilibrage en position soustracteur. Pour cela,
on cherche \`{a} annuler la modulation dans le signal fourni par le
soustracteur. Il est en fait possible de r\'{e}duire cette modulation
jusqu'\`{a} une valeur tr\`{e}s faible \`{a} une fr\'{e}quence donn\'{e}e.
Nous avons plut\^{o}t cherch\'{e} \`{a} obtenir un bon \'{e}quilibrage sur
une large plage de fr\'{e}quence. Pour la valeur optimale de la
r\'{e}sistance variable, on obtient finalement un \'{e}quilibrage en
configuration soustracteur de l'ordre de $0.5\%$ sur une plage de
fr\'{e}quence allant jusqu'\`{a} $20~MHz$.

Le r\'{e}sultat de l'\'{e}quilibrage est repr\'{e}sent\'{e} sur la figure 
\ref{Fig_4equili2}, o\`{u} la courbe repr\'{e}sente l'\'{e}cart relatif $%
20~Log\left| \left( V_{t}-V_{r}\right) /V_{t}\right| $ lorsqu'on balaye la
fr\'{e}quence de modulation du faisceau lumineux. 
\begin{figure}[tbp]
\centerline{\psfig{figure=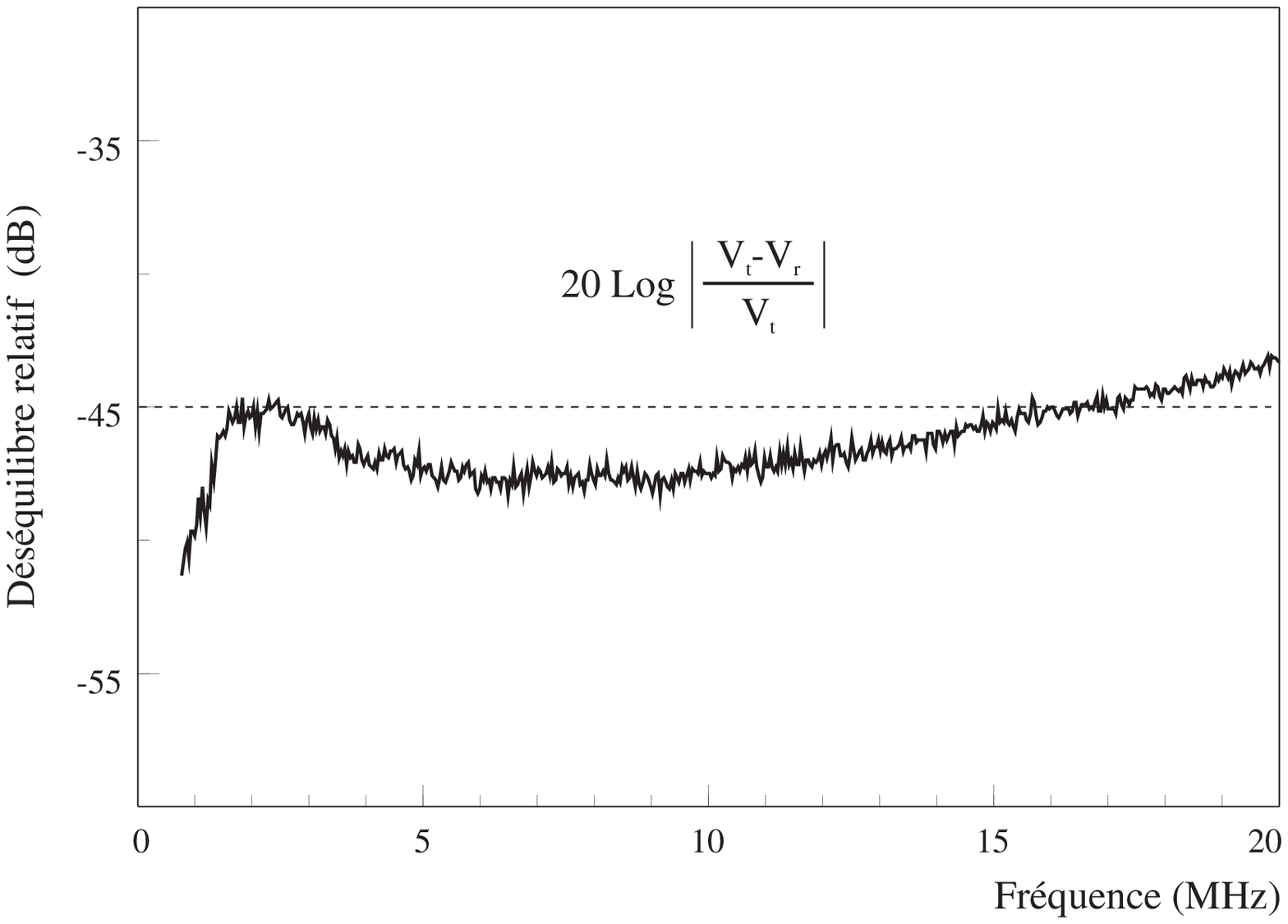,height=9cm}}
\caption{Evolution en fr\'{e}quence du d\'{e}s\'{e}quilibre relatif $\left(
V_{t}-V_{r}\right) /V_{t}$ en $dB$ \`{a} la sortie du soustracteur lorsqu'on
varie la fr\'{e}quence de modulation d'intensit\'{e}}
\label{Fig_4equili2}
\end{figure}
Cette courbe repr\'{e}sente en fait l'\'{e}quilibrage de l'ensemble de la
d\'{e}tection \'{e}quilibr\'{e}e, incluant les d\'{e}fauts d'\'{e}quilibrage
de la s\'{e}paration optique du faisceau \`{a} l'aide de la $\lambda /2$ et
du cube $CP2$, des photodiodes, des pr\'{e}amplificateurs et du dispositif
soustracteur-sommateur. On voit que le d\'{e}s\'{e}quilibre global de tous
les \'{e}l\'{e}ments de la cha\^{\i }ne est tr\`{e}s petit puisqu'il est de
l'ordre de $-45~dB$, soit $0.5\%$, et cela sur une large bande de
fr\'{e}quence d'environ $20~MHz$. La remont\'{e}e que l'on peut observer
\`{a} haute fr\'{e}quence est sans doute due \`{a} une l\'{e}g\`{e}re
diff\'{e}rence des r\'{e}ponses en fr\'{e}quence des voies $HF$, ou \`{a} un
d\'{e}phasage entre les deux voies d'entr\'{e}e du dispositif
soustracteur-sommateur. Quoiqu'il en soit, l'\'{e}quilibrage global du
syst\`{e}me de d\'{e}tection est tr\`{e}s largement suffisant pour pouvoir
mesurer le bruit quantique de la lumi\`{e}re avec une pr\'{e}cision de
l'ordre de $1\%$.

\subsubsection{Caract\'{e}ristiques puissance-tension de la d\'{e}tection%
\label{IV-3-3-5}}

Les deux voies de d\'{e}tection doivent pr\'{e}senter un comportement
lin\'{e}aire vis-\`{a}-vis de l'intensit\'{e} lumineuse arrivant sur les
photodiodes pour des puissances les plus \'{e}lev\'{e}es possibles. En
effet, l'efficacit\'{e} de la mesure homodyne repose en partie sur le fait
que l'intensit\'{e} de l'oscillateur local doit \^{e}tre grande devant celle
du faisceau r\'{e}fl\'{e}chi par la cavit\'{e} \`{a} miroir mobile. Il est
donc important d'\'{e}viter tout effet de saturation aussi bien au niveau
des photodiodes que des amplificateurs. C'est pourquoi nous avons
cherch\'{e} \`{a} caract\'{e}riser le comportement de la tension fournie par
les voies $DC$ et $HF$ de la d\'{e}tection en fonction de la puissance
incidente.

En ce qui concerne les voies $DC$, on mesure \`{a} l'aide d'un voltm\`{e}tre
la somme $V_{+}$ des tensions des voies $DC$ (figure \ref{Fig_4interse}) en
fonction de la puissance incidente que l'on mesure en utilisant un
microwattm\`{e}tre plac\'{e} devant le syst\`{e}me de d\'{e}tection
\'{e}quilibr\'{e}e (avant la lame demi-onde). Les carr\'{e}s sur la figure 
\ref{Fig_4lineint} repr\'{e}sentent les tensions mesur\'{e}es pour des
puissances incidentes allant de $0$ \`{a} $20~mW$. On voit que ces point
sont parfaitement ajust\'{e}s par une droite, ce qui indique que l'ensemble
des \'{e}l\'{e}ments constituant les deux voies $DC$ de la d\'{e}tection ont
un comportement parfaitement lin\'{e}aire sur cette plage de puissance.
D'autre part, la pente de la droite permet de d\'{e}terminer le rendement
quantique $\eta $ des photodiodes, d\'{e}fini comme le rapport entre le flux
de charges \'{e}lectriques $i_{pd}/e$ fourni par la photodiode et le flux de
photons $P_{pd}/\left( \hbar \omega _{L}\right) $: 
\end{subequations}
\begin{equation}
\eta =\frac{hc}{\lambda e}~\frac{i_{pd}}{P_{pd}}  \label{4.3.11}
\end{equation}
\begin{figure}[tbp]
\centerline{\psfig{figure=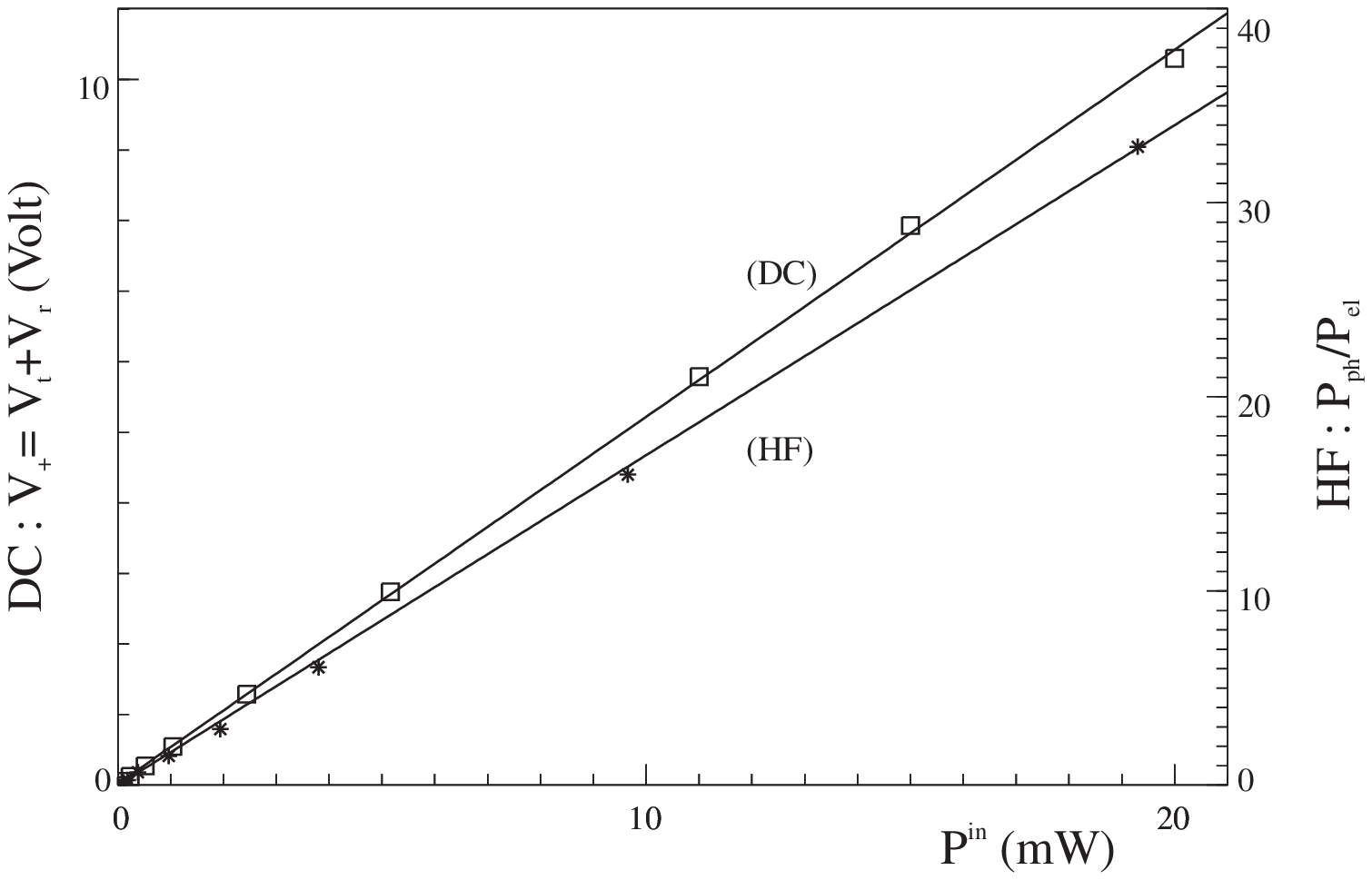,height=9cm}}
\caption{Caract\'{e}ristique puissance-tension du dispositif de
d\'{e}tection \'{e}quilibr\'{e}. La droite ($DC$) d\'{e}crit le comportement
de la tension $V_{+}$ qui est \'{e}gale \`{a} la somme des tensions \`{a} la
sortie des voies $DC$ des deux d\'{e}tecteurs. La droite ($HF$)
repr\'{e}sente le rapport des puissances de bruit de photon $P_{ph}$ et de
bruit \'{e}lectronique $P_{el}$ \`{a} une fr\'{e}quence d'analyse de $2~MHz$}
\label{Fig_4lineint}
\end{figure}
Nous pouvons en effet relier le photocourant $i_{pd}$ \`{a} la tension de
sortie de la voie $DC$ puisque le taux de conversion courant-tension $R_{DC}$
est \'{e}gal \`{a} $1~V/mA$. Les intensit\'{e}s et les deux d\'{e}tecteurs
\'{e}tant \'{e}quilibr\'{e}s, le rendement quantique $\eta $ est reli\'{e}
\`{a} la pente $V_{+}/P^{in}$ de la droite ($DC$) sur la figure \ref
{Fig_4lineint}: 
\begin{equation}
\eta =\frac{hc}{\lambda e}~\frac{V_{+}}{P^{in}}~\frac{1}{R_{DC}}\approx 80\%
\label{4.3.12}
\end{equation}
Pour conna\^{\i }tre le rendement quantique des photodiodes uniquement,
cette valeur doit \^{e}tre corrig\'{e}e des pertes optiques au niveau de la
lame $\lambda /2$ et du cube $CP2$, qui sont de l'ordre de $4\%$. D'autre
part, ces mesures ont \'{e}t\'{e} r\'{e}alis\'{e}es sans les miroirs de
renvoi (figure \ref{Fig_4blocpd}) qui permettent de r\'{e}cup\'{e}rer $9\%$
de l'intensit\'{e} lumineuse incidente. On trouve ainsi que le rendement
quantique des photodiodes est de l'ordre de $90\%$.

En ce qui concerne le comportement \`{a} haute fr\'{e}quence du syst\`{e}me
de d\'{e}tection, nous avons test\'{e} la lin\'{e}arit\'{e} de la puissance
du bruit de photon standard vis-\`{a}-vis de la puissance lumineuse
incidente. Pour cela, on mesure la puissance de bruit sur l'analyseur de
spectre \`{a} une fr\'{e}quence d'analyse fix\'{e}e \`{a} $2~MHz$, le
soustracteur-sommateur \'{e}tant en configuration soustracteur. On mesure en
fait la somme $P_{ph}+P_{el}$ de la puissance $P_{ph}$ du bruit de photon et
du bruit \'{e}lectronique $P_{el}$, dont on peut d\'{e}terminer la valeur en
masquant le faisceau incident. Les \'{e}toiles sur la figure \ref
{Fig_4lineint} repr\'{e}sentent le rapport des puissances de bruit $%
P_{ph}/P_{el}$ pour diff\'{e}rentes valeurs de la puissance lumineuse
incidente. L'alignement de ces \'{e}toiles montre que le comportement des
diff\'{e}rents \'{e}l\'{e}ments des voies $HF$ de nos deux d\'{e}tecteurs
est parfaitement lin\'{e}aire. On peut d'autre part d\'{e}terminer la
puissance incidente minimale que l'on peut mesurer. Pour un rapport signal
\`{a} bruit $P_{ph}/P_{el}$ \'{e}gal \`{a} $1$, c'est-\`{a}-dire pour un
bruit total $3~dB$ au dessus du bruit \'{e}lectronique, la puissance
incidente est de l'ordre de $500~\mu W$.

\section{Excitation optique du r\'{e}sonateur\label{IV-4}}

\bigskip

Afin d'\'{e}tudier les modes acoustiques du r\'{e}sonateur et la fa\c{c}on
dont ils sont coupl\'{e}s au faisceau lumineux, nous avons d\'{e}velopp\'{e}
une technique originale d'excitation optique du r\'{e}sonateur. Les
m\'{e}thodes utilis\'{e}es pour \'{e}tudier les modes acoustiques des
r\'{e}sonateurs m\'{e}caniques en quartz font g\'{e}n\'{e}ralement appel aux
propri\'{e}t\'{e}s pi\'{e}zo\'{e}lectriques du mat\'{e}riau : des
\'{e}lectrodes sont plac\'{e}es sur le r\'{e}sonateur de fa\c{c}on \`{a}
l'exciter \'{e}lectriquement. Une telle approche n'est bien s\^{u}r pas
applicable \`{a} notre cas o\`{u} le r\'{e}sonateur est en silice. On
utilise alors la force de pression de radiation pour exciter les modes du
r\'{e}sonateur. Plus pr\'{e}cis\'{e}ment, un faisceau laser intense est
envoy\'{e} par l'arri\`{e}re de la cavit\'{e} sur le miroir mobile. Ce
faisceau est modul\'{e} en intensit\'{e} \`{a} l'aide d'un modulateur
acousto-optique, \`{a} une fr\'{e}quence fix\'{e}e par un synth\'{e}tiseur
haute fr\'{e}quence. En se r\'{e}fl\'{e}chissant sur le miroir mobile, le
faisceau exerce une force de pression de radiation modul\'{e}e \`{a} la
fr\'{e}quence choisie. En variant l'amplitude et la fr\'{e}quence de
modulation, on peut ainsi d\'{e}terminer la r\'{e}ponse m\'{e}canique du
r\'{e}sonateur \`{a} une force ext\'{e}rieure. Pour cela, on d\'{e}tecte le
mouvement de la face plane du r\'{e}sonateur en mesurant la modulation
induite par celui-ci sur la phase du faisceau r\'{e}fl\'{e}chi par la
cavit\'{e} \`{a} miroir mobile.

Cette excitation optique s'av\`{e}re moins efficace qu'une excitation
pi\'{e}zo\'{e}lectrique, car la force de pression de radiation est moins
intense que l'effet pi\'{e}zo\'{e}lectrique. Elle permet n\'{e}anmoins
d'acc\'{e}der \`{a} des param\`{e}tres autres que la fr\'{e}quence et le
facteur de qualit\'{e} des modes acoustiques. On peut par exemple faire
varier la section du faisceau lumineux et obtenir ainsi des informations sur
la structure spatiale des modes acoustiques.

Nous allons pr\'{e}senter dans cette partie l'ensemble des \'{e}l\'{e}ments
constituant ce dispositif. Comme nous le verrons, un point
particuli\`{e}rement d\'{e}licat est l'isolation du faisceau arri\`{e}re du
reste du montage (cavit\'{e} \`{a} miroir mobile et syst\`{e}me de
d\'{e}tection homodyne). En particulier, le faisceau arri\`{e}re ne doit pas
entrer dans la cavit\'{e}. Sinon la lumi\`{e}re r\'{e}siduelle transmise
vers l'avant peut perturber la d\'{e}tection de phase du faisceau
r\'{e}fl\'{e}chi en rajoutant au signal un terme de bruit provenant du
battement de la lumi\`{e}re r\'{e}siduelle avec l'oscillateur local. 
\begin{figure}[tbp]
\centerline{\psfig{figure=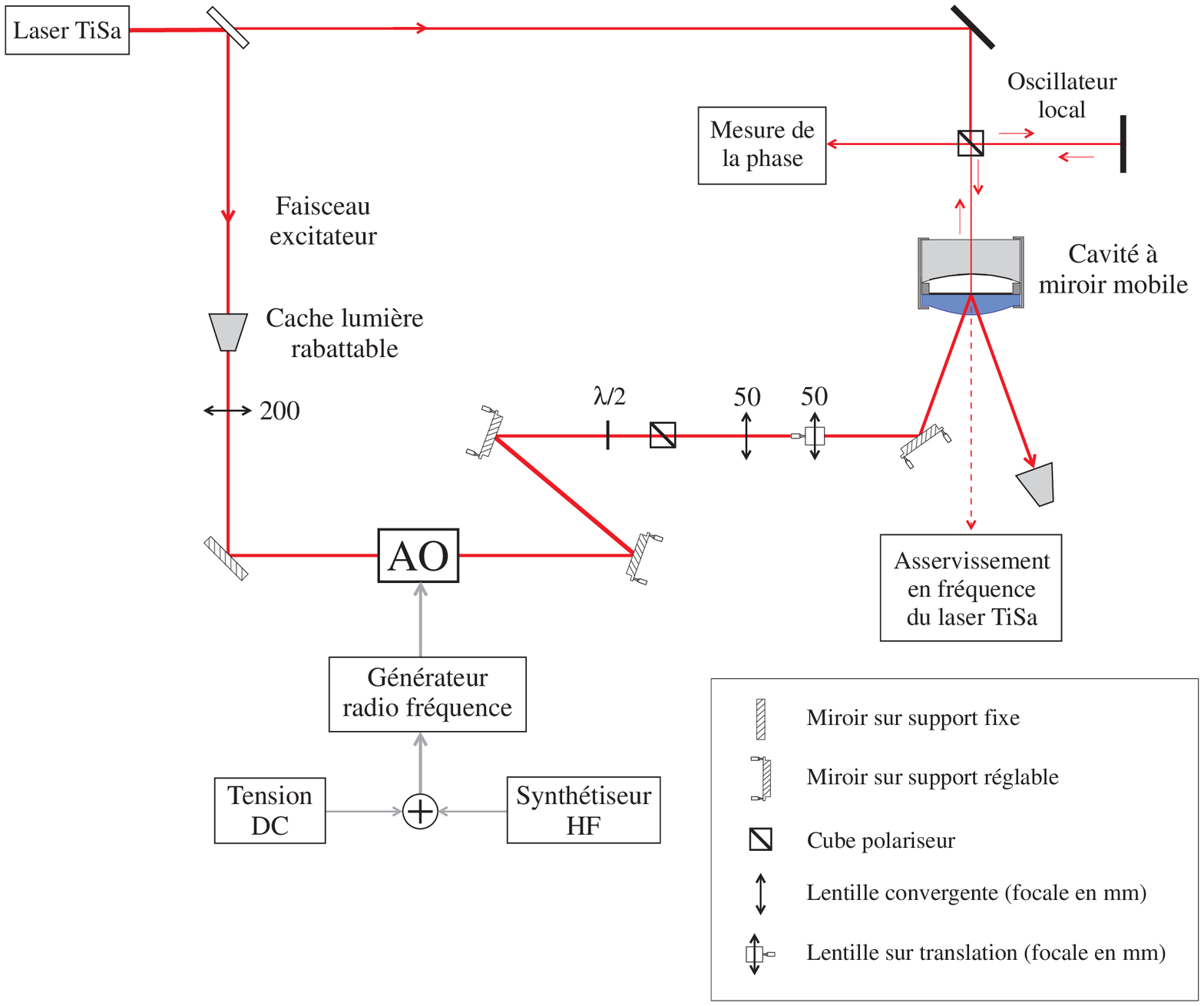,height=13cm}}
\caption{Repr\'{e}sentation sch\'{e}matique du faisceau arri\`{e}re qui sert
\`{a} l'excitation optique du r\'{e}sonateur m\'{e}canique. Ce faisceau est
modul\'{e} en intensit\'{e} \`{a} l'aide d'un acousto-optique (AO)
pilot\'{e} par un synth\'{e}tiseur. Il se r\'{e}fl\'{e}chit ensuite sur le
miroir mobile et il exerce une force de pression de radiation modul\'{e}e
\`{a} la fr\'{e}quence du synth\'{e}tiseur. Le dispositif de mesure homodyne
d\'{e}tecte le mouvement du miroir}
\label{Fig_4excitop}
\end{figure}

La figure \ref{Fig_4excitop} montre le dispositif d'excitation optique du
r\'{e}sonateur. Le faisceau arri\`{e}re est pr\'{e}lev\'{e} \`{a} la sortie
du laser titane saphir, \`{a} l'aide du cube s\'{e}parateur de polarisation
plac\'{e} juste apr\`{e}s l'isolateur optique (voir figure \ref{Fig_4tisasta}%
, page \pageref{Fig_4tisasta}). En fait, l'essentiel de la puissance du
laser est r\'{e}fl\'{e}chi par ce cube afin de constituer le faisceau
arri\`{e}re, l'intensit\'{e} du faisceau allant vers la cavit\'{e} \'{e}tant
beaucoup plus faible (typiquement quelques dizaines de milliwatts). On
dispose ainsi d'une puissance de l'ordre de $700~mW$ pour exciter le
r\'{e}sonateur m\'{e}canique, ce qui repr\'{e}sente une puissance suffisante
pour obtenir une excitation efficace. Sur le trajet du faisceau (figure \ref
{Fig_4excitop}) nous avons plac\'{e} un cache mont\'{e} sur un support
rabattable afin de pouvoir couper le faisceau arri\`{e}re. Le faisceau
traverse ensuite une lentille convergente de focale \'{e}gale \`{a} $200~mm$
qui permet de focaliser le faisceau au niveau du modulateur acousto-optique,
avec un col de l'ordre de $70~\mu m$.

Pour moduler l'intensit\'{e} du faisceau, on utilise un modulateur
acousto-optique plut\^{o}t qu'un \'{e}lectro-optique suivi d'un polariseur.
Dans un modulateur acousto-optique, le faisceau lumineux interagit avec une
onde acoustique qui cr\'{e}e un r\'{e}seau sur lequel se diffracte l'onde
lumineuse. Lorsque l'angle d'incidence du faisceau correspond \`{a} l'angle
de Bragg, l'intensit\'{e} de l'onde diffract\'{e}e dans l'ordre $1$
d\'{e}pend essentiellement de la puissance de l'onde acoustique. Celle-ci
est produite par l'interm\'{e}diaire d'un g\'{e}n\'{e}rateur radio
fr\'{e}quence dont la puissance de sortie est contr\^{o}l\'{e}e par un
signal basse tension (typiquement $0$ \`{a} $5~Volts$). On peut ainsi
moduler en tout ou rien l'intensit\'{e} du faisceau transmis dans l'ordre $1$
jusqu'\`{a} des fr\'{e}quences de modulation de quelques m\'{e}gahertz.
L'obtention d'un r\'{e}sultat \'{e}quivalent avec un \'{e}lectro-optique
n\'{e}cessiterait une tension de commande de plusieurs centaines de volts.

Un autre avantage de l'acousto-optique est qu'il d\'{e}cale la fr\'{e}quence
optique du faisceau diffract\'{e} d'une valeur \'{e}gale \`{a} la
fr\'{e}quence de l'onde acoustique. En pratique, on a utilis\'{e} un
acousto-optique $AA$-$MP10$ (Automates et Automatismes) qui fonctionne \`{a} 
$200~MHz$. Le faisceau arri\`{e}re ayant \`{a} l'origine la m\^{e}me
fr\'{e}quence que le faisceau incident sur la cavit\'{e} \`{a} miroir
mobile, ce d\'{e}calage permet de placer le faisceau arri\`{e}re hors
r\'{e}sonance par rapport \`{a} la cavit\'{e}. Comme ce d\'{e}calage est
grand compar\'{e} \`{a} la bande passante de la cavit\'{e} (de l'ordre de $%
2~MHz$), le faisceau arri\`{e}re ne peut pas se propager dans la cavit\'{e}.
Ceci permet donc de limiter les perturbations induites par le faisceau
arri\`{e}re sur la mesure homodyne du faisceau r\'{e}fl\'{e}chi par la
cavit\'{e}.

Notons que ce filtrage en fr\'{e}quence n'est toutefois pas suffisant pour
assurer une isolation totale entre le faisceau arri\`{e}re et le syst\`{e}me
de d\'{e}tection homodyne. En effet, le filtrage induit par la cavit\'{e}
correspond \`{a} une att\'{e}nuation par un facteur de l'ordre de $10^{4}$
de l'intensit\'{e} transmise. Etant donn\'{e} la dissym\'{e}trie entre les
transmissions des deux miroirs de la cavit\'{e} \`{a} miroir mobile
(quelques $ppm$ pour le miroir mobile et $50~ppm$ pour le coupleur Newport),
le coefficient de transmission de la cavit\'{e} pour le faisceau arri\`{e}re
est de l'ordre de $10^{-6}$ (voir \'{e}quation \ref{4.2.5}). La puissance
r\'{e}siduelle du faisceau arri\`{e}re qui interagit avec le syst\`{e}me de
d\'{e}tection est donc de l'ordre de $0.1~\mu W$, soit environ cent fois
moins que le faisceau r\'{e}fl\'{e}chi par la cavit\'{e} dont la puissance
est de l'ordre de $10~\mu W$. D'autre part, le faisceau arri\`{e}re
pr\'{e}sente une forte modulation d'intensit\'{e}, puisque la profondeur de
modulation peut atteindre $100\%$, alors que la modulation de phase attendue
sur le faisceau r\'{e}fl\'{e}chi est tr\`{e}s petite : les d\'{e}placements
du miroir mobile devraient induire une modulation de phase dont la
profondeur ne d\'{e}passe pas $1\%$. Il appara\^{\i }t ainsi que le faisceau
arri\`{e}re constitue toujours, malgr\'{e} le filtrage en fr\'{e}quence, une
perturbation importante pour le syst\`{e}me de mesure. Nous verrons dans la
suite comment cet effet a pu \^{e}tre \'{e}limin\'{e} gr\^{a}ce \`{a} un
filtrage spatial.

\begin{figure}[tbp]
\centerline{\psfig{figure=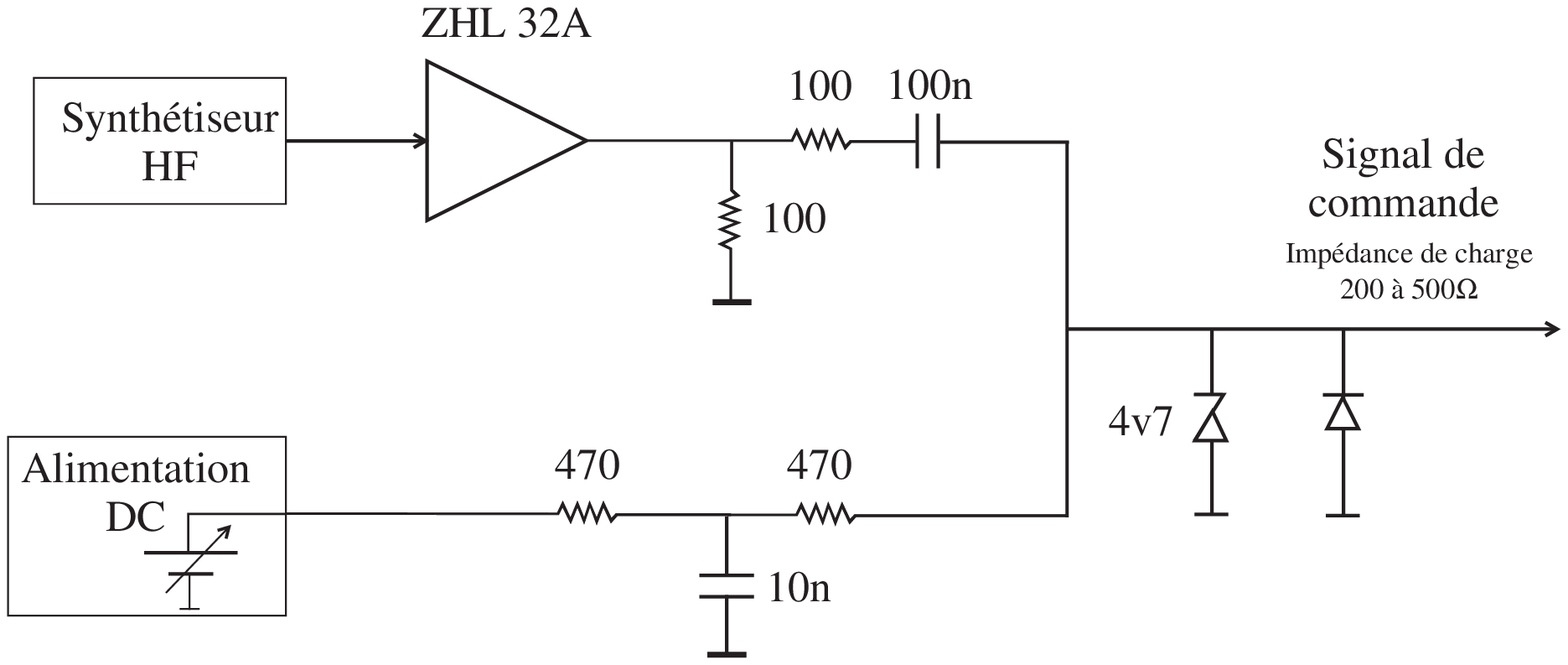,height=6cm}}
\caption{Electronique de commande de l'acousto-optique permettant de moduler
l'intensit\'{e} du faisceau arri\`{e}re}
\label{Fig_4aopcomm}
\end{figure}

L'intensit\'{e} lumineuse diffract\'{e}e par l'acousto-optique est
d\'{e}termin\'{e}e par un signal de commande pilotant le g\'{e}n\'{e}rateur
radio fr\'{e}quence. Lorsque l'amplitude de ce signal varie de $0$ \`{a} $%
5~Volts$, l'intensit\'{e} lumineuse diffract\'{e}e dans l'ordre $1$ varie de 
$0$ \`{a} une intensit\'{e} maximale qui correspond \`{a} l'efficacit\'{e}
de l'acousto-optique (intensit\'{e} lumineuse diffract\'{e}e dans l'ordre $1$%
, rapport\'{e}e \`{a} l'intensit\'{e} incidente). La modulation
d'intensit\'{e} est obtenue en appliquant sur cette entr\'{e}e de commande
une tension sinuso\"{\i }dale issue d'un synth\'{e}tiseur haute
fr\'{e}quence. La figure \ref{Fig_4aopcomm} montre le sch\'{e}ma de
l'\'{e}lectronique de commande. Un amplificateur $ZHL~32A$ permet d'obtenir
une modulation haute fr\'{e}quence avec une amplitude sup\'{e}rieure \`{a} $%
5~Volts$. Un ensemble de filtres $RC$ m\'{e}lange cette modulation avec une
tension continue de fa\c{c}on \`{a} fournir une tension modul\'{e}e comprise
entre $0$ et $5~Volts$. Une diode zener et une diode germanium permettent de
limiter l'excursion du signal de commande. Notons la pr\'{e}sence d'un
filtre en $T$ dans la voie $DC$. Il permet d'\'{e}viter tout retour de la
modulation haute fr\'{e}quence dans l'alimentation continue. Sans ce filtre,
l'alimentation \'{e}mettait un fort rayonnement \`{a} la fr\'{e}quence de
modulation, rayonnement qui \'{e}tait capt\'{e} par le syst\`{e}me de
d\'{e}tection homodyne. 

L'efficacit\'{e} de l'acousto-optique d\'{e}pend beaucoup de l'alignement du
r\'{e}seau acoustique par rapport au faisceau lumineux incident. Pour
optimiser cette efficacit\'{e}, nous avons fix\'{e} l'acousto-optique sur un
support r\'{e}glable \`{a} trois axes de rotation, mont\'{e} sur un ensemble
de translations horizontale et verticale. Lorsque le faisceau incident est
parfaitement focalis\'{e} au niveau du r\'{e}seau et que l'alignement est
optimis\'{e}, on arrive \`{a} une efficacit\'{e} de l'ordre de $95\%$.

Les \'{e}l\'{e}ments plac\'{e}s apr\`{e}s l'acousto-optique permettent de
d\'{e}terminer la position et la g\'{e}om\'{e}trie du faisceau arri\`{e}re
par rapport au miroir mobile (voir figure \ref{Fig_4excitop}). On trouve
tout d'abord un ensemble de deux miroirs mont\'{e}s sur supports
microm\'{e}triques qui permettent de positionner le faisceau au centre du
miroir mobile, puis un att\'{e}nuateur variable constitu\'{e} d'une lame
demi-onde et d'un cube s\'{e}parateur de polarisation qui sert \`{a}
contr\^{o}ler l'intensit\'{e} moyenne du faisceau arri\`{e}re. On
contr\^{o}le enfin la taille du faisceau arri\`{e}re en utilisant un jeu de
deux lentilles comme nous l'avons d\'{e}j\`{a} fait pour nos diff\'{e}rentes
cavit\'{e}s (FPE, FPF et cavit\'{e} \`{a} miroir mobile). Il n'est pas
n\'{e}cessaire ici d'adapter en taille et en position le col du faisceau
arri\`{e}re \`{a} celui de la cavit\'{e} : la pression de radiation ne
d\'{e}pend pas de la courbure du faisceau gaussien et seule importe la
section du faisceau au niveau du miroir mobile, que l'on veut pouvoir faire
varier dans des proportions importantes de fa\c{c}on \`{a} \'{e}tudier
l'influence sur le couplage optom\'{e}canique de l'adaptation spatiale. Nous
avons choisi une premi\`{e}re lentille de focale \'{e}gale \`{a} $50~mm$
plac\'{e}e \`{a} $650~mm$ de l'acousto-optique. La seconde lentille de
m\^{e}me focale est plac\'{e}e $120~mm$ plus loin sur une platine de
translation qui permet de varier la distance entre les deux lentilles. Avec
ce dispositif, la taille du col image est toujours tr\`{e}s petite,
inf\'{e}rieure \`{a} $50~\mu m$, mais sa position peut \^{e}tre
d\'{e}plac\'{e}e par rapport au miroir mobile qui se trouve \`{a} environ $%
350~mm$ de la seconde lentille. Du fait de la divergence importante du
faisceau, on fait varier de cette mani\`{e}re la section du faisceau au
niveau du miroir mobile. Nous avons mesur\'{e} cette section \`{a} l'aide
d'une barette de photodiodes. En variant la distance entre les deux
lentilles d'environ $12~mm$, la taille du faisceau au niveau du miroir
mobile varie de moins de $50~\mu m$ \`{a} plus de $400~\mu m$.

Enfin, comme le montre le sch\'{e}ma de la figure \ref{Fig_4excitop}, le
faisceau excitateur est envoy\'{e}, \`{a} l'aide d'un miroir mont\'{e} sur
un support microm\'{e}trique, sur le miroir mobile avec un angle d'incidence 
$\theta $ non nul, de l'ordre de $10{{}^\circ}$. C'est en effet un moyen
simple d'exciter le r\'{e}sonateur m\'{e}canique sans perturber la
transmission r\'{e}siduelle de la cavit\'{e} \`{a} miroir mobile qui sert
\`{a} maintenir le faisceau de mesure \`{a} r\'{e}sonance avec la
cavit\'{e}. D'autre part, on r\'{e}alise ainsi un filtrage spatial tr\`{e}s
efficace du faisceau arri\`{e}re. En effet, l'int\'{e}grale de recouvrement
spatial entre le faisceau arri\`{e}re et le mode fondamental de la
cavit\'{e} est proportionnelle \`{a} $e^{-\left( \theta /\theta _{0}\right)
^{2}}$ o\`{u} $\theta _{0}=0.2{{}^\circ}$ est l'angle de divergence du mode
fondamental ($\theta _{0}=\lambda /\left( \pi w_{0}\right) $). Le faisceau
arri\`{e}re n'est donc pratiquement pas coupl\'{e} au mode fondamental de la
cavit\'{e}, et on \'{e}limine ainsi toute lumi\`{e}re parasite qui pourrait
perturber le fonctionnement du dispositif de mesure homodyne.

\section{Cavit\'{e} de filtrage de grande finesse\label{IV-5}}

\bigskip

Comme nous l'avons vu dans la section \ref{IV-2-3}, le faisceau issu du
laser titane saphir pr\'{e}sente d'importantes fluctuations d'intensit\'{e}
\`{a} des fr\'{e}quences inf\'{e}rieures \`{a} $2~MHz$ (voir figure \ref
{Fig_4shot}, page \pageref{Fig_4shot}). Cette caract\'{e}ristique ne permet
pas d'observer les effets quantiques du couplage optom\'{e}canique pour des
fr\'{e}quences d'analyse inf\'{e}rieures \`{a} $2~MHz$. Le dispositif de
stabilisation d'intensit\'{e} plac\'{e} dans la source laser permet de
r\'{e}duire efficacement les fluctuations d'intensit\'{e} en dessous de
quelques dizaines de kilohertz (figure \ref{Fig_4asintdb}, page \ref
{Fig_4asintdb}). Il semble toutefois tr\`{e}s difficile de r\'{e}aliser une
boucle d'asservissement \'{e}lectronique qui puisse agir efficacement \`{a}
des fr\'{e}quences plus \'{e}lev\'{e}es sans rajouter un exc\`{e}s de bruit
au dessus de $2~MHz$. Nous avons alors envisag\'{e} une m\'{e}thode de
r\'{e}duction du bruit d'intensit\'{e} bas\'{e}e sur une technique optique,
qui utilise l'effet de filtrage en transmission d'une cavit\'{e} Fabry-Perot
r\'{e}sonnante. En effet, lorsque le faisceau incident est \`{a}
r\'{e}sonance avec la cavit\'{e}, la fonction de transfert de celle-ci
filtre toutes les composantes de bruit pour des fr\'{e}quences
sup\'{e}rieures \`{a} sa bande passante: plus la bande passante est
\'{e}troite et plus le filtrage est efficace. En utilisant une cavit\'{e}
ayant une faible bande passante et donc une grande finesse, on devrait ainsi
\^{e}tre capable de r\'{e}duire l'exc\`{e}s de bruit technique du laser pour
des fr\'{e}quences comprises entre $1$ et $2~MHz$. On notera enfin que cette
cavit\'{e} joue aussi un r\^{o}le de filtrage spatial comme nous l'avons
d\'{e}j\`{a} vu pour la cavit\'{e} FPF : \`{a} terme, cette cavit\'{e} de
grande finesse remplacera dans le montage exp\'{e}rimental la cavit\'{e} FPF.

\subsection{Principe du filtrage\label{IV-5-1}}

La cavit\'{e} que nous utilisons est sym\'{e}trique, ce qui permet d'avoir
une transmission de la cavit\'{e} maximale \`{a} r\'{e}sonance. Un
traitement complet des effets d'une telle cavit\'{e} sur les fluctuations du
champ doit tenir compte des fluctuations du vide $\delta \alpha ^{v}$ qui
sont coupl\'{e}es au champ intracavit\'{e} par la deuxi\`{e}me entr\'{e}e
(voir figure \ref{Fig_4cavreo}). Dans le cas d'une cavit\'{e} sans perte et
de grande finesse, les relations d'entr\'{e}e-sortie pour les fluctuations
sont similaires \`{a} celles obtenues dans la section \ref{II-3-4}. En
tenant compte du couplage avec les fluctuations du vide dans l'\'{e}quation
d'\'{e}volution du champ intracavit\'{e} (2.58a), on peut d\'{e}duire
l'expression \`{a} r\'{e}sonance des fluctuations $\delta \alpha ^{t}\left[
\Omega \right] $ du champ transmis par la cavit\'{e} en fonction des
fluctuations $\delta \alpha ^{in}\left[ \Omega \right] $ du champ entrant et
des fluctuations $\delta \alpha ^{v}\left[ \Omega \right] $ du vide: 
\begin{figure}[tbp]
\centerline{\psfig{figure=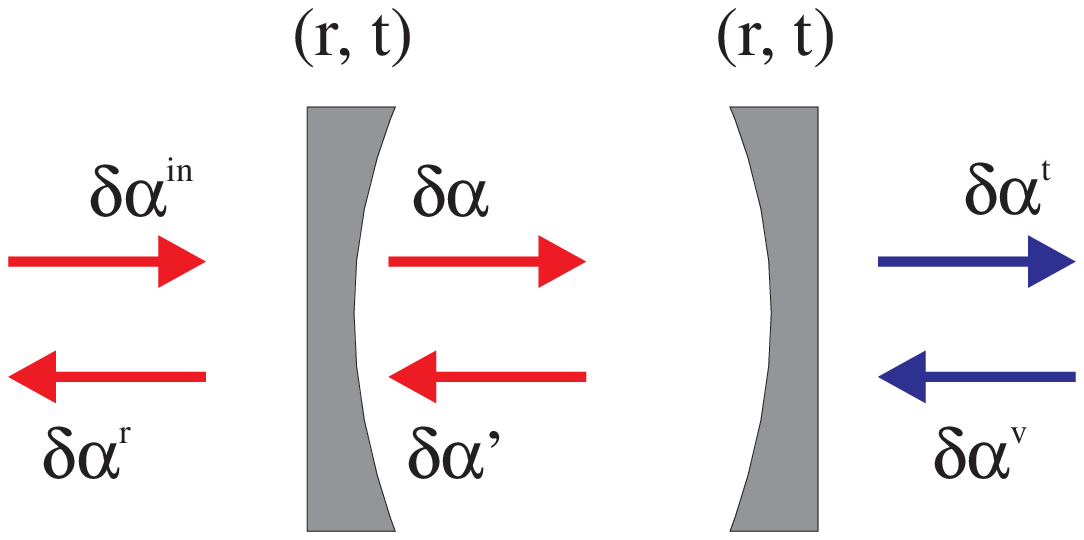,height=4cm}}
\caption{Cavit\'{e} Fabry-Perot sym\'{e}trique \`{a} deux
entr\'{e}es-sorties et sans pertes}
\label{Fig_4cavreo}
\end{figure}
\begin{equation}
\delta \alpha ^{t}\left[ \Omega \right] =t\left[ \Omega \right] ~\delta
\alpha ^{in}\left[ \Omega \right] +r\left[ \Omega \right] ~\delta \alpha
^{v}\left[ \Omega \right]  \label{4.4.1}
\end{equation}
o\`{u} $t\left[ \Omega \right] $ et $r\left[ \Omega \right] $ sont les
coefficient de transmission et de r\'{e}flexion en amplitude de la
cavit\'{e}. Pour une cavit\'{e} r\'{e}sonnante sans pertes et de grande
finesse, ces coefficients s'expriment simplement en fonction du coefficient
de transmission $t$ des miroirs: 
\begin{equation}
t\left[ \Omega \right] =\frac{t^{2}}{t^{2}+i\Omega \tau }\qquad ,\qquad
r\left[ \Omega \right] =-\frac{i\Omega \tau }{t^{2}+i\Omega \tau }
\label{4.4.2}
\end{equation}
Puisque les fluctuations $\delta \alpha ^{in}\left[ \Omega \right] $ et $%
\delta \alpha ^{v}\left[ \Omega \right] $ qui appara\^{\i }ssent dans
l'expression (\ref{4.4.1}) sont ind\'{e}pendantes, le spectre $%
S_{I}^{t}\left[ \Omega \right] $ des fluctuations d'intensit\'{e} du
faisceau transmis peut s'\'{e}crire en fonction du spectre de bruit
d'intensit\'{e} du faisceau incident $S_{I}^{in}\left[ \Omega \right] $: 
\begin{equation}
S_{I}^{t}\left[ \Omega \right] =\bar{I}^{t}~\left\{ {\rm T}\left[ \Omega
\right] ~\frac{S_{I}^{in}\left[ \Omega \right] }{\bar{I}^{in}}+{\rm R}\left[
\Omega \right] \right\}  \label{4.4.3}
\end{equation}
o\`{u} ${\rm T}\left[ \Omega \right] =\left| t\left[ \Omega \right] \right|
^{2}$ et ${\rm R}\left[ \Omega \right] =\left| r\left[ \Omega \right]
\right| ^{2}$ correspondent respectivement aux fonctions de transfert en
transmission et en r\'{e}flexion de la cavit\'{e}, et $\bar{I}^{in}$, $\bar{I%
}^{t}$ sont les intensit\'{e}s moyennes incidente et transmise. On notera
que la conservation de l'\'{e}nergie dans la cavit\'{e} est v\'{e}rifi\'{e}e
puisque la somme des fonctions ${\rm T}$ et ${\rm R}$ est \'{e}gale \`{a} $1$
quelque soit la fr\'{e}quence d'analyse.

Si le faisceau incident pr\'{e}sente un exc\`{e}s de bruit d'intensit\'{e}
par rapport au bruit de photon standard, nous avons vu dans la section \ref
{IV-2-3} (\'{e}quation 4.21) que l'on peut tenir compte de cette
caract\'{e}ristique \`{a} l'aide du facteur de Mandel $Q^{in}\left[ \Omega
\right] $: 
\begin{equation}
S_{I}^{in}\left[ \Omega \right] =\bar{I}^{in}~\left( 1+Q^{in}\left[ \Omega
\right] \right)  \label{4.4.4}
\end{equation}
En substituant cette expression dans la relation (\ref{4.4.3}) et en
utilisant la condition de conservation de l'\'{e}nergie v\'{e}rifi\'{e}e par
les fonctions de transfert, on obtient l'expression suivante pour le spectre
d'intensit\'{e} $S_{I}^{t}\left[ \Omega \right] $: 
\begin{equation}
S_{I}^{t}\left[ \Omega \right] =\bar{I}^{t}~\left( 1+{\rm T}\left[ \Omega
\right] ~Q^{in}\left[ \Omega \right] \right)  \label{4.4.5}
\end{equation}
Le second terme de cette relation montre que le faisceau transmis
pr\'{e}sente toujours un exc\`{e}s de bruit mais qu'il est filtr\'{e} par la
fonction de transfert en transmission ${\rm T}\left[ \Omega \right] $ de la
cavit\'{e}. La cavit\'{e} se comporte donc comme un filtre passe-bas de
forme lorentzienne et de fr\'{e}quence de coupure \'{e}gale \`{a} sa bande
passante $\nu _{BP}=1/\left( 2\tau {\cal F}\right) $ : pour des
fr\'{e}quences tr\`{e}s sup\'{e}rieures \`{a} la bande passante, les
fluctuations incidentes ne sont plus coupl\'{e}es au champ intracavit\'{e}.
Ainsi, les fluctuations du faisceau laser sont directement
r\'{e}fl\'{e}chies par le miroir d'entr\'{e}e, alors que les fluctuations du
faisceau transmis reproduisent les fluctuations du vide r\'{e}fl\'{e}chies
par le miroir de sortie.

Il appara\^{\i}t ainsi que l'efficacit\'{e} du filtrage de la cavit\'{e} est
d'autant plus grande que la bande passante est faible. La valeur de la bande
passante est d\'{e}termin\'{e}e par la longueur de la cavit\'{e} et par sa
finesse. Pour des longueurs raisonnables, il est n\'{e}cessaire d'utiliser
une cavit\'{e} de grande finesse, c'est-\`{a}-dire des miroirs ayant une
grande r\'{e}flectivit\'{e} et de faibles pertes. Dans ces conditions, la
valeur minimale de la bande passante est limit\'{e}e par la tenue au flux
des miroirs et par la puissance maximale du faisceau que l'on veut envoyer
dans la cavit\'{e}.

\subsection{Caract\'{e}ristiques de la cavit\'{e}\label{IV-5-2}}

La cavit\'{e} de filtrage que nous avons test\'{e}e est constitu\'{e}e d'un
miroir d'entr\'{e}e plan mont\'{e} sur une cale pi\'{e}zo\'{e}lectrique et
d'un miroir de sortie courbe de rayon de courbure \'{e}gal \`{a} $1~m$. Les
miroirs que nous avons utilis\'{e}s ont \'{e}t\'{e} fabriqu\'{e}s par $REO$
(Research Electro-Optics). Comme nous l'avons vu dans la section
pr\'{e}c\'{e}dente, les caract\'{e}ristiques des miroirs d\'{e}terminent
\`{a} la fois l'effet de filtrage et la puissance lumineuse admissible par
la cavit\'{e}. Nous avons choisi des miroirs ayant un coefficient de
transmission \'{e}gal \`{a} $2000\pm 50~ppm$ et des pertes inf\'{e}rieures
\`{a} $30~ppm$. La finesse th\'{e}orique de la cavit\'{e} est ainsi
\'{e}gale \`{a} $1600$. Etant donn\'{e} la tenue au flux des miroirs qui est
au moins de $10~kW/cm^{2}$, la puissance du faisceau lumineux peut \^{e}tre
sup\'{e}rieure \`{a} $50~mW$. Cette valeur est suffisante pour que cette
cavit\'{e} puisse remplacer la cavit\'{e} FPF plac\'{e}e dans la source
laser. Par ailleurs, la longueur de la cavit\'{e} est d\'{e}termin\'{e}e par
un barreau cylindrique en invar, de $210~mm$ de long, sur lequel sont
mont\'{e}s les miroirs. On obtient ainsi une bande passante th\'{e}orique de 
$230~kHz$. Enfin, la taille des miroirs est relativement petite ($7.75~mm$
de diam\`{e}tre et $4~mm$ d'\'{e}paisseur), ce qui permet d'optimiser la
dynamique de la cale pi\'{e}zo\'{e}lectrique qui sert \`{a} l'asservissement
de la cavit\'{e}.

Pour tester la cavit\'{e}, on a utilis\'{e} l'un des deux faisceaux
r\'{e}fl\'{e}chis par la lame de verre qui pr\'{e}l\`{e}ve le faisceau
envoy\'{e} vers le lambdam\`{e}tre (voir figure \ref{Fig_4tisasta}, page 
\pageref{Fig_4tisasta}). Pour adapter spatialement ce faisceau sur le mode
fondamental $TEM_{00}$ de la cavit\'{e}, on utilise comme pour les
cavit\'{e}s pr\'{e}c\'{e}dentes un jeu de deux lentilles convergentes. Nous
avons choisi une premi\`{e}re lentille de focale $200~mm$ plac\'{e}e \`{a} $%
1.7~m$ du col du laser. La seconde lentille de focale $100~mm$ est
plac\'{e}e $340~mm$ plus loin sur une platine de translation qui permet
d'ajuster le col image. On obtient ainsi un col de $0.3~mm$ \`{a} $460~mm$
de la seconde lentille, l\`{a} o\`{u} se trouve le miroir plan de la
cavit\'{e}. L'alignement du faisceau sur la cavit\'{e} est r\'{e}alis\'{e}
\`{a} l'aide de deux miroirs mont\'{e}s sur des supports microm\'{e}triques. 
\begin{figure}[tbp]
\centerline{\psfig{figure=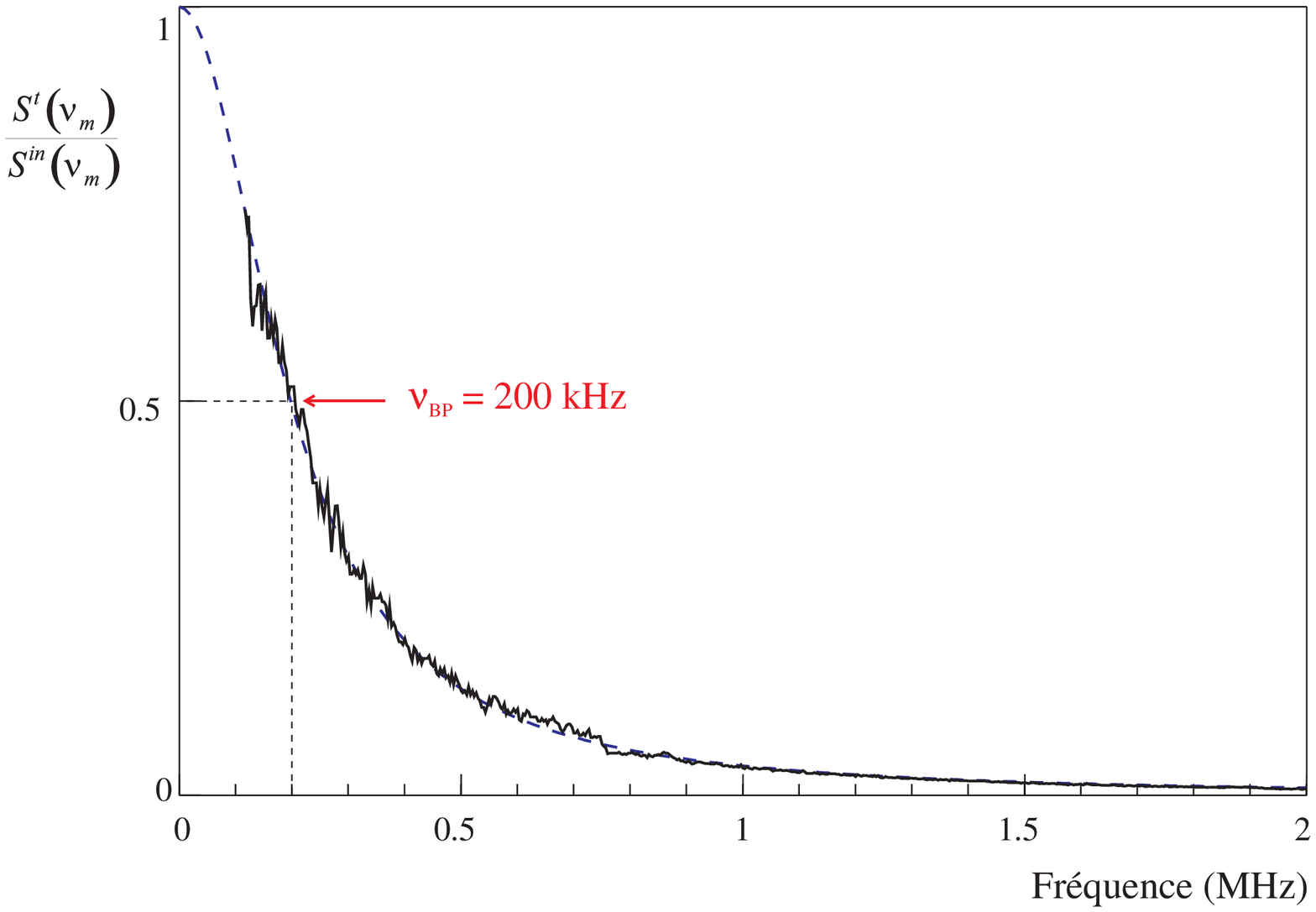,height=9cm}}
\caption{Fonction de transfert de la cavit\'{e} de filtrage de grande
finesse. L'ajustement Lorentzien en tirets permet de d\'{e}terminer la bande
passante $\nu _{BP}$ de la cavit\'{e} qui est \'{e}gale \`{a} $200~kHz$}
\label{Fig_4bpasreo}
\end{figure}

Nous avons mesur\'{e} la bande passante de cette cavit\'{e} en utilisant la
m\^{e}me technique de modulation d'intensit\'{e} du faisceau incident que
pour la cavit\'{e} FPF (voir section 4.2.4.3). On obtient la fonction de
transfert ${\rm T}\left[ \Omega \right] $ en faisant le rapport des spectres
de modulation d'intensit\'{e} transmis $S^{t}\left[ \Omega _{m}\right] $ et
incident $S^{in}\left[ \Omega _{m}\right] $, la cavit\'{e} \'{e}tant
maintenue \`{a} r\'{e}sonance avec le faisceau. Le r\'{e}sultat est
repr\'{e}sent\'{e} sur la figure \ref{Fig_4bpasreo}. Un ajustement
Lorentzien permet de d\'{e}terminer la bande passante $\nu _{BP}$ de la
cavit\'{e}, que l'on trouve \'{e}gale \`{a} $200~kHz$. Cette mesure
exp\'{e}rimentale de la bande passante permet par ailleurs d'estimer la
finesse de la cavit\'{e} ${\cal F}=\nu _{ISL}/2\nu _{BP}$, sachant que
l'intervalle spectral libre $\nu _{ISL}$ est \'{e}gal \`{a} $710~MHz$. On
trouve une finesse \'{e}gale \`{a} $1780$, en bon accord avec la valeur
th\'{e}orique de $1600$.

\subsection{Asservissement et effet de filtrage\label{IV-5-3}}

L'efficacit\'{e} du filtrage du bruit d'intensit\'{e} par la cavit\'{e}
Fabry-Perot d\'{e}pend aussi de la qualit\'{e} de l'asservissement qui sert
\`{a} maintenir la cavit\'{e} \`{a} r\'{e}sonance avec le faisceau incident.
L'asservissement que nous avons mis au point repose sur la technique des
bandes lat\'{e}rales que nous avons d\'{e}j\`{a} utilis\'{e}e pour la
stabilisation en fr\'{e}quence du laser sur la cavit\'{e} FPE (voir section 
\ref{IV-2-2}). Le sch\'{e}ma de principe de cet asservissement est tout
\`{a} fait similaire \`{a} celui repr\'{e}sent\'{e} sur la figure \ref
{Fig_4fpeasse} page \pageref{Fig_4fpeasse}. La seule diff\'{e}rence
r\'{e}side dans le fait que l'asservissement agit sur la cale
pi\'{e}zo\'{e}lectrique de la cavit\'{e} de filtrage et non sur la
fr\'{e}quence du laser titane saphir.

L'\'{e}lectro-optique utilis\'{e} pour moduler la phase du faisceau est un
mod\`{e}le New Focus $4001M$ r\'{e}sonnant \`{a} $12~MHz$, mont\'{e} sur une
platine de positionnement New Focus $9071$. Le choix de cette fr\'{e}quence
de modulation est li\'{e} aux caract\'{e}ristiques de la cavit\'{e} de
filtrage : elle est grande par rapport \`{a} la bande passante de la
cavit\'{e} et petite devant l'intervalle spectral entre modes transverses ($%
110~MHz$). D'autre part, il est pr\'{e}f\'{e}rable de choisir une
fr\'{e}quence assez \'{e}loign\'{e}e de la fr\'{e}quence de $20~MHz$
utilis\'{e}e pour l'asservissement en fr\'{e}quence du laser titane saphir,
afin d'\'{e}viter d'\'{e}ventuelles interf\'{e}rences. Le synth\'{e}tiseur
qui pilote l'\'{e}lectro-optique, le m\'{e}langeur et le bloc photodiode
sont similaires \`{a} ceux utilis\'{e}s pour la cavit\'{e} FPE, les filtres
\'{e}tant adapt\'{e}s \`{a} la fr\'{e}quence de modulation de $12~MHz$. Le
signal d'erreur obtenu \`{a} la sortie du m\'{e}langeur est utilis\'{e} pour
agir sur la longueur de la cavit\'{e} de filtrage de fa\c{c}on \`{a}
maintenir celle-ci \`{a} r\'{e}sonance avec la fr\'{e}quence du faisceau
incident. L'\'{e}lectronique de commande qui pilote la cale
pi\'{e}zo\'{e}lectrique de la cavit\'{e} est constitu\'{e}e d'un premier
int\'{e}grateur de pente $-6~dB/octave$, suivi par un deuxi\`{e}me
int\'{e}grateur pour des fr\'{e}quences inf\'{e}rieures \`{a} $700~Hz$, afin
d'augmenter le gain \`{a} basse fr\'{e}quence. Le signal est enfin
envoy\'{e} sur un amplificateur haute tension ($0$-$1000~V$) qui pilote la
cale pi\'{e}zo\'{e}lectrique. La r\'{e}sistance de sortie de l'amplificateur
haute tension ($460~k\Omega $) est choisie de mani\`{e}re \`{a} constituer
avec la capacit\'{e} de la cale pi\'{e}zo\'{e}lectrique ($10nF$) un filtre
passe-bas de fr\'{e}quence \'{e}gale \`{a} $35~Hz$. 
\begin{figure}[tbp]
\centerline{\psfig{figure=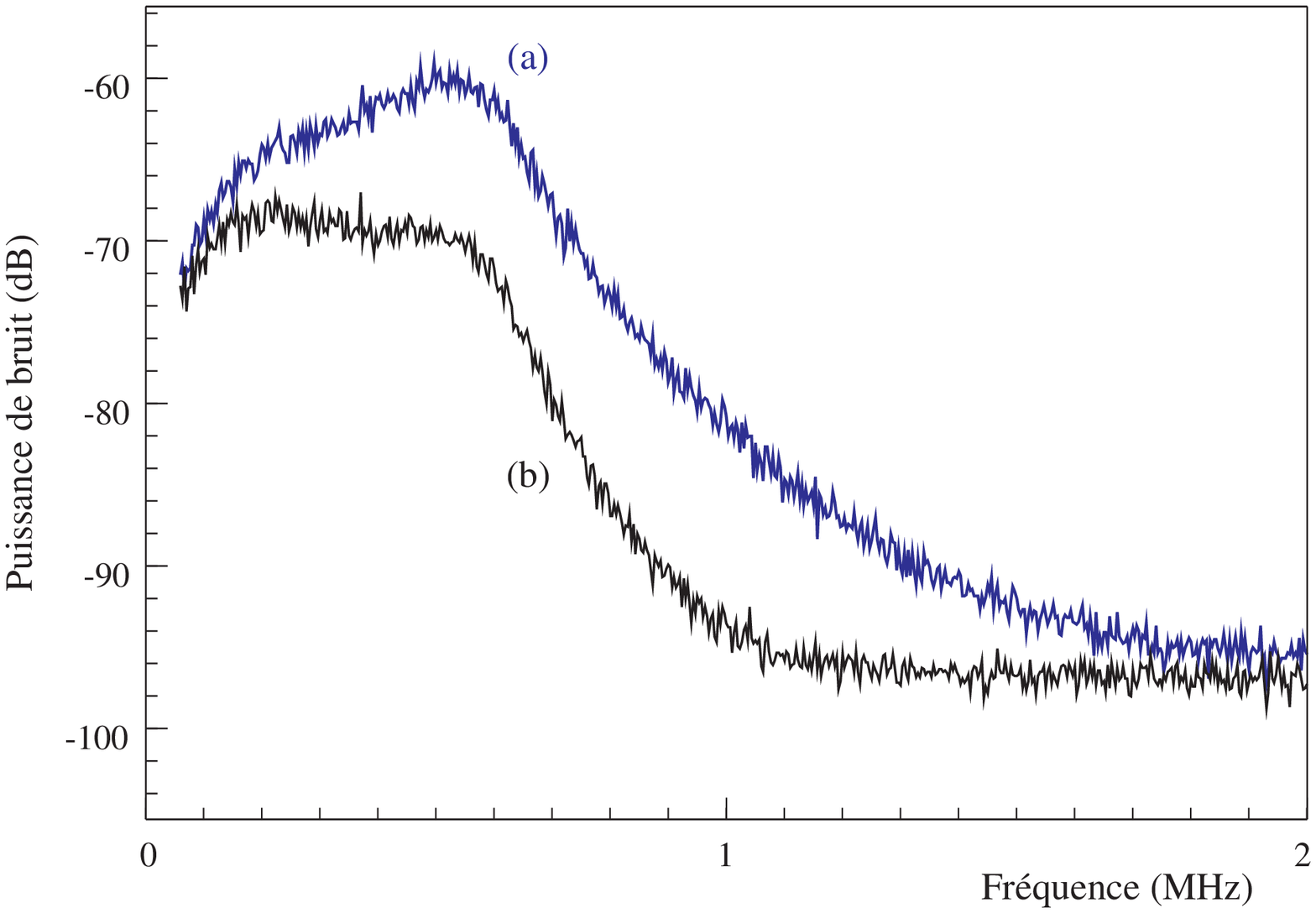,height=8cm}}
\caption{Spectre de bruit d'intensit\'{e} du faisceau avant (a) et apr\`{e}s
(b) le filtrage par la cavit\'{e} de grande finesse, pour une puissance
lumineuse de $4~mW$}
\label{Fig_4filtreo}
\end{figure}

Lorsque l'asservissement est activ\'{e}, l'intensit\'{e} transmise devrait
pr\'{e}senter un bruit technique r\'{e}duit par rapport \`{a} celui du
faisceau incident. Pour observer cet effet de filtrage du bruit
d'intensit\'{e}, nous avons mesur\'{e} les spectres de bruit d'intensit\'{e}
incident et transmis, pour la m\^{e}me intensit\'{e} moyenne. Les spectres
sont obtenus en pla\c{c}ant un bloc photodiode rapide ($FND100$ et
pr\'{e}amplificateur rapide transimp\'{e}dance) avant et apr\`{e}s la
cavit\'{e}. La figure \ref{Fig_4filtreo} montre le r\'{e}sultat de la
mesure. La trace (a) repr\'{e}sente le spectre de bruit d'intensit\'{e} du
faisceau incident pour une puissance de $4~mW$. Comme nous l'avons
d\'{e}j\`{a} vu (figure \ref{Fig_4shot}), on retrouve ici un bruit technique
important pour des fr\'{e}quences inf\'{e}rieures \`{a} $1~MHz$ et le bruit
d'intensit\'{e} du faisceau n'atteint le bruit de photon standard qu'\`{a}
des fr\'{e}quences sup\'{e}rieures \`{a} $2~MHz$. La trace (b)
repr\'{e}sente le spectre de bruit d'intensit\'{e} du faisceau transmis par
la cavit\'{e} pour la m\^{e}me intensit\'{e} moyenne. Ce spectre montre bien
l'effet du filtrage de la cavit\'{e} sur le bruit technique du faisceau
incident. On observe en particulier une r\'{e}duction de l'ordre de $-3~dB$
\`{a} la fr\'{e}quence de coupure ($200~kHz$) de la fonction de transfert de
la cavit\'{e}. Au-del\`{a}, l'effet de filtrage devient de plus en plus
important jusqu'\`{a} atteindre $-12~dB$ pour des fr\'{e}quences voisines de 
$1~MHz$. On obtient ainsi \`{a} la sortie de la cavit\'{e} de filtrage un
faisceau de puissance \'{e}gale \`{a} $4~mW$ et dont le bruit
d'intensit\'{e} rejoint le bruit de photon standard \`{a} partir de $1~MHz$.
Notons enfin que ce r\'{e}sultat d\'{e}pend de l'intensit\'{e} moyenne du
faisceau (voir \'{e}quation 4.23) et que l'on peut atteindre la limite du
bruit quantique \`{a} des fr\'{e}quences inf\'{e}rieures au m\'{e}gahertz
pour des puissances plus faibles. \newpage

%% file: chapter5.tex
\chapter{RESULTATS EXPERIMENTAUX}

\bigskip \bigskip

Nous avons d\'{e}crit dans le chapitre pr\'{e}c\'{e}dent les diff\'{e}rents
\'{e}l\'{e}ments du montage exp\'{e}rimental et leurs principales
caract\'{e}ristiques. Comme nous l'avons indiqu\'{e} dans la partie \ref
{IV-1}, la finesse de la cavit\'{e} \`{a} miroir mobile n'est pas suffisante
pour mettre en \'{e}vidence les effets quantiques du couplage
optom\'{e}canique. Par ailleurs, une telle mise en \'{e}vidence
n\'{e}cessite de r\'{e}duire le bruit thermique du miroir mobile. Un
cryostat, sp\'{e}cifiquement adapt\'{e} \`{a} notre cavit\'{e}, est en cours
de r\'{e}alisation. Notre montage nous a cependant permis de r\'{e}aliser la
premi\`{e}re \'{e}tape de l'exp\'{e}rience qui consiste \`{a}
caract\'{e}riser le couplage optom\'{e}canique dans la cavit\'{e}. Cette
\'{e}tude est importante afin d'optimiser les caract\'{e}ristiques optiques
et m\'{e}caniques de la cavit\'{e}.

Nous avons aussi pu mettre en \'{e}vidence l'extr\^{e}me sensibilit\'{e} de
notre dispositif \`{a} des petits d\'{e}placements du miroir mobile. Nous
avons en particulier observ\'{e} le mouvement Brownien du miroir mobile. Ce
r\'{e}sultat d\'{e}montre qu'il est possible d'utiliser une cavit\'{e} de
grande finesse pour mener une \'{e}tude quantitative du bruit thermique d'un
r\'{e}sonateur m\'{e}canique. Etant donn\'{e} la sensibilit\'{e} atteinte,
un tel dispositif constitue aussi une alternative int\'{e}ressante aux
dispositifs capacitifs plac\'{e}s sur les barres de Weber\cite{optique weber}%
.

Nous pr\'{e}sentons dans ce chapitre l'ensemble de ces r\'{e}sultats
exp\'{e}rimentaux. Nous commencerons par pr\'{e}senter le spectre de bruit
de phase du faisceau r\'{e}fl\'{e}chi par la cavit\'{e} \`{a} miroir mobile.
Ce spectre nous a permis de mettre en \'{e}vidence le mouvement Brownien du
r\'{e}sonateur (partie \ref{V-1}). Nous pr\'{e}senterons ensuite les
r\'{e}sultats concernant la caract\'{e}risation de la r\'{e}ponse
m\'{e}canique du r\'{e}sonateur (partie \ref{V-2}). La derni\`{e}re partie
de ce chapitre est consacr\'{e}e \`{a} la d\'{e}termination de la
sensibilit\'{e} de la cavit\'{e} \`{a} des petits d\'{e}placements du miroir
mobile (partie \ref{V-3}).

\section{Observation du bruit thermique\label{V-1}}

\bigskip

Comme nous l'avons expliqu\'{e} dans le chapitre 2 (sections \ref{II-2-3} et
\ref{II-4-2}), une application de notre montage consiste \`{a} mesurer le
spectre des fluctuations thermiques du r\'{e}sonateur m\'{e}canique \`{a}
temp\'{e}rature ambiante. La sensibilit\'{e} de la cavit\'{e} devrait nous
permettre d'observer non seulement les pics de bruit thermique associ\'{e}s
aux r\'{e}sonances acoustiques du r\'{e}sonateur mais aussi le bruit de fond
entre ces r\'{e}sonances. Pour r\'{e}aliser cette \'{e}tude, le faisceau
laser est maintenu \`{a} r\'{e}sonance avec la cavit\'{e} \`{a} l'aide de
l'asservissement d\'{e}crit dans la section \ref{IV-2-5}. L'intensit\'{e} du
faisceau entrant dans la cavit\'{e}, qui est \'{e}gale \`{a} $100~\mu W$,
est suffisamment faible pour pouvoir n\'{e}gliger les effets de pression de
radiation du champ intracavit\'{e}. On mesure alors les fluctuations de
phase du faisceau r\'{e}fl\'{e}chi induites par les variations de longueur
de la cavit\'{e}, en verrouillant l'asservissement de l'oscillateur local de
mani\`{e}re \`{a} ce que les deux faisceaux soient en quadrature de phase :
le signal \`{a} la sortie du syst\`{e}me de d\'{e}tection \'{e}quilibr\'{e}e
en configuration soustracteur est proportionnel aux fluctuations de phase du
faisceau r\'{e}fl\'{e}chi (\'{e}quation \ref{4.3.5}, page \pageref{4.3.5}).
Le spectre de bruit ainsi obtenu est la somme du bruit de photon standard
correspondant au bruit de phase du faisceau incident et de la contribution
associ\'{e}e au d\'{e}placement du miroir mobile (voir \'{e}quation \ref{4.7}%
, page \pageref{4.7}).

\subsection{Acquisition des spectres\label{V-1-1}}

L'acquisition de spectres que nous devons r\'{e}aliser pour observer les
pics de bruit thermique est relativement d\'{e}licate. Ces pics sont en
effet \'{e}troits : pour une fr\'{e}quence de r\'{e}sonance de $2~MHz$ et un
facteur de qualit\'{e} de $10^{5}$, la largeur du pic est \'{e}gale \`{a} $%
20~Hz$. Par ailleurs, on veut pouvoir explorer syst\'{e}matiquement une
large plage de fr\'{e}quence, sup\'{e}rieure \`{a} la bande passante de la
cavit\'{e} \`{a} miroir mobile. Ainsi une simple acquisition directe par un
analyseur de spectre n'est pas adapt\'{e}e \`{a} cette mesure. Le nombre de
points fournis par l'analyseur de spectre que nous utilisons ($HP~8560E$)
est limit\'{e} \`{a} $600$, et cela quelque soit sa bande d'analyse ({\it %
span}) : dans une plage de fr\'{e}quence de $4~MHz$ ceci correspond \`{a}
une r\'{e}solution spectrale d'environ $1$ point tous les $7~kHz$.

Pour am\'{e}liorer cette r\'{e}solution, on peut diviser la largeur totale
d'analyse en plusieurs intervalles plus petits. Mais il faudrait environ $%
200 $ intervalles de $20~kHz$ de large pour atteindre une r\'{e}solution
d'un point tous les $30~Hz$. C'est pourquoi nous avons pr\'{e}f\'{e}r\'{e}
utiliser une m\'{e}thode d'acquisition plus \'{e}labor\'{e}e, qui permet de
r\'{e}aliser une acquisition sur une plage de fr\'{e}quence de $500~kHz$ de
large avec la r\'{e}solution souhait\'{e}e. Le principe de cette m\'{e}thode
consiste \`{a} r\'{e}cup\'{e}rer le signal fourni par l'analyseur de spectre
sur un oscilloscope digital ($TDS420$) synchronis\'{e} avec l'analyseur de
spectre. Le nombre de points \'{e}lev\'{e} de l'oscilloscope ($15000$
points) permet de r\'{e}duire consid\'{e}rablement le nombre d'intervalles,
puisque l'on passe \`{a} seulement $8$ intervalles de largeur \'{e}gale
\`{a} $500~kHz$, pour parcourir la plage de fr\'{e}quence de $0$ \`{a} $%
4~MHz $ tout en gardant une r\'{e}solution d'un point tous les $30~Hz$.

Pour g\'{e}rer cette acquisition, on utilise un programme informatique qui
pilote \`{a} la fois l'analyseur de spectre et l'oscilloscope, connect\'{e}s
\`{a} un ordinateur PC par une liaison GPIB. Le signal issu du
soustracteur-sommateur est envoy\'{e} dans l'analyseur de spectre qui
fonctionne en mode d'acquisition d\'{e}clench\'{e} ({\it single sweep}), ce
qui permet de contr\^{o}ler le d\'{e}clenchement du balayage \`{a} partir du
programme. On utilise la sortie vid\'{e}o de l'analyseur de spectre pour
r\'{e}cup\'{e}rer le spectre sur l'oscilloscope. Pour une r\'{e}solution
spectrale ({\it rbw}) de l'analyseur de spectre sup\'{e}rieure \`{a} $300~Hz$%
, cette sortie vid\'{e}o fournit un signal analogique qui repr\'{e}sente le
spectre de bruit. La fr\'{e}quence d'analyse est balay\'{e}e de mani\`{e}re
lin\'{e}aire en fonction du temps, la plage de fr\'{e}quence totale ($%
500~kHz $) \'{e}tant parcourue durant le temps de balayage de l'analyseur de
spectre. Le signal fourni \`{a} un instant donn\'{e} correspond \`{a} la
puissance de bruit \`{a} la fr\'{e}quence d'analyse associ\'{e}e. Ce signal
est compris entre $0$ et $1~Volt$, avec un taux de conversion de $9.6~mV/dB$%
, le niveau $1~Volt$ correspondant au niveau de r\'{e}f\'{e}rence ({\it %
reference level}) de l'analyseur.

L'acquisition par l'oscilloscope de ce signal est r\'{e}alis\'{e}e de
mani\`{e}re synchrone en d\'{e}clenchant ce dernier \`{a} partir de la voie
de sortie {\it trigger} de l'analyseur de spectre. Le temps de balayage de
l'oscilloscope est choisi un peu plus grand que celui de l'analyseur de
spectre, ce qui permet de r\'{e}cup\'{e}rer le spectre sur toute la plage de
fr\'{e}quence. D\`{e}s que l'oscilloscope termine son acquisition, le
programme d\'{e}clenche \`{a} nouveau le balayage de l'analyseur de spectre
et un nouveau cycle d'acquisition commence. Cette proc\'{e}dure est
r\'{e}p\'{e}t\'{e}e un certain nombre de fois de fa\c{c}on \`{a} moyenner la
trace au niveau de l'oscilloscope (qui fonctionne en mode {\it average}). Le
programme sauvegarde ensuite les donn\'{e}es dans un fichier, apr\`{e}s
avoir converti le signal fourni par l'oscilloscope en $dB$. Enfin, le
programme relance toute la proc\'{e}dure d'acquisition pour l'intervalle de
fr\'{e}quence suivant, en augmentant de $500~kHz$ la fr\'{e}quence centrale
de l'analyseur. La boucle globale d'acquisition se termine lorsque toute la
bande de fr\'{e}quence (de $0$ \`{a} $4~MHz$) a \'{e}t\'{e} balay\'{e}e.

En pratique, la configuration des param\`{e}tres de l'analyseur de spectre
est la suivante : r\'{e}solution spectrale ({\it rbw}) de $300~Hz$, bande de
fr\'{e}quence d'analyse ({\it span}) de $500~kHz$, filtre vid\'{e}o ({\it %
video filter}) de $3~MHz$, temps de balayage ({\it sweep time}) de $14~${\it %
secondes}. En ce qui concerne l'oscilloscope, on choisit une base de temps
de $50~ms/div$, ce qui permet d'acqu\'{e}rir les $15000$ points en $15~${\it %
secondes}, c'est \`{a} dire en un temps un peu plus long que le temps de
balayage de l'analyseur de spectre. L'oscilloscope fonctionne d'autre part
en mode moyennage ({\it average}). En choisissant un nombre de moyennes
\'{e}gal \`{a} $60$, on obtient un spectre de $500~kHz$ de large en $15$
minutes. Pour parcourir l'ensemble de l'intervalle de $0$ \`{a} $4~MHz$ par
tranches de $500~kHz$, il est n\'{e}cessaire d'acqu\'{e}rir $8$ spectres, ce
qui correspond \`{a} un temps total d'acquisition de $2$ heures. Notons que
tous les asservissements sont suffisamment efficaces pour assurer une
compensation des d\'{e}rives lentes et un fonctionnement stable du montage
durant ce laps de temps.

\subsection{Spectre de bruit de phase du faisceau r\'{e}fl\'{e}chi\label%
{V-1-2}}

\begin{figure}[tbp]
\centerline{\psfig{figure=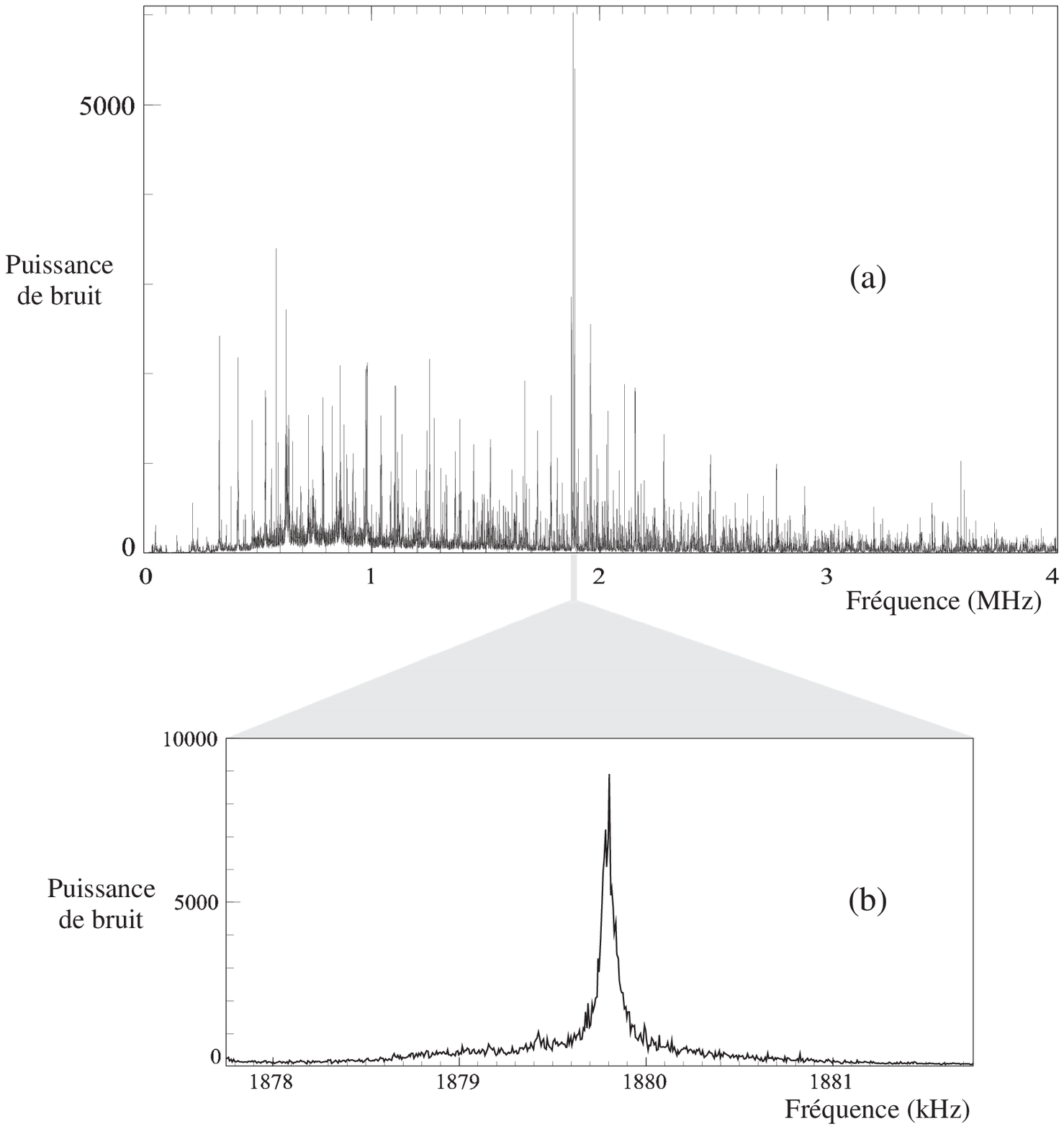,height=16cm}}
\caption{Spectres normalis\'{e}s au bruit de photon standard ({\it shot noise%
}) des fluctuations de phase du faisceau r\'{e}fl\'{e}chi par la cavit\'{e}
\`{a} miroir mobile $\left\{ N1,~M32\right\}$. Le spectre (a) a \'{e}t\'{e}
obtenu par la m\'{e}thode d'acquisition d\'{e}crite dans la section
pr\'{e}c\'{e}dente alors que le spectre (b) a \'{e}t\'{e} obtenu en
utilisant uniquement l'analyseur de spectre sur une petite bande de
fr\'{e}quence ($4~kHz$) autour de la fr\'{e}quence du mode acoustique
fondamental du r\'{e}sonateur}
\label{Fig_5m32ther}
\end{figure}

Un point important pour une mesure de spectre de bruit consiste \`{a}
calibrer le bruit de photon standard. Dans une mesure homodyne, ce bruit
correspond au bruit de photon de l'oscillateur local, auquel on a acc\`{e}s
en masquant le faisceau r\'{e}fl\'{e}chi par la cavit\'{e} \`{a} miroir
mobile (voir \'{e}quation \ref{4.3.5}, page \pageref{4.3.5}). Par ailleurs,
nous avons v\'{e}rifi\'{e} que le bruit de phase du faisceau
r\'{e}fl\'{e}chi, lorsque la cavit\'{e} n'est pas \`{a} r\'{e}sonance avec
le laser, est \'{e}gal au bruit de photon standard. Ainsi toute
d\'{e}viation par rapport au bruit de photon standard que l'on observe dans
le bruit de phase du faisceau r\'{e}fl\'{e}chi par la cavit\'{e} \`{a}
r\'{e}sonance est li\'{e}e \`{a} l'interaction du faisceau avec la
cavit\'{e}.

Le spectre de bruit de phase normalis\'{e} au bruit de photon standard,
obtenu \`{a} r\'{e}sonance avec la cavit\'{e} \`{a} miroir mobile $\left\{
N1,~M32\right\} $ est repr\'{e}sent\'{e} sur la figure \ref{Fig_5m32ther}a.
On constate la pr\'{e}sence de nombreux pics de bruit thermique dans la
plage de fr\'{e}quence \'{e}tudi\'{e}e, de $0$ \`{a} $4~MHz$. On peut noter
aussi la tr\`{e}s grande sensibilit\'{e} de la mesure, puisque les pics sont
plusieurs milliers de fois plus grands que le bruit de photon standard. Ce
point sera \'{e}tudi\'{e} plus en d\'{e}tail dans la partie \ref{V-3}.

Ce spectre pr\'{e}sente de nombreux pics, en particulier pour des
fr\'{e}quences inf\'{e}rieures \`{a} la fr\'{e}quence de r\'{e}sonance du
modes acoustique fondamental du miroir mobile, dont la valeur th\'{e}orique
est de l'ordre de $1.9~MHz$. Cela signifie que le faisceau intracavit\'{e}
est sensible au bruit thermique d'\'{e}l\'{e}ments autres que le miroir
mobile. On peut incriminer essentiellement deux \'{e}l\'{e}ments dont les
vibrations m\'{e}caniques sont susceptibles de modifier la longueur de la
cavit\'{e} : il s'agit du coupleur Newport d'entr\'{e}e et de l'espaceur
entre les deux miroirs. Nous avons cherch\'{e} \`{a} d\'{e}terminer quels
pics sont dus \`{a} chacun de ces \'{e}lements, en comparant les spectres
obtenus avec diff\'{e}rentes cavit\'{e}s. Nous avons ainsi utilis\'{e} la
cavit\'{e} $\left\{ N1,~M32\right\} $ avec diff\'{e}rents espaceurs (cuivre,
silice et plexiglass) sans noter de diff\'{e}rence appr\'{e}ciable entre les
spectres. Par contre le spectre obtenu avec la cavit\'{e} $\left\{
N1,~N2\right\} $ reproduit la plupart des pics du spectre de la figure \ref
{Fig_5m32ther}a, mais le pic le plus \'{e}lev\'{e}, dont la fr\'{e}quence
est voisine de $1.9~MHz$, a disparu. Ces r\'{e}sultats semblent indiquer que
la plupart des pics sont dus au mouvement Brownien du coupleur d'entr\'{e}e,
tandis que le pic aux alentours de $1.9~MHz$ est d\^{u} au mode fondamental
du miroir mobile. Une confirmation de ce r\'{e}sultat sera pr\'{e}sent\'{e}e
dans la partie suivante, gr\^{a}ce \`{a} l'\'{e}tude de la r\'{e}ponse
m\'{e}canique \`{a} une excitation optique du miroir mobile et du coupleur
d'entr\'{e}e.

Le spectre de la figure \ref{Fig_5m32ther}b repr\'{e}sente le pic de bruit
thermique dont la fr\'{e}quence est voisine de $1.9~MHz$. Il a \'{e}t\'{e}
obtenu directement \`{a} l'aide de l'analyseur de spectre, dont la plage de
balayage ainsi que la r\'{e}solution spectrale ont \'{e}t\'{e} r\'{e}duites
afin d'am\'{e}liorer la d\'{e}finition de la r\'{e}sonance. Ainsi, pour une
plage de fr\'{e}quence de $4~kHz$ autour de la r\'{e}sonance et une
r\'{e}solution spectrale \'{e}gale \`{a} $10~Hz$, on peut visualiser la
forme de cette r\'{e}sonance. Ce spectre montre par ailleurs qu'il est
possible d'observer le bruit thermique m\^{e}me tr\`{e}s loin sur les ailes
de la r\'{e}sonance puisque ce bruit est toujours bien sup\'{e}rieur au
bruit de photon standard. La sensibilit\'{e} atteinte dans notre montage
devrait donc permettre de mener une \'{e}tude exhaustive du bruit thermique
et des m\'{e}canismes de dissipation des modes acoustiques internes du
r\'{e}sonateur.

\section{R\'{e}ponse m\'{e}canique du r\'{e}sonateur\label{V-2}}

\bigskip

Afin d'am\'{e}liorer la caract\'{e}risation des modes acoustiques des
miroirs de la cavit\'{e}, nous avons d\'{e}velopp\'{e} une m\'{e}thode
d'excitation optique utilisant le faisceau arri\`{e}re dont
l'impl\'{e}mentation dans l'exp\'{e}rience a \'{e}t\'{e} d\'{e}crite dans la
partie \ref{IV-4}. Ce faisceau est modul\'{e} en intensit\'{e} \`{a} une
fr\'{e}quence variable, ce qui permet d'exciter s\'{e}lectivement les
diff\'{e}rents modes acoustiques du miroir arri\`{e}re de la cavit\'{e}. On
d\'{e}tecte alors l'effet du mouvement du miroir sur la phase du faisceau
r\'{e}fl\'{e}chi par la cavit\'{e}, lorsque celle-ci est r\'{e}sonnante avec
le faisceau incident. L'effet du mouvement appara\^{\i }t au niveau du
spectre de phase sous la forme d'un pic de modulation lorsqu'on fait varier
la fr\'{e}quence du synth\'{e}tiseur qui pilote le modulateur
acousto-optique autour d'une fr\'{e}quence de r\'{e}sonance acoustique. Nous
pr\'{e}sentons dans cette partie les r\'{e}sultats obtenus pour des
fr\'{e}quences voisines du pic de bruit thermique observ\'{e} dans la partie
pr\'{e}c\'{e}dente (section \ref{V-2-1}). Nous d\'{e}crivons ensuite les
r\'{e}sultats concernant les r\'{e}ponses m\'{e}caniques du miroir mobile et
du coupleur d'entr\'{e}e, \'{e}tudi\'{e}es sur une large plage de
fr\'{e}quence (section \ref{V-2-2}).

\subsection{Le mode acoustique fondamental\label{V-2-1}}

\begin{figure}[tbp]
\centerline{\psfig{figure=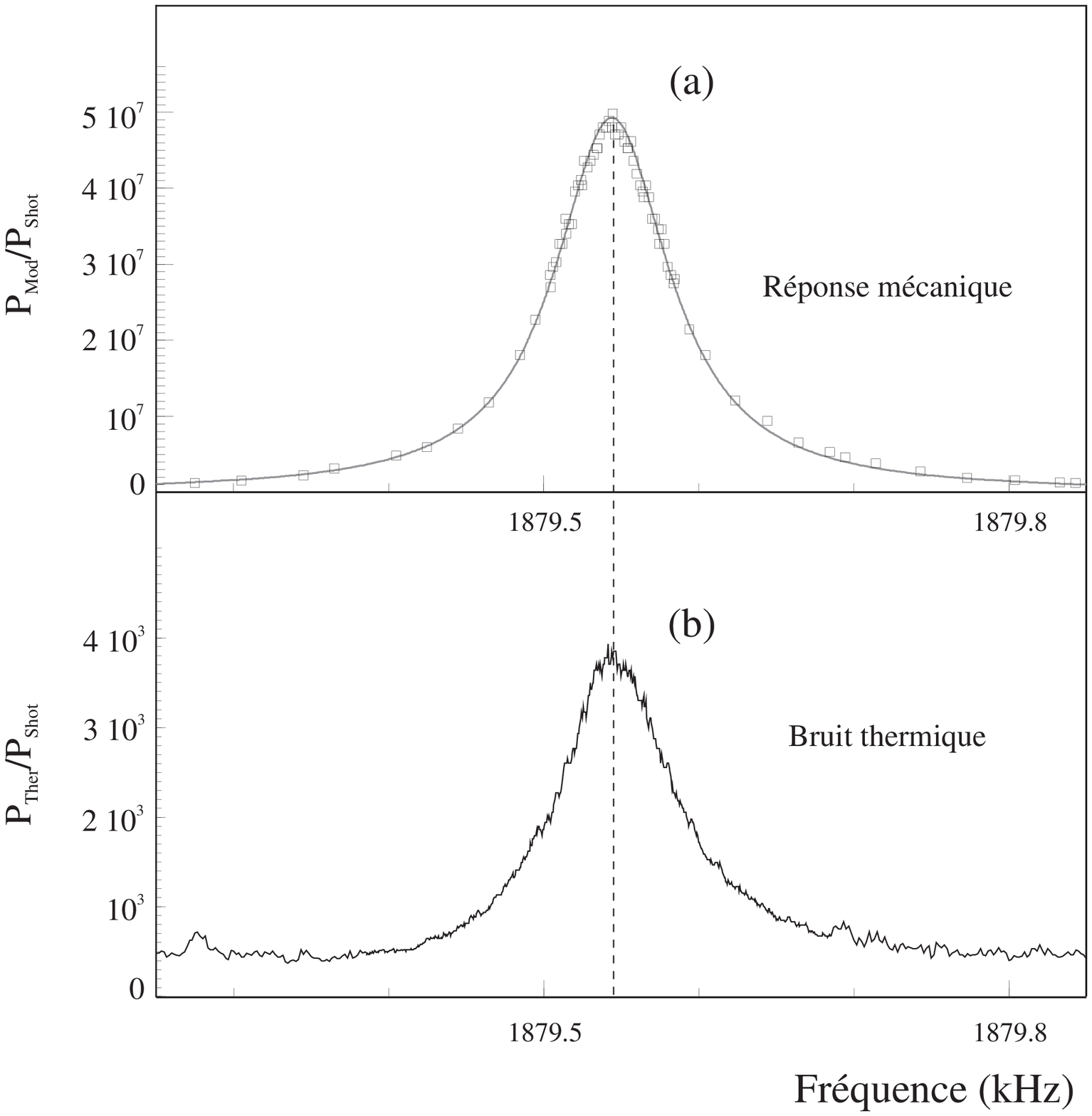,height=12cm}}
\caption{Comparaison entre la r\'{e}ponse m\'{e}canique (a) et le bruit
thermique (b) du mode acoustique fondamental du r\'{e}sonateur.
L'\'{e}chelle verticale repr\'{e}sente les puissances de bruit
normalis\'{e}es au bruit de photon standard. Chaque carr\'{e} correspond
\`{a} la puissance de modulation mesur\'{e}e sur la phase du faisceau
r\'{e}fl\'{e}chi, pour une fr\'{e}quence donn\'{e}e de la modulation
d'intensit\'{e} du faisceau excitateur. L'ajustement Lorentzien (trait
plein) permet de d\'{e}terminer la fr\'{e}quence de r\'{e}sonance ($%
1879.54~kHz$) et le facteur de qualit\'{e} ($21000$)}
\label{Fig_5mod1879}
\end{figure}

Pour d\'{e}terminer avec pr\'{e}cision la r\'{e}ponse m\'{e}canique du mode
acoustique fondamental du miroir mobile, on mesure \`{a} l'aide du
syst\`{e}me de d\'{e}tection homodyne l'amplitude de modulation pour
diff\'{e}rentes valeurs de la fr\'{e}quence de modulation, que l'on fait
varier dans une petite bande de fr\'{e}quence autour de la r\'{e}sonance
acoustique. Le r\'{e}sultat obtenu est repr\'{e}sent\'{e} sur la figure \ref
{Fig_5mod1879}a o\`{u} chaque carr\'{e} repr\'{e}sente la puissance $P_{Mod}$
de la modulation de la phase du faisceau r\'{e}fl\'{e}chi par la cavit\'{e},
sur une plage de fr\'{e}quence de $\pm 300~Hz$ autour de la r\'{e}sonance.
Les valeurs obtenues sont normalis\'{e}es par rapport au bruit de photon
standard $P_{Shot}$ de l'oscillateur local. La pr\'{e}sence de ce pic de
modulation r\'{e}v\`{e}le l'existence d'une r\'{e}sonance m\'{e}canique du
miroir mobile et confirme le fait que le pic de bruit thermique observ\'{e}
sur la figure \ref{Fig_5m32ther}b est bien d\^{u} au miroir mobile.

Le spectre de la figure \ref{Fig_5mod1879}b, reproduit ce pic de bruit
thermique, normalis\'{e} au bruit de photon standard. Compar\'{e} au spectre
de la figure \ref{Fig_5m32ther}b, la m\'{e}thode d'acquisition a \'{e}t\'{e}
am\'{e}lior\'{e}e. En particulier, le nombre de moyennage a \'{e}t\'{e}
augment\'{e} jusqu'\`{a} $1000$, gr\^{a}ce \`{a} un programme qui pilote
l'analyseur de spectre et qui effectue un moyennage sur l'ordinateur. La
comparaison des deux r\'{e}sonances de la figure \ref{Fig_5mod1879} indique
clairement qu'elles correspondent au m\^{e}me mode acoustique du miroir
mobile, puisqu'elles se situent \`{a} la m\^{e}me fr\'{e}quence de
r\'{e}sonance et elles ont des largeurs comparables.

Ces r\'{e}sultats permettent d'autre part d'obtenir des informations
pr\'{e}cises sur les caract\'{e}ristiques du mode acoustique. La courbe en
trait continu de la figure \ref{Fig_5mod1879}a est un ajustement lorentzien
des points exp\'{e}rimentaux. On trouve alors que la r\'{e}ponse
m\'{e}canique est caract\'{e}ris\'{e}e par une fr\'{e}quence de
r\'{e}sonance \'{e}gale \`{a} $1879.54~kHz$ et une largeur \`{a} mi-hauteur
\'{e}gale \`{a} $90~Hz$, ce qui correspond \`{a} un facteur de qualit\'{e}
de $21000$.

\subsection{Spectre de modulation \label{V-2-2}}

Comme nous venons de le voir, le dispositif d'excitation optique associ\'{e}
au syst\`{e}me de d\'{e}tection homodyne permet d'observer avec une grande
efficacit\'{e} la r\'{e}ponse m\'{e}canique du r\'{e}sonateur pour des
fr\'{e}quences voisines de la fr\'{e}quence de r\'{e}sonance d'un mode
acoustique. Nous avons g\'{e}n\'{e}ralis\'{e} cette m\'{e}thode afin
d'observer la r\'{e}ponse m\'{e}canique du r\'{e}sonateur sur une plage de
fr\'{e}quence beaucoup plus large. Ceci permet de rechercher
syst\'{e}matiquement les fr\'{e}quences des modes acoustiques du
r\'{e}sonateur. Pour cela on utilise un programme informatique pour piloter
\`{a} la fois le synth\'{e}tiseur HF et l'analyseur de spectre. Ce programme
permet d'automatiser l'\'{e}tude de la r\'{e}ponse du r\'{e}sonateur \`{a}
une excitation optique en contr\^{o}lant la fr\'{e}quence de modulation du
faisceau arri\`{e}re et la fr\'{e}quence de l'analyseur de spectre. En
pratique, le programme fait varier la fr\'{e}quence du synth\'{e}tiseur
d'une valeur minimale \`{a} une valeur maximale, avec un pas donn\'{e}. Pour
chaque fr\'{e}quence, il lance l'acquisition de l'analyseur de spectre en
mode {\it zero span}, tout d'abord avec la modulation activ\'{e}e, puis sans
modulation. Le programme r\'{e}cup\`{e}re ces deux spectres et en
d\'{e}duit, par moyennage des deux traces, la puissance de bruit thermique
et la puissance de modulation. Le processus est r\'{e}p\'{e}t\'{e} pour la
fr\'{e}quence suivante, jusqu'\`{a} parcourir l'ensemble de la plage de
fr\'{e}quence d\'{e}sir\'{e}e. On obtient finalement deux spectres qui
correspondent au bruit thermique et \`{a} la puissance de modulation.
\begin{figure}[tbp]
\centerline{\psfig{figure=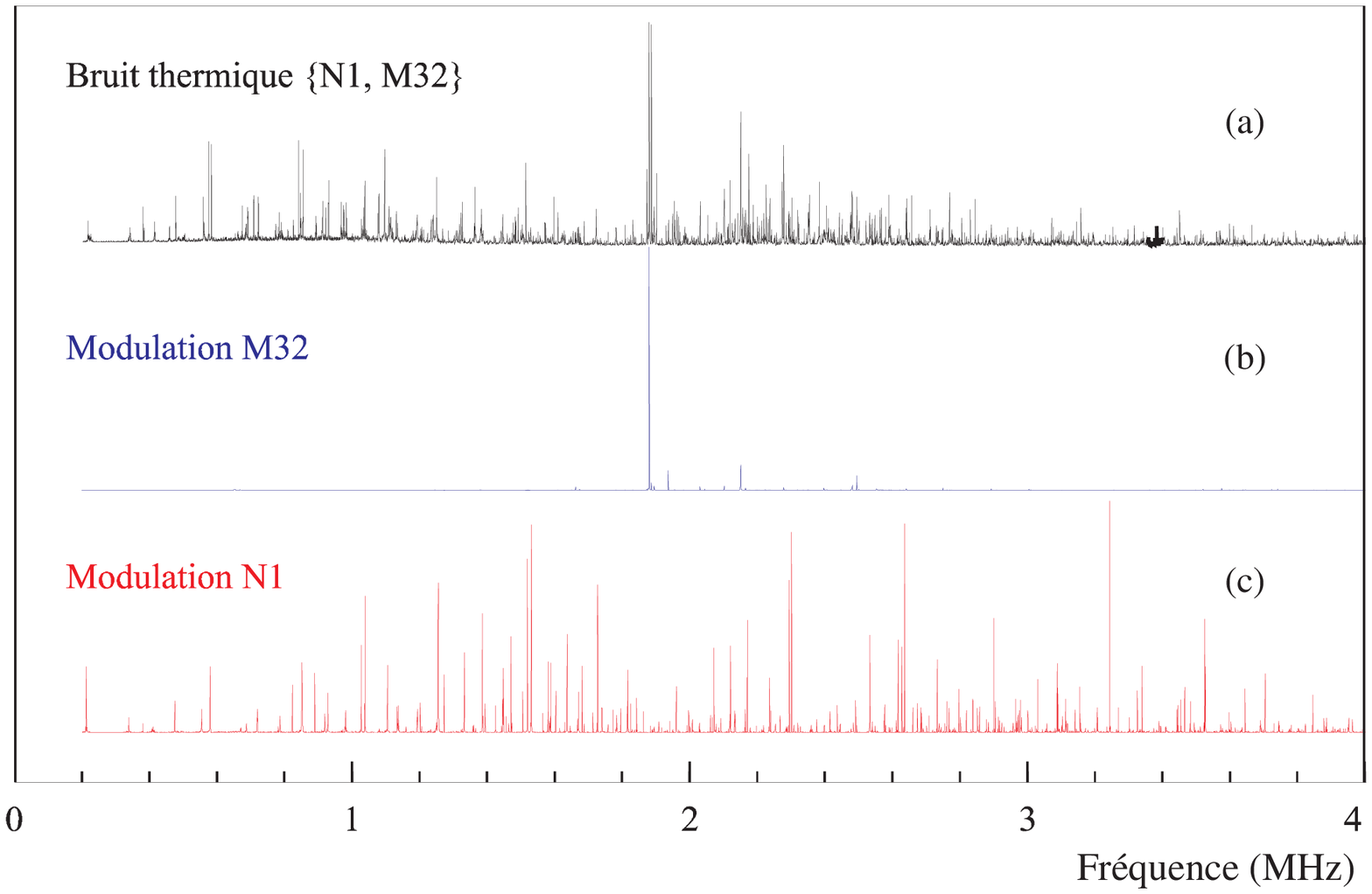,width=155mm}}
\caption{Comparaison entre le spectre du bruit thermique de la cavit\'{e}
\`{a} miroir mobile $\left\{ N1,~M32\right\} $ et les spectres de modulation
du miroir mobile $M32$ et du coupleur d'entr\'{e}e Newport $N1$}
\label{Fig_5new-mac}
\end{figure}

Les spectres (a) et (b) de la figure \ref{Fig_5new-mac} repr\'{e}sentent le
r\'{e}sultat obtenu pour la cavit\'{e} \`{a} miroir mobile $\left\{
N1,~M32\right\} $ pour des fr\'{e}quences allant de $200~kHz$ \`{a} $4~MHz$
avec un pas de $100~Hz$. On notera que tous les pics qui apparaissent dans
le spectre de modulation du miroir mobile $M32$ sont aussi pr\'{e}sent dans
le spectre de bruit thermique, ce qui permet d'identifier les pics de bruit
thermique associ\'{e}s au miroir mobile. Nous avons aussi \'{e}tudi\'{e} les
modes acoustiques du coupleur d'entr\'{e}e en utilisant la cavit\'{e} $%
\left\{ N2,~N1\right\} $ o\`{u} le miroir $N1$ est plac\'{e} \`{a}
l'arri\`{e}re de la cavit\'{e}. Le faisceau arri\`{e}re excite alors les
modes du miroir $N1$. Le spectre de modulation r\'{e}sultant est
repr\'{e}sent\'{e} sur la figure \ref{Fig_5new-mac}c. Il appara\^{\i }t
ainsi que la plupart des pics de bruit thermique du spectre (a) sont en fait
dus au coupleur d'entr\'{e}e Newport de la cavit\'{e}.

En conclusion, l'excitation optique par le faisceau arri\`{e}re permet de
r\'{e}aliser une \'{e}tude d\'{e}taill\'{e}e des modes acoustiques des
miroirs de la cavit\'{e}. Cette \'{e}tude est compl\'{e}mentaire de
l'observation directe du bruit thermique. En ce qui concerne notre
cavit\'{e} \`{a} miroir mobile, cette \'{e}tude a montr\'{e} que la plupart
des pics de bruit thermique sont en fait li\'{e}s aux modes acoustiques du
coupleur d'entr\'{e}e. D'autre part, le mouvement Brownien du miroir mobile
ne semble pas correspondre \`{a} celui pr\'{e}vu par le calcul th\'{e}orique
pr\'{e}sent\'{e} dans la section \ref{III-4-2}. Seul le mode acoustique
fondamental a une influence notable sur le champ intracavit\'{e} et son
facteur de qualit\'{e} n'est pas aussi \'{e}lev\'{e} que la valeur
pr\'{e}vue pour un r\'{e}sonateur en silice pure. Ceci pourrait \^{e}tre
d\^{u} \`{a} la fa\c{c}on dont sont tenus les miroirs de la cavit\'{e}. La
fixation peut jouer un r\^{o}le important dans le comportement m\'{e}canique
des miroirs, en modifiant la structure des modes et en induisant des
processus de dissipation par l'interm\'{e}diaire du support. En vue de
l'observation des effets quantiques du couplage optom\'{e}canique, il est
n\'{e}cessaire d'am\'{e}liorer le support de la cavit\'{e}, par exemple en
ne tenant les miroirs qu'en trois points situ\'{e}s \`{a} $120{{}^{\circ }}$
l'un de l'autre sur la circonf\'{e}rence du miroir. Quoi qu'il en soit, les
m\'{e}thodes que nous avons d\'{e}velopp\'{e}es (observation du bruit
thermique et excitation optique du r\'{e}sonateur) s'av\`{e}rent des outils
performants pour optimiser les caract\'{e}ristiques m\'{e}caniques de la
cavit\'{e}.

\section{D\'{e}termination de la sensibilit\'{e} \label{V-3}}

\bigskip

Comme nous l'avons soulign\'{e} dans les chapitres pr\'{e}c\'{e}dents, la
grande finesse de la cavit\'{e} \`{a} miroir mobile permet d'atteindre une
tr\`{e}s grande sensibilit\'{e} aux variations de longueur de la cavit\'{e}.
Nous avons cherch\'{e} \`{a} mesurer avec pr\'{e}cision cette
sensibilit\'{e} en utilisant une m\'{e}thode de modulation de fr\'{e}quence
du faisceau incident sur la cavit\'{e} \`{a} miroir mobile. Cette modulation
de fr\'{e}quence se traduit par une modulation de phase du faisceau
r\'{e}fl\'{e}chi \'{e}quivalente \`{a} l'effet produit par une variation de
longueur de la cavit\'{e}. La phase du faisceau r\'{e}fl\'{e}chi est en
effet sensible au d\'{e}phasage $\Psi =2kx$ du champ dans la cavit\'{e}.
Ainsi, une modulation $\delta \nu _{m}$ de la fr\'{e}quence du laser est
\'{e}quivalente \`{a} un d\'{e}placement $\delta x_{m}$ du miroir mobile:
\begin{equation}
\frac{\lambda }{c}~\delta \nu _{m}=\frac{\delta x_{m}}{L}  \label{5.1bis}
\end{equation}
o\`{u} $L$ est la longueur de la cavit\'{e}. Le principe de la mesure de la
sensibilit\'{e} consiste donc \`{a} comparer les d\'{e}placements
mesur\'{e}s \`{a} l'effet d'une modulation de fr\'{e}quence. Cette
modulation de fr\'{e}quence est \'{e}talonn\'{e}e \`{a} l'aide d'une
cavit\'{e} de r\'{e}f\'{e}rence qui n'est autre que la cavit\'{e} de
filtrage FPF de la source laser.

\subsection{Etalonnage de la modulation de fr\'{e}quence \label{V-3-1}}

Pour moduler la fr\'{e}quence du laser nous utilisons un g\'{e}n\'{e}rateur
HF ($HP$-$8648A$) qui fournit une tension sinuso\"{\i }dale d'amplitude $%
V_{m}$ et de fr\'{e}quence $\Omega _{m}$ variables. Ce signal pilote
l'\'{e}lectro-optique interne du laser titane saphir par l'interm\'{e}diaire
de l'entr\'{e}e modulation rapide pr\'{e}vue \`{a} cet effet dans le
syst\`{e}me de pr\'{e}amplification de l'asservissement en fr\'{e}quence de
la source laser (voir figure \ref{Fig_4alhp}, page \pageref{Fig_4alhp}). On
module ainsi la longueur optique de la cavit\'{e} laser, ce qui se traduit
\`{a} la sortie du laser titane saphir par une modulation de la
fr\'{e}quence du faisceau avec une amplitude $\delta \nu _{m}$ qui
d\'{e}pend lin\'{e}airement de la tension $V_{m}$ (dans la plage de tension
o\`{u} le fonctionnement de l'\'{e}lectro-optique est lin\'{e}aire).

La premi\`{e}re \'{e}tape de la mesure de la sensibilit\'{e} de la
cavit\'{e} \`{a} miroir mobile consiste \`{a} d\'{e}terminer le coefficient
de proportionnalit\'{e} $p_{\nu }$ qui relie l'amplitude de modulation $%
\delta \nu _{m}$ \`{a} la tension $V_{m}$ du g\'{e}n\'{e}rateur HF. Afin de
r\'{e}aliser cet \'{e}talonnage, on utilise comme cavit\'{e} de
r\'{e}f\'{e}rence la cavit\'{e} Fabry-Perot FPF dont on conna\^{\i }t
pr\'{e}cis\'{e}ment la bande passante (voir section \ref{IV-2-4}). Lorsque
la fr\'{e}quence du faisceau incident est modul\'{e}e, on observe en
transmission une modulation de l'intensit\'{e} $\delta I_{m}^{t}$ dont
l'amplitude d\'{e}pend de l'amplitude $\delta \nu _{m}$, du point de
fonctionnement de la cavit\'{e} et de l'effet de filtrage li\'{e} \`{a} la
bande passante $\Omega _{cav}$ de la cavit\'{e}. Plus pr\'{e}cis\'{e}ment,
le d\'{e}saccord $\Psi =2kL$ du champ dans la cavit\'{e} FPF est modul\'{e}
\`{a} la fr\'{e}quence $\Omega _{m}$ selon la relation:
\begin{subequations}
\label{5.1}
\begin{eqnarray}
\Psi \left( t\right)  &=&\bar{\Psi}+\delta \Psi _{m}\left( t\right)
\label{5.1a} \\
\delta \Psi _{m}\left( t\right)  &=&4\pi \frac{L}{c}~\delta \nu _{m}~\cos
\left( \Omega _{m}t\right)   \label{5.1b}
\end{eqnarray}
\end{subequations}
o\`{u} $L$ est la longueur de la cavit\'{e} et $\delta \Psi _{m}$ est
l'amplitude de modulation \`{a} la fr\'{e}quence $\Omega _{m}$ du
d\'{e}saccord $\Psi \left( t\right) $ autour du d\'{e}saccord moyen $\bar{%
\Psi}$ qui d\'{e}finit le point de fonctionnement de la cavit\'{e}. Le champ
intracavit\'{e} est lui-m\^{e}me modul\'{e} autour du champ moyen $\bar{%
\alpha}$. Cette modulation $\delta \alpha _{m}$ ob\'{e}it \`{a}
l'\'{e}quation diff\'{e}rentielle suivante, d\'{e}duite des \'{e}quations
(2.40a) et (\ref{5.1a}):
\begin{equation}
\tau ~\frac{d}{dt}\delta \alpha _{m}\left( t\right) =\left( -\gamma +i\bar{%
\Psi}\right) ~\delta \alpha _{m}\left( t\right) +i\bar{\alpha}~\delta \Psi
_{m}\left( t\right)  \label{5.3}
\end{equation}
o\`{u} $2\gamma $ repr\'{e}sente les pertes totales de la cavit\'{e} ($\sqrt{%
\gamma }$ est \'{e}gal \`{a} la transmission en amplitude pour chacun des
miroirs dans le cas d'une cavit\'{e} sym\'{e}trique et sans perte). Cette
\'{e}quation permet de d\'{e}terminer la modulation du champ transmis $%
\delta \alpha _{m}^{t}\left( t\right) =\sqrt{\gamma }~\delta \alpha
_{m}\left( t\right) $ et celle de l'intensit\'{e} transmise $\delta
I_{m}^{t}\left( t\right) $:
\begin{equation}
\delta I_{m}^{t}\left( t\right) =-\bar{I}^{t}\frac{2\pi \delta \nu _{m}}{%
\Omega _{cav}}\frac{2\bar{\Psi}/\gamma }{\sqrt{\left( 1+\bar{\Psi}%
^{2}/\gamma ^{2}-\Omega _{m}^{2}/\Omega _{cav}^{2}\right) ^{2}+\left(
2\Omega _{m}/\Omega _{cav}\right) ^{2}}}~\cos \left( \Omega _{m}t+\varphi
\right) \medskip  \label{5.4}
\end{equation}

avec
\[
\tan \left( \varphi \right) =-\frac{2\Omega _{m}/\Omega _{cav}}{1+\bar{\Psi}%
^{2}/\gamma ^{2}-\Omega _{m}^{2}/\Omega _{cav}^{2}}
\]
o\`{u} $\tau =\gamma /\Omega _{cav}$ est le temps d'aller et retour dans la
cavit\'{e}. Cette relation permet de d\'{e}terminer l'amplitude de
modulation $\delta \nu _{m}$ de la fr\'{e}quence du laser, en fonction de la
profondeur de modulation $\delta I_{m}^{t}/\bar{I}^{t}$ du faisceau transmis
et du point de fonctionnement $\bar{\Psi}$. La pente $p_{\nu }=\delta \nu
_{m}/V_{m}$ reliant la modulation de fr\'{e}quence \`{a} la tension de
modulation appliqu\'{e}e sur le laser est donn\'{e}e par:
\begin{equation}
p_{\nu }=\frac{\Omega _{cav}}{2\pi V_{m}}\frac{\delta I_{m}^{t}}{\bar{I}^{t}}%
\frac{\sqrt{\left( 1+\bar{\Psi}^{2}/\gamma ^{2}-\Omega _{m}^{2}/\Omega
_{cav}^{2}\right) ^{2}+\left( 2\Omega _{m}/\Omega _{cav}\right) ^{2}}}{2\bar{%
\Psi}/\gamma }  \label{5.5}
\end{equation}

En pratique, la cavit\'{e} est maintenue \`{a} mi-transmission ($\bar{\Psi}%
\approx \gamma $). Pour cela, nous avons modifi\'{e} l\'{e}g\`{e}rement le
fonctionnement de la source laser : l'asservissement de la cavit\'{e} FPF
\`{a} r\'{e}sonance est d\'{e}sactiv\'{e}. On d\'{e}sactive aussi
l'asservissement d'intensit\'{e} et on utilise son \'{e}lectronique (la
photodiode $Pd3$ plac\'{e}e \`{a} la sortie de la cavit\'{e} FPF et les
pr\'{e}amplificateurs, voir figures \ref{Fig_4assint} et \ref{Fig_4tisasta})
pour piloter la cale pi\'{e}zo\'{e}lectrique de la cavit\'{e} FPF. Cet
asservissement contr\^{o}le donc la longueur de la cavit\'{e} de sorte que
l'intensit\'{e} transmise soit \'{e}gale \`{a} la moiti\'{e} de sa valeur
\`{a} r\'{e}sonance. Comme l'intensit\'{e} incidente est relativement
stable, on assure ainsi la stabilit\'{e} du point de fonctionnement de la
cavit\'{e}. Nous avons v\'{e}rifi\'{e} qu'au cours du processus de mesure,
aussi bien l'intensit\'{e} transmise $\bar{I}^{t}$ (mesur\'{e}e sur $Pd3$)
que le point de fonctionnement $\bar{\Psi}$ ne varient pas de plus d'un pour
cent.

Nous avons effectu\'{e} les mesures pour une fr\'{e}quence de modulation de $%
2~MHz$. La modulation r\'{e}sultante $\delta I_{m}^{t}$ sur le faisceau
transmis est mesur\'{e}e sur l'analyseur de spectre depuis la photodiode $%
Pd3 $ qui sert aussi \`{a} mesurer le niveau continu $\bar{I}^{t}$. Nous
avons \'{e}talonn\'{e} sa r\'{e}ponse en fr\'{e}quence en la comparant \`{a}
celle d'une photodiode FND100 suivie par un amplificateur transimp\'{e}dance
CLC425 (montage similaire \`{a} la voie HF de la figure \ref{Fig_4amphdio},
sans capacit\'{e} de fa\c{c}on \`{a} avoir un gain plat entre $0$ et $2~MHz$%
). On obtient ainsi un gain \`{a} $2~MHz$ environ $1.42$ fois plus grand que
le gain en continu. Le tableau suivant montre les r\'{e}sultats obtenus pour
diff\'{e}rentes tensions de modulation $V_{m}$:\medskip \bigskip

\centerline{\psfig{figure=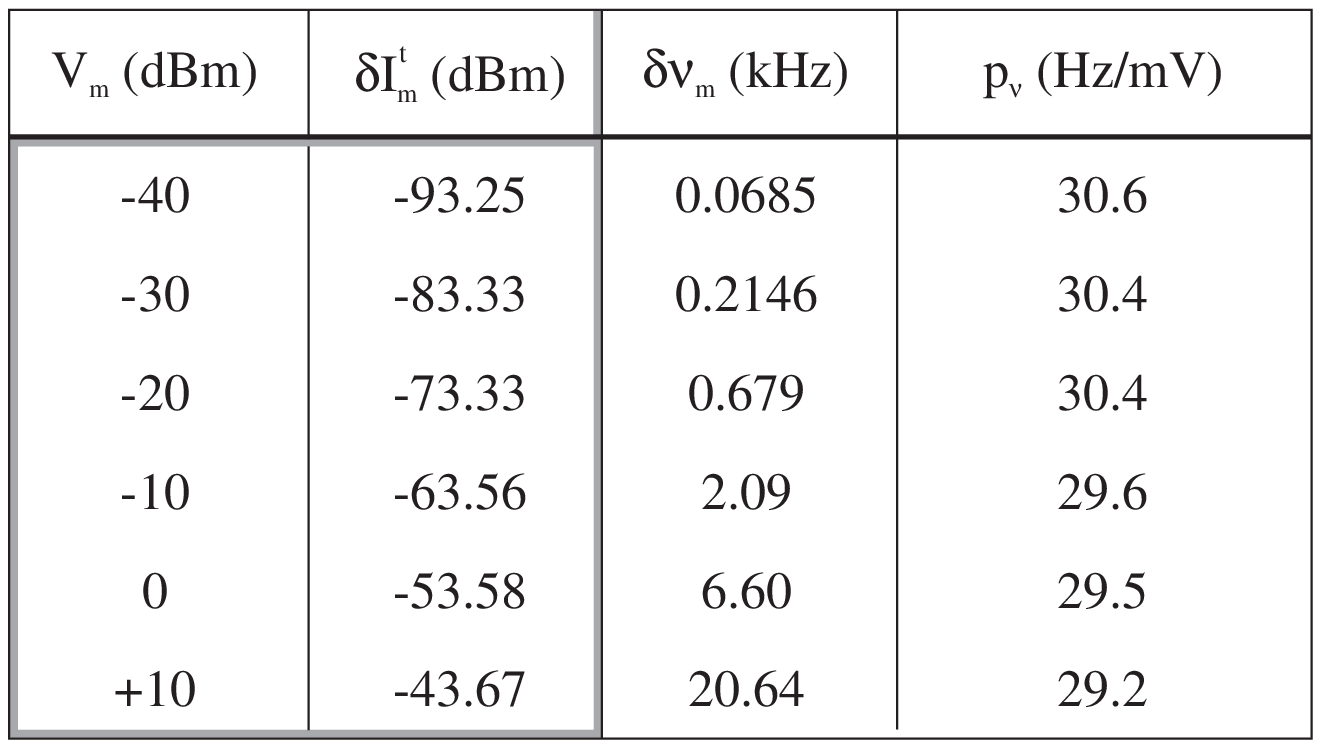,width=12cm}}\label{5tab2dbm}\bigskip
\medskip

La modulation de fr\'{e}quence $\delta \nu _{m}$ et la pente $p_{\nu
}=\delta \nu _{m}/V_{m}$ sont calcul\'{e}es \`{a} partir de l'\'{e}quation (%
\ref{5.5}), sachant que l'intensit\'{e} transmise moyenne $\bar{I}^{t}$
correspond \`{a} $291~mV$ en sortie de $Pd3$ et que le point de
fonctionnement $\bar{\Psi}$ de la cavit\'{e} est \'{e}gal \`{a} $1.05$. Les
tensions en $dBm$ sont converties en $Volt$ gr\^{a}ce \`{a} la formule:
\begin{equation}
V_{dBm}=10~Log\left[ 20~V_{Volt}^{2}\right]  \label{5.5bis}
\end{equation}
puisque $0~dBm$ correspond \`{a} une puissance d'un milliwatt dans une
r\'{e}sistance de charge de $50~\Omega $. Les valeurs mesur\'{e}es de $%
\delta I_{m}^{t}$ sont en plus multipli\'{e}es par le facteur correctif de $%
1.42$ entre les gains \`{a} $2~MHz$ et en continu.

\begin{figure}[tbp]
\centerline{\psfig{figure=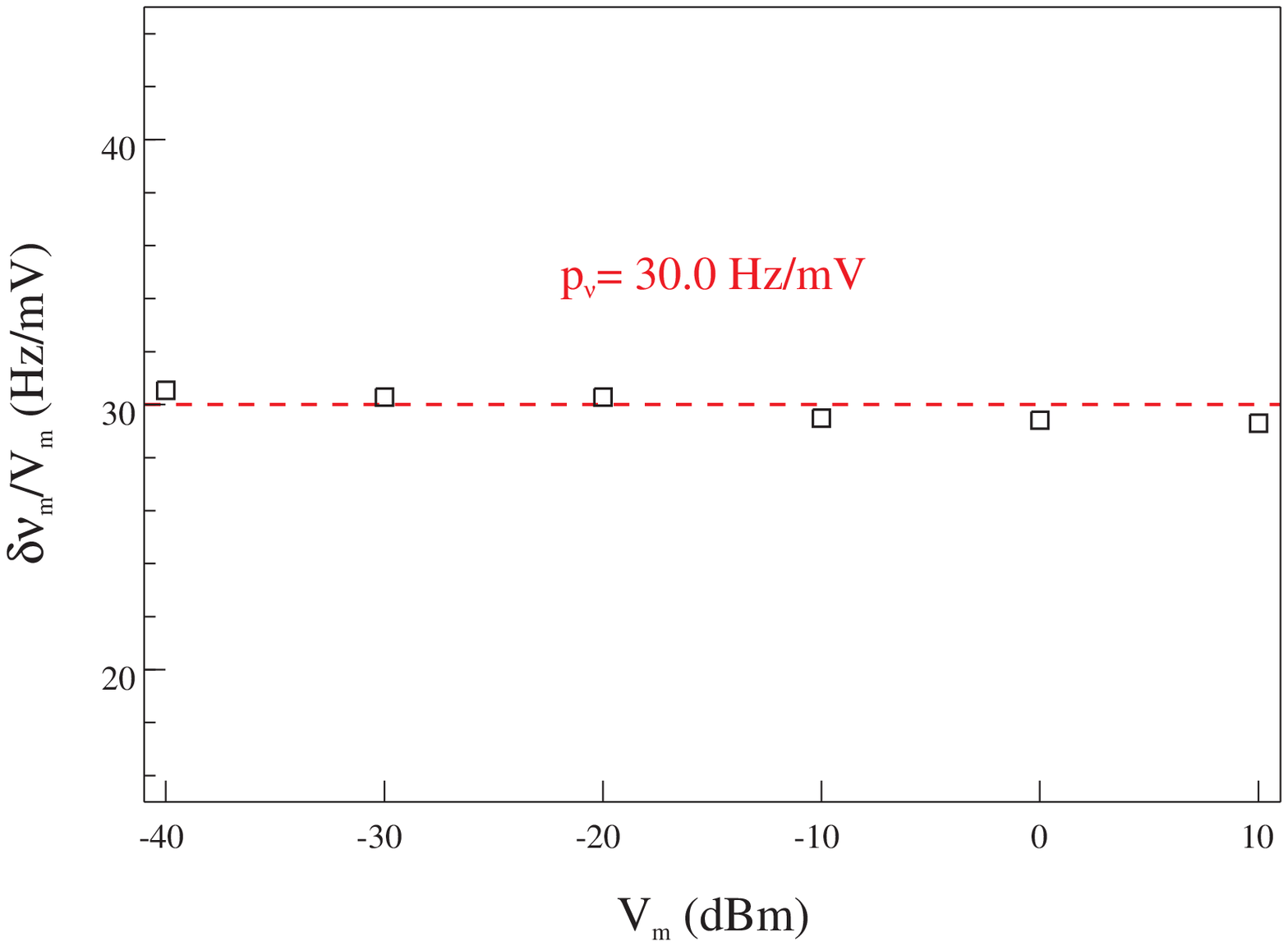,height=8cm}}
\caption{Variation de l'amplitude de modulation $\delta \nu _{m}$ de la
fr\'{e}quence du laser, en fonction de l'amplitude $V_{m}$ de la tension
appliqu\'{e}e sur le laser titane saphir. Les carr\'{e}s repr\'{e}sentent
les pentes $p_{\nu }=\delta \nu _{m}/V_{m}$ obtenues pour diff\'{e}rentes
amplitudes $V_{m}$. La droite correspond \`{a} la valeur moyenne}
\label{Fig_5mod2dbm}
\end{figure}

La figure \ref{Fig_5mod2dbm} montre le r\'{e}sultat des mesures. Les
carr\'{e}s repr\'{e}sentent les valeurs de la pente $p_{\nu }$ pour les
diff\'{e}rentes valeurs de l'amplitude $V_{m}$. On voit que la modulation de
fr\'{e}quence $\delta \nu _{m}$ est pratiquement lin\'{e}aire en fonction de
l'amplitude $V_{m}$, avec une pente moyenne $p_{\nu }$ \'{e}gale \`{a} $%
30.0~Hz/mV$. On notera que la pr\'{e}cision de ces mesures est tr\`{e}s
bonne puisque l'\'{e}cart des points obtenus par rapport \`{a} cette valeur
moyenne est au plus de $2\%$. Cet \'{e}talonnage permet donc de conna\^{\i
}tre avec pr\'{e}cision l'amplitude de modulation de fr\'{e}quence $\delta
\nu _{m}$ de la source laser produite par une tension quelconque $V_{m}$
appliqu\'{e}e \`{a} la source laser.

\subsection{Sensibilit\'{e} de la cavit\'{e} de grande finesse\label{V-3-2}}

Pour \'{e}valuer la sensibilit\'{e} de la cavit\'{e} aux variations de sa
longueur, on utilise la modulation de fr\'{e}quence dont l'amplitude a
\'{e}t\'{e} pr\'{e}alablement \'{e}talonn\'{e}e. Cette modulation de
fr\'{e}quence du faisceau incident se traduit par une modulation de la phase
du faisceau r\'{e}fl\'{e}chi que l'on peut mesurer gr\^{a}ce au syst\`{e}me
de d\'{e}tection homodyne. On peut de cette fa\c{c}on d\'{e}terminer
l'\'{e}volution de l'amplitude $V_{\varphi }$ de cette modulation en
fonction de la tension $V_{m}$ appliqu\'{e}e au laser. La connaissance de la
pente $p_{\varphi }=V_{\varphi }/V_{m}$ permet alors d'associer une
amplitude de modulation de phase $V_{\varphi }$ du faisceau r\'{e}fl\'{e}chi
\`{a} une tension $V_{m}$ qui correspond, comme nous l'avons vu dans la
section pr\'{e}c\'{e}dente, \`{a} une amplitude de modulation de
fr\'{e}quence $\delta \nu _{m}$ bien d\'{e}finie. Nous savons par ailleurs
que cette modulation de fr\'{e}quence est \'{e}quivalente \`{a} une
amplitude de modulation $\delta x_{m}$ de la longueur $L$ de la cavit\'{e}
(\'{e}quation \ref{5.1bis}):
\begin{equation}
\delta x_{m}=\frac{\lambda L}{c}~\delta \nu _{m}=\frac{\lambda L}{c}~\frac{%
p_{\nu }}{p_{\varphi }}~V_{\varphi }  \label{5.6}
\end{equation}
Il est donc possible, connaissant $p_{\nu }$ et $p_{\varphi }$, de
d\'{e}terminer la variation de longueur \'{e}quivalente associ\'{e}e \`{a}
un niveau de bruit de phase quelconque du faisceau r\'{e}fl\'{e}chi par la
cavit\'{e}.

La pente $p_{\varphi }$ caract\'{e}rise la sensibilit\'{e} de la cavit\'{e}
\`{a} une modulation de fr\'{e}quence du laser. Elle d\'{e}pend donc de la
puissance du faisceau incident et des caract\'{e}ristiques de la cavit\'{e},
telles que la finesse et la bande passante. Nous avons mesur\'{e} $%
p_{\varphi }$ pour une cavit\'{e} constitu\'{e}e par un miroir Newport plan
\`{a} l'entr\'{e}e (not\'{e} $Np$) et le miroir Newport $N1$, d\'{e}j\`{a}
utilis\'{e} comme coupleur pour la cavit\'{e} \`{a} miroir mobile. Les
caract\'{e}ristiques de cette cavit\'{e} ont \'{e}t\'{e} d\'{e}termin\'{e}es
en suivant les proc\'{e}dures d\'{e}crites dans la section \ref{IV-1-4}. La
bande passante $\nu _{BP}$ est \'{e}gale \`{a} $1.9~MHz$, le coefficient de
r\'{e}flexion \`{a} r\'{e}sonance ${\cal \tilde{R}}_{0}$ est \'{e}gal \`{a} $%
10.1\%$ et l'intervalle spectral libre $\Delta \lambda $ vaut $3.11~\AA $.
Ces valeurs permettent de d\'{e}terminer la finesse de la cavit\'{e} ${\cal F%
}=37240$, sa longueur $L=1.06~mm$, le coefficient de transmission du miroir
plan $T_{p}=60~ppm$ et les pertes totales de la cavit\'{e} $%
T_{p}+T_{1}+P_{p}+P_{1}=169~ppm$. Le tableau suivant pr\'{e}sente les
mesures r\'{e}alis\'{e}es avec cette cavit\'{e}, pour une puissance
incidente de $100~\mu W$:\medskip \bigskip

\centerline{\psfig{figure=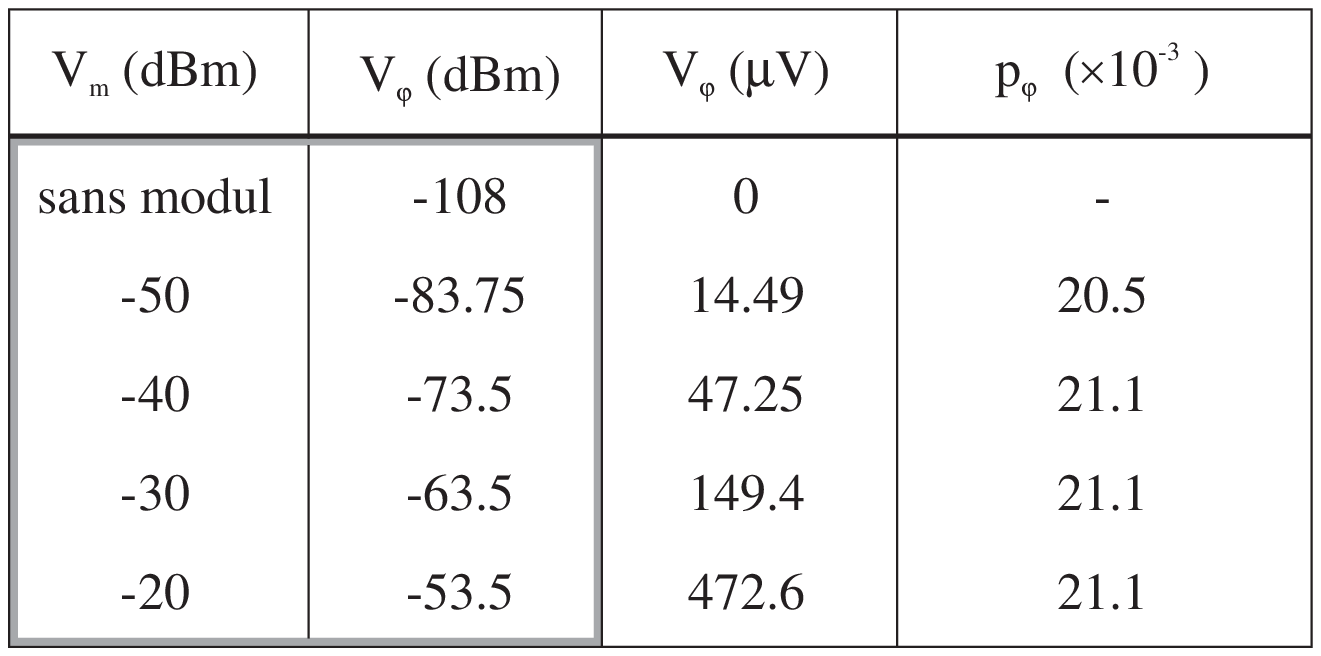,width=12cm}}\label{5tab2mod}\bigskip
\medskip

Les valeurs de $V_{\varphi }$ en $dBm$ sont mesur\'{e}es \`{a} l'aide de
l'analyseur de spectre, \`{a} la fr\'{e}quence $\Omega _{m}$ de modulation
de $2~MHz$, avec une r\'{e}solution spectrale de $1~Hz$. Ces valeurs sont
converties en $Volt$ (troisi\`{e}me colonne du tableau) \`{a} partir de
l'\'{e}quation (\ref{5.5bis}), en retranchant la puissance de bruit
thermique donn\'{e}e par la premi\`{e}re ligne (sans modulation).

\begin{figure}[tbp]
\centerline{\psfig{figure=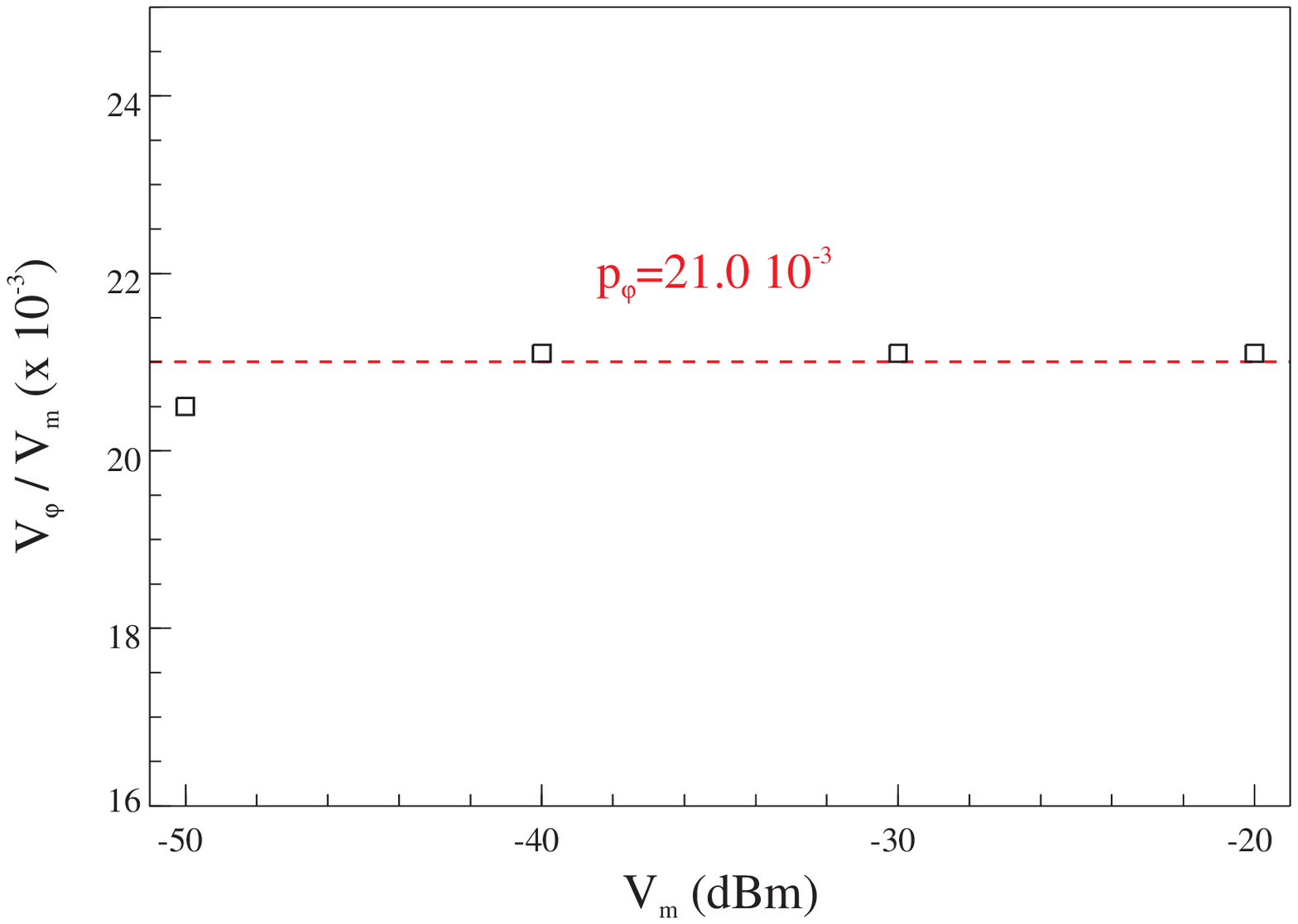,height=8cm}}
\caption{Variation en fonction de la tension de modulation $V_{m}$ du
rapport entre l'amplitude $V_{\varphi }$ de la modulation de phase du
faisceau r\'{e}fl\'{e}chi par la cavit\'{e} et la tension de modulation $%
V_{m}$. Les carr\'{e}s repr\'{e}sentent les valeurs obtenues pour
diff\'{e}rentes tensions $V_{m}$. La droite correspond \`{a} la valeur
moyenne}
\label{Fig_5fpm2mhz}
\end{figure}

Les valeurs r\'{e}sultantes de la pente $p_{\varphi }=V_{\varphi }/V_{m}$
sont repr\'{e}sent\'{e}es par des carr\'{e}s sur la figure \ref{Fig_5fpm2mhz}%
. A partir de ces valeurs on peut d\'{e}duire la pente $p_{\varphi }$ qui
correspond \`{a} la valeur moyenne repr\'{e}sent\'{e}e par la droite en
tirets. On obtient une pente \'{e}gale \`{a} $p_{\varphi }=21.0~10^{-3}$.
Ces r\'{e}sultats permettent de calibrer les mesures effectu\'{e}es par
l'analyseur de spectre. D'apr\`{e}s l'\'{e}quation (\ref{5.6}), une tension $%
V_{\varphi }$ sur l'analyseur correspond \`{a} un d\'{e}placement $\delta
x_{m}$ des miroirs de la cavit\'{e} donn\'{e}e par:
\begin{equation}
\delta x_{m}/V_{\varphi }=4.09~10^{-12}m/V  \label{5.6bis}
\end{equation}

Gr\^{a}ce \`{a} cette calibration, il est maintenant possible de
d\'{e}terminer la sensibilit\'{e} de la mesure du bruit thermique
pr\'{e}sent\'{e}e dans la partie \ref{V-1}. Cette mesure est en effet
limit\'{e}e par le bruit de photon standard de l'oscillateur local. Nous
avons mesur\'{e} la tension de bruit $V_{\varphi }^{\min }$ correspondant
\`{a} ce bruit quantique en pla\c{c}ant la cavit\'{e} hors r\'{e}sonance.
L'analyseur de spectre ayant une r\'{e}solution spectrale de $1~Hz$, la
valeur mesur\'{e}e sur l'analyseur en $dBm$ correspond en fait \`{a} une
tension de bruit qui s'exprime en $Volt/\sqrt{Hz}$. On trouve ainsi une
tension de bruit $V_{\varphi }^{\min }$ \'{e}gale \`{a} $67.1~nV/\sqrt{Hz}$.
A partir de l'\'{e}quation (\ref{5.6bis}), on en d\'{e}duit que la limite de
sensibilit\'{e} de la mesure du bruit thermique correspond \`{a} un
d\'{e}placement des miroirs:
\begin{equation}
\delta x_{\min }=2.74~10^{-19}m/\sqrt{Hz}  \label{5.7bis}
\end{equation}
Cette sensibilit\'{e} est pratiquement comparable \`{a} celle pr\'{e}vue
pour l'interf\'{e}rom\`{e}tre gravitationnel VIRGO. On peut noter par
ailleurs qu'il est possible d'am\'{e}liorer encore cette sensibilit\'{e} en
utilisant des miroirs de qualit\'{e} optique sup\'{e}rieure \`{a} celle des
miroirs commerciaux Newport afin d'augmenter la finesse de la cavit\'{e}, ou
en augmentant la puissance lumineuse incidente.

\subsection{Comparaison avec la th\'{e}orie\label{V-3-3}}

On peut comparer la valeur mesur\'{e}e de $\delta x_{\min }$ \`{a} celle
obtenue th\'{e}oriquement en utilisant le r\'{e}sultat des calculs
effectu\'{e}s dans la partie \ref{II-4}. Les calculs th\'{e}oriques que nous
avons men\'{e}s supposent cependant que la cavit\'{e} est sans perte, ce qui
n'est pas le cas exp\'{e}rimentalement. La pr\'{e}sence de pertes r\'{e}duit
l'influence des d\'{e}placements des miroirs dans le spectre de bruit de
phase du faisceau r\'{e}fl\'{e}chi. Les pertes peuvent \^{e}tre
mod\'{e}lis\'{e}es par une transmission non nulle du miroir arri\`{e}re de
la cavit\'{e}. Les \'{e}quations d'entr\'{e}e-sortie des fluctuations du
champ (\'{e}quations \ref{2.58a} et \ref{2.58b}, page \pageref{2.58})
s'\'{e}crivent alors \`{a} r\'{e}sonance:
\begin{subequations}
\label{5.15}
\begin{eqnarray}
\left( \gamma _{1}+\gamma _{2}-i\Omega \tau \right) ~\delta \alpha \left[
\Omega \right]  &=&\sqrt{2\gamma _{1}}~\delta \alpha ^{in}\left[ \Omega
\right] +\sqrt{2\gamma _{2}}~\delta \alpha ^{v}\left[ \Omega \right] +2ik%
\overline{\alpha }~\delta x\left[ \Omega \right]   \label{5.15a} \\
\delta \alpha ^{out}\left[ \Omega \right]  &=&\sqrt{2\gamma _{1}}~\delta
\alpha \left[ \Omega \right] -\delta \alpha ^{in}\left[ \Omega \right]
\label{5.15b}
\end{eqnarray}
\end{subequations}
o\`{u} $2\gamma _{1}=T_{p}$ est la transmission du coupleur d'entr\'{e}e, $%
2\gamma _{2}=T_{1}+P_{p}+P_{1}$ repr\'{e}sente les pertes dans la
cavit\'{e}, et $\delta \alpha ^{v}$ sont les fluctuations du vide
coupl\'{e}es \`{a} la cavit\'{e} par l'interm\'{e}diaire des pertes. Le
champ intracavit\'{e} moyen $\bar{\alpha}$ est reli\'{e} au champ incident $%
\bar{\alpha}^{in}$ par l'\'{e}quation:
\begin{equation}
\bar{\alpha}=\frac{\sqrt{2\gamma _{1}}}{\gamma _{1}+\gamma _{2}}~\bar{\alpha}%
^{in}  \label{5.16}
\end{equation}
Les fluctuations de la quadrature de phase $\delta q^{out}$ du champ
r\'{e}fl\'{e}chi (\'{e}quation \ref{4.2}, page \pageref{4.2}) sont
reli\'{e}es au d\'{e}placement $\delta x$ et aux fluctuations incidentes $%
\delta q^{in}$ et $\delta q^{v}$:
\begin{equation}
\left( \gamma _{1}+\gamma _{2}-i\Omega \tau \right) ~\delta q^{out}=\left(
\gamma _{1}-\gamma _{2}+i\Omega \tau \right) ~\delta q^{in}+\sqrt{4\gamma
_{1}\gamma _{2}}~\delta q^{v}+4\frac{2\gamma _{1}}{\gamma _{1}+\gamma _{2}}~%
\overline{\alpha }^{in}k~\delta x  \label{5.17}
\end{equation}
Cette \'{e}quation permet de d\'{e}terminer le spectre de phase $S_{q}^{out}$
du champ r\'{e}fl\'{e}chi, sachant que les spectres $S_{q}^{in}$ et $%
S_{q}^{v}$ des fluctuations incidentes $\delta \alpha ^{in}$ et $\delta
\alpha ^{v}$ sont tous deux \'{e}gaux \`{a} $1$:
\begin{equation}
S_{q}^{out}\left[ \Omega \right] =1+\frac{256}{1+\left( \Omega /\Omega
_{cav}\right) ^{2}}\frac{{\cal F}^{2}\bar{I}^{in}}{\lambda ^{2}}\left( \frac{%
\gamma _{1}}{\gamma _{1}+\gamma _{2}}\right) ^{2}~S_{x}\left[ \Omega \right]
\label{5.7}
\end{equation}
La sensibilit\'{e} $\delta x_{\min }$ \`{a} une fr\'{e}quence d'analyse $%
\Omega $ correspond au d\'{e}placement qui fournit un signal du m\^{e}me
ordre de grandeur que le bruit de photon standard (terme $1$ dans
l'\'{e}quation \ref{5.7}):
\begin{equation}
\delta x_{\min }\left[ \Omega \right] =\frac{\lambda }{16{\cal F}\sqrt{\bar{I%
}^{in}}}\sqrt{1+\left( \Omega /\Omega _{cav}\right) ^{2}}\left( \frac{%
T_{p}+T_{1}+P_{p}+P_{1}}{T_{p}}\right)  \label{5.8}
\end{equation}
Compar\'{e}e \`{a} l'expression obtenue pour la cavit\'{e} sans perte
(\'{e}quation \ref{4.9}, page \pageref{4.9}), la sensibilit\'{e} est
diminu\'{e}e dans la proportion de la transmission du coupleur d'entr\'{e}e
compar\'{e}e aux pertes totales de la cavit\'{e}.

En utilisant les valeurs exp\'{e}rimentales des param\`{e}tres de la
cavit\'{e}, on obtient une sensibilit\'{e} optimale $\delta x_{\min }$ \`{a}
une fr\'{e}quence de $2~MHz$ \'{e}gale \`{a} $2.76~10^{-19}~m/\sqrt{Hz}$. Si
on compare cette valeur th\'{e}orique \`{a} celle obtenue
exp\'{e}rimentalement (\'{e}quation \ref{5.7bis}), on constate un \'{e}cart
de moins d'un pourcent. Ceci montre que l'ensemble de nos mesures
exp\'{e}rimentales, que ce soit la mesure de la sensibilit\'{e} ou celle des
param\`{e}tres de la cavit\'{e}, sont tr\`{e}s pr\'{e}cises.

Notons enfin que la bande passante de la cavit\'{e} r\'{e}duit la
sensibilit\'{e} \`{a} haute fr\'{e}quence (\'{e}quation \ref{5.8}) : \`{a} $%
2~MHz$, la sensibilit\'{e} est environ $\sqrt{2}$ fois moins bonne qu'\`{a}
basse fr\'{e}quence. Nous avons donc repris l'ensemble de la proc\'{e}dure
de mesure de la sensibilit\'{e} \`{a} une fr\'{e}quence de modulation
\'{e}gale \`{a} $500~kHz$. D'apr\`{e}s l'expression (\ref{5.8}), on devrait
obtenir \`{a} cette fr\'{e}quence une sensibilit\'{e} \'{e}gale \`{a} $%
\delta x_{\min }=1.96~10^{-19}~m/\sqrt{Hz}$. L'\'{e}talonnage sur la
cavit\'{e} FPF donne une pente $p_{\nu }$ \'{e}gale \`{a} $84.6~Hz/mV$. La
mesure de la modulation de phase r\'{e}fl\'{e}chie par la cavit\'{e} $%
\left\{ Np,~N1\right\} $ donne une pente $p_{\varphi }$ \'{e}gale \`{a} $%
58.4~10^{-3} $, tandis que le niveau du bruit de photon standard $V_{\varphi
}^{\min }$ est \'{e}gal \`{a} $46.8~nV/\sqrt{Hz}$. On trouve ainsi que la
sensibilit\'{e} $\delta x_{\min }$ mesur\'{e}e \`{a} $500~kHz$ est \'{e}gale
\`{a}:
\begin{equation}
\delta x_{\min }=1.95~10^{-19}m/\sqrt{Hz}  \label{5.9}
\end{equation}
On remarque ici encore l'excellent accord avec la valeur th\'{e}orique, qui
prouve la pr\'{e}cision de nos mesures exp\'{e}rimentales.

Notons pour terminer qu'afin de faciliter la comparaison entre la
th\'{e}orie et l'exp\'{e}rience, la tension de bruit $V_{\varphi }^{\min }$
correspond au bruit de photon seulement. En fait, le bruit du bloc de
d\'{e}tection intervient aussi dans la d\'{e}termination de la
sensibilit\'{e} exp\'{e}rimentale. A $2~MHz$, ce bruit \'{e}lectronique
correspond \`{a} une tension de $22.4~nV/\sqrt{Hz}$. La tension de bruit $%
V_{\varphi }^{\min }$ correspondant au bruit total (\'{e}lectronique plus
bruit de photon standard) est donc de $70.7~nV/\sqrt{Hz}$. Ceci a pour effet
d'augmenter l\'{e}g\`{e}rement la valeur du d\'{e}placement $\delta x_{\min
} $ correspondant \`{a} la sensibilit\'{e} exp\'{e}rimentale. On trouve
ainsi, pour les deux fr\'{e}quences \'{e}tudi\'{e}es:
\begin{subequations}
\label{5.14}
\begin{eqnarray}
\delta x_{\min }\left[ 500kHz\right]  &=&2.09~10^{-19}m/\sqrt{Hz}\medskip
\label{5.14a} \\
\delta x_{\min }\left[ 2MHz\right]  &=&2.89~10^{-19}m/\sqrt{Hz}
\label{5.14b}
\end{eqnarray}
\end{subequations}
La prise en compte du bruit \'{e}lectronique du bloc de d\'{e}tection
modifie de moins de $10\%$ la valeur mesur\'{e}e de la sensibilit\'{e}. Il
est par ailleurs possible de r\'{e}duire l'influence du bruit
\'{e}lectronique en augmentant l'intensit\'{e} de l'oscillateur local. Ceci
a pour effet d'augmenter le niveau du bruit de photon et de rendre la
contribution du bruit \'{e}lectronique encore plus faible.\newpage

%% file: conclu.tex
\chapter{CONCLUSION}

\bigskip \bigskip

Nous avons pr\'{e}sent\'{e} une \'{e}tude th\'{e}orique et exp\'{e}rimentale
du couplage optom\'{e}canique dans une cavit\'{e} de grande finesse dont
l'un des miroirs est susceptible de se d\'{e}placer sous l'effet de la
pression de radiation du champ intracavit\'{e}. Nous avons tout d'abord
men\'{e} une \'{e}tude th\'{e}orique de ce couplage dans le cadre d'un
mod\`{e}le monodimensionnel, o\`{u} le champ est d\'{e}crit comme une onde
plane et le miroir mobile comme un oscillateur harmonique (syst\`{e}me
pendulaire). Nous avons montr\'{e} qu'un tel dispositif peut \^{e}tre
utilis\'{e} pour mettre en \'{e}vidence les effets quantiques dus \`{a} la
pression de radiation. Il est ainsi possible de contr\^{o}ler les
fluctuations de la lumi\`{e}re en produisant des \'{e}tats comprim\'{e}s, ou
encore de cr\'{e}er des corr\'{e}lations quantiques entre la position du
miroir mobile et l'intensit\'{e} lumineuse. Nous avons d'autre part
montr\'{e} la grande sensibilit\'{e} d'une telle cavit\'{e} \`{a} des petits
d\'{e}placements du miroir mobile. Une application directe de cette
sensibilit\'{e} consiste \`{a} mesurer le bruit thermique du miroir ou
encore \`{a} r\'{e}aliser une mesure quantique non destructive de
l'intensit\'{e} lumineuse.

Cette \'{e}tude nous a permis de d\'{e}gager les param\`{e}tres essentiels
qui caract\'{e}risent l'efficacit\'{e} du couplage optom\'{e}canique. Nous
avons ainsi montr\'{e} que les effets li\'{e}s \`{a} la pression de
radiation sont significatifs lorsque le d\'{e}placement moyen du miroir
mobile produit par la pression de radiation moyenne est de l'ordre de la
largeur de la r\'{e}sonance optique. Cette condition fait intervenir \`{a}
la fois les caract\'{e}ristiques optiques et m\'{e}caniques de la
cavit\'{e}. Elle impose en particulier une grande finesse, une puissance
lumineuse incidente \'{e}lev\'{e}e, et une masse du miroir aussi petite que
possible. La contribution des effets thermiques au d\'{e}placement du miroir
mobile doit aussi \^{e}tre n\'{e}gligeable compar\'{e}e \`{a} celle li\'{e}e
aux fluctuations quantiques de la pression de radiation. Afin de r\'{e}duire
le bruit thermique, il est n\'{e}cessaire de travailler \`{a} basse
temp\'{e}rature, avec un miroir mobile dont le facteur de qualit\'{e} et la
fr\'{e}quence de r\'{e}sonance sont \'{e}lev\'{e}s.

L'ensemble de ces contraintes nous a amen\'{e} \`{a} choisir comme miroir
mobile un r\'{e}sonateur m\'{e}canique constitu\'{e} d'un substrat en silice
de structure plan-convexe de $1.5~mm$ d'\'{e}paisseur. Nous avons
pr\'{e}sent\'{e} dans ce m\'{e}moire les caract\'{e}ristiques du couplage
optom\'{e}canique avec un tel r\'{e}sonateur, en d\'{e}veloppant un
mod\`{e}le th\'{e}orique qui tient compte de la pr\'{e}sence des
diff\'{e}rents modes acoustiques internes du r\'{e}sonateur et de la
structure tridimensionnelle du r\'{e}sonateur et du faisceau gaussien. Nous
avons montr\'{e} qu'il est possible de se ramener \`{a} une description
monodimensionnelle, en int\'{e}grant la structure spatiale dans une
susceptibilit\'{e} effective qui d\'{e}crit l'effet sur le champ de la
r\'{e}ponse m\'{e}canique \`{a} la pression de radiation intracavit\'{e}.
Nous avons mis en \'{e}vidence l'importance de l'adaptation spatiale entre
les modes optique et acoustique, et nous avons d\'{e}fini la masse effective
du r\'{e}sonateur qui d\'{e}crit l'amplitude du couplage optom\'{e}canique
\`{a} basse fr\'{e}quence. Cette masse d\'{e}pend de la section du faisceau
et peut \^{e}tre beaucoup plus petite que la masse totale du miroir.

Au cours de ce travail de th\`{e}se, nous avons r\'{e}alis\'{e} une
exp\'{e}rience qui doit permettre de mettre en \'{e}vidence les effets
quantiques du couplage optom\'{e}canique. Le montage exp\'{e}rimental est
compos\'{e} d'une cavit\'{e} de grande finesse \`{a} miroir mobile, d'une
source laser tr\`{e}s stable construite autour d'un laser titane saphir,
d'un syst\`{e}me de d\'{e}tection homodyne qui permet de mesurer le bruit
quantique de n'importe quelle quadrature du faisceau r\'{e}fl\'{e}chi, et
enfin d'un dispositif permettant d'exciter optiquement les modes acoustiques
du r\'{e}sonateur. Un ensemble de mesures nous a permis de d\'{e}terminer
avec une grande pr\'{e}cision les caract\'{e}ristiques optiques de la
cavit\'{e}. Nous avons ainsi mesur\'{e} la finesse de la cavit\'{e}, la
transmission et les pertes du coupleur d'entr\'{e}e et les pertes totales du
miroir mobile. Les finesses obtenues avec les diff\'{e}rents miroirs que
nous avons utilis\'{e}s sont comprises entre $30000$ et $47000$.

Nous avons pu d\'{e}montrer l'extr\^{e}me sensibilit\'{e} de notre
cavit\'{e} pour la mesure de petits d\'{e}placements du miroir mobile. Nous
avons en particulier observ\'{e} le bruit thermique du miroir mobile. Nous
avons aussi mis en \'{e}vidence la pr\'{e}sence de nombreux pics de bruit
thermique qui indiquent que le champ intracavit\'{e} est sensible au
mouvement Brownien d'\'{e}l\'{e}ments autres que le miroir mobile. En
comparant les spectres obtenus avec diff\'{e}rentes cavit\'{e}s, nous avons
montr\'{e} que la plupart des pics sont dus au mouvement Brownien du
coupleur d'entr\'{e}e. Pour d\'{e}terminer avec certitude quels pics sont
dus au miroir mobile et au coupleur d'entr\'{e}e, nous avons r\'{e}alis\'{e}
une \'{e}tude de la r\'{e}ponse m\'{e}canique des deux miroirs \`{a} une
excitation optique. En excitant s\'{e}lectivement les diff\'{e}rents modes
acoustiques du miroir, on obtient des informations pr\'{e}cises sur les
caract\'{e}ristiques de ces modes. Nous avons en particulier
d\'{e}termin\'{e} la fr\'{e}quence de r\'{e}sonance du mode acoustique
fondamental du miroir mobile ainsi que son facteur de qualit\'{e}.

Nous avons enfin d\'{e}termin\'{e} la sensibilit\'{e} de la mesure du bruit
thermique. Pour cela, nous avons utilis\'{e} une modulation de fr\'{e}quence
du faisceau incident dont l'amplitude a \'{e}t\'{e} pr\'{e}alablement
\'{e}talonn\'{e}e. Les r\'{e}sultats obtenus permettent d'associer un
d\'{e}placement \'{e}quivalent (en $m/\sqrt{Hz}$) \`{a} une tension de bruit
mesur\'{e}e sur la phase du faisceau r\'{e}fl\'{e}chi. On trouve que le plus
petit d\'{e}placement observable, dans une plage de fr\'{e}quence comprise
dans la bande passante de la cavit\'{e}, est de $2~10^{-19}~m/\sqrt{Hz}$.
Cette sensibilit\'{e} est pratiquement comparable a celle pr\'{e}vue pour
les interf\'{e}rom\`{e}tres gravitationnels. Elle peut par ailleurs \^{e}tre
am\'{e}lior\'{e}e en utilisant des miroirs de meilleure qualit\'{e} optique.
Nous avons aussi compar\'{e} la valeur mesur\'{e}e de la sensibilit\'{e}
\`{a} celle obtenue th\'{e}oriquement en tenant compte des
caract\'{e}ristiques de la cavit\'{e}. Le tr\`{e}s bon accord entre ces deux
valeurs montre que l'ensemble de nos mesures exp\'{e}rimentales sont
tr\`{e}s pr\'{e}cises.

Les m\'{e}thodes exp\'{e}rimentales que nous avons d\'{e}velopp\'{e}es
s'av\`{e}rent des outils performants pour d\'{e}terminer les
caract\'{e}ristiques optiques et m\'{e}caniques de la cavit\'{e} \`{a}
miroir mobile. Les r\'{e}sultats obtenus permettent d'identifier les
param\`{e}tres \`{a} optimiser en vue d'une mise en \'{e}vidence
exp\'{e}rimentale des effets quantiques du couplage optom\'{e}canique. Les am%
\'{e}liorations \`{a} apporter au montage exp\'{e}rimental concernent
essentiellement la qualit\'{e} optique et la r\'{e}ponse m\'{e}canique des
miroirs.

La finesse de la cavit\'{e} est limit\'{e}e par les pertes des miroirs,
qu'il s'agisse des coupleurs commerciaux ou des miroirs mobiles dont
l'\'{e}tat de surface ne permet pas d'atteindre le niveau de perte
souhait\'{e} de quelques $ppm$. L'utilisation de miroirs de meilleurs
qualit\'{e} optique devrait permettre d'atteindre des finesses de l'ordre de 
$10^{5}$. D'autre part, la puissance lumineuse incidente est limit\'{e}e
\`{a} quelques centaines de microwatts, du fait de la faible tenue au flux
des coupleurs d'entr\'{e}e. Les traitements multidi\'{e}lectriques
r\'{e}cents ont une tenue au flux bien meilleure, de l'ordre de plusieurs
dizaines de kilowatts par centim\`{e}tre carr\'{e}. Ces am\'{e}liorations
devraient non seulement augmenter les effets de pression de radiation mais
aussi la sensibilit\'{e} de la cavit\'{e} aux d\'{e}placements du miroir.

En ce qui concerne les caract\'{e}ristiques m\'{e}caniques de la cavit\'{e},
les \'{e}tudes r\'{e}alis\'{e}es sur le bruit thermique du r\'{e}sonateur
montrent que la r\'{e}ponse m\'{e}canique du miroir mobile ne correspond pas
\`{a} celle pr\'{e}vue par la th\'{e}orie, aussi bien en ce qui concerne la
pr\'{e}sence des diff\'{e}rents modes acoustiques que pour le facteur de
qualit\'{e} du mode fondamental. La fixation du miroir joue certainement un
r\^{o}le important dans le comportement m\'{e}canique du r\'{e}sonateur
puisqu'elle modifie la structure des modes et elle induit des processus de
dissipation par l'interm\'{e}diaire du support. C'est pourquoi il semble
n\'{e}cessaire de modifier le syst\`{e}me de fixation du miroir mobile. On
peut par exemple tenir le miroir sur sa circonf\'{e}rence par trois point
situ\'{e}s \`{a} $120{{}^{\circ }}$ l'un de l'autre.

L'observation du bruit thermique pour les diff\'{e}rentes cavit\'{e}s que
nous avons r\'{e}alis\'{e}es montre par ailleurs que le fond thermique est
pour l'essentiel d\^{u} au coupleur Newport. On peut d\'{e}terminer
th\'{e}oriquement l'apport relatif des deux miroirs au fond thermique. Le
bruit thermique \`{a} basse fr\'{e}quence est en effet proportionnel \`{a}
la susceptibilit\'{e} effective. Les r\'{e}sultats du chapitre 3 montrent
que la susceptibilit\'{e} effective du miroir mobile est \'{e}gale \`{a} $%
\chi _{eff}\left[ \Omega \approx 0\right] =1/M_{eff}\Omega _{M}^{2}\approx
3~10^{-8}~m/N$ alors que celle du coupleur d'entr\'{e}e est \'{e}gale \`{a} $%
3.5~10^{-10}~m/N$\cite{cypres}. En supposant l'angle de perte identique pour
les deux miroirs, le fond thermique d\^{u} au miroir mobile devrait donc
\^{e}tre environ $100$ fois plus grand que celui induit par le coupleur
Newport. Ce d\'{e}saccord avec l'exp\'{e}rience traduit ici encore le fait
que les modes acoustiques du miroir mobile sont modifi\'{e}s par le
syst\`{e}me de fixation. L'am\'{e}lioration de la fixation devrait donc
permettre d'accro\^{\i}tre la r\'{e}ponse m\'{e}canique du miroir mobile, et
en comparaison de rendre celle du coupleur d'entr\'{e}e n\'{e}gligeable.
Notons enfin qu'il est possible d'\'{e}liminer compl\`{e}tement les
probl\`{e}mes caus\'{e}s par le coupleur d'entr\'{e}e en utilisant une
cavit\'{e} constitu\'{e}e de deux miroirs mobiles, dont la face plane aurait
un rayon de courbure de l'ordre du m\`{e}tre de fa\c{c}on \`{a} obtenir une
cavit\'{e} optique stable. Ceci aurait pour effet de supprimer tous les pics
de bruit thermique d\^{u}s actuellement au coupleur Newport.

Gr\^{a}ce \`{a} ces am\'{e}liorations, il devrait \^{e}tre possible de r\'{e}%
aliser une \'{e}tude plus pr\'{e}cise du bruit thermique et des
m\'{e}canismes de dissipation, en excitant optiquement le r\'{e}sonateur
tr\`{e}s loin sur les ailes des r\'{e}sonances acoustiques. On pourra en
particulier avoir acc\`{e}s \`{a} l'\'{e}volution de l'angle de perte $\Phi
\left[ \Omega \right] $ en fonction de la fr\'{e}quence. Enfin,
l'am\'{e}lioration de la qualit\'{e} optique des miroirs et de la
r\'{e}ponse m\'{e}canique du r\'{e}sonateur devrait permettre de rendre
observable, \`{a} basse temp\'{e}rature, les effets quantiques de la
pression de radiation.